\documentclass[
10pt, 
oneside, 
english, 
singlespacing, 
headsepline, 
consistentlayout, 
]{MastersDoctoralThesis} 

\usepackage[utf8]{inputenc} 
\usepackage[T1]{fontenc} 

\usepackage{mathpazo} 

\usepackage[autostyle=true]{csquotes} 

\usepackage{mathrsfs}
\usepackage{amsmath}
\usepackage{cite}
\usepackage{bm}
\usepackage{notoccite}
\usepackage{subcaption}
\usepackage{rotating}
\usepackage{tikz}
\usepackage{contour}
\usetikzlibrary{shapes,arrows,positioning,automata,backgrounds,calc,er,patterns}
\usepackage[compat=1.1.0]{tikz-feynman}
\usepackage{slashed}
\usepackage{diagbox}
\usepackage{multirow}
\usepackage{caption}
\usepackage{mathabx}
\usepackage[makeroom]{cancel}

\usepackage{afterpage}

\newcommand\blankpage{%
    \null
    \thispagestyle{empty}%
    \addtocounter{page}{-1}%
    \newpage}
    
\thesistitle{Study of the Transverse-Momentum-Dependent \\ structure of the nucleon \\ in Semi-Inclusive DIS} 
\supervisor{Prof. Anna \textsc{Martin}} 
\examiner{} 
\author{Andrea \textsc{Moretti}} 
\addresses{} 

\university{\href{http://www.units.it}{Università degli Studi di Trieste}} 

\AtBeginDocument{
\hypersetup{pdftitle=\ttitle} 
\hypersetup{pdfauthor=\authorname} 
\hypersetup{pdfkeywords=\keywordnames} 
}

\begin{document}

\frontmatter 

\pagestyle{plain} 


\begin{titlepage}
\begin{center}

\vspace*{.06\textheight}
\includegraphics[width=0.2\textwidth]{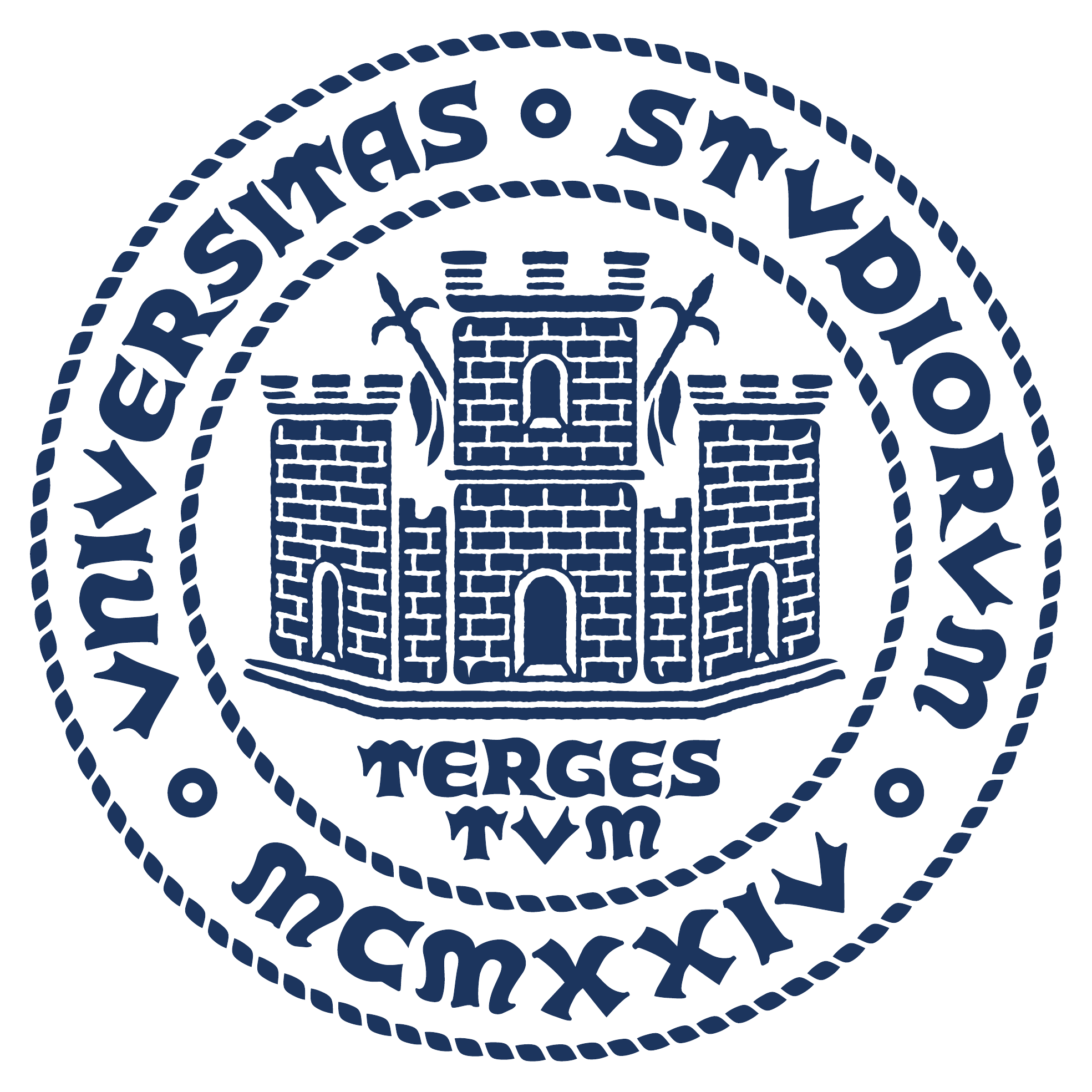} \\
{\scshape\LARGE \univname\par}\vspace{1.0cm} 
\textsc{\Large XXXIII ciclo del Dottorato di Ricerca in Fisica }\\[0.5cm] 

\bigskip 

\HRule \\[0.4cm] 
{\huge \bfseries \ttitle\par}\vspace{0.4cm} 
\HRule \\[0.8cm] 
 
\textsc{Settore Scientifico-Disciplinare: FIS/04} 

\vspace{1.5cm}
 
\begin{minipage}[t]{0.6\textwidth}
\begin{center} \large
\emph{Candidate}\\
{\authorname} \\ 
\emph{}\\

\emph{Coordinator} \\
{Prof. Francesco \textsc{Longo}} \\ 
\emph{}\\

\emph{Supervisor} \\
{\supname} %
\end{center}
\end{minipage}


 
\vspace{2.5cm}

{\large Academic Year 2019/2020}\\[4cm] 

\vfill
\end{center}
\end{titlepage}

\afterpage{\blankpage}
\dedicatory{Alla mia famiglia}



 
 
\afterpage{\blankpage}
\cleardoublepage


\vspace*{0.2\textheight}

\begin{flushleft}
\textit{Sic a principiis ascendit motus et exit \\ paulatim nostros ad sensus, ut moveantur \\ illa quoque, in solis quae lumine cernere quimus \\ nec quibus id faciant plagis apparet aperte.} 
\end{flushleft} 

\vspace{0.1cm}
\begin{flushright}
Thus motion ascends from the primevals on, \\ and stage by stage emerges to our sense, \\ until those objects also move which we \\ can mark in sunbeams, though it not appears \\ what blows do urge them. \\
\bigskip 
(Lucretius, De Rerum Natura)
\end{flushright}

\vspace{3cm}
\clearpage


\afterpage{\blankpage}

\begin{extraAbstract}
Lo studio della struttura interna del nucleone è un tema di grande rilevanza in fisica adronica. Negli ultimi anni è stato intrapreso un grande sforzo, sia dal punto di vista teorico che sperimentale, per ottenere una descrizione completa della struttura del nucleone al di là dell'approccio collineare in Cromodinamica Quantistica, dove i gradi di libertà trasversi non vengono considerati. D'altronde, sempre più risultati sperimentali non possono essere spiegati senza ricorrere al momento trasverso del partone $k_{T}$, al suo spin trasverso $s_{T}$ e alla loro correlazione con il nucleone. Nel formalismo dipendente dal momento trasverso (TMD), la struttura del nucleone è descritta da otto funzioni di distribuzione partonica (TMD PDFs), che emergono come generalizzazione delle tre PDF collineari (densità, elicità e trasversità). La nostra conoscenza di queste funzioni è molto limitata. Di particolare interesse è la funzione di Boer-Mulders $h_{1}^{\perp}$, che descrive la correlazione fra il momento trasverso e lo spin trasverso di un partone in un nucleone non polarizzato. \\

Lo scopo di questa Tesi è contribuire alla comprensione della struttura del nucleone mediante lo studio di due osservabili TMD accessibili nella diffusione semi-inclusiva profondamente inelastica (SIDIS) di leptoni di alta energia su protoni non polarizzati, ossia le distribuzioni di momento trasverso e le ampiezze delle modulazioni nell'angolo azimutale (o asimmetrie azimutali) degli adroni nello stato finale. Queste forniscono informazioni importanti sul momento trasverso dei partoni e sulla funzione di Boer-Mulders. In questa Tesi è riassunto il lavoro fatto lunga tale linea di ricerca durante il dottorato, consistito nell'analisi completa di parte dei dati raccolti a COMPASS nel 2016. COMPASS è un esperimento a bersaglio fisso, situato al CERN, che utilizza un fascio di muoni di 160 GeV e, nell'anno considerato, un bersaglio di idrogeno liquido. La qualità dei dati, la stabilità dei rivelatori e la consistenza dei dati ricostruiti sono state studiate in modo approfondito, in parallelo alla implementazione e validazione delle simulazioni Monte Carlo, necessarie per la correzione per l'accettanza e per la stima della contaminazione da processi diffrattivi. È stata condotta un'analisi dei possibili effetti sistematici e del loro contributo all'incertezza complessiva delle misure. Le varie dipendenze cinematiche sono state trattate in dettaglio e dei risultati, confrontati con quelli già ottenuti in misure precedenti a COMPASS, è stata condotta un'analisi fenomenologica. \\

Questa Tesi è così organizzata. La struttura del nucleone nell'approccio TMD è introdotta nel Capitolo 1, con particolare attenzione al processo SIDIS e alle osservabili ad esso associate. Il Capitolo 2 è dedicato alla descrizione dell'esperimento COMPASS e del setup sperimentale per la presa dati del 2016. Il Capitolo 3 è dedicato all'analisi dei dati, in termini di selezione degli eventi DIS, degli adroni prodotti nel SIDIS e delle correzioni per l'accettanza. La procedura per l'estrazione delle distribuzioni in momento trasverso e delle asimmetrie azimutali, la stima delle incertezze sistematiche ad esse associate, i risultati e la loro interpretazione sono presentati nei Capitoli 4 e 5. Il Capitolo 6 tratta della produzione diffrattiva dei mesoni vettoriali, con particolare attenzione al mesone $\rho^0$ e alla distribuzione angolare dei suoi prodotti di decadimento. Le conclusioni sono tratte nel Capitolo 7.
\end{extraAbstract}

\afterpage{\blankpage}
\begin{abstract}
The study of the internal structure of the nucleon is a hot topic in hadron physics. In recent years, a huge effort has been undertaken, both on the theoretical and on the experimental side, to provide a comprehensive description of the nucleon structure beyond the collinear Quantum Chromo Dynamics (QCD) approach, where all transverse degrees of freedom are assumed negligible. As a matter of fact, many experimental results can not be explained without considering the parton transverse momentum $k_{T}$, its transverse spin $s_{T}$ and the way how they correlate with the nucleon. In the recently developed QCD formalism, the nucleon structure is described by eight Transverse-Momentum-Dependent (TMD) Parton Distribution Functions (PDFs), which are generalizations of the three collinear ones (number density, helicity and transversity). To a large extent, our knowledge of these functions is still very poor. \\

The aim of this Thesis is to contribute to the understanding of the nucleon structure through the study of two observables accessible in Semi-Inclusive Deep Inelastic Scattering (SIDIS) of high energy leptons off unpolarized protons: the transverse-momentum distributions and the amplitudes of the modulations in the azimuthal angle of the final state hadrons, the latter referred to as "azimuthal asymmetries". They give relevant information on the transverse momentum of the partons inside the nucleon and on the Boer-Mulders TMD PDF. This Thesis summarizes the work done in this direction during the Ph.D., which consisted in a complete analysis of part of the data collected in 2016 in COMPASS, a fixed target experiment at the CERN SPS using 160 GeV/$c$ $\mu^+$ and $\mu^-$ beams and a liquid hydrogen target. The data quality and the detector stability have been investigated, as well as the stability and the consistency of the reconstructed data. A significant effort has been put in the validation of the Monte Carlo simulations, necessary for the evaluation of the acceptance of the spectrometer and to estimate the contamination to the SIDIS sample by the hadrons produced in the decay of diffractively produced vector mesons (particularly $\rho^0$ and $\phi$), which give strong contributions both to the transverse-momentum distributions and to the azimuthal asymmetries of the inclusive hadrons. This diffractive process had also to be studied in detail and implemented in dedicated Monte Carlo simulations, reducing the systematic uncertainties affecting previous measurements. Other systematic effects have also been investigated, and the corresponding systematic uncertainties evaluated. A deep inspection of the various kinematic dependences has been performed. A phenomenological analysis of the new results is also presented, with a comparison of the current COMPASS findings with the previous ones obtained on a deuteron target. \\

This Thesis is organized as follows. In Chapter 1 the transverse-momentum-dependent structure of the nucleon is introduced, with a focus on the SIDIS process and the related observables. The second Chapter is dedicated to the description of the COMPASS experiment with details on the 2016 setup. The third Chapter is devoted to the data analysis, with a description of the selection of the DIS events and of the hadron samples and of the acceptance corrections. The procedure for the extraction of the transverse-momentum distributions and of the azimuthal asymmetries, the estimation of their systematic uncertainties, the results and their interpretation are presented in Chapter 4 and 5 respectively. Chapter 6 hosts a study of the exclusive diffractive vector meson production, with a focus on the $rho^0$ vector meson. The conclusions are given in Chapter 7.
\end{abstract}

\afterpage{\blankpage}

\tableofcontents 

\mainmatter 

\pagestyle{thesis} 



\newcommand{\Conv}{\mathscr{C}}
\newcommand{\f}{f_{1}}
\newcommand{\fq}{f_{1}^{q}}
\newcommand{\D}{D_{1}}
\newcommand{\Dq}{D_{1}^{h/q}}
\newcommand{\eps}{\varepsilon}
\newcommand{\fih}{\phi_{h}}
\newcommand{\kt}{k_{T}}
\newcommand{\ktsq}{k_{T}^{2}}
\newcommand{\aktsq}{\langle k_{T}^{2} \rangle}
\newcommand{\Faktsq}{\langle k_{T,~q}^{2} \rangle}
\newcommand{\aktsqbm}{\langle k_{T}^{2} \rangle_{BM}}
\newcommand{\Faktsqbm}{\langle k_{T,~q}^{2} \rangle_{BM}}
\newcommand{\vkt}{\bm{k}_{T}}
\newcommand{\pperp}{p_{\perp}}
\newcommand{\pperpsq}{p_{\perp}^{2}}
\newcommand{\apperpsq}{\langle p_{\perp}^{2} \rangle}
\newcommand{\Fapperpsq}{\langle p_{\perp,~h/q}^{2} \rangle}
\newcommand{\apperpsqcoll}{\langle p_{\perp}^{2} \rangle_{C}}
\newcommand{\apperpsqfav}{\langle p_{\perp,~fav}^{2} \rangle}
\newcommand{\apperpsqunf}{\langle p_{\perp,~unf}^{2} \rangle}
\newcommand{\apperpsqcollfav}{\langle p_{\perp,~fav}^{2} \rangle_{C}}
\newcommand{\apperpsqcollunf}{\langle p_{\perp,~unf}^{2} \rangle_{C}}
\newcommand{\vpperp}{\bm{p}_{\perp}}
\newcommand{\Pt}{{P}_{T}}
\newcommand{\vPt}{{\bm{P}_{T}}}
\newcommand{\Ptsq}{P_{T}^{2}}
\newcommand{\aPtsq}{\langle P_{T}^{2} \rangle}
\newcommand{\FaPtsq}{\langle P_{T,~h/q}^{2} \rangle}
\newcommand{\aPtsqbm}{\langle P_{T}^{2} \rangle_{BM}}
\newcommand{\vt}{\bm{t}}
\newcommand{\hh}{\hat{\bm{h}}}
\newcommand{\bom}{h_{1}^{\perp}}
\newcommand{\bomq}{h_{1}^{\perp~q}}
\newcommand{\coll}{H_1^{\perp}}
\newcommand{\collq}{H_1^{\perp~h/q}}
\newcommand{\Qsq}{Q^{2}}
\newcommand{\mbm}{M_{BM}}
\newcommand{\mcoll}{M_{C}}
\newcommand{\qt}{q_{T}}

\newcommand{\vmphi}{\phi}
\newcommand{\vmrho}{\rho^0}
\newcommand{\vbeta}{\bm{\beta}}
\newcommand{\vp}{\bm{p}}
\newcommand{\vbetahat}{\bm{\hat{\beta}}}
\newcommand{\gevc}{GeV/\textit{c}}

\newcommand{\op}{\left(}
\newcommand{\cp}{\right)}
\newcommand{\opc}{\left\{}
\newcommand{\cpc}{\right\}}
\newcommand{\qq}{\qquad \qquad}
\newcommand{\qqq}{\qquad \qquad \qquad}

\makeatletter
\providecommand*{\diff}%
	{\@ifnextchar^{\DIfF}{\DIfF^{}}}
\def\DIfF^#1{%
	\mathop{\mathrm{\mathstrut d}}%
	\nolimits^{#1}\gobblespace}
\def\gobblespace{%
	\futurelet\diffarg\opspace}
\def\opspace{%
	\let\DiffSpace\!%
	\ifx\diffarg(%
		\let\DiffSpace\relax
	\else
		\ifx\diffarg[%
			\let\DiffSpace\relax
		\else
			\ifx\diffarg\{%
				\let\DiffSpace\relax
			\fi\fi\fi\DiffSpace}

\chapter{Transverse-Momentum-Dependent structure of the nucleon} 

\label{Chapter1_SIDIS} 

\section{Introduction}
The study of the three-dimensional structure of the nucleon is a fascinating and challenging topic. In recent years, it has been at the center of a very intense research activity, with a flourishing of theoretical developments and experimental investigations. Still, our knowledge about this subject is  limited. 

\begin{figure}[h!]
\captionsetup{width=\textwidth}
    \centering
    \includegraphics[width=0.85\textwidth]{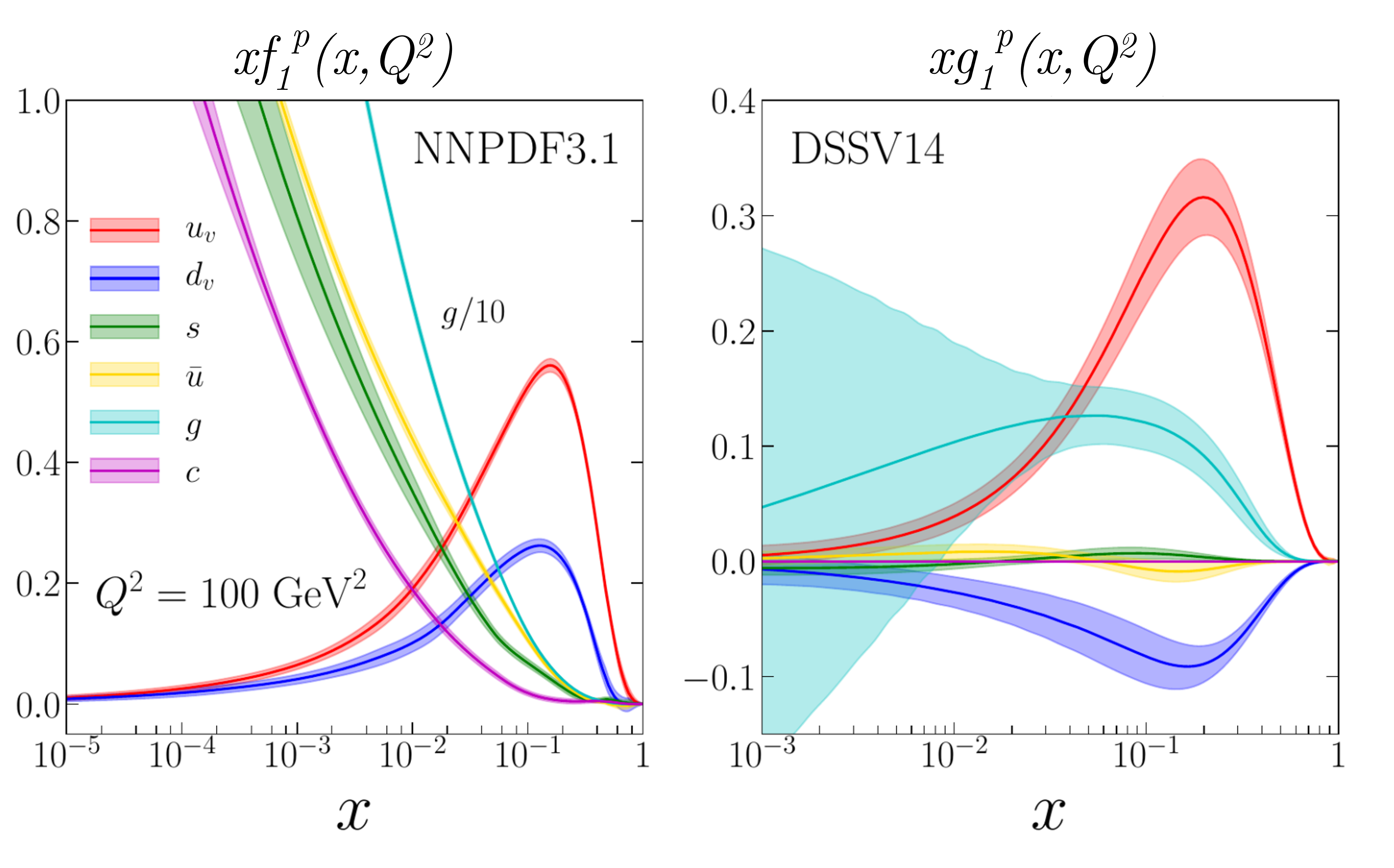}
    \caption{Examples of unpolarized (left) and helicity proton PDFs (right) from the NNPDF3.1 (NNLO) and DSSV14 (NLO) extractions, respectively. Uncertainty bands correspond to Monte Carlo 68\% confidence levels. Figure adapted from Ref.~\cite{Ethier:2020way}. }
    \label{fig:pdfs}
\end{figure}

In its collinear approach, where the nucleon and parton transverse degrees of freedom are neglected, the quantum theory of the strong interaction (Quantum Chromo Dynamics $-$ QCD), is successful in explaining a large amount of experimental data. The collinear structure functions and the unpolarized Parton Distribution Functions (PDF) are well known from the measurements performed in Deep Inelastic Scattering (DIS), unpolarized Semi-Inclusive DIS (SIDIS) and Drell-Yan (see Ref.~\cite{Ball:2014uwa} for a review). As can be seen in Fig.~\ref{fig:pdfs} (left), the most recent extractions are characterized by a very high level of accuracy. 

Also the longitudinal spin structure of the nucleon, accessed in DIS and SIDIS with longitudinally polarized targets \cite{deFlorian:2019zkl}, is well known (Fig.~\ref{fig:pdfs}, right)

However, since the pioneering measurements of the EMC experiment at CERN \cite{Ashman:1987hv}, it is also known that the spin of the valence quarks is not sufficient to generate the total nucleon spin: the contributions from the sea quarks and the gluons as well as from the orbital angular momentum of the partons are also to be taken into account. The orbital angular momentum contribution, which can not be accounted for in a collinear framework, can be accessed in an indirect way by measuring a set of multidimensional partonic distribution functions, the Generalized Parton Distributions (GPDs) \cite{Diehl:2003ny} through the Deeply Virtual Compton Scattering (DVCS) and the Hard Exclusive Meson Production (HEMP) mechanisms. Important contributions in this respect have been produced by HERMES \cite{HERMES:2011bou,HERMES:2001bob,HERMES:2006pre,HERMES:2009cqe,HERMES:2012gbh,HERMES:2012idp,HERMES:2010hnl,HERMES:2010dsx,HERMES:2008abz,HERMES:2009xsg,HERMES:2009bqn}, H1 \cite{Adloff:2001cn,Aktas:2005ty,Aaron:2007ab,Aaron:2009ac} and ZEUS \cite{Chekanov:2003ya,Schoeffel:2007ng} at DESY, by the COMPASS experiment at CERN \cite{COMPASS:2017mvk,COMPASS:2019fea} and by the Hall A \cite{JeffersonLabHallA:2006prd,JeffersonLabHallA:2007jdm,JeffersonLabHallA:2015dwe} and CLAS experiments \cite{CLAS:2001wjj,CLAS:2006krx,CLAS:2007clm,CLAS:2008ahu,CLAS:2014qtk,CLAS:2015bqi,CLAS:2015uuo} at the Jefferson Lab.

\bigskip

\begin{figure}
\captionsetup{width=\textwidth}
    \centering
    \includegraphics[width=0.37\textwidth]{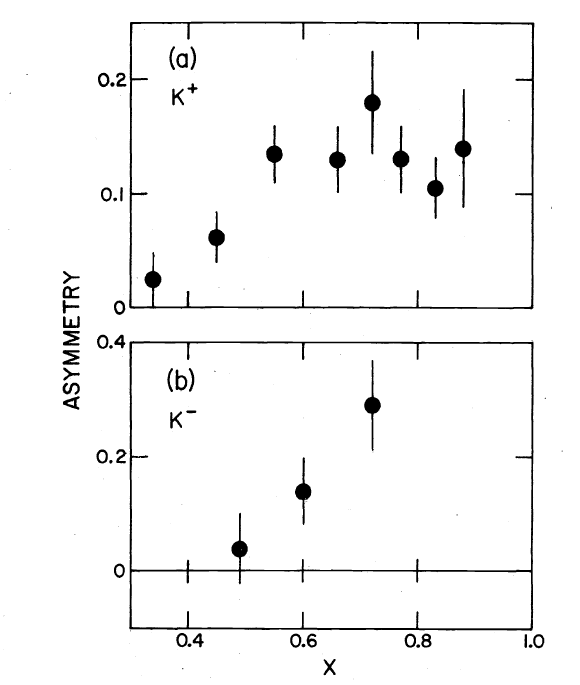}
    \includegraphics[width=0.62\textwidth]{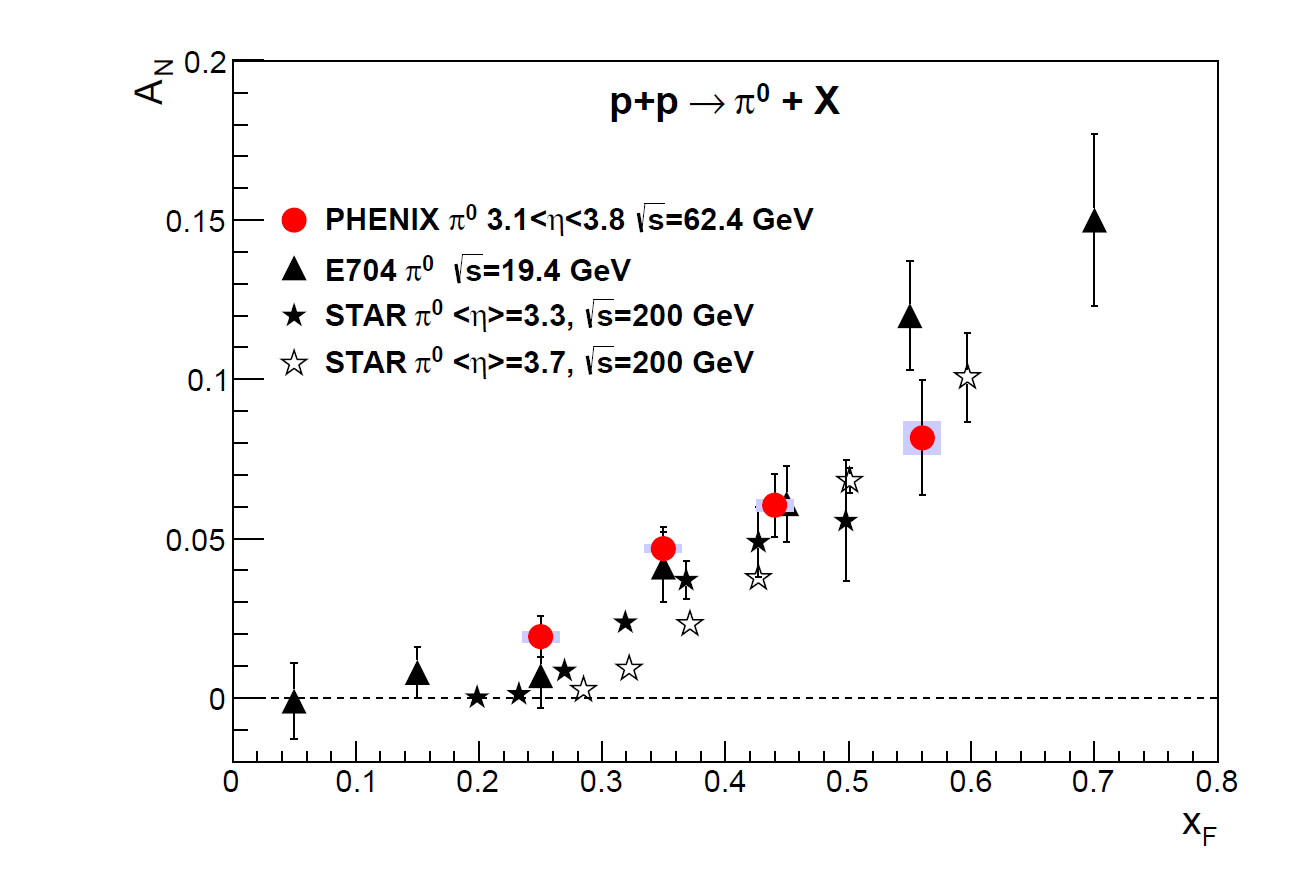}
    \caption{On the left: on the top (bottom) plot, the asymmetry for inclusive $K^+$ ($K^-$) production from hydrogen as a function of $x$, measured at Argonne~\cite{Dragoset:1978gg}. On the right: neutral pion $A_N$ as function of $x_F$ at various center-of-mass energies, measured at RICH \cite{Adare:2013ekj}.}
    \label{fig:dragoset}
\end{figure}

In terms of transverse momentum and transverse spin, the nucleon structure is less known. The first indications that the transverse spin effects are sizable came in the 1970s, with the observation of large single spin asymmetries in inclusive hadroproduction from polarized $pp$ collisions at Argonne \cite{Dick:1975ty,Klem:1976ui,Dragoset:1978gg} (Fig.~\ref{fig:dragoset}, left), later confirmed by the E704 Collaboration at Fermilab \cite{Adams:1991rw,Adams:1991cs,Adams:1991ru,Bravar:1996ki} at different energies and, more recently, at RHIC \cite{Adams:2003fx,Adams:2006uz,Adler:2003pb,Adler:2005in,Arsene:2007jd,Abelev:2008af,Arsene:2008aa,Adare:2013ekj}. Just as an example, Fig.~\ref{fig:dragoset} (left) shows the measured left-right asymmetry, as a function of the Bjorken variable $x$, for identified positive and negative Kaons at Argonne \cite{Dragoset:1978gg}: the positive sign of the asymmetry indicates that more Kaons are produced to the left than to the right, when the beam has a transverse up polarization.  Fig.~\ref{fig:dragoset} (right) shows instead the asymmetry for identified $\pi^0$, whose trend is linear as a function of the Feynman variable $x_F$ at various center-of-mass energies. 

\begin{figure}
    \captionsetup{width=\textwidth}
    \centering
    \includegraphics[width=0.49\textwidth]{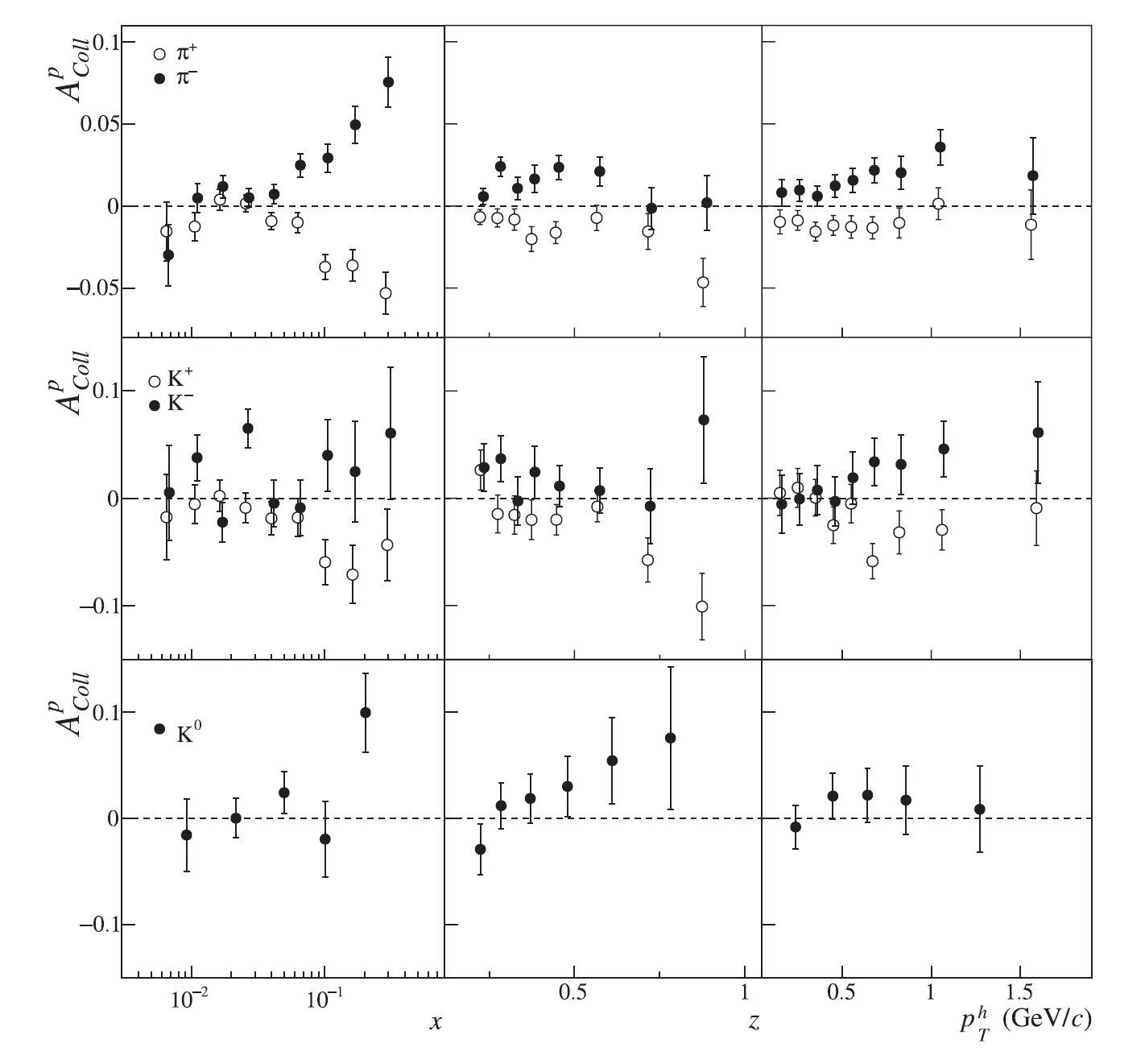}
    \includegraphics[width=0.49\textwidth]{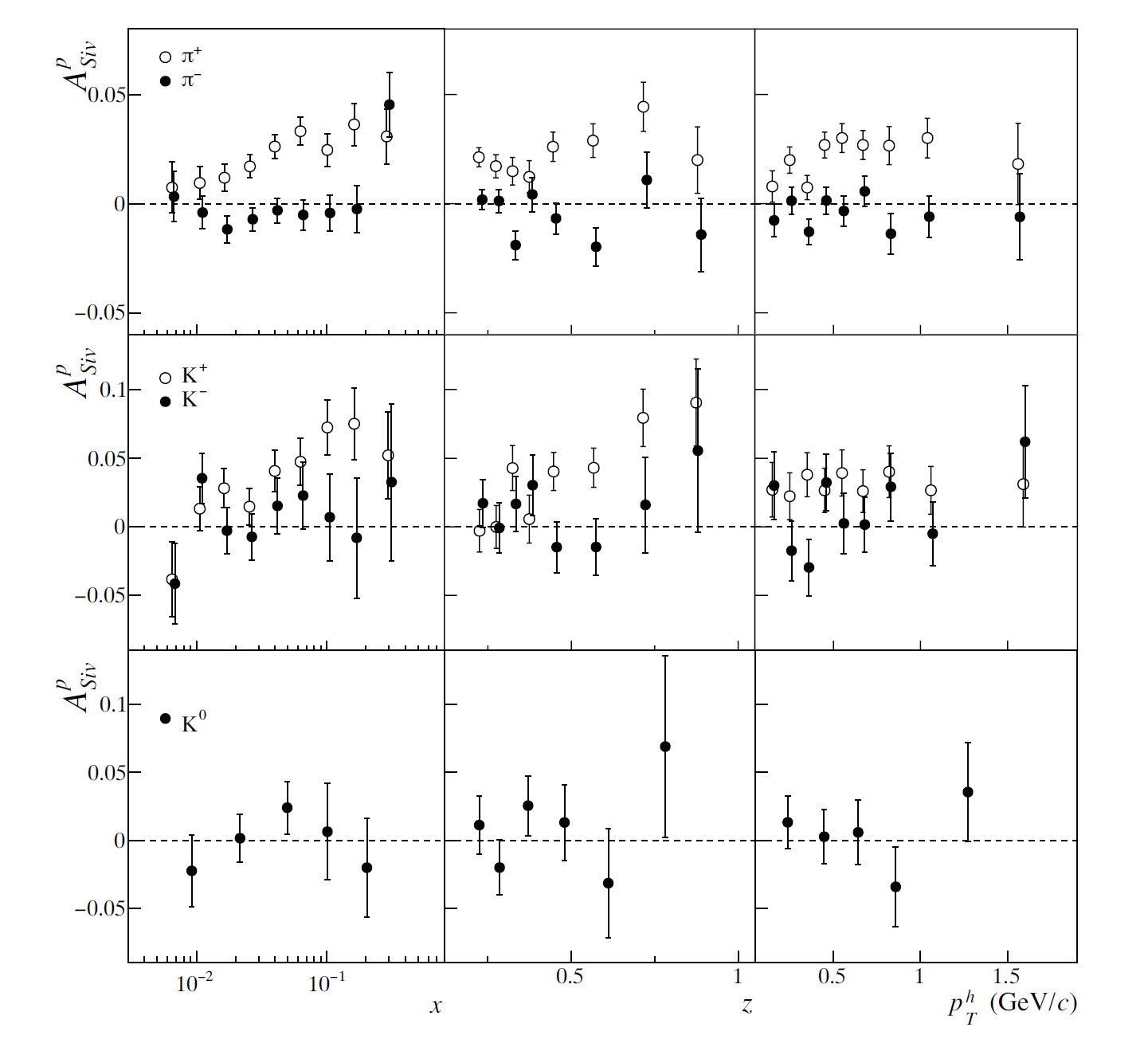}
    \caption{Left: The Collins asymmetry on proton for charged pions (top), charged Kaons (middle) and neutral Kaons (bottom) as a function of $x$, $z$ and $\Pt$ measured at COMPASS \cite{COMPASS:2012ozz}. Right: same, for the Sivers asymmetry \cite{COMPASS:2012ozz}.}
    \label{fig:apcollsiv}
\end{figure}

The inclusion of the transverse spin in the QCD theoretical framework led, in the Nineties, to the formal definition of the chiral-odd transversity function $h_1$, first suggested by Ralston and Soper in 1979 \cite{Ralston:1979ys} and later by Artru \cite{Artru:1989zv}. Several experimental channels were originally proposed to access transversity function in SIDIS off transversely polarized nucleons. Two of them, the measurement of the Collins asymmetries (in which $h_1$ couples to the chiral-odd Collins fragmentation function $\coll$) and of the hadron-pair azimuthal asymmetries on transversely polarized protons, both measured to be sizable and with clear kinematic dependences both by HERMES and COMPASS \cite{Airapetian:2004tw,Airapetian:2010ds,COMPASS:2010hbb,COMPASS:2012ozz, Airapetian:2008sk,COMPASS:2012bfl,COMPASS:2014ysd}, provided convincing evidence that transversity is experimentally accessible and different from zero. This is evident, e.g., from the $x$-dependence of the Collins asymmetry of positive and negative pions measured in COMPASS on transversely polarized protons (Fig.~\ref{fig:apcollsiv} (left), from Ref.~\cite{COMPASS:2012ozz}), which is a clear manifestation of the transversity PDF. Combining the Collins asymmetries measured with proton and deuteron targets with the information coming from the $e^+e^-$ annihilation data \cite{Seidl:2008xc}, the transversity functions for the $u$ and $d$ quarks have been extracted point-by-point and found to be different from zero at large $x$, where $h_1^u(x)$ and $h_1^d(x)$ are almost of the same size but opposite in sign, while $h_1^{\bar{u}}$ and $h_1^{\bar{d}}$ have been found compatible with zero (see, e.g., Ref.~\cite{Martin:2014wua}). A third channel to access transversity in SIDIS, i.e. the measurement of the spin transfer from transversely polarized nucleon to $\Lambda$ hyperons, has been recently investigated in COMPASS \cite{COMPASS:2021bws}. 

The measurement of non-zero Sivers asymmetries, in addition to the Collins asymmetries, marked a further important step in the study of the transverse spin effects in SIDIS. The Sivers asymmetry, shown in Fig.~\ref{fig:apcollsiv} (right) as measured in COMPASS \cite{COMPASS:2012ozz} for identified hadrons as a function of $x$, $z$ and $\Pt$, is proportional to the Sivers function $f_{1T}^{\perp}$ \cite{Sivers:1989cc,Sivers:1990fh}, which encodes the correlation between the nucleon spin and the parton transverse momentum $\vkt$ for an unpolarized parton in a transversely polarized nucleon.

\bigskip
The Sivers asymmetry is not the only manifestation of the intrinsic transverse momentum of the parton, $\vkt$. It was predicted by Cahn \cite{Cahn:1978se} that its presence gives rise to a cosine modulation of the azimuthal angle $\fih$ of the final-state hadron produced in SIDIS off an unpolarized target. The transverse momentum $\vkt$ also contributes to the transverse momentum vector $\Pt$ of the final-state hadrons, and consequently has a primary role in defining the shape of the SIDIS cross-section as a function of the hadron transverse momentum. Several measurements exist of these transverse momentum effects: they will be described later in this Chapter.

\bigskip     

The transverse momentum and the transverse spin of the partons, their correlation and their correlation with the nucleon spin have been encoded in a set of Transverse-Momentum-Dependent Parton-Distribution-Functions (TMD-PDFs) which, in addition to the dependence on the Bjorken variable $x$ and on the scale $\Qsq$, acquire a dependence on the intrinsic transverse momentum of the parton, $\kt$. At leading order, a complete description of the nucleon structure is given by eight independent TMD-PDFs, of which only three (number density, helicity and transversity) survive to the integration upon the parton transverse momentum. Currently, only the number density and the Sivers function have been measured different from zero, while the other TMDs have not been extracted yet. Specularly to the Sivers function, the Boer-Mulders function $\bom$ \cite{Boer:1997nt} encodes the correlation between the parton spin and the parton momentum for a transversely polarized parton in an unpolarized nucleon. Also because of this similarity with $f_{1T}^{\perp}$, the $\bom$ function is particularly interesting. Its extraction from the experimental data is however challenging, and several attempts have not been conclusive so far. Analogously to the PDF case, it is possible to define eight independent TMD Fragmentation Functions (FFs) at leading twist. Generally, these functions are still poorly known.

\bigskip

This thesis aims at contributing to the study of the transverse-momentum-dependent structure of the nucleon with the analysis of part of the SIDIS data collected in COMPASS with a high-energy muon beam and a liquid hydrogen target. After a brief overview of the TMD-PDFs, this Chapter hosts a description of the SIDIS process, with the indication of the kinematic variables best suited for its definition. Two classes of observables (the transverse-momentum distributions and the azimuthal asymmetries) are then introduced and indicated as particularly interesting, due to their proportionality to the unpolarized SIDIS structure functions, in turn related to the TMD-PDFs and FFs. The last part of this Chapter summarizes the existing unpolarized SIDIS measurements and describes some phenomenological works performed to interpret the results.

\section{Transverse-Momentum-Dependent distribution functions}
In Leading Order (LO) QCD, when neglecting the parton transverse momentum, the structure of the nucleon is encoded in three independent parton distribution functions (PDFs): the number density $f_1^q(x)$, the helicity $g_1^q(x)$ and the transversity $h_1^q(x)$ \cite{Ralston:1979ys,Artru:1989zv}. All of them are characterized by a probabilistic interpretation:
\begin{itemize}
    \item $f_1^q(x)$ is the probability of finding, in a nucleon with momentum $P$, a parton $q$ with longitudinal momentum equal to $xP$;
    \item $g_1^q(x)$ is the difference of the probability that the parton $q$ has its helicity aligned or anti-aligned with that of the parent nucleon;
    \item $h_1^q(x)$ is the difference of the probability that the parton $q$ has its spin aligned or anti-aligned with that of the transversely polarized parent nucleon.
\end{itemize}

Despite the similar definitions, $h_1(x)$ and $g_1(x)$ are different and independent functions, due to the breaking of the rotational symmetry that occurs in a relativistic motion along the longitudinal direction. In addition, at variance with $g_1(x)$, $h_1(x)$ is chiral-odd. In other words, the eigenstates of the transverse polarization are orthogonal to the helicity eigenstates: this means that the struck quark must flip its helicity at the photon vertex. In order the reaction to be observable the quark must flip back its helicity, and this can only occur in the fragmentation process, when a final-state hadron is produced. For this reason, transversity can not be measured in DIS, but only in a Semi-Inclusive measurement of the DIS, where it can couple to another chiral-odd function (like the Collins function $\coll$). These three collinear PDFs are subject to the so-called Soffer bound \cite{Soffer:1994ww}, according to which the absolute value of the transversity PDF cannot exceed half the sum, in absolute value, of the number density and helicity PDFs:

\begin{equation}
    |h_1(x)| \leq \frac{1}{2} | f_1(x) + g_{1}(x) |.
\end{equation}
Related to the transversity PDF is the tensor charge $\delta q$, defined as:
\begin{equation}
    \delta q = \int_0^1 \diff x \op h_1^{q}(x) - h_1^{\bar{q}}(x) \cp.
\end{equation}
In a non-relativistic quark model, $h_1^q$ would be equal to $g_{1}^q$ and $\delta q$ would be equal to the valence quark contribution to the nucleon spin. The tensor charge, which reflects the actual difference between helicity and transversity, provides important constraints to any model of the nucleon. The quark tensor charge is in turn related to the isovector nucleon tensor charge, defined as the difference of the tensor charges for the $u_v$ and $d_v$ quarks:
\begin{equation}
    g_T = \delta u_v -\delta d_v.
\end{equation}
which, together with the vector and axial charges, characterizes the nucleon as a whole and whose magnitude can put limits on observables related to the Physics Beyond the Standard Model (BSM) \cite{Courtoy:2015haa}.

\bigskip

In general, the partons are not collinear to the parent nucleon, their transverse motion being characterized by the intrinsic transverse momentum $\vkt$. Taking into account the parton transverse motion, the three leading order collinear partonic distributions $f_1$, $g_1$ and $h_1$ are generalized as Transverse Momentum Dependent (TMD) parton distributions in $x$ and $\vkt$. The three-dimensional structure of the nucleon is however not exhausted by these three TMD-PDFs: the complete description of the nucleon at leading twist, namely at the leading order in the hard scale which characterizes the interaction between the probe and the target, requires eight TMD-PDFs \cite{Mulders:1995dh,Bacchetta:2006tn}, which describe all possible correlations among the nucleon spin, the parton spin and the parton transverse momentum: they are given in Tab.~\ref{tab:tmds}.

Along the diagonal of Tab.~\ref{tab:tmds} there are the only three TMD-PDFs (number density, helicity and transversity) that do not vanish upon integration over the transverse momentum $\vkt$, just reducing to their collinear counterparts: for example, for the number density TMD-PDF $f_1(x,\kt)$, one has that:

\begin{equation}
    f_1^{q}(x) = \int \diff^2 \vkt~f_1^{q}(x,\kt).
\end{equation}
and similarly for the helicity PDF and for the transversity PDFs. The other five TMD-PDFs are zero in the collinear limit.

\bigskip

The Sivers function $f_{1T}^{\perp}(x,\kt)$ \cite{Sivers:1989cc,Sivers:1990fh} and the Boer-Mulders function $\bom(x,\kt)$ \cite{Boer:1997nt}, are T-odd: they change sign upon a time-reversal transformation. In particular, the Sivers function encodes the correlation between the transverse momentum $\vkt$ of an unpolarized quark  in a transversely polarized nucleon and the nucleon polarization vector, while the Boer-Mulders function encodes the correlation between the transverse momentum $\vkt$ and the transverse polarization of a quark inside an unpolarized nucleon.

The pretzelosity TMD $h_{1T}^{\perp}$ is somehow related to the nucleon shape: if different from zero, it would suggest that the nucleon is not exactly spherical \cite{Miller:2007ae}. The two worm-gear functions, $g_{1T}$ \cite{Kotzinian:1995cz} and $h_{1L}^{\perp}$, describe the distributions of longitudinally polarized quarks inside a transversely polarized nucleon ($g_{1T}$)  and of transversely polarized quarks in a longitudinally polarized nucleon ($h_{1L}^{\perp}$).

\begin{table}
\captionsetup{width=\textwidth}
    \centering
        \begin{tabular}{|c|*{3}{c|}}\hline
        \backslashbox{\textbf{\Large{N}}}{\textbf{\Large{q}}}
        &\makebox[3em]{\textbf{\large{U}}}&\makebox[3em]{\textbf{\large{L}}}&\makebox[3em]{\textbf{\large{T}}} \\ \hline
        \multirow{4}{*}{\textbf{\large{U}}} &  & &  \\
        & $f_1$ && $h_1^{\perp}$ \\
        & \footnotesize{number density} & & \footnotesize{Boer-Mulders} \\
        & & &  \\\hline
        \multirow{4}{*}{\textbf{\large{L}}} && &  \\
         & & $g_{1}$ & $h_{1L}^\perp$ \\
         && \footnotesize{helicity} & \footnotesize{Kotzinian-Mulders} \\
         &&  & \footnotesize{worm-gear L} \\ \hline
        \multirow{4}{*}{\textbf{\large{T}}} &  &  & $h_{1}$ \\
         & $f_{1T}^{\perp}$ & $g_{1T}$ & \footnotesize{transversity} \\
         & \footnotesize{Sivers} & \footnotesize{Kotzinian-Mulders} & $h_{1T}^\perp$ \\
         &  & \footnotesize{worm-gear T} & \footnotesize{pretzelosity} \\ \hline
        \end{tabular}
    \caption{The eight leading-twist TMD-PDFs. The indices U, L and T indicate unpolarized, longitudinally or transversely polarized quarks (columns) or nucleons (rows). }
    \label{tab:tmds}
\end{table}

\bigskip
Also the fragmentation of a parton of given polarization into a final state hadron can be formalized in a transverse-momentum-dependent description, not integrating over the outgoing quark transverse momentum. At leading twist, there are 8 TMD Fragmentation Functions (TMD-FFs). Among them, there is the unpolarized FF $D_1(z,\pperp)$, which describes the fragmentation of an unpolarized quark into an unpolarized hadron. Here, $\pperp$ is the transverse momentum acquired by the fragmenting parton during the hadronization process. One of the most interesting fragmentation functions is the T-odd Collins function $\coll(z,\pperp)$ \cite{Collins:1992kk}, which encodes the correlation of the produced hadron transverse momentum and the spin of the fragmenting quark.

\bigskip 
The PDFs and FFs are thought to be universal, namely independent on the process in which they are observed. The same is true for the TMD-PDFs, but for the T-odd ones, (the Sivers and Boer-Mulders functions). An important prediction based on the T-oddity of these functions is that they should have an opposite sign when measured in SIDIS and in Drell-Yan, where the final state interactions are replaced by initial state interaction between the colliding hadrons \cite{Collins:2002kn}. Recent measurements done in COMPASS, the only experiment that can run both SIDIS and Drell-Yan measurements with the same apparatus and a similar kinematic coverage, favor the hypothesis of the sign change for the Sivers function \cite{COMPASS:2017jbv}. \\

\section{Semi-Inclusive Deep Inelastic Scattering}
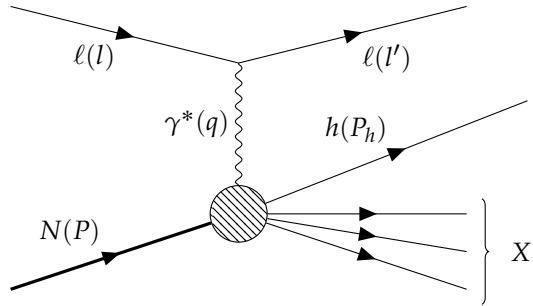
\begin{figure}[h!]
\captionsetup{width=\textwidth}
\centering
\begin{tikzpicture}
    \begin{feynman}
        \vertex[blob] (m) at ( 0, -1) {};
        \vertex (a) at (-3, 1.75);
        \vertex (b) at (-3,-2);
        \vertex (c) at ( 3, 1.75);
        \vertex (d) at ( 3, -2);
        \vertex (e) at ( 3, -1.5);
        \vertex (f) at ( 3, -1);
        \vertex (g) at (0,1);
        \vertex (dd) at ( 3.2, -2.2);        
        \vertex (ff) at ( 3.2, -0.8);        
        \vertex (hh) at ( 3.8, 0.5);       
        \diagram*{
        (a) -- [fermion,edge label'=$\ell(l)$] (g) -- [fermion,edge label'=$\ell(l^\prime)$] (c),
        (b) -- [fermion, very thick, edge label=$N(P)$] (m) -- [fermion] (d),
        (m) -- [fermion] (e),
        (m) -- [fermion, edge label=$h(P_h)$] (hh),
        (m) -- [fermion, label=right:$X$] (f),
        (g) -- [photon, edge label'=$\gamma^{*}(q)$] (m),
        };
        \draw [decoration = {brace} , decorate] (ff.north west) -- (dd.south west) node [pos = 0.5 , right = 0.25cm] {$X$};
    \end{feynman}
\end{tikzpicture}
\caption{Schematic representation of the Semi-Inclusive Deep Inelastic Scattering in the one-photon-exchange approximation. }
\label{fig:sidis_scheme}
\end{figure}

In the SIDIS process:
\begin{equation}
    \ell(l) + N(P) \rightarrow \ell(l^\prime) + h(P_h) + X(P_X)
\end{equation}
a high energy lepton $\ell$ scatters off a target nucleon $N$, probing its internal structure, and at least one final-state hadron $h$ is detected in coincidence with the scattered lepton $\ell^\prime$. The quantities in parentheses denote the four-momenta and $X$ represents the unobserved part of the final state. In the one-photon-exchange approximation, depicted in Fig.~\ref{fig:sidis_scheme}, the electromagnetic interaction between the lepton and the nucleon is mediated by a virtual photon $\gamma^*$ of momentum $q=l-l^\prime$ and virtuality $\Qsq = -q^2$.  

\bigskip 
In the DIS regime (typically $\Qsq \gg M^2$, where $M$ is the target nucleon mass) the energy of the virtual photon is high enough to probe the inner parton $q$ of the nucleon via the elementary interaction $\gamma^* q \rightarrow q^\prime$. Moreover, the time scale of the process ($\sim1/\Qsq$) is so short that the parton can be considered free from the binding forces due to the strong interaction.
The final state requires at least two variables to be fully described. Along with $\Qsq$, useful kinematic variables are:

\begin{enumerate}
    \item the Bjorken scaling variable $x$, which can be identified with the fraction of the longitudinal nucleon momentum carried by the parton, in a system (the Breit frame) where the nucleon has infinite momentum. It is given by:
    \begin{equation}
    x = \frac{Q^2}{2P\cdot q} \stackrel{\text{lab}}{=} \frac{Q^2}{2M(E-E^{\prime})}
    \end{equation}
    where $E$ ($E^\prime$) is the energy, measured in the laboratory frame, of the incoming (scattered) lepton;
    \item the inelasticity $y$, that can be identified in the laboratory system as the ratio of the virtual photon energy with the incoming lepton energy:
    \begin{equation}
    y = \frac{P\cdot q}{P\cdot l} \stackrel{\text{lab}}{=} \frac{E-E^\prime}{E}
    \end{equation}
    \item the virtual photon energy in the laboratory system, $\nu$:
    \begin{equation}
        \nu = \frac{P\cdot q}{M} \stackrel{\text{lab}}{=} E-E^\prime
    \end{equation}
    \item the invariant mass squared of the hadronic final state $W^2$:
    \begin{equation}
        W^2 = \op P+q\cp^2 \stackrel{\text{lab}}{=} M^2-Q^2+2M(E-E^\prime)
    \end{equation}
\end{enumerate}
Note that not all these variables are independent. In practice, two of them (e.g. $x$ and $\Qsq$) completely identify the process.
In addition to the DIS variables, a full description of the SIDIS process also requires to introduce the relative energy of the final state hadron $z$: 
\begin{equation}
z = \frac{P\cdot P_h}{P\cdot q} \stackrel{\text{lab}}{=} \frac{E_h}{E-E^{\prime}}
\end{equation}
and the hadron transverse momentum $\vPt$ with respect to the virtual photon:

\begin{equation}
    \vPt = \textbf{\textit{P}}_h - \frac{\op\textbf{\textit{P}}_h \cdot \textbf{\textit{q}}\cp \textbf{\textit{q}} }{|\textbf{\textit{q}}^2|}.
\end{equation}

\begin{figure}
\captionsetup{width=\textwidth}
    \centering
    \includegraphics[scale=0.38]{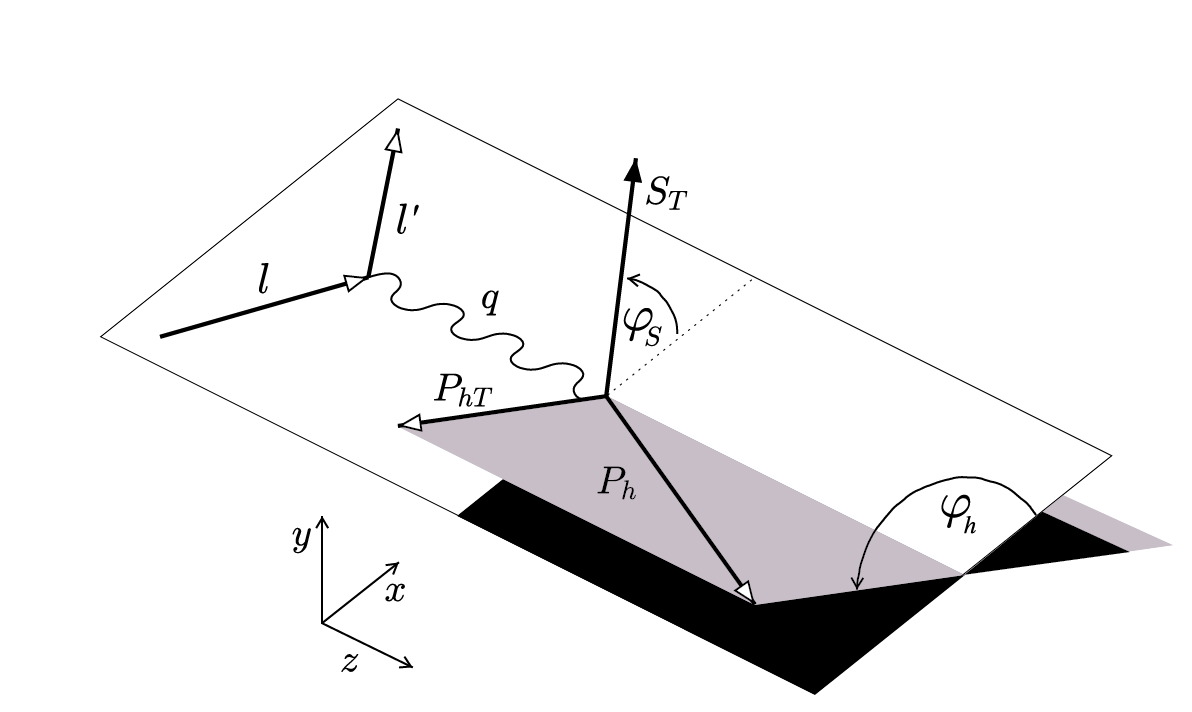}
    \caption{The Gamma-Nucleon System (GNS), where the z-axis is fixed by the direction of the virtual photon, the xz-plane coincides with the lepton scattering plane and the y-axis is defined in order to have an orthogonal right-handed system, with the positive x-axis direction fixed according to the scattered muon direction. Plot by J.~Matousek (COMPASS). }
    \label{fig:gns}
\end{figure}
Denoting by $\fih$ and $\phi_S$ the azimuthal angles of $\vPt$ and of the nucleon spin in the Gamma Nucleon System (GNS, defined in Fig.~\ref{fig:gns}) and considering the possible polarization of the beam (that can be either unpolarized or longitudinally polarized) and of the target nucleon (unpolarized, longitudinally polarized or transversely polarized) the differential cross-section in the one-photon-exchange approximation can be written in a model-independent way in terms of 18 independent structure functions \cite{Kotzinian:1994dv,Mulders:1995dh,Diehl:2005pc,Bacchetta:2006tn}:

\begin{equation}
    \begin{split}
    & \frac{\diff^6\sigma}{\diff x \diff y \diff \psi \diff z \diff \fih \diff \Ptsq} = \frac{\alpha^2}{xy\Qsq}\frac{y^2}{2\op 1-\eps\cp}\op 1+\frac{\gamma^2}{2x}\cp \cdot \\
    & \qqq \bigg\{ F_{UU,T} + \eps F_{UU,L} + \sqrt{2\eps\op 1+\eps\cp} F_{UU}^{\cos\fih}\cos\fih + \eps F_{UU}^{\cos2\fih}\cos2\fih \\ 
    & \qqq + \lambda \bigg( \sqrt{2\eps\op 1-\eps\cp} F_{LU}^{\sin\fih}\sin\fih \bigg) \\
    & \qqq + S_L \bigg( \sqrt{2\eps\op 1+\eps\cp} F_{UL}^{\sin\fih}\sin\fih + \eps F_{UL}^{\sin2\fih}\sin2\fih \bigg) \\
    & \qqq + S_L\lambda \bigg( \sqrt{1-\eps^2} F_{LL} + \sqrt{2\eps\op 1-\eps\cp} F_{LL}^{\cos\fih}\cos\fih \bigg) \\
    & \qqq + S_T \bigg( \op F_{UT,T}^{\sin\op \fih-\phi_S\cp} + \varepsilon F_{UT,L}^{\sin\op \fih-\phi_S\cp}\cp \sin\op\fih-\phi_S \cp \\
    & \qqq \qq + \eps F_{UT}^{\sin\op \fih+\phi_S\cp}\sin\op\fih+\phi_S\cp + \eps F_{UT}^{\sin\op 3\fih-\phi_S\cp}\sin\op3\fih-\phi_S\cp \\
    & \qqq \qq + \sqrt{2\eps\op 1+\eps\cp} F_{UT}^{\sin\phi_S}\sin\phi_S + \sqrt{2\eps\op 1+\eps\cp} F_{UT}^{\sin\op2\fih-\phi_S\cp}\sin\op2\fih-\phi_S\cp \bigg) \\
    & \qqq +S_T\lambda \bigg( \sqrt{1-\eps^2}F_{LT}^{\cos\op \fih-\phi_S\cp}\cos\op \fih-\phi_S\cp \\ 
    & \qqq \qq + \sqrt{2\eps\op 1-\eps\cp} F_{LT}^{\cos\phi_S}\cos\phi_S + \sqrt{2\eps\op 1-\eps\cp} F_{LT}^{\cos\op2\fih-\phi_S\cp}\cos\op2\fih-\phi_S\cp \bigg) \bigg\}.
    \end{split}
\label{eq:sidis_xsec}
\end{equation}
where $\psi$ is the azimuthal angle of the nucleon spin vector in a system where the z-axis is chosen as the lepton beam momentum \cite{Diehl:2005pc}, well approximated by $\phi_S$; $\lambda$ is the longitudinal polarization of the incoming lepton, while $S_L$ ($S_T$) is the nucleon longitudinal (transverse) component of the nucleon spin in the GNS. The kinematic factor $\eps$, defined as:

\begin{equation}
    \eps = \frac{1-y-\frac{1}{4}\gamma^2y^2}{1-y+\frac{1}{2}y^2+\frac{1}{4}\gamma^2y^2}
\end{equation}
corresponds to the ratio of the longitudinal and transverse photon flux; $F_{XY}^{f(\fih,\phi_S)}$ indicates the structure function associated to the modulation described by $f(\fih,\phi_S)$, for a given beam polarization $X$ and a target polarization $Y$. The structure functions $F_{UU,T}$, $F_{UU,L}$ and $F_{LL}$ are not associated to any angular modulation. The third subscript indicates the photon polarization, that can be either transverse ($T$) or longitudinal ($L$). In general, all the structure functions in Eq.~\ref{eq:sidis_xsec} depend on the hadron type and charge, as well as on the kinematic variables ($x$, $\Qsq$, $z$, $\vPt$), omitted here.

\subsection{SIDIS structure functions in TMD formalism}
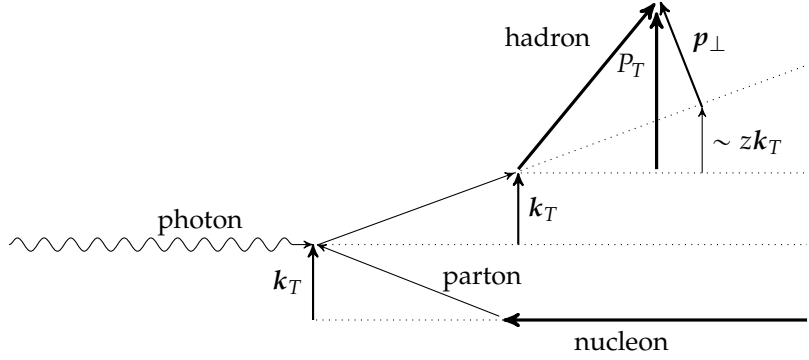
\begin{figure}
\captionsetup{width=\textwidth}
\centering
\begin{tikzpicture}[
  >=stealth',
  pos=.8,
  photon/.style={decorate,decoration={snake,post length=2mm}}
]
  \draw[->,photon] (-6,0) -- node[above left] {photon} (-2,0);
  \draw[->,very thick] (4.5,-1) -- node[below right] {nucleon} (0.5,-1);
  \draw[->] (0.45,-0.95) -- node[below right] {~~~~~~~~~~~~parton} (-1.95,0);
  \draw[->] (-1.93,0) -- node[below right] {} (0.65,0.95);
  \draw[->,very thick] (0.70,1.00) -- node[left] {hadron~~~~~} (2.5,3.2);
  \draw[->,very thick] (2.52,1.0) -- node[below left] {~~~~~~~~~$P_T$} (2.52,3.08);
  \draw[->,thick] (3.12,1.82) -- node[below right] {~~$\vpperp$} (2.57,3.22);
  \draw[->] (3.12,0.95) -- node[below right] {$\sim z\vkt$} (3.12,1.82);
  \draw[->,thick] (0.7,0) -- node[below right] {$\vkt$} (0.7,0.95);
  \draw[dotted] (-2,0) -- node[below right] {} (4.5,0);
  \draw[dotted] (0.65,0.95) -- node[below right] {} (4.5,0.95);
  \draw[dotted] (0.65,0.95) -- node[below right] {} (4.5,2.38);
  \draw[dotted] (0.5,-1) -- node[below right] {} (-2,-1);
  \draw[->,thick] (-2,-1) -- node[below left] {$\vkt$} (-2,0.0);

\end{tikzpicture}
    \caption{Diagram describing the momenta involved in a SIDIS event, in the GNS: a virtual photon interacts with a parton inside the nucleon. The parton transverse momentum is indicated as $\vkt$. The struck parton fragments into a final-state hadron, which acquires a further transverse momentum $\vpperp$. The total measured hadron transverse momentum is $\vPt$. When $Q^2$ is large, $\vPt \approx z\vkt + \vpperp$.}
\label{fig:sidis_momenta}
\end{figure}

At small transverse momentum, the factorization theorems \cite{Collins:2011zzd,Collins:1989gx,Ji:2004wu,Echevarria:2012js,Rogers:2015sqa} suggest that the structure functions of Eq.~\ref{eq:sidis_xsec} can be written as convolutions of TMD-PDFs and TMD-FFs:

\begin{equation}
    F(x,\Qsq,z,\Pt) = \mathcal{C}[wf^qD^{q\to h}]
\end{equation}
where $w=w(\vkt,\vpperp)$ is a weight, $f$ and $D$ are TMD-PDFs and FFs and the convolution $\mathcal{C}$ is defined as:

\begin{equation}
\begin{split}
    \mathcal{C}[wf^qD^{q\to h}]  = x \sum_q e_q^2 & \int \diff^2 \vkt \diff^2 \vpperp \delta^{(2)} \op \vPt - z\vkt - \vpperp \cp \cdot \\
    & \cdot w(\vkt,\vpperp) f^q(x,\kt,\Qsq) D^{q\to h} (z,\pperp,\Qsq).
\end{split}
\label{eq:convolution}
\end{equation}
Here, the sum runs over the quark flavors $q$, weighted with their electric charge squared $e_q^2$. The integration of TMD-PDFs and FFs is performed over the intrinsic transverse momentum of the quark $\vkt$ and over the transverse momentum acquired during the fragmentation $\vpperp$, both not experimentally accessible (Fig. \ref{fig:sidis_momenta}). At large $Q^2$, the total transverse momentum of the hadron $\vPt$ is related to $\vkt$ and $\vpperp$ through the relation:
\begin{equation}
    \vPt = z\vkt + \vpperp.
\end{equation}
As $\vkt$ and $\vpperp$ are independent, the mean value of the angle $\theta$ between them is zero; hence, a linear relation holds among the mean values of the three transverse momenta:
\begin{equation}
\begin{split}
    \aPtsq & = \langle (z\vkt + \vpperp)^2\rangle  \\
    & = z^2\aktsq + \apperpsq +2z \langle k_{T} p_{\perp} \cos\theta \rangle \\
    & = z^2\aktsq + \apperpsq 
\end{split}
\end{equation}
Making explicit the flavor dependence of $\aktsq$ and $\apperpsq$, one has:
\begin{equation}
    \FaPtsq = z^2\Faktsq + \Fapperpsq.
\end{equation}

Out of the 18 structure functions appearing in the cross-section, only eight are non-vanishing at leading twist (that is twist-2, according to the definition by Jaffe \cite{Jaffe:1997vlv}). They are \cite{Bacchetta:2006tn,Anselmino:2011ch}:

\begin{equation}
\begin{split}
    F_{UU,T} = \mathcal{C} & \left[ f_1D_1 \right] \\
    F_{UU}^{\cos2\fih} = \mathcal{C} &\left[\frac{2(\hh\cdot\vkt)(\hh\cdot\vpperp)-\vkt\cdot\vpperp}{zMM_h}\bom\coll\right] \\
    F_{UL}^{\sin2\fih} = \mathcal{C} &\left[\frac{2(\hh\cdot\vkt)(\hh\cdot\vpperp)-\vkt\cdot\vpperp}{zMM_h}h_{1L}^{\perp}\coll\right] \\
    F_{LL} = \mathcal{C} &\left[g_{1L}D_1\right] \\
    F_{UT,T}^{\sin(\fih-\phi_S)} = \mathcal{C} &\left[-\frac{\hh\cdot\vkt}{M}f_{1T}^{\perp}D_1\right] \\
    F_{UT}^{\sin(\fih+\phi_S)} = \mathcal{C} &\left[\frac{\hh\cdot\vpperp}{zM_h}h_1\coll\right] \\
    F_{UT}^{\sin(3\fih-\phi_S)} = \mathcal{C} &\left[-\frac{2(\hh\cdot\vkt)(\vkt\cdot\vpperp) + \kt^2(\hh\cdot\vpperp) -4\op\hh\cdot\vkt\cp^2\op\hh\cdot\vpperp\cp }{2zM^2M_h}h_{1T}^{\perp}\coll\right] \\
    F_{LT}^{\cos(\fih-\phi_S)} = \mathcal{C} &\left[\frac{\hh\cdot\vkt}{M}g_{1T}D_1\right] \\
\end{split}
\end{equation}
where $\hh = \frac{\vPt}{|\vPt|}$. In particular, associated to the structure functions $ F_{UT,T}^{\sin(\fih-\phi_S)}$ and $ F_{UT}^{\sin(\fih+\phi_S)}$, we recognize the Sivers function $f_{1T}^\perp$ and the Collins function $\coll$, the most famous TMDs. We also see that, at twist-2, the only azimuthal modulation expected from an unpolarized target comes from the $F_{UU}^{\cos2\fih}$ structure function, in which the Boer-Mulders function $\bom$ appears coupled to the Collins function. Let's now concentrate on the SIDIS off an unpolarized nucleon target, the subject of this thesis.

\section{SIDIS off unpolarized nucleons}
\label{sect:ch1_sidis_unpol}
If the target nucleon is not polarized, \textit{on average} the $S_L$ and $S_T$-dependent terms give no contribution to the cross-section. Integrating over $\diff \psi$ the cross-section of Eq.~\ref{eq:sidis_xsec}, one gets the expression for the SIDIS production of a hadron $h$ on an unpolarized nucleon target: 

\begin{equation}
    \begin{split}
    & \frac{\diff^5\sigma}{\diff x \diff y \diff z \diff \fih \diff \Ptsq} = \frac{2\pi \alpha^2}{xy\Qsq}\frac{y^2}{2\op 1-\eps\cp}\op 1+\frac{\gamma^2}{2x}\cp \cdot \\
    & \qqq \bigg\{ F_{UU,T} + \eps F_{UU,L} + \sqrt{2\eps\op 1+\eps\cp} F_{UU}^{\cos\fih}\cos\fih + \eps F_{UU}^{\cos2\fih}\cos2\fih \\ 
    & \qqq + \lambda \bigg( \sqrt{2\eps\op 1-\eps\cp} F_{LU}^{\sin\fih}\sin\fih \bigg) \bigg\}
    \end{split}
\end{equation}
or, analogously (using the relation $Q^2 = xys$): 
\begin{equation}
    \begin{split}
    & \frac{\diff^5\sigma}{\diff x \diff \Qsq \diff z \diff \fih \diff \Ptsq} = \frac{2\pi \alpha^2}{xQ^4}\frac{y^2}{2\op 1-\eps\cp}\op 1+\frac{\gamma^2}{2x}\cp \cdot \\
    & \qqq \bigg\{ F_{UU,T} + \eps F_{UU,L} + \sqrt{2\eps\op 1+\eps\cp} F_{UU}^{\cos\fih}\cos\fih + \eps F_{UU}^{\cos2\fih}\cos2\fih \\ 
    & \qqq + \lambda \bigg( \sqrt{2\eps\op 1-\eps\cp} F_{LU}^{\sin\fih}\sin\fih \bigg) \bigg\}.
    \end{split}
\label{eq:sidis_unp_Q2}
\end{equation}

At twist-3 (that is, order $1/Q$) the $F_{UU,L}$ structure function is still expected to be zero, while for $F_{UU}^{\cos\fih}$ and $F_{LU}^{\sin\fih}$ one has \cite{Bacchetta:2006tn}:

\begin{equation}
\begin{split}
    F_{UU}^{\cos\fih} = \frac{2M}{Q}\mathcal{C} &\left[\frac{\hh\cdot\vpperp}{zM_h} \op xh\coll + \frac{M_h}{M}f_1\frac{\tilde{D}^\perp}{z}\cp - \frac{\hh\cdot\vkt}{M} \op xf^\perp D_1 + \frac{M_h}{M}\bom\frac{\tilde{H}}{z}\cp\right] \\
    F_{LU}^{\sin\fih} = \frac{2M}{Q}\mathcal{C} &\left[\frac{\hh\cdot\vpperp}{zM_h} \op xe\coll + \frac{M_h}{M}f_1\frac{\tilde{G}^\perp}{z}\cp + \frac{\hh\cdot\vkt}{M} \op xg^\perp D_1 + \frac{M_h}{M}\bom\frac{\tilde{E}}{z}\cp\right],
\end{split}
\end{equation}
where $h$, $f^{\perp}$, $e$ and $g^{\perp}$ on one hand, and $\tilde{D}^\perp$, $\tilde{H}$, $\tilde{G}^\perp$ and $\tilde{E}$ on the other hand are new twist-3 TMD-PDFs and FFs.
The expression for $F_{UU}^{\cos\fih}$ is often simplified in the context of the so-called Wandzura-Wilczek approximation \cite{Bastami:2018xqd}, according to which the terms originating from the quark-gluon-quark correlator are neglected. In particular, $ xh  \approx - \frac{\ktsq}{M^2}\bom$ and $xf^\perp  \approx f_1$, so that $F_{UU}^{\cos\fih}$ reduces to:
\begin{equation}
    F_{UU}^{\cos\fih} \simeq -\frac{2M}{Q}\mathcal{C} \left[\frac{\hh\cdot\vpperp}{zM_h}  \frac{\ktsq}{M^2}\bom\coll  + \frac{\hh\cdot\vkt}{M} f_1 D_1\right],
\end{equation}
where one recognizes a contribution proportional to the convolution of the Boer-Mulders function with the Collins function, suppressed of a factor $\kt/M$ with respect to the second term, expected to be dominant and proportional to the convolution of the unpolarized TMDs $f_1$ and $D_1$. The presence of this term was suggested long ago by Cahn \cite{Cahn:1978se,Cahn:1989yf} as arising from the non-coplanarity of the virtual photon and parton momenta, i.e. due to the intrinsic transverse momentum $\vkt$ of the parton. Cahn showed that the interaction probability between parton and lepton depends on the relative orientation of the leptonic and parton planes. This results in more hadrons being produced on the side opposite the scattered muon in the current fragmentation region, giving a negative modulation in $\cos\fih$ in the forward direction. As the target remnants must balance the transverse momentum $\Pt$ of the produced hadron, a positive modulation is expected in the backward direction. 
Also the emission of a hard gluon by the quark before or after the interaction can give rise to a similar asymmetry, as suggested by Georgi and Politzer in the Seventies \cite{Georgi:1977tv}. On the other hand, this contribution is expected to be significant at large $\Pt$, while being small for $\Pt<1$ GeV/$c$.

The Cahn effect also contributes to the $F_{UU}^{\cos2\fih}$ structure function at twist-4 (order 1/$\Qsq$):

\begin{equation}
    F_{UU|Cahn}^{\cos2\fih} = \frac{2M^2}{Q^2}\mathcal{C} \left[\frac{2 (\hh\cdot\vkt)^2 - \ktsq}{M^2} f_1D_1 \right].
\end{equation}
However, this is just one of the possible contributions arising at the same twist, whose size is still not known \cite{Barone:2015ksa}. 

The production of pions in the semi-exclusive (i.e. high-$z$) regime has been investigated in Ref.~\cite{Brandenburg:1994mm}. There, for various choices of the pion distribution amplitudes and of the intrinsic transverse momentum of the parton, the amplitudes of the modulations in $\cos\fih$ and $\cos2\fih$ are predicted to have opposite sign with respect to the one expected from the Cahn mechanism.

\bigskip
\subsection{Transverse momentum distributions in the Gaussian Ansatz}
\label{ssect:TMDGA}
To solve the convolutions $\mathcal{C}$, one needs to know the transverse-momentum dependence of the TMDs, i.e. the dependence on $\kt$ of the PDFs and that on $\pperp$ of the FFs. Usually, such dependence is factorized and the TMDs are written as the product of a collinear term, dependent on $x$ (or $z$) and $\Qsq$, and a function describing the dependence on the transverse momentum. The range of validity of such factorized approach is presently the subject of intense discussions, whose description goes beyond the scope of this work. Here the factorization is assumed to hold true. In this context, the Gaussian Ansatz, in which the transverse-momentum-dependent parts of the TMDs are assumed to be Gaussian distributions, is the most commonly used. A power-law approximation is also often considered: it has been used, e.g., in the description of the $e^+e^-$ annihilation data from TASSO \cite{Boglione:2017jlh}.

In the Gaussian approximation, that we adopt in this work, the unpolarized PDF and FF can be written as:

\begin{equation}
	f_1^q(x,\kt,\Qsq) = f_1^q(x,\Qsq)\frac{e^{-\ktsq/\Faktsq}}{\pi\Faktsq}
\end{equation}

\begin{equation}
	D_1^{h/q}(z,\pperp,\Qsq) = D_1^{h/q}(z,\Qsq)\frac{e^{-\pperpsq/\Fapperpsq}}{\pi\Fapperpsq}
\end{equation}
where the subscripts $q$ and $h/q$ indicate the possible flavor-dependence of the transverse momenta. Just as an example, it is interesting to derive the expression for the $F_{UU}$ structure function. From the definitions given before, it follows that: 
\begin{equation}
\begin{split}
   	F_{UU} & = \sum_{q} e_q^2 xf_1^q(x,\Qsq) D_1^{h/q}(z,\Qsq) \int \diff^2\vkt \diff^2\vpperp \delta^2 \op\vPt-z\vkt-\vpperp\cp \frac{e^{-\ktsq/\Faktsq}}{\pi\Faktsq}\frac{e^{-\pperpsq/\Fapperpsq}}{\pi\Fapperpsq} \\
	 & =  \sum_{q} e_q^2 xf_1^q(x,\Qsq) D_1^{h/q}(z,\Qsq) \frac{1}{\pi^2\Faktsq\Fapperpsq} \underbrace{\int \diff^2\vkt e^{-\left(\frac{\ktsq}{\Faktsq}+\frac{(\vPt-z\vkt)^2}{\Fapperpsq}\right)}}_{I_0}.
\end{split}
\end{equation}
Let's now focus on the integral $I_0$, omitting all flavor indices for the sake of clarity. By expanding the square and rearranging the terms, and inserting the identity $\aPtsq = z^2\aktsq+\apperpsq$, the exponent can be rewritten as:
\begin{equation}
\begin{split}
   \frac{\ktsq}{\aktsq}+\frac{\op\vPt-z\vkt\cp^2}{\apperpsq} & =  \frac{\ktsq}{\aktsq} + \frac{\Ptsq}{\apperpsq}-\frac{2z\vPt\cdot\vkt}{\apperpsq} + \frac{z^2\ktsq}{\apperpsq} \\
   & = \left(\frac{1}{\aktsq}+\frac{z^2}{\apperpsq}\right)\ktsq + \frac{\Ptsq}{\apperpsq} - \frac{2z\vPt\cdot\vkt}{\apperpsq} \\
   & = \frac{\aPtsq}{\aktsq\apperpsq}\left(\ktsq + \frac{\aktsq\Ptsq}{\aPtsq} - \frac{2z\aktsq\vPt\cdot\vkt}{\aPtsq} \right) \\
   & = \frac{\aPtsq}{\aktsq\apperpsq}\left(\vkt-\frac{z\aktsq\vPt}{\aPtsq}\right)^2 + \frac{\Ptsq}{\aPtsq}.
\end{split}
\end{equation}
The integral $I_0$ can then be solved to give: 
\begin{equation}
I_0  = e^{-\frac{\Ptsq}{\aPtsq}} \int \diff^2\vkt e^{-\frac{\aPtsq}{\aktsq\apperpsq}\left(\vkt-\frac{z\aktsq\vPt}{\aPtsq}\right)^2} = e^{-\frac{\Ptsq}{\aPtsq}} \frac{\pi\aktsq\apperpsq}{\aPtsq},
\end{equation}
so that the flavor-dependent structure function $F_{UU}$ finally reads:
\begin{equation}
\begin{split}
   F_{UU} & = \sum_{q}e_q^2 xf_1^q(x,\Qsq) D_1^{h/q}(z,\Qsq) \frac{e^{-\Ptsq/\FaPtsq}}{\pi\FaPtsq}.
\end{split}
\end{equation}
Neglecting instead the possible flavor dependence of $\aktsq$ and $\aPtsq$, one gets:
\begin{equation}
   F_{UU}  = \sum_{q}e_q^2 xf_1^q(x,\Qsq) D_1^{h/q}(z,\Qsq) \cdot \frac{e^{-\Ptsq/\aPtsq}}{\pi\aPtsq}
\end{equation}
The structure function at small $\Pt$ is thus predicted to have an exponential dependence on $\Ptsq$, with an \textit{inverse slope} equal to $\aPtsq = z^2\aktsq + \apperpsq$. The SIDIS cross-section, integrated over the hadron azimuthal angle, can thus be written as:

\begin{equation}
\begin{split}
    \frac{\diff^4\sigma}{\diff x \diff \Qsq \diff z \diff \Ptsq} & = \frac{4\pi^2 \alpha^2}{x^3y^2s^2}\frac{y^2}{2\op 1-\eps\cp}\op 1+\frac{\gamma^2}{2x}\cp \cdot F_{UU} \\
    &  = \frac{4\pi^2 \alpha^2}{x^3y^2s^2}\frac{y^2}{2\op 1-\eps\cp}\op 1+\frac{\gamma^2}{2x}\cp \cdot \sum_{q}e_q^2 xf_1^q(x,\Qsq) D_1^{h/q}(z,\Qsq) \cdot \frac{e^{-\Ptsq/\aPtsq}}{\pi\aPtsq} 
\end{split}
\end{equation}
and its measurement as a function of $\Ptsq$ allows getting information on $\aPtsq$. Usually however it is easier to measure the hadron multiplicities, namely:
\begin{equation}
\begin{split}
    \frac{\diff^2 M^{h}}{\diff z \diff \Ptsq} & = \left. \op \frac{\diff^4\sigma}{\diff x \diff \Qsq \diff z \diff \Ptsq} \cp \right/ \op \frac{\diff^2\sigma^{DIS}}{\diff x \diff \Qsq} \cp \\
    & = N\frac{e^{-\Ptsq/\aPtsq}}{\pi\aPtsq}.
\end{split}
\label{eq:mult}
\end{equation}
where $N$, which corresponds to the transverse-momentum-independent part of the ratio of SIDIS and DIS cross-sections, can be calculated from known PDFs and FFs\footnote{Note that this LO expression is correct at given values of $x$, $\Qsq$ and $z$. When integrating over the experimental bin widths, correction factors could be needed. Also, other phenomena, like radiative effects, could affect differently the DIS and the SIDIS cross-sections, requiring again correction factors depending on the kinematic region.}.
The measurement of the multiplicities thus allows extracting $\aPtsq$. From Eq.~\ref{eq:mult} it is also clear that information on $\aPtsq$ can be obtained even by measuring the shape of the $\Ptsq$-distributions with arbitrary normalization, as done in this work (Ch.~\ref{Chapter4_PT-distributions}).

\bigskip
\subsection{Azimuthal asymmetries in the Gaussian Ansatz}
\label{sect:ch1_aa_definition}
The unpolarized SIDIS cross-section can be written as:
\begin{equation}
     \frac{\diff^5\sigma}{\diff x \diff \Qsq \diff z \diff \fih \diff \Ptsq} \propto \op 1 + \eps_1 A_{UU}^{\cos\fih}\cos\fih + \eps_2 A_{UU}^{\cos2\fih}\cos2\fih  + \lambda \eps_3 A_{LU}^{\sin\fih}\sin\fih \cp
\end{equation}
where the quantities $A_{XY}^{f(\fih)} = F_{XY}^{f(\fih)}/F_{UU}$, which can be measured in SIDIS experiments, are here referred to as \textit{azimuthal asymmetries} and give independent access to the intrinsic transverse momentum $\aktsq$ and to the Boer-Mulders function $\bom$. They have also been measured in this work (Ch.~\ref{Chapter5_Azimuthal_asymmetries}).

With a calculation similar to the one presented in Sect.~\ref{ssect:TMDGA}, one can derive the expression for the Cahn contribution to the $F_{UU}^{\cos\fih}$ structure function:

\begin{equation}
\begin{split}
   F_{UU|Cahn}^{\cos\fih} & = -\frac{2}{Q}\sum_{q} e_q^2 xf_1^q(x,\Qsq) D_1^{h/q}(z,\Qsq) \frac{e^{-\Ptsq/\FaPtsq}}{\pi^2\Faktsq\Fapperpsq}\frac{\pi z\Pt\Faktsq^2\Fapperpsq}{\FaPtsq^2}\\
   & = -\frac{2z\Pt}{Q}\sum_{q} e_q^2 xf_1^q(x,\Qsq) D_1^{h/q}(z,\Qsq) \frac{e^{-\Ptsq/\FaPtsq}}{\pi\FaPtsq}\frac{\Faktsq}{\FaPtsq}.\\
\end{split}
\end{equation}
which reduces, assuming no flavor-dependence, to:
\begin{equation}
   F_{UU|Cahn}^{\cos\fih} = -\frac{2z\Pt\aktsq}{Q\aPtsq} \op \sum_{q} e_q^2 xf_1^q(x,\Qsq) D_1^{h/q}(z,\Qsq) \frac{e^{-\Ptsq/\aPtsq}}{\pi\aPtsq}\cp = -\frac{2z\Pt\aktsq}{Q\aPtsq} F_{UU}
\end{equation}
so that the \textit{Cahn asymmetry} reads:
\begin{equation}
   A_{UU|Cahn}^{\cos\fih} = \frac{F_{UU|Cahn}^{\cos\fih}}{F_{UU}} = -\frac{2z\Pt\aktsq}{Q\aPtsq}.
\end{equation}

The full calculations for $F_{UU|Cahn}^{\cos\fih}$, as well as for the other structure functions appearing in the unpolarized SIDIS cross-section, are given in Appendix~\ref{AppendixA}. There, it is shown that the Cahn contribution to $F_{UU}^{\cos2\fih}$ and the Boer-Mulders contributions to $F_{UU}^{\cos\fih}$ and $F_{UU}^{\cos2\fih}$ can be written as:

\begin{equation}
   F_{UU|Cahn}^{\cos2\fih} = \frac{2z^2\Ptsq}{Q^2}\sum_{q} e_q^2 xf_1^q(x,\Qsq) D_1^{h/q}(z,\Qsq) \frac{e^{-\Ptsq/\FaPtsq}}{\pi\FaPtsq} \frac{\Faktsq^2}{\FaPtsq^2},
\end{equation}

\begin{equation}
\begin{split}
   F_{UU|BM}^{\cos\fih} & = -\frac{2}{zQMM_h} \sum_{q} e_q^2x\bomq(x,\Qsq) \collq(z,\Qsq) \frac{e^{-\frac{\Ptsq}{\FaPtsq}}}{\pi \FaPtsq} \cdot \\
   & ~~~\cdot \frac{\Faktsq\Fapperpsq\Pt}{\FaPtsq^3}\op\Fapperpsq\FaPtsq + z^2\Faktsq\op\Ptsq-\FaPtsq\cp\cp
\end{split}
\end{equation}

\begin{equation}
   F_{UU|BM}^{\cos2\fih} = \frac{1}{zMM_h}\sum_{q} e_q^2 x\bomq(x,\Qsq) \collq(z,\Qsq)
   \frac{e^{-\frac{\Ptsq}{\FaPtsq}}}{\pi\FaPtsq}\frac{ z \Ptsq \Faktsq\Fapperpsq}{\FaPtsq^2}
\end{equation}
which respectively reduce, assuming no flavor-dependence, to the asymmetries:
\begin{equation}
\begin{split}
   A_{UU|Cahn}^{\cos2\fih} & = \frac{F_{UU|Cahn}^{\cos2\fih}}{F_{UU}} = \frac{2z^2\aktsq^2\Ptsq}{Q^2\aPtsq^2}. 
\end{split}
\end{equation}

\begin{equation}
\begin{split}
   A_{UU|BM}^{\cos\fih} = \frac{F_{UU|BM}^{\cos\fih}}{F_{UU}} & = -\frac{2}{zQMM_h} \frac{\aktsq\apperpsq\Pt}{\aPtsq^3}\op\apperpsq\aPtsq + z^2\aktsq\op\Ptsq-\aPtsq\cp\cp \cdot \\
   & ~~~\cdot \frac{\sum_{q} e_q^2x\bomq(x,\Qsq) \collq(z,\Qsq)}{\sum_{q} e_q^2 xf_1^q(x,\Qsq) D_1^{h/q}(z,\Qsq)} 
\end{split}
\end{equation}

\begin{equation}
   A_{UU|BM}^{\cos2\fih}  = \frac{F_{UU|BM}^{\cos2\fih}}{F_{UU}}= \frac{ \Ptsq \aktsq\apperpsq}{MM_h\aPtsq^2}\frac{\sum_{q} e_q^2 x\bomq(x,\Qsq) \collq(z,\Qsq)}{\sum_{q} e_q^2 xf_1^q(x,\Qsq) D_1^{h/q}(z,\Qsq)} 
\end{equation}

\section{Existing measurements and their interpretation}
Several measurements of the $\Ptsq$-dependent multiplicities and of the azimuthal asymmetries exist. They are summarized in the following, together with a short review of the corresponding interpretation work.

\subsection{$\Ptsq$-dependent distributions}
\begin{figure}
\captionsetup{width=\textwidth}
    \centering
    \includegraphics[width=\textwidth]{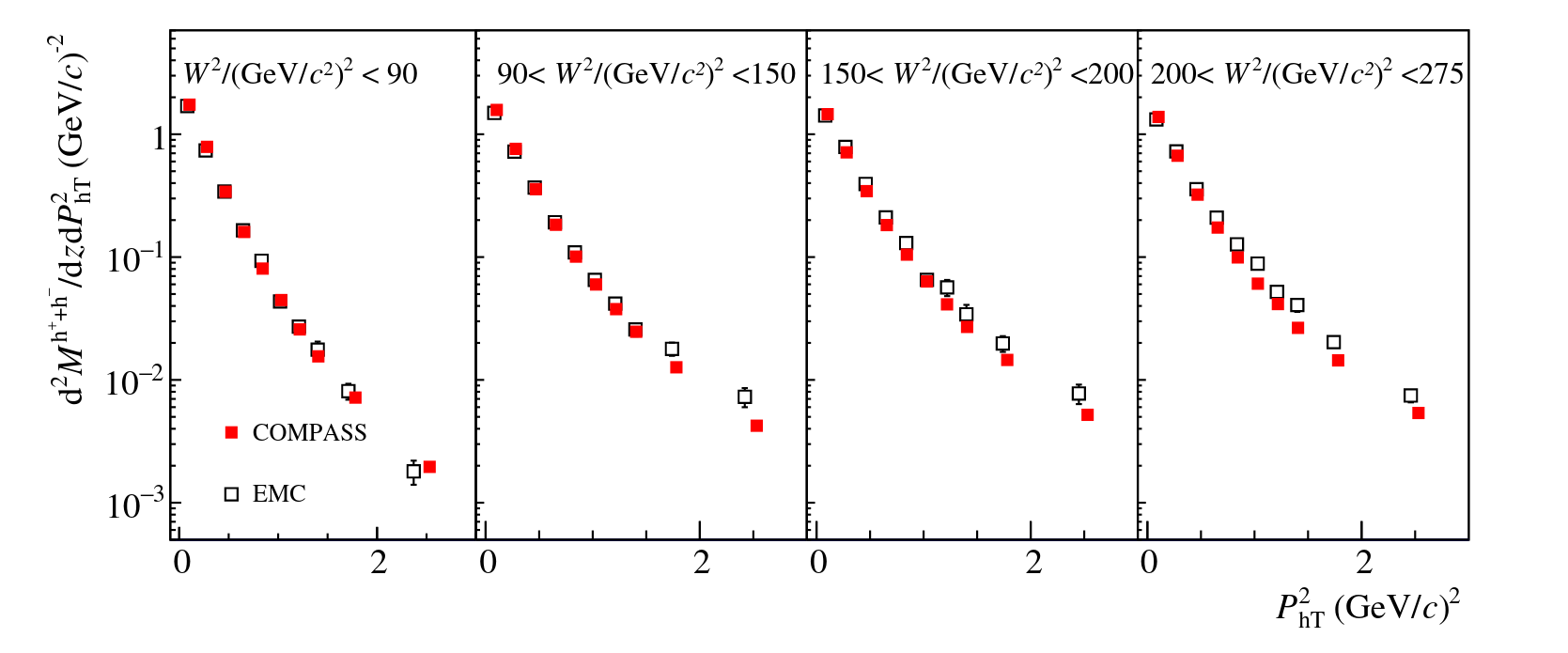}
    \caption{Charged hadron multiplicities from COMPASS (closed points) compared to the EMC results (open points), shown in four bins of $W^2$ \cite{COMPASS:2017mvk}.}
    \label{fig:comp_compass_emc}
\end{figure}

The first measurement of the $\Ptsq$-distributions, normalized to the number of DIS events ($\Ptsq$-multiplicities), was performed by the EMC Collaboration at CERN \cite{Ashman:1991cj}, using muon beams with energies between 100 and 280~GeV scattering off proton and deuteron targets. More recently, these measurements have been performed by the HERMES experiment at DESY \cite{Airapetian:2012ki} and by the COMPASS experiment at CERN \cite{COMPASS:2013bfs,COMPASS:2017mvk}. HERMES used an electron (or positron) beam of 27.6~GeV/$c$ and a proton or deuteron target and measured the $\Pt$-multiplicities for identified $\pi^\pm$ and $K^\pm$ as a function of $\Pt$, $x$ and $\Qsq$ in bins of $z$, with an average value of $\Qsq$ ranging from 1 to 10~(GeV/$c$)$^2$. COMPASS measured the $\Ptsq$-multiplicities using a muon beam of 160~GeV/$c$, thus in a kinematic range complementary to HERMES, and a deuteron target. Lower energy measurements have been performed at Jefferson Lab Hall-C (E00-108) \cite{Asaturyan:2011mq} using a 5.5 GeV/$c$ electron beam scattering off proton and deuteron targets and more recently at CLAS \cite{Osipenko:2008aa}. Some of the results are compared in Ref.~\cite{COMPASS:2017mvk}. Just as an example, the comparison of the COMPASS and EMC results is shown also here (Fig.~\ref{fig:comp_compass_emc}) in four $W$ bins and summing over positive and negative hadrons multiplicities in the lowest $z$ bin ($0.20<z<0.40$): a good agreement between the two experiments can be seen, in particular in the lowest two $W$ bins; at higher $W$, the ratio of the EMC and COMPASS multiplicities is larger than one for $P_T>1$~GeV/$c$. 

In Ref.~\cite{COMPASS:2017mvk}, the COMPASS and HERMES multiplicities are found to agree in size at small $\Ptsq$ and $z<0.60$; elsewhere, and likely due to the different kinematic range, the agreement is less good, with a smaller $\aPtsq$ in the HERMES case. Similarly, a difference both in size and shape is observed in the comparison of the COMPASS multiplicities and the E00-108 results, the latter showing a steeper decrease in $\Ptsq$. Once again, the differences could be motivated by the different kinematic range, particularly the $\Qsq$ range, lower in the E00-108 case.

\bigskip

A fit of the multiplicities of charged hadrons produced in SIDIS measured at HERMES and COMPASS has been performed by the Torino Group \cite{Anselmino:2013lza}, in the context of the Gaussian Ansatz, obtaining:
\begin{equation}
    \aktsq = 0.57\pm0.08~\mathrm{(GeV/} c \mathrm{)} ^2, \hspace{2cm} \apperpsq = 0.12\pm0.01~\mathrm{(GeV/} c \mathrm{)} ^2 
\end{equation}
from the HERMES data \cite{Airapetian:2012ki} and 
\begin{equation}
    \aktsq = 0.61\pm0.20~\mathrm{(GeV/} c \mathrm{)} ^2, \hspace{2cm} \apperpsq = 0.19\pm0.02~\mathrm{(GeV/} c \mathrm{)} ^2 
\end{equation}
from the COMPASS data \cite{COMPASS:2013bfs}. No $\Qsq$-dependence has been assumed, in addition to the DGLAP evolution typically considered for the collinear PDFs and FFs. However, no clear indication for a $\Qsq$ dependence of $\aPtsq$ was found, for the HERMES data, using a fit expression based on the Collins-Soper-Sterman resummation scheme \cite{Collins:1984kg}, according to which one would expect:

\begin{equation}
    \langle P_T \rangle(Q^2) = \langle P_T \rangle(Q_0^2) + a z^2 \ln{\frac{Q^2}{Q_0^2}}.
\end{equation}

The values of the HERMES fit have been compared with the shape of the cross-section as a function of $\Ptsq$ as measured in JLab-Hall C \cite{Asaturyan:2011mq}, with good agreement at small $\Ptsq$, and with the CLAS results \cite{Osipenko:2008aa} with a good overall agreement. The parameters extracted from the fit of the COMPASS data have been found adequate to describe the EMC multiplicities as well.

\bigskip

The identified-hadron multiplicities on proton and deuteron target measured in HERMES have been analyzed by the Pavia Group \cite{Signori:2013mda} taking into account a possible flavor dependence of $\aktsq$ and $\apperpsq$. In particular, the multiplicities $M^h$ have been modeled as:

\begin{equation}
    M^h(x,z,\Ptsq) = \frac{\pi}{\sum_q e_q^2 f_1^q(x)}\sum_q e_q^2 f_1^q(x)D_1^{h/q}(z) \frac{e^-\Ptsq/(z^2\Faktsq + \Fapperpsq)}{\pi(z^2\Faktsq + \Fapperpsq) }
\end{equation}
with the average transverse momentum squared $\Faktsq$ assumed to depend on $x$ according to the form:
\begin{equation}
    \Faktsq(x) = \Faktsq(x_0) \frac{(1-x)^\alpha x^\sigma}{(1-x_0)^\alpha x_0^\sigma}
\label{eq:signori1}
\end{equation}
where $\alpha$ and $\sigma$ have been taken as flavor-independent, while three options were allowed for $\Faktsq$: $\aktsq_{u_v}$, $\aktsq_{d_v}$ and $\aktsq_{sea}$. As for $\Fapperpsq$, the considered functional form was:

\begin{equation}
    \Fapperpsq(z) = \Fapperpsq(z_0) \frac{(z^\beta + \delta)(1-z)^\gamma}{(z_0^\beta + \delta)(1-z_0)^\gamma}
\label{eq:signori2}
\end{equation}
where the $\beta$, $\delta$ and $\gamma$ parameters have been assumed to be flavor-independent. On the other hand, a distinction has been made among favored and unfavored fragmentation, also keeping distinct the fragmentation into Kaons, thus writing the hadron multiplicities in terms of $\langle p_{\perp,fav}^2 \rangle$, $\langle p_{\perp,unf}^2 \rangle$, $\langle p_{\perp,uK}^2 \rangle$ and $\langle p_{\perp,sK}^2 \rangle$.

The fit of the multiplicities returned the following values for $\aktsq$: $\aktsq_{u_v}=0.36\pm0.14$~(GeV/$c$)$^2$, $\aktsq_{d_v}=0.30\pm0.17$~(GeV/$c$)$^2$ and $\aktsq_{sea}=0.41\pm0.16$~(GeV/$c$)$^2$. Well compatible results were found for $\aktsq_{u_v}$ and $\aktsq_{d_v}$ selecting the data at $\Qsq>1.6$~((GeV/$c$)$^2$ or considering only pions, while $\aktsq_{sea}$ was observed to decrease. The flavor-independent fit returned: $\aktsq=0.30\pm0.10$~(GeV/$c$)$^2$. 

As for $\Fapperpsq$, it was found that: $\langle p_{\perp,fav}^2 \rangle = 0.15 \pm 0.04$~(GeV/$c$)$^2$, while $\langle p_{\perp,unf}^2 \rangle \approx \langle p_{\perp,sK}^2 \rangle \approx \langle p_{\perp,uK}^2 \rangle \approx 0.19 \pm 0.05$~(GeV/$c$)$^2$, with a flavor-independent result equal to: $\langle p_{\perp}^2 \rangle = 0.18 \pm 0.03$~(GeV/$c$)$^2$.

\bigskip

Recently, several phenomenological analyses of the transverse-momentum-distributions in SIDIS and Drell-Yan have been carried out using the most up-to-date TMD evolution frameworks. In parallel, a review has been performed of the kinematic ranges in which the TMD factorization is expected to hold (see e.g. Ref.~\cite{Boglione:2016bph}). 

\begin{figure}
\captionsetup{width=\textwidth}
    \centering
    \includegraphics[width=0.85\textwidth]{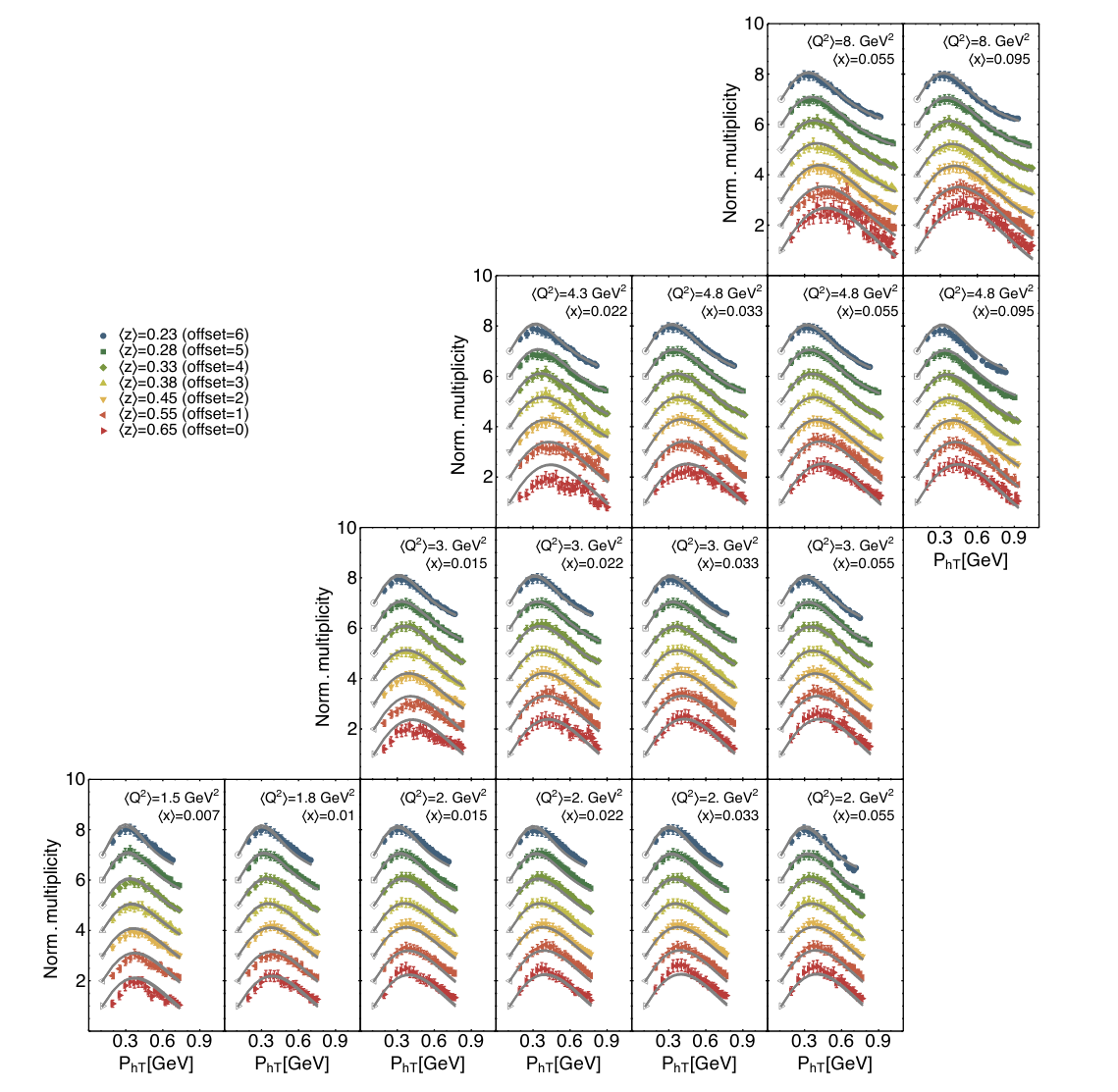}
    \caption{COMPASS unpolarized $\Pt$-distributions for positive hadrons produced in SIDIS off a deuteron target, in bins of $x$, $z$ and $\Qsq$, normalized to the first bin in $\Pt$ for each $z$ bin \cite{Bacchetta:2017gcc}.}
    \label{fig:bacchetta}
\end{figure}

\begin{figure}
\captionsetup{width=\textwidth}
    \centering
    \includegraphics[width=0.85\textwidth]{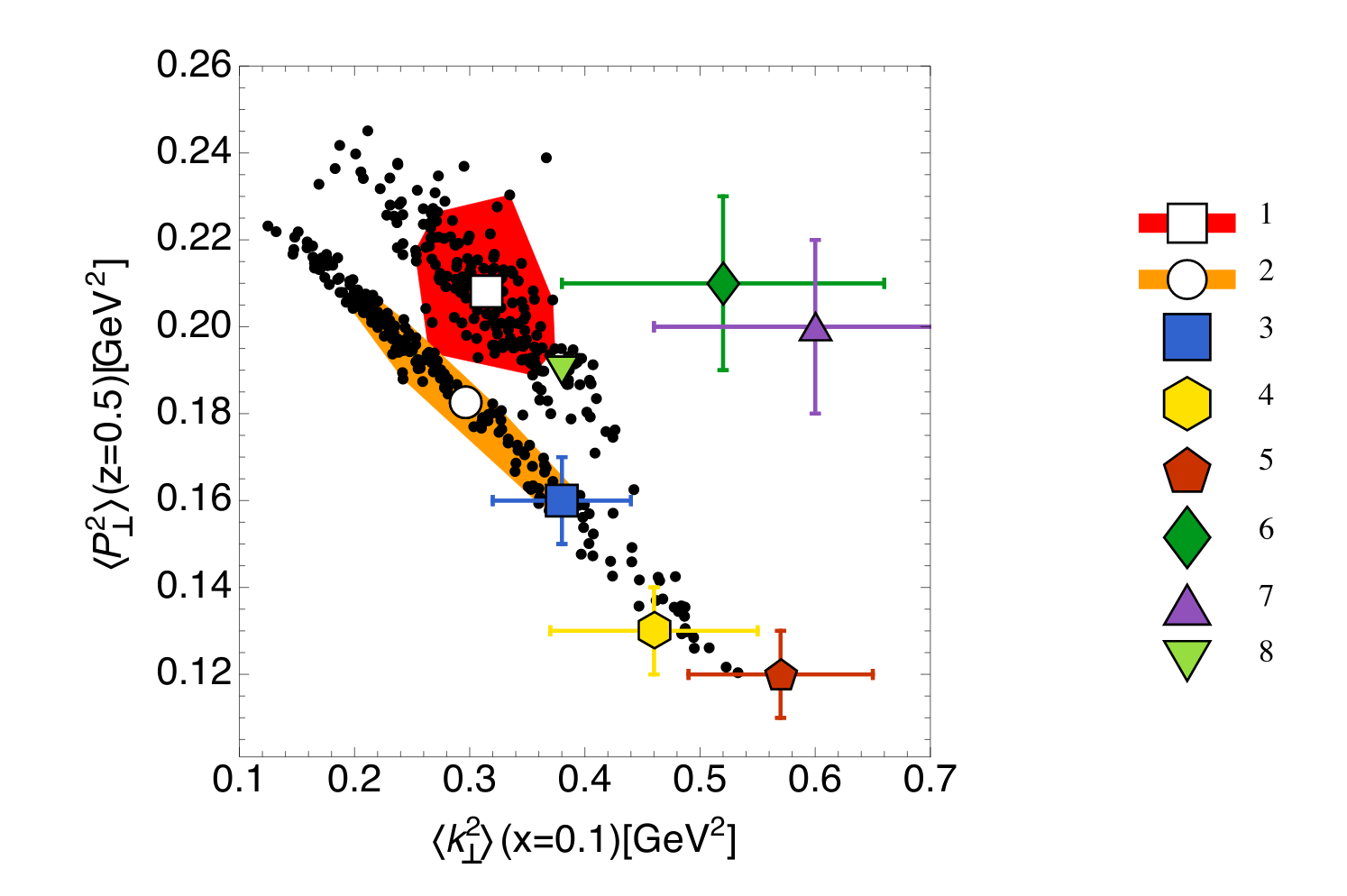}
    \caption{Correlation between $\aktsq$ (horizontal axis) and $\apperpsq$ (vertical axis) as from different phenomenological extractions. (1): mean values obtained from the replica method; the red region indicates the 68\% C.L., the black dots the results of the replicas. Similarly for (2), the orange region and the black dots around it, obtained from Ref.~\cite{Signori:2013mda}. (3): results from Ref.~\cite{Schweitzer:2010tt}. (4): results from Ref.~\cite{Anselmino:2013lza} (HERMES data). (5): same, at high $z$. (6): same, for normalized COMPASS data. (7): same, at high $z$. (8): results from Ref.~\cite{Echevarria:2014xaa}.}
    \label{fig:bacchetta2}
\end{figure}

In Ref.~\cite{Bacchetta:2017gcc} the first simultaneous fit of SIDIS data (including COMPASS), low-energy Drell-Yan and $Z$-boson production data at a next-to-leading order (NLO) accuracy has been presented. Allowing for a free normalization constant, the description of the COMPASS $\Ptsq$-distributions \cite{COMPASS:2013bfs} is very good, as can be seen in Fig.~\ref{fig:bacchetta} for the case of positive hadrons. In this work, the average transverse momenta $\aktsq$ and $\apperpsq$ were assumed to have a dependence on $x$ and $z$ similar to the ones of Eq.~\ref{eq:signori1} and Eq.~\ref{eq:signori2}. Using the replica method, a set of values was obtained for $\aktsq$ and $\apperpsq$, shown in Fig.~\ref{fig:bacchetta2} together with the estimates from other phenomenological analyses. Naturally, the distributions in the transverse momentum are not sufficient to disentangle the two transverse momenta, which appear strongly anti-correlated.

\begin{figure}
\captionsetup{width=\textwidth}
    \centering
    \includegraphics[width=0.85\textwidth]{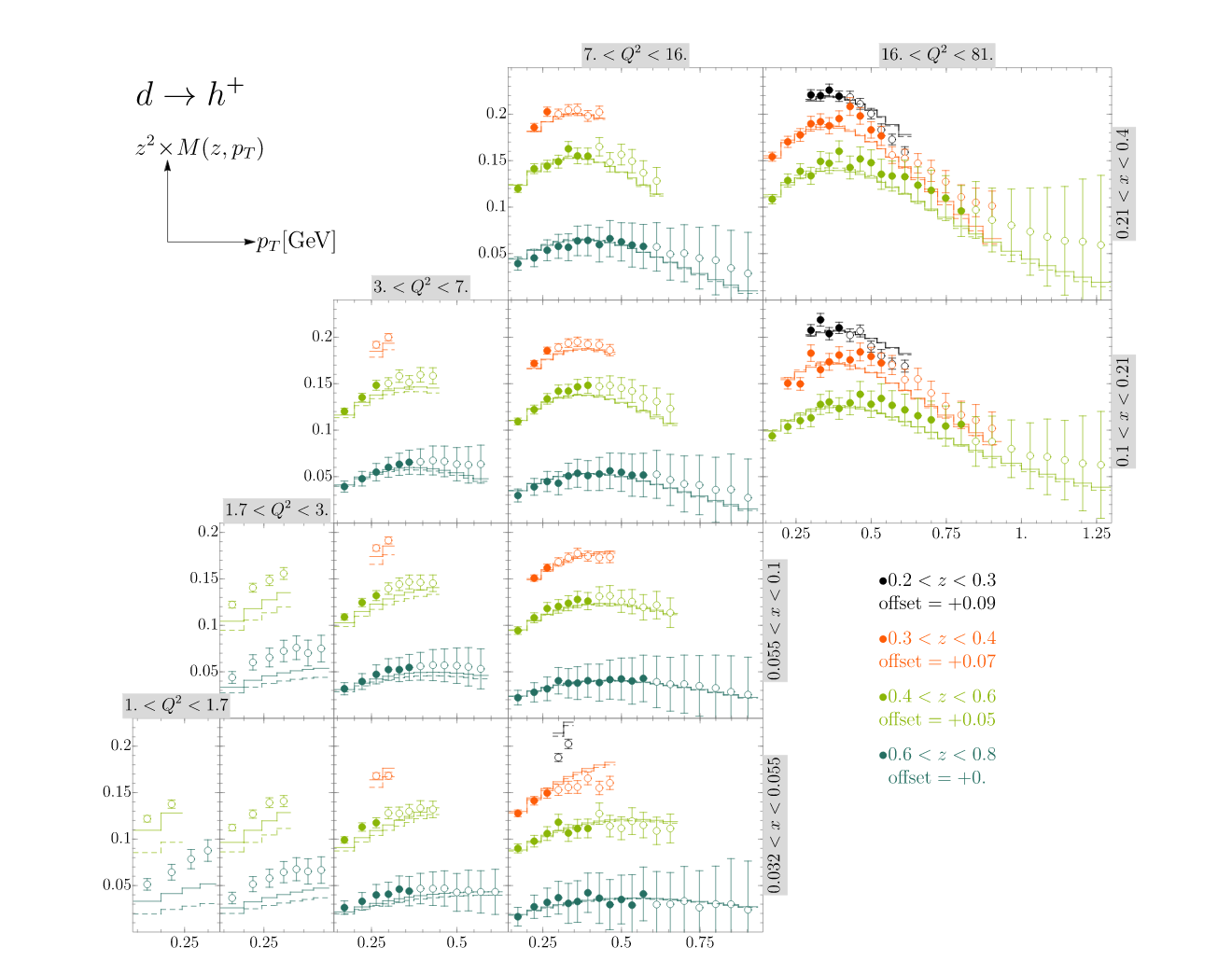}
    \caption{COMPASS unpolarized multiplicities, multiplied by $z^2$ and in bins of $x$, $\Qsq$, $z$ and $\Pt$, for positive hadrons produced in SIDIS off a deuteron target. The closed (open) points have been included (not included) in the phenomenological analysis. Solid and dashed lines correspond to the theory predictions at different approximation levels: NNLO and N3LO \cite{Scimemi:2019cmh}.}
    \label{fig:scimemi}
\end{figure}

\bigskip
Strictly related to the transverse momentum $\Pt$ is the ratio $q_T=P_T/z$, about which a growing interest has risen in recent times. It is often indicated as a good quantity to identify the region of validity of the TMD formalism. For example, in Ref.~\cite{Scimemi:2019cmh}, such region has been selected by requiring $q_T<0.25~Q$, based on the value of the reduced $\chi^2$ obtained from the comparison of the theory predictions and the data points from SIDIS and Drell-Yan \cite{Scimemi:2017etj}. A lively debate is ongoing in the community about the application of such $q_T$ cut, which has been shown to be effective in getting a good global description of SIDIS and Drell-Yan data with no evidence, in the SIDIS case (Fig.~\ref{fig:scimemi}), for normalization issues observed elsewhere \cite{Bacchetta:2017gcc,Gonzalez-Hernandez:2018ipj}. We will come back to this point in Sect.~\ref{sect:ch4_qt}.

\bigskip

To conclude, it is important to underline how useful complementary information on $\apperpsq$ comes from the $e^+e^-$ annihilation experiments. Here, the $\Ptsq$-distributions measured with respect to the thrust axis (chosen as the direction which maximizes the sum of the longitudinal hadron momenta in the center-of-mass) give direct information on the transverse momentum $\pperp$ acquired by the hadrons in the fragmentation process. Such measurements were performed by TASSO  \cite{Althoff:1983ew,Braunschweig:1990yd} and PLUTO \cite{Berger:1983yp} at DESY and by MARK-II at SLAC \cite{Petersen:1987bq}. In particular, the TASSO data were collected at four different center-of-mass energies, from $\sqrt{s}=14$~GeV to $\sqrt{s}=44$~GeV; the resulting $\pperp$ distributions were produced summing over the hadron charge and type and integrating over $z=\frac{2E_h}{\sqrt{s}}$, where $E_h$ is the energy of the hadron in the $e^+e^-$ center-of-mass. The MARK-II data were collected at $\sqrt{s}=29$~GeV, while the PLUTO center-of-mass energy ranged from 7.7~GeV to 27.6~GeV. The TASSO results have been analyzed in Ref.~\cite{Boglione:2017jlh}, both in the context of the Gaussian Ansatz and using a power-law parametrization of the $\pperp$ distributions. The Gaussian parametrization was observed to work well only at small $\Pt$, with no evidence for a $\Qsq$-dependence of $\apperpsq$. Allowing for an arbitrary normalization of the distributions, the Gaussian approach gave $\apperpsq = 0.098 \pm 0.005$~(GeV/$c$)$^2$. The choice of a power-law fit function proved to be more effective in giving a good description of the TASSO data up to $\pperp = 2$~GeV/$c$.

\begin{figure}
\captionsetup{width=\textwidth}
    \centering
    \includegraphics[width=0.85\textwidth]{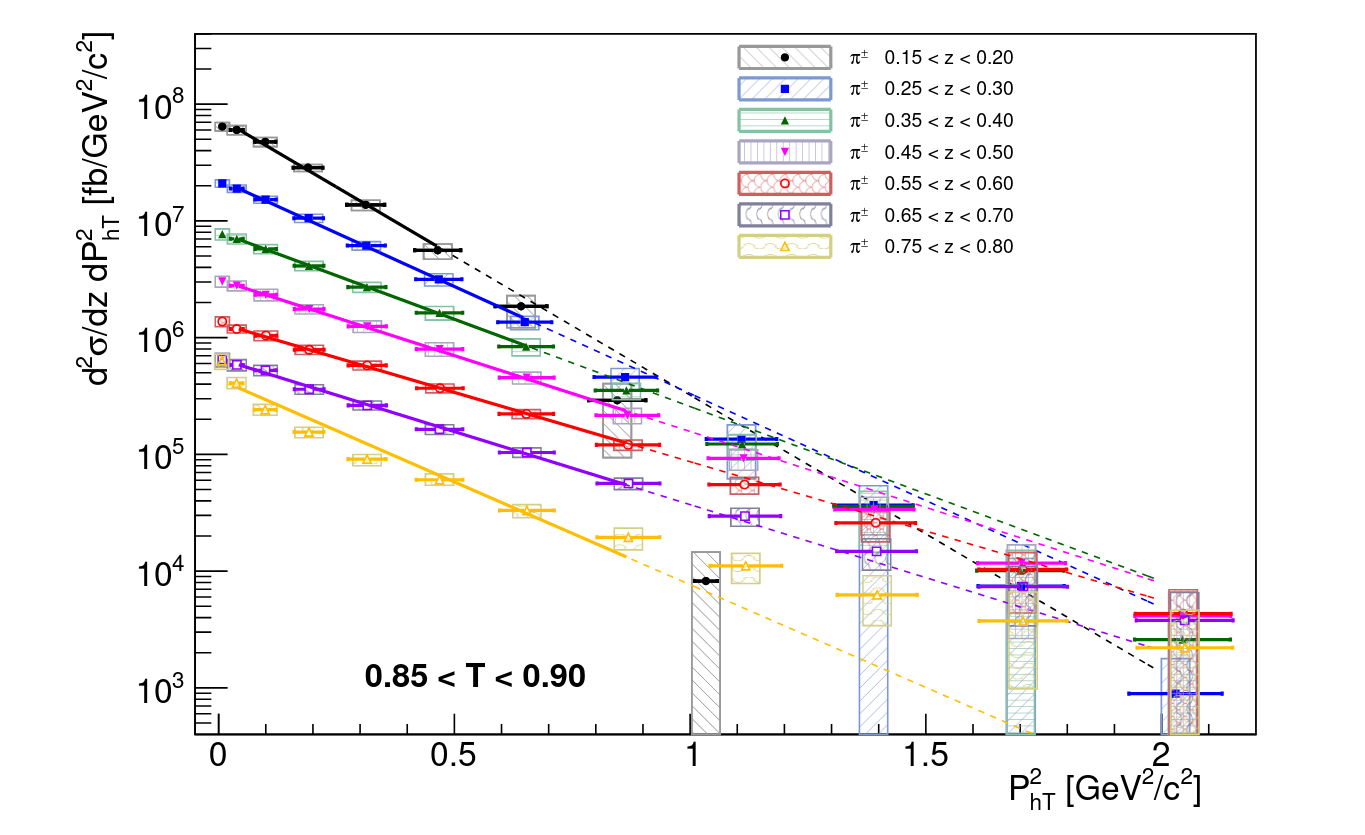}
    \caption{Single charged pion cross-sections as a function of the hadron transverse momentum squared in bins of $z$ at high thrust $T$ ($0.85< T <0.90$). The full lines at low transverse momenta correspond to the Gaussian fits to the data, extended as dotted lines to larger transverse momenta not included in the fit. The error boxes represent the  systematic uncertainties \cite{Seidl:2019jei}.}
    \label{fig:belle_apt2}
\end{figure}

The Belle Collaboration has published a set of transverse-momentum-dependent cross-sections measured at $\sqrt{s} = 10.58$~GeV for identified pions, Kaons and protons in different $z$ and thrust bins \cite{Seidl:2019jei}. These new experimental data, of great importance for the TMD studies, show an exponential trend at low $\Pt$, thus confirming the Gaussian approximation in that range: an example is shown in Fig.~\ref{fig:belle_apt2} for pions at high thrust in bins of $z$. The publication of these data has also stimulated flourishing developments and discussions on the theoretical side. In particular, the authors of Ref.~\cite{Boglione:2020cwn} discuss how to reconcile the (single-hadron) Belle data with the cross-section for the hadron-pair production in $e^+e^-$ annihilation, where the TMD factorization can be applied, indicating that the presence of process-dependent soft factors in the former case prevents a direct comparison of the two channels. The predicted cross-sections, whose calculation has been performed in the context of a novel factorization scheme, are found to be in very good agreement with the Belle experimental results. A good agreement between predictions and experimental results is found also by the authors of Ref.~\cite{Kang:2020yqw}, who propose a TMD factorization and resummation formula for the unpolarized transverse momentum distributions of single hadron production produced in $e^+e^-$ annihilation in different kinematic regions.

\subsection{Azimuthal asymmetries}
The study of the transverse-momentum-dependent structure of the nucleon via the measurement of the unpolarized azimuthal asymmetries started in the European Muon Collaboration (EMC) \cite{Aubert:1983cz}. There, the first measurement of the azimuthal asymmetries was performed considering all charged hadron at $\Qsq>5$~(GeV/$c$)$^2$, in three bins of $W$ and as a function of $z$ and $\Pt$. The analyzed sample consisted of about 100~000 hadrons produced in the Deep Inelastic Scattering of 280 and 120~GeV muons off a liquid hydrogen target.
 
\begin{figure}
\captionsetup{width=\textwidth}
    \centering
    \includegraphics[width=0.45\textwidth]{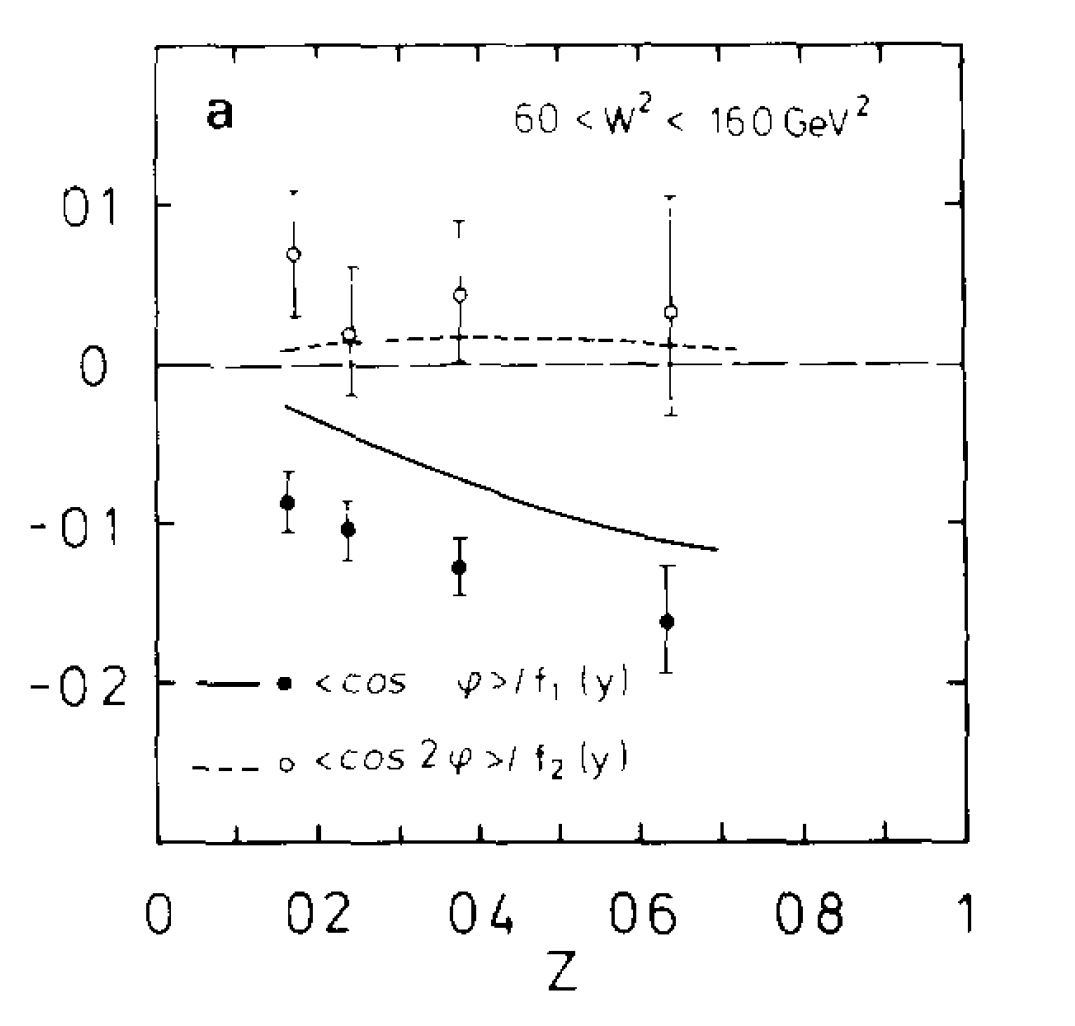}
    \includegraphics[width=0.45\textwidth]{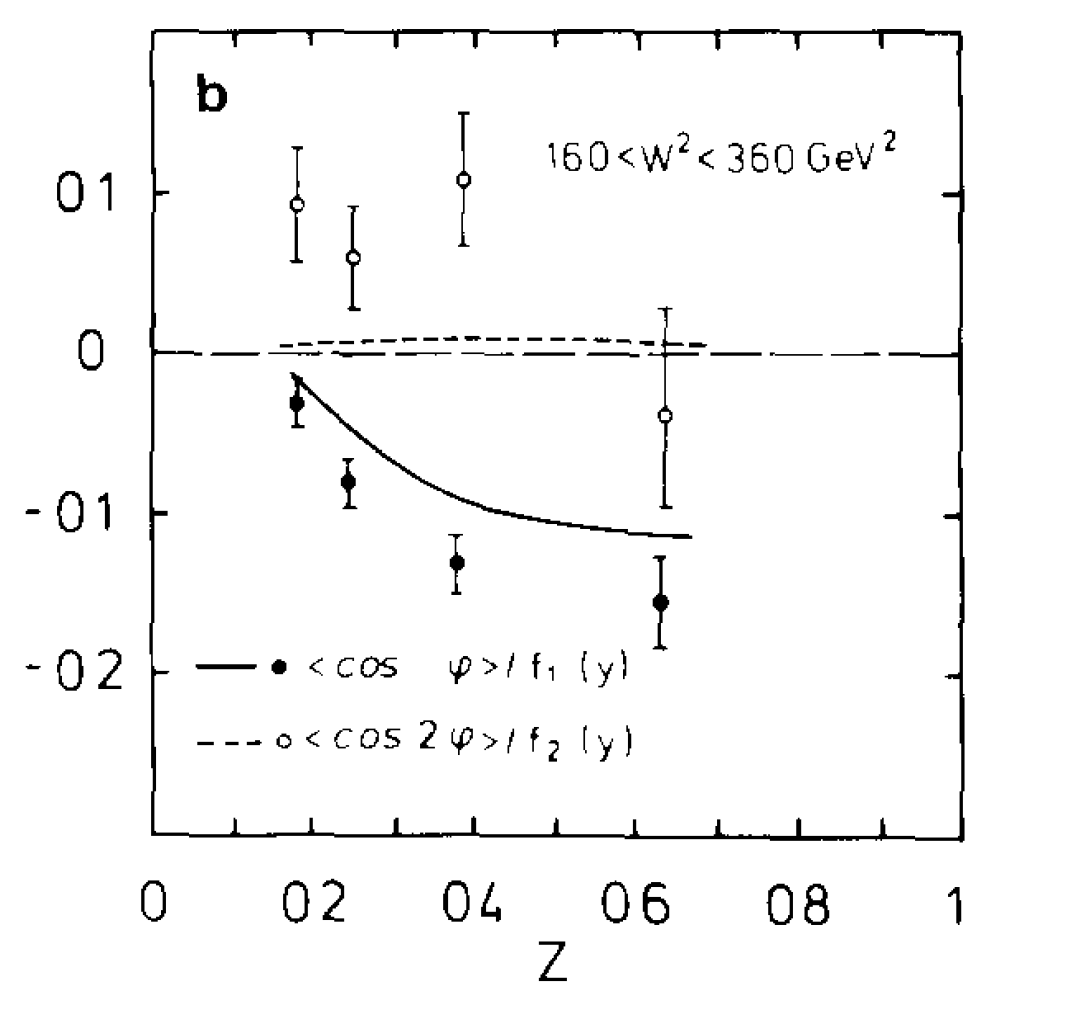}
    \includegraphics[width=0.60\textwidth]{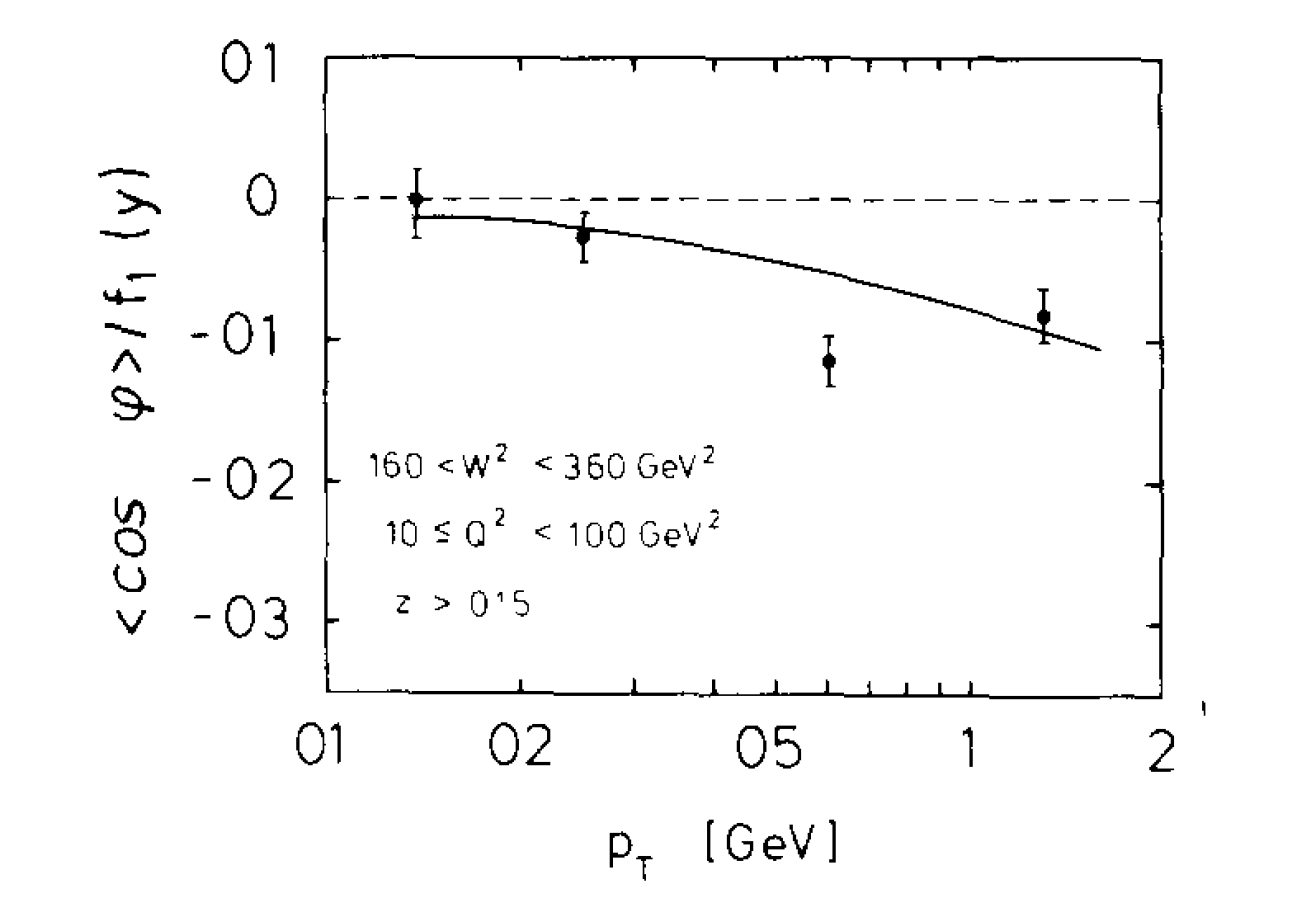}
    \caption{Top plots: $z$-dependence of the $A_{UU}^{\cos\fih}$ (closed points) and $A_{UU}^{\cos2\fih}$ asymmetry (open points) in the first (left) and second $W$ bin (right), as measured by the EMC Collaboration at CERN \cite{Aubert:1983cz}. Bottom plot: $\Pt$-dependence of  $A_{UU}^{\cos\fih}$ in the second $W$ bin. The reference curves are from the model calculations of Ref.~\cite{Konig:1982uk}. }
    \label{fig:emc_aa_1}
\end{figure}

The asymmetry in $\cos\fih$ was found sizable, with a mean value ranging from $\langle A_{UU}^{\cos\fih} \rangle = -0.11\pm0.02$ at $W^2=100$~GeV/$c^2$ to $\langle A_{UU}^{\cos\fih} \rangle = -0.18\pm0.02$ at $W^2=325$~GeV/$c^2$. The $z$-dependence, shown in Fig.~\ref{fig:emc_aa_1} (closed points) after the correction for the kinematic factor, was found to be linear in the first $W$ bin, with a different trend at intermediate $z$ at higher $W$. A decreasing trend in $\Pt$ could also be observed, as expected from the Cahn mechanism. The $A_{UU}^{\cos2\fih}$ asymmetry was found positive and smaller than 0.1 at small $W$, with a weak $W$-dependence. The $A_{LU}^{\sin\fih}$ asymmetry was found compatible with zero. In Fig.~\ref{fig:emc_aa_1}, taken from Ref.~\cite{Aubert:1983cz}, the measured asymmetries have been compared to a model prediction by Konig and Kroll \cite{Konig:1982uk}, done at first order in perturbative QCD assuming  $\aktsq=0.7$~(GeV/$c$)$^2$ and $\apperpsq=0.3$~(GeV/$c$)$^2$.

\bigskip

\begin{figure}
\captionsetup{width=\textwidth}
    \centering
    \includegraphics[width=0.50\textwidth,angle=-0.75]{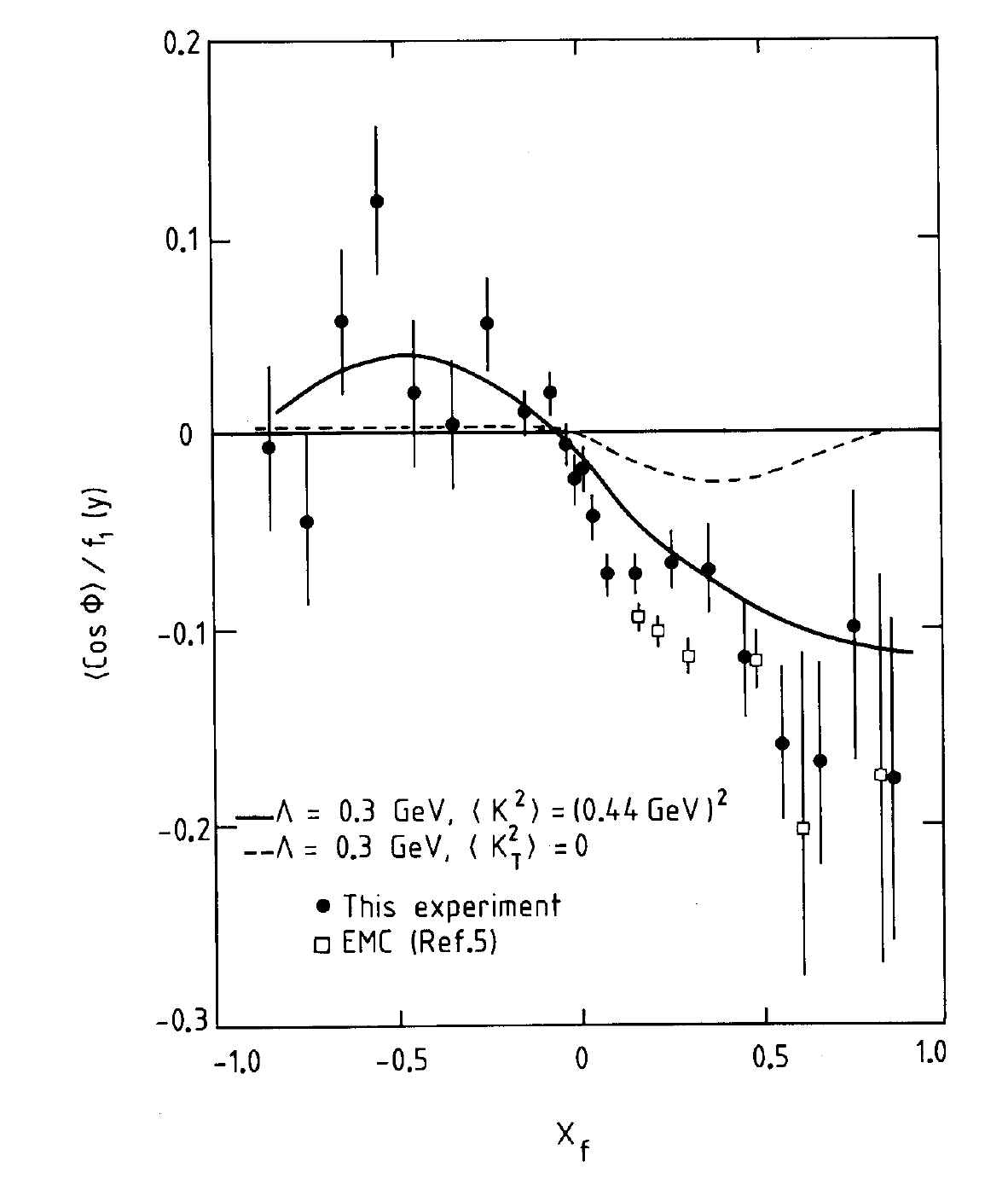}
    \caption{$x_F$-dependence of the $A_{UU}^{\cos\fih}$ (closed points) as measured by the EMC Collaboration at CERN \cite{Arneodo:1986cf}, compared to the previous measurement in the forward region (open points) \cite{Aubert:1983cz}. The reference curves are from the model calculations of Ref.~\cite{Konig:1982uk}. }
    \label{fig:emc_aa_2}
\end{figure}

A few years after the first publication, the EMC Collaboration produced new results for the azimuthal asymmetries using 27~000 hadrons produced in the scattering of a 280~GeV muon beam off a liquid hydrogen target \cite{Arneodo:1986cf}. The azimuthal asymmetries were inspected as a function of the Feynman variable $x_F=2p_L/W$, where $p_L$ is the hadron longitudinal momentum in the virtual photon-nucleon center-of-mass, complementing the previous results with the backward region at $x_F<0$. The $A_{UU}^{\cos\fih}$ was found to change sign, becoming positive, at $x_F\approx 0$: a comparison between the first and second sets of result, corresponding to the forward region ($z>0.2$) and to the full $x_F$ range, is shown in Fig.~\ref{fig:emc_aa_2}. The reference curves were produced assuming a smaller value of the intrinsic transverse momentum: $\aktsq=0.44$~(GeV/$c$)$^2$.

\bigskip
In the early Nineties, the E665 Collaboration at Fermilab \cite{Adams:1993hs} studied the azimuthal asymmetries of hadrons produced in the scattering of a 490~GeV muon beam off a proton or deuteron target. 
At least four hadrons were required to be produced in the DIS event. A negative modulation in $\cos\fih$ was found for the hadrons produced in events with large value of $\Pi = \sum_h |\Pt^h|$, dominated by hard QCD effects (gluon radiation and photon-gluon fusion) while being small at small $\Pi.$

\bigskip

A phenomenological analysis \cite{Anselmino:2005nn} of both azimuthal asymmetries and $\Ptsq$-multiplicities from the EMC experiment and E665 experiments, performed in the Gaussian approximation, suggests the following values for the average transverse momenta:

\begin{equation}
    \aktsq \sim 0.25~\mathrm{(GeV/} c \mathrm{)} ^2, \hspace{2cm} \apperpsq \sim 0.20~\mathrm{(GeV/} c \mathrm{)} ^2.
\end{equation}
The fit of the azimuthal asymmetries has been performed with both the exact kinematics (i.e. including all orders in $\kt/\ Q$) and considering only the twist-3 term, with good agreement between the two.
These results have also been checked to be in agreement with the differential cross-section $\diff \sigma /\ \diff \Ptsq$ measured in EMC \cite{Ashman:1991cj} on proton and deuteron targets. It is to be noted, however, that the data were collected in the two experiments at different energies, different $\Qsq$, $x$ and $z$ and no dependence on these kinematic variables was assumed in the estimation of $\aktsq$ and $\apperpsq$.

\begin{figure}[h!]
    \centering
    \includegraphics[width=0.75\textwidth]{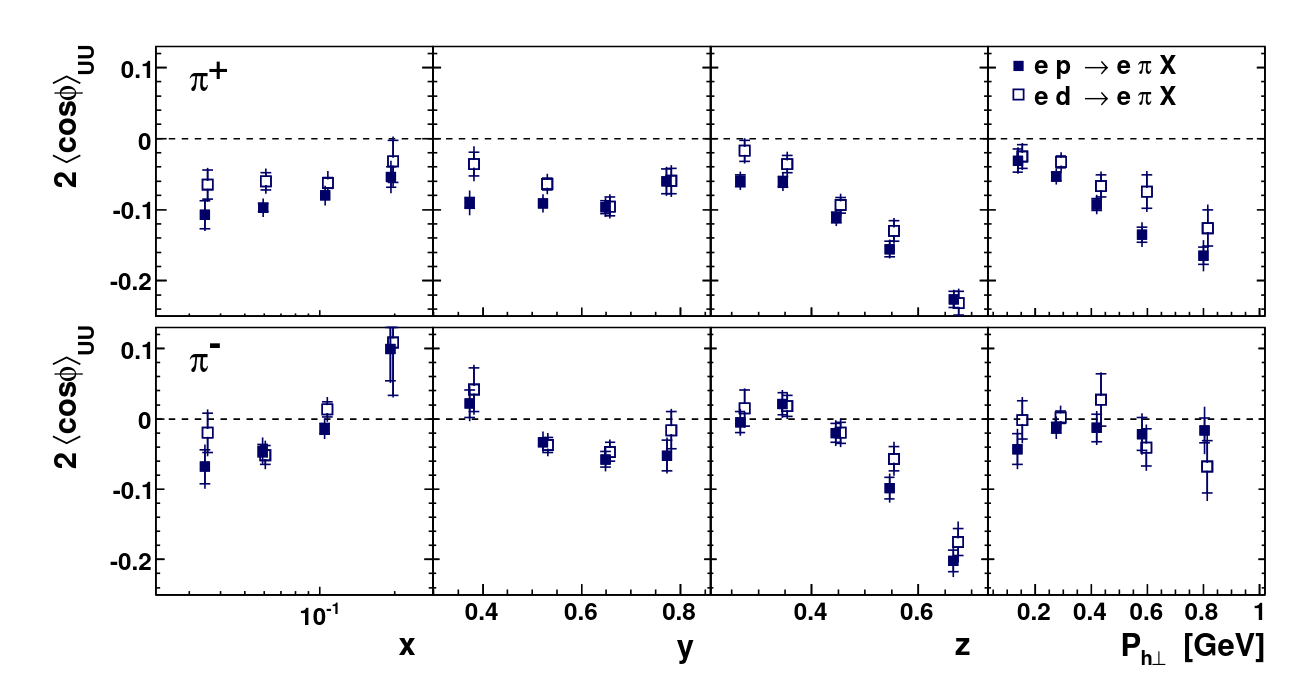}
    \includegraphics[width=0.75\textwidth]{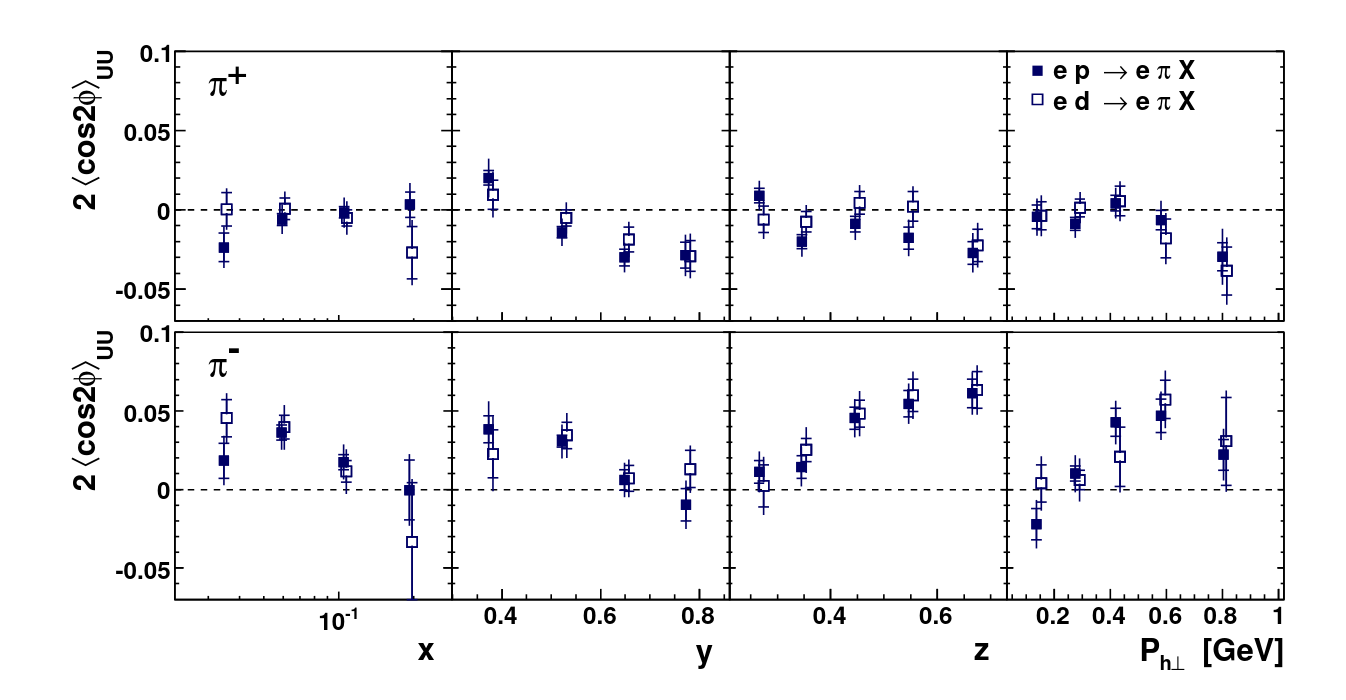}
    \captionsetup{width=\textwidth}
    \caption{Azimuthal asymmetries in $\cos\fih$ (top) and $\cos2\fih$ (bottom) for positive and negative pions, as a function of $x$, $y$, $z$ and $\Pt$, as measured by the HERMES Collaboration \cite{Airapetian:2012yg}.}
    \label{fig:hermes}
\end{figure}

\bigskip
In more recent years, new measurements of the azimuthal asymmetries in unpolarized SIDIS have been performed by HERMES \cite{Airapetian:2012yg}, COMPASS \cite{COMPASS:2014kcy} and CLAS (later CLAS12) \cite{Osipenko:2008aa,Diehl:2021rnj}. Using the data collected from 2000 to 2006 with both deuteron and proton targets, HERMES produced the azimuthal asymmetries of identified hadrons in bins of $x$, $y$, $z$ and $\Pt$. The kinematic region for the pion analysis was selecting by requiring $0.023<x<0.270$, $0.30<y<0.85$, $0.20<z<0.75$ and $0.05~\mathrm{GeV/}c\mathrm{}<\Pt<1.00~\mathrm{GeV/}c\mathrm{}$. As shown in Fig.~\ref{fig:hermes}, an overall agreement was found between the proton and deuteron results. In particular, the $\cos2\fih$ asymmetries was observed to be close to zero for positive pions and clearly positive for negative pions. As for the $\cos\fih$ case, the asymmetries were found large and negative, with a clear linear dependence upon $z$ and $\Pt$, being larger in size for positive pions.

A description of the azimuthal asymmetries of charged hadron measured by HERMES has been proposed in Ref.~\cite{Giordano:2008zzc}, where the transverse momenta are estimated to be:
\begin{equation}
    \aktsq \sim 0.18~\mathrm{(GeV/} c \mathrm{)} ^2, \hspace{2cm} \apperpsq = 0.42~z^{0.37}~(1-z)^{0.54} \sim 0.20~\mathrm{(GeV/} c \mathrm{)} ^2.
\end{equation}
These values, in line with the one derived from EMC and E665, have been used as input for a later extraction, unfortunately not conclusive, of the Boer-Mulders function in \cite{Barone:2009hw} from the $\cos2\fih$ asymmetry measured by HERMES and COMPASS. \\

As said, also the COMPASS Collaboration produced important results for the azimuthal asymmetries in $\cos\fih$, $\cos2\fih$ and $\sin\fih$, measured on a deuteron target \cite{COMPASS:2014kcy}. The results were produced both in a one-dimensional and a three-dimensional approach, i.e. binning the data either as a function of $x$, or $z$ or $\Pt$ (integrating over the other two), or performing a simultaneous binning in the three variables. Clear and strong kinematic dependences were observed, particularly as a function of $z$ and $\Pt$ in the $A_{UU}^{\cos\fih}$ asymmetry. The contribution to the asymmetries originating from the hadrons produced in the decay of diffractively produced vector mesons has been recently estimated to be sizable \cite{COMPASS:2019lcm}. These data will be referred to and compared to the new results, obtained from a proton target, all throughout this Thesis. 

\bigskip

A phenomenological analysis of the COMPASS and HERMES multidimensional data on azimuthal asymmetries and multiplicities has been performed at twist-3, suggesting a small value of $\aktsq$ \cite{Barone:2015ksa}:
\begin{equation}
    \aktsq \simeq 0.03-0.04~\mathrm{(GeV/} c \mathrm{)} ^2,
\end{equation}
mainly driven by the $\cos\fih$ asymmetry, with a marginal role of the Boer-Mulders contribution to this asymmetry. The difference between the $A_{UU}^{\cos\fih}$ asymmetries for positive and negative hadrons indicate a flavor-dependence that should originate from the Boer-Mulders term or from higher-twist terms. As a possible explanation of the difference with the previously-obtained values of $\aktsq$, it was pointed that the twist-3 contribution could be not negligible in the $\cos\fih$ term. 

\bigskip

A measurement of the semi-inclusive electroproduction of positive pions has been performed at CLAS \cite{Osipenko:2008aa} at small $\Qsq$ ($1.4< \Qsq/\mathrm{(GeV/}c\mathrm{)}^2 <5.7$. The precision of the data was not sufficient to obtain information about the contribution of the Boer-Mulders function, while a disagreement was found between the structure function associated to the $\cos\fih$ modulation and the prediction from the Cahn effect. Recently, very interesting results have been produced at CLAS12 for the $A_{LU}^{\sin\fih}$ beam-spin asymmetry of positive pions \cite{Diehl:2021rnj} in the context of a high-precision multidimensional analysis. This asymmetry, already investigated in COMPASS and HERMES, is compatible with zero in the COMPASS kinematics.


\chapter{The COMPASS Experiment} 

\label{Chapter2_COMPASS}

COMPASS (COmmon Muon Proton Apparatus for Structure and Spectroscopy) is a fixed-target experiment located at CERN in the North Area, along the M2 SPS beamline. The experiment, in operation since 2002, was approved in 1997 with a broad scientific program \cite{Baum:1996yv} and it is designed, by concept, as a multi-purpose facility.\\

Along with hadron spectroscopy (see Ref.~\cite{Ketzer:2019wmd} for a review) and chiral dynamics studies \cite{COMPASS:2014eqi,COMPASS:2011vvh}, one of the main scientific goals of the "Phase-1" (2002-2011) was the study of the nucleon structure using the high-energy, naturally polarized muon beam of positive charge and either a proton or a deuteron longitudinally or transversely polarized target. In the quest for a solution to the \textit{spin puzzle}, the flagship measurement at the proposal time was the first direct measurement  of the gluon contribution to the nucleon spin $\Delta g$. It has been inferred considering the photon-gluon fusion (PGF) mechanism, in which the virtual photon is absorbed by a gluon producing a quark-antiquark pair, tagged either by the detection of $D^0$ mesons in the final state \cite{COMPASS:2012mpe}, or looking at hadron pairs at high $\Pt$ \cite{COMPASS:2012pfa}. The results for $\Delta g /\ g$, obtained from the data collected on a deuteron ($6$LiD) target in 2002-2004 and 2006, indicate a small contribution of the gluon to the nucleon spin and a reasonable agreement between the two channels.\\

DIS and SIDIS studies could also be made using the same data; in addition, they were also performed using the longitudinally polarized proton (NH$_3$) target in 2007 and 2011. Relevant measurements have been performed in COMPASS of the longitudinal double spin asymmetries $A_1^p$ and $A_1^d$ \cite{COMPASS:2005xxc,COMPASS:2006mhr,Ageev:2007du,COMPASS:2010wkz}, related to the structure functions $g_1^p$ and $g_1^d$ of proton and deuteron respectively. The measurement of the same asymmetries for identified hadrons in the final state allowed for a flavor separation of the helicity PDFs \cite{COMPASS:2009kiy}.  \\

In parallel to the study of the structure of the nucleon with a longitudinally polarized target, SIDIS measurements with a transversely polarized target were performed. While most of the deuteron data were collected in 2002-2004, the proton data were collected in 2007 (half of the data-taking) and 2010. Apart from the already quoted results for the Collins \cite{COMPASS:2006mkl, COMPASS:2008isr,COMPASS:2010hbb,COMPASS:2012ozz,COMPASS:2014bze} and Sivers asymmetries \cite{COMPASS:2006mkl,COMPASS:2008isr,COMPASS:2010hbb,COMPASS:2012dmt,COMPASS:2014bze,COMPASS:2016led}, clearly different from zero on proton, a lot of different measurements were performed, in particular all the azimuthal asymmetries expected in the single-hadron cross-section \cite{Parsamyan:2013fia}, the $\Pt$-weighted Sivers asymmetry \cite{COMPASS:2018ofp} and the di-hadron asymmetries \cite{COMPASS:2012bfl,COMPASS:2014ysd,COMPASS:2015drt}. These results have been very important to access transversity and the Sivers function. Also, the need for more deuteron data is nowadays very clear, and a dedicated run will take place in 2021/2022. In addition, new exploratory measurements were done using the same data, like the asymmetries for high-$\Pt$ hadrons \cite{COMPASS:2017ezz}, $J/\Psi$ \cite{Matousek:2018nco}, $\omega$ \cite{COMPASS:2016ium} and $\vmrho$ \cite{COMPASS:2012ngo,COMPASS:2013fsk}, and the $\Lambda$ polarization \cite{COMPASS:2021bws}. Mixing up the data collected with opposite deuteron polarization (either longitudinal or transverse) the $\Pt$-multiplicities \cite{COMPASS:2013bfs,COMPASS:2017mvk} and the azimuthal asymmetries in SIDIS off unpolarized deuteron \cite{COMPASS:2014kcy} could also be measured: these results will be discussed in the following of this Thesis. \\

In 2012, the experiment entered a "Phase-2", expanding further the original physics goals \cite{Gautheron:2010wva}. In 2015 and 2018 COMPASS collected Drell-Yan data with a transversely polarized proton target and a negative hadron beam, to compare the Sivers asymmetry measured in Drell-Yan and in SIDIS \cite{COMPASS:2017jbv}. \\

The 2016 and 2017 data taking were dedicated to the measurement of the Deeply Virtual Compton Scattering (DVCS) and of Hard Exclusive Meson Production (HEMP) as a way to access the Generalized Parton Distributions (GPDs). A pilot DVCS run took place in 2012; some results from those data will be presented in Ch.~\ref{Chapter6_SDMEs}. Positive and negative muon beams scattered off an unpolarized proton target allow the measurement of the sum and difference of cross-sections, with the possibility to access different azimuthal modulations \cite{COMPASS:2018pup} and the transverse extension of partons in the proton \cite{COMPASS:2019fea}. In parallel to DVCS, SIDIS data were also collected, which are the subject of this Thesis. As already mentioned, a new data taking to be held in 2021/2022 has been approved for the measurement of the $d-$quark transversity \cite{Friedrich:2286954}, and this will complete the COMPASS program. \\

\section{The experimental apparatus}

\begin{figure}[h!]
    \captionsetup{width=\textwidth}
    \centering 
    \hspace{0.85cm}
    \includegraphics[width=\textwidth]{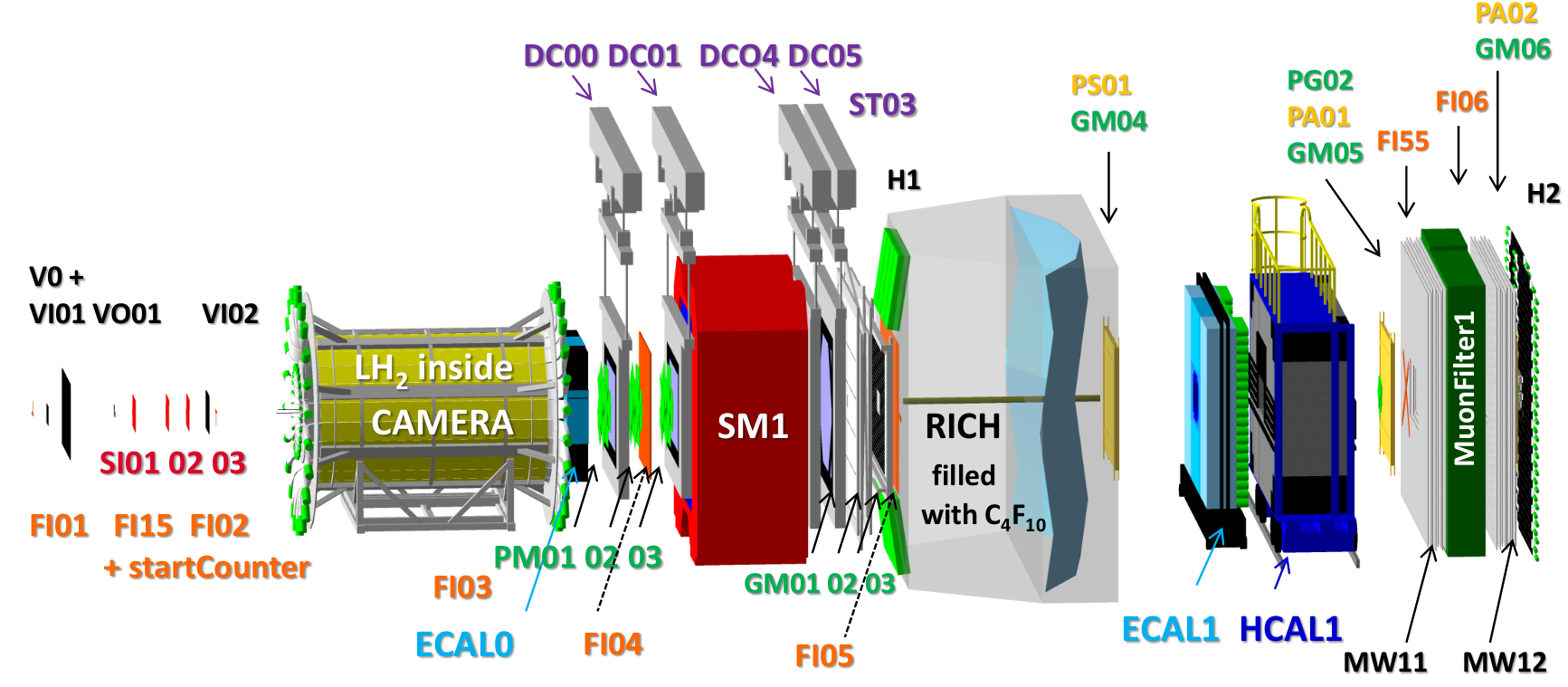} \\
    \includegraphics[width=\textwidth]{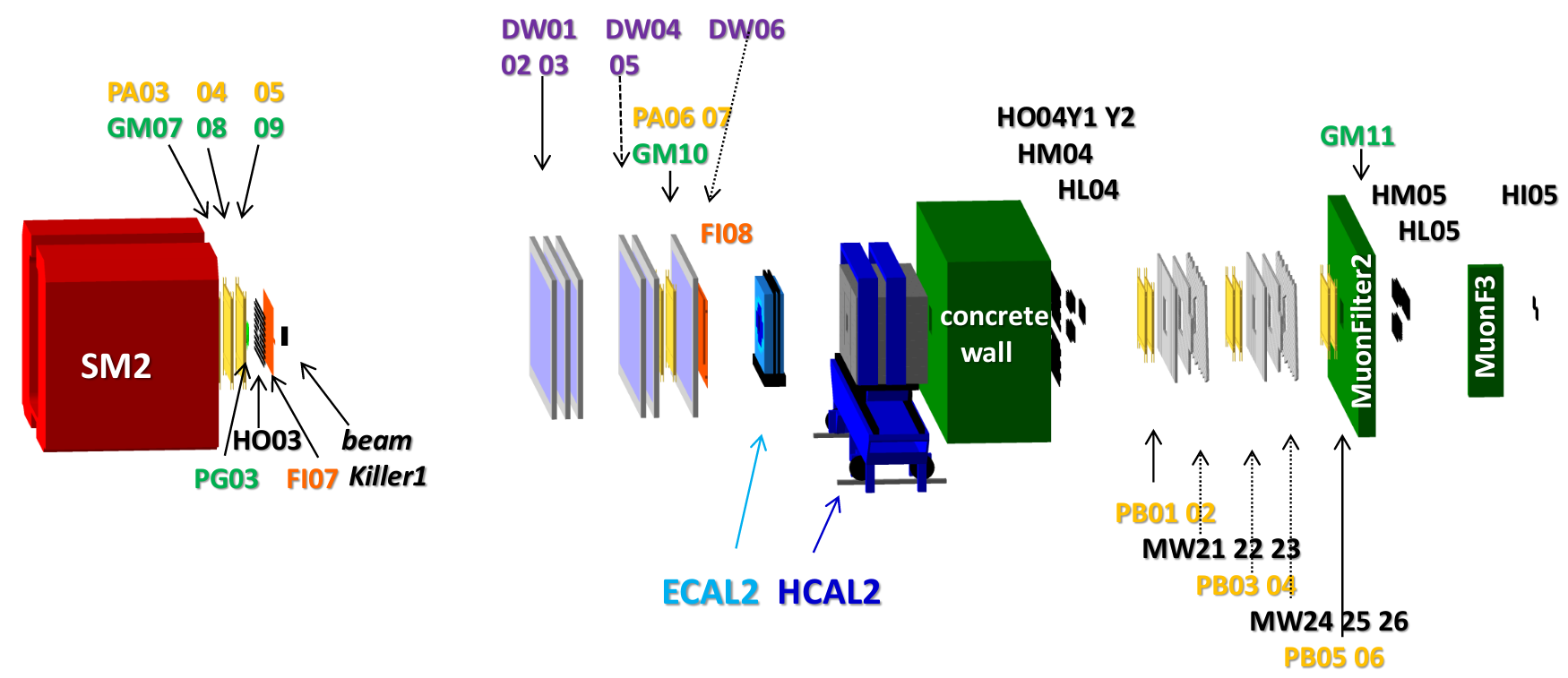}
    \caption{The COMPASS setup in 2016/2017: the Large Angle Spectrometer on the top part, the Small Angle Spectrometer on the bottom. }
    \label{fig:apparatus}
\end{figure}

Along the years, the experimental apparatus has been adapted to the various measurements with different beams and targets and improved with several upgrades. This Chapter is dedicated to a brief description of the COMPASS apparatus as it was set up for the 2016 data taking (shown in Fig.~\ref{fig:apparatus}), in which the data analyzed in this Thesis have been collected.  \\

The COMPASS apparatus consists of a 50~m long, two-stage spectrometer equipped with both trackers and detectors used for the particle identification. The fundamental requirements imposed to the setup are: the widest possible angular and kinematic acceptance, high resolution and precise track reconstruction down to small angles. In 2016 (and 2017) the apparatus was optimized for DVCS measurements, with an electromagnetic calorimeter placed close to the target: this resulted in a reduced acceptance for the SIDIS measurement with respect to previous SIDIS runs. The main parts of the apparatus, described in the following, are:

\begin{itemize}
    \item the beam line;
    \item the target region;
    \item the Large Angle Spectrometer (LAS);
    \item the Small Angle Spectrometer (SAS).
\end{itemize}

\subsection{The M2 beamline and the muon beam}
The beam used in the COMPASS experiment is delivered by the SPS M2 beamline \cite{COMPASS:2014cka}, which was commissioned in 1978 to serve the EMC and the BCDMS experiments and subsequently used by the NMC experiment. From 1991 onward, the beam has been used by the SMC Collaboration. In 2002 the beam was somewhat upgraded and since then it has been used by COMPASS. The beam properties, described in Ref.~\cite{Doble:1994np} are very well reproduced by a full Monte Carlo code based on the TRANSPORT charged particle transport optics program. The M2 beamline can provide hadron or muon beams with high intensity and a momentum up to 280~GeV/$c$. The secondary beams are obtained from the collision of an intense primary proton beam which is first accelerated in the SPS to more than 400 GeV/$c$ and then extracted from the SPS onto a Beryllium target of adjustable thickness (T6). For the DVCS data taking in 2016 (2017), which was conducted using both $\mu^+$ and $\mu^-$ beams, the proton beam flux on T6 was about 100 (150)$\cdot$ 10$^{11}$ protons/spill, for a typical spill length of 4.8 seconds. Two spills were delivered every 36 seconds. The $\mu^{\pm}$ beams are obtained by sending the charged pions into a decay section, where they decay into muons and neutrinos. Being the production of positive pions favored in the proton collision at T6, using the same T6 setup would result in a $\mu^+$ beam intensity about 2.7 times larger than the  $\mu^-$ one. To ensure the best experimental conditions for the DVCS measurement, the thickness of the T6 production target was set at 100 (500) mm for the $\mu^+$ ($\mu^-$) beam, thus getting typical $\mu^+$ ($\mu^-$) fluxes of 7.6 (6.3) $\cdot$ 10$^7$ muons per spill (about 1.5 $\cdot$ 10$^7$ muons per second). \\

\begin{figure}
    \captionsetup{width=\textwidth}
    \centering 
    \hspace{0.85cm}
    \includegraphics[width=\textwidth]{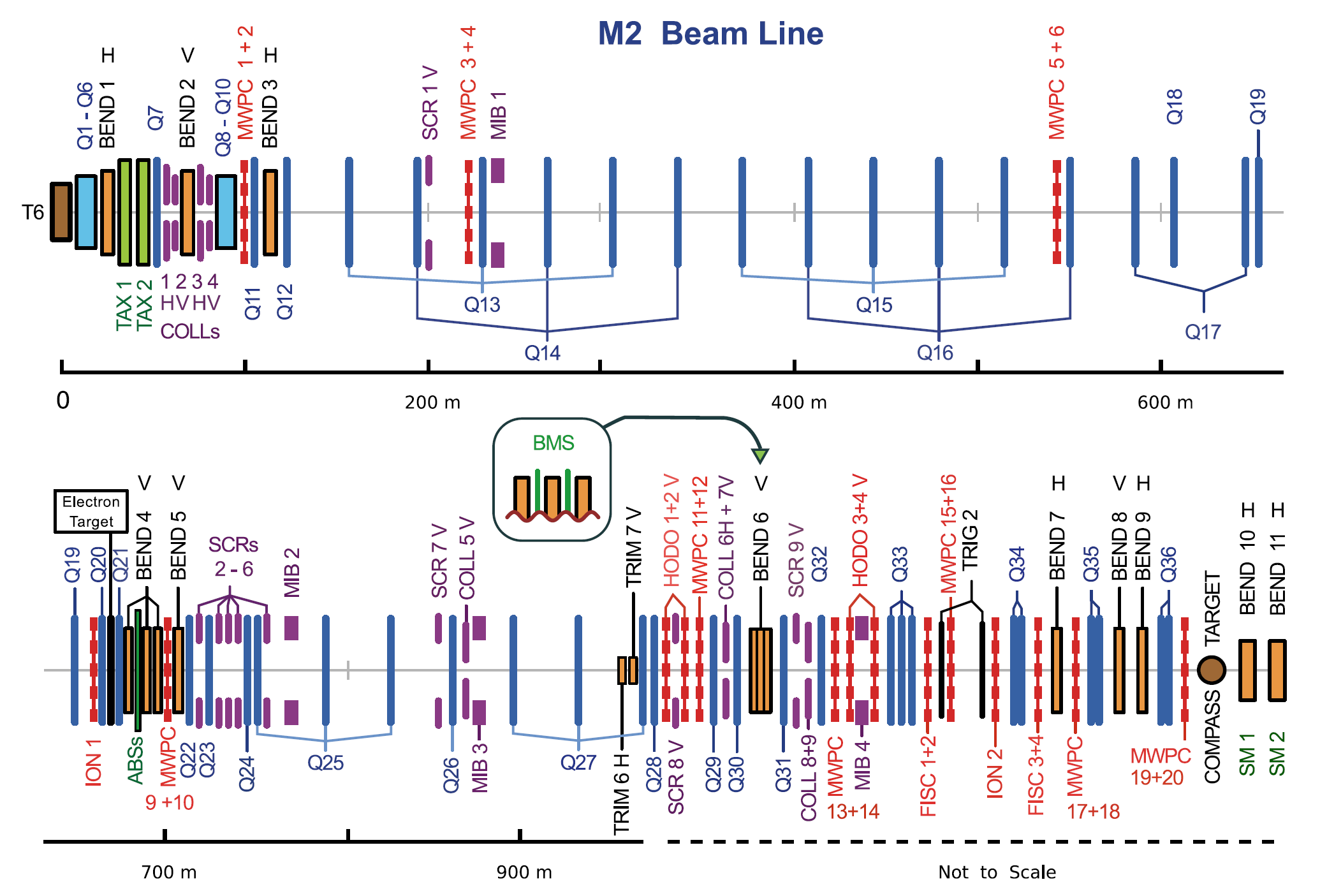} 
    \caption{Overview of the M2 beamline, including quadrupoles, bending magnets, absorbers, scrapers and collimators \cite{COMPASS:2014cka}.}
    \label{fig:beamline}
\end{figure}

An overview of the M2 beamline is given in Fig.~\ref{fig:beamline}. The beamline is constituted of many elements including dipole magnets, quadrupole magnets, collimators, scrapers and absorbers. The dipole magnets are used to bend the beam direction, while the quadrupoles are used to focus the beam. As a quadrupole can only focus on one plane, while defocusing in the perpendicular plane, at least two quadrupoles are needed to effectively focus the beam. The collimators are fundamental in order to define the beam momentum. The scrapers are used to reduce the halo component and the absorbers are used to reduce the hadron contamination in the muon beam. \\

The particles produced in the collision at T6 are mostly pions, with a small Kaon contamination. A sequence of quadrupole and bending magnets is used to collect them within a large angular acceptance and to select their momentum at 225 GeV/$c$ $\pm$ 10\% before they reach the long decay section, equipped with alternating focusing and defocusing quadrupoles (FODO). There, a large fraction of the pions undergoes a weak decay into muons and neutrinos \cite{COMPASS:2007rjf}. The muons, naturally fully polarized in the rest frame of the parent hadron, are further bent and selected with Beryllium absorbers. A second FODO section is used to perform the final focusing for the experiment. Before entering the experimental hall, the beam passes through a set of dipole magnets that bend the beam to the horizontal plane.\\

\begin{figure}
    \captionsetup{width=\textwidth}
    \centering
    \includegraphics[scale=0.8]{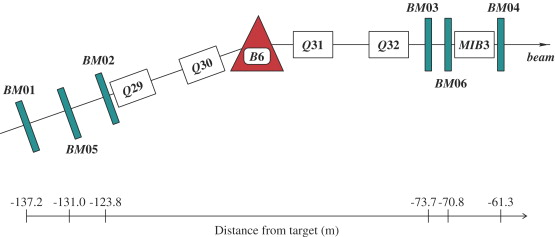}
    \caption{Layout of the BMS for the COMPASS muon beam \cite{COMPASS:2007rjf}.}
    \label{fig:bms}
\end{figure}

The momentum of each incoming particles is measured by the Beam Momentum Station (BMS, Fig.~\ref{fig:bms}) which consists of a bending dipole magnet (B6) surrounded by four quadrupoles (Q29-Q32) and six scintillator hodoscopes (BM01-BM06). The beam momentum is parametrized based on the coordinates of the track passing through these detectors. The precision of the momentum measurement is better than 1\%. Close to the target position, the beam direction is measured with scintillating fibres and silicon detectors and traced back to the BMS. The momentum spread is typically $\sigma_p/p$ = 0.05 and the spatial spread is about 7 mm at the target position. Scintillator veto counters allow one to separate the beam from the halo, composed by particles not passing through the target but giving spurious triggers and interfering with the reconstruction of the events. \\

The polarization of the beam in the laboratory depends on the mass $m_h$ of the decaying hadron (pion or Kaon) and on the ratio of the energies $E_h /\ E_\mu$ in the laboratory frame, according to the formula \cite{Adeva:1993kp}:

\begin{equation}
    P_{\mu^{\pm}} = \mp \frac{m_h^2 + \op 1-2\frac{E_h}{E_\mu}\cp m_\mu^2}{m_h^2-m_\mu^2}.
\end{equation}
Thus, the polarization for a positive muon beam is maximum, and equal to -100\%, if $E_h= E_\mu$, while a polarization of +100\% can be reached for:

\begin{equation}
    \frac{E_\mu}{E_h} = \op \frac{m_\mu}{m_h} \cp^2 \stackrel{\text{$h=\pi$}}{\approx} 0.57.
\end{equation}
This case is however less favorable, as the polarization of the muons decaying from Kaons would have opposite sign with respect to the one produced in the pion decay. For this reason, a high energy ratio is preferred: the optimal value for the beam polarization is obtained for a momentum ratio $p_{\mu} /\ p_{h}$ between 0.9 and 0.95 \cite{Doble:1994np}. A beam polarization $P_{\mu^+} = - 80\%$ is expected from the pion decay by selecting $p_{\mu^+} /\ p_{\pi^+} = 0.93$. Two direct measurements of the muon beam polarization \cite{SpinMuonSMC:1993coa,SpinMuon:1999nzf} have been found to agree well with the expected value.

\subsection{The target region}
\begin{figure}
    \captionsetup{width=\textwidth}
    \centering
    \includegraphics[scale=0.4]{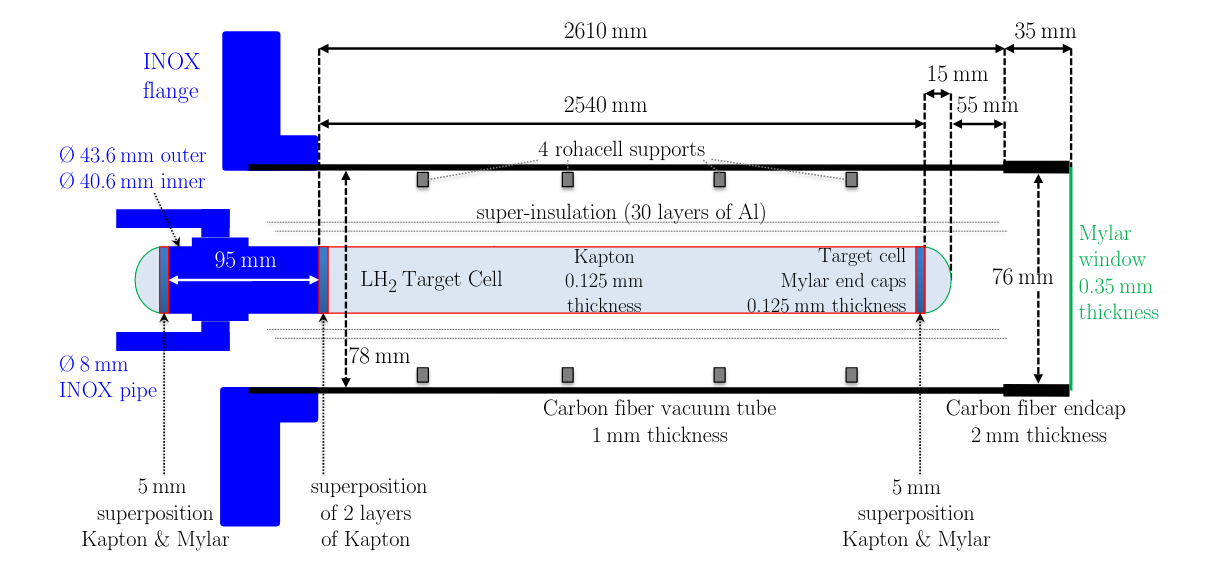}
    \caption{A schematic view of the liquid hydrogen target \cite{Vidon:2019dbm}.} 
    \label{fig:target}
\end{figure}

In 2016 and 2017 the COMPASS spectrometer was equipped with a liquid hydrogen target \cite{Bielert:2014jfa} of cylindrical shape, more than 2.5~m long, with a diameter of 40~mm and a volume of 3.3~l. A schematic view of the target is given in Fig.~\ref{fig:target}. In order to allow for an accurate detection of the recoil proton at small momentum transfer, required by the DVCS measurement, the material budget had to be minimal: the target cell was made of a 0.125 mm thick Kapton sheet with a Mylar end cap of the same thickness. The Kapton tube was glued to a stainless steel cylinder in the upstream part and to a carbon fiber end cap in the downstream part, its whole length being insulated with 30 aluminum layers. Several rohacell supports ensured the tube to be kept in place inside the vacuum tube. 
Thanks to a complex refrigerator system, the target was operated at a temperature $T\approx$ 20~K, corresponding to a hydrogen density of 70.3~kg/m$^3$ at 1150~mbar. \\

\begin{figure}
    \captionsetup{width=\textwidth}
    \centering
    \includegraphics[width=\textwidth]{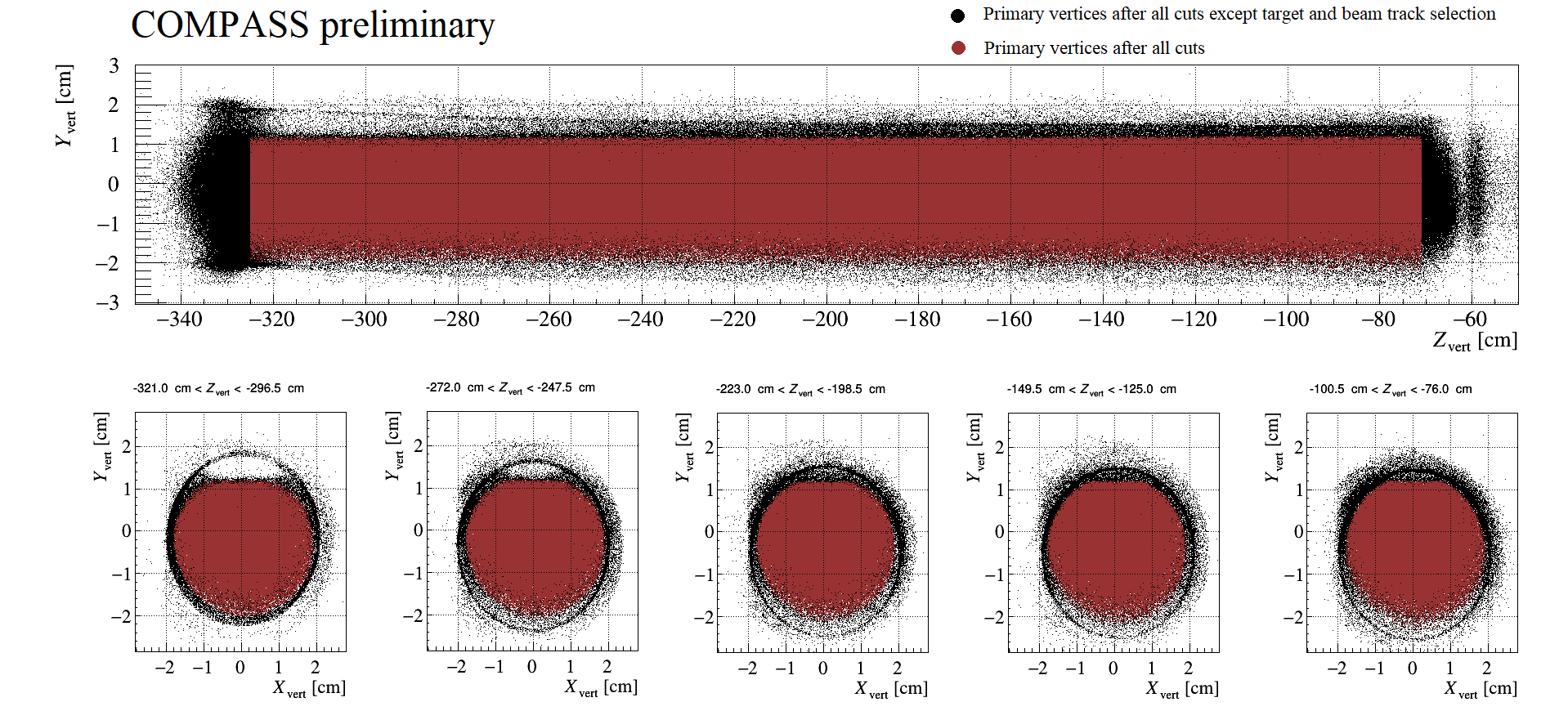}
    \caption{Side-view of the target (top), as obtained from the interaction vertex distribution in the data: a fiducial region, indicated in red, is determined by rejecting the vertices originating from the target walls, from the end-cap and from the empty part located upstream. Five transverse sections of the target are given in the bottom part, where the walls are well visible as dense rings. Plot by J. Matousek (COMPASS).} 
    \label{fig:targetv}
\end{figure}

The fiducial target volume for the 2016 target configuration (in red in Fig.~\ref{fig:targetv}) can be identified by considering different projections and slices of the three-dimensional distribution of the interaction vertex. From the position of the target walls, well visible as dense rings in the bottom part of Fig.~\ref{fig:targetv} due to the higher interaction probability, it is possible to reconstruct the position of the center and the radius of the target along the longitudinal axis. This reconstruction procedure consists first in a conversion of the (x,y) Cartesian coordinates of the target wall into a polar equation of the form:

\begin{equation}
    r^2 -2r r_0 \cos(\phi - \phi_0) + r_0^2 = a^2
\end{equation}
where ($r_0,\phi_0$) is the position of the target center, ($r,\phi$) corresponds to the generic point on the wall and $a$ is the radius of the target cell. This equation can be solved for $r$ obtaining:
\begin{equation}
    r = r_0 \cos(\phi - \phi_0) \pm \sqrt{a^2 - r_0^2\sin^2(\phi - \phi_0)}.
\end{equation}
This expression has been used to fit points on the target wall, obtaining the best values of the parameters $r_0$, $\phi_0$ and $a$, from which the position of the target center could be converted back to Cartesian coordinates in the laboratory system.
As can be observed in Fig.~\ref{fig:targetv}, the target cell was slightly tilted, the center position being higher in the upstream part, and partially filled there.

\subsubsection{CAMERA}
\begin{figure}[h!]
\captionsetup{width=\textwidth}
\captionsetup[subfigure]{labelfont=rm}
  \begin{subfigure}{0.45\textwidth}
    \includegraphics[scale=0.29]{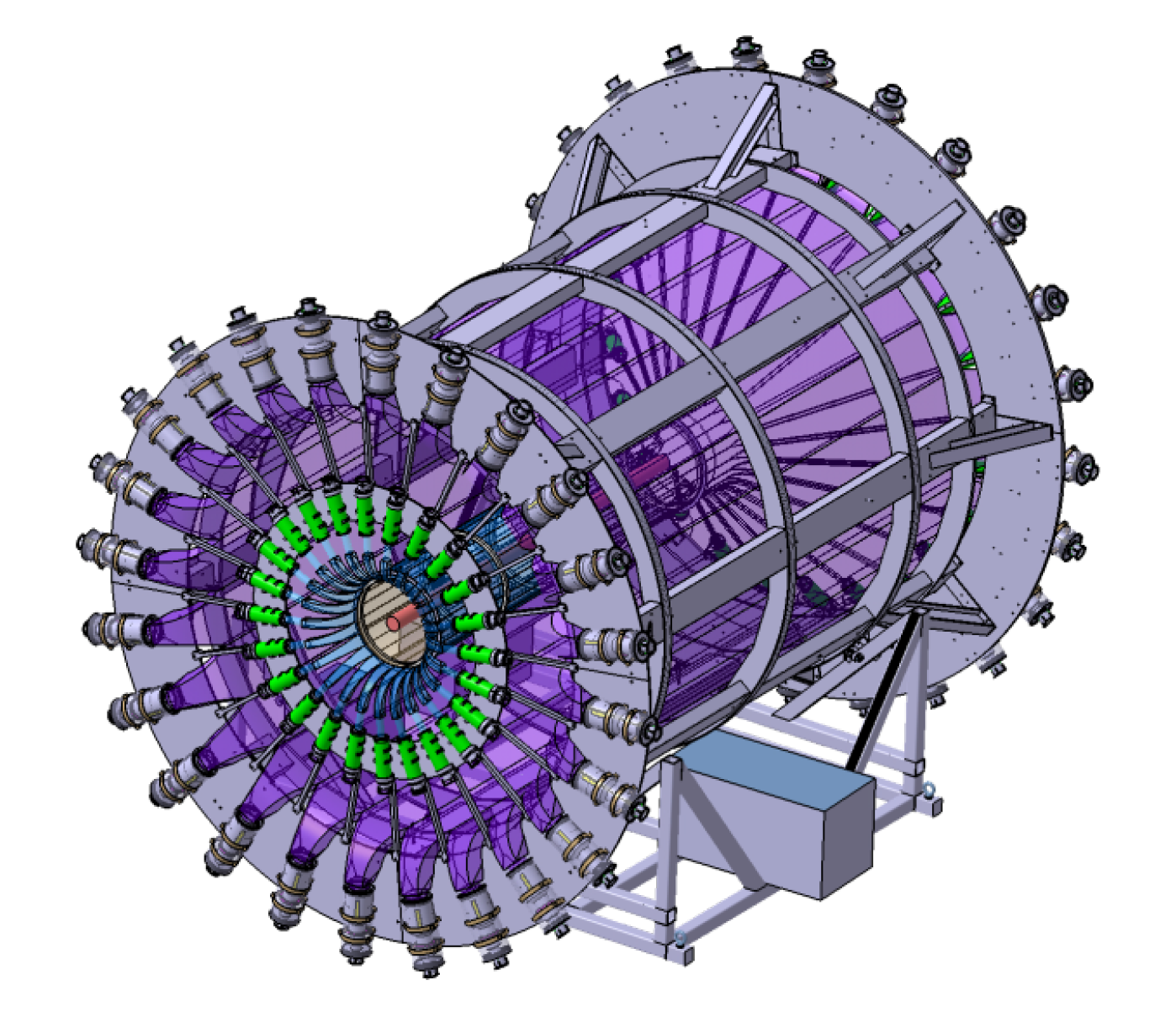}
    \caption{} \label{fig:camera_a}
  \end{subfigure}%
  \hspace*{\fill}   
  \begin{subfigure}{0.45\textwidth}
    \includegraphics[scale=0.36]{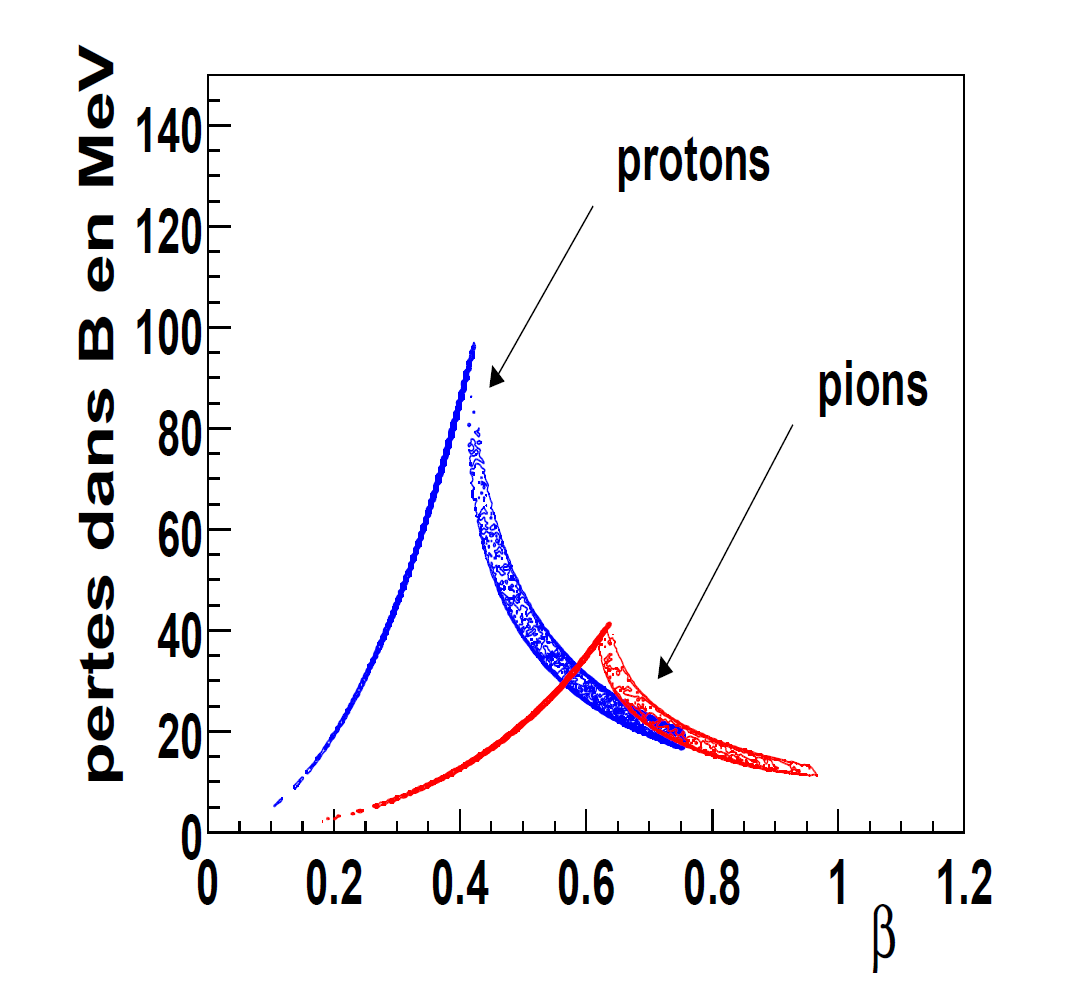}
    \caption{} \label{fig:camera_b}
  \end{subfigure}%
 \caption{(a):~Schematic view of the CAMERA detector, upstream view. (b):~Energy loss in Ring~B as a function of the reconstructed velocity $\beta$ \cite{Vidon:2019dbm,Mosse:2002hka}. }
 \label{fig:camera}
\end{figure}

The direct reconstruction of the recoil proton in an exclusive reaction was made possible thanks to a Time-of-Flight (ToF) detector named CAMERA. Such detector, shown in Fig.~\ref{fig:camera_a}, was placed around the target and consisted of two concentric cylinders (\textit{Ring A} and \textit{Ring B}) placed at a radial distance of 25~cm and 110~cm from the target axis, respectively. Both rings were made of 24 scintillating slabs, covering 15$^\circ$ each. To increase the resolution, the Ring~B was rotated of 7.5$^\circ$ with respect to Ring~A. Each end of the scintillators was connected to Photo-Multiplier-Tubes (PMTs) via long light guides. The bending of the light guides ensured that the PMTs lied out of acceptance. The principle of the CAMERA detector is to allow for a simultaneous measurement of the Time-of-Flight and of the Distance-of-Flight of the tracks leaving hits in coincidence in the two rings, thus giving access to the velocity $\beta$ of the particles. To identify the recoil particle as a proton, the peculiar correlation between the amplitude of the signal collected in the PMTs of Ring~B (proportional to the energy loss) and $\beta$ (Fig.~\ref{fig:camera_b}, from simulation) is used. The energy loss distribution shows two regimes: from small values of $\beta$ up to the peak position, the curve is populated by the particles that are stopped in the Ring~B, while the decrease at larger $\beta$, proportional to $1/\beta^2$, is dictated by the Bethe-Bloch formula for the energy loss in a medium. This second regime is also more dependent on the angle of the particle crossing the Ring~B: this is the reason for the widening of the curve at large $\beta$.

\subsection{The spectrometer}
\subsubsection{\underline{The spectrometer magnets}}
The two stages of the COMPASS spectrometer (LAS and SAS) are built around two large dipole magnets, called SM1 and SM2. Both dipoles feature a magnetic field along the vertical direction: bending charged particles in the horizontal plane. SM1, which is used for the reconstruction of the momenta of the particles produced at large angles, has a bending power $\int B \diff l$ = 1~Tm, with $l=110$~cm; its central aperture is 229~cm wide and 152~cm high. To reconstruct the momentum of the more energetic particles at small angles, the SM2 magnet is used instead, thanks to its large bending power of $\int B \diff l$ = 4.4~Tm, where $l=4$~m. Its central aperture is smaller than the one of SM1, being 2~m wide and 1~m high.

\subsubsection{\underline{The tracking detectors}}
The COMPASS spectrometer hosts numerous tracking stations. They are distributed along the full spectrometer length. Each station comprises a set of detectors of the same type, in order to have several projections of the particle trajectory, thus increasing the resolution of the station and solving ambiguities due to the high flux of particles. \\

Close to the beam axis, the tracking detectors have to be characterized by a very good spatial resolution and must show a high radiation hardness. Moreover, in order to limit multiple scattering and secondary interactions, the amount of material has to be small. Detector technologies fulfilling these requirements are Silicon Microstrip detectors (SIs) and Scintillating Fibres (SciFis). At intermediate distance from the beam, a high rate capability and a good spatial resolution is assured by the Micro MEsh Gaseous detectors or MicroMegas (MMs) and by the Gas Electron Multiplier detectors (GEMs). At larger distances, the particle flux is lower and the region to be covered is larger: MultiWire Proportional Chambers (MWPCs), Drift Chambers (DCs) and Straw tube detectors are used. Their position is shown in Fig.~\ref{fig:apparatus}. A brief description of the various detector technologies is given in the following. \\

\paragraph{Silicon Microstrip detectors (SI)}
Silicons Microstrip detectors are placed before and after the target. They were originally designed for the use at HERA \cite{Abt:2000ws} and optimized for high fluxes. They consist of a 300~$\mu$m thick n-type wafer with an active area of 5$\times$7~cm$^2$. The signals are read out from strips on both sides. The number of strips is 1280 on one side and 1024 on the other, where the strips are orthogonal to the first ones. In this way, the two-dimensional position of the hit is obtained with half the material amount of a common single-side wafer. One detector station is formed by two of those detectors mounted back to back, where the second one is rotated by 5$^\circ$ for a better overall spatial resolution. The silicons are generally operated at low temperatures (around 130~K) to reduce noise and to improve their performance.
In the best scenario, these detectors have a spatial resolution of 4–6~$\mu$m and a time resolution of 2.5~ns.

\begin{figure}[h!]
    \captionsetup{width=\textwidth}
    \centering
    \includegraphics[width=0.55\textwidth]{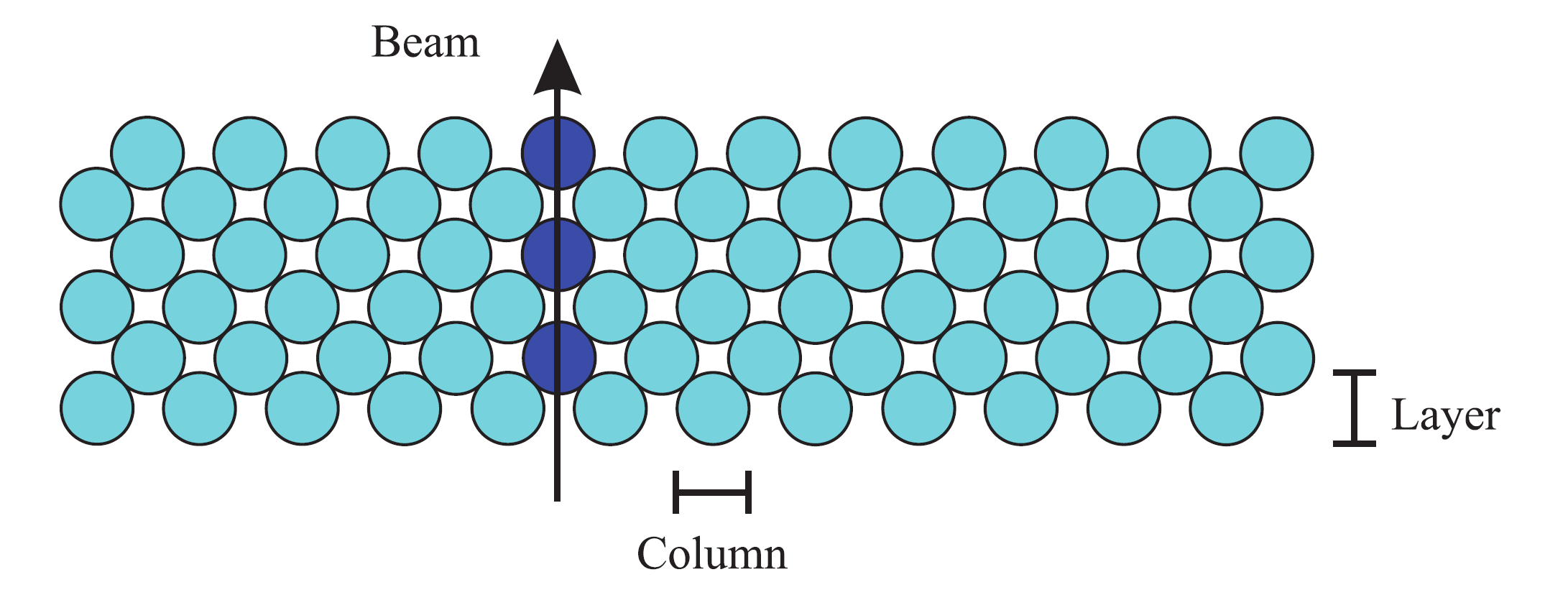}
    \caption{Illustration of a Scintillating Fibres detector \cite{COMPASS:2007rjf}.} 
    \label{fig:scifi}
\end{figure}

\paragraph{Scintillating Fibres detectors (SciFi)}
As illustrated in Fig.~\ref{fig:scifi}, these detectors consist of several layers of scintillating fibres. They are placed upstream and downstream of the target, with different orientation, in order to measure more than one projection and thus maximize the spatial resolution. Upstream the target, the diameter of the fibres is 0.5~mm and the active area is 4~cm$^2$; the downstream stations are generally larger, with a fibre diameter up to 1~mm. The spatial resolution can be as good as 130~$\mu$m, with a time resolution of 350~ps.

\begin{figure}[h!]
\captionsetup{width=\textwidth}
\captionsetup[subfigure]{labelfont=rm}
  \begin{subfigure}{0.35\textwidth}
    \includegraphics[width=\textwidth]{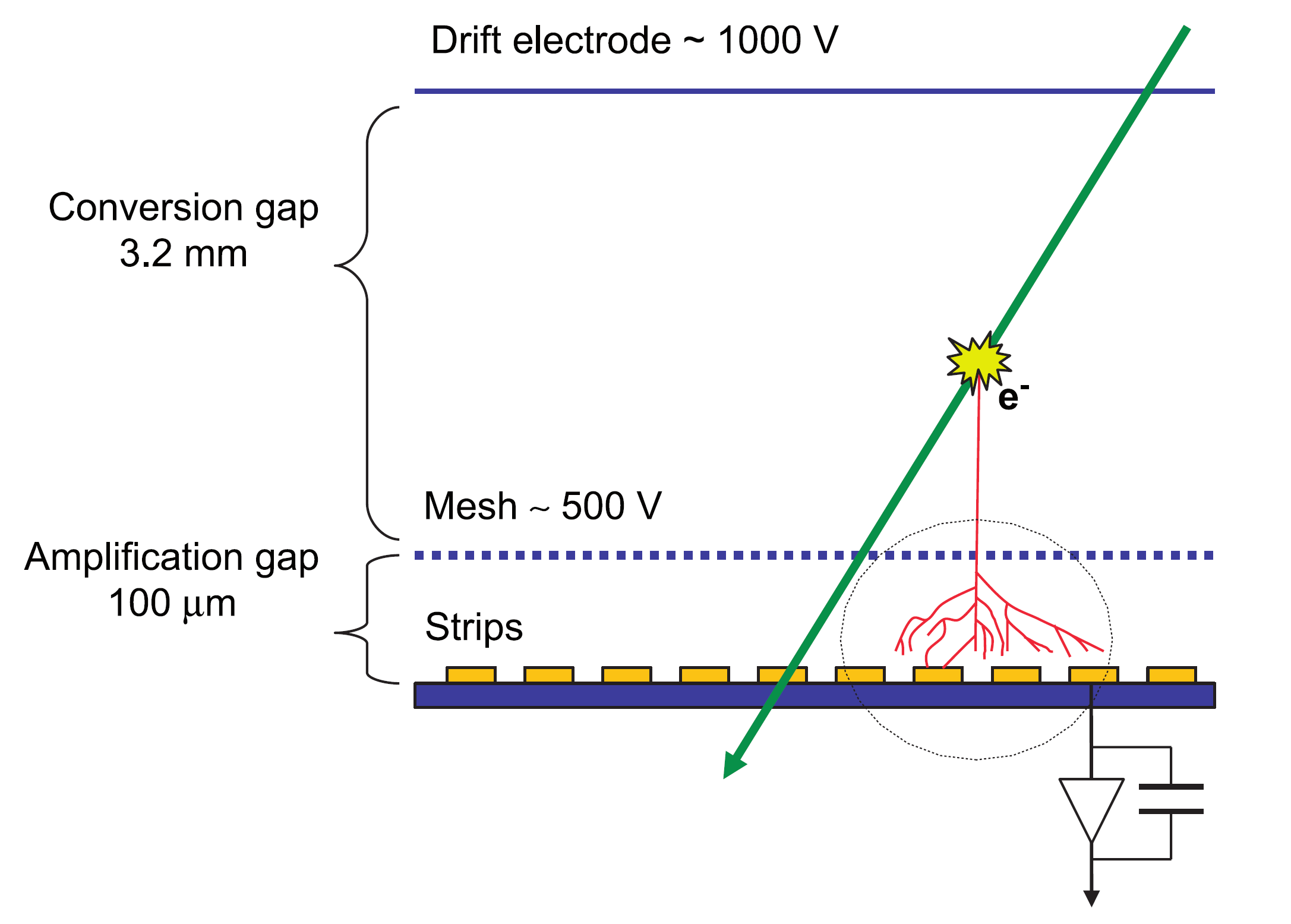}
    \caption{} \label{fig:gemms_a}
  \end{subfigure}%
  \hspace*{\fill}   
  \begin{subfigure}{0.55\textwidth}
    \includegraphics[width=\textwidth]{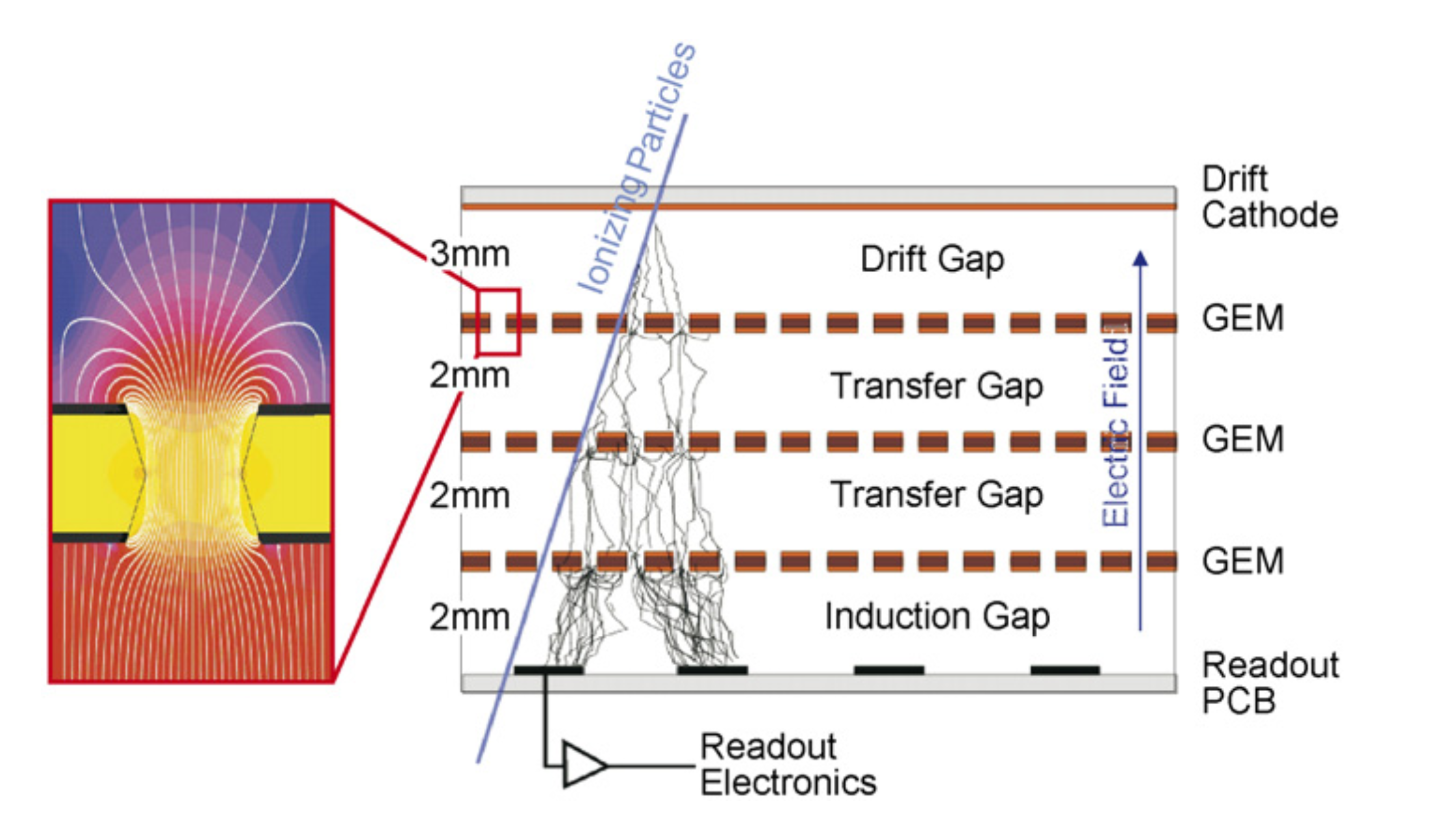}
    \caption{} \label{fig:gemms_b}
  \end{subfigure}%
 \caption{Illustration of a MicroMegas detector (a) and of a GEM detector (b)\cite{COMPASS:2007rjf}. }
 \label{fig:gemms}
\end{figure}

\paragraph{MicroMegas (MMs)}
The MicroMegas detectors (Fig.~\ref{fig:gemms_a}) have a parallel-plate electrode structure. The space between the electrodes is divided by a metallic micro-mesh in a conversion gap with a moderate electric field (1~kV/cm) and an amplification gap with a high field (40~kV/cm). The mesh captures most of the positive ions created in the amplification gap, and the small size of the gap reduces the diffusion of the electrons. The signal generated by the avalanche in the amplification gap is collected at the anode, which is divided into strips. In COMPASS, the first high-energy experiment to employ this technology, each detector has an active area of 40$\times$40~cm$^2$ with a central dead zone with a diameter of 5~cm. These detectors are placed in the LAS between the target and SM1, and each station consists of two doublets. Each doublet is formed by two perpendicular MMs; the second doublet is rotated of 45$^\circ$ with respect to the first, so to increase the spatial resolution ($\approx$ 90~$\mu$m). The time resolution is 9.3~ns. Between 2010 and 2015, some stations have been upgraded and their readout in the central part is since then pixelized.

\paragraph{Gas Electron Multipliers (GEMs)}
Similarly to MMs, GEM detectors (Fig.~\ref{fig:gemms_b}) consist of two parallel electrodes. In this case, however, the amplification does not take place near the strip. Several thin polyimide foils, cladded on both sides with copper, are itched with microscopic holes and a potential difference of about 100~V is applied on their two sides. The avalanche multiplication of electrons induced by the passing charged particles takes place in the holes. The electric field of the two electrodes collects the electrons until they are detected by a readout anode segmented in two perpendicular layers of strips. The detectors have an active area of 31$\times$31~cm$^2$ with a dead zone of 5~cm in diameter in the center. A GEM station is formed by two of these detectors, mounted back-to-back with an inclination of 45$^\circ$. The spatial resolution of a GEM detector is 70~$\mu$m; the time resolution is 12~ns. Since 2008, smaller GEMs detectors with pixelized readout in the central part and no dead zone have been added to the setup. 

\paragraph{MultiWire Proportional Chambers (MWPCs)}
MWPCs are used for the reconstruction of large angle particles in the SAS. They consist of several parallel anode wires in between cathode foils. A gas mixture Ar/CO$_2$/CF$_4$~(74/6/20) fills the detector volume. When a charged particle traverses the detectors, it ionizes the gas along its path and generates an electron avalanche due to the high potential difference. The signal is collected by 1~m-long wires, which have a diameter of 20~$\mu$m and a pitch of 2~mm. Three different kinds of MWPCs are used at COMPASS. The A-type ones consist of a X-, U- and V-plane, where the U- and V-planes are inclined by $\pm$ 10.14$^\circ$. The A*-type ones feature an additional Y-plane. The size of these chambers is 178$\times$120~cm$^2$ with a central dead zone of 16 to 20~cm in diameter. The B-type chambers have a smaller active area (178$\times$90~cm$^2$) with a central dead zone of 22~cm in diameter. They consist of an X-plane and a U- or V-plane, inclined by $\pm$ 10.14$^\circ$. The spatial resolution of these detectors is 1.6~mm.

\begin{figure}[h!]
    \captionsetup{width=\textwidth}
    \centering
    \includegraphics[width=0.55\textwidth]{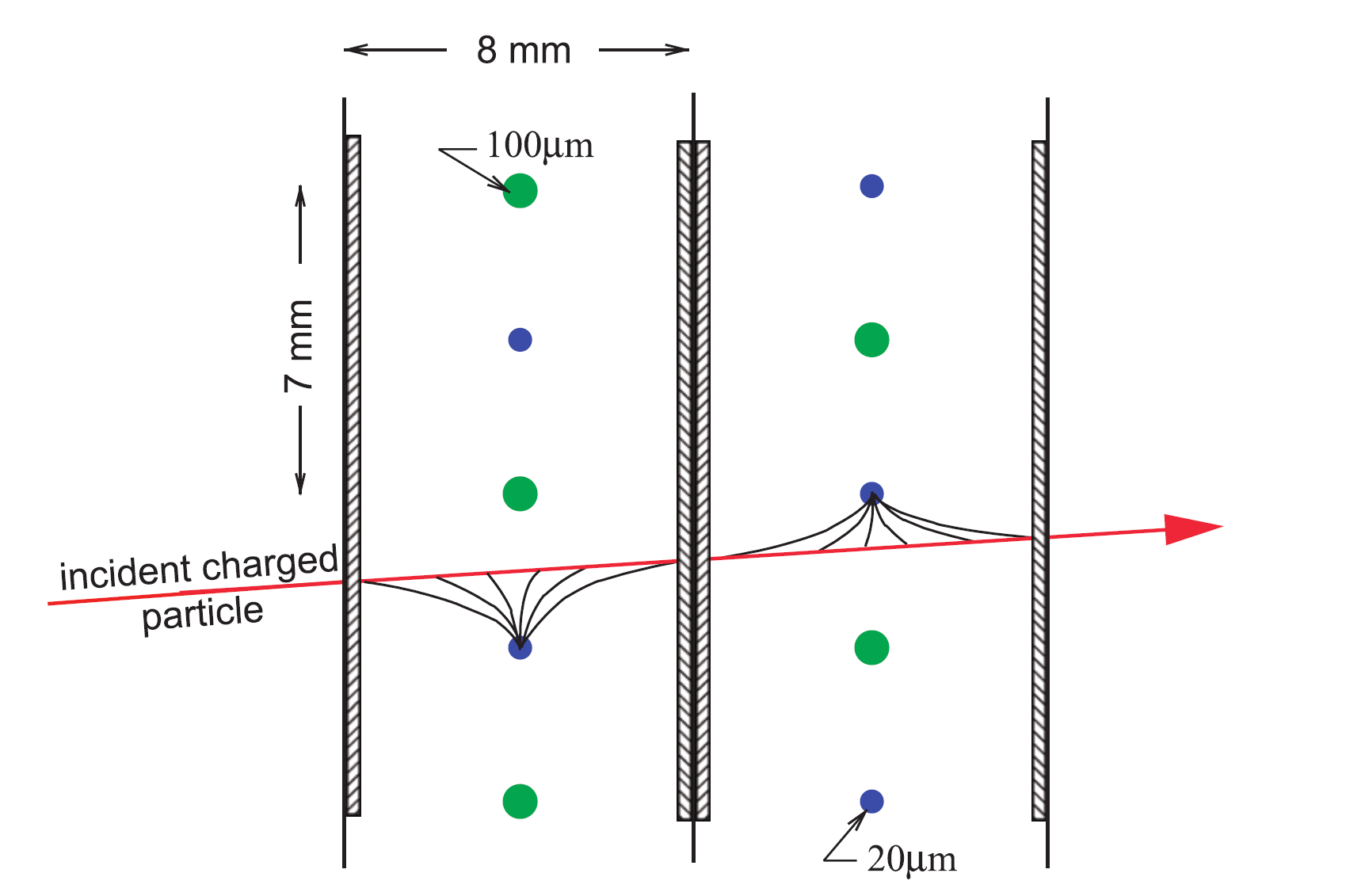}
    \caption{Drift cell geometry used at COMPASS \cite{COMPASS:2007rjf}.} 
    \label{fig:dc}
\end{figure}

\paragraph{Drift Chambers (DCs)}
In the Drift Chambers (Fig.~\ref{fig:dc}), the drift time of the avalanche to the anode wire is measured in addition to the charge collected, ensuring a good spatial resolution. The DCs consist of two cathodes foils together with anode and potential wires. The cathode and the potential wires are kept at -1700~V and the anode wires are at 0~V. In order to solve left-right ambiguities, two drift cells are staggered with shifted wires. The detector volume is filled with the gas mixture Ar/CO$_2$/CF$_4$~(85/5/10). Several types of DCs are in the setup. Two of them are installed in the LAS with active areas of 180$\times$127~cm$^2$, 30~cm diameter dead zone and spatial resolutions of 110~$\mu$m in the horizontal direction and of 170~$\mu$m in the vertical one (due to the SM1 fringe field). Two large ones, covering 248$\times$208~cm$^2$, are positioned after SM1. One of them was produced only before the 2015 run to replace an aging Straw Tube detector. Finally, six large area DCs with an active area of 5$\times$2.5~m$^2$ and less fine resolution of about 0.5~mm are used also in the SAS.

\bigskip

\paragraph{Straw Tubes}
With an active area of 9~m$^2$ and a resolution up to 200~$\mu$m, the Straw Tubes are used in the LAS downstream SM1 for the tracking of large angle particles. Each detector is made of two staggered layers of drift tubes, glued together and mounted on an aluminum frame. A single drift tube works similarly to a small drift cell and consists of a gold-plated tungsten anode wire surrounded with a thin foil consisting of two layers, where the inner foil is aluminium-cladded and serves as a cathode. The tubes have a diameter ranging from 6 to 9~mm. The gas mixture in use is Ar/CO$_2$/CF$_4$~(74/6/20).

\subsubsection{\underline{Particle identification}}
The particle identification is performed in COMPASS thanks to several different detectors. The muon identification is performed with several muon walls and filters both in the LAS and in the SAS. Two hadron calorimeters measure the energy of hadrons and provide complementary trigger signals, while three electromagnetic calorimeters determine the energy of photon and electron in a wide angular range. Finally, a RICH detector in the LAS can be used to separate pions, Kaons and protons in a wide momentum range. In addition, the reconstruction of the exclusive events (important for the DVCS cross-section measurement) is made possible also thanks to the CAMERA detector. 

\paragraph{Muon detectors}
An efficient strategy to identify muons is to exploit the difference in the penetration range that the muons show with respect to hadrons. This is done in COMPASS using a combination of absorbers and tracking detectors both in the LAS and in the SAS. The radiation length of the absorber is large enough to surely identify as muons the particles detected behind the absorber. In the LAS there are the Muon Wall1 (MW1) and the Muon Filter1 (MF1); in the SAS, the Muon Wall2 (MW2) in combination with the Muon Filter2 (MF2); at the very end of the spectrometer, the Muon Filter3 (MF3). The three muon filters are made of iron or concrete. \\
The MW1 system consists of Mini Drift Tubes. The tubes are made of 0.6~mm thick aluminum tubes surrounding a 50~$\mu$m thick tungsten wire and they are filled with a gas mixture of Ar/CO$_2$~(70/30). The muon filter surrounded by the MW1 system is made of 60~cm of iron.\\
The MW2 system in the SAS has two identical stations of layers of drift tubes. Each of the two stations consists of 6 layers with an active area of 447$\times$202 cm$^2$. The stainless steel drift tubes have an inner diameter of 29~mm and a wall thickness of 0.5~mm and the wires are 50~$\mu$m thick. They are filled with a gas mixture of Ar/CH$_4$~(75/25).

\paragraph{Calorimeters}
Three electromagnetic calorimeters are used to measure the energy deposit of electrons and photons. ECAL0, which is placed just after the target, is used for a large angle detection, while ECAL1 and ECAL2, respectively placed after SM1 and SM2, are used for intermediate to small angle detection. The presence of ECAL0, required for the DVCS measurements, reduces the angular acceptance of the spectrometer with respect to the SIDIS measurements performed from 2006 to 2011. The structure of these calorimeters is similar, as they are mainly made of lead glass and \textit{shashlik} modules. Inside the lead-glass modules, photons radiate showers of $e^+e^-$ and the emitted Cherenkov light is collected by Photo-Multiplier Tubes (PMTs). The shashlik modules, instead, are alternating layers of lead and scintillating material. The lead layers produce $e^+e^-$ pairs which radiate visible light within the scintillating material. The light is collected through optical fibres and detected by the PMTs. \\

The two hadron calorimeters (HCAL1 and HCAL2) are located before the muon filters MF1 and MF2. They are sampling calorimeters: their structure is modular, with iron or lead plates alternated to scintillating material. Both hadron calorimeters measure the energy of the hadrons, released in the hadron shower. While a hadron is expected to stop in the calorimeter, a muon just releases a small fraction of its energy: for this reason, the hadron calorimeters participate in the trigger decision for events at small muon angle. \\

\begin{figure}
\captionsetup{width=\textwidth}
\captionsetup[subfigure]{labelfont=rm}
  \begin{subfigure}{0.4\textwidth}
     \includegraphics[scale=0.35]{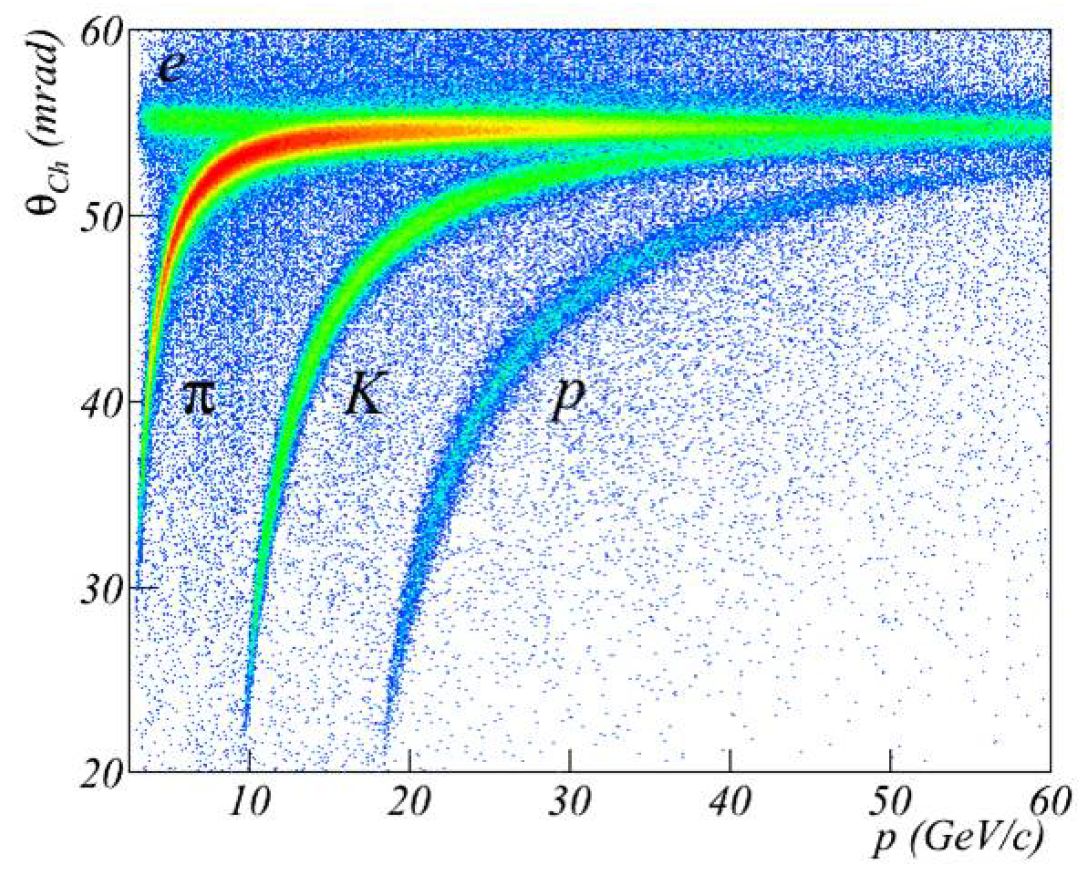}
    \caption{} \label{fig:rich_a}
  \end{subfigure}%
  \hspace*{\fill}   
  \begin{subfigure}{0.5\textwidth}
    \includegraphics[scale=0.32]{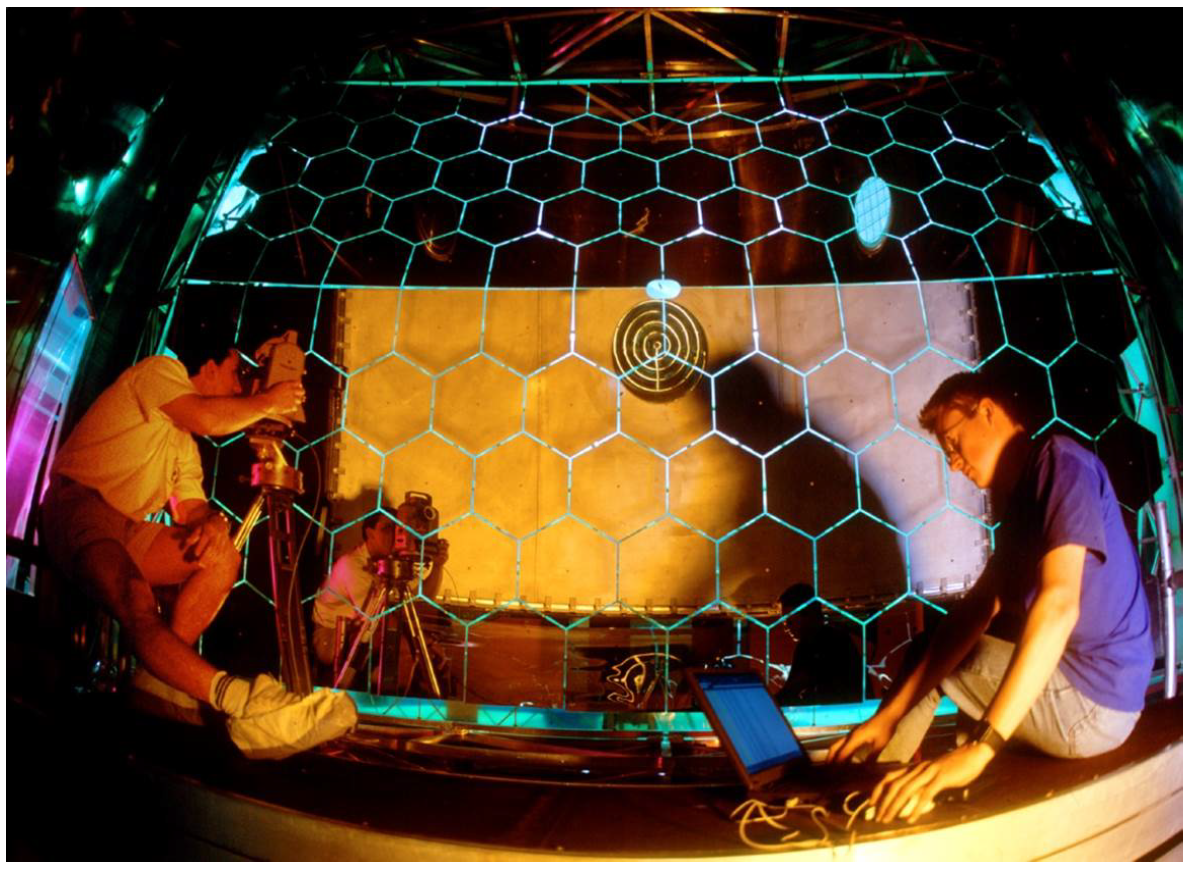}
    \caption{} \label{fig:rich_b}
  \end{subfigure}%
 \caption{(a):~The measured ring Cherenkov angle $\theta_{Ch}$ as a function of the particle momentum $p$ \cite{Tessarotto:2014lia}. (b):~The RICH optical detector system.}
 \label{fig:rich}
\end{figure}

\paragraph{Ring-Imaging Cherenkov Detector (RICH)}
Installed at the end of the LAS, the COMPASS RICH \cite{Albrecht:2005ew,Tessarotto:2014lia} is a large size Ring Imaging Cherenkov Detector that allows for the identification of pions, Kaons and protons from the threshold ($\sim$ 2.5~GeV/$c$ for pions, $\sim$ 10~GeV/$c$ for Kaons and $\sim$ 18~GeV/$c$ for protons in the $\mathrm{C_4F_{10}}$ radiator gas) up to more than 50~GeV/$c$ (Fig.~\ref{fig:rich_a}). Thanks to its large dimensions, it covers the whole angular acceptance of the LAS, ranging $\pm$ 180~mrad on the vertical axis and $\pm$ 250~mrad on the horizontal axis. Two spherical mirror systems \ref{fig:rich_b}, placed one above and one below the beam line, reflect the Cherenkov photons emitted by the charged particles that travel through the gas. The reflected photons are collected and converted to electrons by a system of detectors placed outside the spectrometer acceptance. Three photodetection technologies are used to this end: Multi-Wire Proportional Chambers (MWPCs) with CsI photocathodes, Multi-Anode Photo-Multipliers Tubes (MAPMTs) and Micro-Pattern Gaseous Detectors (MPGDs) \cite{Agarwala:2018bgm}. In its original version (in operation from 2002 to 2004) only the first technology was adopted and 16 MWPCs were used to cover the entire detection surface; then, to cope with the high particle flux in the central region, 4 of them have been replaced by an array of MAPMT. Finally, in 2016, other 4 of the original MWPCs were replaced by the novel hybrid detectors based on MicroMegas (MM) and ThickGEMs.

\subsection{The trigger system}
\begin{figure}
\captionsetup{width=\textwidth}
\captionsetup[subfigure]{labelfont=rm}
  \begin{subfigure}{0.5\textwidth}
     \includegraphics[scale=0.2]{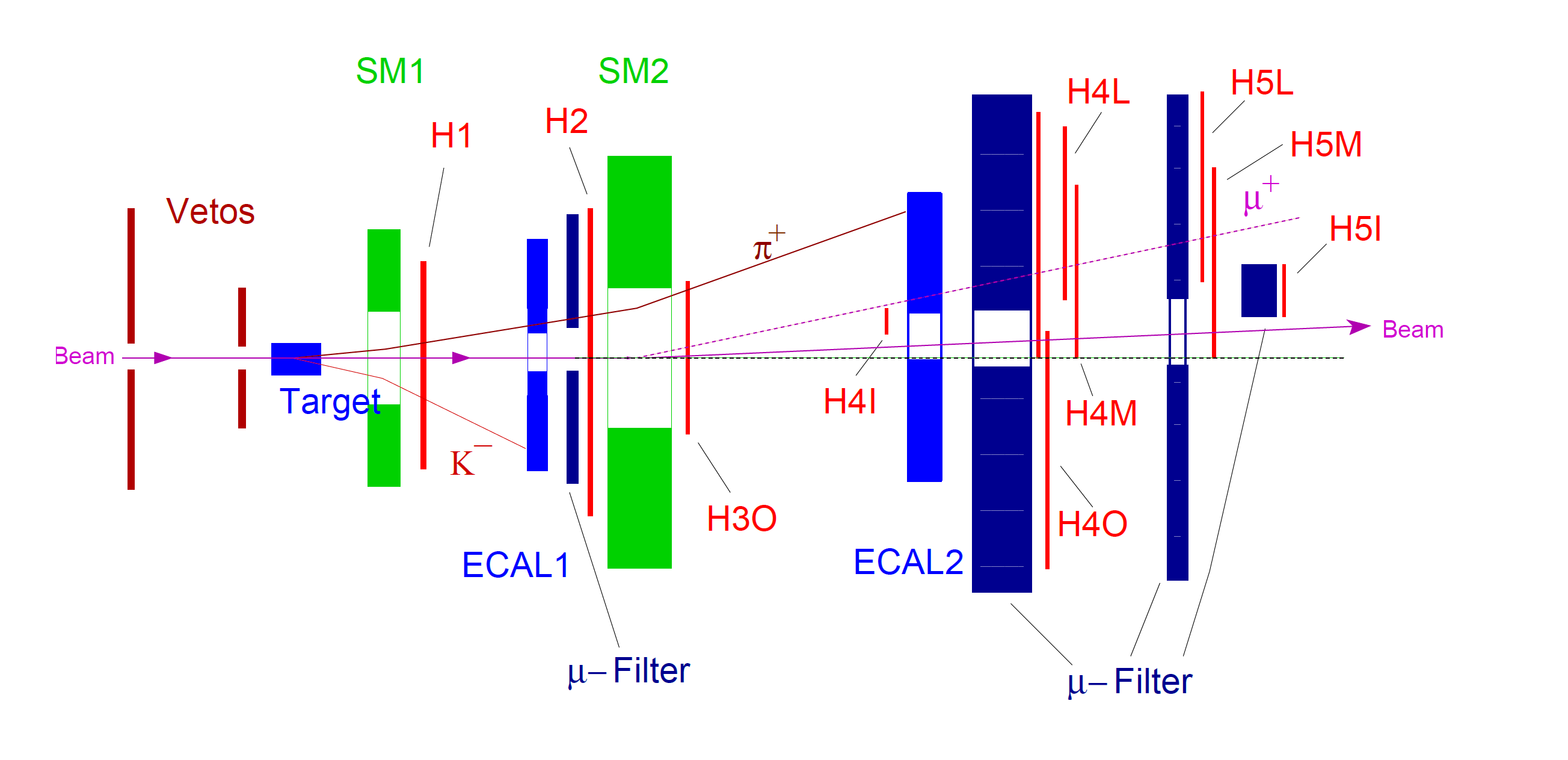}
    \caption{} \label{fig:trg_a}
  \end{subfigure}%
  \hspace*{\fill}   
  \begin{subfigure}{0.4\textwidth}
    \includegraphics[scale=0.2]{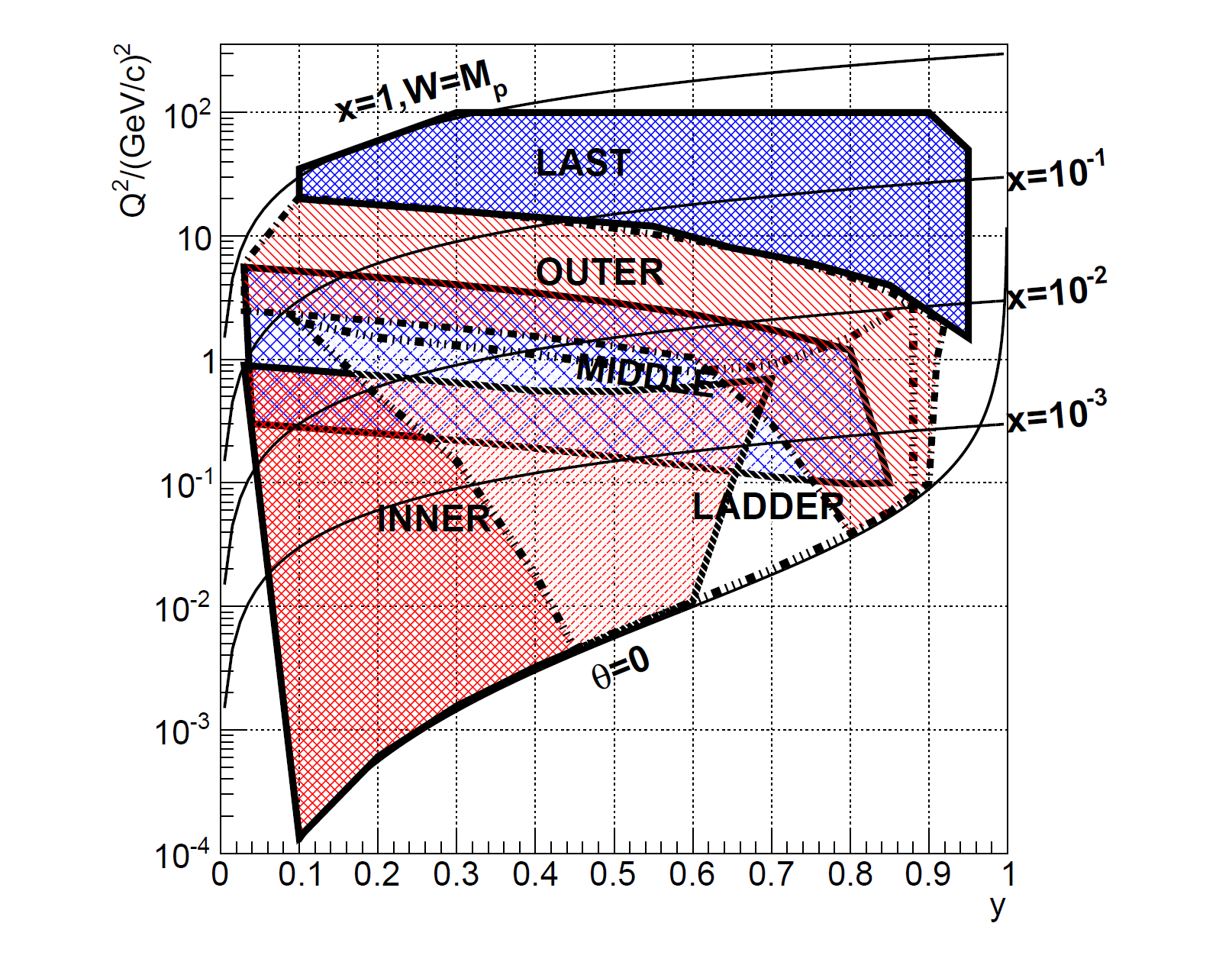}
    \caption{} \label{fig:trg_b}
  \end{subfigure}%
 \caption{(a):~A schematic overview of the muon trigger system \cite{Vidon:2019dbm,Bernet:2005yy}. (b):~The kinematic acceptance of the COMPASS muon trigger (source: COMPASS note 2016-4). }
 \label{fig:trg}
\end{figure}

The COMPASS data are recorded on an event-by-event basis. The readout process thus needs to be activated by an efficient trigger system, which has to rapidly identify good event candidates in a high-rate environment, and with a large kinematic coverage. Here we will concentrate on the muon trigger system \cite{Bernet:2005yy}, shown in Fig.~\ref{fig:trg_a}, not discussing the CAMERA trigger and the random trigger, also in use for the DVCS measurements of 2016 and 2017. 

The muon-based trigger decision is taken using hodoscope signals, energy deposition in the hadron calorimeters and a veto system. Its kinematic coverage in the $y,\Qsq$ plane is shown in Fig.~\ref{fig:trg_b}. At least two signals from different hodoscopes, of which at least one placed behind an absorber, are required. According to the scattered muon angle, two different trigger logics are used: target-pointing and energy-loss.

\subsubsection{Target-pointing trigger}
If the angle of the scattered muon in the vertical direction is large enough, it is possible to check the compatibility of the trajectory with the target position without reconstructing its momentum. This check is performed by using pairs of scintillating hodoscopes with horizontal strips, placed at different z positions along the beam axis. Only certain combinations of elements from both stations correspond to a possible interaction inside the target: to verify this, the signals from the two hodoscopes are sent to a coincidence matrix. A muon track that has not undergone an interaction will fail the coincidence. This method, illustrated in Fig.~\ref{fig:trig_mat1}, is used for the Middle Trigger (MT, plane H4M and H5M), the Outer Trigger (OT, planes H3O and H4O) and the Large Angle Spectrometer Trigger (LAST, planes H1 and H2), which cover different kinematic regions.

\subsubsection{Energy-loss trigger}
\begin{figure}[h!]
\captionsetup{width=\textwidth}
\captionsetup[subfigure]{labelfont=rm}
  \begin{subfigure}{0.45\textwidth}
     \includegraphics[width=\textwidth]{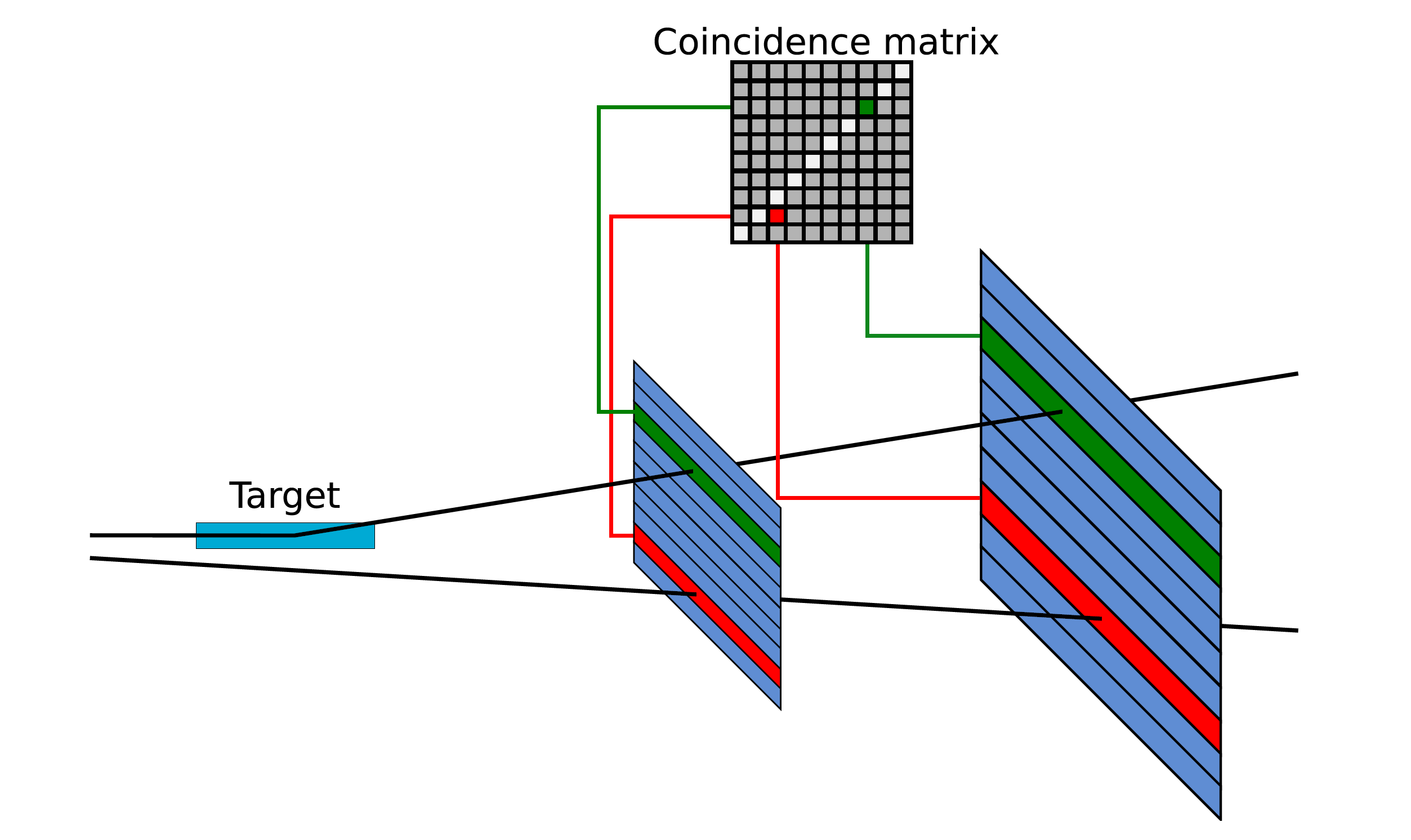}
    \caption{} \label{fig:trig_mat1}
  \end{subfigure}%
  \hspace*{\fill}   
  \begin{subfigure}{0.45\textwidth}
    \includegraphics[width=\textwidth]{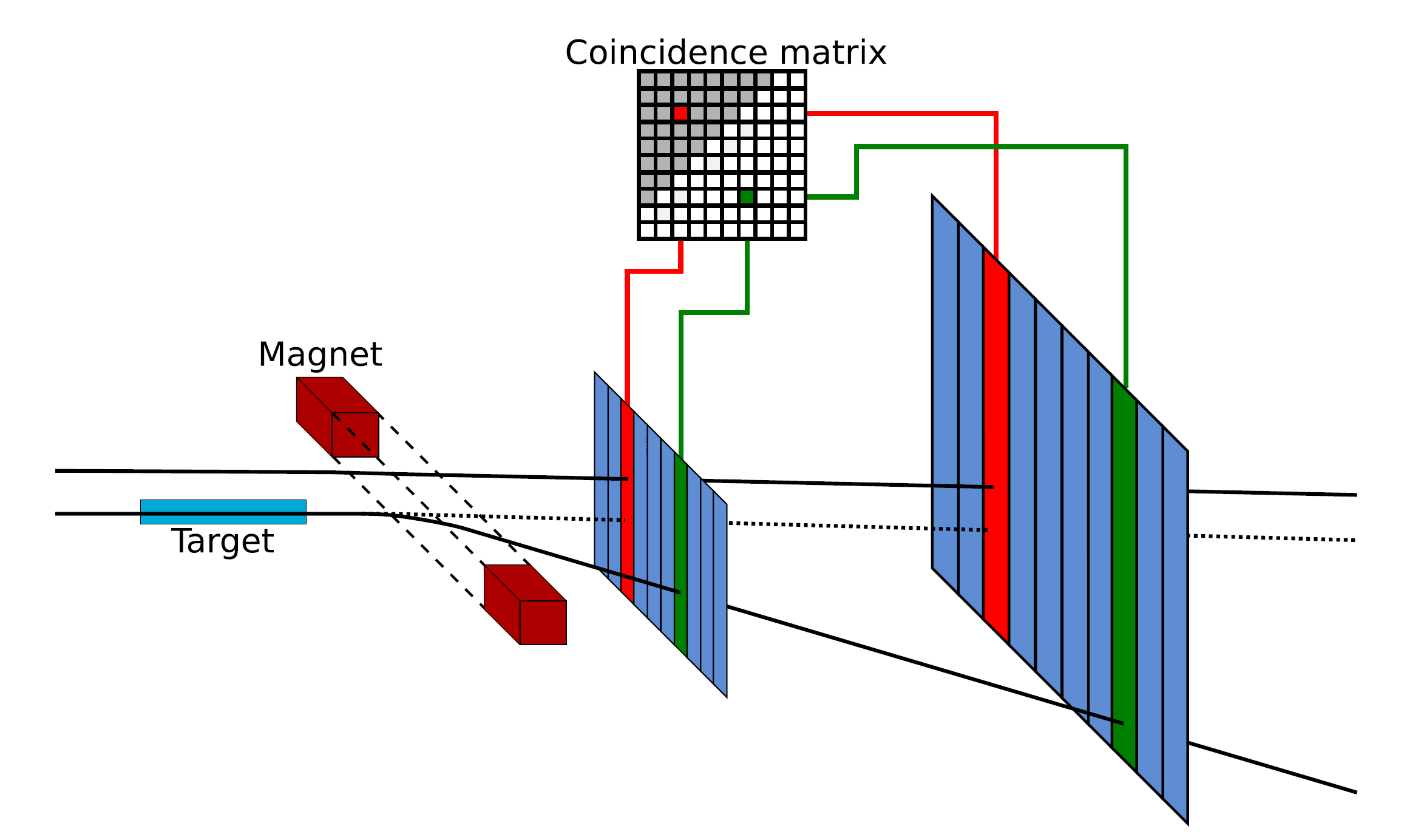}
    \caption{} \label{fig:trig_mat2}
  \end{subfigure}%
 \caption{Illustration of the target-pointing (left) and of the energy-loss method (right). A scattered muon results in a coincidence between the two strips of the hodoscopes in agreement with an interaction inside the target. A halo muon fails to produce such a coincidence. \cite{COMPASS:2007rjf}.}
 \label{fig:trig_mat}
\end{figure}

At small scattered muon angles, the target pointing is not accurate enough. In this case, the trigger decision is based on the muon candidate energy loss. The principle is illustrated in Fig.~\ref{fig:trig_mat2}: two vertical scintillator hodoscopes are used and all possible combinations between the two planes are used in a coincidence matrix, this time of triangular shape. This method is used for the Inner Trigger (IT, planes HI4 and HI5) and for the Ladder Trigger (LT, planes H4L and H5L). The large background coming from processes like the $\mu-e$ elastic scattering and the (quasi-)elastic radiative scattering off nuclei is suppressed, requiring signals in coincidence in the hadronic calorimeters. \\

Except for the Inner Trigger, a veto system is added to the trigger conditions to prevent triggers from halo muons. The veto systems consists of five scintillator hodoscope stations with a central hole for the beam. They are positioned at different distances to the beam to suppress different unwanted contributions. Two large veto hodoscopes suppress halo components further away from the beam, and three veto hodoscopes are build to suppress the part of the beam that does not go through the target. At least one signal from the five stations is needed to have a final veto signal. \\

A further trigger mechanism is implemented, not based on muons. It is the \textit{pure} calorimeter trigger (CT), which is based on the energy deposition of particles inside the hadronic calorimeters HCAL1 and HCAL2, and inside ECAL1. This trigger is used to extend the kinematic range of the trigger system towards larger photon virtualities and to trigger on events with a scattered muon outside the acceptance of the scintillating hodoscopes. This CT signal can also be used in coincidence with the signals from a hodoscope trigger to form a \textit{semi-inclusive} trigger. This is the case for the IT, LT and LAST triggers.

\section{Data acquisition}
\label{sect:data_acq}
The COMPASS data taking is organized in \textit{runs}, i.e. a collection of subsequent \textit{spills}. The time interval corresponding to a single run depends on the spill length; generally it is not larger than 30 minutes. Conventionally, a full run is made of 200 spills. The acquisition of the information from the detectors in coincidence with a trigger signal and the creation of a corresponding \textit{event} is done by the Data Acquisition system (DAQ). \\

The original COMPASS DAQ was designed to read the large number of detector channels (approximately 300 000) with an event rate up to 100 kHz and with the smallest possible deadtime: this required a dedicated design of the readout electronics and the readout-driver modules. A comprehensive description of the original system can be found elsewhere \cite{COMPASS:2007rjf,Schmitt:2004bh}. More recently (before the 2016 data taking), the old system has been upgraded \cite{Bodlak:2016vdt}. \\ 

In the original DAQ system, the front-end electronics (the lowest \textit{layer}) continuously preprocessed and digitized analog data from the detectors. Data from multiple channels were readout and assembled by the concentrator modules called CATCH, GeSiCA, and GANDALF. These modules also received the signals from the time and trigger system, so that the readout was performed at the arrival of the trigger signal. By adding the timestamp and the event identification to the data, a \textit{sub-event} was created. Up to this point, the new and original DAQ share the same logic. The next layer of the original DAQ was the \textit{event building network}, composed of \textit{readout buffers} and \textit{event builders}. The readout buffers were standard servers, equipped with custom PCI cards that allowed to distribute the data load through the full cycle of the SPS accelerator. Finally, subevents were sent over the Gigabit Ethernet to the event builders that assembled full events. Assembled events were stored temporarily on event builder’s local disks before being transferred to the CERN Advanced STORage manager (CASTOR). \\

\begin{figure}
    \captionsetup{width=\textwidth}
    \centering
    \includegraphics[width=\textwidth]{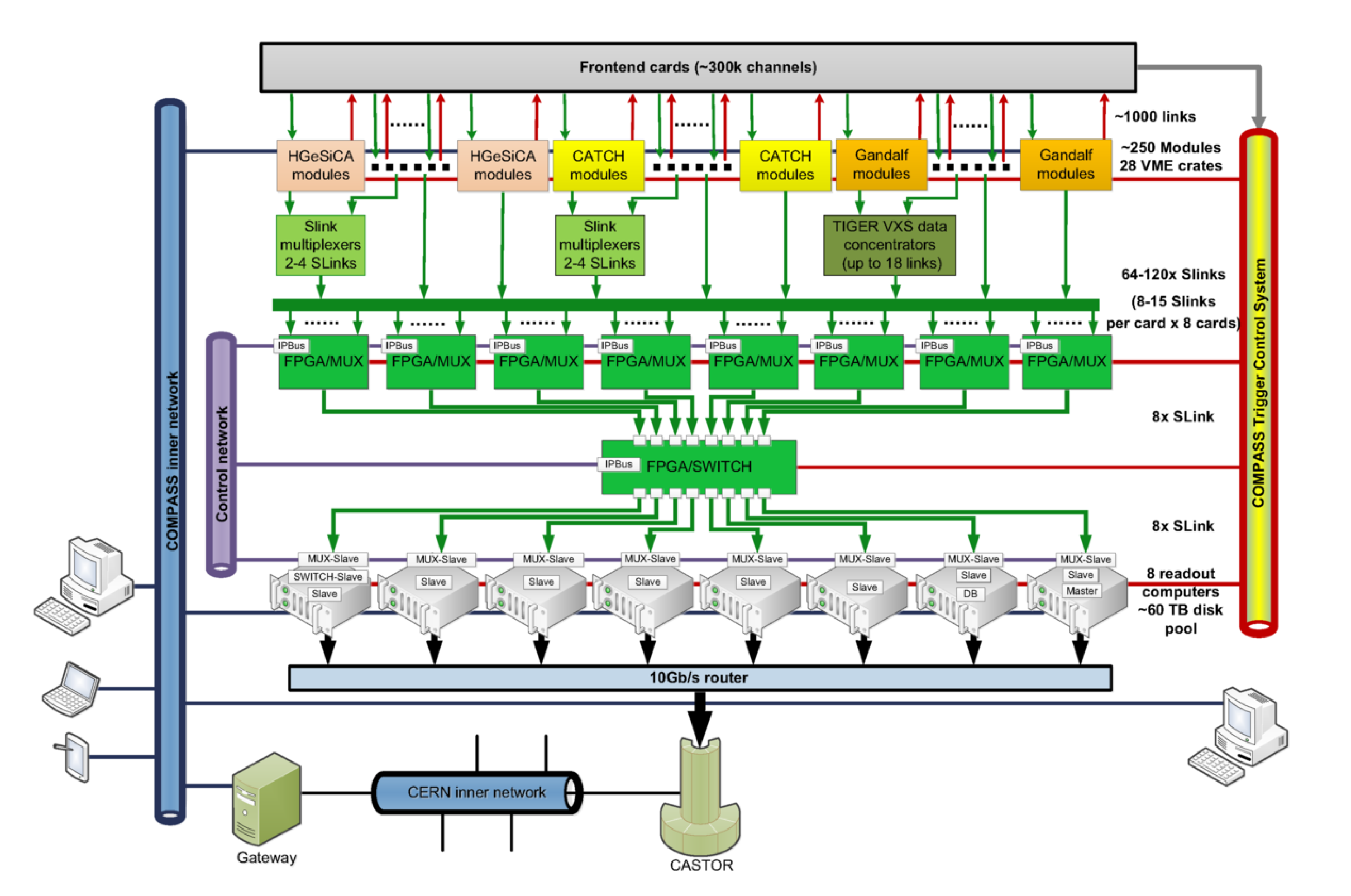}
    \caption{The COMPASS DAQ architecture \cite{Bodlak:2016vdt}.}
    \label{fig:daq}
\end{figure}

The event building network has been replaced with two layers of special FPGA Data Handling Cards (DHC), allowing also for online data consistency check and error recovery algorithms with the desired error tolerance level. From these layers, the full events are transferred to eight readout engine computers, where they are received and temporarily stored before being transferred to CASTOR. A schematic view of the new DAQ architecture is shown in Fig.~\ref{fig:daq}. Up to the DAQ level, an event is a collection of raw data collected at the trigger time. The physics content of these raw data is obtained with the event reconstruction.

\subsection{Online data quality monitoring}
The data taking conditions and the stability of the apparatus are constantly monitored during the data collection. Two software infrastructures are used to this end: the Detector Control System (DCS) and the COMPASS Object-Oriented OnLine (COOOL) software tool. \\

\subsubsection{The COMPASS Detector Control System}
The main aim of the DCS \cite{Bordalo:2012qc} is to provide the control of all the COMPASS setup parameters during the data taking. This includes the control of the high- and low-voltage systems, gas supplies, racks and crates with electronics, as well as the monitoring of slowly varying parameters (pressure, temperature, humidity etc.) both in the hall and in the vicinity of important parts of the spectrometer. Also, the status of experiment-wide infrastructure, like the cooling water system, is monitored in the DCS. The DCS comprises a device layer, a front-ends layer and a supervision layer, each characterized by a dedicated communication technology. In the supervision layer, the DCS uses a commercial SCADA (Supervisory Controls and Data Acquisition) system, called PVSS. Each device communicates with the PVSS through an OPC (Object Linking and Embedding for Process Control) server or through the Distributed Information Management (DIM) system, developed at CERN. To allow for an easier navigation in the system, the integration of new devices, the changing of setup parameters etc., other frameworks are implemented in the DCS as upper layers to the SCADA system. 

\begin{figure}[h!]
    \captionsetup{width=\textwidth}
    \centering
    \includegraphics[width=\textwidth]{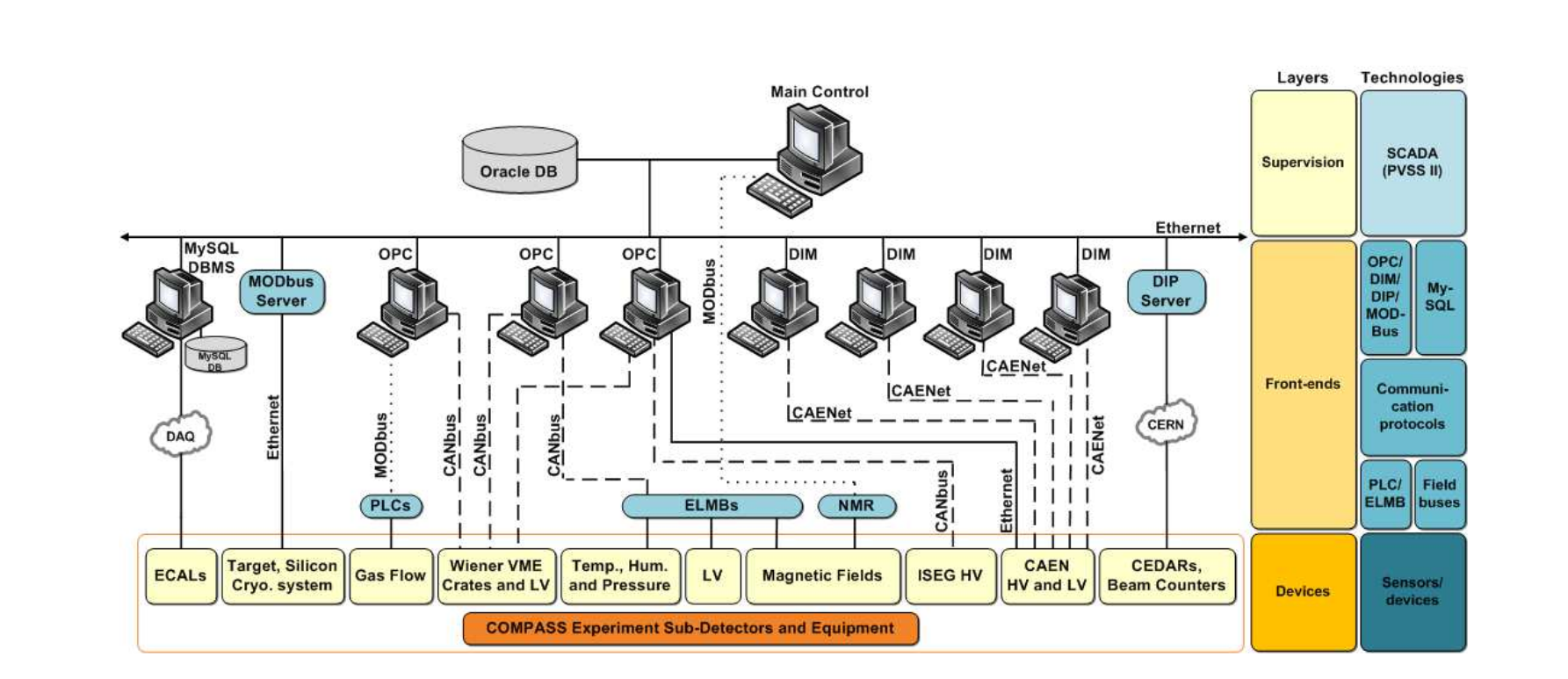}
    \caption{The COMPASS Detector Control System architecture, comprising a devices layer, a front-ends layer and a supervision layer. The technologies used in each layer are indicated in the rightmost column \cite{Bordalo:2012qc}.}
    \label{fig:dcs}
\end{figure}

\subsubsection{The COMPASS Object Oriented OnLine (COOOL) software tool}
The stability of the detectors can be monitored thanks to the COOOL software \cite{Bernet:2004vla}, which performs an online, run-by-run decoding of part of the raw data coming from the DAQ and produces numerous histograms, from which the good detector functioning can be judged. They are mostly detector plane profiles, which allow identifying, for instance, the following problems:
\begin{enumerate}
    \item the failure of the high-voltage supply of a detector, which can be identified in the temporal distribution or in the amplitude distribution of the collected hits. Indeed, a non-powered detector can produce only electronic noise, which shows a small signal amplitude and no correlation with the trigger;
    \item the failure of the low-voltage system, which powers the front-end electronics of a given detector, can be observed as a total absence of data coming from the detector;
    \item the failure of a front-end electronic card can be inferred from the absence of signal from a block of neighboring channels;
    \item the malfunctioning of a single electronic channel can be inferred by comparison with the neighboring ones.
\end{enumerate}
The histograms are continuously checked during the data taking and compared to those of the previous runs, in order to identify and solve as soon as possible emerging issues. The analysis of the COOOL histograms is very relevant offline too, before starting the complete data processing and the Monte Carlo simulation. In order to ensure stable conditions, it may be convenient to exclude from the reconstruction, for entire data taking periods, the unstable or noisy detector planes (or parts of them). Correspondingly, it is of primary relevance to exclude them in the simulations. 
 \\

\begin{figure}[h!]
    \captionsetup{width=\textwidth}
    \centering
    \includegraphics[width=0.53\textwidth]{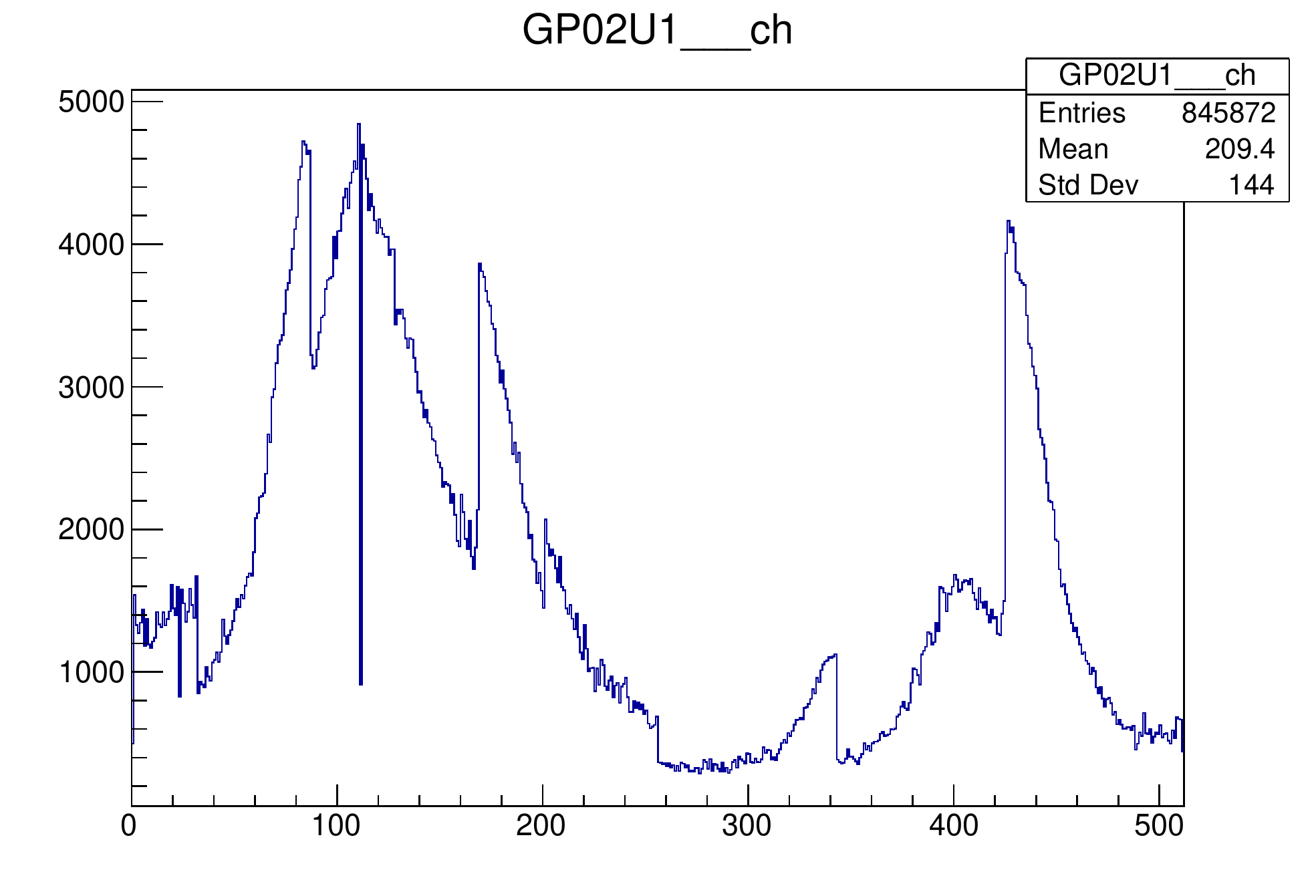}
    \includegraphics[width=0.45\textwidth]{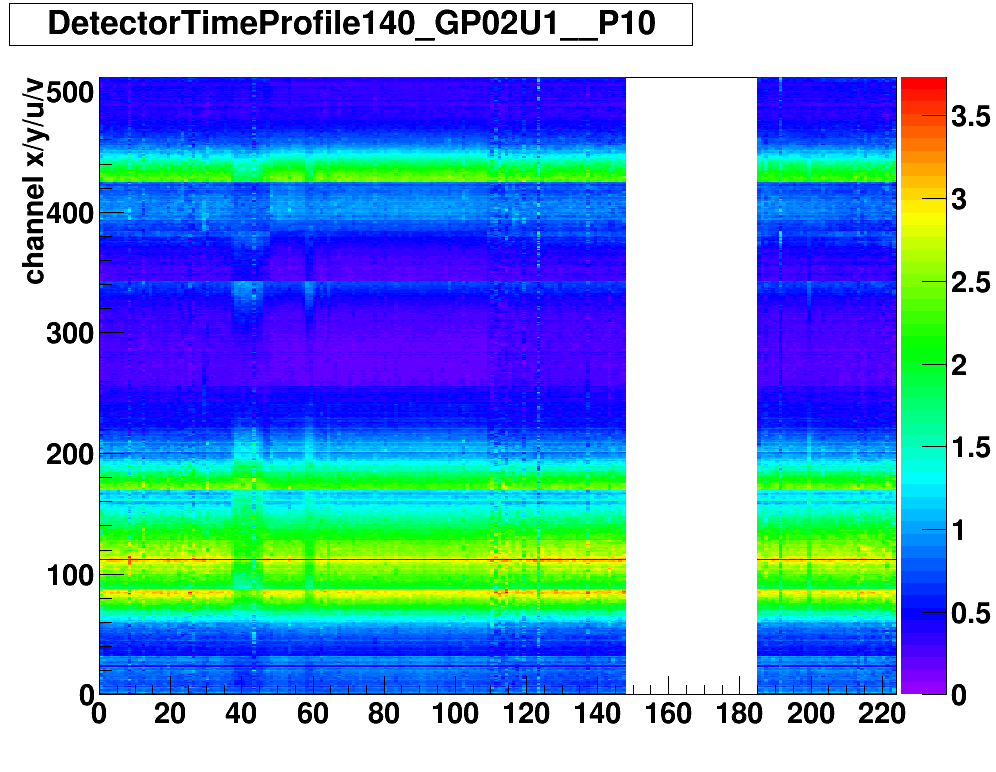}
    \caption{Left: an example of detector profile, for the pixel-GEM GP02, plane U1, period P10-2016, run 276405 (the 100th run in the period). Right: corresponding detector time profile, for the same plane, for the full period P10-2016. }
    \label{fig:coool1}
\end{figure}

A complete analysis of the COOOL histograms has been performed for the tracking detectors in use in 2016 and 2017, allowing to redact a period-based list of bad channels to be excluded in the reconstruction of the real data, of the Monte Carlo simulated data, or both of them. More than 330 planes have been inspected with a dedicated software in order to spot instabilities and malfunctioning along the data taking, according to the following recipe:
\begin{enumerate}
    \item for a given detector plane with a number of channels equal to $N_c$, the COOOL histogram with the number of hits per channel (the \textit{detector profile}) is analyzed; this is done for each of the $N_r$ runs in the considered period. Generally, a period amounts to a few hundred runs.
    \item the average detector profile is obtained from the sum of the $N_r$ profiles;
    \item its integral is used as normalization constant for each of the $N_r$ profiles: this allows to correct the histogram for the run time length, not always the same;
    \item the $N_r$ normalized profiles are drawn in a two-dimensional plot with $N_r$ on the horizontal axis and the channel number on the vertical axis (as shown in Fig.~\ref{fig:coool1}, see later for details), with a color scale for the number of hits per channel.
\end{enumerate}

An example of (not typical) \textit{detector profile} is given in Fig.~\ref{fig:coool1} (left) for the pixel-GEM GP02, plane U1, run 276405 (the 100th run of period P10-2016): each bin corresponds to an electronic channel and the bin content, here not normalized, corresponds to the signal collected in that channel. This specific profile enters, once properly normalized, in the \textit{detector time profile} of the period P10, given in the same Figure on the right. On the horizontal axis there is the number of runs ($N_r=224$) and on the vertical axis the number of channels ($N_c=512$). It appears clearly that the full detector was off for a block of runs (149-185). This has been taken into account in the Monte Carlo simulations. Some structures can be observed around run 40-45 and 60, where the channel content looks blurred: this has been investigated and associated to a beam instability. A malfunctioning (\textit{hot}) channel can also be observed (channel 117).

\section{Event reconstruction and data quality}
\label{sec:evt_rec}
\begin{figure}
    \captionsetup{width=\textwidth}
    \centering
    \includegraphics[width=0.65\textwidth]{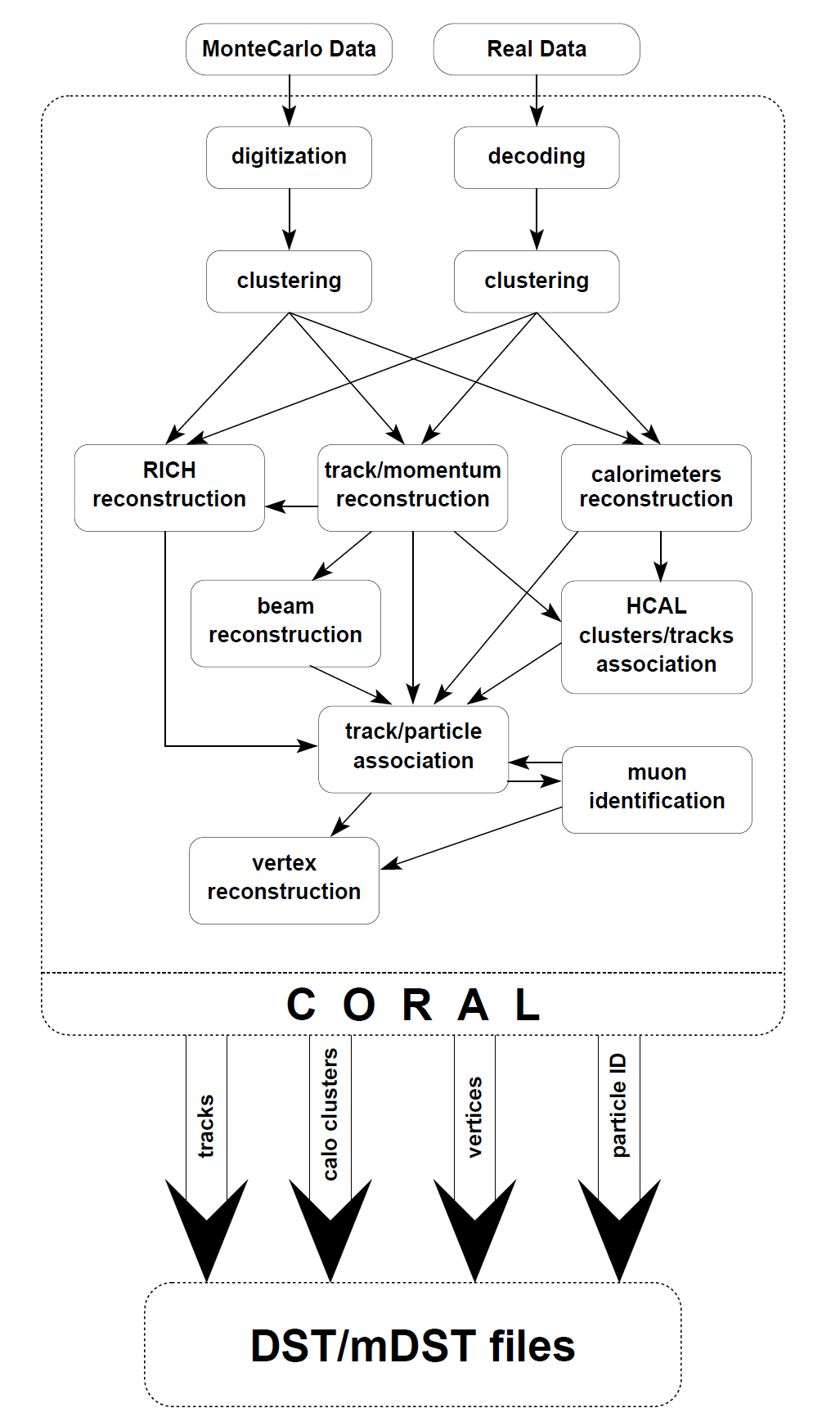}
    \caption{Schematic representation of the COMPASS reconstruction software \cite{COMPASS:2007rjf}.}
    \label{fig:coral}
\end{figure}

The raw data, stored in CASTOR by the DAQ, are processed in order to reconstruct the physics events. This process, also called \textit{production}, is performed offline using an object-oriented package called CORAL \cite{COMPASS:2007rjf, website:CORAL}. The input for CORAL are the raw data from the detectors, the information to decode them, the detector positions and calibrations, the map of the fields in the magnets and the maps of all the material present in the apparatus. It is also possible to provide CORAL with the information coming from the COOOL analysis about malfunctioning detectors. In turn, CORAL fits the particle tracks and the interaction vertices, calculates the number of radiation lengths passed by the particles in the spectrometer and performs the association of the tracks to the observed calorimeter clusters. A schematic representation of the CORAL functioning is given in Fig.~\ref{fig:coral}. As can be seen, similar steps are taken for the reconstruction of real data and Monte Carlo data. In the following, the RICH and calorimeters reconstruction will not be covered, as it is not needed for the analysis presented in this Thesis.
In the following subsections \ref{subs:det_pos} and  \ref{subs:mat_maps} the detector position and the material maps, which are necessary input to CORAL, are discussed; the track and vertex reconstructions are covered in \ref{subs:track_rec} and  \ref{subs:vert_rec}. \\
The output of CORAL is a set of ROOT trees \cite{Brun:1997pa} called mini Data Summary Tapes (mDSTs), which contain all the information of the reconstructed events. The information stored in the mDSTs is then analyzed with the software PHAST (PHysics Analysis Software Tools \cite{website:PHAST}), which also provides a set of algorithms to compute the relevant physics variables for each event.

\subsection{Detector position}
\label{subs:det_pos}
A precise knowledge of the position of the detectors in the experimental hall is required, and passed to CORAL as input: needless to say, if a detector is assumed to be in a wrong position, the momentum of the tracks of the particles can be biased and the reconstruction efficiency lowered. Since each plane has to be aligned with an uncertainty smaller than the detector resolution, the optical measurement of the detector positions (the \textit{survey}) is not enough, and it is followed by an \textit{alignment} procedure. 

Given the large number of detector planes, the alignment is a difficult task. The alignment procedure, which is also implemented and distributed as a part of the CORAL package, is based on the survey measurements and on the data collected during special alignment runs with low intensity and dedicated trigger and beam settings. It is a complex iterative procedure based on the minimization of a global $\chi^2$, calculated from the difference between the measured and estimated hit positions, where the estimation is done using reconstructed tracks excluding the plane under study.  using tracks. There are at least four free parameters for each plane: the rotation and the translation in the plane orthogonal to the nominal beam line and the translation along the beam. Once the alignment is completed on the special data, it is repeated on real data in order to check, in particular, that the magnetic fields do not introduce misalignment.

Several criteria can help in evaluating the goodness of the alignment results. The most relevant are: the number of reconstructed tracks per event, the $\chi^2$ distribution of these tracks, the number of reconstructed vertices per event, the number of tracks per vertices, the width of the mass distributions. Also, the $\chi^2$ associated to the vertex (explained later) can be useful, as it may point to a wrong resolution $\sigma$ for the detector involved in the reconstruction of the tracks associated to that vertex or to the need for a revision of the track model (as, e.g., regarding the multiple scattering).

\subsection{Material maps}
\label{subs:mat_maps}
\begin{figure}
    \captionsetup{width=\textwidth}
    \centering
    \includegraphics[width=\textwidth]{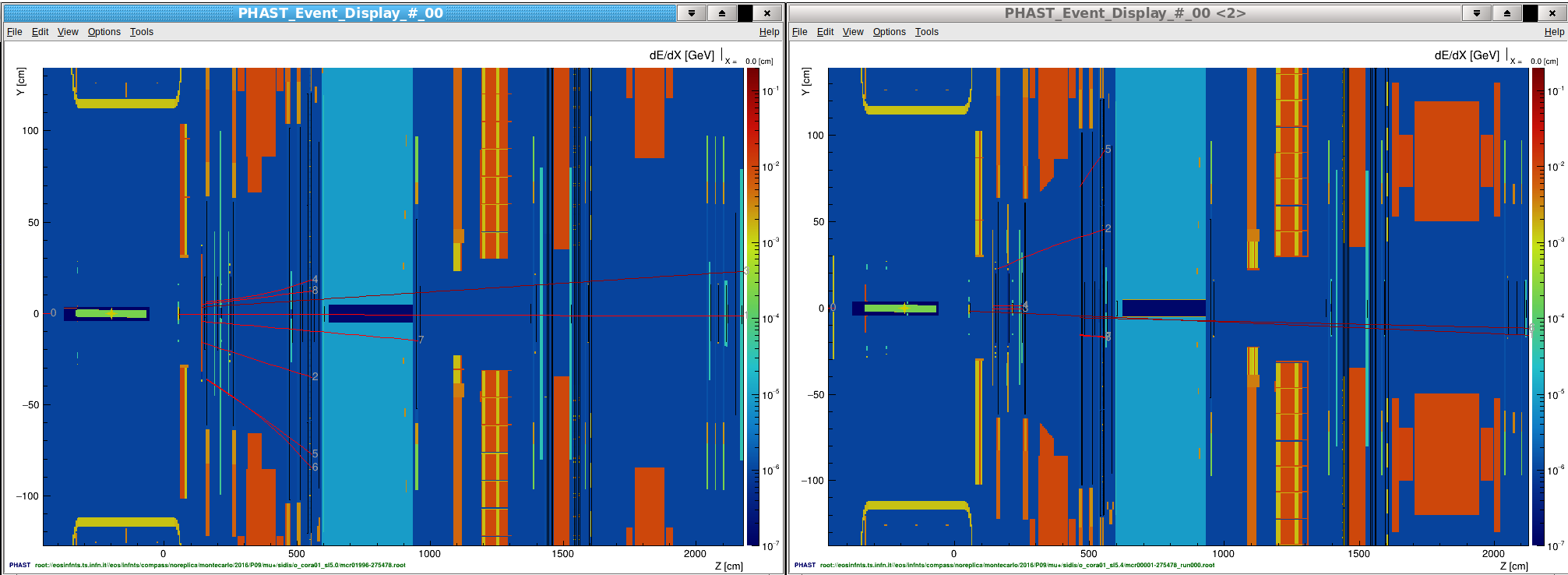}
    \caption{Examples of material maps from PHAST. Plot by A.~Bressan (COMPASS). }
    \label{fig:matmap}
\end{figure}

Another input to the CORAL reconstruction algorithm are the material maps, used to identify the scattered muons and to take into account the multiple scattering in the fitting procedure. They encode the three-dimensional distribution of the material of which the various elements of the spectrometer are composed. Assuming that the traversing particles are minimum-ionizing, the material map can be converted into an energy loss map, as the one shown in Fig.~\ref{fig:matmap}, where the red color indicates high energy loss and blue indicates negligible energy loss.
The material maps used in the reconstruction are stored in the mDST and can be accessed with PHAST.

\subsection{Track reconstruction}
\label{subs:track_rec}
The reconstruction of the tracks is done in three steps. In the first step, the hits observed in the spectrometer detectors are grouped in three regions according to their position (before SM1, between SM1 and SM2, after SM2). A pattern recognition is then used to find clusters that are consistent with track segments, expected to be straight lines in each of the three regions. The track finding algorithm starts using projections over given directions. The reconstructed path is given an uncertainty due to the detector resolution; the path is then used to find further hits along a possible track. The information from all projections is then combined to determine the track in three dimensions. In the second step, the segments are connected through the magnetic fields in a procedure called \textit{bridging}; usually, to have a good reconstruction, the last hit associated to a track is required to be after SM1. The combination of track segments is based on the $\chi^2$ of the pair; combinations with a bad $\chi^2$, or containing segments already used for an accepted combination, are rejected. In the third and last step, a fit of the tracks is performed using a Kalman filter \cite{Fruhwirth:1987fm}, using the magnetic field and the material maps to estimate the best parameters of the track.

\subsection{Vertex reconstruction}
\label{subs:vert_rec}
Two kinds of vertices are reconstructed, i.e. \textit{primary} and \textit{secondary} vertices. Primary vertices have an incoming muon track (reconstructed in the beam telescope) and one or more outgoing tracks; secondary vertices correspond to decays of neutral particles into two oppositely charged particles, thus no incoming track is observed in this case. 
The search for the primary vertex starts from the identification of the position in which the event tracks are at minimum distance. If a track is too far from the point, it is not considered in this procedure. A $\chi^2$ value is associated to the point and the contribution of each track to the overall $\chi^2$ is computed with a Kalman filter; tracks with a too large $\chi^2$ contribution can be rejected, and the procedure is repeated until the reduced $\chi^2$ has the desired value. In general, there can be more than one incoming muon tracks: as a consequence, more than one primary vertices can be reconstructed per event. The choice of the \textit{best} primary vertex is done at the PHAST level, by comparing the $\chi^2$ values of the various vertices and the number of tracks associated to each of them.
The secondary vertices are reconstructed by combining all possible track pairs with opposite charge. A Kalman filter technique is used to find the decay position as the point in which the tracks are at minimum distance. In general, a track can be associated to both a primary and a secondary vertex.

\subsection{Data quality}
The online quality monitoring, discussed in Sect.~\ref{sect:data_acq}, does not require the production of the data, as it is performed in real-time on the data being collected by the DAQ. After the production, however, it is necessary to ensure the reliability and stability of the data with further inspections, first on a spill-by-spill- and then on a run-by-run basis.

\subsubsection{Bad spill rejection}
The spill quality is inspected by comparing each spill to the neighboring ones. Among the quantities that are compared for SIDIS measurements there are:
\begin{itemize}
    \item the number of primary vertices per event and per trigger;
    \item the number of tracks per primary vertex and trigger;
    \item the number of triggers (normalizing to the flux integrated over the spill). 
\end{itemize}
If a spill is observed to behave differently with respect to the neighboring spills, this can be an due to beam instabilities, or to instabilities in the detector efficiencies: in this case, the spill is marked as \textit{bad} and excluded from the analysis. Several other rejection criteria can be introduced for specific purposes and detectors: this is the case for the RICH and for the calorimeters, for which dedicated badspill lists can be produced.

\subsubsection{Bad run rejection}
The inspection of the run quality is done after the rejection of the bad spills, and it is based on the comparison of a set of kinematic distributions, produced for each run. For the 2016 data, the considered variables were: the x, y and z position of the primary vertex; the energy of the beam, $x$, $y$, $Q^2$, the polar and azimuthal angles of the scattered muon $\theta_{\mu^\prime}$ and $\phi_{\mu^\prime}$, the energy $E_h$ and the polar and azimuthal angles $\theta_{h}$ and $\phi_{h}$ of the hadron, its transverse momentum, the mass of the reconstructed $K_0^s$. Energies and angles are those measured in the laboratory system. Based on these variables, each run has been compared to all the other runs in the same period, with the same beam charge. The tag of \textit{bad run} is not always easily associated to a malfunctioning of detector planes as observed with COOOL. In general, the impact of the detector planes on the bad run definition depends on the time interval in which a detector plane was not performing well and on the amount of tracks whose reconstruction was strongly dependent on that plane.

For the 2016 data considered in this work, the rejection of bad spills and bad runs corresponded to a $\sim 10\%$ decreaseof the available statistics. 

\section{Event simulations}
\label{sec:evt_sim}
The measurements performed with an unpolarized target require an extensive use of Monte Carlo (MC) simulations for the determination of the spectrometer acceptance and for the background evaluation. Here, the COMPASS MC chain is briefly presented. \\

The simulation of the apparatus, as well as the simulation of the passage of particles through it, is done within the TGEANT package \cite{szameitat}, based on GEANT4 \cite{website:GEANT4}. The events of the physics process of interest are provided to TGEANT by an event generator. For the generation of DIS events, the common choice in COMPASS is LEPTO, v.~6.5.1 \cite{Ingelman:1996mq}, which is based on the LUND string fragmentation model for the quark fragmentation; it refers to PYTHIA-JETSET \cite{Sjostrand:1995iq} for the hadronization process and has access to the LHAPDF library \cite{Giele:2002hx} for the PDFs. HEPGEN \cite{Sandacz:2012at} is used to generate hard exclusive electroproduction events; the $Q^2$-dependence of the diffractive cross-section implemented in HEPGEN is based on NMC results \cite{NewMuon:1994xtx} tuned to COMPASS data. DJANGOH \cite{Aschenauer:2013iia}, a DIS generator based on LEPTO integrating HERACLES \cite{Kwiatkowski:1990es} for electroweak radiative effects and SOPHIA \cite{Mucke:1999yb} for low hadronic masses (instead of JETSET), is being tested for an estimate of the impact of radiative effects.

TGEANT uses the same geometry of the apparatus as CORAL and contains all the information on the detectors dead zones. In addition, the two software tools utilize the same material maps and magnetic field maps. TGEANT performs the transportation of the particles taking into account the selected physical processes. The digitization of the signals generated in the detectors is also provided and constitutes, together with the information on the generated particles, its main output. With the current graphic interface, the user has almost full control on all the ingredients of the TGEANT simulation: it is possible to set the beam particle type and its momentum distribution, to indicate the active detectors and their setting (the position of the trackers, in particular), to import efficiency maps and alignment files. CORAL analyzes the simulated data using the same reconstruction code used for the real data: only the decoding of the raw data is different: as for the real data, also Monte Carlo reconstructed (and generated) events are stored in mDST, that can be accessed and analyzed with the aforementioned software tools.

\subsection{Efficiency- and Pseudo-efficiency maps}
The efficiency of each detector plane is a fundamental quantity that has to be taken into account in the Monte Carlo simulation. In the simplest approach, a mean value of the detector efficiency can be provided to the reconstruction algorithm via the \texttt{detectors.dat} file. However, this may be not enough, if the efficiency is not uniform over the detector active area. For this reason, two-dimensional efficiency maps are produced from data samples and later used in the reconstruction of the Monte Carlo events.

As is the case for the alignment procedure, the production of efficiency maps is a complex task. The efficiency of each detector plane has to be evaluated by comparing the number of expected and observed hits, with an accuracy that should be better than the detector resolution. Also, the efficiency of each plane is to be evaluated, not considering the contribution to the track reconstruction given by the plane. A hit is said to be expected whenever the considered track has crossed the detector: this is the case, for example, when one hit is recorded upstream and another hit is recorded downstream of the chosen detector, with the line connecting the two hits crossing the detector active area. 

\begin{figure}[h!]
    \captionsetup{width=\textwidth}
    \centering
    \includegraphics[width=0.9\textwidth]{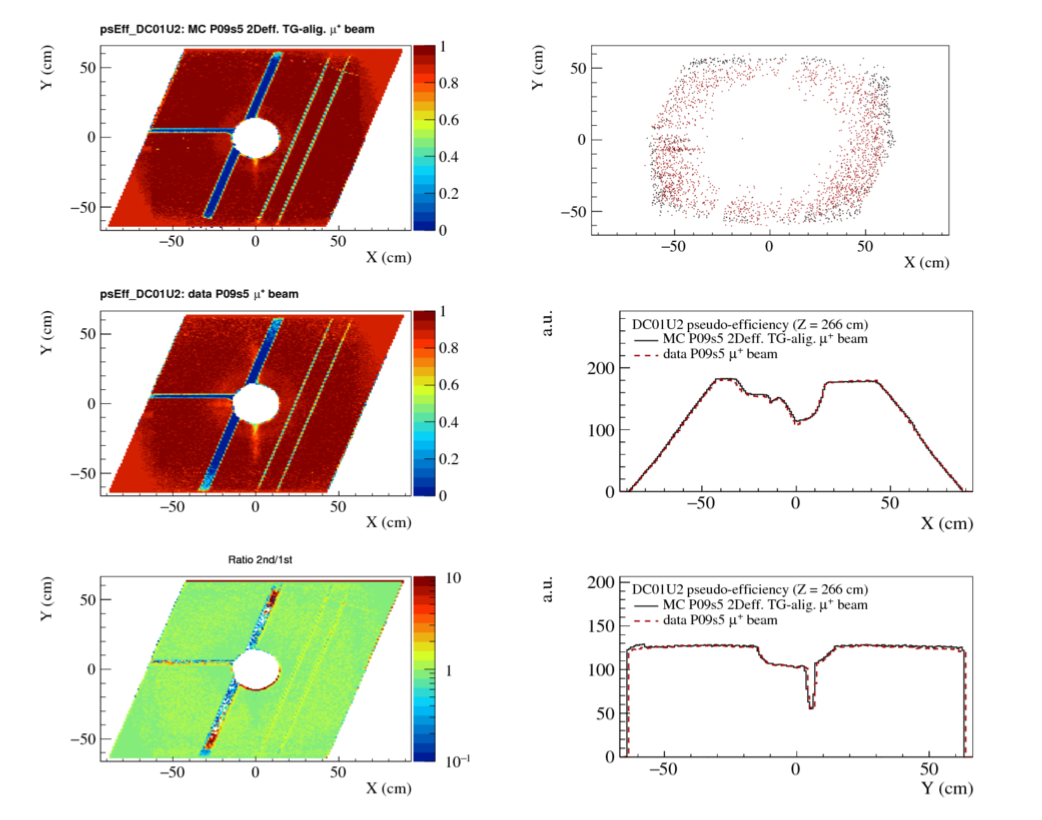}
    \caption{Examples of pseudo-efficiency comparison for Monte Carlo and real data: the case of DC01U2 plane, period P9. Plot by J.~Matousek (COMPASS).}
    \label{fig:pseudoeff}
\end{figure}

If the efficiency of the detector planes is correctly taken into account in the Monte Carlo, the pseudo-efficiencies for data and Monte Carlo are expected to agree. By pseudo-efficiency one means, again, the ratio of the expected and observed hits in a detector plane. This time, however, the detector at stake is not removed from the reconstruction of the track: this biases the determination of the efficiency in an unknown way, but this does not prevent a comparison of different samples. For the 2016 data, a comparison of the pseudo-efficiencies of real data and Monte Carlo has been performed on the same mDST format also used in analysis. An example of comparison is given in Fig.~\ref{fig:pseudoeff} for the DC01U2 plane, period P9, production slot5: the Monte Carlo (top left) and the data (central left) agree very well as for both the x- (central right) and y-projections (bottom right).

\chapter{Data analysis} 

\label{Chapter3_Data_analysis}

This Chapter summarizes the analysis steps that are common to both the measurement of the transverse-momentum distribution and of the azimuthal asymmetries, starting from the reconstructed events on mDSTs. The strategy of the analysis is: first, the Deep Inelastic Scattering events and the final-state hadrons are selected with proper kinematic cuts. Secondly, the hadrons produced in the decay of diffractive vector mesons are removed or subtracted using the Monte Carlo. Then, the acceptance corrections are applied, and the results are finally obtained. The results will be presented in dedicated Chapters (Ch.~\ref{Chapter4_PT-distributions} for the $\Ptsq$-distributions and Ch.~\ref{Chapter5_Azimuthal_asymmetries} for the azimuthal asymmetries). In this Chapter, the data samples are presented in Sect.~\ref{sect:ch3_samples}, the event and hadron selections are described in Sect.~\ref{sect:ch3_ev_had_sel} and some kinematic distributions are shown in Sect.~\ref{sect:ch3_kinematic}. The contribution of exclusive hadrons to the selected samples is discussed in Sect.~\ref{sect:ch3_exclusive}. The acceptance corrections are defined in Sect.~\ref{sect:ch3_acceptance}, while possible systematic effects are discussed in Sect.~\ref{sect:ch3_systematics}.

\section{Data samples}
\label{sect:ch3_samples}
The data used in this analysis have been collected during three periods (denoted as P08, P09 and P10) of the 2016 data taking, corresponding to about 11\% of the whole statistics collected in 2016 and 2017. Each of these three periods is divided into two sub-periods of balanced statistics, according to the charge of the muon beam. The samples have been pre-filtered into micro-DSTs ($\mu$DSTs) asking for at least one primary vertex with measured incoming muon momentum, at least one outgoing muon with the same charge of the incoming one and for $Q^2>0.8$~(GeV/$c$)$^2$. Bad spills and bad runs have been removed from the $\mu$DST samples before all the other cuts. The resulting statistics is given in the first line of Tab.~\ref{tab:devcuts}. Given the stability of the data taking and of the detector performances, the three periods are expected to give compatible results. \\

Four sets of Monte Carlo events have been used, having as event generators:
\begin{enumerate}
    \item LEPTO \cite{Ingelman:1996mq}, for the simulation of SIDIS data;    
    \item HEPGEN-$\rho$ \cite{Sandacz:2012at}, for the simulation of the diffractive production and decay of the $\vmrho$ vector mesons ($\vmrho \to \pi^+ \pi^-$, $BR\sim100\%$ \cite{Zyla:2020zbs}) and the angular modulations of the decay pions, induced by the Spin Density Matrix Elements (SDMEs, discussed in detail in Ch.~\ref{Chapter6_SDMEs});
    \item HEPGEN-$\phi$ to describe, similarly to the $\vmrho$ case, the  diffractive production and decay of the $\vmphi$ meson ($\vmphi \to K^+K^-$, $BR=49.2\%$ \cite{Zyla:2020zbs});
    \item HEPGEN-$\omega$ to account for the $\omega$ diffractive production and decay ($\omega \longrightarrow \pi^+\pi^-\pi^0$, $BR=89.2\%$ \cite{Zyla:2020zbs}).
\end{enumerate}
For each set, two samples have been produced (one per each beam charge), for a total of eight samples. Given the stability of the apparatus in the three considered periods, the four sets have been generated according to the geometry of the central period (P09). The reconstruction of the Monte Carlo events has been performed on the Trieste computing farm in January 2020, using the same CORAL version as for the data production and taking into account the 2D-efficiency maps extracted from the data. The number of reconstructed events for each Monte Carlo sample is given in the first line of Tab.~\ref{tab:MCevcuts}. At variance with LEPTO, the events generated with HEPGEN are characterized by a \textit{weight}, which encodes information on the phase space and on various kinematic dependences of the diffractive cross-section, and with an \textit{s-weight} (\textit{s} for SDME) for the angular dependences. The statistics of Tab.~\ref{tab:MCevcuts} is given assuming both weight and s-weight equal to 1, while the properly weighted statistics is given in Tab.~\ref{tab:MCevcuts2}.

\section{Event and hadron selection}
\label{sect:ch3_ev_had_sel}
The requirements, applied to both real- and Monte Carlo data to select events produced in Deep Inelastic Scattering, are:
\begin{itemize}
    \item \textbf{primary vertex:} the primary vertex (PV) is required to be the ``best'' according to the PHAST definition, based on the vertex $\chi^2$ and on the number of outgoing tracks; the position of the PV must be inside the fiducial target volume.
    \item \textbf{incoming muon $\mu$:} the beam muon must be reconstructed in the BMS and its momentum must satisfy the condition $140<P_{\mu}/(\mathrm{GeV}/c)<180$; the reconstructed track must cross the whole target length and have a good $\chi^2$.
    \item \textbf{scattered muon $\mu^\prime$:} to be identified as the scattered muon, an outgoing track must have crossed material corresponding to a large number of radiation lengths ($X/X_0>15$). Moreover, it is required to have the first (last) hit before (after) the SM1 magnet and a small $\chi^2$. The unambiguous definition of the scattered muon is also ensured by the rejection of all the events with more than one muon candidate and of those in which at least one track, different from the $\mu^\prime$ candidate and with the same charge, is reconstructed after the muon filter MF2.
    \item \textbf{kinematic cuts}. The DIS event selection has been done asking for $Q^2>1$~(GeV/$c$)$^{2}$ and $W>5$~GeV/$c^2$, to ensure a large enough photon virtuality and to avoid the hadron resonance region. In addition, the Bjorken variable $x$ has been selected to be in the range $0.003<x<0.130$ (the upper limit being fixed by the reduced acceptance of the apparatus). Events with $y<0.1$ or $y>0.9$ have been removed, where the lower value has been chosen in order to ensure a precise measurement of the kinematic variables and the upper one to limit the impact of the radiative effects. One last requirement was $\theta_{\gamma^*}<60$~mrad, where $\theta_{\gamma^*}$ is the polar angle of the virtual photon calculated in the laboratory system with respect to the incoming muon direction. This condition, reproducible in phenomenological analyses, was already introduced in Ref.~\cite{COMPASS:2014kcy} and it has been replicated here, to reduce the corrections due to the spectrometer acceptance. Its effect is to discard hadrons with large polar angle in the laboratory, for which large acceptance corrections would be needed. 
    \item \textbf{trigger selection}: at least one of the among the Middle Trigger (MT), the Ladder Trigger (LT), the Outer Trigger (OT) and the Large Area Spectrometer Trigger (LAST) must have been fired.\\
\end{itemize}
The effect on the sample statistics of the various blocks of cuts is shown in Tab.~\ref{tab:devcuts} for the three periods P08, P09 and P10, separately for the $\mu^+$ and $\mu^-$ cases (real data). The percentages of events passing the cuts, given in italic, indicate a good agreement among the periods. Summing over $\mu^+$ and $\mu^-$, the number of selected events in the real data is about 5$\cdot$10$^6$. Analogously, Tab.~\ref{tab:MCevcuts} and Tab.~\ref{tab:MCevcuts2} illustrate the impact of the various cuts on the number of events and on the number of weighted HEPGEN events, respectively. The Monte Carlo samples are statistically limited: in particular, for both $\mu^+$ and $\mu^-$, the ratio $R$ of selected events (LEPTO/data) is $R \approx 1.3$.
\newline

\begin{table}[h!]
\small
\captionsetup{width=\textwidth}
\centering
\begin{tabular}{c|rrr|rrr}
	\hline
	& \multicolumn{3}{c|}{$\mu^+$} & \multicolumn{3}{c}{$\mu^-$}\\
    & \multicolumn{1}{c}{P08} & \multicolumn{1}{c}{P09} & \multicolumn{1}{c|}{P10} & \multicolumn{1}{c}{P08} & \multicolumn{1}{c}{P09} & \multicolumn{1}{c}{P10}\\ \hline
 
All events in $\mu$DSTs 
        & 7530495 & 5790426 & 6224393 & 7326779 & 5551404 & 5449564 \\
        & \it{100.00} & \it{100.00} & \it{100.00} & \it{100.00} & \it{100.00} & \it{100.00} \\
Primary vertex
        & 4038922 & 3100141 & 3327774 & 3924362 & 2990704 & 2919125 \\
        & \it{53.63} & \it{53.54} & \it{53.46} & \it{53.56} & \it{53.87} & \it{53.57} \\
Incoming muon $\mu$
        & 3907032 & 2999056 & 3216331 & 3796819 & 2896812 & 2825161 \\
        & \it{51.88} & \it{51.79} & \it{51.67} & \it{51.82} & \it{52.18} & \it{51.84} \\
Scattered muon $\mu^\prime$
        & 3462480 & 2667380 & 2827736 & 3227142 & 2467859 & 2404080 \\
        & \it{45.98} & \it{46.07} & \it{45.43} & \it{44.05} & \it{44.45} & \it{44.12} \\ 
Kinematic cuts
        & 1055756 & 812872 & 885585 & 975147 & 747146 & 722712 \\
        & \it{14.02} & \it{14.04} & \it{14.23} & \it{13.31} & \it{13.46} & \it{13.26} \\    
Trigger
        & 1006194 & 775981 & 844526 & 932528 & 714634 & 691747 \\
        & \it{13.36} & \it{13.40} & \it{13.57} & \it{12.73} & \it{12.87} & \it{12.69} \\  
    \hline
\end{tabular}
\caption{Number of events after each cut applied to select DIS events, starting from the events on $\mu$DSTs. The values in italic show the percentage of events passing the different block of cuts.}
\label{tab:devcuts}
\end{table}

\begin{table}[!h]
\centering
\small
\captionsetup{width=\textwidth}
\begin{tabular}{c|rrrr|rrrr}
	\hline
	& \multicolumn{4}{c|}{$\mu^+$} & \multicolumn{4}{c}{$\mu^-$}\\
        & LEPTO & HG-$\rho$ & HG-$\phi$ & HG-$\omega$ & LEPTO & HG-$\rho$ & HG-$\phi$ & HG-$\omega$ \\
\hline
All events in $\mu$DSTs 
        & 9949854 & 513943 & 957302 & 191113 & 8360162 & 1935317 & 959754 & 484882 \\
        & \it{100.00} & \it{100.00} & \it{100.00} & \it{100.00} & \it{100.00} & \it{100.00} & \it{100.00} & \it{100.00}\\
Primary vertex
        & 9175031 & 464203 & 862601 & 172462 & 7709180 & 1748109 & 864995 & 437821 \\
        & \it{92.21} & \it{90.32} & \it{90.11} & \it{90.24} & \it{92.21} & \it{90.33} & \it{90.13} & \it{90.29} \\
Incoming muon $\mu$
        & 9019460 & 444250 & 825748 & 165018 & 7619448 & 1675651 & 829407 & 419650 \\
        & \it{90.65} & \it{86.44} & \it{86.26} & \it{86.35} & \it{91.14} & \it{86.58} & \it{86.42} & \it{86.55} \\
Scattered muon $\mu^\prime$
        & 8200168 & 348058 & 642590 & 131250 & 6929712 & 1315885 & 648042 & 334620 \\
        & \it{82.41} & \it{67.72} & \it{67.13} & \it{68.68} & \it{82.89} & \it{67.99} & \it{67.52} & \it{69.01} \\
Kinematic cuts
        &  3853909 & 165283 & 303019 & 62228 & 3266473 & 627653 & 308424 & 159091 \\
        & \it{38.73} & \it{32.16} & \it{31.65} & \it{32.56} & \it{39.07} & \it{32.43} & \it{32.14} & \it{32.81} \\
Trigger
        & 3525109 & 136847 & 252411 & 51490 & 2989137 & 519177 & 256823 & 131523 \\
        & \it{35.43} & \it{26.63} & \it{26.37} & \it{26.94} & \it{35.75} & \it{26.83} & \it{26.76} & \it{27.12} \\
    \hline
\end{tabular}
\caption{Number of Monte Carlo events after each cut applied to select DIS events, starting from the events on $\mu$DSTs. Here, each event has a weight equal to 1. The values in italic show the percentage of events passing the corresponding block of cuts.}
\label{tab:MCevcuts}
\end{table}

\begin{table}[!h]
\centering
\small
\captionsetup{width=\textwidth}
\begin{tabular}{c|rrr|rrr}
	\hline
	& \multicolumn{3}{c|}{$\mu^+$} & \multicolumn{3}{c}{$\mu^-$}\\
        & HG-$\rho$ & HG-$\phi$ & HG-$\omega$ & HG-$\rho$ & HG-$\phi$ & HG-$\omega$ \\
\hline
All events in $\mu$DSTs 
        & 220266.6 & 98484.9 & 8326.8 & 846040.1 & 97524.0 & 21214.7 \\
        & \it{100.00} & \it{100.00} & \it{100.00} & \it{100.00} & \it{100.00} & \it{100.00}\\
Primary vertex
        & 197154.6 & 88061.3 & 7412.1 & 754572.0 & 86839.1 & 18900.3 \\
        & \it{89.51} & \it{89.42} & \it{89.01} & \it{89.19} & \it{89.04} & \it{89.09} \\
Incoming muon $\mu$
        & 188739.8 & 84305.1 & 7094.2 & 723066.4 & 83410.3 & 18114.0 \\
        & \it{85.69} & \it{85.60} & \it{85.20} & \it{85.46} & \it{85.53} & \it{85.38} \\
Scattered muon $\mu^\prime$
        & 157636.0 & 70123.3 & 6000.2 & 602770.9 & 69359.8 & 15287.4 \\
        & \it{71.57} & \it{71.20} & \it{72.06} & \it{71.25} & \it{71.12} & \it{72.06} \\
Kinematic cuts
        & 18765.6 & 6779.6 & 707.8 & 71571.2 & 6839.6 & 1807.6 \\
        & \it{8.52} & \it{6.88} & \it{8.50} & \it{8.46} & \it{7.01} & \it{8.52} \\
Trigger
        & 16504.6 & 5956.3 & 619.9 & 62638.6 & 6024.4 & 1585.1 \\
        & \it{7.49} & \it{6.05} & \it{7.44} & \it{7.40} & \it{6.18} & \it{7.47} \\
    \hline
\end{tabular}
\caption{Same as Tab.~\ref{tab:MCevcuts}, but with the proper weight assigned to each HEPGEN event.}
\label{tab:MCevcuts2}
\end{table}

\newpage
\clearpage

The final-state hadrons have been selected with the following criteria:
\begin{itemize}
    \item \textbf{identification as hadron:} based on the number of crossed radiation lengths ($X/X_0<10$). In this first analysis, the RICH and the calorimeters have not been used;
    \item \textbf{track quality:} based on the quality of the reconstructed track and on the position of the first and last hits, respectively before and after SM1;
    \item \textbf{kinematic cuts:} on the fractional energy in the laboratory frame ($0.1<z<1.0$) and on the transverse momentum in the GNS ($0.1<P_{T}<2.0$~GeV/$c$).
\end{itemize}
The number of positive and negative hadrons surviving each block of cuts, for each of the six sub-periods, is given in Tab.~\ref{tab:dhacuts}. There is a good agreement between the three periods and between the $\mu^+$ and $\mu^-$ cases. The total number of selected hadrons is $N_{h^+}\approx 3.6\cdot 10^6$ and $N_{h^-}\approx 3.0\cdot 10^6$. Analogously, Tab.~\ref{tab:MChacuts} gives the number of surviving hadrons for the Monte Carlo samples, where all events are weighted 1; in Tab.~\ref{tab:MChacuts2}, the HEPGEN statistics is properly weighted. Summing over $\mu^+$ and $\mu^-$, the ratio in the number of selected hadrons (LEPTO/data) is $r \approx 1.4$ for both positive and negative hadrons. It is to be noted that the total number of events and hadrons just depends on the different size of the various samples, and that no relative normalization is introduced at this stage. A sizable difference can be observed in the percentage of surviving hadrons between LEPTO and HEPGEN, originating from the \textit{Identification} and \textit{Kinematic cuts} steps. In the first place, a larger number of tracks are rejected in LEPTO due to the number of crossed radiation lengths; then, the cut in $z$ ($z>0.1$, in particular) strongly contributes in reducing the amount of hadron tracks, while this is not the case for HEPGEN due to the peculiar $z$ distribution of the exclusive hadrons (see Sect.~\ref{sect:ch3_exclusive}).

More stringent cuts have been applied in the analyses for the extraction of the transverse-momentum distributions and of the azimuthal asymmetries. As will be explained in Sect.~\ref{sect:ch3_acceptance}, this is the case for the $y$ range, for which a lower cut at $y=0.2$ has been preferred to have small acceptance corrections. Also, as explained in Sect.~\ref{sect:ch3_exclusive}, the events in which exactly two hadrons are observed in the final state, with opposite charge and $z_{h^+}+z_{h^-}>0.95$, are tagged as exclusive and discarded. The selection of events and hadrons is identical for real data and Monte Carlo data.

\bigskip

\begin{table}[h!]
\small
\centering
\captionsetup{width=\textwidth}
\begin{tabular}{c|rrrrrr}
	\hline
    $\mu^+$ &  \multicolumn{2}{c}{P08} & \multicolumn{2}{c}{P09} & \multicolumn{2}{c}{P10} \\
        & $h^+$ & $h^-$ & $h^+$ & $h^-$ & $h^+$ & $h^-$ \\ \hline
Tracks from the PV 
        & 1588846 & 1392100 & 1218944 & 1069158 & 1324050 & 1162604 \\ 
        & \it{100.00} & \it{100.00} & \it{100.00} & \it{100.00} & \it{100.00} & \it{100.00} \\ 
Identification
        & 1566456 & 1358163 & 1202170 & 1043784 & 1305752 & 1135248 \\ 
        & \it{98.59} & \it{97.56} & \it{98.62} & \it{97.63} & \it{98.62} & \it{97.65} \\ 
Track quality
        & 1505537 & 1300087 & 1152168 & 996105 & 1245961 & 1076752 \\ 
        & \it{94.76} & \it{93.39} & \it{94.52} & \it{93.17} & \it{94.10}& \it{92.62} \\ 
Kinematic cuts
        & 730976 & 604044 & 561754 & 465318 & 612294 & 507229 \\ 
        & \it{46.01} & \it{43.39} & \it{46.09} & \it{43.52} & \it{46.24} & \it{43.63} \\ 
    \hline
        $\mu^-$ &  \multicolumn{2}{c}{P08} & \multicolumn{2}{c}{P09} & \multicolumn{2}{c}{P10} \\
        & $h^+$ & $h^-$ & $h^+$ & $h^-$ & $h^+$ & $h^-$ \\ \hline
Tracks from the PV 
        & 1479198 & 1281072 & 1126699 & 976153 & 1083686 & 938026 \\
        & \it{100.00} & \it{100.00} & \it{100.00} & \it{100.00} & \it{100.00} & \it{100.00} \\ 
Identification
        & 1442290 & 1264769 & 1099654 & 964017 & 1057309 & 926124 \\
        & \it{97.50} & \it{98.73} & \it{97.60} & \it{98.76} & \it{97.57} & \it{98.73} \\         
Track quality
        & 1384531 & 1210306 & 1051875 & 919887 & 1007839 & 881177 \\
        & \it{93.60} & \it{94.48} & \it{93.36} & \it{94.24} & \it{93.00} & \it{93.94} \\  
Kinematic cuts
        & 675011 & 560829 & 515620 & 427366 & 495458 & 411039 \\
        & \it{45.63} & \it{43.78} & \it{45.76} & \it{43.78} & \it{45.72} & \it{43.82} \\          
    \hline    
\end{tabular}
\caption{Number of hadrons after each block of cuts, starting from the tracks associated to the selected DIS events, not identified as muons (data, $\mu^+$ and $\mu^-$ beam). The values in italic show the percentage of tracks passing each block of cuts.}
\label{tab:dhacuts}
\end{table}

\begin{table}[h!]
\small
\centering
\captionsetup{width=\textwidth}
\begin{tabular}{c|rrrrrrrr}
	\hline
    \multicolumn{1}{c|}{$\mu^+$} &  \multicolumn{2}{c}{LEPTO} & \multicolumn{2}{c}{HEPGEN-$\rho$} & \multicolumn{2}{c}{HEPGEN-$\phi$} &  \multicolumn{2}{c}{HEPGEN-$\omega$} \\
        & $h^+$ & $h^-$ & $h^+$ & $h^-$ & $h^+$ & $h^-$ & $h^+$ & $h^-$ \\ \hline
Tracks from the PV 
        & 11493220 & 11422024 & 138497 & 134394 & 255772 & 246479 & 64367 & 62972 \\
        & \it{100.00} & \it{100.00} & \it{100.00} & \it{100.00} & \it{100.00} & \it{100.00} & \it{100.00} & \it{100.00} \\
Identification
        & 6458268 & 5263227 & 137881 & 128683 & 254817 & 237266 & 64200 & 61152 \\
        & \it{56.19} & \it{46.08} & \it{99.56} & \it{95.75} & \it{99.63} & \it{96.26} & \it{99.74} & \it{97.11} \\
Track quality
        & 6220844 & 5054937 & 135143 & 125685 & 249831 & 231555 & 62230 & 59106 \\
        & \it{54.13} & \it{44.26} & \it{97.58} & \it{93.52} & \it{97.68} & \it{93.95} & \it{96.68} & \it{93.86} \\
Kinematic cuts
        & 2738348 & 2225741 & 103819 & 100640 & 179769 & 172917 & 36106 & 35277 \\
        & \it{23.83} & \it{19.49} & \it{74.96} & \it{74.88} & \it{70.28} & \it{70.15} & \it{56.09} & \it{56.02} \\
    \hline
    \multicolumn{1}{c|}{$\mu^-$} &  \multicolumn{2}{c}{LEPTO} & \multicolumn{2}{c}{HEPGEN-$\rho$} & \multicolumn{2}{c}{HEPGEN-$\phi$} &  \multicolumn{2}{c}{HEPGEN-$\omega$} \\
        & $h^+$ & $h^-$ & $h^+$ & $h^-$ & $h^+$ & $h^-$ & $h^+$ & $h^-$ \\ \hline
Tracks from the PV 
        & 10040456 & 10040456 & 531721 & 507172 & 262921 & 248554 & 165328 & 159910 \\
        & \it{100.00} & \it{100.00} & \it{100.00} & \it{100.00} & \it{100.00} & \it{100.00} & \it{100.00} & \it{100.00}\\ 
Identification
        & 5420082 & 4503392 & 510318 & 505155 & 252812 & 247649 & 160400 & 159525 \\
        & \it{53.98} & \it{44.85} & \it{95.97} & \it{99.60} & \it{96.16} & \it{99.64} & \it{97.02} & \it{99.76}\\ 
Track quality
        & 5217762 & 4316880 & 500994 & 491784 & 247888 & 240886 & 155539 & 153605 \\
        & \it{51.97} & \it{42.99} & \it{94.22} & \it{96.97} & \it{94.28} & \it{96.91} & \it{94.08} & \it{96.06}\\  
Kinematic cuts
        & 2303864 & 1902172 & 384551 & 392267 & 177807 & 179807 & 90256 & 91936 \\
        & \it{22.95} & \it{18.95} & \it{72.32} & \it{77.34} & \it{67.63} & \it{72.34} & \it{54.59} & \it{57.49}\\ 
    \hline
\end{tabular}
\caption{Effect of the cuts applied to the hadrons produced in the selected DIS events (Monte Carlo samples, $\mu^+$ and $\mu^-$ beams). Here, each event is assigned a weight equal to 1. The values in italic show the percentage of events passing the corresponding block of cuts.}
\label{tab:MChacuts}
\end{table}

\begin{table}[h!]
\small
\centering
\captionsetup{width=\textwidth}
\begin{tabular}{c|rrrrrr}
	\hline
    $\mu^+$ & \multicolumn{2}{c}{HEPGEN-$\rho$} & \multicolumn{2}{c}{HEPGEN-$\phi$} &  \multicolumn{2}{c}{HEPGEN-$\omega$}\\
        & $h^+$ & $h^-$ & $h^+$ & $h^-$ & $h^+$ & $h^-$ \\ \hline
Tracks from the PV 
        & 15829.3 & 15501.9 & 5726.5 & 5575.2 & 731.5 & 721.1 \\
        & \it{100.00} & \it{100.00} & \it{100.00} & \it{100.00} & \it{100.00} & \it{100.00} \\
Identification
        & 15787.7 & 14929.5 & 5712.0 & 5403.6 & 729.9 & 704.5 \\
        & \it{99.74} & \it{96.31} & \it{99.75} & \it{96.92} & \it{99.78} & \it{97.70} \\
Track quality
        & 15462.3 & 14565.7 & 5588.2 & 5258.0 & 708.1 & 681.2 \\
        & \it{97.68} & \it{93.96} & \it{97.58} & \it{94.31} & \it{96.80} & \it{94.47} \\
Kinematic cuts
        & 12133.8 & 11791.7 & 4235.4 & 4075.0 & 443.4 & 432.3 \\
        & \it{76.65} & \it{76.07} & \it{73.96} & \it{73.09} & \it{60.62} & \it{59.95} \\
    \hline
    $\mu^-$ & \multicolumn{2}{c}{HEPGEN-$\rho$} & \multicolumn{2}{c}{HEPGEN-$\phi$} &  \multicolumn{2}{c}{HEPGEN-$\omega$}\\
        & $h^+$ & $h^-$ & $h^+$ & $h^-$ & $h^+$ & $h^-$ \\ \hline
Tracks from the PV 
        & 60994.6 & 59202.2 & 5868.8 & 5629.9 & 1889.8 & 1835.5 \\
        & \it{100.00} & \it{100.00} & \it{100.00} & \it{100.00} & \it{100.00} & \it{100.00}\\ 
Identification
        & 58979.0 & 59013.1 & 5690.3 & 5616.4 & 1844.2 & 1832.5 \\
        & \it{96.70} & \it{99.68} & \it{96.96} & \it{99.76} & \it{97.59} & \it{99.84}\\ 
Track quality
        & 57771.3 & 57335.7 & 5559.1 & 5444.4 & 1789.9 & 1763.4 \\
        & \it{94.72} & \it{96.85} & \it{94.72} & \it{96.71} & \it{94.71} & \it{96.07}\\  
Kinematic cuts
        & 45345.5 & 45986.4 & 4186.1 & 4190.2 & 1106.1 & 1109.1 \\
        & \it{74.34} & \it{77.68} & \it{71.33} & \it{74.43} & \it{58.53} & \it{60.42}\\ 
    \hline    
\end{tabular}
\caption{Same as Tab.~\ref{tab:MChacuts}, but with the proper weight assigned to each HEPGEN event.}
\label{tab:MChacuts2}
\end{table}

\section{Kinematic distributions}
\label{sect:ch3_kinematic}
This Section hosts a set of kinematic distributions for the selected events and hadrons. All the plots have been produced using the $\mu^+$ beam data, summing over the three periods in the case of real data, and for $Q^2>1~(\mathrm{Gev}/c)^2$, $x>0.003$ and $W>5~\mathrm{GeV}/c^2$. In the $\mu^-$ case, all the distributions are very similar.

\begin{figure}[h!]
\captionsetup{width=\textwidth}
\begin{center}
	\includegraphics[width=0.65\textwidth,angle=90]{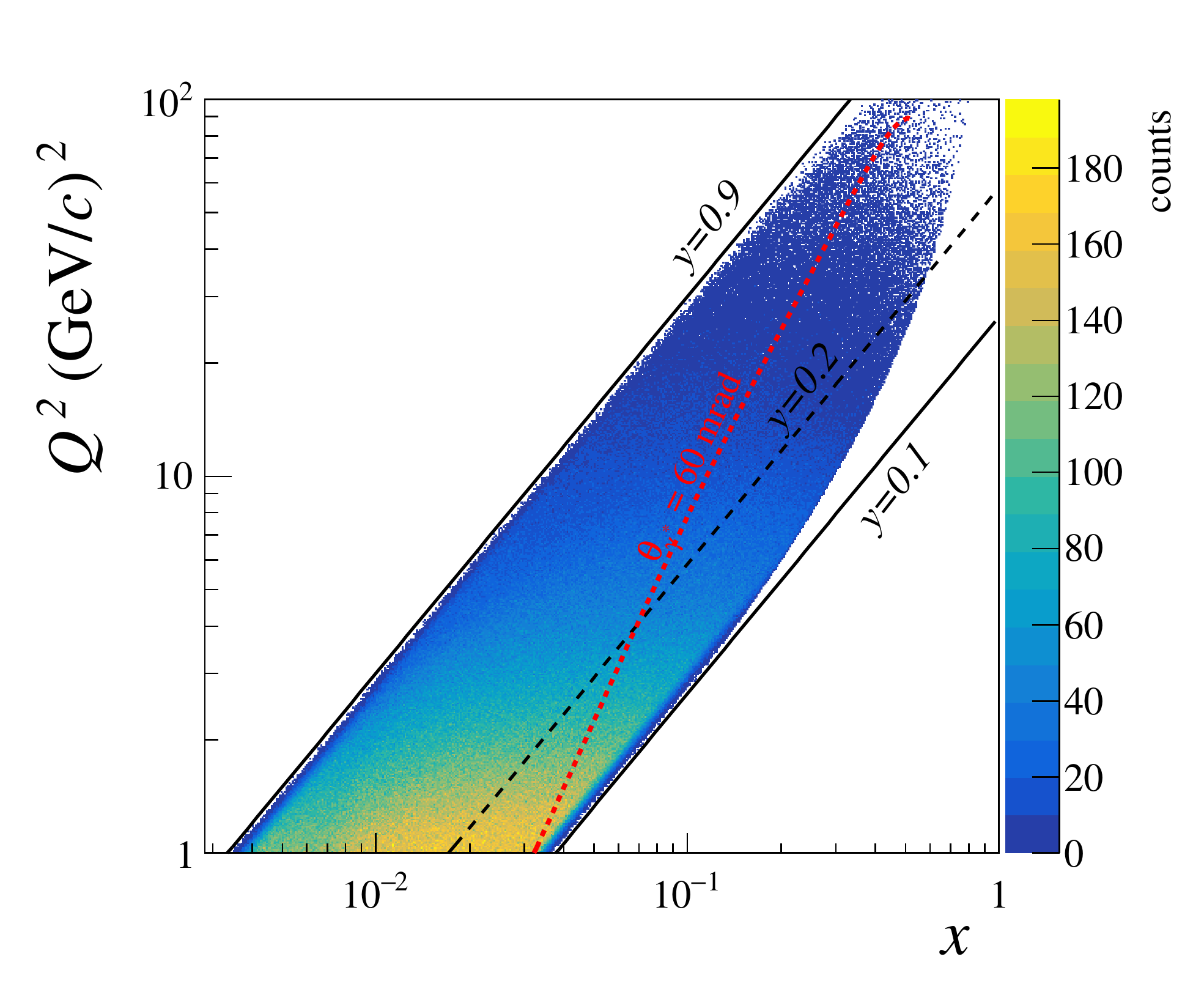}
	\caption{The $x-Q^2$ correlation for the selected DIS events in the data. The two continuous lines indicate the largest $y$ range ($0.1<y<0.9$), while the dashed black line indicate the $y=0.2$ condition. The red dashed line corresponds to the $\theta_{\gamma^*}$ cut.}
\label{fig:xq2data0}
\end{center}
\end{figure}

\begin{figure}[h!]
\captionsetup{width=\textwidth}
\begin{center}
	\includegraphics[width=0.49\textwidth]{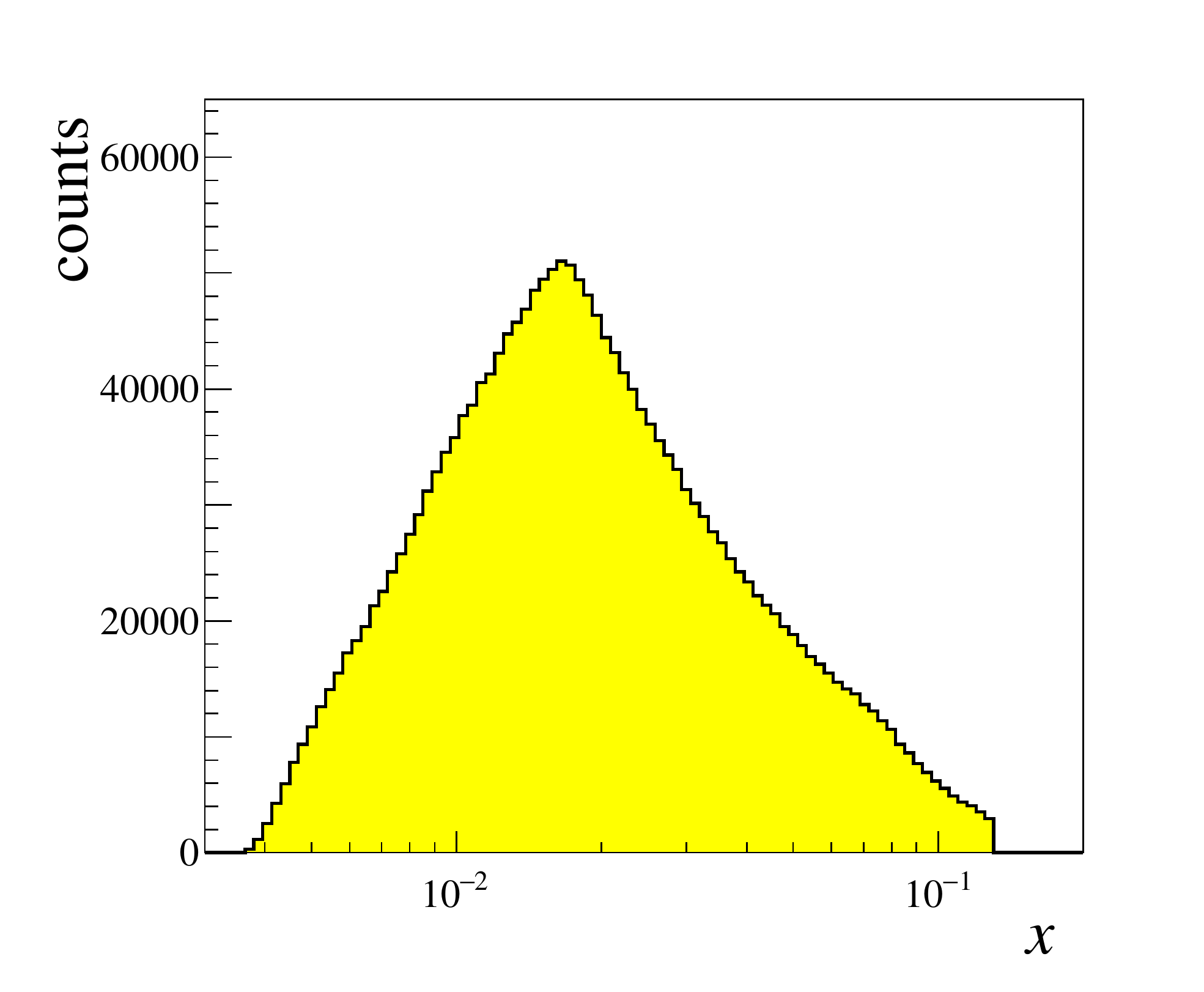}
	\includegraphics[width=0.49\textwidth]{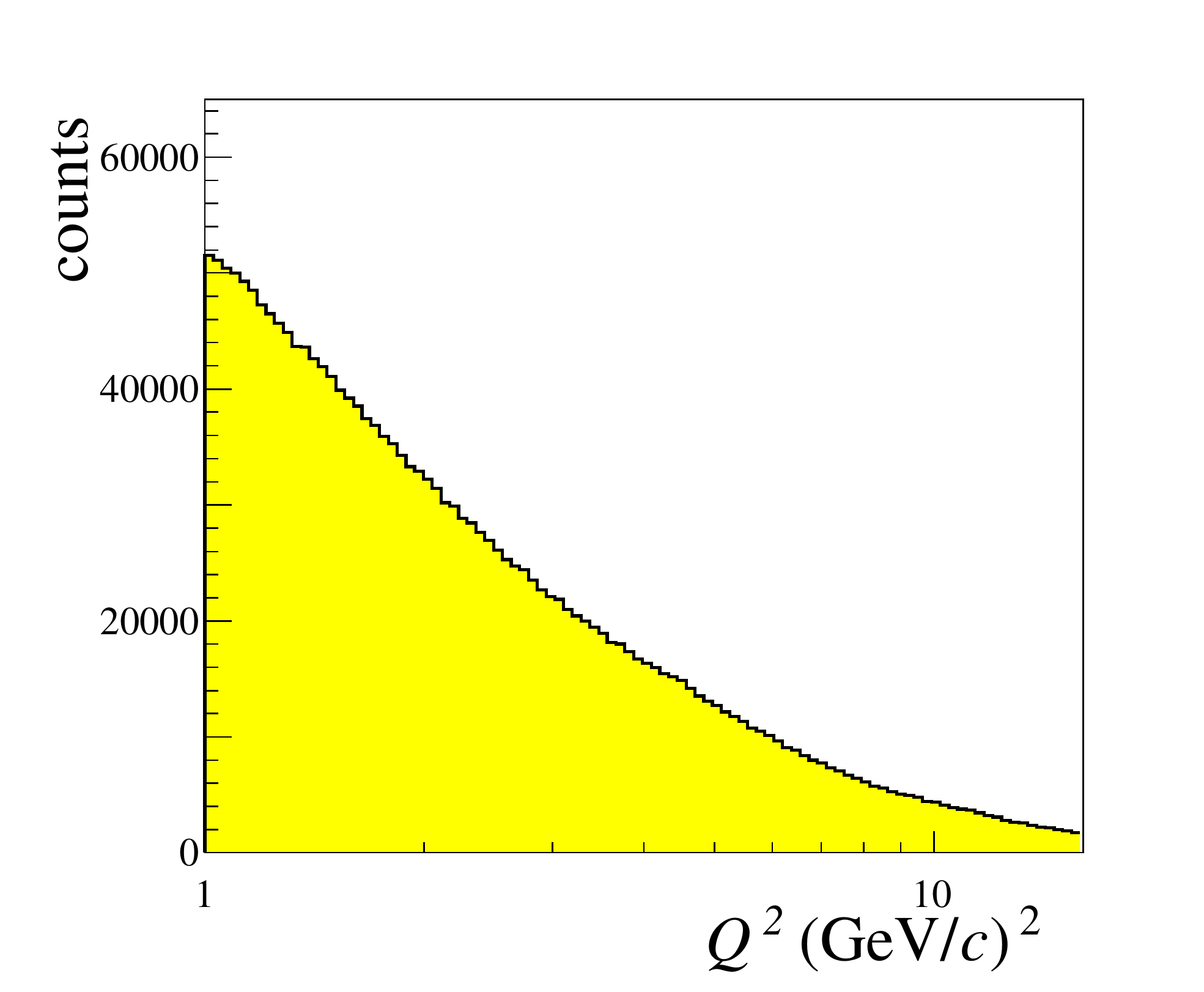}\\
	\caption{The $x$ and $Q^2$ distributions for the $\mu^+$ DIS events in the data for $0.2<y<0.9$.}
\label{fig:kin0}
\end{center}
\end{figure}

\begin{figure}[h!]
\captionsetup{width=\textwidth}
\begin{center}
	\includegraphics[width=0.49\textwidth]{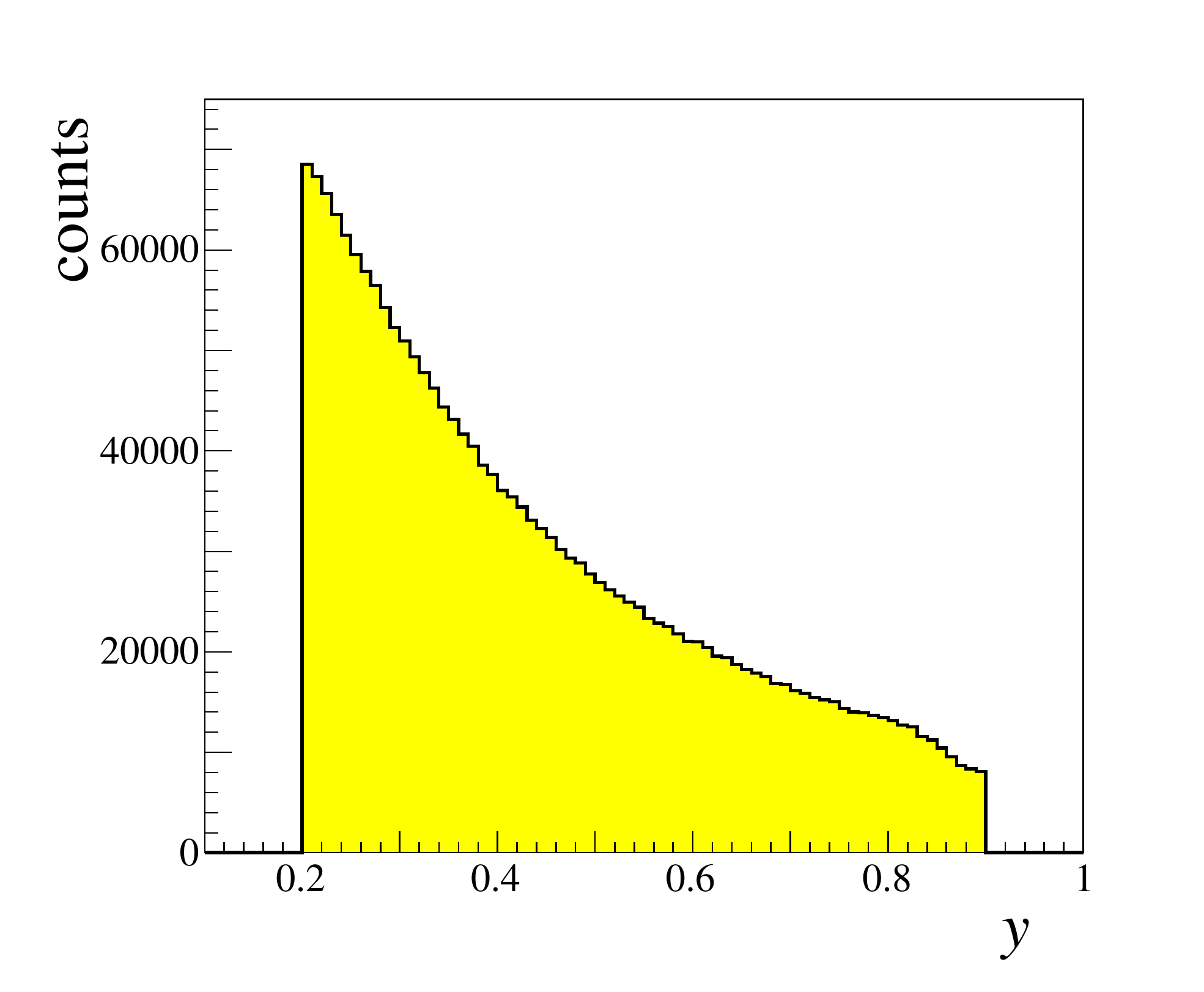}
	\includegraphics[width=0.49\textwidth]{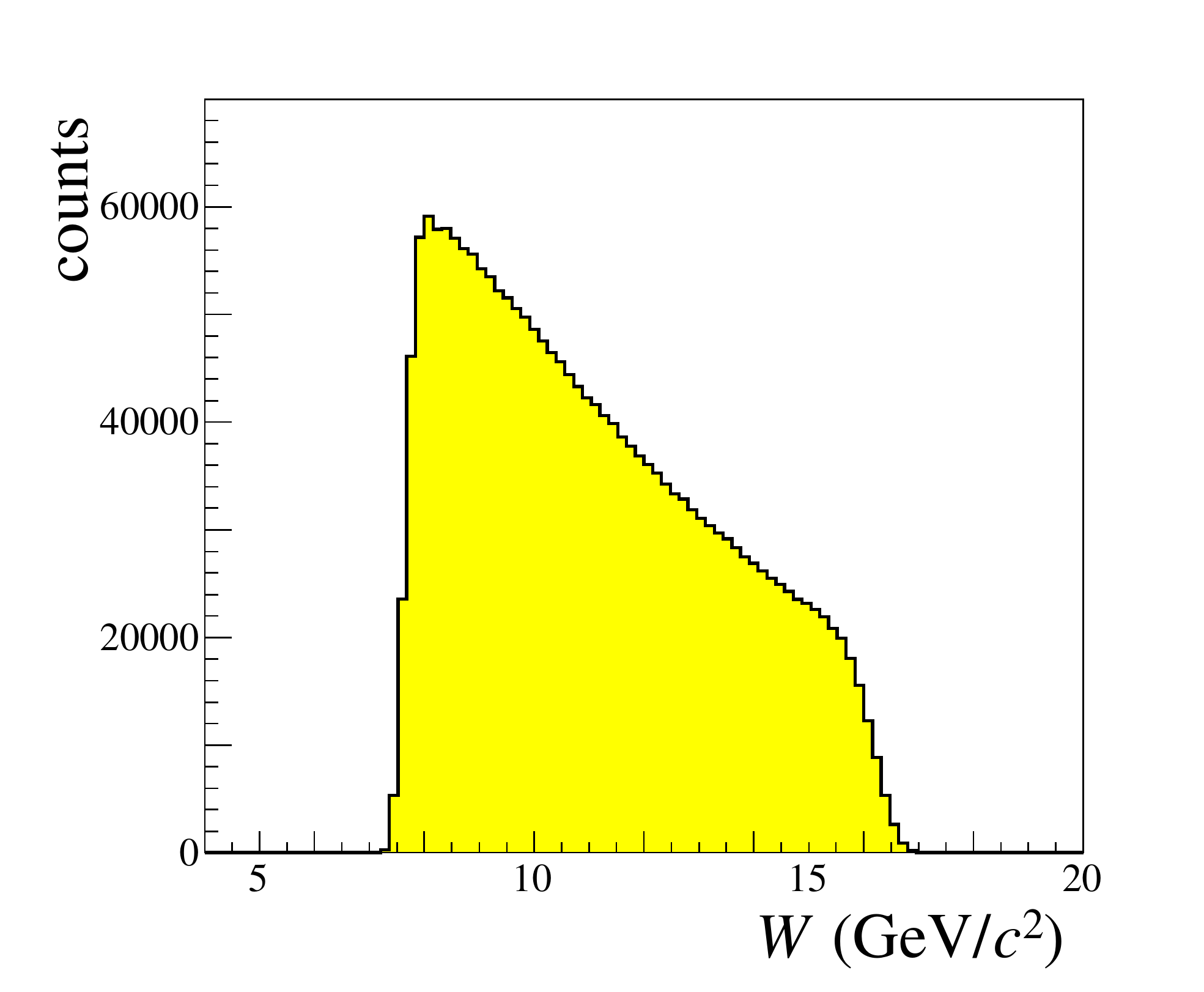}\\
	\caption{The $y$ and $W$ distributions for the $\mu^+$ DIS events in the data for $0.2<y<0.9$.}
\label{fig:kin1}
\end{center}
\end{figure}

The typical $x-Q^2$ correlation of the DIS events is shown in Fig.~\ref{fig:xq2data0}. The two continuous inclined lines correspond to $y=0.1$ and $y=0.9$, and the dashed line to $y=0.2$. As described in Sect.~\ref{sect:ch3_acceptance}, the analysis of the acceptance corrections suggests not to use the region at low $y$ ($0.1<y<0.2$). A dotted red line corresponds to $\theta_{\gamma^*}=60$~mrad. Only the events at the left of this line survive in the DIS selection. The impact of this cut (in terms of reduction in statistics) is not large if $y>0.2$. The one-dimensional distributions of $x$, $Q^2$, $y$ and $W$ are shown in Fig.~\ref{fig:kin0} and Fig.~\ref{fig:kin1} for the selected DIS events in the $0.2<y<0.9$ range. 

\begin{figure}[h!]
\captionsetup{width=\textwidth}
\begin{center}
	\includegraphics[width=0.49\textwidth]{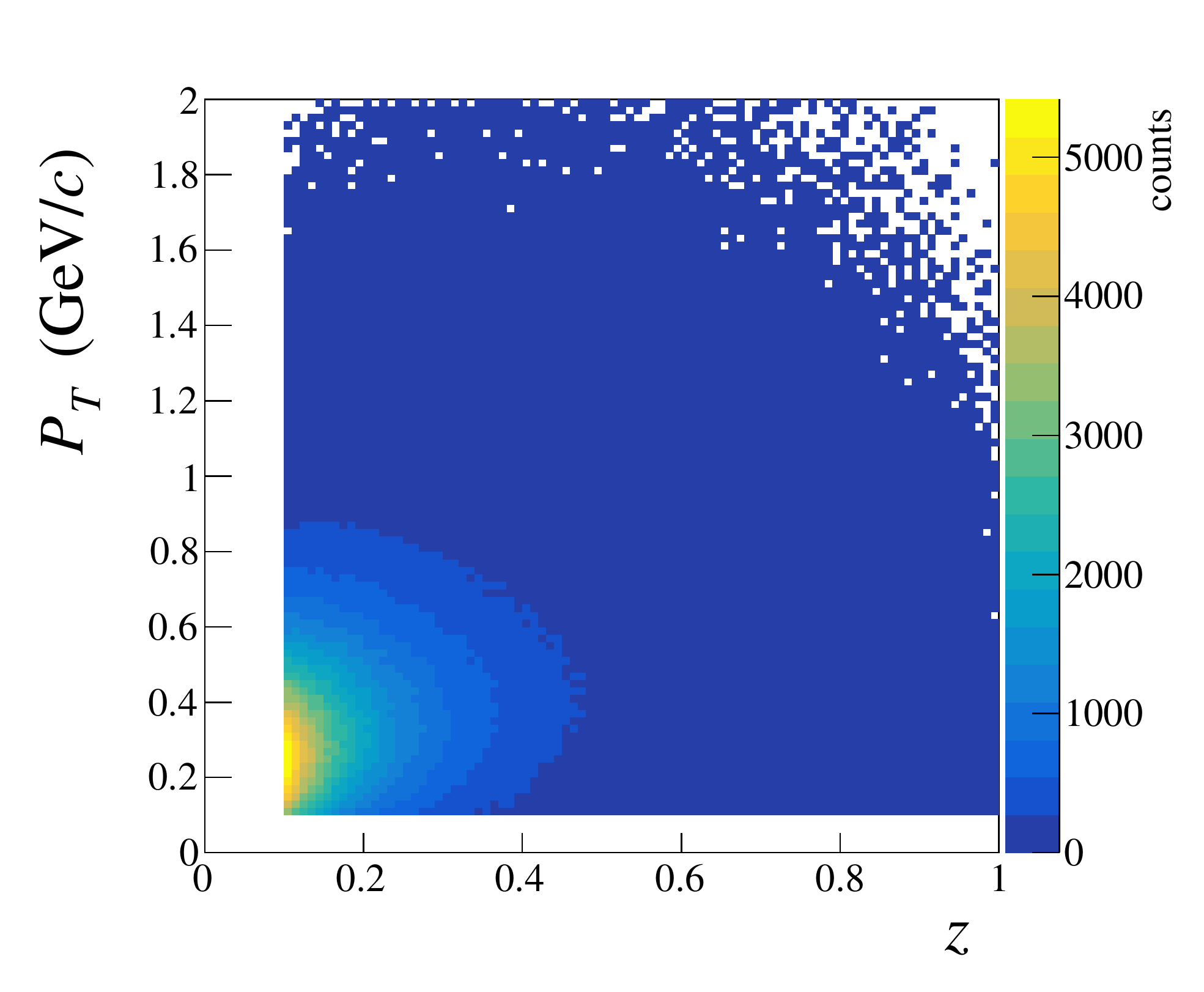}
	\caption{The $z$-$\Pt$ correlation for all selected positive hadrons in the $\mu^+$ data sample.}
\label{fig:zpt}
\end{center}
\end{figure}

\begin{figure}[h!]
\captionsetup{width=\textwidth}
\begin{center}
	\includegraphics[width=0.49\textwidth]{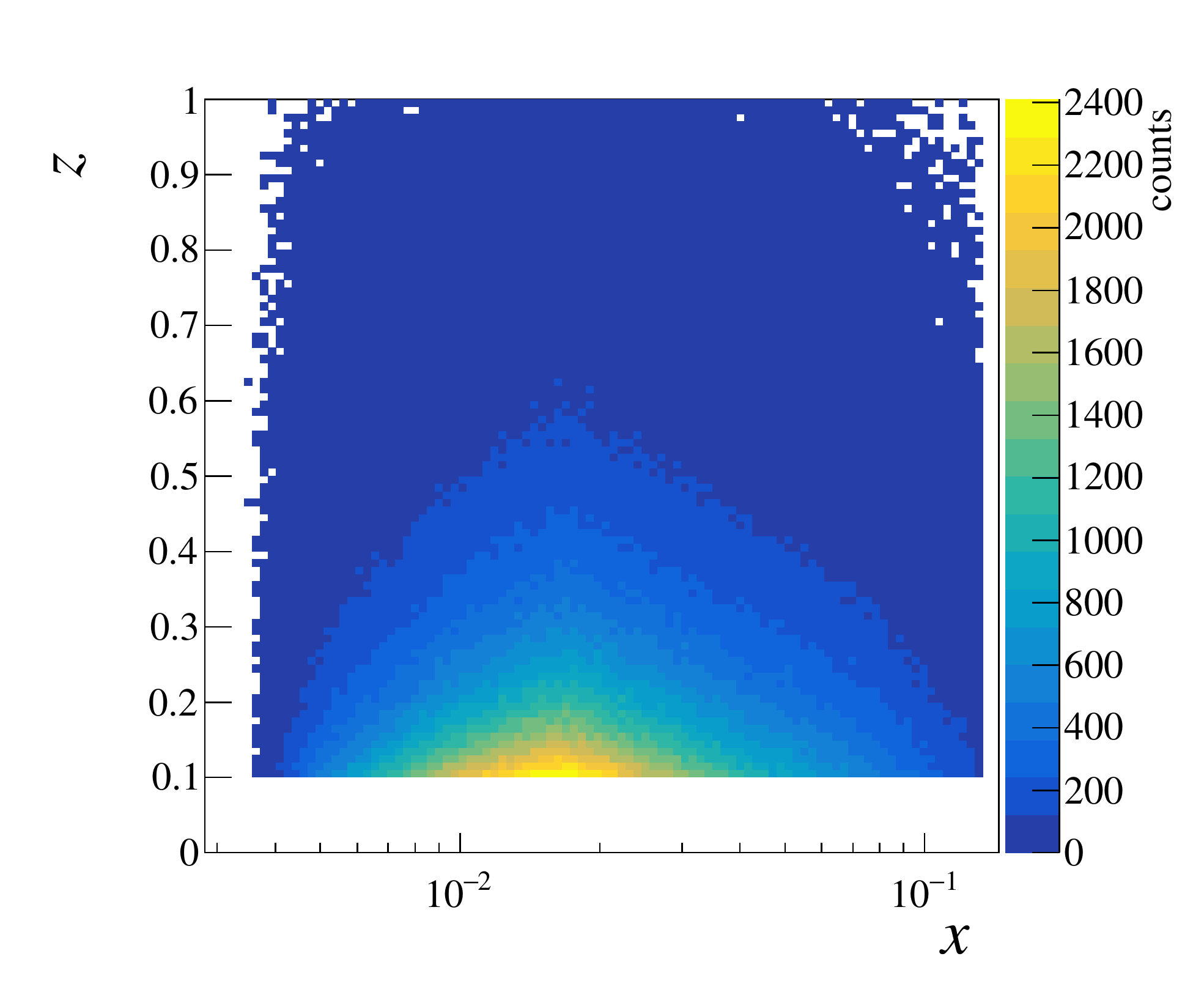}
	\includegraphics[width=0.49\textwidth]{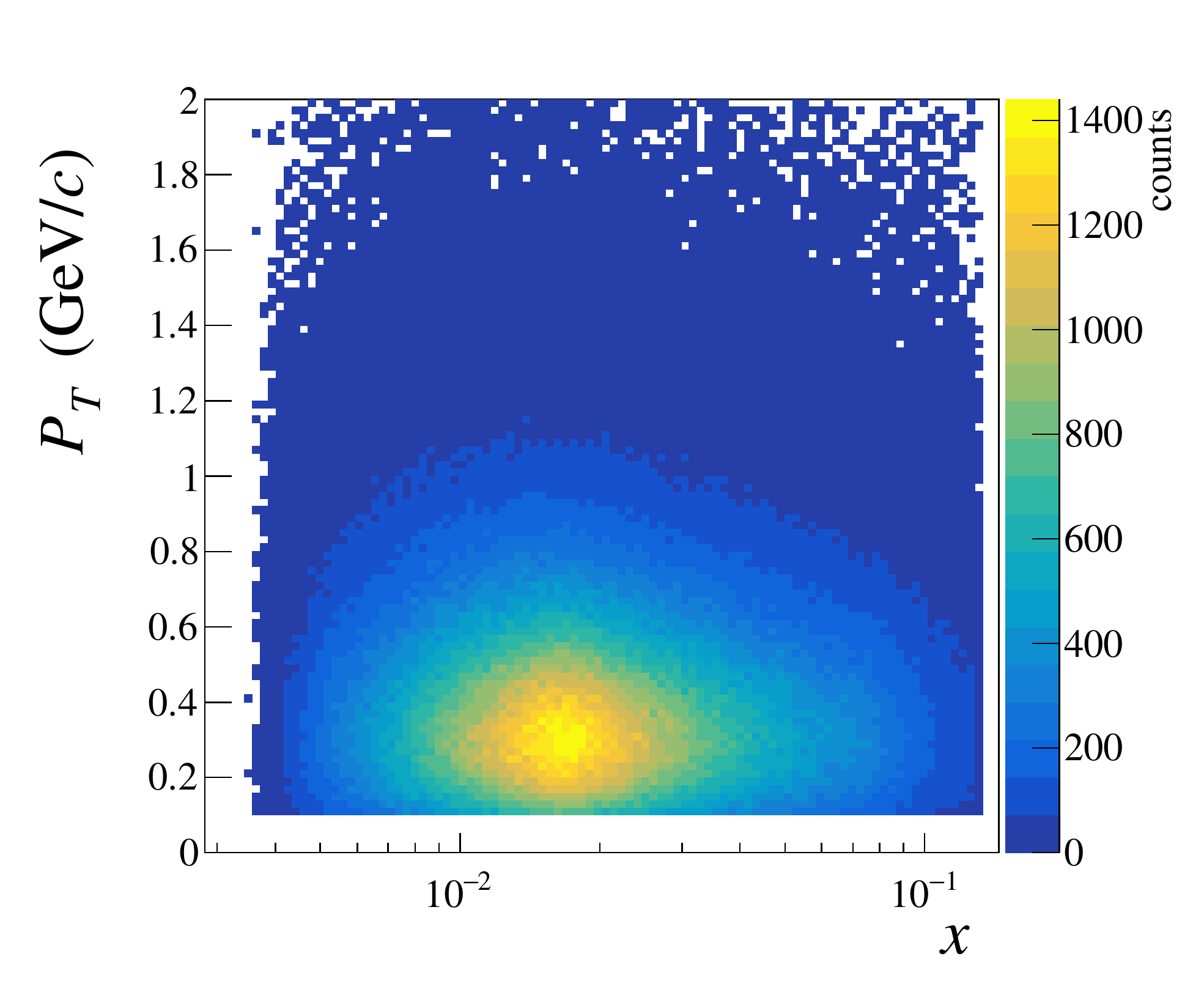}
	\caption{The $x-z$ and $x-\Pt$ correlations for all selected positive hadrons in the $\mu^+$ data sample.}
\label{fig:xzpt}
\end{center}
\end{figure}

Moving to the hadron variables, the $z-\Pt$ correlation for all selected positive hadrons in the same $\mu^+$ sample is shown in Fig.~\ref{fig:zpt}. The $x-z$ and $x-\Pt$ correlations are shown in Fig.~\ref{fig:xzpt}. As can be seen, the correlations are very small, if any.

\begin{figure}[h!]
\captionsetup{width=\textwidth}
\begin{center}
	\includegraphics[width=0.49\textwidth]{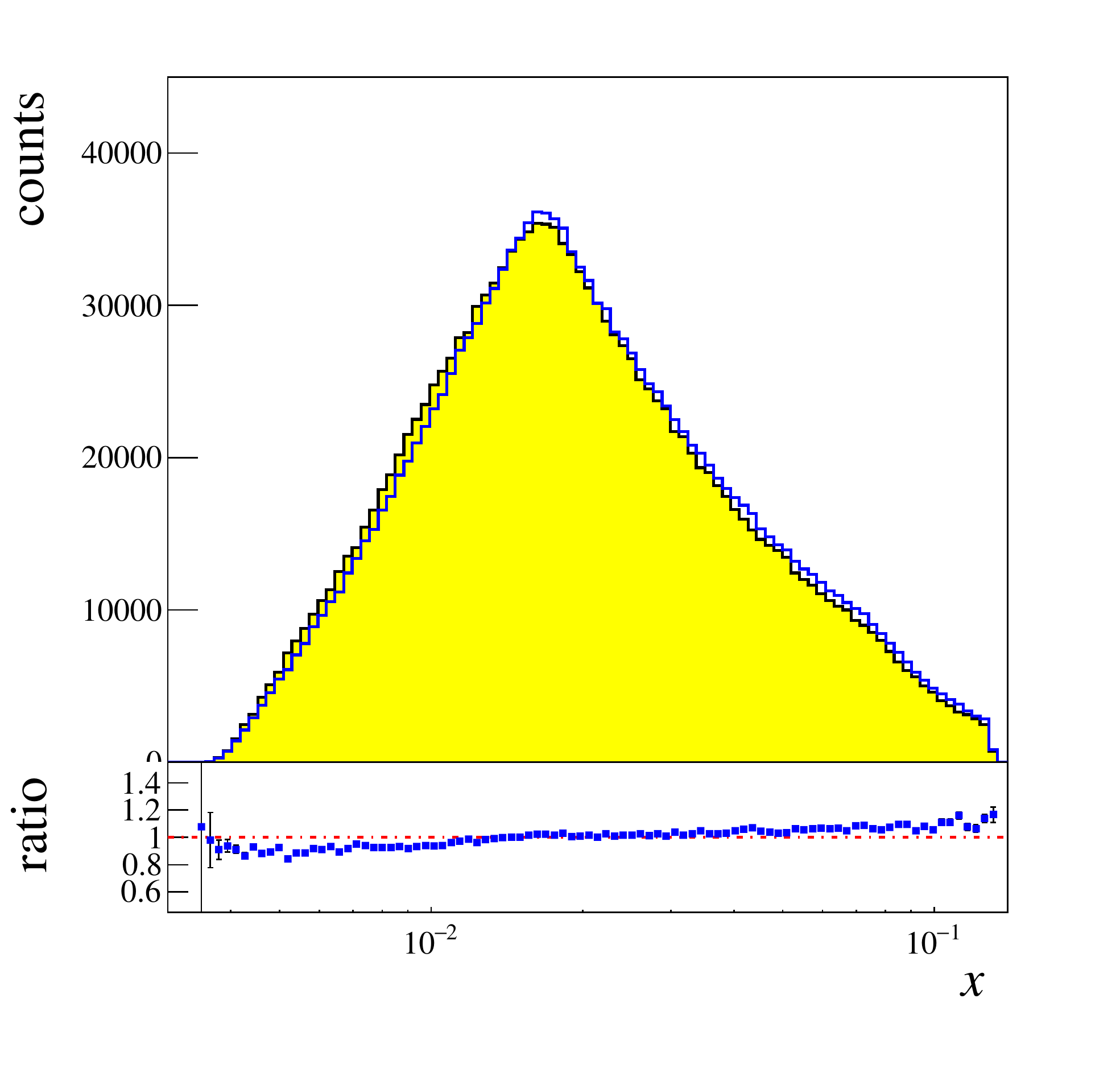}
	\includegraphics[width=0.49\textwidth]{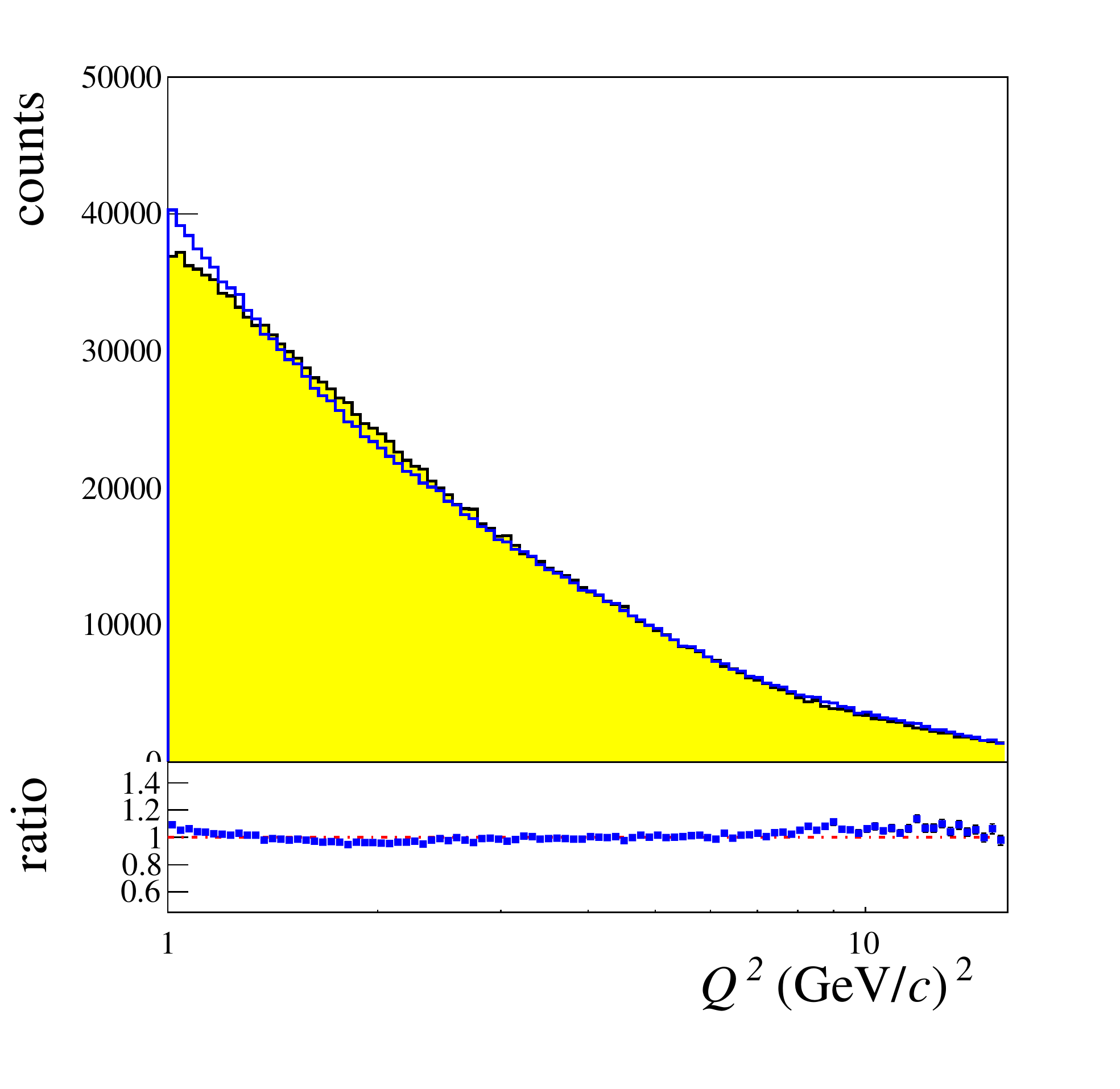}
	\caption{The $x$ and $Q^2$ distributions for positive hadrons are compared with the analogous distributions from the LEPTO Monte Carlo. The bottom panels show the Monte Carlo-over-data ratio.}
\label{fig:cmpMCnorm1}
\end{center}
\end{figure}

\begin{figure}[h!]
\captionsetup{width=\textwidth}
\begin{center}
    \includegraphics[width=0.49\textwidth]{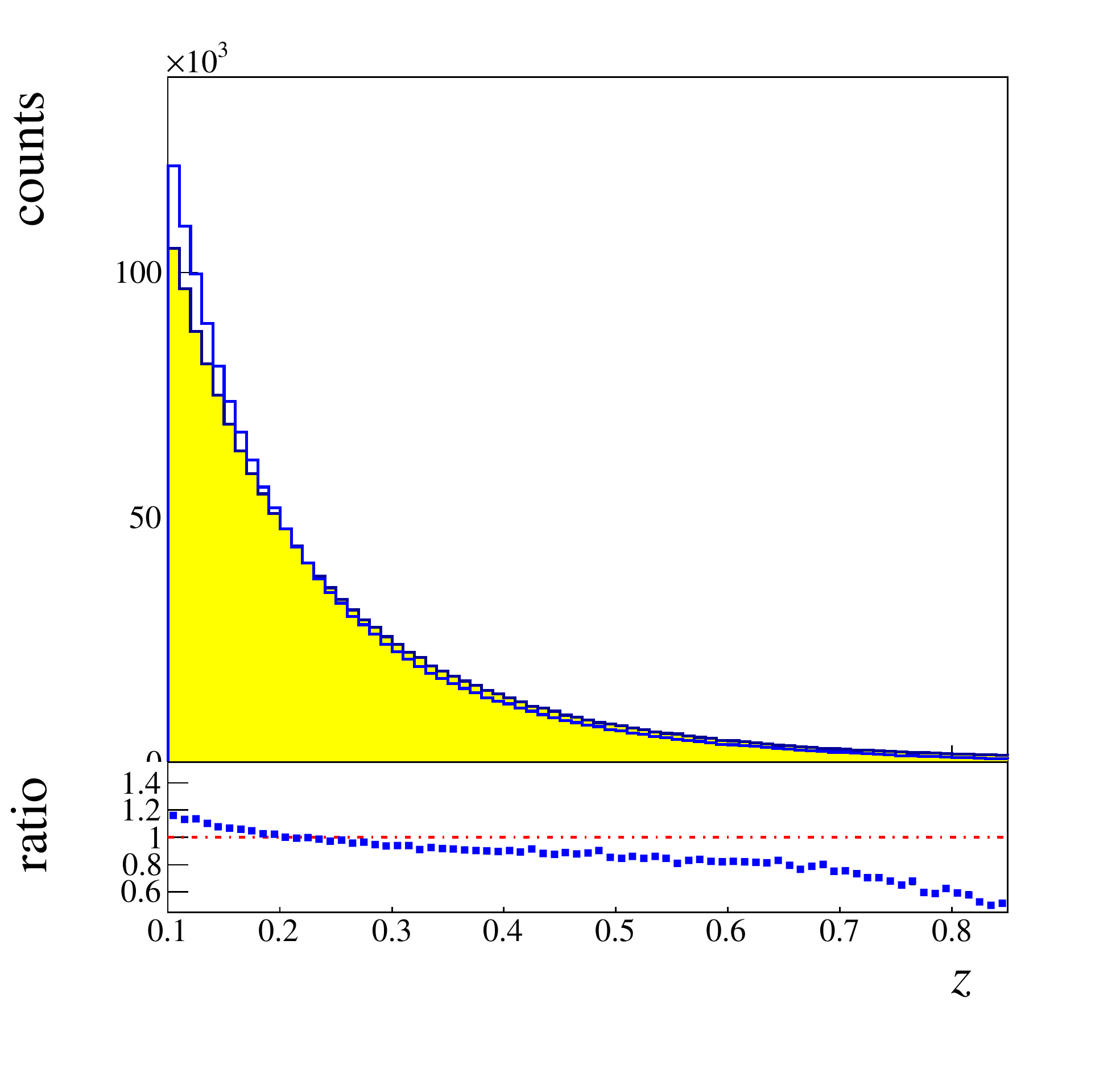}
	\includegraphics[width=0.49\textwidth]{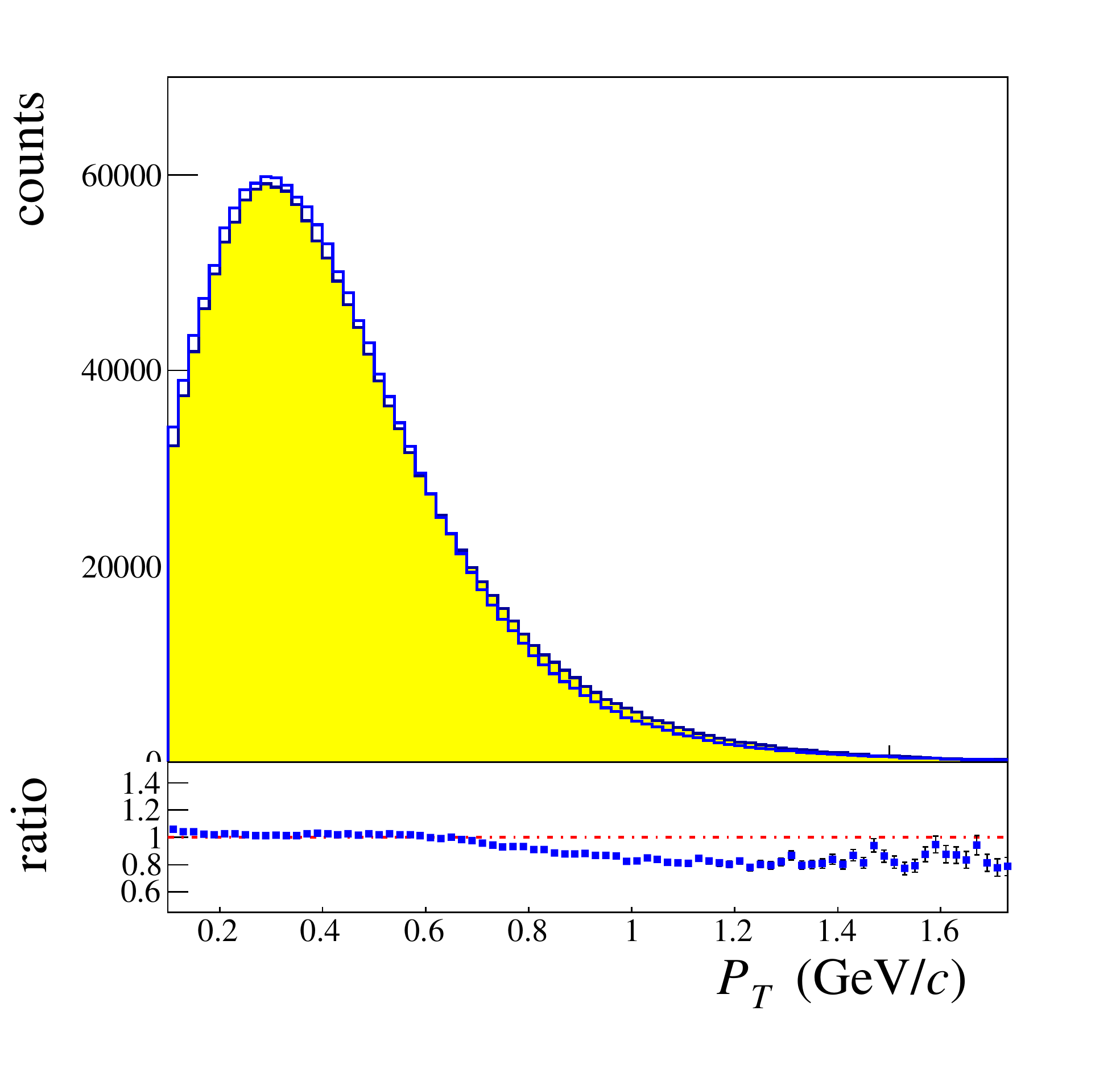}
	\caption{The $z$ and $\Pt$ distributions for positive hadrons are compared with the analogous distributions from the LEPTO Monte Carlo. The bottom panels show the Monte Carlo-over-data ratio.}
\label{fig:cmpMCnorm2}
\end{center}
\end{figure}

The $x$, $Q^2$, $z$ and $\Pt$ distributions for the selected positive hadrons and $\mu^+$ beam are compared with the reconstructed distributions from the LEPTO Monte Carlo in Fig.~\ref{fig:cmpMCnorm1} and \ref{fig:cmpMCnorm2}. As said before, both the data and the Monte Carlo have been processed following the same flow of cuts. The distributions for the Monte Carlo (blue lines) have been scaled in order to match the statistics in the data distributions (in yellow). The plots in the bottom panels show the Monte Carlo-over-data ratio. Despite some deviations from unity (particularly at high $\Pt$ and high $z$, the agreement is satisfactory, since the Monte Carlo events are mainly used to evaluate the acceptance in multi-dimensional analyses.

\newpage
\clearpage

\section{Contribution of exclusive hadrons}
\label{sect:ch3_exclusive}
A non-negligible contribution to the selected hadron sample is constituted by the decay products of diffractively produced vector mesons. The presence of such contribution is known since long ago, and it has been taken into account in several HERMES and COMPASS works. In particular, the collinear multiplicities measured in COMPASS \cite{COMPASS:2016xvm,COMPASS:2016crr} have been corrected for it. The transverse-momentum-dependent multiplicities have been published \cite{COMPASS:2017mvk} with and without the correction, estimated with a dedicated HEPGEN Monte Carlo. The estimation of the fraction of exclusive events and hadrons was done, in that case, by normalizing the LEPTO and HEPGEN samples based on the ratio of the SIDIS and diffractive cross-sections; however, since the latter is affected by a large uncertainty, the correction suffered from a large uncertainty as well. The COMPASS unpolarized azimuthal asymmetries on deuteron \cite{COMPASS:2014kcy}, on the other hand, were not corrected. Only recently it has been realized, by looking directly at the data, that the hadrons produced in the decay of diffractive vector mesons are characterized by large asymmetries: their contribution has been subtracted \cite{COMPASS:2019lcm}, using the same contamination estimates as for the multiplicities paper \cite{COMPASS:2017mvk}. For the analysis presented in this Thesis, a different correction procedure has been used, which depends much less on the Monte Carlo.\\

The diffractive production mechanism, briefly presented in Ch.~\ref{Chapter6_SDMEs}, can not be interpreted in the framework of the parton model. At variance with the DIS process, the virtual photon does not interact with a parton in the nucleon, but rather with the nucleon itself, through the exchange of a Pomeron. Also the kinematic region is in principle different: the DIS is defined for large enough $\Qsq$, while the diffraction mechanism becomes relevant at small $x$. The rapidity distribution of the hadrons produced in Semi-Inclusive DIS is continuous, while large gaps are present between the nucleon and the diffractively produced hadrons. More importantly, the cross-section for the diffractive production of a hadron in the final state can not be written in terms of the standard PDFs, but in terms of GPDs \cite{Collins:2011zzd,Diehl:2003ny}. For these reasons, the impact of the diffractively produced vector mesons on the measured TMD observables has to be taken into account. 

In the diffractive process, the virtual photon oscillates into a $q\bar{q}$ pair which interacts with the target nucleon and then converts into a vector meson. The conservation of the photon quantum numbers ($J^{PC}=1^{--})$ limits the possible vector meson species to $\vmrho(770)$, $\vmphi(1020)$ and $\omega(782)$ \footnote{These three vector mesons are expected to give the largest contributions to the hadron sample in the COMPASS kinematics. Other particles, with the same quantum numbers, but difficult to identify due to combinatorial background, to the small production cross section or characterized by a different decay mechanism, and thus not considered in this work, are: $\omega(1420)$, $\rho(1450)$, $\rho(1570)$, $\omega(1650)$, $\phi(1680)$, $\rho(1700)$, $\rho(1900)$, $\rho(2150)$, $\phi(2170)$. In the charm sector, the same quantum numbers are shared by: $J/\Psi(1S)$, $\Psi(2S)$, $\Psi(3770)$ etc.; in the bottom sector, $\Upsilon(1S)$, $\Upsilon(2S)$ etc.}. The main decay modes of these vector mesons are \cite{Zyla:2020zbs}:

\begin{equation}
\begin{split}
    \vmrho \longrightarrow \pi^+\pi^-~~&~~(BR\sim100\%)\\
    \vspace{0.5cm}\\
    \vmphi \longrightarrow K^+K^-~~&~~(BR=49.2\%)\\
    \vmphi \longrightarrow K_LK_S~~&~~(BR=34.0\%)\\
    \vspace{0.5cm}\\
    \omega \longrightarrow \pi^+\pi^-\pi^0~~&~~(BR=89.2\%)\\
\end{split}
\end{equation}
The $K_L$ and $K_S$ from the decay of the $\vmphi$ do not contribute to the SIDIS sample, since they can not be associated to any primary vertex. The $\vmrho$ decay and the $\vmphi$ decay into charged Kaons can be seen in the data, thanks to the exclusivity of the diffractive production mechanism. In fact, in most of the cases, the interaction of the quark pair with the nucleon leaves the latter intact; thus, the vector meson takes all the virtual photon energy and so do its decay products, that for this reason are here referred to as \textit{exclusive hadrons}. The diffractive production cross-section for the $\vmrho$, $\vmphi$ and $\omega$ vector mesons has been measured at HERA \cite{Cartiglia:1996xv}, where the $\vmrho$ has been found dominant with respect to the others. In general, the ratio of the production cross-section for $\vmrho$ and $\omega$ depends on $R=\sigma_L/\sigma_T$ \cite{Collins:1996fb}; in COMPASS, for the $\Qsq$ range of interest of this analysis, the measured value of $R$ is approximately equal to 1 \cite{COMPASS:2007ogt}, from which the ratio of the $\omega$ and $\vmrho$ cross-sections is calculated to be about 0.1. Thus, given its low cross-section, the contribution of the $\omega$ meson has not been taken into account in this work, as was the case in Ref.~\cite{COMPASS:2017mvk}.

\begin{figure}
\captionsetup{width=\textwidth}
\begin{center}
	\includegraphics[width=0.49\textwidth]{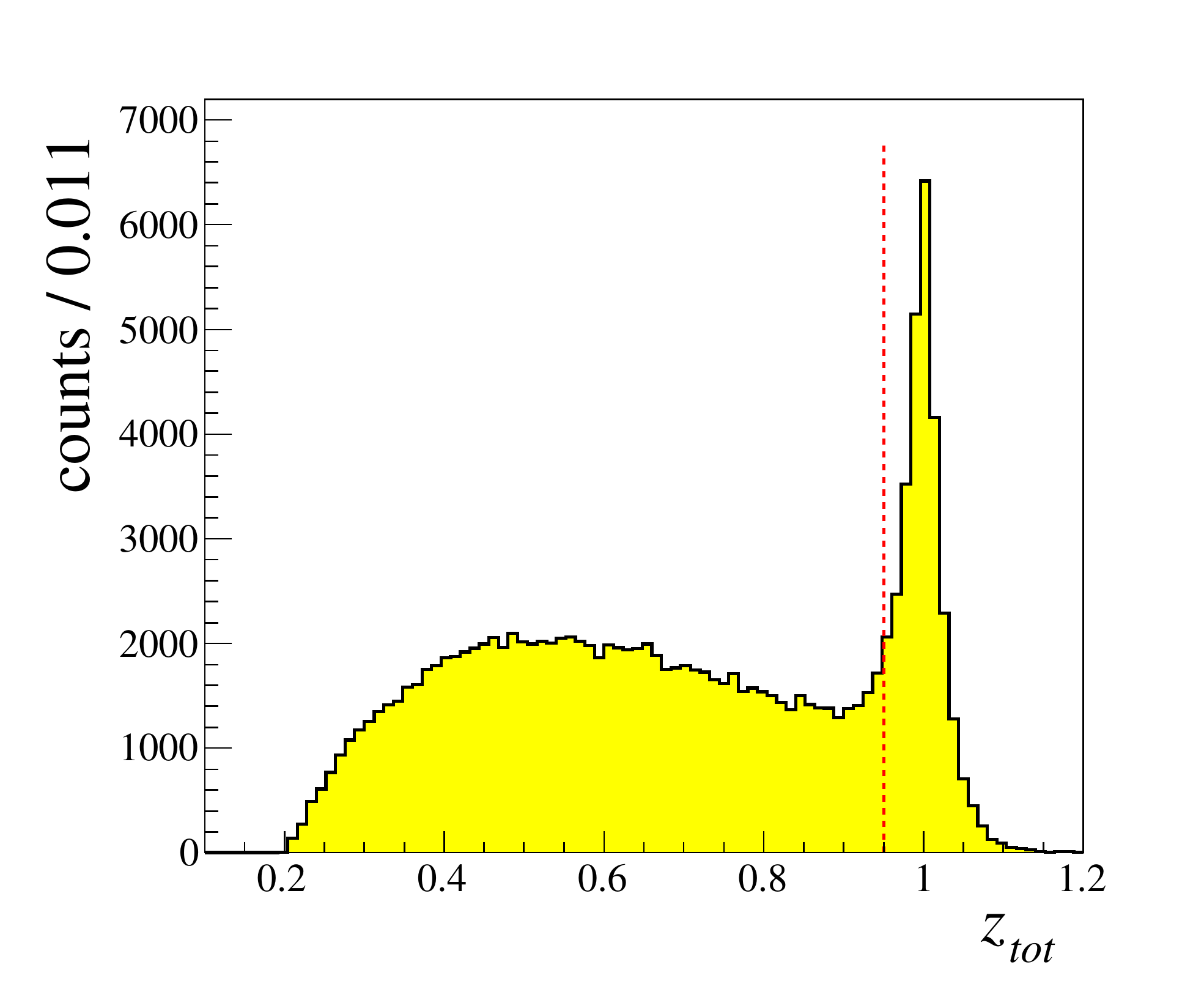}
	\includegraphics[width=0.49\textwidth]{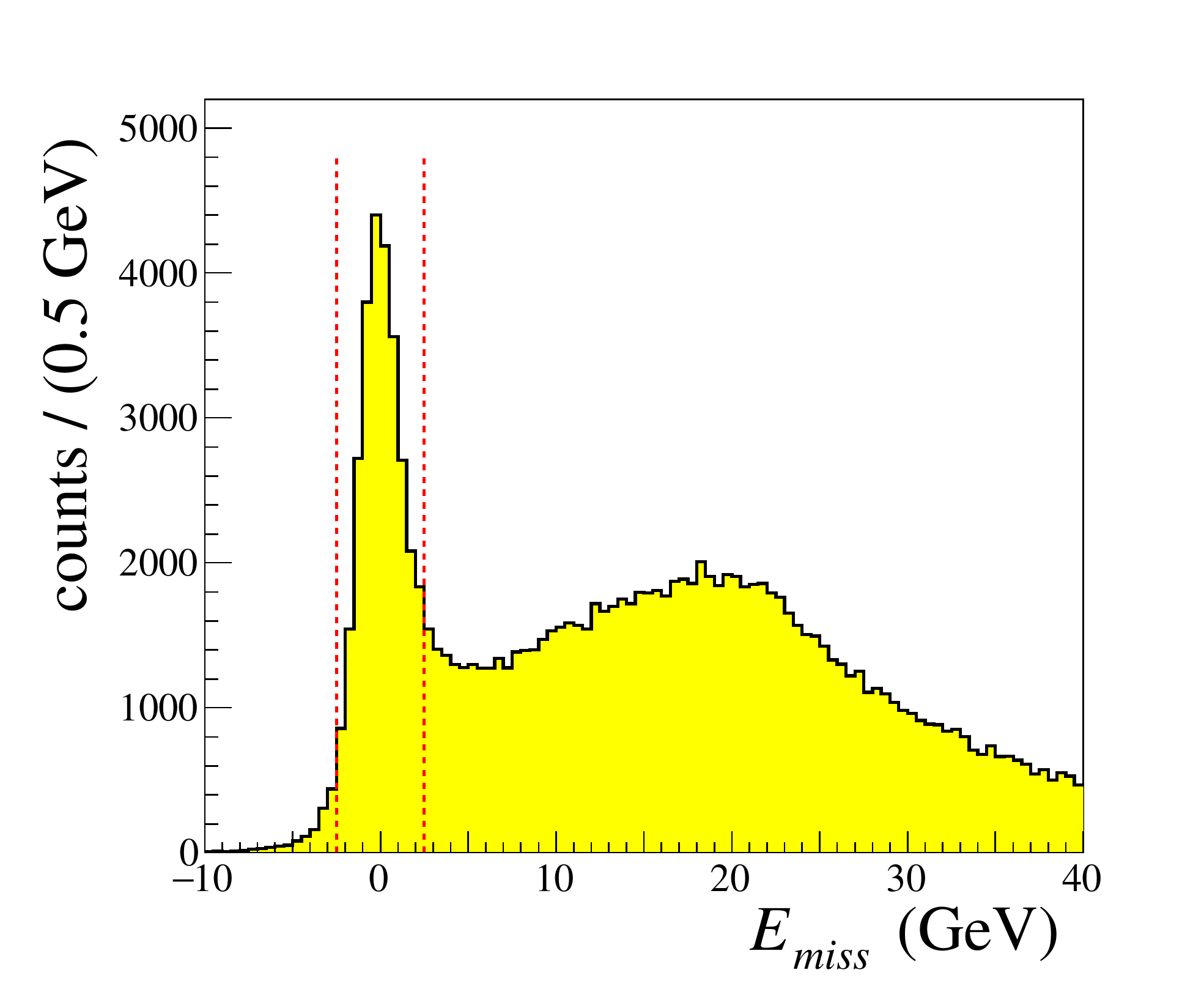}
	\caption{The $z_{tot}$ (left) and the $E_{miss}$ distributions (right). }
\label{fig:ztot_emiss}
\end{center}
\end{figure}

The exclusivity of the event can be observed in the data by looking at two quantities: 
\begin{itemize}
    \item the sum of the fractional energy $z_{tot}=z_{h^+}+z_{h^-}$ of the decay hadrons;
    \item the missing energy $E_{miss}$, defined as:
    \begin{equation}
        E_{miss} = \frac{M_X^2 - M_p^2}{2M_p}
    \end{equation}
where $M_p$ is the proton mass and $M_X^2 = \op p+q-(p_{h^+}+p_{h^-})\cp^2$ is the missing mass squared, calculated from the four-vectors of the proton $p$, of the virtual photon $q$ and of the two decay hadrons $p_{h^+}$ and $p_{h^-}$.
\end{itemize}
The plots of $z_{tot}$ and of $E_{miss}$, for the events with exactly two oppositely charged hadrons in the final state, are shown in Fig. \ref{fig:ztot_emiss}. The peaks, around 1 and 0 respectively, correspond to the exclusive events, the ``background'' under the peak being constituted by the combinatorial of hadrons produced in real SIDIS events. A cut on the $z_{tot}$ ($z_{tot}<0.95$) or $E_{miss}$ peak, indicated in Fig. \ref{fig:ztot_emiss} with the red vertical lines, eliminates most of the \textit{visible component} of the diffractive exclusive contribution to the hadron sample. \\

The \textit{non-visible component}, constituted of all those diffractive events in which one of the two hadrons is not reconstructed, has to be estimated and subtracted. This is done with the HEPGEN Monte Carlo, used to simulate the diffractive production and decay of $\vmrho$, $\vmphi$ and $\omega$ vector mesons. The Monte Carlo reconstructed events, selected with the same procedure as the real data, are normalized to the real data by using the distributions of the visible component, as explained in Sect.~\ref{subsec:pure_sidis}. Then, in each kinematic bin the relevant distributions (azimuthal angle $\fih$ and $\Ptsq$) of the non-visible component, as from the normalized Monte Carlo, are subtracted from the real data distributions. \\
In the next Subsection, some relevant distributions of the exclusive hadrons as from the HEPGEN Monte Carlo samples will be illustrated. The distributions for the reconstructed exclusive events in HEPGEN and in the data have been found in very good agreement. 

\clearpage
\subsection{Exclusive hadrons}
The properties of the exclusive hadrons and their contribution to the measured TMD observables have been investigated and estimated with the HEPGEN Monte Carlo event generator \cite{Sandacz:2012at}. HEPGEN (Hard Exclusive Production GENerator) is dedicated to the study of hard exclusive lepto-production in the COMPASS kinematic regime. Along with the simulation of the hard exclusive production of vector mesons, possibly including the target diffractive dissociation, HEPGEN can be also used to simulate single photon production via the DVCS and the Bethe-Heitler mechanisms. For the work presented in this thesis, the standard vector meson generation has been modified, in order to include the preliminary results of the Spin Density Matrix Elementes (SDMEs) measured in COMPASS and presented in Ch.~\ref{Chapter6_SDMEs}.\\

\begin{figure}[h!]
\captionsetup{width=\textwidth}
\begin{center}
	\includegraphics[width=0.49\textwidth]{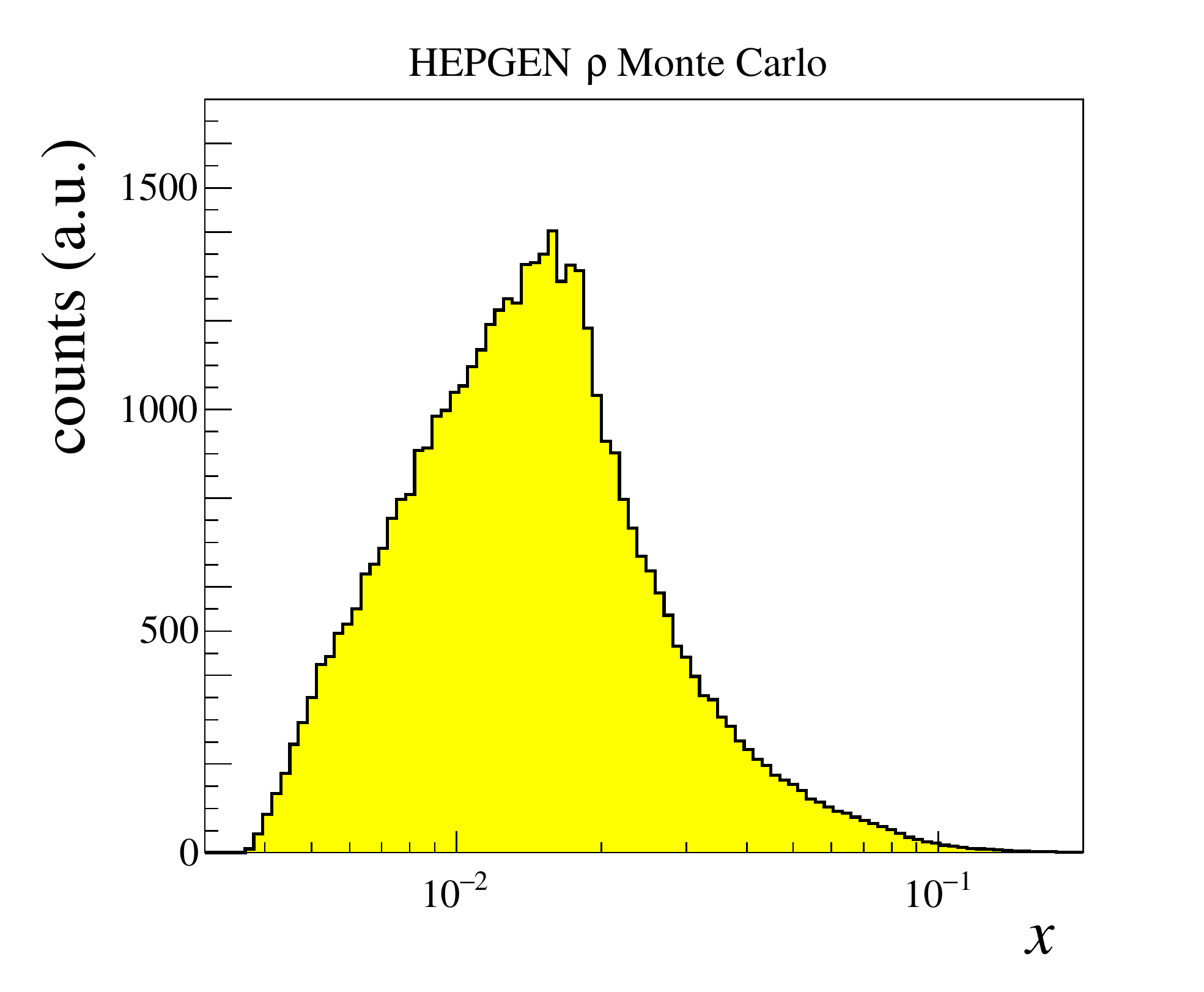}
	\includegraphics[width=0.49\textwidth]{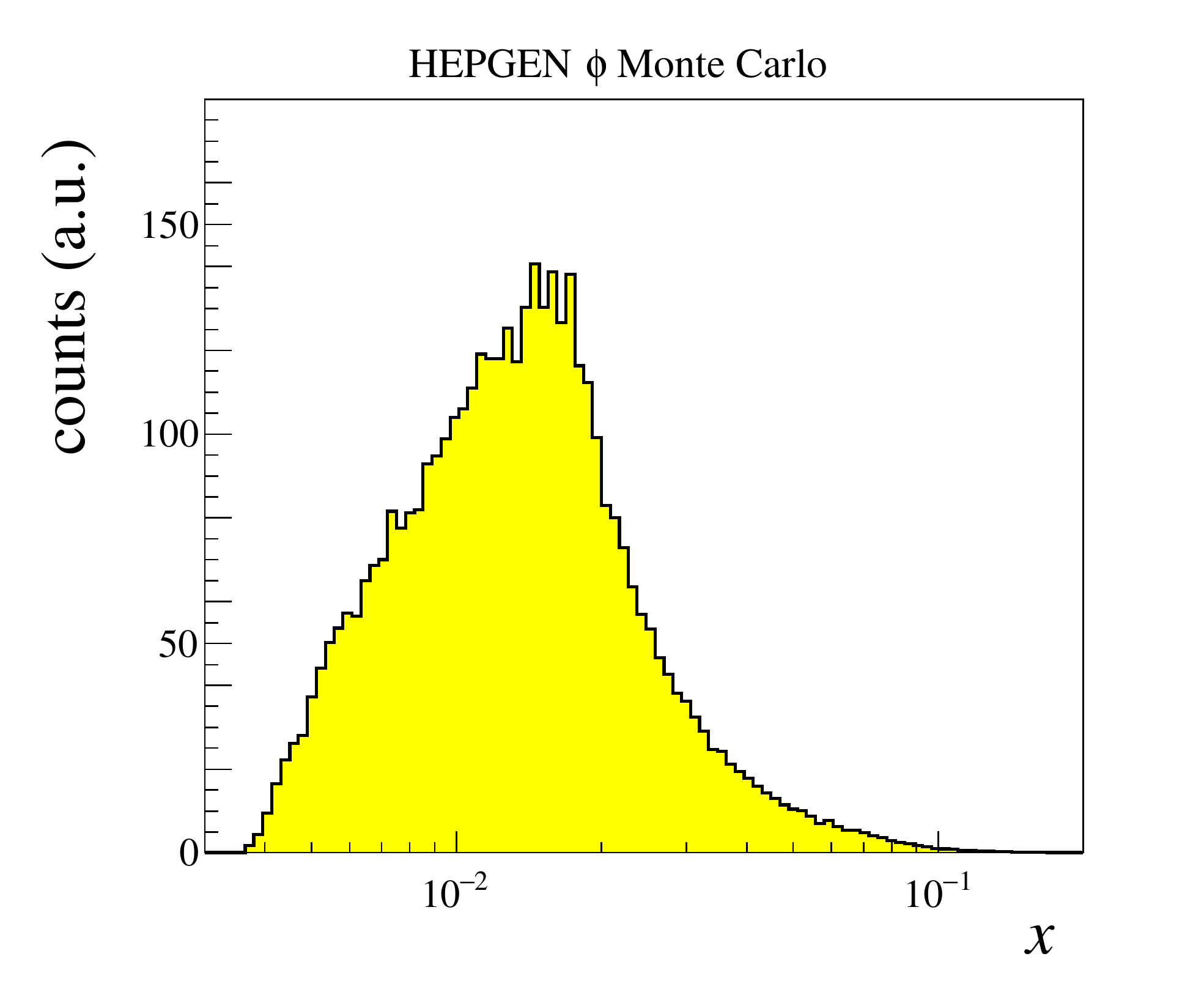}
	\caption{The $x$ distribution of the exclusive events from HEPGEN: $\vmrho$ (left) and $\vmphi$ (right).}
\label{fig:x_excl}
\end{center}
\end{figure}

\begin{figure}[h!]
\captionsetup{width=\textwidth}
\begin{center}
	\includegraphics[width=0.49\textwidth]{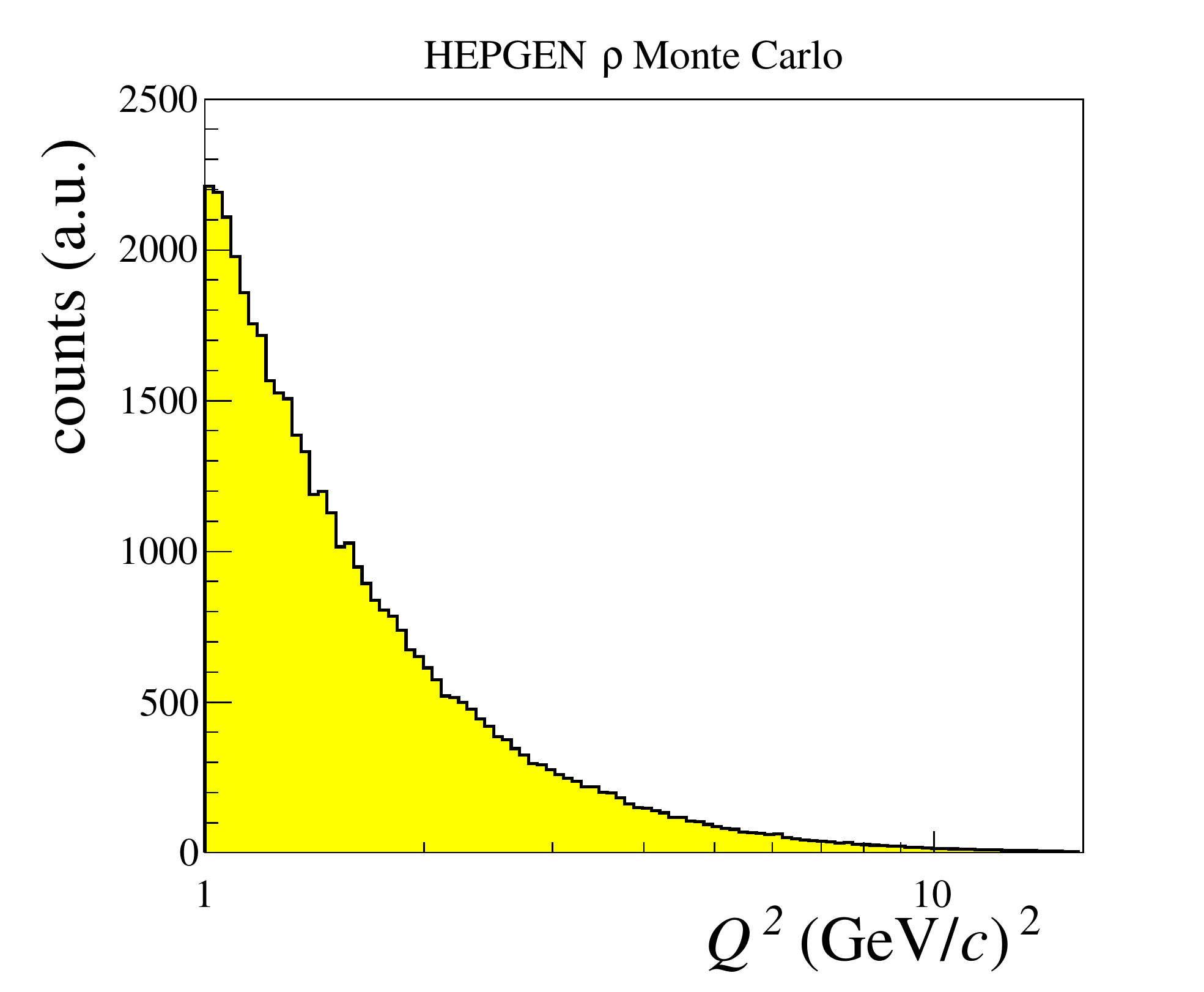}
	\includegraphics[width=0.49\textwidth]{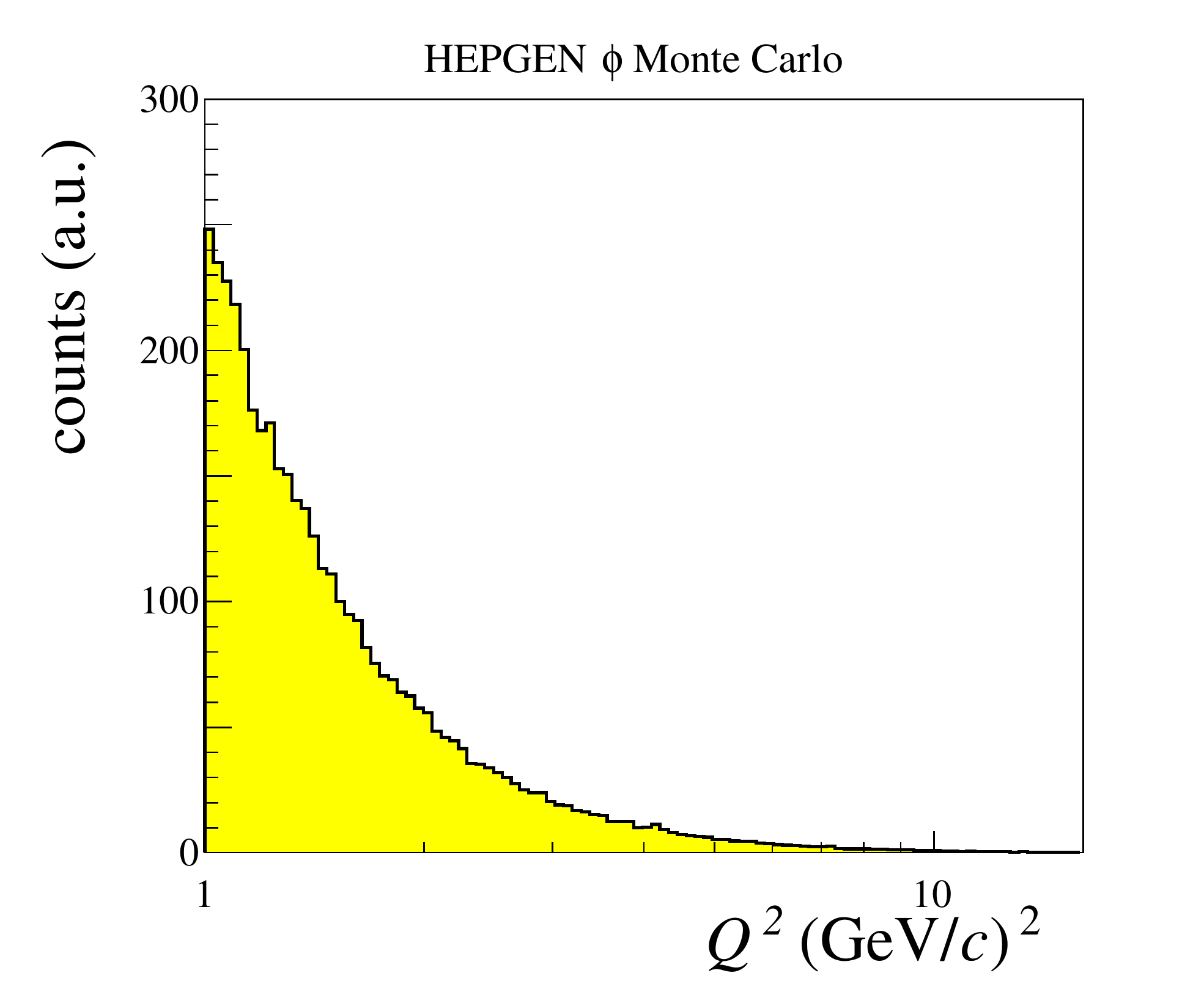}
	\caption{The $\Qsq$ distribution of the exclusive events from HEPGEN: $\vmrho$ (left) and $\vmphi$ (right).}
\label{fig:Q2_excl}
\end{center}
\end{figure}

The exclusive hadrons from the diffractive vector mesons decay are characterized by peculiar kinematics and angular distributions. The $x$ and $\Qsq$ distributions of the reconstructed exclusive events, as from the HEPGEN simulation, are shown in Fig.~\ref{fig:x_excl} and Fig.~\ref{fig:Q2_excl} respectively, for $\vmrho$ and $\vmphi$. Compared to the corresponding distributions in Fig.~\ref{fig:kin0}, it is easy to notice that the exclusive events are characterized by lower values of $x$ and by a much steeper $\Qsq$ distribution.

\begin{figure}[h!]
\captionsetup{width=\textwidth}
\begin{center}
	\includegraphics[width=0.49\textwidth]{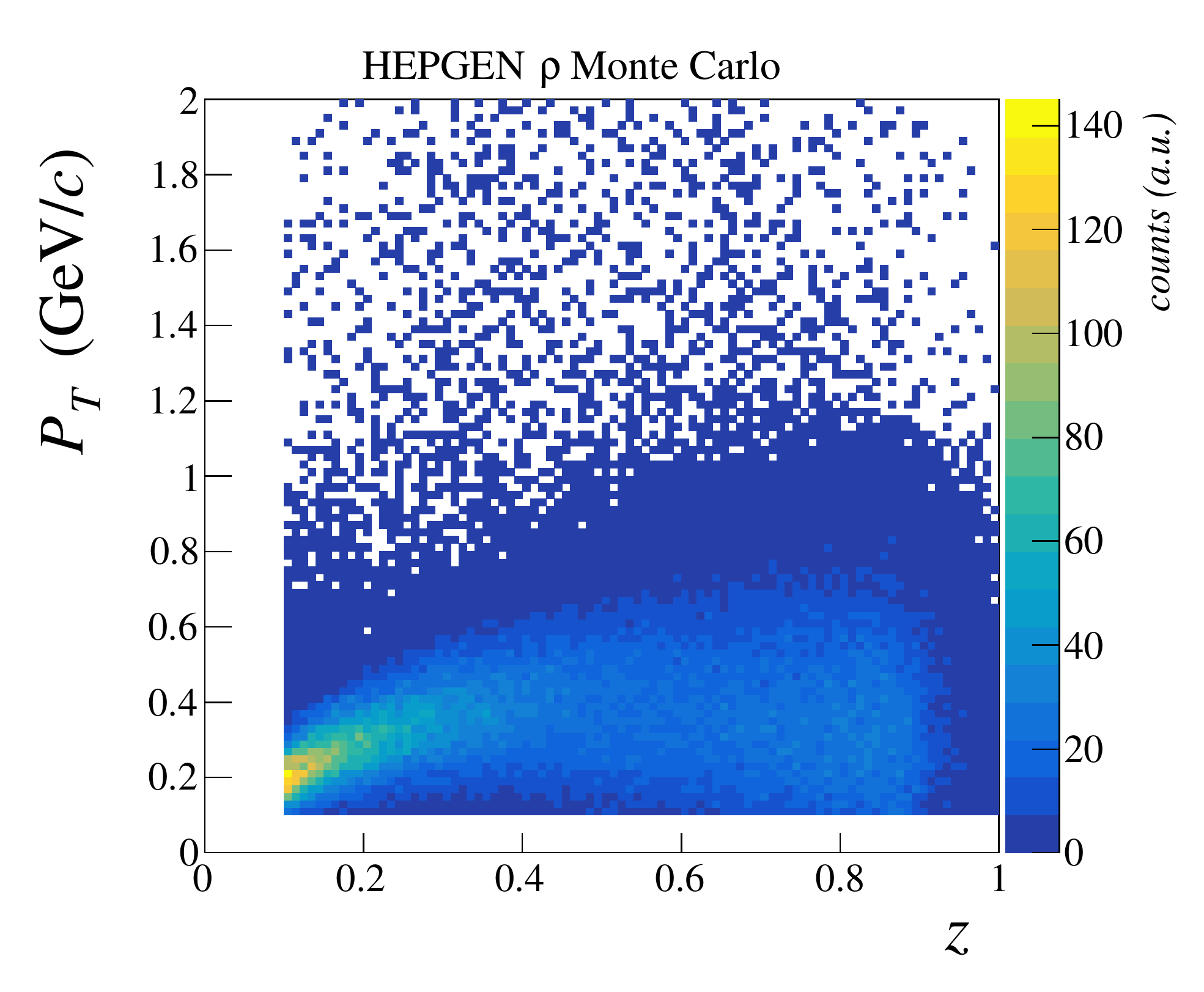}
	\includegraphics[width=0.49\textwidth]{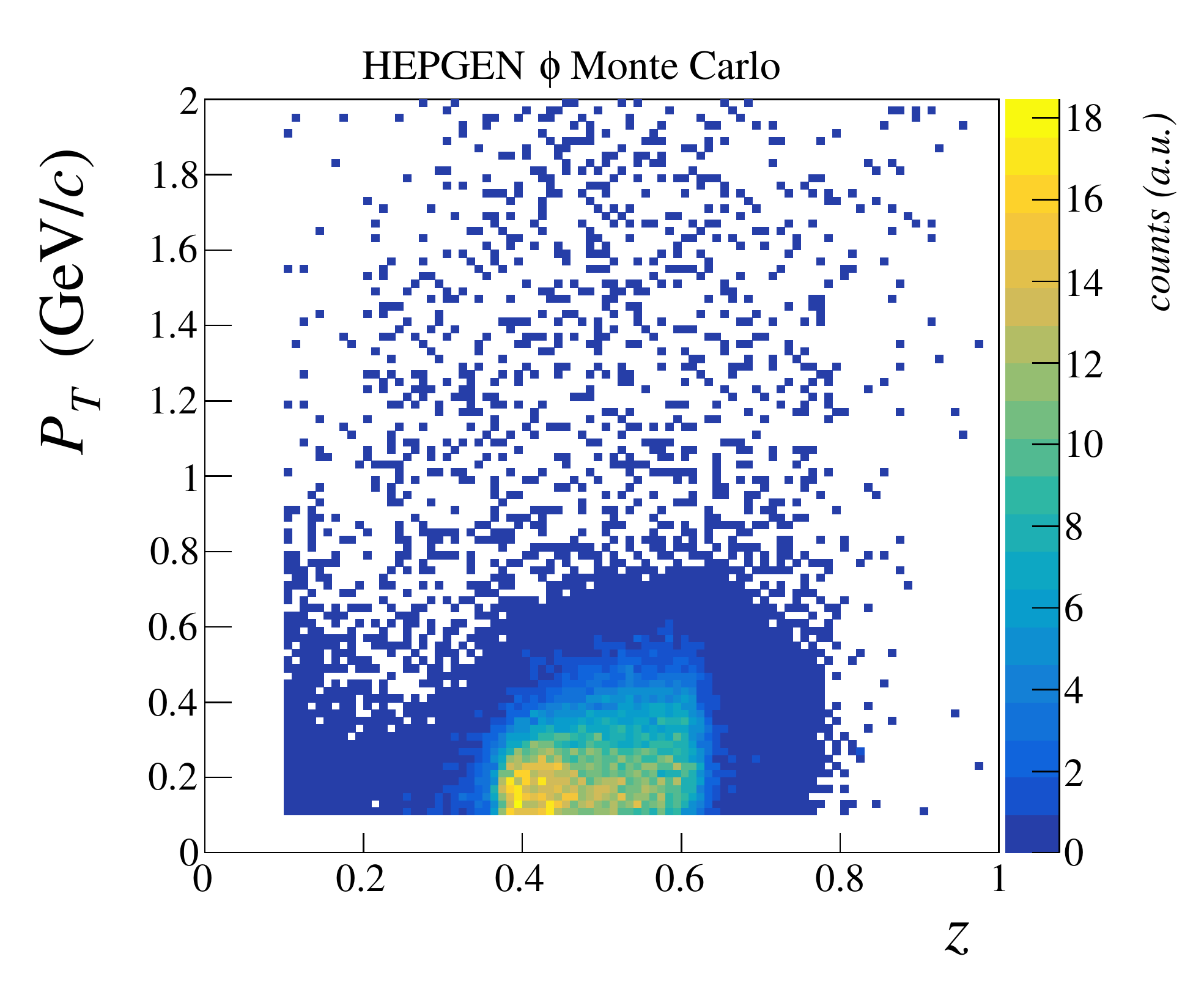}
	\caption{The $z-\Pt$ correlation of the exclusive hadrons from HEPGEN: $\vmrho$ (left) and $\vmphi$ (right).}
\label{fig:zpt_excl}
\end{center}
\end{figure}

The $\Pt$ and $z$ distributions are quite different from those of the hadrons produced in SIDIS, as clear when comparing Fig.~\ref{fig:zpt} with Fig.~\ref{fig:zpt_excl}, which show the $\Pt-z$ correlation for hadrons coming from exclusive $\vmrho$ (left) and $\vmphi$ (right). The $\Pt$-values are much lower, while the $z$ distribution are approximately flat in the range $0.1<z<0.9$ for the $\vmrho$ and in $0.35<z<0.65$ for the $\vmphi$ decay products.

Different, compared to the hadron distributions of Fig.~\ref{fig:zpt}, is also the correlation between $z$ and $\Pt$, as a result of the smaller $\Pt$ range of the exclusive hadrons and, in the case of $\vmphi$, of the $z$ range (Fig.~\ref{fig:zpt_excl}). The difference in the $z$ range between the two vector meson species is due to the different decay kinematics. Indeed, considering a boost along a generic direction $\vbeta$ with $\gamma = (1-\beta^2)^{-1/2}$, the energy $E_1^{lab}$ of one of the two decay products in the laboratory system can be written as:
\begin{equation}
  E_1^{lab} = \gamma \op E_1^{CM} - \vbeta\cdot\vp_1^{CM}\cp =  \gamma \op \frac{M_{VM}}{2} - \vbeta\cdot\vp_1^{CM}\cp  
\end{equation}
where $M_{VM}$ is the mass of the vector meson. For the vector meson, it is $E_{VM}^{lab} = \gamma M_{VM}$ and $\vp_{VM}^{lab} = - \gamma M_{VM}\vbeta$, so that:
\begin{equation}
  E_1^{lab} = \frac{E_{VM}^{lab}}{M_{VM}} \op \frac{M_{VM}}{2} + \frac{\vp_{VM}^{lab}\cdot \vp_1^{CM}}{\gamma M_{VM}}\cp = \frac{E_{VM}^{lab}}{2} + \frac{\vp_{VM}^{lab}\cdot \vp_1^{CM}}{M_{VM}}.
\end{equation}
Since the process is exclusive, $E_{VM}$ is equal to the total energy $\nu$ available to the hadrons, so that:
\begin{equation}
  z_1 = \frac{E_1^{lab}}{E_{VM}^{lab}} = \frac{1}{2} + \frac{\vp_{VM}^{lab}\cdot \vp_1^{CM}}{M_{VM}E_{VM}^{lab}}.
\end{equation}
The same relation hold, with opposite sign, for the second hadron produced in the decay. Then, in general:
\begin{equation}
  z_{1,2} = \frac{1}{2} \pm \frac{\vp_{VM}^{lab}\cdot \vp^{CM}}{M_{VM}E_{VM}^{lab}} \approx \frac{1}{2} \pm \underbrace{\frac{p^{CM}\cos\theta}{M_{VM}}}_{\delta}.
\end{equation}
where the approximation holds if $p_{VM}^{lab}$ is large enough and where $\theta$ is the angle between $\vp_{VM}^{lab}$ and $\vp^{CM}$. The limits of the $z$ ranges are obtained for $|\cos\theta|=1$, which gives $\delta_{\pi}^{\vmrho} = 0.47$ and $\delta_{K}^{\vmphi}=0.14$. We can thus conclude that $z_{\pi}^{\rho^0}$ is basically unconstrained and can range from 0 to 1, while $z_{K}^\vmphi$ is limited to a smaller range, approximately from 0.35 to 0.65, where most of the hadrons in Fig.~\ref{fig:zpt_excl} (right) are.

\begin{figure}[h!]
\captionsetup{width=\textwidth}
\begin{center}
	\includegraphics[width=0.49\textwidth]{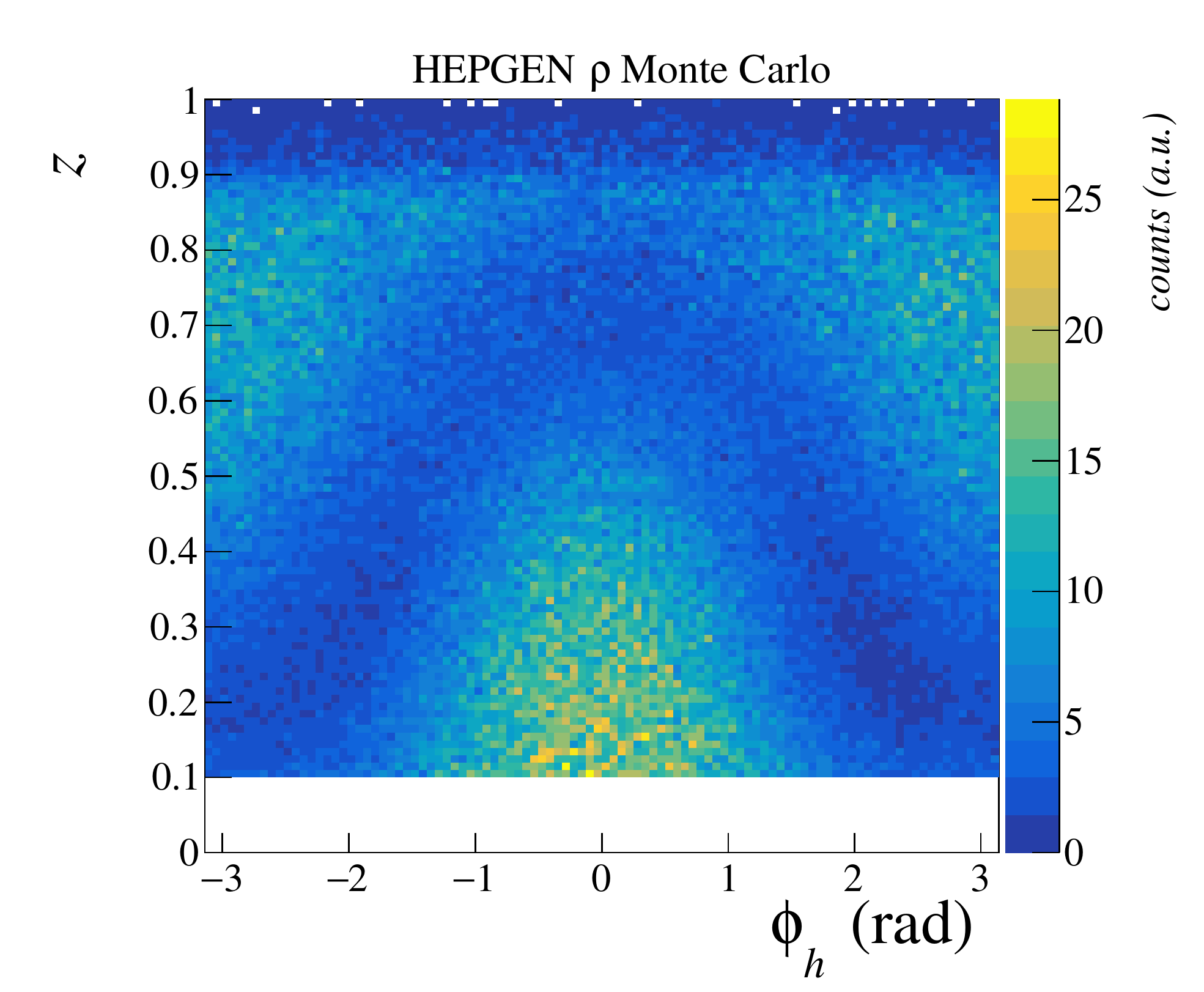}
	\includegraphics[width=0.49\textwidth]{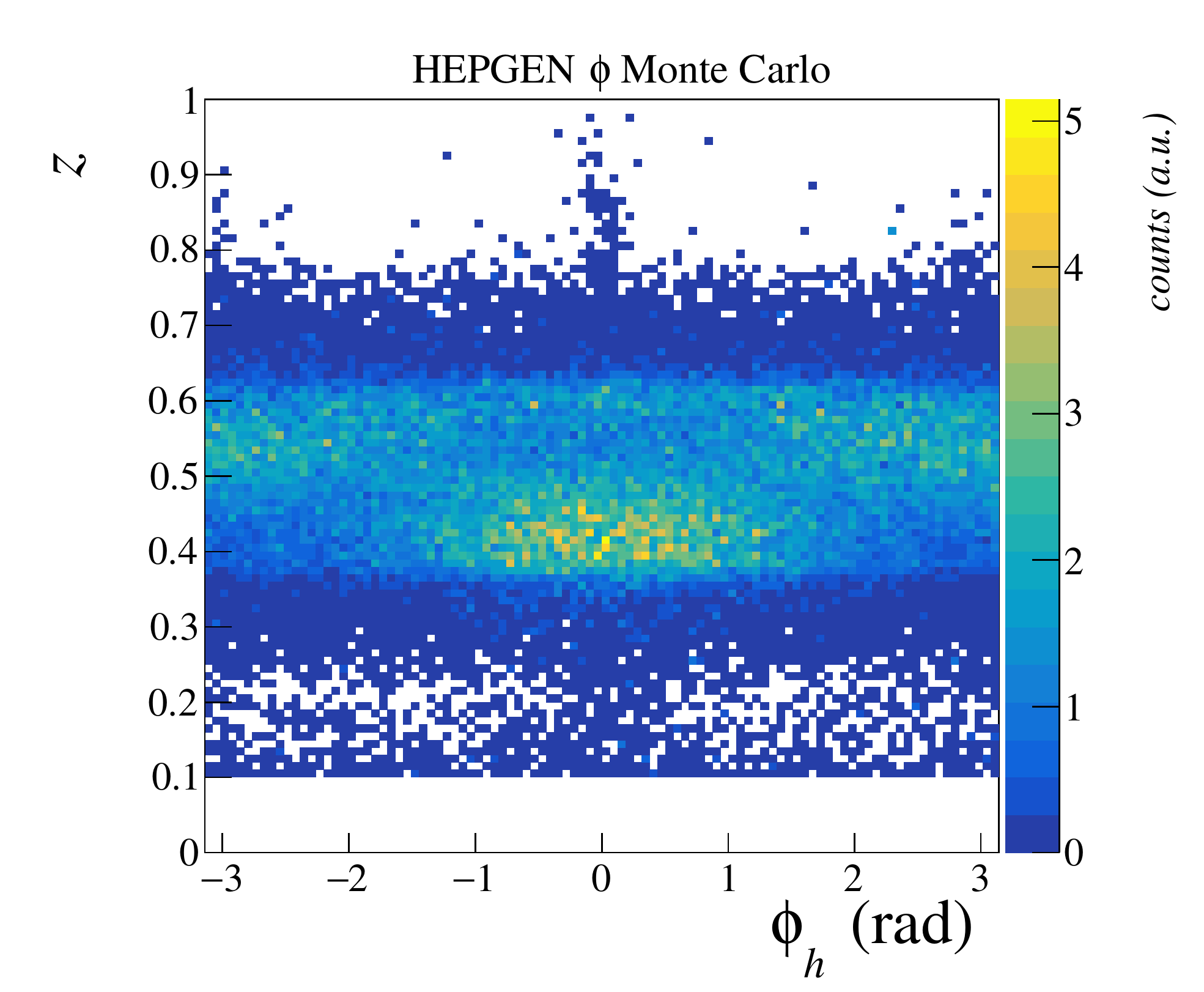}
	\caption{The $z-\fih$ correlation of the exclusive events from HEPGEN: $\vmrho$ (left) and $\vmphi$ (right).}
\label{fig:zphi_excl}
\end{center}
\end{figure}

The $z-\fih$ correlation, shown in Fig.~\ref{fig:zphi_excl}, indicates the presence of strong $z$-dependent modulations in the azimuthal angle. In particular, in the $\vmrho$ case one expects a positive modulation in $\cos\fih$ at small $z$, a negative modulation at large $z$ with a change of sign at $z\approx 0.6$. The correlation has the same features in the $\vmphi$ case, but in a limited $z$ range ($0.35<z<0.65$).

\subsection{Final SIDIS sample}
\label{subsec:pure_sidis}
\begin{figure}[h!]
\captionsetup{width=\textwidth}
\begin{center}
	\includegraphics[width=0.49\textwidth]{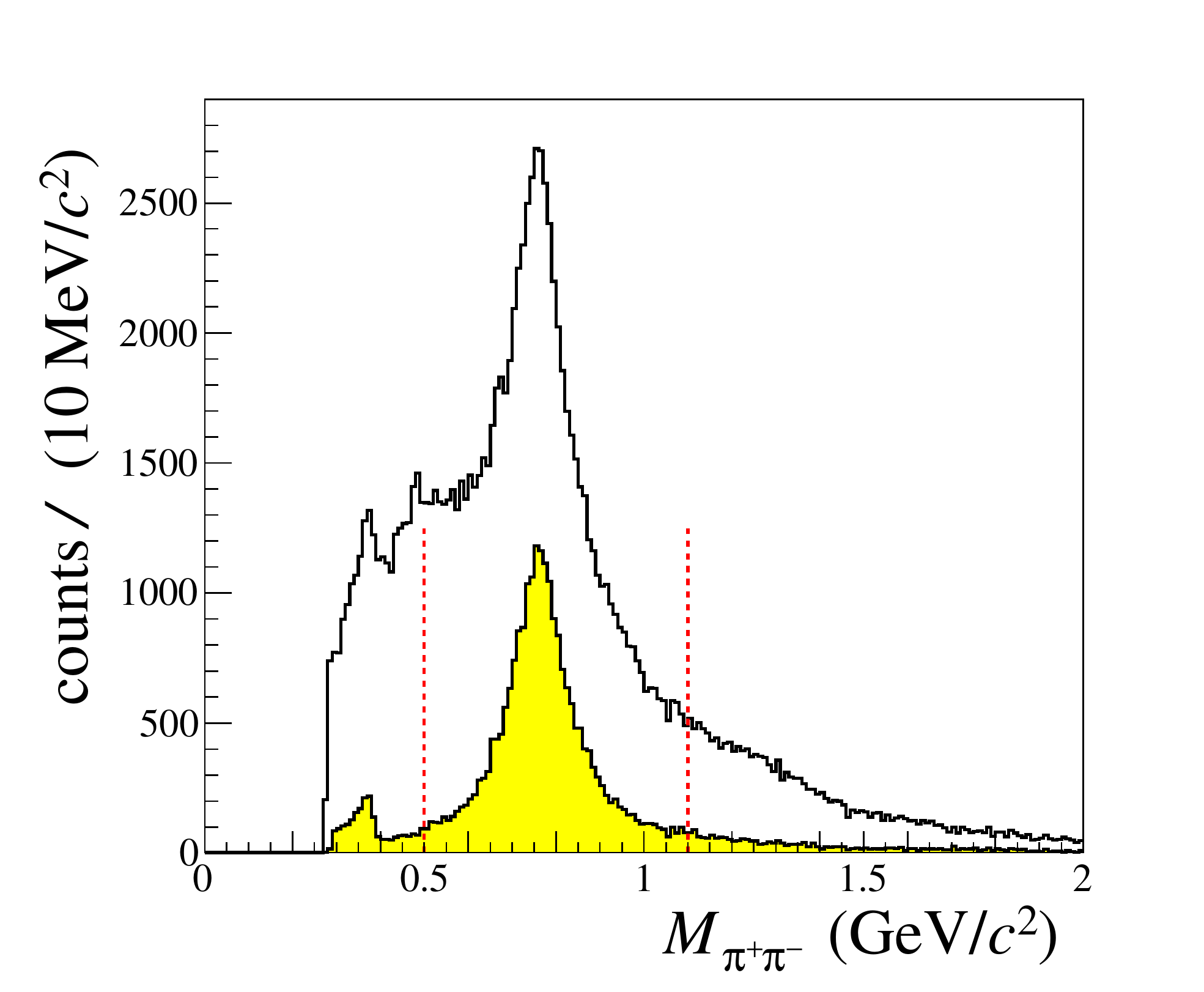}
	\includegraphics[width=0.49\textwidth]{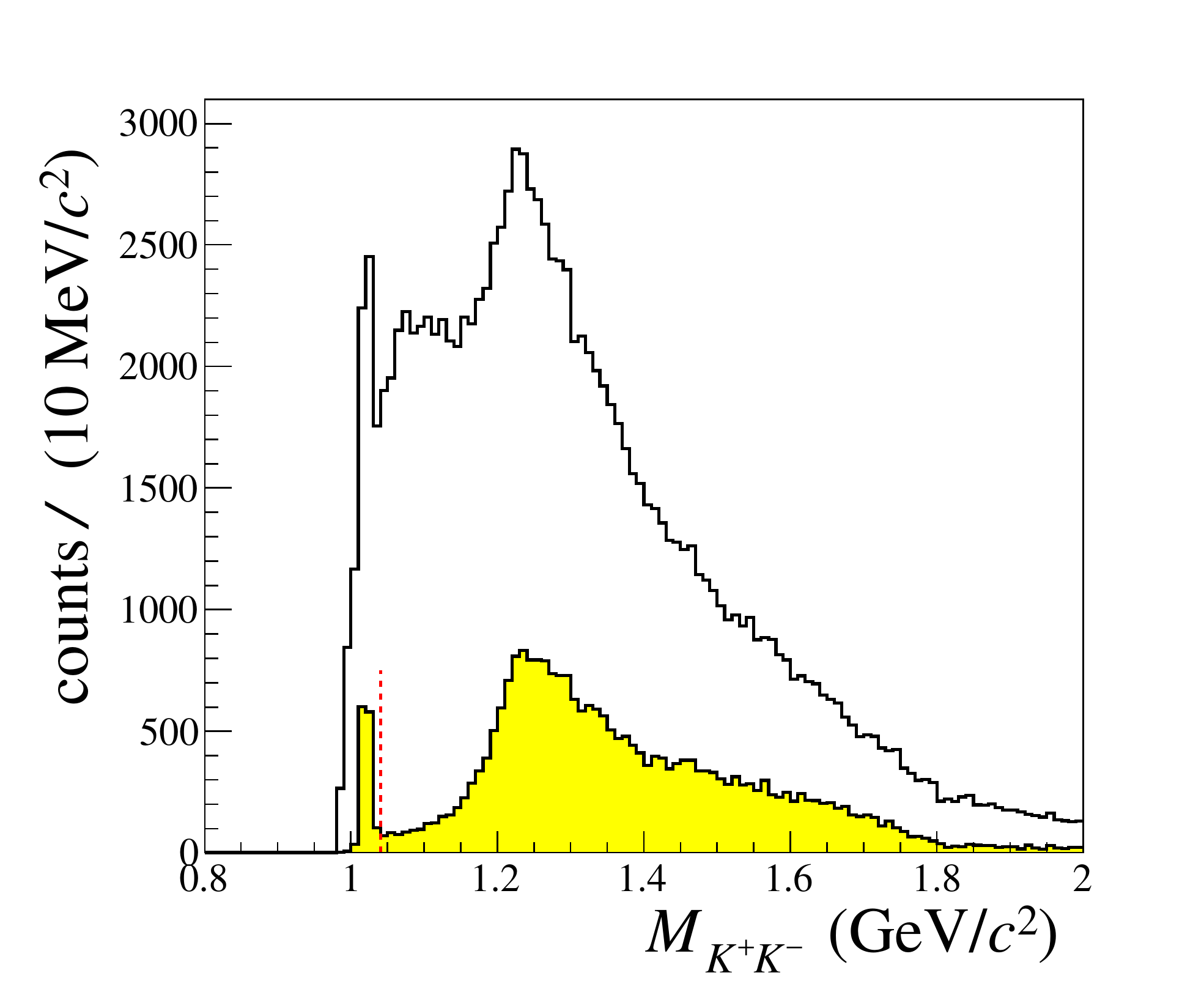}
	\caption{The invariant mass distributions of the pairs in the pion mass hypothesis ($M_{\pi^+\pi^-}$, left) and in the Kaon mass hypothesis ($M_{K^+K^-}$, right) from real data after the cuts explained in the text. The vertical lines indicate the cuts applied in order to select the normalization regions.}
\label{fig:inv_mass}
\end{center}
\end{figure}

As previously said, if an exclusive event is fully reconstructed (that is, exactly two hadrons are observed, with opposite charge and large $z_{tot}$, e.g. $z_{tot}>0.95$), it gets discarded from the data sample and not included in further analysis steps. However, such events are useful to normalize the $\vmrho$ and $\vmphi$ HEPGEN samples. The normalization is done separately for the two species and for $\mu^+$ and $\mu^-$ beams. The identification of the vector mesons is performed by looking at the invariant mass distribution of the hadron pair, assuming alternatively the pion or Kaon mass hypothesis for the two hadrons in the pair. The invariant mass distribution of the reconstructed pairs, from the data for the two mass hypotheses, is shown in Fig.~\ref{fig:inv_mass} (white histograms). A cut on $E_{miss}$ ($-2.5~\mathrm{GeV}<E_{miss}<2.5~\mathrm{GeV}$), whose effect on the mass distributions is also shown (as the yellow histograms), allows getting rid of most of the SIDIS background and in selecting the exclusive component of the distributions. Both in the pion (left) and Kaon hypothesis (right), two components can be observed. In the $M_{\pi^+\pi^-}$ spectrum, the highest peak corresponds to the $\vmrho$, while the smaller peak at $M<0.3$ GeV/$c^2$ is given by the $\vmphi$ candidates, in which the two decay Kaons have been assigned the pion mass. Similarly, in the $M_{K^+K^-}$ spectrum the $\vmphi$ peak at 1 GeV/$c^2$ is accompanied by a broad contribution of $\vmrho$ candidates, in which the two decay pions have been assigned the Kaon mass. The vertical lines in the plots indicate the cuts on the invariant masses used to separate the $\vmrho$ and $\vmphi$ mesons in the normalization procedure.

The estimation of the HEPGEN normalization factors has been performed using:
\begin{itemize}
    \item real data: two samples, corresponding to the two possible beam charges, each obtained from the sum of three sub-periods (P08$\mu^+$, P09$\mu^+$ and P10$\mu^+$ on one side, P08$\mu^-$, P09$\mu^-$ and P10$\mu^-$ on the other);
    \item HEPGEN: four samples, for the two considered vector meson species ($\vmrho$ and $\vmphi$) and the two beam charges;
    \item LEPTO, to take into account the contribution of SIDIS events in the visible component: two samples, one per beam charge. 
\end{itemize}
For each sample, the events have first been filtered according to the selection presented in Sect.~\ref{sect:ch3_ev_had_sel}, also requiring $0.2<y<0.9$; then, the following additional cuts have been applied:
\begin{itemize}
    \item $\Qsq<7.0$~(GeV/$c$)$^2$;
    \item invariant mass $M_{K^+K^-}<1.04$~GeV/$c^2$ in the $\vmphi$ case;
    \item invariant mass $0.5~\mathrm{(GeV/}c^2\mathrm{)}<M_{\pi^+\pi^-}<1.1~\mathrm{(GeV/}c^2\mathrm{)}$ with $M_{K^+K^-}>1.04$~GeV/$c^2$ in the $\vmrho$ case.
\end{itemize}
The first cut has been introduced to select the kinematic region where the exclusive contamination is non-negligible. The cuts on the invariant masses are in line with the ones of Ref.~\cite{COMPASS:2013fsk}, for which an extensive study of the mass spectra had been performed. In particular, in the chosen mass ranges, the contribution originating from the interference of the non-resonant pion pairs production, known to change sign at $M_{\pi^+\pi^-}=M_{\rho}$ \cite{Ross:1965qa,Soding:1965nh}, is minimized.

\begin{figure}[h!]
\captionsetup{width=\textwidth}
\begin{center}
	\includegraphics[width=0.49\textwidth]{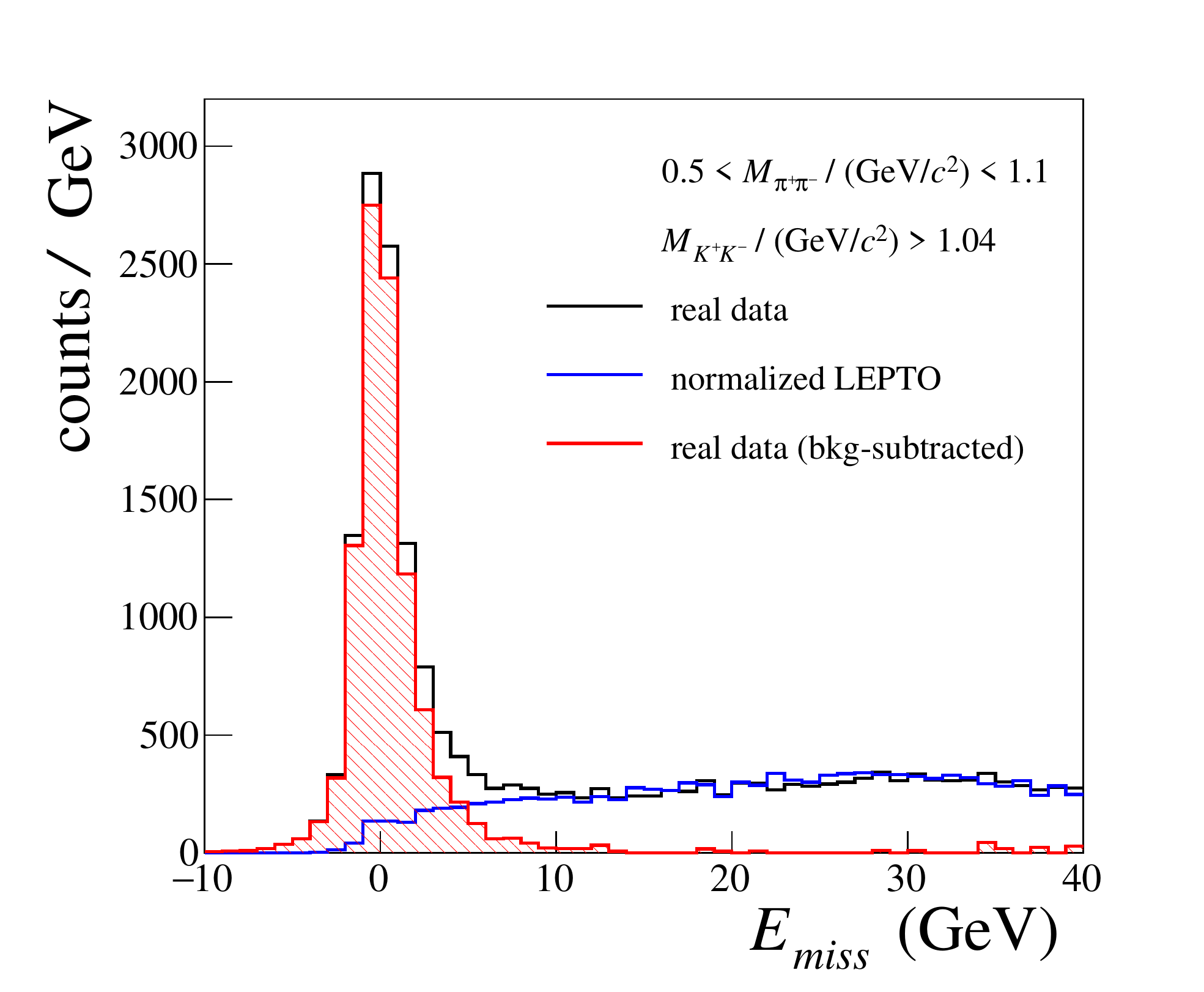}
	\includegraphics[width=0.49\textwidth]{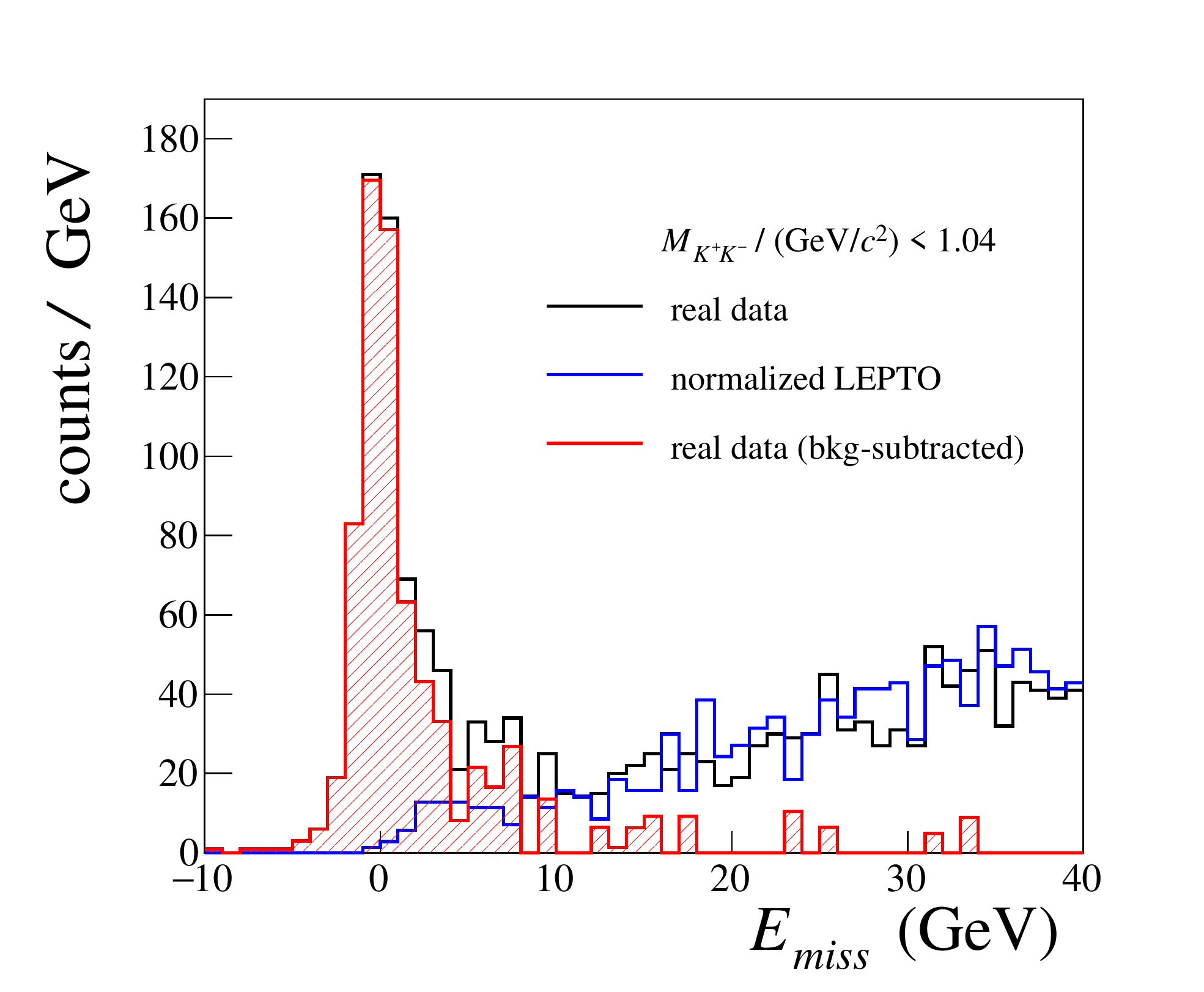}
	\caption{The missing energy distribution $E_{miss}$ for the selected hadron pairs from the data (black), the normalized LEPTO sample (blue) and their difference (shaded red), for the $\mu^+$ case and for the $\vmrho$ (left) and $\vmphi$ vector mesons (right).}
\label{fig:emiss}
\end{center}
\end{figure}

For each data sample, the SIDIS background under the exclusive peak has been estimated with the LEPTO Monte Carlo, which has been first normalized to the data in the range $10~\mathrm{GeV}<E_{miss}<20~\mathrm{GeV}$ and then subtracted from the data. Then, the subtracted distributions have been used to normalize the HEPGEN Monte Carlo in the range $-2.5~\mathrm{GeV}<E_{miss}<2.5~\mathrm{GeV}$. The $E_{miss}$ distributions for the data sample, the normalized LEPTO sample and their difference are shown, for the $\mu^+$ case and the $\vmrho$ and $\vmphi$ case separately, in Fig.~\ref{fig:emiss}. The normalization factors have been found to be: $ n_{\rho^0}^{\mu^+} = 4.97 \pm 0.10$, $n_{\rho^0}^{\mu^-} = 1.00 \pm 0.01$,  $n_{\phi}^{\mu^+} = 0.80 \pm 0.04$ and $n_{\phi}^{\mu^-} = 0.68 \pm 0.04$, where the uncertainties are statistical only. These factors just represent a scaling factor to be applied to the Monte Carlo distributions in the subtraction of the non-visible exclusive component from the data. For the aim of this work, no attempt has been made to convert these normalization values into a ratio of diffractive and SIDIS cross-sections, for which a dedicated study is planned. A systematic uncertainty of $\sim10\%$ has been assigned to the normalization factors, based on a check of the normalization procedure in the kinematic bins used for the extraction of the transverse-momentum distributions and of the azimuthal asymmetries.

Once the HEPGEN normalization values have been fixed, the non-visible component in the HEPGEN samples has been analyzed and the $\Pt$- and $\fih$-distributions subtracted from those of the real data in each kinematic bin of the measurements. As it will be shown in Ch.~\ref{Chapter5_Azimuthal_asymmetries}, the hadrons from the non-visible component are about the 15\% of the total exclusive hadrons. As said before, the diffractive dissociation of the proton is also implemented in HEPGEN: in addition to the two decay hadrons, other hadrons (pions) can be generated from the target proton. These events have been taken into account in the normalization procedure.

\section{Acceptance corrections}
\label{sect:ch3_acceptance}
The acceptance corrections for the azimuthal asymmetries and for the $\Ptsq$-distributions, which include both the geometrical acceptance and the reconstruction efficiency, have been evaluated using the LEPTO Monte Carlo samples, separately for the data collected with $\mu^+$ and $\mu^-$ beams. They have been applied to the measured $\fih$- and $\Ptsq$-distributions in each kinematic bin after subtracting the exclusive diffractive contributions. The same definition of acceptance correction has been used in both cases:

\begin{equation}
    Acc(X) = \frac{N_h^{rec}(X^{rec})}{N_h^{gen}(X^{gen})}
    \label{eq:acc}
\end{equation}
that is, the acceptance correction has been taken as the ratio of the number of reconstructed and generated hadrons in the Monte Carlo samples in a particular bin of the variable $X$, where $X=\fih$ or $X=\Ptsq$. The uncertainty on the acceptance has been calculated as:

\begin{equation}
    \sigma_{Acc}(X) = \frac{\sqrt{N_h^{rec}(X^{rec})}}{N_h^{gen}(X^{gen})},
\end{equation}
thus assuming no uncertainty on the number of generated hadrons and Poissonian statistics for the reconstructed. This definition has been preferred over the Binomial option, which would have given a smaller uncertainty:

\begin{equation}
    \sigma_{Acc}^{Bin}(X) = \frac{\sqrt{N_h^{rec}(X^{rec}) \op 1- \frac{N_h^{rec}(X^{rec})}{N_h^{gen}(X^{gen})}\cp }}{N_h^{gen}(X^{gen})},
\end{equation}
in order to take into account some observed non-statistical fluctuations in the Monte Carlo distributions.

The acceptance corrections also depends on $x$, $Q^2$ and $z$. As already stressed, the $x$ and $y$ ranges of the measurements have been chosen in order to have small acceptance corrections and, when the measurements are performed integrating over one (or more) variables, it is important that the agreement between their distributions from data and Monte Carlo samples are in good agreement, as in our case. \\

Having a 2.5~m long target, the acceptance depends also on the primary vertex position. To investigate this effect, the target has been divided into four \textit{quarters}:
\begin{itemize}
    \item upstream quarter: $-325~\mathrm{cm} < z_{vtx} < -251~\mathrm{cm}$;
    \item central-upstream quarter: $-251~\mathrm{cm} < z_{vtx} < -191~\mathrm{cm}$;
    \item central-downstream quarter: $-191~\mathrm{cm} < z_{vtx} < -131~\mathrm{cm}$;
    \item downstream quarter: $-131~\mathrm{cm} < z_{vtx} < -71~\mathrm{cm}$;
\end{itemize}
and the acceptance correction has been inspected in all the kinematic bins. \\

\begin{figure}
\captionsetup{width=\textwidth}
    \centering
	\includegraphics[width=0.95\textwidth]{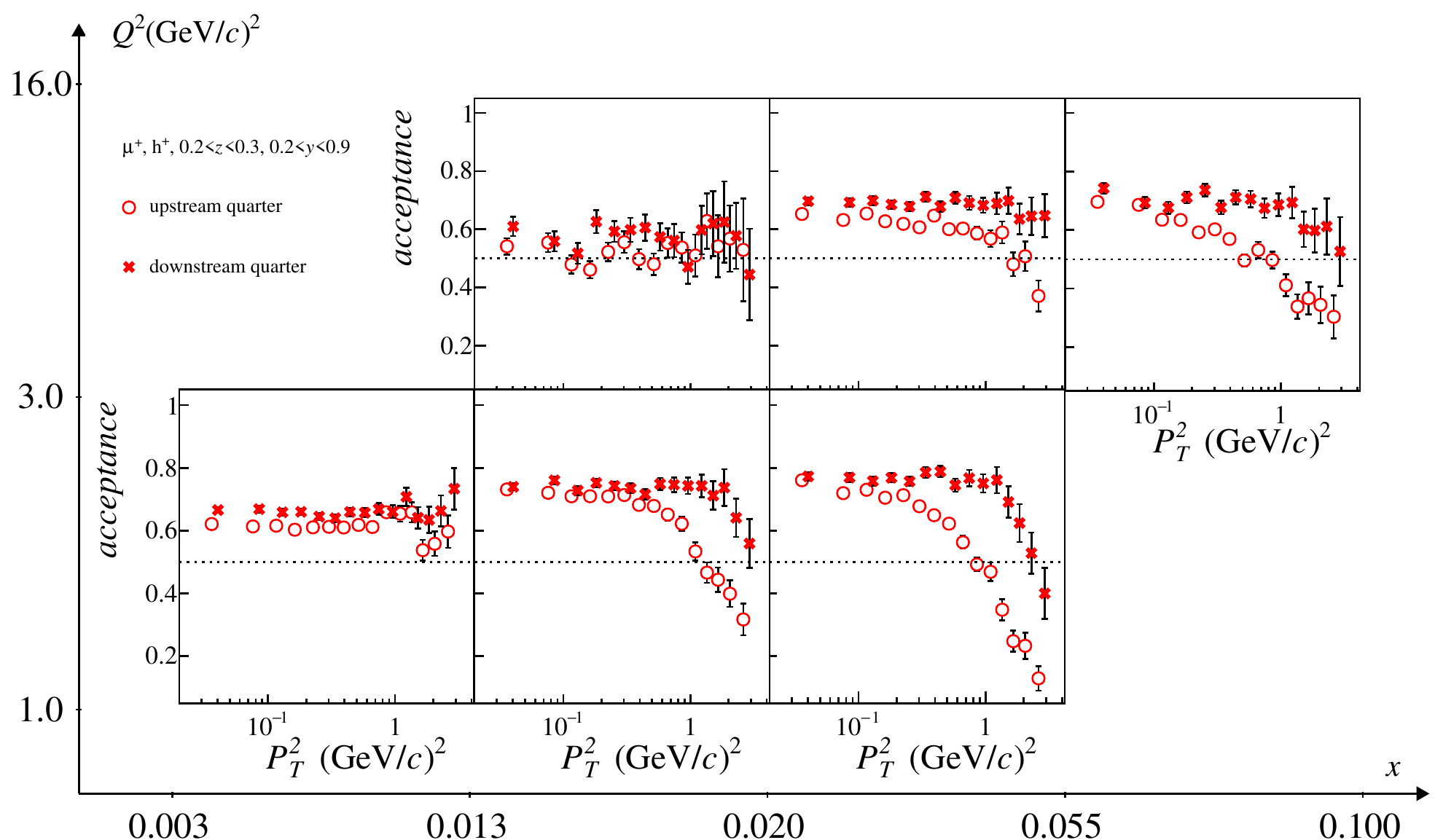}
    \caption{Acceptance as a function of $\Ptsq$ for positive hadrons and $\mu^+$ beam, in bins of $x$ and $Q^2$ for $0.2<y<0.9$ and $0.2<z<0.3$. The two sets of points refer to the upstream and downstream target quarters.}
    \label{fig:ptacc-cells}
\end{figure}

The acceptance as a function of $\Ptsq$ in the upstream and downstream target quarters is shown in Fig.~\ref{fig:ptacc-cells} for positive hadrons collected with a $\mu^+$ beam in bins of $x$ and $\Qsq$ for $0.20<z<0.30$ (the worst case). The acceptance, almost flat at small $\Ptsq$ and weakly dependent on $x$ and $\Qsq$, shows a clear dependence on the primary vertex position at larger values of $\Ptsq$, as expected from geometrical considerations. In particular, in the kinematic bin defined by $0.020<x<0.055$ and $1~\mathrm{(GeV/}c\mathrm{)}^2<Q^2<3~\mathrm{(GeV/}c\mathrm{)}^2$, the acceptance is reduced by more than 50\% at $\Pt=1$ GeV/$c$ for the upstream part of the target only (open points). 

\begin{figure}[!h]
\captionsetup{width=\textwidth}
    \centering
   	\includegraphics[width=0.95\textwidth]{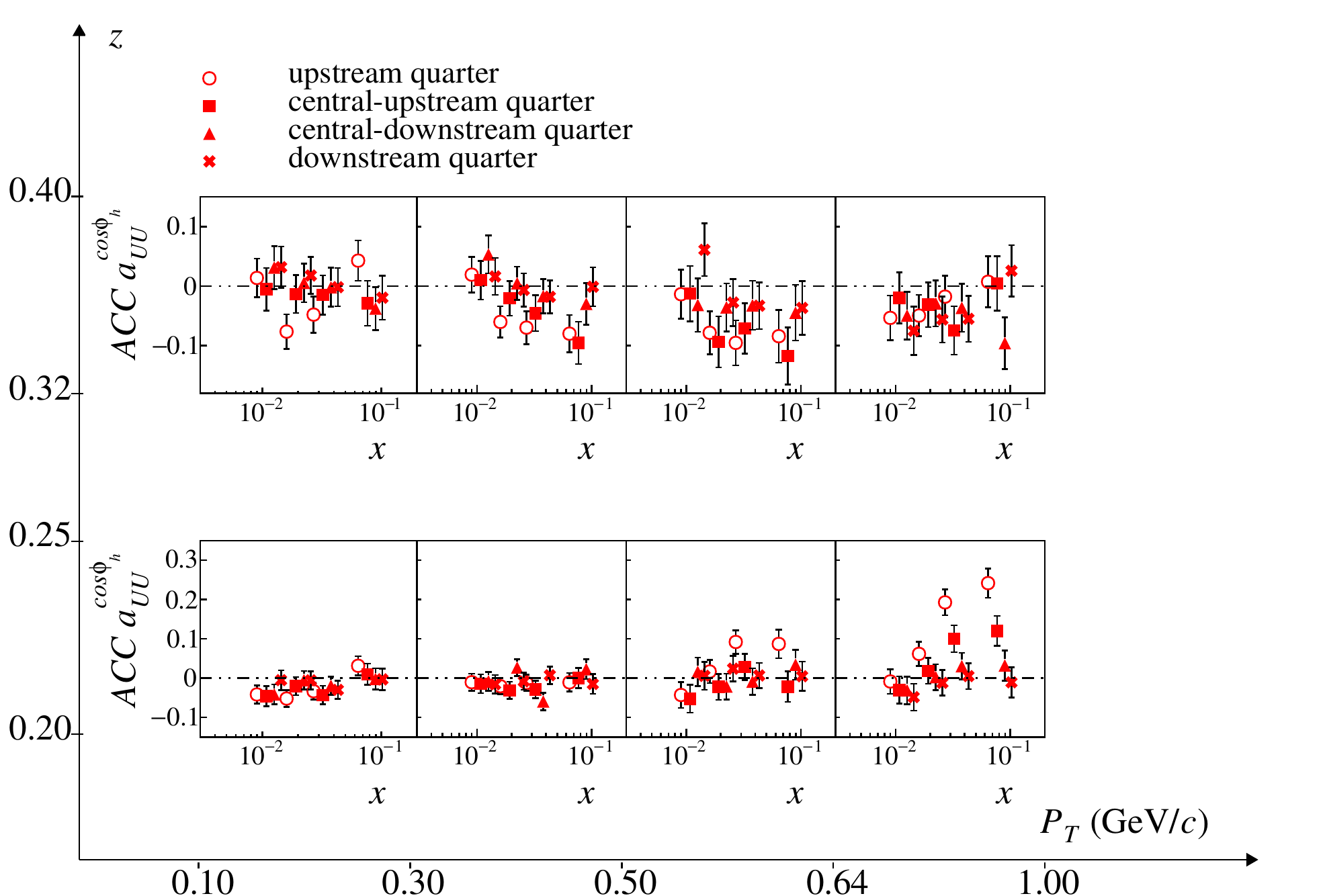}
    \caption{Amplitude of the $\cos\fih$ acceptance modulation for positive hadrons and $\mu^+$ beam, as a function of $x$ in two bins of $z$ ($0.20<z<0.25$ and $0.32<z<0.40$) and four bins in $\Pt$. The four sets of points refer to the four target quarters.}
    \label{fig:aaacc-cells3D}
\end{figure}

In the case of the azimuthal asymmetries, our request is that the modulations of the acceptance corrections are not large compared to the measured azimuthal asymmetries. To have an idea of the effect of the acceptance, we fitted the ratio of the $\fih$ distributions of reconstructed and generated hadrons in the Monte Carlo, using the same fit function as the one used for the data \ref{eq:asymfit}. At first order, the asymmetries after the acceptance corrections are equal to the asymmetries before correction, subtracted of the amplitudes of the acceptance modulations. The amplitude of the $\cos\fih$ acceptance modulations, indicated as $ACC~a_{UU}^{\cos\fih}$, for positive hadrons and $\mu^+$ beam, when simultaneously binning in $x$, $z$ and $\Pt$, is shown in Fig.~\ref{fig:aaacc-cells3D} in two bins of $z$ ($0.20<z<0.25$ and $0.32<z<0.40$, note the different scale). Also in this case, the expected positive acceptance asymmetries are larger for the upstream quarter, even if a clear systematic trends can be seen only at low $z$ and high $\Pt$. \\

To reduce the acceptance correction, the final results for both observables have been produced discarding the events with primary vertex in the upstream quarter. In other words, in addition to the kinematic constraints introduced in Sect. \ref{sect:ch3_ev_had_sel}, a new cut has been applied to the events:

\begin{equation}
 -251~\mathrm{cm} <z_{vtx} <-71~\mathrm{cm}.   
\end{equation}
It has been checked that the number of events as a function of $z_{vtx}$, observed in Monte Carlo, compares well with the distribution obtained for the data, their ratio being flat along $z_{vtx}$. For this reason, the vertex position is integrated over.

\section{Systematic effects}
\label{sect:ch3_systematics}
To conclude, here is the list of the tests performed to estimate the systematic uncertainty:
\begin{itemize}
    \item compatibility between the results from the data collected in the three data taking periods P08, P09 and P10. As said, the spectrometer acceptance and efficiency were stable during these periods, thus no dependence of the results on the period is expected.
    \item compatibility between $\mu^+$ and $\mu^-$ results. The results are not expected to depend on the beam charge. The only observable for which the beam charge plays a role is the $A_{LU}^{\sin\fih}$ asymmetry through the beam polarization, which has opposite sign in the two cases.
    \item compatibility of the results for different target quarters. If the Monte Carlo description of the apparatus is satisfactory, both azimuthal asymmetries and $\Ptsq$-distributions should not show a dependence on the position of the primary vertex, once corrected for acceptance. 
    \item uncertainty on the HEPGEN normalization in the subtraction of the exclusive vector mesons contribution. The statistical uncertainty on the normalization of HEPGEN to the data is $\sim$10\%. The impact on the final results of the normalization has been tested by varying it by 20\%.
    \item uncertainty on the acceptance. This test was done for the $\Ptsq$-distributions only, in order to check the impact on the final results of the $\Ptsq$-slope of the acceptance.
    \item compatibility of results obtained with different Monte Carlo samples for acceptance correction. Several Monte Carlo with (slightly) different geometries, detector efficiency and CORAL versions have been tested. We checked that the acceptance corrections give compatible results.
\end{itemize}

\begin{figure}
\captionsetup{width=\textwidth}
    \centering
    \includegraphics[scale=0.35]{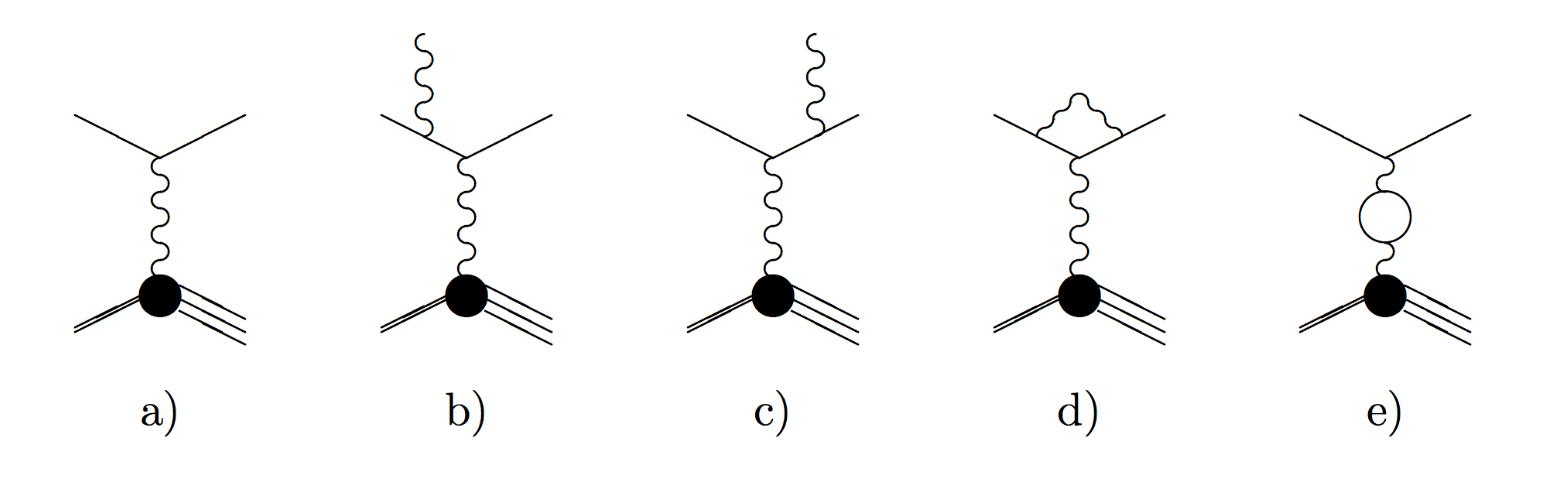}
    \caption{Feynman diagrams for: a) tree level (Born) interaction, b) emission of real photon by the incoming lepton, c) same, by the scattered lepton, d) vertex correction, e) vacuum polarization \cite{Pierre:2019nry}.}
    \label{fig:rad_corr}
\end{figure}

The reconstruction of the kinematics of a Deep Inelastic Scattering event is based the one-photon approximation: the difference of the incoming and scattered lepton reconstructed momenta is taken as the virtual photon momentum. However, radiative effects can change this picture and spoil the reconstruction of the kinematics. If no particle is emitted other than the virtual photon, the cross-section is referred to as \textit{Born cross-section} $\sigma_0$, to which the observed experimental cross-section $\sigma_{exp}$ can be related to via the expression:

\begin{equation}
    \sigma_{exp} = \op 1+\delta_{RC} \cp \sigma_0,
\end{equation}
$\delta_{RC}$ being an overall correction term for radiative effects. At first order in $\alpha$, the radiative corrections arising from QED comprise leptonic radiation, hadronic radiation, two-photon exchange corrections, vacuum polarization and weak corrections. Among these, the leptonic radiation and the vacuum polarization are of particular relevance; they are depicted in Fig.~\ref{fig:rad_corr}. On the other hand, the weak corrections ($Z$-boson exchange and interference with the virtual photon) are expected to be small in the COMPASS regime. \\

The impact of the radiative corrections to the measured kinematic variables can be studied with proper Monte Carlo simulations, as done with RADGEN \cite{Akushevich:1998ft} or DJANGOH \cite{Aschenauer:2013iia}. The latter has been used to get preliminary estimates of their impact on the measured SIDIS observables considered in this Thesis. As the validation of the Monte Carlo results was still ongoing at the time of writing, no correction has been directly applied to the measured observables: as discussed in the Sections dedicated to the systematic uncertainties affecting the measurements, only some estimates of their possible impact will be given. Such estimates have been preliminarily derived by comparing the output of the generator with the radiative effects on and off. 

DJANGOH was born as a Monte Carlo event simulation tool for the treatment of neutral- and charged current $ep$ interactions 
at HERA. Only recently it has been modified in order to simulate also the $\mu p$ interactions occurring at COMPASS. DJANGOH is built over the LEPTO generator, and it coincides with it if the radiative effects are not active. When active, the radiative effects include the single-photon emission from the (incoming or outgoing) lepton lines, the self energy correction, and a complete set of one-loop weak corrections. The background from
radiative elastic scattering is also included.

\bigskip 
The numerical values of each of the quoted systematic tests, as well as a comment on the possible impact of the radiative effects, will be given in the following Ch.~\ref{Chapter4_PT-distributions} and Ch.~\ref{Chapter5_Azimuthal_asymmetries}. \\

\chapter{Measurement of the $\Ptsq$~$-$~distributions} 

\label{Chapter4_PT-distributions}
The event and hadron selection, the kinematic region that can be covered in the present measurements, as well as the general aspects of the corrections applied to the selected hadron samples (namely, exclusive hadrons subtraction, acceptance and radiative effects corrections) have been described in Ch.~\ref{Chapter3_Data_analysis}. Here we focus on the specific aspects related to the measurement of the $\Ptsq$-distributions.\\

The first part of this Chapter is dedicated to the measurement of the $\Ptsq$-distributions in what we call the \textit{standard} binning. Different choices for the binning have also been made, and the results are summarized in Sect.~\ref{sect:ch4_kin_dep}. In all cases, the $\Ptsq$-distributions have been measured separately for positive and negative hadrons from the data collected with $\mu^+$ and $\mu^-$ beams, subtracting the exclusive hadrons contributions and correcting for acceptance. After merging the results obtained with $\mu^+$ and $\mu^-$ beams, all the measured $\Ptsq$-distributions have been normalized to the first bin in $\Ptsq$: this is an arbitrary choice motivated by the fact that, in the present work, we are mainly interested in the shape of the distributions and not in their absolute normalization. The measurement of the $\Ptsq$-dependent cross-section will be the subject of a future analysis. \\

Sections~\ref{sect:ch4_binning} to \ref{sect:ch4_results} are dedicated to the measurements of the $\Ptsq$-distributions in the standard binning, defined in Sect.~\ref{sect:ch4_binning}. Details on the exclusive hadron subtraction and on the acceptance corrections are given in Sect.~\ref{sect:ch4_exclusive} and \ref{sect:ch4_acceptance}, while the systematic uncertainties are discussed in Sect.~\ref{sect:ch4_systematics}. The results for the $\Ptsq$-distributions are presented and discussed in Sect.~\ref{sect:ch4_results}, where they are also interpreted in terms of Leading-Order (LO) TMD formalism and compared to the most recent COMPASS measurements of the $\Ptsq$-distributions for SIDIS off an isoscalar (deuteron) target \cite{COMPASS:2017mvk}. To conclude this part, the $q_T=\Pt/z$ distributions are presented in Sect.~\ref{sect:ch4_qt}. All these results have been released by the COMPASS Collaboration, and they have already been presented at international conferences \cite{Moretti:2019lkb,Moretti:CPHI,Moretti:ICNFP,Matousek:IWHSS,Moretti:DIS}. \\

Further investigations of the kinematic dependences of the $\Ptsq$-distributions are described in Sect.~\ref{sect:ch4_kin_dep}. In particular, the $\Ptsq$-distributions have been measured in more $z$ bins, in $W$ bins and in more $\Qsq$ bins. Most of these results have already been released by the COMPASS Collaboration and are available upon request. Finally, a first extraction of the $\Qsq$-dependence of $\aktsq$ from these data is presented. \\

\section{The \textit{standard} binning}
\label{sect:ch4_binning}
\begin{figure}[h!]
\captionsetup{width=\textwidth}
\begin{center}
	\includegraphics[width=0.65\textwidth]{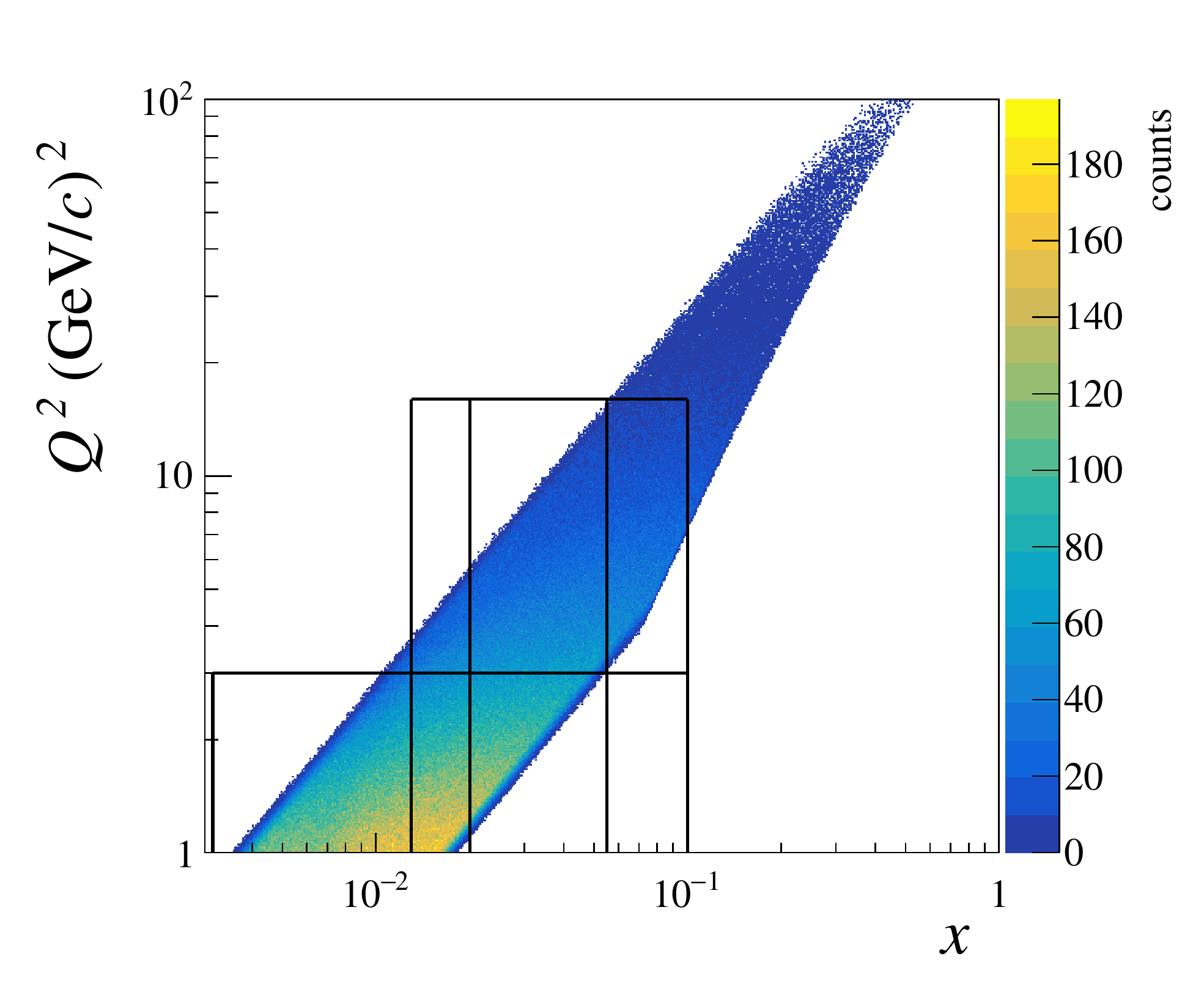}
	\caption{The standard binning used for the measurement of the $\Ptsq$-distributions, drawn on top of the $x-\Qsq$ correlation plot.}
\label{fig:xQ2_dist}
\end{center}
\end{figure}

The transverse-momentum distributions have been measured in bins of $x$, $\Qsq$ and $z$. The first choice for the binning has been to keep the same intervals as for the published COMPASS multiplicities on deuteron \cite{COMPASS:2017mvk} in the region common to the two analyses (\textit{standard} binning). The $x$ and $\Qsq$ bins are shown in Fig.~\ref{fig:xQ2_dist}. The $\Ptsq$-distributions have been obtained for $0.2<y<0.9$, but the low-$y$ region ($0.1<y<0.2$), where the acceptance is smaller, has also been inspected. In each ($x,\Qsq$) bin the $z$ range has been divided into four intervals and in each of them the distributions have been measured in 15 bins from $\Ptsq=$ 0.02~(GeV/$c$)$^2$ to $\Ptsq=$~3.00~(GeV/$c$)$^2$. The $x$, $\Qsq$ and $z$ binning is given in Tab.~\ref{tab:dist_stat}, where the total number of positive and negative hadrons in each useful bin, for $0.2<y<0.9$ and summing over the beam charge, is also given. \\

In each $x$, $\Qsq$ and $z$ bin, the $\Ptsq$-distributions have been corrected for the remaining exclusive hadrons and for the acceptance, as explained in the next two Sections.

\begin{table}[tbh!]
\captionsetup{width=\textwidth}
\small
    \centering
    \begin{tabular}{c|c|cc|cc}
     & & \multicolumn{2}{c|}{$1.0<\Qsq<3.0$} & \multicolumn{2}{c}{$3.0<\Qsq<16.0$} \\
    \hline
     $x$ & $z$ & $N_{h^+}$ & $N_{h^-}$ & $N_{h^+}$ & $N_{h^-}$\\
    \hline
    \multirow{4}{*}{0.003 - 0.013} & 0.20 - 0.30 & 134728 & 119023 & & \\
    & 0.30 - 0.40 & 69069 & 58466 & &   \\
    & 0.40 - 0.60 & 60397 & 49646 & &   \\
    & 0.60 - 0.80 & 22191 & 17824 & &  \\ 
    \hline
    
    \multirow{4}{*}{0.013 - 0.020} & 0.20 - 0.30 & 100792 & 85294 & 10341 & 8921 \\
    & 0.30 - 0.40 & 53100 & 43159 & 5253 & 4256 \\
    & 0.40 - 0.60 & 47584 & 35908 & 4300 & 3291 \\
    & 0.60 - 0.80 & 17590 & 12823 & 1397 & 1049 \\
    \hline
    
    \multirow{4}{*}{0.020 - 0.055} & 0.20 - 0.30 & 85698 & 70818 & 78882 & 64766 \\ 
    & 0.30 - 0.40 & 46714 & 35628 & 40671 & 30577 \\ 
    & 0.40 - 0.60 & 41785 & 29550 & 34560 & 23380 \\ 
    & 0.60 - 0.80 & 15207 & 10124 & 11611 & 6893 \\ 
    \hline
    
    \multirow{4}{*}{0.055 - 0.100} & 0.20 - 0.30 &  &  & 38442 & 30079 \\ 
    & 0.30 - 0.40 &  &  & 20871 & 13988 \\ 
    & 0.40 - 0.60 &  &  & 18188 & 10658 \\ 
    & 0.60 - 0.80 &  &  & 5915 & 2925 \\ 
    \hline
    
    \end{tabular}
    \caption{Number of positive ($N_{h^+}$) and negative hadrons ($N_{h^-}$) in the bins of $x$, $z$ and $\Qsq$ used in the measurement of the $\Ptsq$-distributions. $\Qsq$ is given in units of (GeV/$c$)$^2$.}
        \label{tab:dist_stat}
\end{table}

\section{Exclusive hadron contribution}
\label{sect:ch4_exclusive}
\begin{figure}[h!]
\captionsetup{width=\textwidth}
    \centering
    \includegraphics[width=0.95\textwidth]{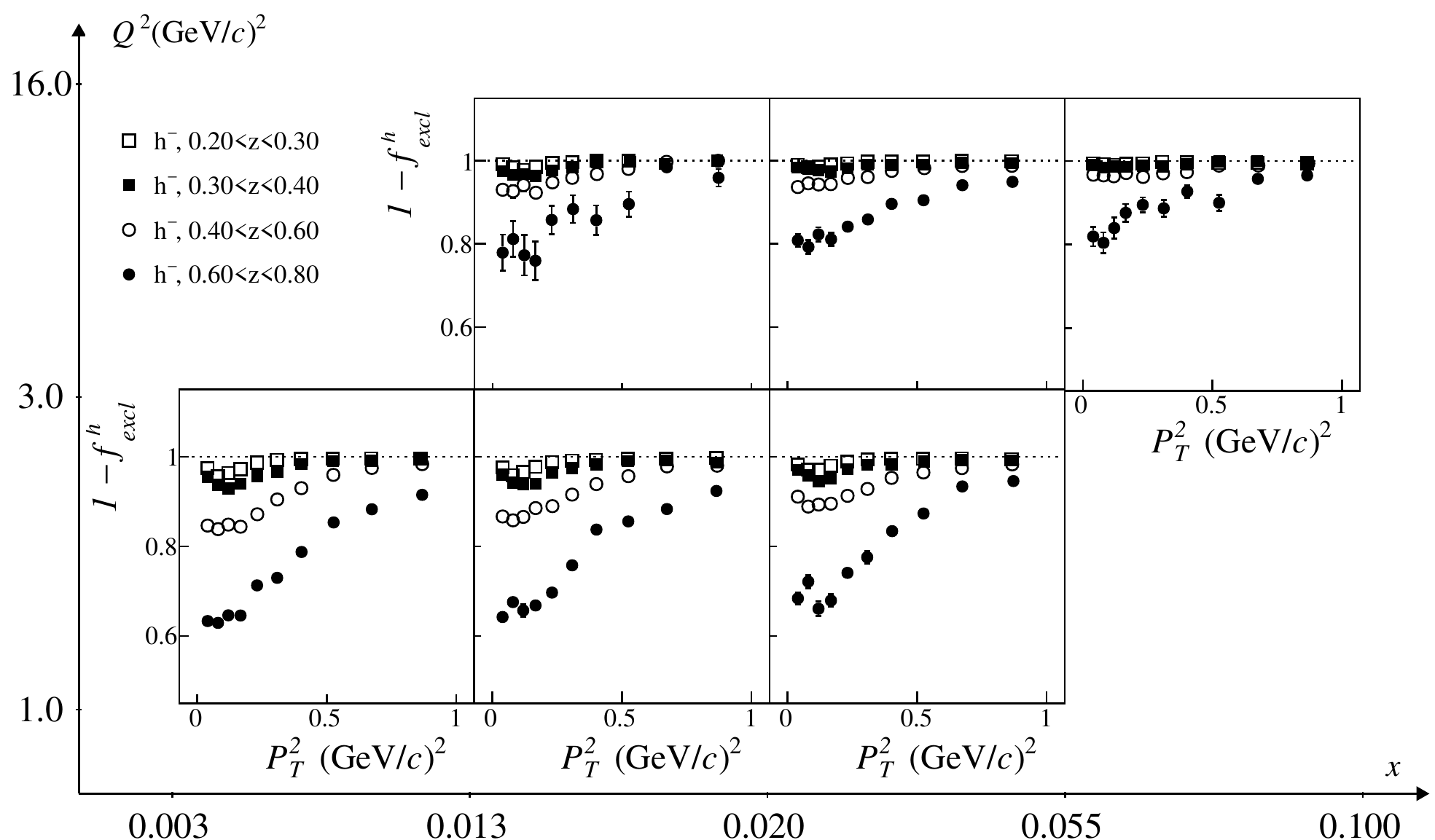}
    \caption{The estimated fraction of negative SIDIS hadrons over the total number of negative hadrons in the data, for the average of $\mu^+$ and $\mu^-$ data.}
    \label{fig:ptds_cont}
\end{figure}

The amount of exclusive hadrons in the selected hadron sample is kinematic-dependent and, in general, not negligible. This can be seen in Fig.~\ref{fig:ptds_cont}, which shows the estimated fraction of hadrons from SIDIS $f_{SIDIS}^h$
\begin{equation}
    f_{SIDIS}^h = 1 - f_{excl}^h = \frac{N_{tot}^h-N_{excl}^h}{N_{tot}^h}
\label{eq:fract_h}    
\end{equation}
as a function of $\Ptsq$ in the various $x,\Qsq,z$ bins, for the average of the $\mu^+$ and $\mu^-$ beam cases and for negative hadrons. In Eq.~\ref{eq:fract_h}, $N_{excl}^{h}$ is the total number of exclusive hadrons, that is the sum of the exclusive hadrons reconstructed in the data and the non-visible exclusive hadrons, estimated from the Monte Carlo sample, normalized to the data following the procedure described in Ch.~\ref{Chapter3_Data_analysis}. As can be seen, there is a weak dependence on $x$, while the dependences on $\Qsq$ and $z$ are strong. In particular, for $\Qsq<3$~(GeV/$c$)$^2$ and $0.60<z<0.80$, about 40\% of the hadrons are decay products of diffractively produced vector mesons.

For positive hadrons, the trends are similar; the exclusive hadron percentage is however smaller in this case, due to the higher probability of producing a positive SIDIS hadron in the final state. What remains after this cut is the non-visible component only. The $\Ptsq$-distributions of these hadrons, as obtained from the normalized HEPGEN samples (see Ch.~\ref{Chapter3_Data_analysis}) are subtracted in each bin of $x$, $\Qsq$ and $z$ from the corresponding measured distributions. The exclusive hadrons from the non-visible components are at most 7\% of the total number of hadrons: this fraction, observed at low $\Pt$, at low $x$ and $\Qsq$, and in the fourth $z$ bin, approximately corresponds to one quarter of the total fraction of exclusive hadrons as from Fig.~\ref{fig:ptds_cont}.

The uncertainty on the HEPGEN normalization and its impact on the measured $\Ptsq$-distributions are discussed in Sect.~\ref{sect:ch4_systematics}.

\newpage
\clearpage
\section{Acceptance correction}
\label{sect:ch4_acceptance}
\begin{figure}
\captionsetup{width=\textwidth}
    \centering
	\includegraphics[angle=0,width=0.75\textwidth]{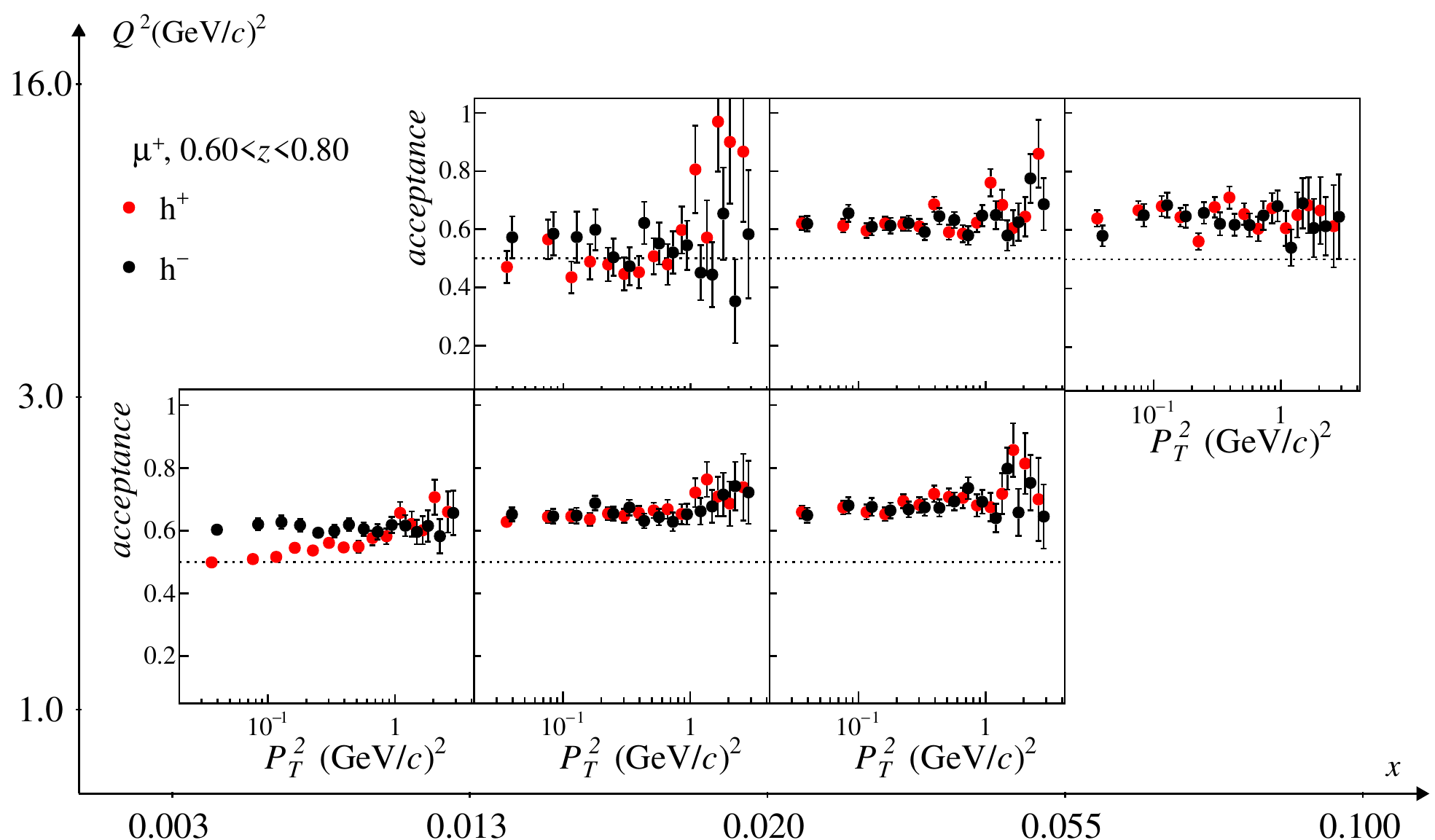}\\
	\includegraphics[angle=0,width=0.75\textwidth]{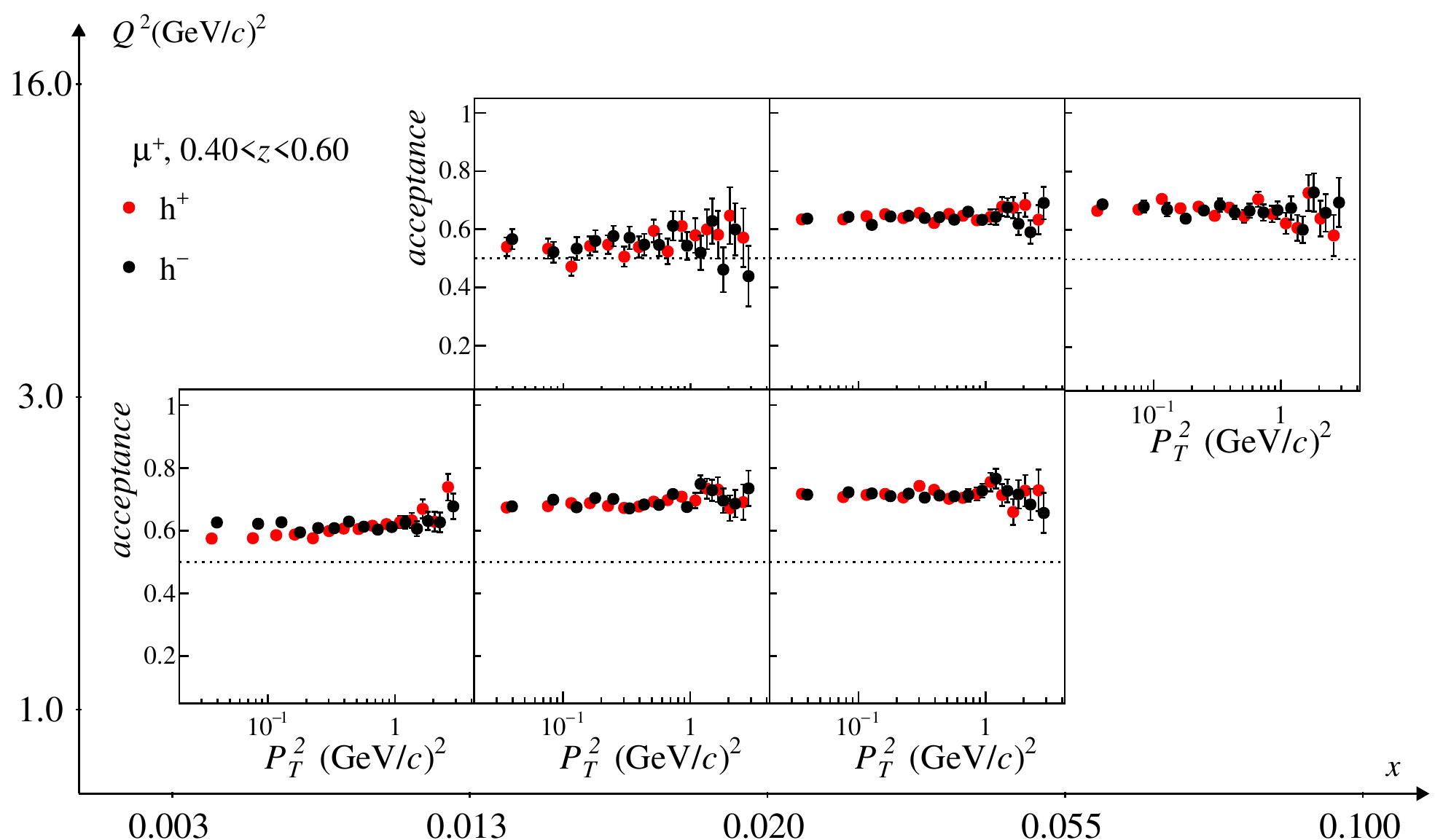}\\
	\includegraphics[angle=0,width=0.75\textwidth]{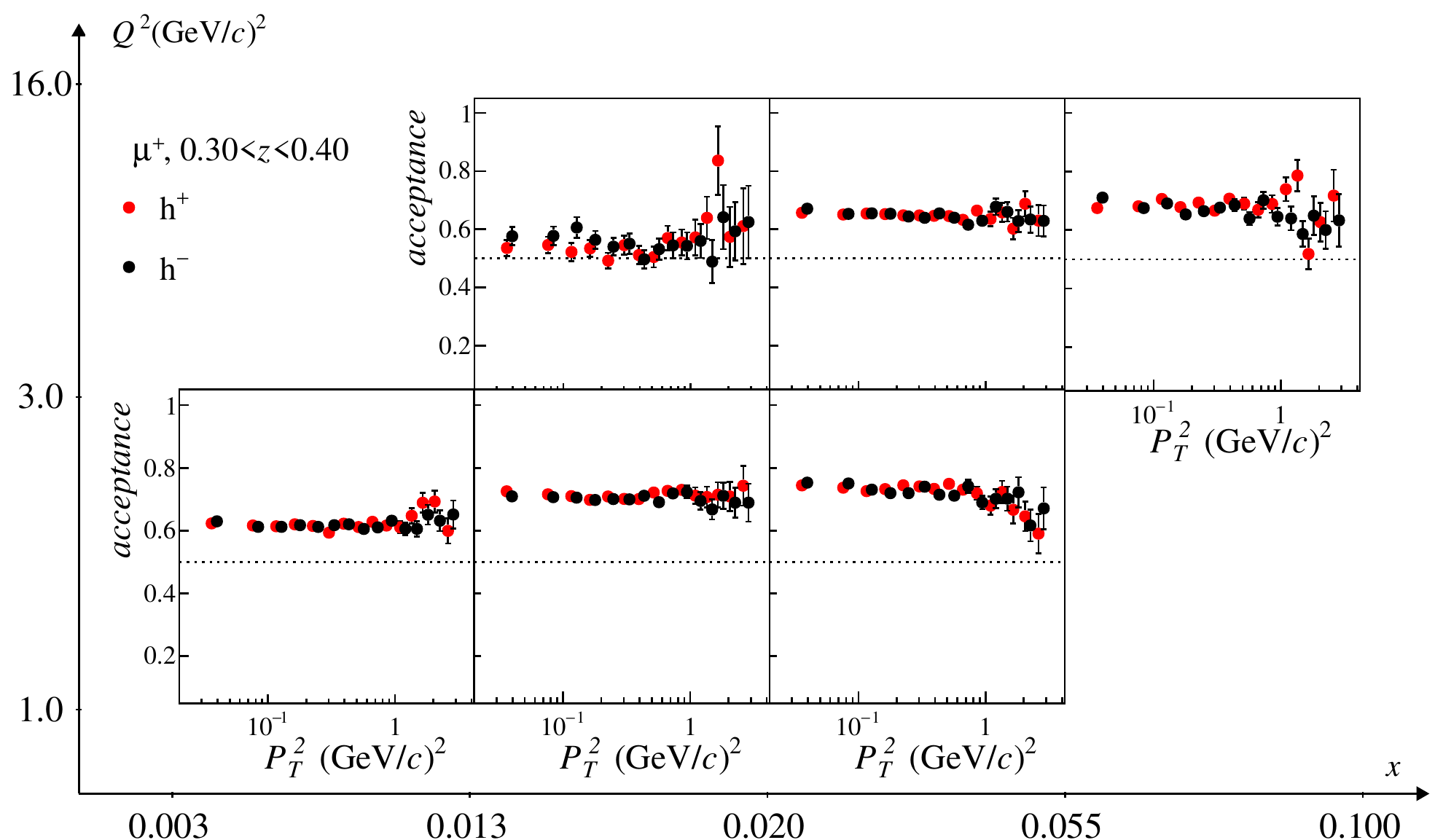}\\
	\includegraphics[angle=0,width=0.75\textwidth]{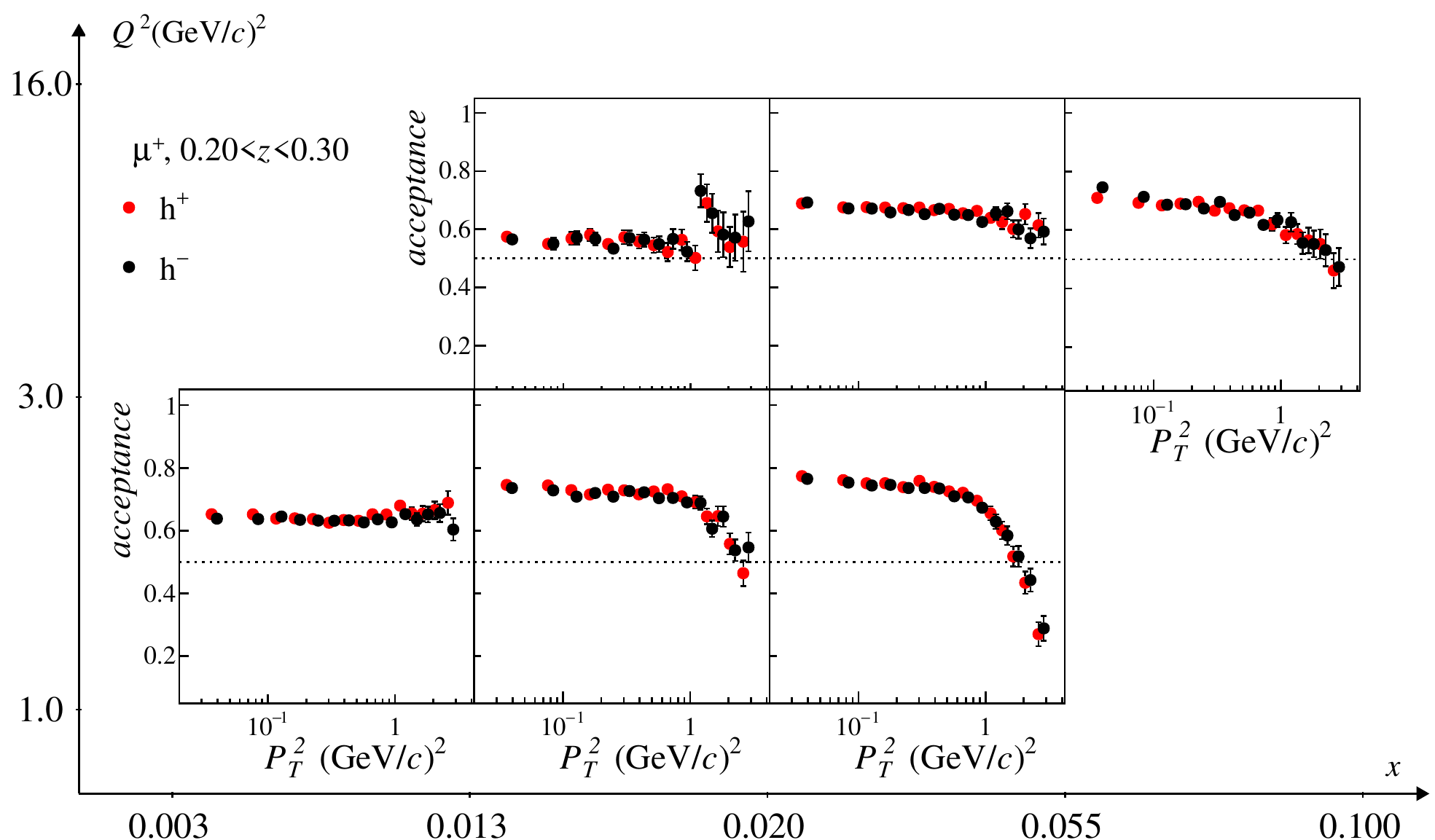}
	\caption{Acceptance as a function of $\Ptsq$, for $\mu^+$ beam, for positive (red) and negative hadrons (black). The different plots refer to the different $z$ bins and in each plot the panels refer to the different $x, Q^2$ bins.}
    \label{fig:pt2acc_mup}
\end{figure}

The acceptance, calculated according to Eq.~\ref{eq:acc}, is shown in Fig.~\ref{fig:pt2acc_mup} as a function of $\Ptsq$ in the different $x$, $Q^2$ and $z$ bins and for the $\mu^+$ beam. As described in Ch.~\ref{Chapter3_Data_analysis}, it has been evaluated from the simulated TGEANT samples produced for $\mu^+$ and $\mu^-$ beam using LEPTO as event generator. The acceptance is very similar for positive and negative hadrons, but at low $x$ and high $z$. There, the acceptance for negative hadrons is higher than for the positive ones: this effect is due to the cut on the tracks passing through the absorber holes (explained in Sect.~\ref{sect:ch3_ev_had_sel}), applied to the reconstructed events in order to prevent an ambiguous definition of the scattered muon. The acceptance for the $\mu^-$ beam (not shown) is similar to the one for $\mu^+$, with the positive and negative hadrons exchanging their roles. As can be seen, the acceptance is almost flat everywhere for $\Pt<1.00$~GeV/$c$ and becomes almost 0.20 at $\Ptsq=3.00$~(GeV/$c$)$^2$ only for the lowest $z$ and $\Qsq$ bins at $0.020<x<0.055$. \\
After the subtraction of the non-visible component of the exclusive hadrons, the $\Ptsq$-distributions have been corrected for the acceptances, evaluated separately in each kinematic bin for positive and negative hadrons and for $\mu^+$ and $\mu^-$ beams. A study on systematic uncertainty due to the acceptance correction is described in Sect.~\ref{sect:ch4_systematics}.

\begin{figure}
\captionsetup{width=\textwidth}
    \centering
	\includegraphics[angle=0,width=0.95\textwidth]{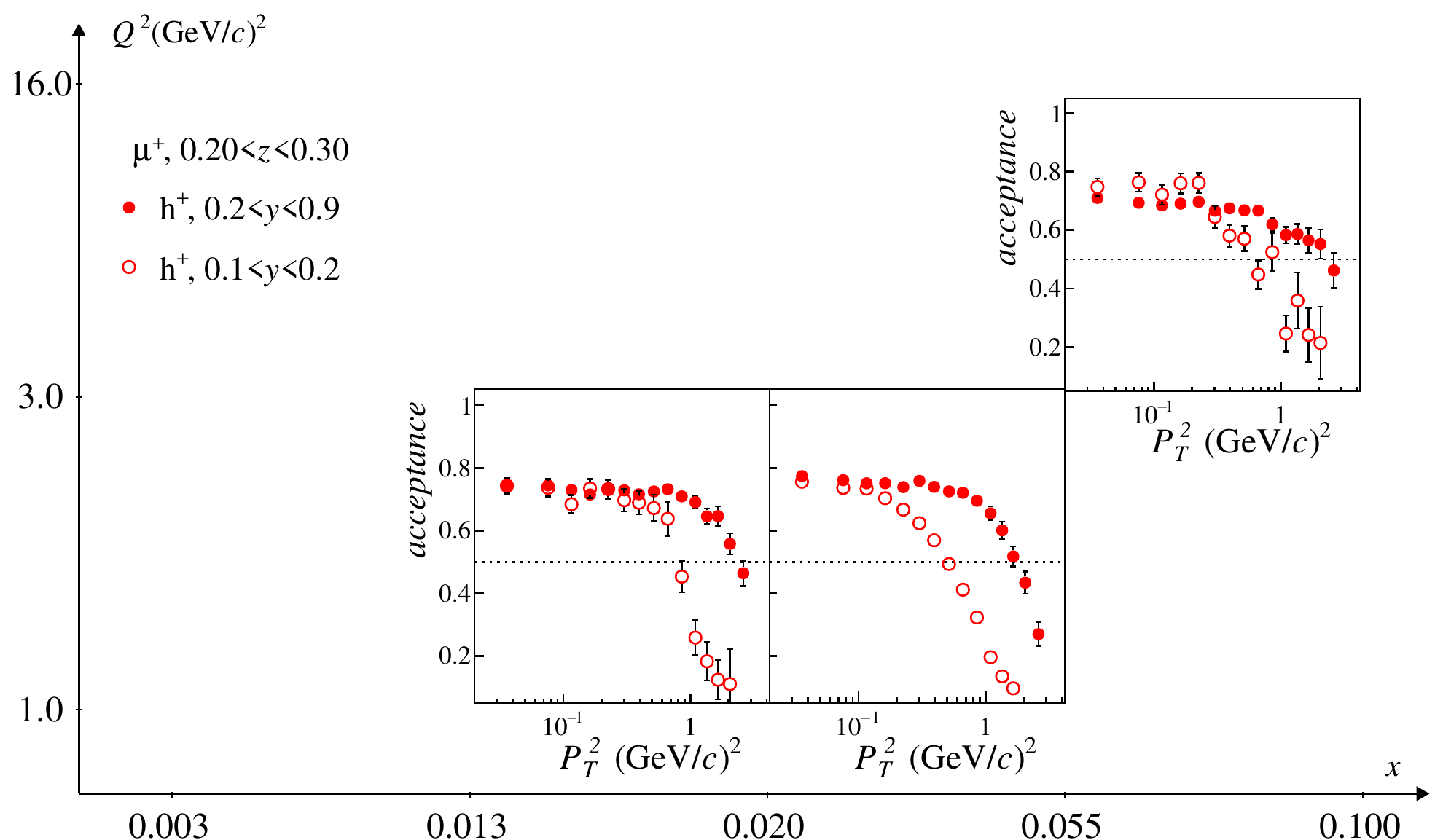}
	\caption{Acceptance as a function of $\Ptsq$, for $\mu^+$ beam and positive hadrons with $0.20<z<0.30$. The acceptance for low-$y$ hadrons ($0.1<y<0.2$, empty markers) is compared to the acceptance in the standard $y$ range ($0.2<y<0.9$, full markers). The panels refer to the different $x, Q^2$ bins where the statistics allows calculating the acceptance also for $0.1<y<0.2$.}
    \label{fig:pt2acc_mum}
\end{figure}

The acceptances for $0.1<y<0.2$ and for $0.2<y<0.9$ are compared in Fig.~\ref{fig:pt2acc_mum}, for the $\mu^+$ beam and for positive hadrons with $0.20<z<0.30$. It is clear that the low-$y$ contribution is relevant in one bin only, that is $0.020<x<0.055$ at low $\Qsq$. There, the low-$y$ acceptance shows a sharp decrease at large $\Ptsq$. For this reason, the low-$y$ range is not included in the present results. 

\newpage
\section{Systematic uncertainty}
\label{sect:ch4_systematics}
Several tests have been performed to investigate possible systematic effects. They are described in the following and summarized at the end of this Section.

\subsection{Period compatibility}
\begin{figure}
\captionsetup{width=\textwidth}
    \centering
	\includegraphics[width=\textwidth]{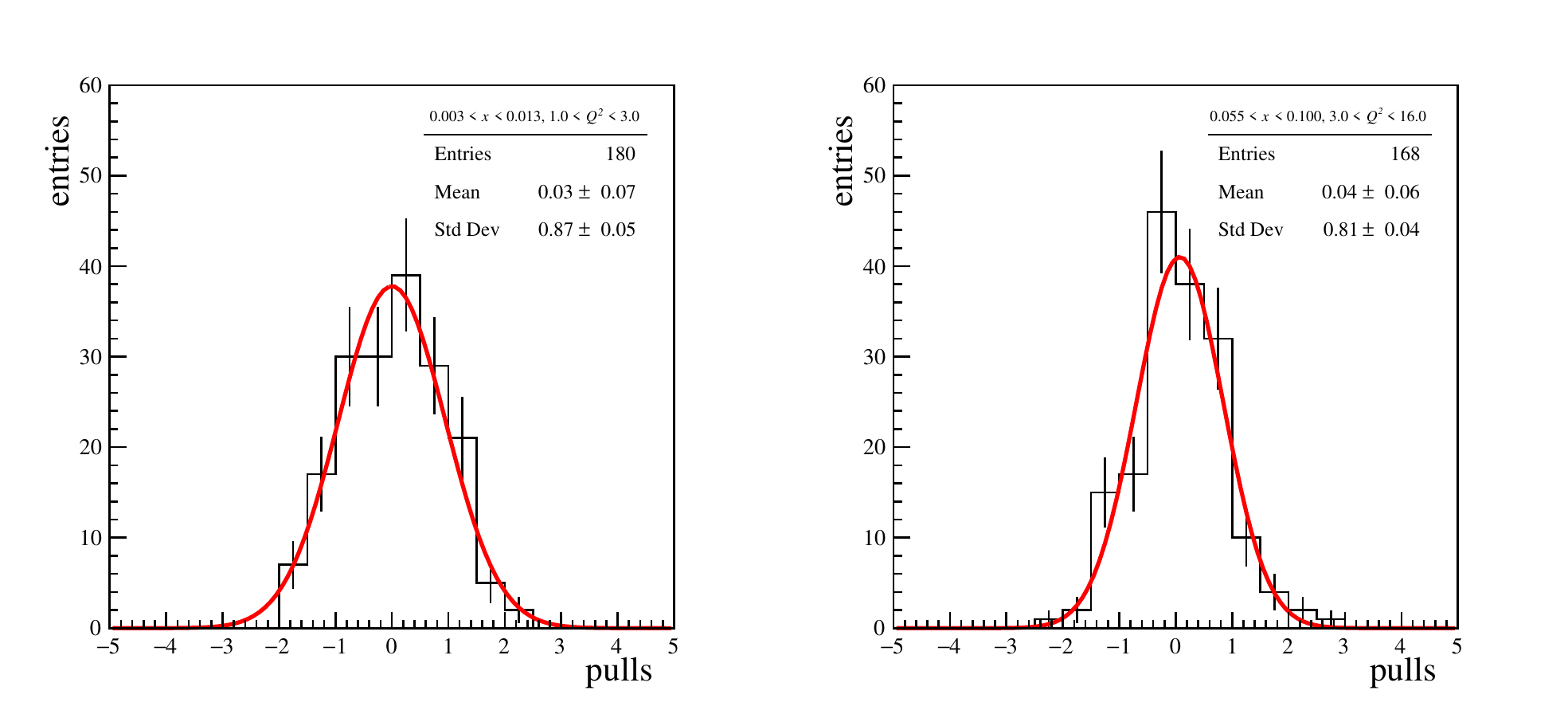}
	\caption{Pulls distributions used for the evaluation of the period compatibility, for $0.003<x<0.013$ with $1.0<\Qsq<3.0$ (left) and $0.055<x<0.100$ with $3.0<\Qsq<16.0$ (right), both in the first $z$ bin ($0.20<z<0.30$). The values of $\Qsq$ are in units of (GeV/$c$)$^2$.}
    \label{fig:pulls_pt}
\end{figure}

The compatibility of the results, as obtained from the data collected in the three data taking periods used in this work, has been studied in order to investigate possible systematic effects due to the apparatus instabilities. It has been estimated by evaluating the distribution of the pulls with respect to the average. The raw $\Ptsq$-distributions, normalized to the value in the first bin in $\Ptsq$, have been evaluated for each of the three periods separately; then, for each value of the beam and hadron charge and in each bin of $x$, $Q^2$, $z$ and $\Ptsq$ the distributions have been compared by looking at the distribution of the pulls $p$:

\begin{equation}
    p_{ij} = \frac{D_{ij} - \langle D \rangle_i }{\sqrt{\sigma_{D_{ij}}^2 - \sigma_{\langle D \rangle_i}^2}}
\end{equation}
where $D_{ij}$ represents the value of the $\Ptsq$-distribution in the kinematic bin $i$ and for the period $j$, while $\langle D \rangle$ represents its average over the three periods. The pulls are expected to be normally distributed: a deviation from this expectation hints at possible systematic effects. The distributions of the pulls have been merged in order to have one of them in each of the useful $x$, $\Qsq$ and $z$ bins, which makes 24 distributions. In each of them, the maximum number of entries is 180 = 2 (beam charge) $\times$ 2 (hadron charge) $\times$ 15 (bins in $\Ptsq$) $\times$ 3 (periods). The pulls have not been calculated if the raw number of hadrons was smaller than 10 in at least one of the three periods. No indications of systematic effects could be drawn from this test, due to fact that in all bins the mean value and the standard deviation were found compatible with zero (the former) and not exceeding one (the latter). This is shown e.g. in Fig.~\ref{fig:pulls_pt} for two cases: the lowest and highest bins in $x$ and $\Qsq$, in the first $z$ bin. 

\subsection{$\mu^{+}-\mu^{-}$ compatibility}
The compatibility of the results from the $\mu^+$ and $\mu^-$ data has been checked by calculating a global $\chi^2$: 
\begin{equation}
    \chi^2=\sum_i \frac{(r_i-\bar{r})^2}{\sigma_{r_i}^2}    
\end{equation}
where the sum runs over the kinematic bins in $x$, $\Qsq$, $z$ and $\Ptsq$, for positive and negative hadrons; $r_i$ is the ratio of the number of positive or negative hadrons in the bin $i$ measured with positive and negative muons, $\sigma_{r_i}$ its statistical uncertainty, and $\bar{r}$ is the mean value of the $r_i$ ratios. The ratios have been calculated after the corrections for the exclusive hadrons and for the acceptance and neglecting the bins with less than 10 hadrons (before corrections). Starting with 960 bins, 742 bins have been used finding $\bar{r}=1.13\pm0.03$ and $\chi^2=745$. For positive hadrons, with 394 points, it is $\chi^2=401$, while for negative hadrons, with 390 points, $\chi^2=384$. As these $\chi^2$s are very good, there is no indication for systematic effects.  \\

In addition, a similar test has been performed considering the $\Ptsq$-distributions normalized to the first bin in $\Ptsq$ (and not, as before, the number of hadrons in each bin). In this case, the $\chi^2$ has been written as:

\begin{equation}
    \chi^2_1=\sum_i \frac{(D_i^+-D_i^-)^2}{\sigma_{D_i^+}^2 + \sigma_{D_i^-}^2}    
\end{equation}
where $D_{i}^{\pm}$ represents the value of the $\Ptsq$-distribution in the kinematic bin $i$, measured with a $\mu^{\pm}$ beam.  This test gave a value of $\chi^2_1=781$ over 742 bins, again giving no indication of systematic effects.\\

\subsection{Target quarter compatibility}
The dependence of the final $\Ptsq$-distributions on the primary vertex position along the z-axis has also been investigated. Being the acceptance correction different for the different target regions, the $\Ptsq$-distributions could be different if the acceptance is not correctly evaluated.
The studies have been performed only for the $\mu^+$ case (having already checked the $\mu^+-\mu^-$ compatibility), by comparing the $\Ptsq$-distributions obtained in the six ($x,Q^2$) bins, in the four $z$ bins, and in the four target quarters, before the normalization to the first bin. The distributions measured in the downstream quarter (indicated with D) have been used as a reference for the other three quarters (U for upstream, CU and CD for central-upstream and central-downstream respectively), even if the most upstream one has not been included in the final results. Similarly to the test done for the $\mu^{+}-\mu^{-}$ compatibility, the U/D, CU/D and CD/D ratios of the distributions have been evaluated point-by point and their average calculated summing over the $x$, $\Qsq$, $z$ and $\Ptsq$ bins. Then, a global $\chi^2$ has been calculated, using the bins with at least ten hadrons before correction, to evaluate the compatibility of each point with respect to the mean. The mean values of the ratios and the $\chi^2$ in the different target quarters are given in Tab.~\ref{tab:ptsyst_cells}. The large U/D ratio reflects the fact that the U region is longer than the others. The $\chi^2$ values are good, and no indication of a systematic uncertainty can be drawn from this test.

\begin{table}[h!]
\captionsetup{width=\textwidth}
\centering
\small
\begin{tabular}{llccc}
	\hline
       &          & U/D & CU/D & CD/D \\
    \hline
\multirow{2}{*}{mean}   & $h^+$    & $1.253\pm0.006$ & $1.006\pm0.005$ &  $1.003\pm0.005$ \\  
       & $h^-$    & $1.253\pm0.006$ & $0.996\pm0.005$ &  $0.998\pm0.005$ \\    \hline
       
\multirow{2}{*}{$\chi^2$ / $N_{bins}$}   & $h^+$    &  323.8/360  & 334.0/355    & 304.3/359    \\
       & $h^-$    &  325.5/353  & 325.3/343    & 307.3/349    \\
\hline
\end{tabular}
\caption{Test of compatibility of the $\Ptsq$-distributions, for events with the primary vertex in different target quarters (upstream U, central-upstream CU, central-downstream CD and downstream D).}
\label{tab:ptsyst_cells}
\end{table}

\subsection{Uncertainty on the acceptance correction}
\begin{figure}
\captionsetup{width=\textwidth}
    \centering
    \includegraphics[width=0.49\textwidth]{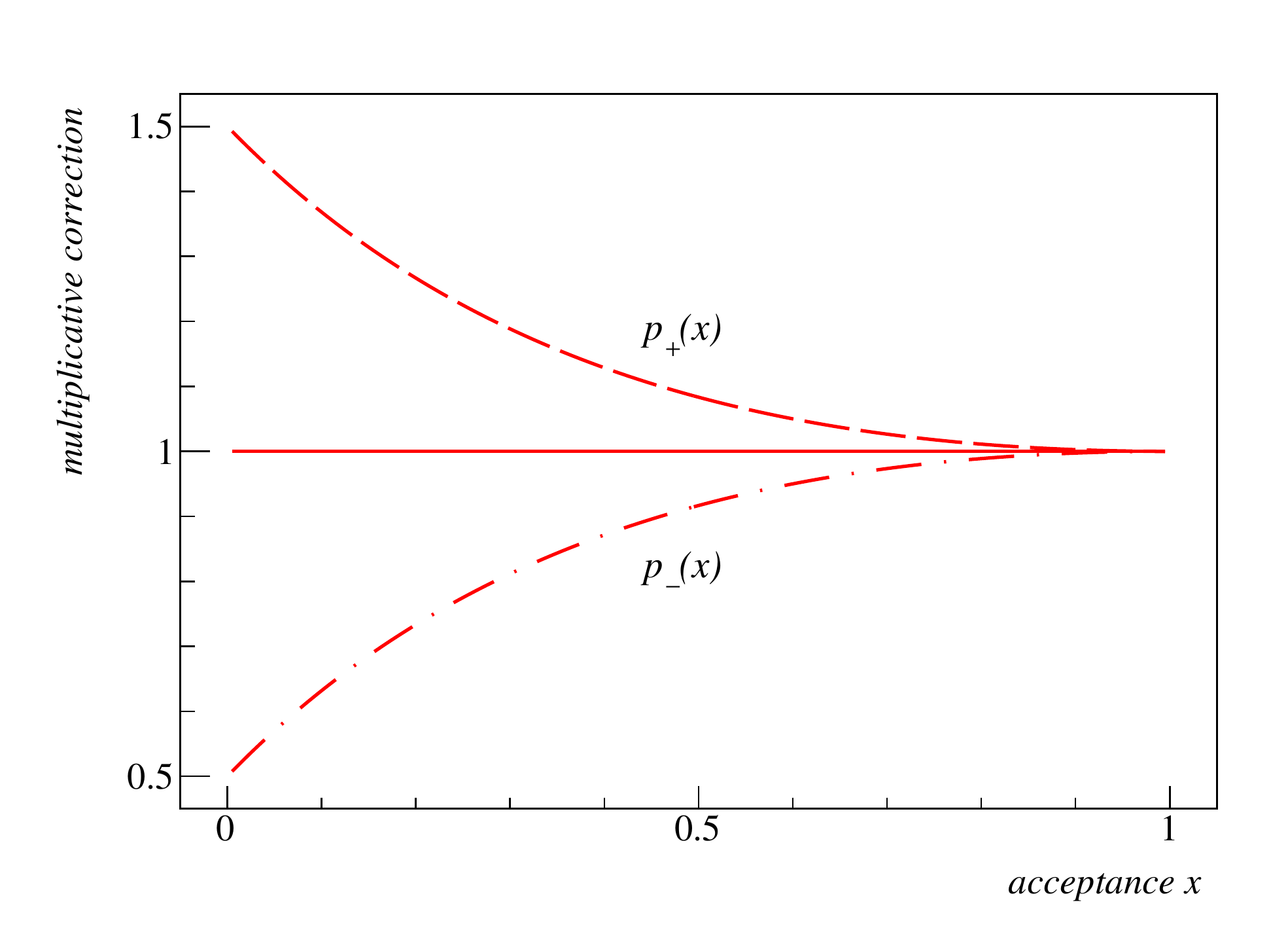}
    \includegraphics[width=0.49\textwidth]{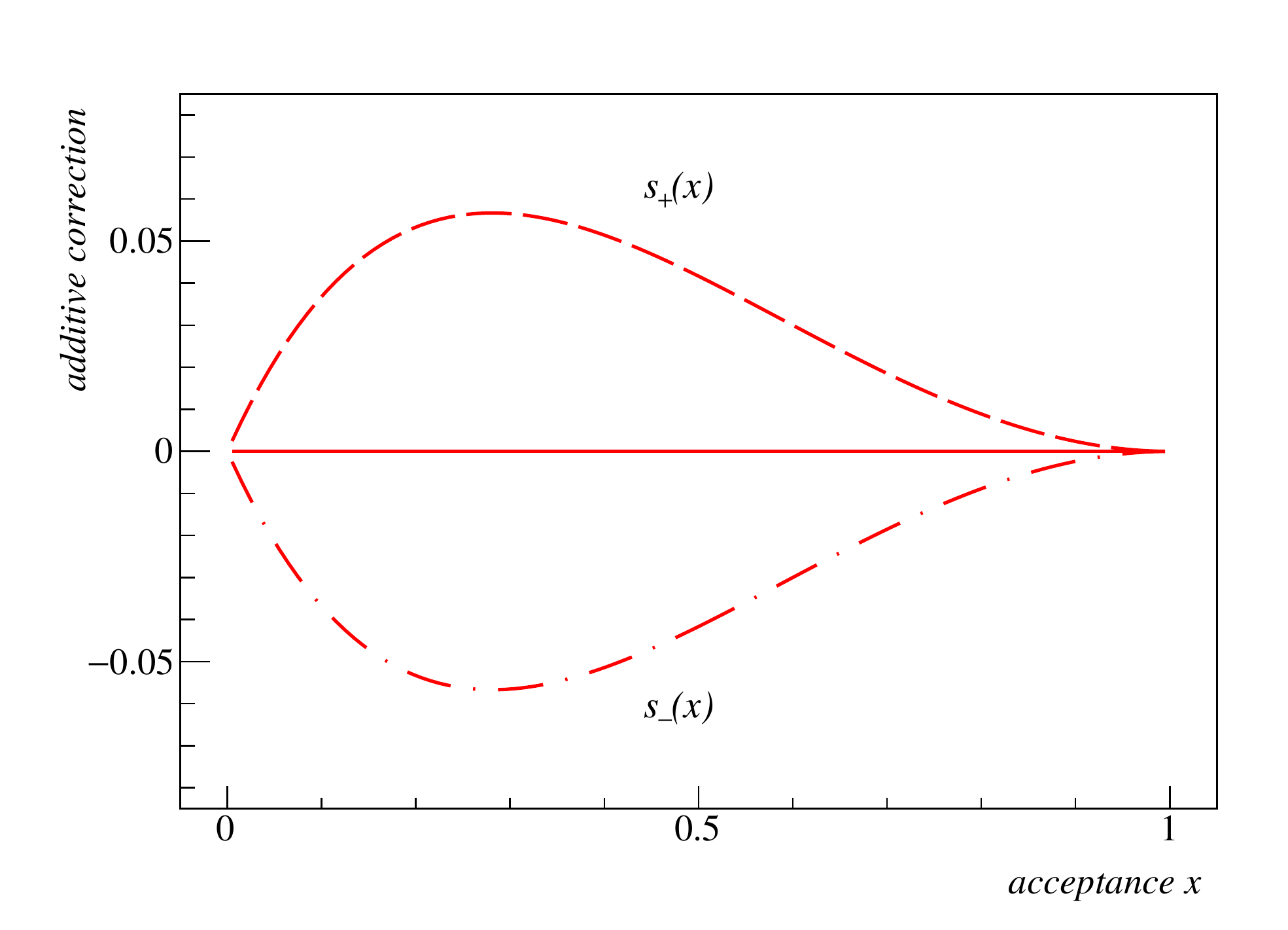}
    \caption{Multiplicative (left) and additive corrections (right) applied to the acceptance in order to evaluate the impact of a different acceptance to the final results.}
    \label{fig:syst_acc}
\end{figure}

\begin{figure}
\captionsetup{width=\textwidth}
    \centering
    \includegraphics[width=0.95\textwidth]{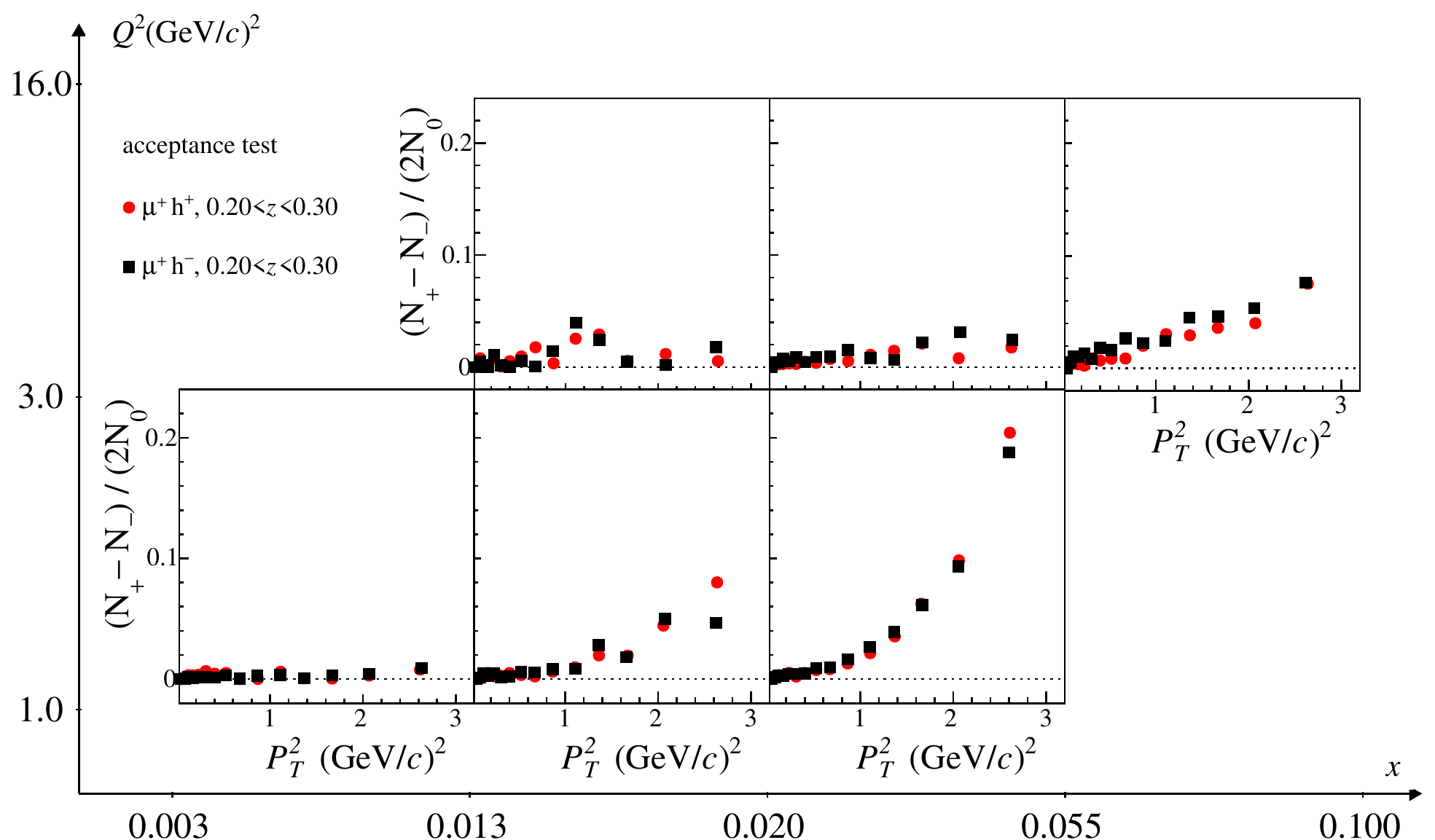}
    \caption{Effect of a modified acceptance on the $\Ptsq$-distributions for positive (red) and negative hadrons (black) compared to the nominal distributions, for $0.20<z<0.30$. The panels refer to the different $x, Q^2$ bins.}
    \label{fig:syst_acc2}
\end{figure}

In this analysis, the $\Ptsq$-distributions have an arbitrary normalization they are normalized to the value in the first bin in $\Ptsq$), thus the absolute value of the acceptance has no relevance. On the contrary, the shape of the acceptance as a function of $\Ptsq$ has an impact on the results. In order to check this, the acceptance has been modified by introducing two multiplicative correction factors $p_{\pm}$ defined, for an acceptance $x\in$ [0,1], as:

\begin{equation}
\begin{split}
    p_{+}(x) & = 2- \op \frac{2x}{1+x} + \frac{1-x}{2} \cp \\
    p_{-}(x) & = \frac{2x}{1+x} + \frac{1-x}{2} 
\end{split}
\end{equation}
so that the correction is larger for smaller values of the acceptance: in particular, $p_{\pm} \to 1 \pm \frac{1}{2}$ (i.e. 50$\%$ correction) when the acceptance goes to zero, while it is equal to 1 (i.e. no correction) when the acceptance goes to 1. Such multiplicative corrections can be translated into additive ones: in this case the corrections take the form $s_{\pm}(x) = x(p_{\pm}(x)-1)$, which have their maximum (minimum) at $x\approx 0.28$, close to the minimum value of the calculated acceptance (low $z$, high $x$, low $\Qsq$). There, the additive correction amounts to $|s_{\pm}|\approx 0.057$, corresponding to the multiplicative corrections $p_{+} \approx 1.2$ and $p_{-} \approx 0.8$. Both the multiplicative corrections and the additive ones are shown in Fig.~\ref{fig:syst_acc}. Taking as reference the set of results produced with the nominal acceptance (indicated here as $N_0$ distributions), other two sets of results have been produced scaling the nominal acceptance by $p_{+}$ in one case, and by $p_{-}$ in the second case (or equivalently, adding $s_{+}$ in one case and $s_{-}$ in the other). These two sets of distributions are indicated as $N_{+}$ and $N_{-}$. The point-by-point calculated half-difference of these two sets of distributions, divided by the statistical uncertainty of the nominal distribution, is taken here as an estimate of systematic uncertainty $\sigma_{syst}^{acc}$ due to the acceptance correction:

\begin{equation}
        \sigma_{syst}^{acc} = \frac{N_+ - N_-}{2\sigma_{N_0}}.
\end{equation}
Generally, it is $\sigma_{stat}^{acc}<0.3~\sigma_{stat}$ where $\sigma_{stat}$ is the statistical uncertainty on the $\Ptsq$-distributions obtained with the nominal acceptance. Only in one bin (low $z$, high $x$, low $\Qsq$) it amounts to $\sigma_{syst}^{acc}\approx\sigma_{stat}$. As expected from the trend of the acceptance (shown in Fig.~\ref{fig:pt2acc_mup}), $\sigma_{syst}^{acc}$ is smaller at higher values of $z$. The impact of such modified acceptances in the lowest $z$ bin can be appreciated from the plot in Fig.~\ref{fig:syst_acc2}, where $N_+-N_-$ has been divided by $2N_0$. 

It has been checked that there is agreement between the values of the acceptance calculated for the $\Ptsq$-distributions and those given in Sect.~\ref{sect:ch5_acceptance}. Also, the correction to the acceptance due to the integration over $\phi_h$ has been evaluated using the values of azimuthal acceptances and asymmetries given in Ch.~\ref{Chapter5_Azimuthal_asymmetries}, and found to be negligible with respect to the quoted systematic uncertainty.

\subsection{Uncertainty on the HEPGEN normalization}	
\begin{figure}[h!]
\captionsetup{width=\textwidth}
    \centering
    \includegraphics[width=0.95\textwidth]{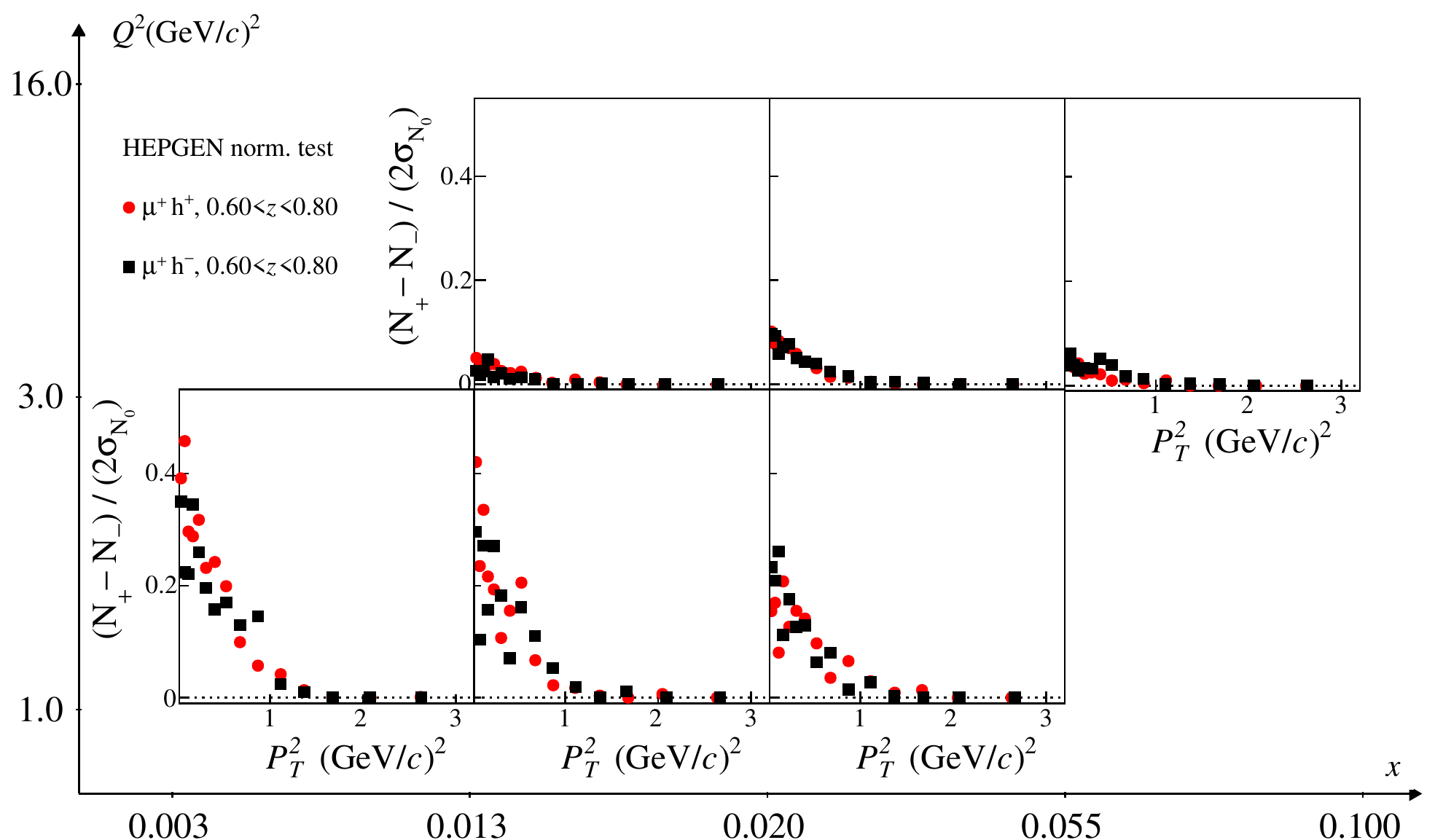}
	\caption{Effect of a modified HEPGEN normalization on the $\Ptsq$-distributions for positive (red) and negative hadrons (black) in terms of statistical uncertainty, for $0.60<z<0.80$. The panels refer to the different $x, Q^2$ bins.}
   \label{fig:pt2systnorm}
\end{figure}

The relative uncertainty on the HEPGEN normalization has been estimated to be 10\% by comparing the HEPGEN normalization, by default calculated at low $x$ and low $\Qsq$, with a set of normalization values calculated in different kinematic bins. However, due to lack of Monte Carlo statistics, this check could not be made in all the bins used for the analysis of the $\Ptsq$-distributions. For this reason, and to be on the safe side, the systematic uncertainty associated to the HEPGEN normalization has been evaluated by checking the impact on the final results after modifying the normalization values of Sect.~\ref{subsec:pure_sidis} by $\pm$ 20\%. 
Similarly to the previous test, an estimate of the systematic uncertainty due to the normalization has been calculated as:

\begin{equation}
    \sigma_{syst}^{norm} = \frac{N_+ - N_-}{2\sigma_{N_0}}.
\end{equation}
where $N_0$ is a measured $\Ptsq$-distribution, obtained with the nominal value of the HEPGEN normalization and where $N_+$ (respectively $N_-$) is the corresponding distribution, obtained with an increased (lowered) normalization.
Apart from a few $\Ptsq$ bins, the quantity $\sigma_{syst}^{norm}$, shown in Fig.~\ref{fig:pt2systnorm} for positive and negative hadrons with $0.60<z<0.80$, is smaller than 0.3, being almost negligible at low $z$, where the exclusive contamination is negligible.

\subsection{Summary}
Only two contributions to the systematic uncertainty have been found not to be negligible for the $\Ptsq$-distributions, namely the uncertainty on the acceptance and on the HEPGEN normalization. As they show an opposite trend as a function of $z$, the value $\sigma_{syst} = 0.3~\sigma_{stat}$ can be regarded as an upper limit of the overall systematic uncertainty in all kinematic bins and at all values of $\Ptsq$. 

As said before, the radiative effects could modify the final results shown here; however, in view of more accurate simulations of these effects at the COMPASS kinematics, no correction has been directly applied to the results, and only and estimate of the possible impact is given here. Preliminary tests done with the DJANGOH Monte Carlo indicate that the ratio of the $\Ptsq$-distributions with and without the radiative effects accounted for has a linear trend in $\Ptsq$, with a slope smaller or equal to 0.1~(GeV/$c$)$^{-2}$. Assuming for simplicity that the $\Ptsq$-distributions, without and with the radiative effects in, can be modeled with two single-exponential functions $f_1(\Ptsq) = e^{-\Ptsq/\aPtsq_1}$ and $f_2(\Ptsq) = e^{-\Ptsq/\aPtsq_2}$, with $\aPtsq_2 \gtrsim \aPtsq_1$, one has that:

\begin{equation}
\frac{f_2(\Ptsq)}{f_1(\Ptsq)} = e^{-\Ptsq \op \frac{1}{\aPtsq_2} - \frac{1}{\aPtsq_1} \cp} \approx 1 +\Ptsq \op \frac{1}{\aPtsq_1} - \frac{1}{\aPtsq_2}  \cp = 1 + 0.1~\Ptsq
\end{equation}
from which $\aPtsq_2< 1.07~\aPtsq_1$ for a typical value of $\aPtsq_1$ between 0.2 and 0.6. The radiative effects are thus expected to have a minor impact on the estimated $\aPtsq$.

\newpage
\section{Results for the $\Ptsq$-distributions}
\label{sect:ch4_results}
In this Section we show the final results for the $\Ptsq$-distributions, after the correction for the exclusive hadrons contamination and for the acceptance, as obtained by making the weighted average of the results from the $\mu^+$ and $\mu^-$ data. The distributions have been normalized by dividing the values and the uncertainties in each kinematic bins for $\mu^+, \mu^-, h^+, h^-$ by the corresponding value of the first bin in $\Ptsq$. Then, the values in each $\Ptsq$ bin has been calculated as the weighted average of $\mu^+$ and $\mu^-$ results, considering only the statistical uncertainty.

\begin{figure}
\captionsetup{width=\textwidth}
    \centering
	\includegraphics[angle=0,width=0.75\textwidth]{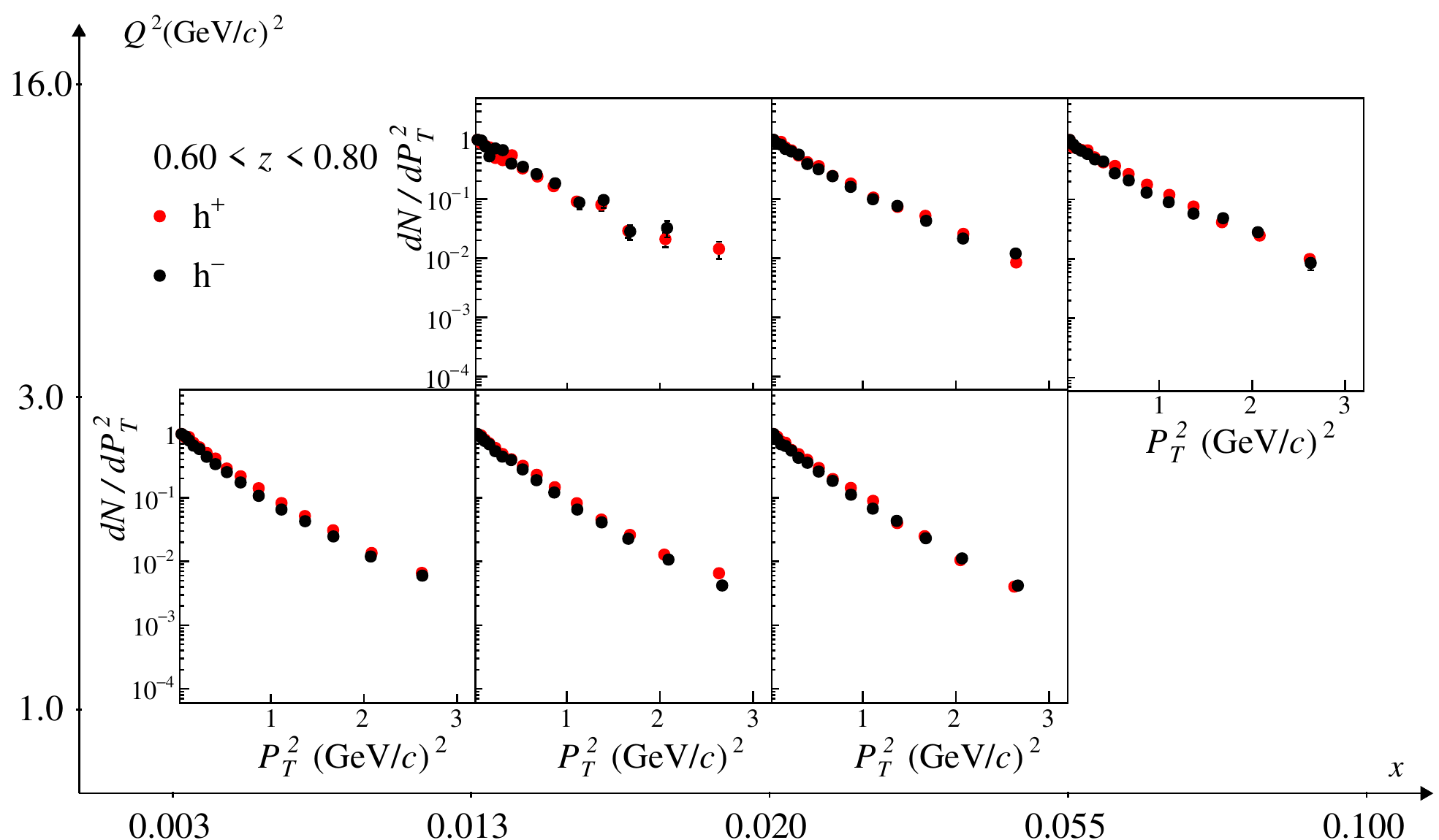}\\
	\includegraphics[angle=0,width=0.75\textwidth]{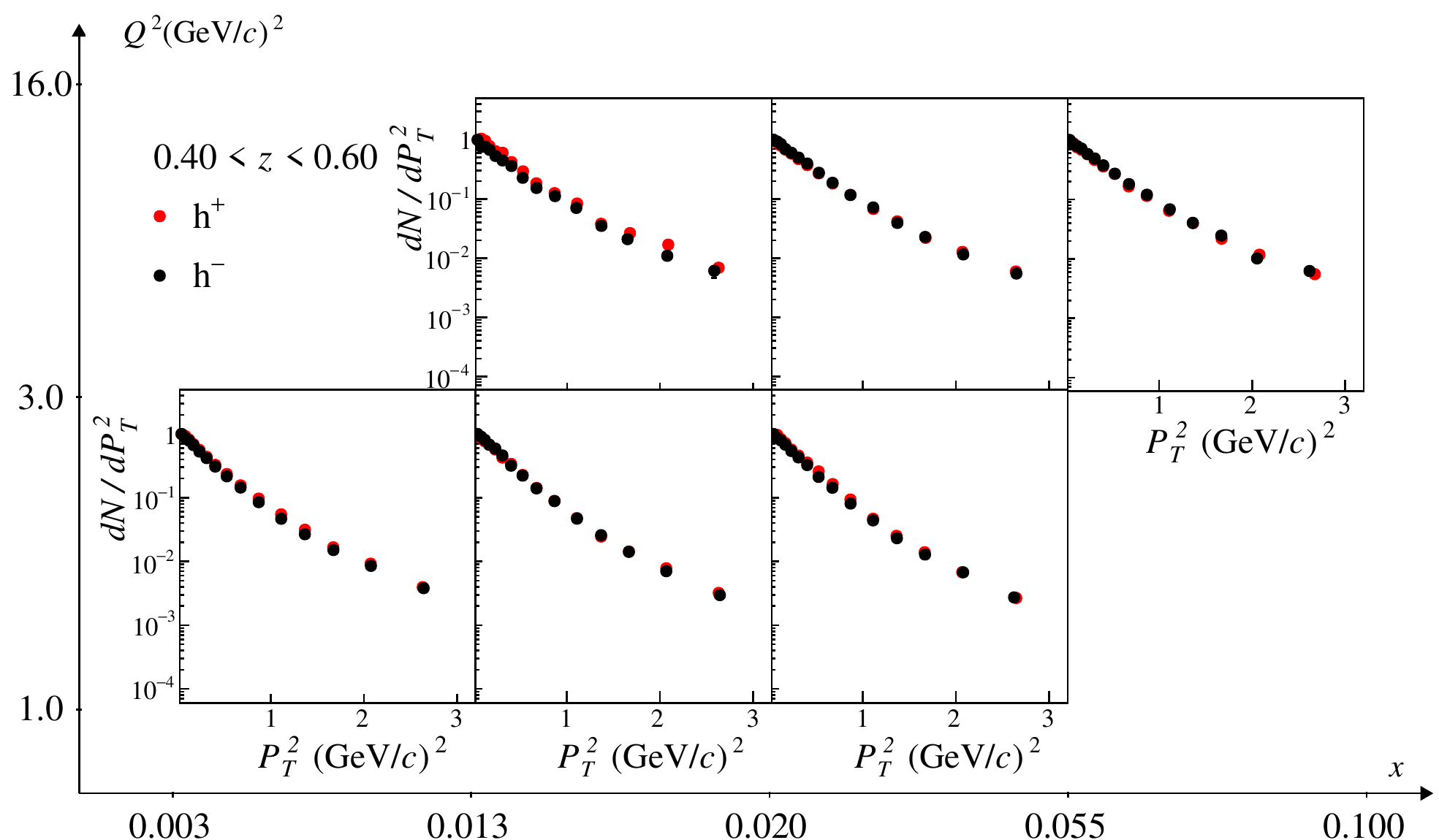}\\
	\includegraphics[angle=0,width=0.75\textwidth]{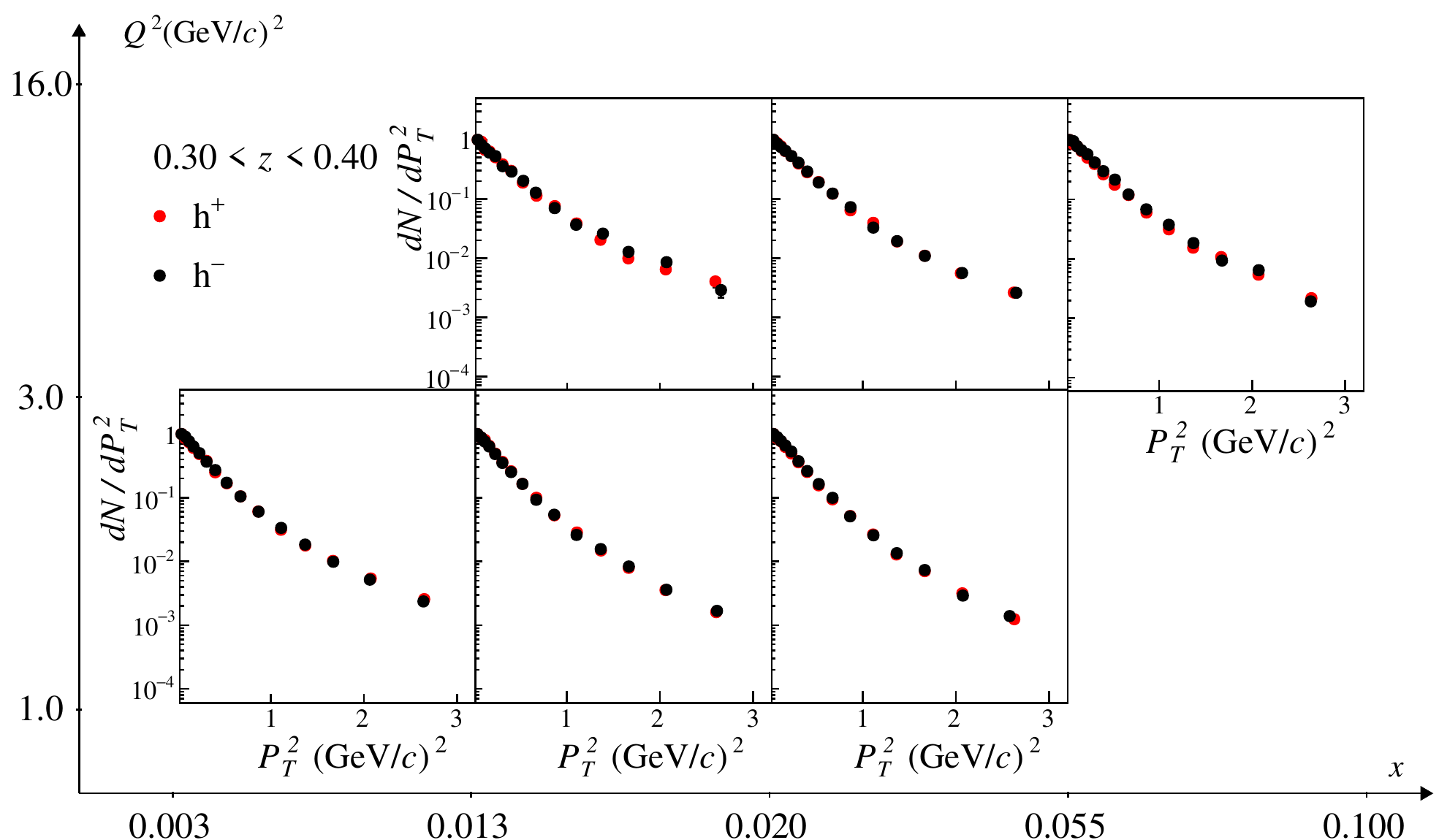}\\	\includegraphics[angle=0,width=0.75\textwidth]{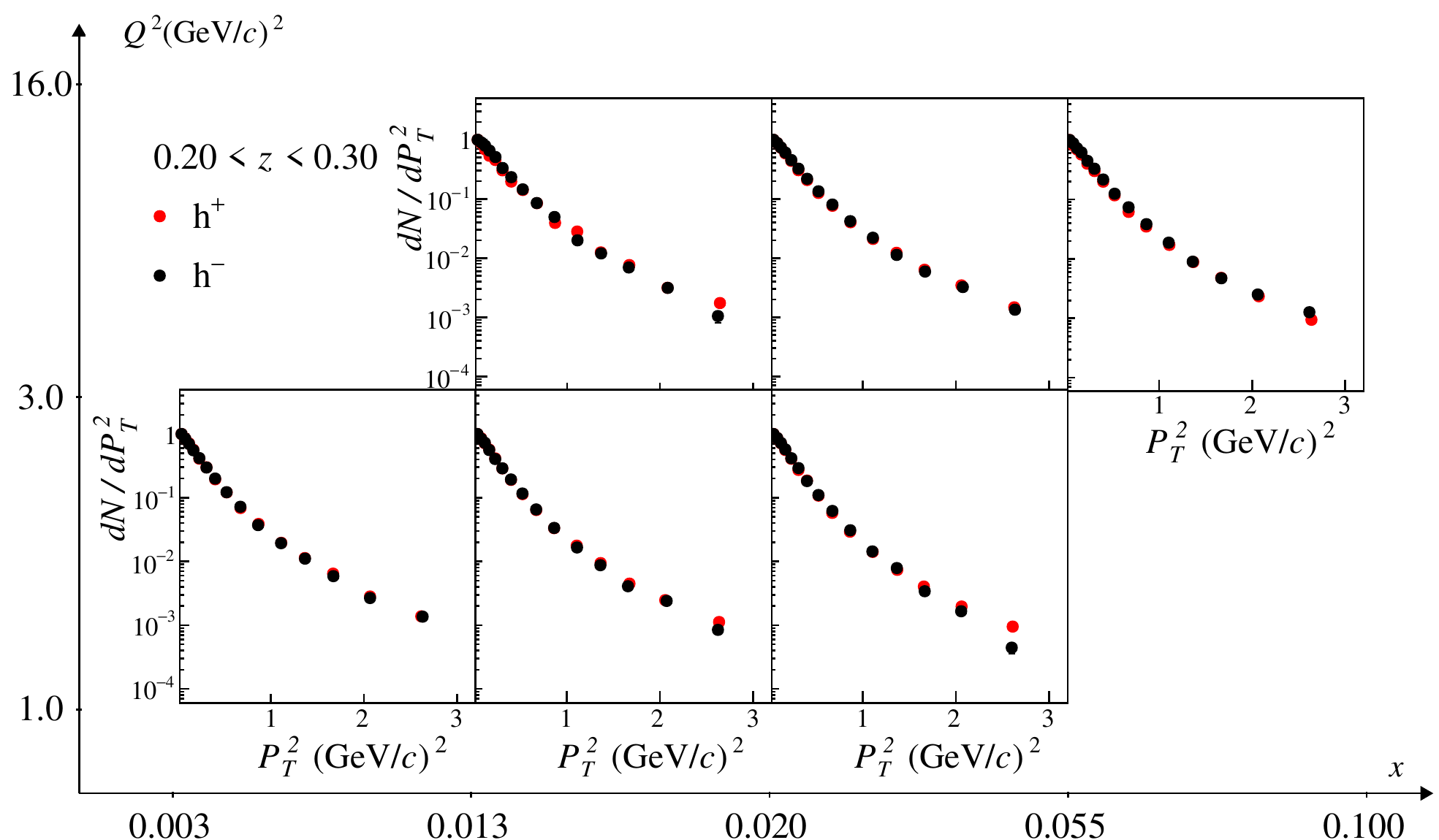}    
	\caption{$\Ptsq$-distributions for positive (red) and negative (black) hadrons. The different plots refer to the different $z$ bins and in each plot the panels refer to the different $x, Q^2$ bins. }
    \label{fig:pt2res3}
\end{figure}

Figure~\ref{fig:pt2res3} shows the $\Ptsq$-distributions up to $\Ptsq = 3$~(\gevc)$^2$ in the four $z$ bins for positive (red) and negative hadrons (black), obtained in the range $0.2<y<0.9$. In this Figure, and in all the Figures of this Chapter, the error bars, often not visible, indicate the statistical uncertainties only. 

The exponential trend expected at low $\Ptsq$ is quite clear. A second component, emerging at higher values of $\Ptsq$, where higher-order perturbative QCD effects may be relevant \cite{Anselmino:2006rv,Georgi:1977tv,Kniehl:2004hf,Daleo:2004pn}, can also be seen at low $z$. Recent calculations (e.g. Ref.~\cite{Gonzalez-Hernandez:2018ipj}) however suggest that the contribution of the perturbative component may not be enough to explain the data.

The dependence on $x$ of the distributions is weak; for fixed $x$, a stronger dependence on $\Qsq$ can instead be observed, with a larger inverse slope at higher $\Qsq$. The dependence on $z$ is remarkable: moving from low to high $z$, both the slope at small $\Ptsq$ and the relative contribution of the small- and high-$\Ptsq$ components show an interesting evolution.\\

\begin{figure}
\captionsetup{width=\textwidth}
    \centering
	\includegraphics[angle=0,width=0.75\textwidth]{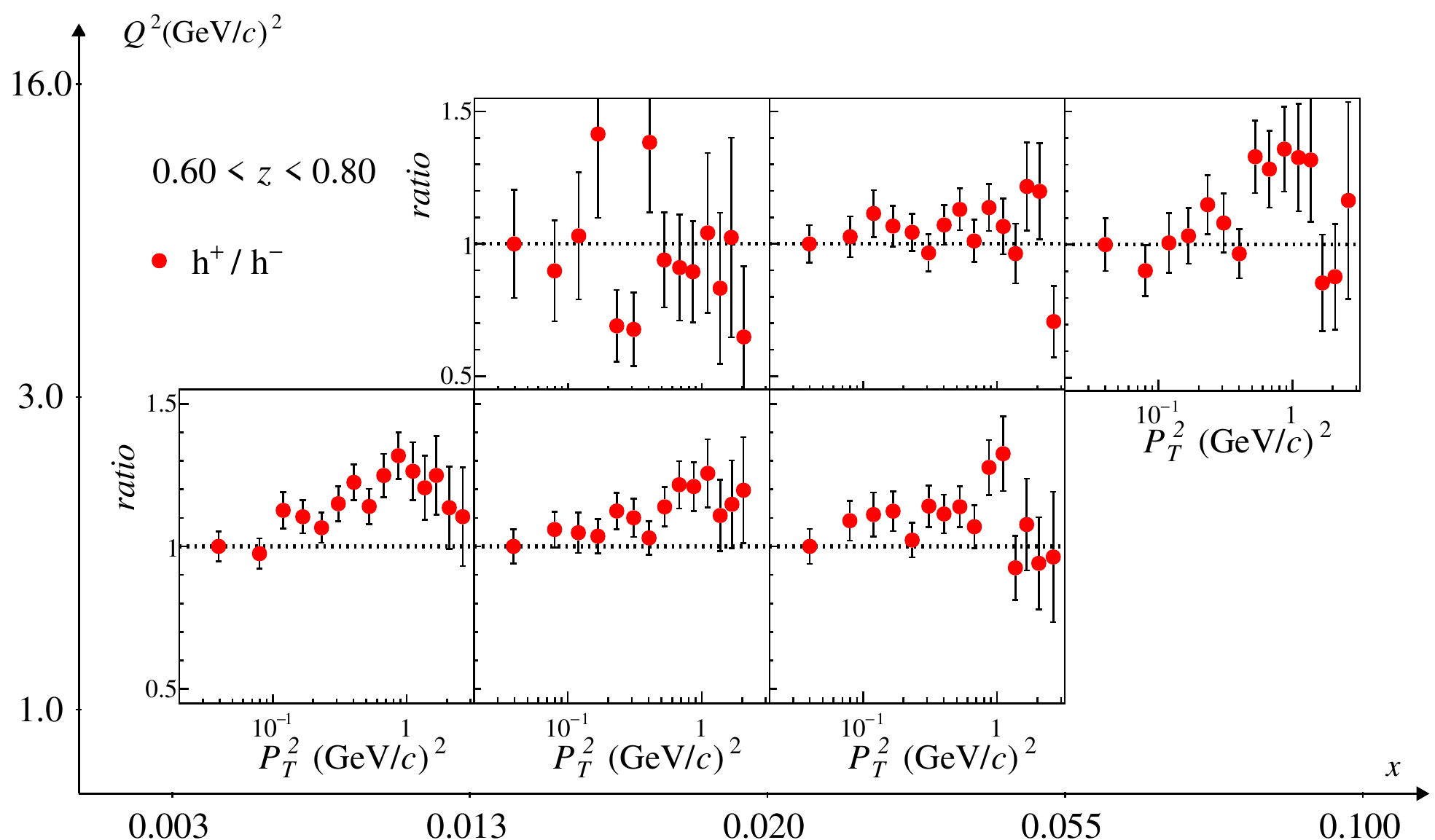}\\
	\includegraphics[angle=0,width=0.75\textwidth]{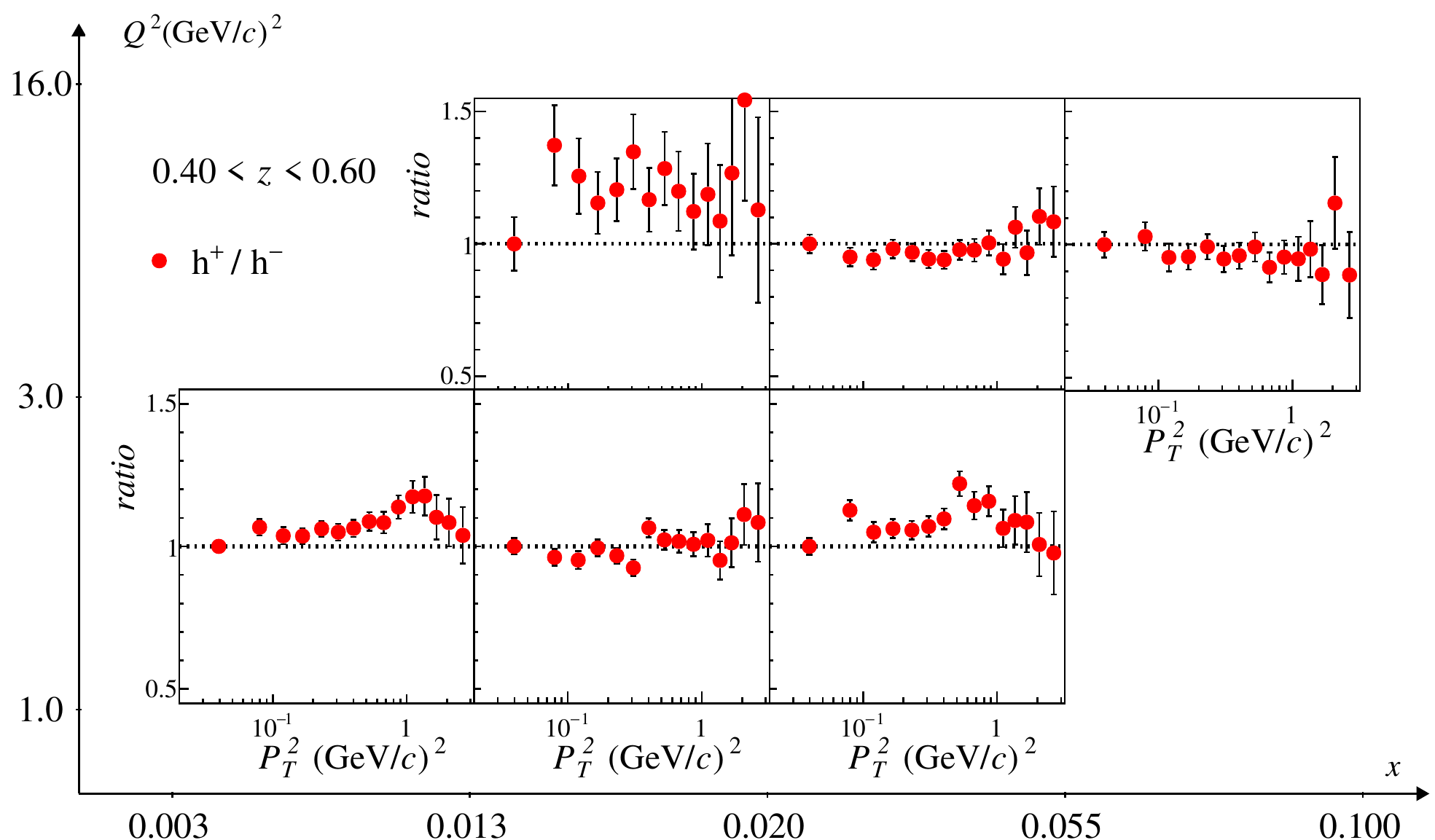}\\
	\includegraphics[angle=0,width=0.75\textwidth]{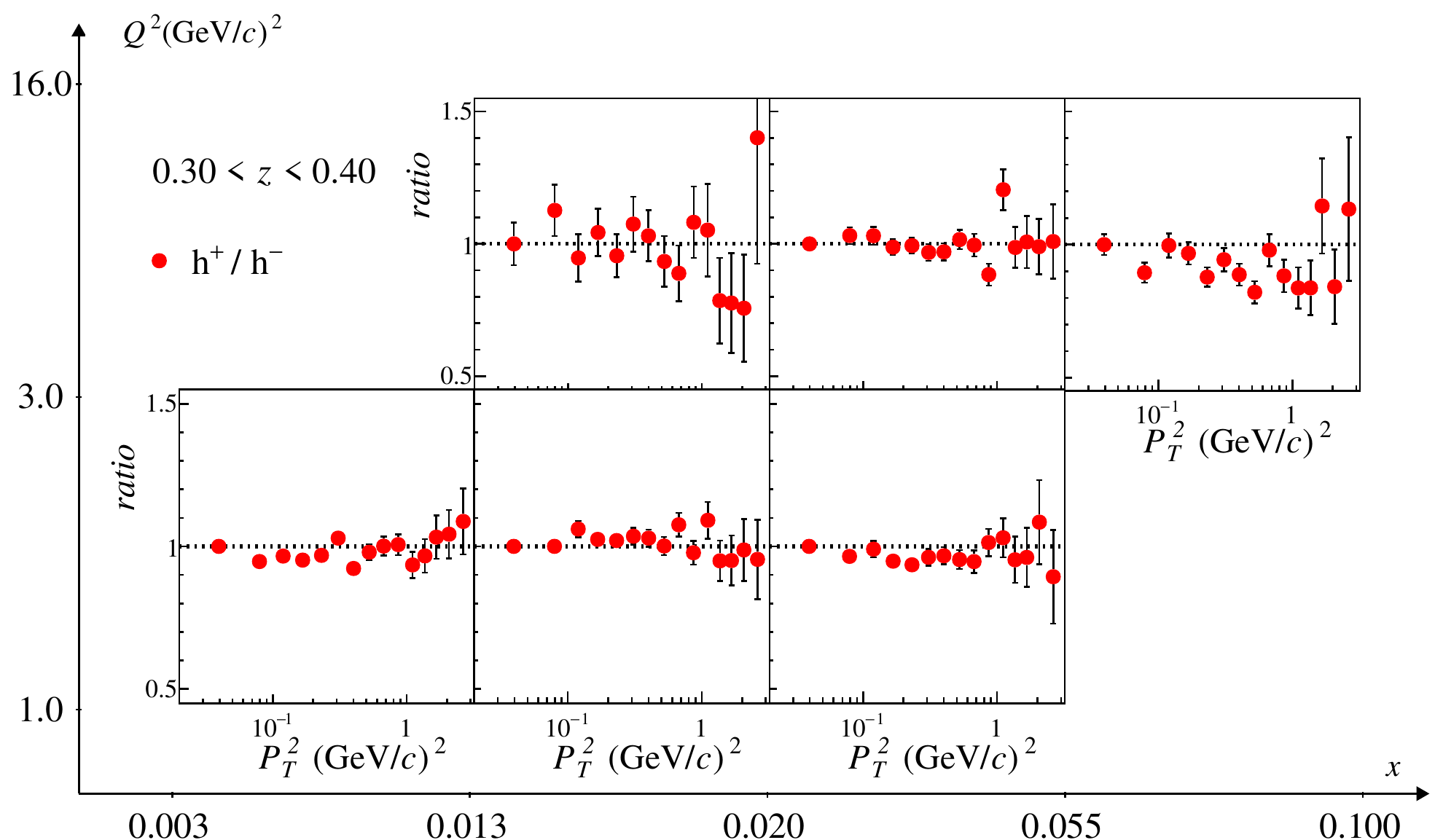}\\	\includegraphics[angle=0,width=0.75\textwidth]{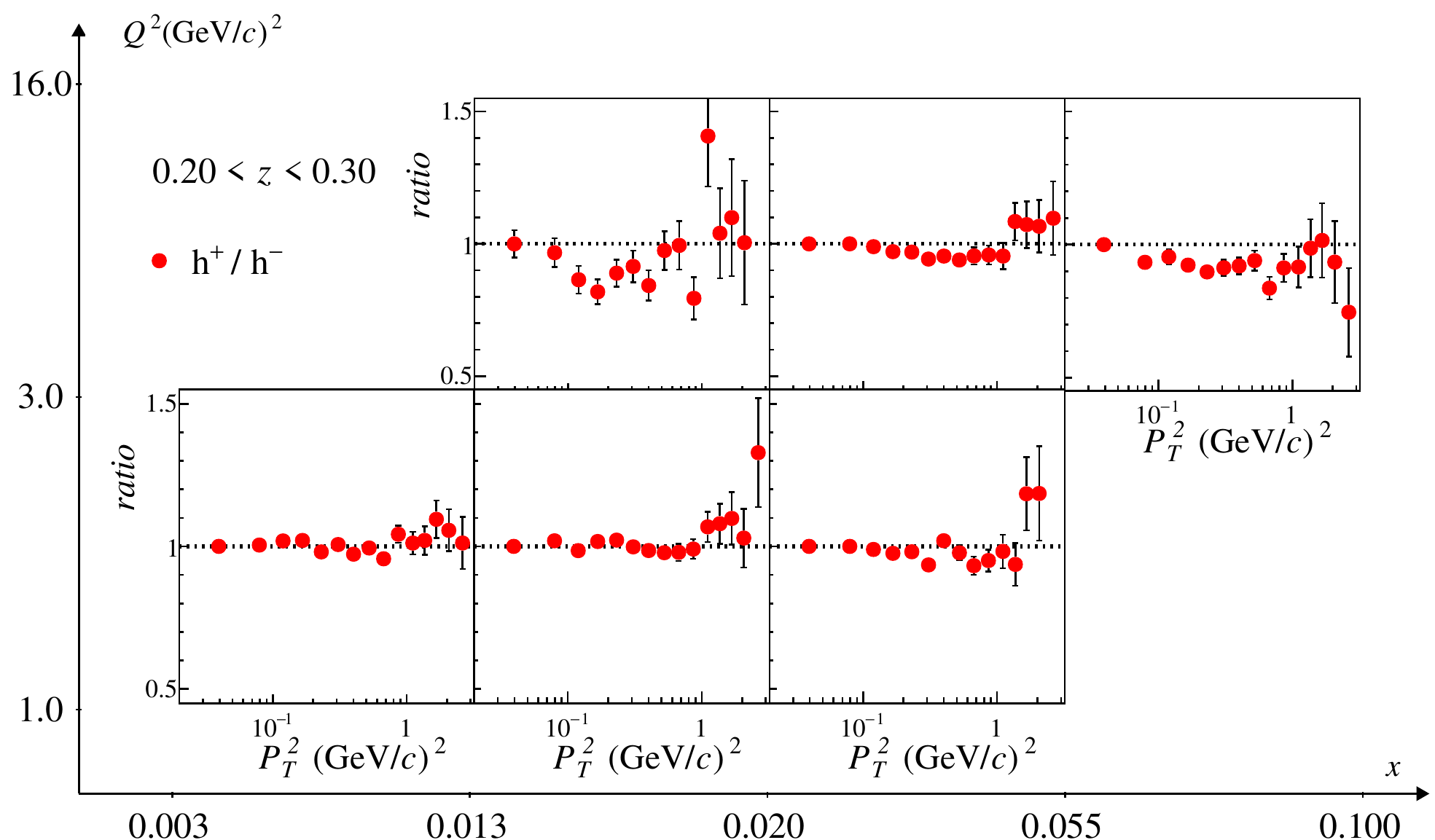}
	\caption{Ratio of the $\Ptsq$-distributions (positive over negative hadrons). The different plots refer to the different $z$ bins and in each plot the panels refer to the different $x, Q^2$ bins. }
    \label{fig:pt2ratio}
\end{figure}

No difference between positive and negative hadrons can be seen in Fig.~\ref{fig:pt2res3}; it is to be reminded, however, that we are comparing here the $\Ptsq$-dependence only, and not the absolute normalization. The ratio of the distributions for positive and negative hadrons is given in Fig.~\ref{fig:pt2ratio}. As expected, for each bin in $x$, $\Qsq$ and $z$, the ratio is equal to one in the first $\Ptsq$ bin. At larger values of $\Ptsq$, the different trends of the distributions for the two hadron charges can be observed to emerge at growing $z$, while being compatible with one at small $z$. This could indicate a flavor dependence of $\aktsq$, but the effect seems to be small, since the ratio reaches 1.3 at most, with a non-obvious $x$ and $\Qsq$ dependence. A different choice for the normalization point would have resulted in ratio plots with a possibly different absolute value of the ratio, but with the same shape.

\newpage
\clearpage

\subsection{Evaluation of $\aPtsq$ and $z^2$-dependence}
\label{sec:ptdist_interpr}
In order to compare our results with the Leading Order expectations, we calculated the average transverse momentum $\aPtsq$ in each $x$, $\Qsq$ and $z$ bin fitting the $\Ptsq$-distributions with three different functions:

\begin{itemize}
    \item a single-exponential function:
        \begin{equation}
            f_0(\Ptsq) = A_0~e^{-\Ptsq/a_0}
        \end{equation}
        up to $\Ptsq=1$ GeV/$c^2$, i.e. using 10 out of the 15 available data points. Being the distributions normalized to their value in the first $\Ptsq$ bin, $A_0$ is expected to be compatible with one, while $a_0$ gives the inverse slope of the distributions at low $\Pt$: $\aPtsq_{f_{0}} = a_0$. 
    \item a double-exponential function:
        \begin{equation}
            f_1(\Ptsq) = B_0~e^{-\Ptsq/b_0} + B_1~e^{-\Ptsq/b_1}
        \end{equation}
        up to $\Ptsq=3$ GeV/$c^2$. The second exponential is introduced to describe the change in the slope occurring at $\Ptsq\approx 1$ GeV/$c^2$. The parameters $B_0$ and $B_1$ are proportional to the relative weight of the two exponential, while the global $\aPtsq$, obtained integrating $\Ptsq f_1(\Ptsq)$ in $[0,+\infty)$, is given by:
        \begin{equation}
            \aPtsq_{f_{1}} = \frac{B_0 b_0^2 + B_1 b_1^2}{B_0b_0 + B_1b_1}
        \end{equation}
    \item a power-law (\textit{Tsallis-like}) function:
        \begin{equation}
            f_2(\Ptsq) = c_0 \op 1+c_1 \Ptsq\cp^{-c_2}
        \end{equation}
        whose form has been borrowed from the expression of the Tsallis entropy \cite{Cleymans:2013tna}, already considered in a previous COMPASS publication of the hadron multiplicities on deuteron \cite{COMPASS:2017mvk} as well as in papers of some LHC experiments \cite{Aamodt:2010my,Khachatryan:2010us}. The $\aPtsq$, obtained integrating $\Ptsq f_2(\Ptsq)$ in $[0,+\infty)$, reads in this case:
        \begin{equation}
            \aPtsq_{f_{2}} = \frac{1}{c_1(c_2-2)}
        \end{equation}
\end{itemize}

\begin{figure}[h!]
    \centering
    \captionsetup{width=\textwidth}
    \includegraphics[width=0.85\textwidth]{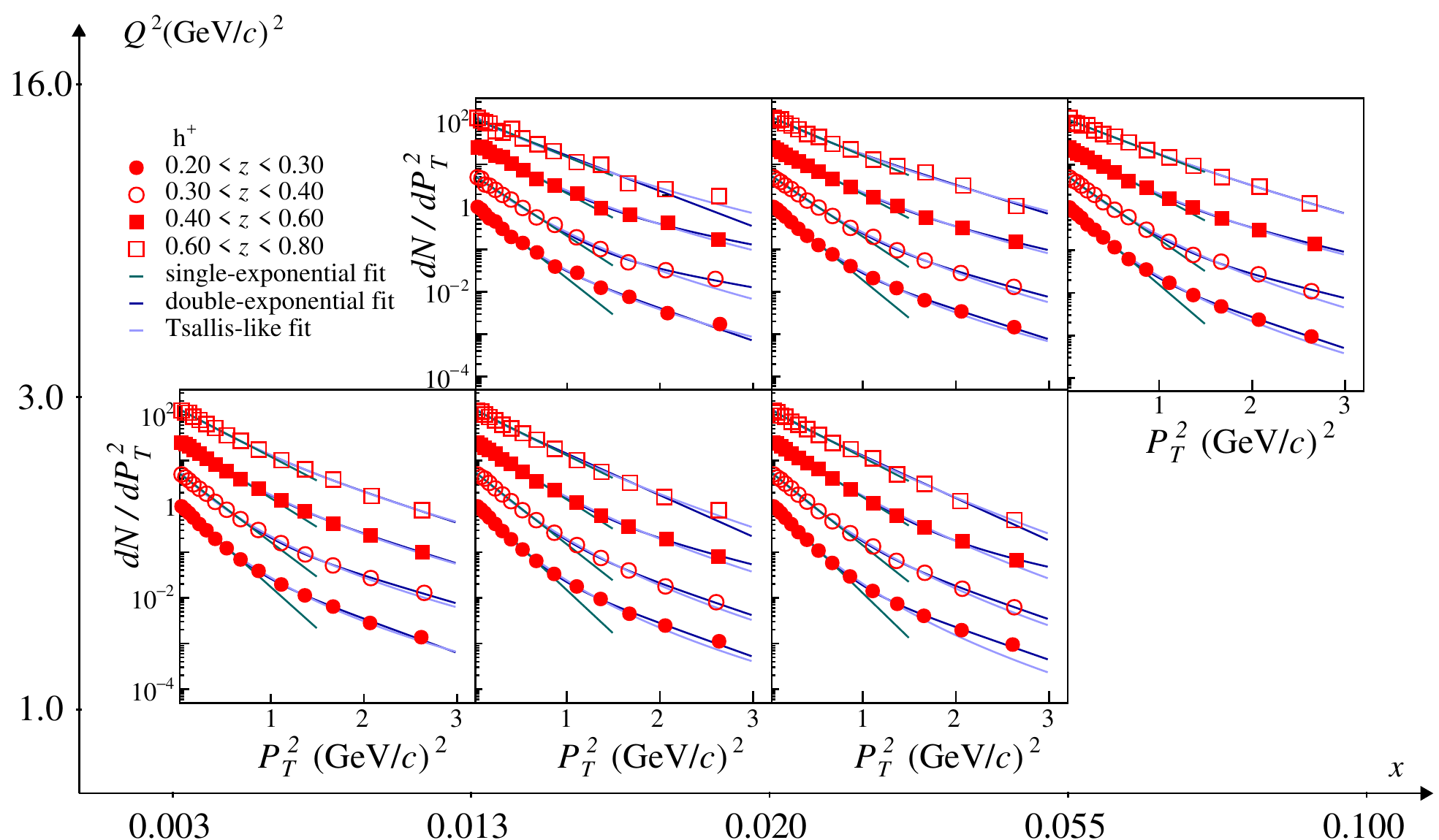}
    \includegraphics[width=0.85\textwidth]{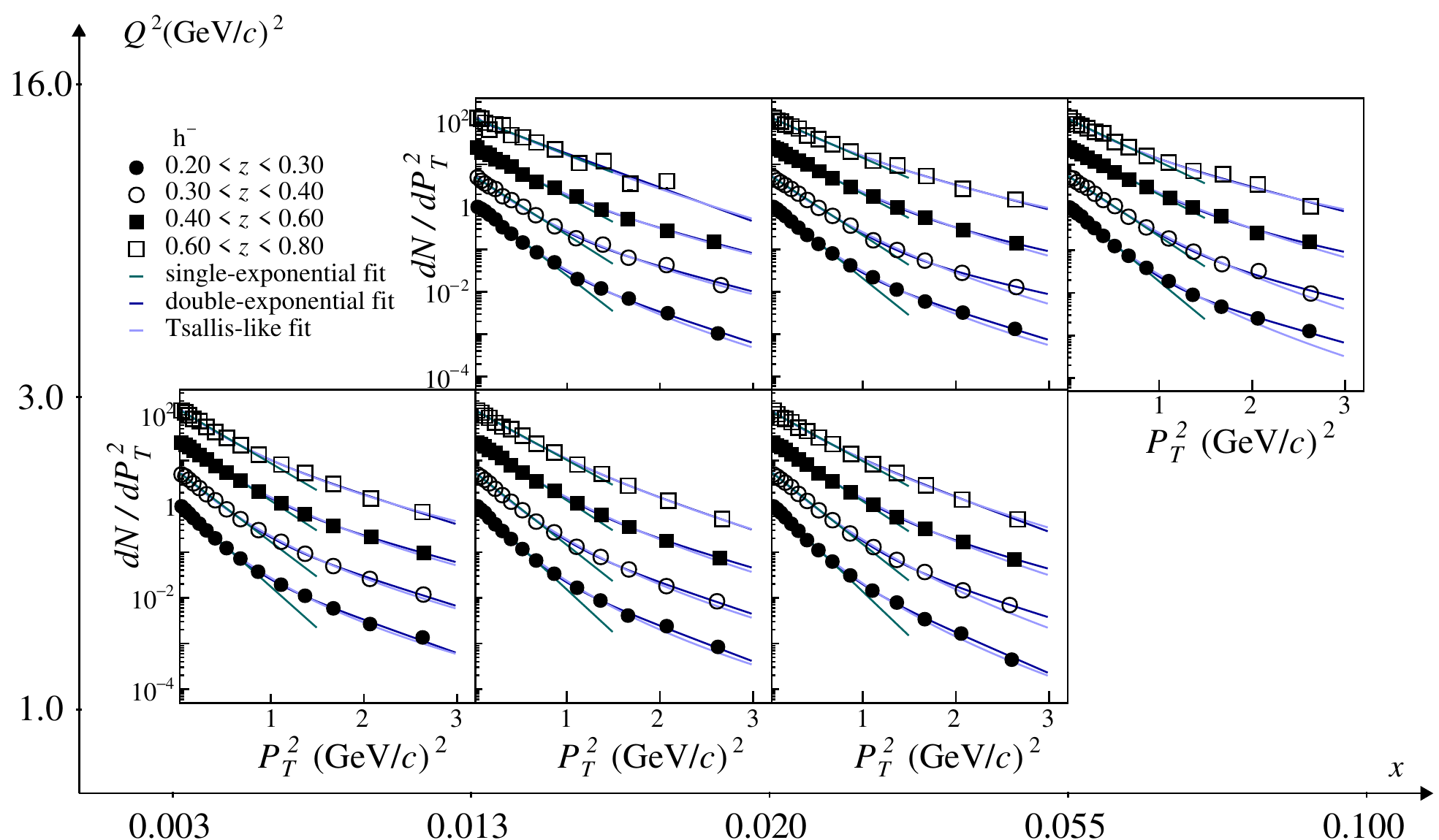}
    \caption{Fit of the $\Ptsq$-distributions in bins of $x$, $\Qsq$ and $z$ with three fit functions: single-exponential (cyan), double-exponential (dark blue) and Tsallis-like function (light-blue). }
    \label{fig:fit_pt2dist_fit}
\end{figure}

\begin{figure}[h!]
\captionsetup{width=\textwidth}
    \centering
    \includegraphics[width=0.95\textwidth]{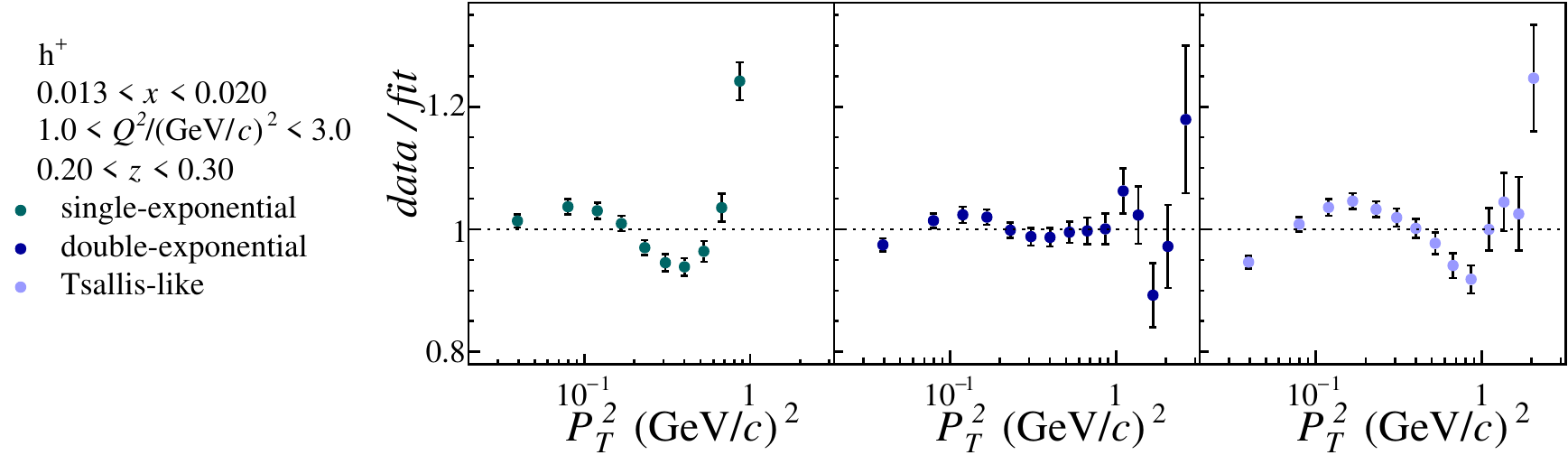}
    \includegraphics[width=0.95\textwidth]{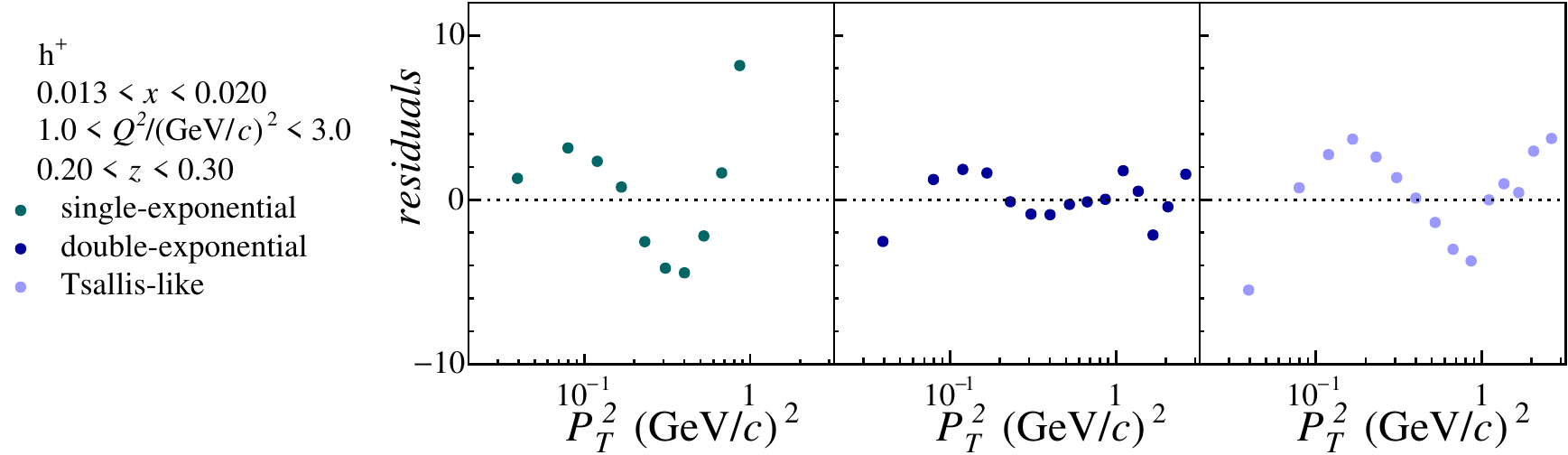}
    \caption{For a particular bin in $x$, $\Qsq$ and $z$ and for positive hadrons, on top: the ratio of the measured distribution over the estimated ones, based on the three fit functions (single exponential, double-exponential and Tsallis-like); on the bottom: the corresponding residuals. }
\label{fig:residuals_pt2}
\end{figure}

\begin{figure}[h!]
\captionsetup{width=\textwidth}
    \centering
    \includegraphics[width=\textwidth]{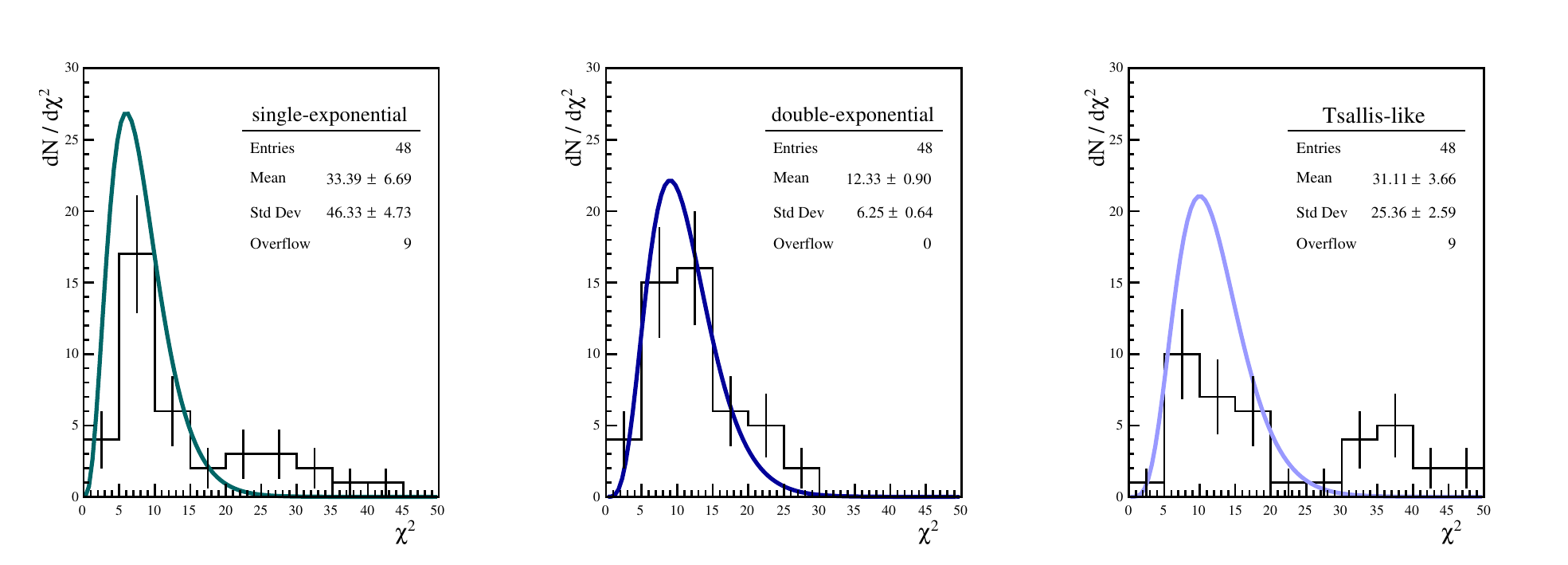}
    \caption{Distributions of the $\chi^2$ values obtained from the fit of the $\Ptsq$-distributions with a single exponential (left), double-exponential (middle) and Tsallis-like function (right). }
\label{fig:chisq_pt2}
\end{figure}

More details on these expressions, and on the way the statistical uncertainties on $\aPtsq$ have been calculated, are given in 
Appendix~\ref{AppendixB}. All the fits have been performed using the \texttt{MINUIT} package \cite{James:1994vla} and \texttt{MIGRAD} and \texttt{HESSE} algorithms.
The $\Ptsq$-distributions in each $x$, $\Qsq$ and $z$ bin, for positive and negative hadrons, are shown in Fig. \ref{fig:fit_pt2dist_fit} together with the three fit functions calculated at the estimated values of their parameters.
The error bars again indicate the statistical uncertainty only, calculated taking into account the correlations among the estimates of the parameters. Of the three functions, the one that best describes the data is the double-exponential one: this is shown, for a particular $x$, $\Qsq$ and $z$ bin and for positive hadrons, in Fig.~\ref{fig:residuals_pt2}. There, on the top plot, the ratio of the measured distribution over the fit function, calculated in each $\Ptsq$ bin, is presented for the single-exponential (left panel), the double-exponential (central panel) and for the Tsallis-like function (right panel). Similarly and for the same kinematic bin, the residuals (i.e. the difference between measured and estimated points, normalized by the uncertainty on the measured points) are shown in the bottom plot. In the single-exponential case, the fit function undershoots the data at low $\Pt$ ($\Ptsq<0.1$~(GeV/$c$)$^2$), while the opposite trend is observed in the intermediate $\Pt$ region, where the measured distribution is lower than the fit function. For $\Pt>1$~GeV/$c$, as expected, the fit function clearly undershoots the data. A similar comparison between data and fit function can be observed, in a different $\Ptsq$ range, also for the Tsallis-like function; differently from the previous case, the fit function overshoots the data in the first $\Ptsq$ bin. As said before, the best description is obtained with the double-exponential function, where still the data-over-fit ratio does not look flat at small $\Pt$. The distributions of the $\chi^2$ values of the fits, for each of the three fit functions, is shown in Fig. \ref{fig:chisq_pt2}, where the reference $\chi^2$ curves for a number of degrees of freedom equal to 8, 11 and 12 are also plotted for comparison. The agreement is satisfactory for the double-exponential only. These features could point to the presence of more complex structures of the $\Ptsq$-distributions which are relevant at small $\Ptsq$ in particular at low $z$. They could be related to the hadrons produced in the decay of inclusive vector mesons \cite{Avakian:2019uzf}.

\begin{figure}[h!]
\captionsetup{width=\textwidth}
    \centering
	\includegraphics[width=\textwidth]{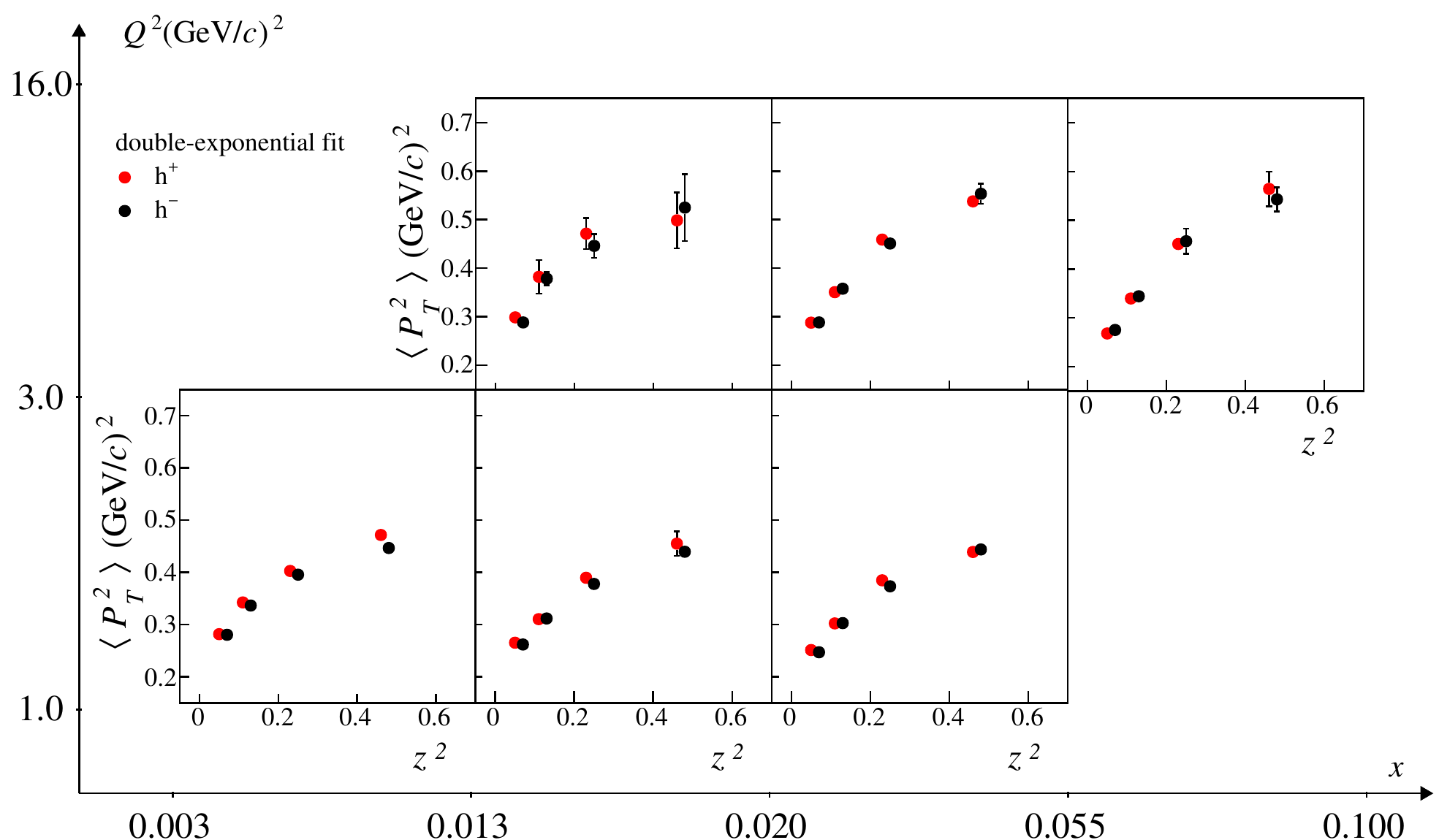}    
	\caption{$\aPtsq$ for positive (red) and negative hadrons (black), in bins of $x$ and $\Qsq$ as a function of $z^2$, as obtained from the double-exponential fit of the $\Ptsq$-distributions.}
    \label{fig:apt2_de}
\end{figure}

The set of estimated $\aPtsq$, in bins of $x$, $\Qsq$ and $z$, for positive and negative hadrons and for the three fit functions, is given in Tab.~\ref{tab:dist_fit}. While $\aPtsq$ is found smaller in the case of the single-exponential, as expected, the agreement between the double-exponential and Tsallis-like estimates is good, despite the different fit quality. The difference between the single-exponential and the other two estimates does not decrease with $z$. The estimates of $\aPtsq$ as from the double-exponential fit are shown in Fig.~\ref{fig:apt2_de}, where they are plotted against $z^2$ in bins of $x$ and $\Qsq$ for positive and negative hadrons. Almost no dependence can be observed upon $x$, while the $\Qsq$-dependence looks stronger, as already observed in the past COMPASS measurements \cite{COMPASS:2013bfs,COMPASS:2017mvk}. A good agreement between positive and negative hadrons can be seen in almost all bins.

\clearpage

\begin{table}[tbh!]
\scriptsize 
\captionsetup{width=\textwidth}
    \centering
    \begin{tabular}{c|c|c|cc|cc|cc}
    \hline
      & & & \multicolumn{2}{c|}{\textbf{single-exponential}} & \multicolumn{2}{c|}{\textbf{double-exponential}} & \multicolumn{2}{c}{\textbf{Tsallis-like}}  \\ \hline 
     $\Qsq$ & $x$ & $z$ & \multicolumn{2}{c|}{$\aPtsq$} & \multicolumn{2}{c|}{$\aPtsq$} & \multicolumn{2}{c}{$\aPtsq$} \\
     & & & $h^{+}$ & $h^{-}$ & $h^{+}$ & $h^{-}$ & $h^{+}$ & $h^{-}$ \\
    \hline
    \multirow{12}{*}{1.0 - 3.0} & \multirow{4}{*}{0.003 - 0.013} & 020 - 0.30 & 0.238 $\pm$ 0.001 & 0.240 $\pm$ 0.001 & 0.282 $\pm$ 0.002 & 0.281 $\pm$ 0.002 & 0.284 $\pm$ 0.002 & 0.281 $\pm$ 0.002  \\
    & & 0.30 - 0.40 & 0.284 $\pm$ 0.002 & 0.283 $\pm$ 0.002 & 0.342 $\pm$ 0.003 & 0.337 $\pm$ 0.003 & 0.342 $\pm$ 0.003 & 0.338 $\pm$ 0.003  \\
    & & 0.40 - 0.60 & 0.342 $\pm$ 0.003 & 0.330 $\pm$ 0.003 & 0.403 $\pm$ 0.004 & 0.396 $\pm$ 0.005 & 0.407 $\pm$ 0.004 & 0.396 $\pm$ 0.004  \\
    & & 0.60 - 0.80 & 0.413 $\pm$ 0.009 & 0.365 $\pm$ 0.008 & 0.472 $\pm$ 0.009 & 0.447 $\pm$ 0.008 & 0.478 $\pm$ 0.008 & 0.466 $\pm$ 0.010  \\
    \cline{2-9}
    
    & \multirow{4}{*}{0.013 - 0.020} & 0.20 - 0.30 & 0.229 $\pm$ 0.001 & 0.231 $\pm$ 0.001 & 0.265 $\pm$ 0.002 & 0.262 $\pm$ 0.002 & 0.263 $\pm$ 0.002 & 0.260 $\pm$ 0.002  \\
    & & 0.30 - 0.40 & 0.272 $\pm$ 0.002 & 0.271 $\pm$ 0.002 & 0.310 $\pm$ 0.003 & 0.312 $\pm$ 0.003 & 0.309 $\pm$ 0.002 & 0.311 $\pm$ 0.044  \\
    & & 0.40 - 0.60 & 0.336 $\pm$ 0.003 & 0.327 $\pm$ 0.004 & 0.390 $\pm$ 0.006 & 0.378 $\pm$ 0.005 & 0.382 $\pm$ 0.004 & 0.376 $\pm$ 0.004  \\
    & & 0.60 - 0.80 & 0.432 $\pm$ 0.011 & 0.394 $\pm$ 0.010 & 0.455 $\pm$ 0.023 & 0.439 $\pm$ 0.010 & 0.463 $\pm$ 0.008 & 0.443 $\pm$ 0.009  \\
    \cline{2-9}
    
    & \multirow{4}{*}{0.020 - 0.055} & 0.20 - 0.30 & 0.220 $\pm$ 0.001 & 0.223 $\pm$ 0.001 & 0.251 $\pm$ 0.002 & 0.247 $\pm$ 0.002 & 0.245 $\pm$ 0.002 & 0.246 $\pm$ 0.002  \\
    & & 0.30 - 0.40 & 0.269 $\pm$ 0.002 & 0.270 $\pm$ 0.003 & 0.302 $\pm$ 0.003 & 0.303 $\pm$ 0.003 & 0.300 $\pm$ 0.003 & 0.299 $\pm$ 0.003  \\
    & & 0.40 - 0.60 & 0.345 $\pm$ 0.004 & 0.325 $\pm$ 0.004 & 0.385 $\pm$ 0.009 & 0.373 $\pm$ 0.006 & 0.371 $\pm$ 0.003 & 0.368 $\pm$ 0.004  \\
    & & 0.60 - 0.80 & 0.414 $\pm$ 0.010 & 0.391 $\pm$ 0.011 & 0.439 $\pm$ 0.007 & 0.444 $\pm$ 0.009 & 0.445 $\pm$ 0.007 & 0.453 $\pm$ 0.196  \\
    \cline{1-9}
    
    \multirow{12}{*}{3.0 - 16.0} & \multirow{4}{*}{0.013 - 0.020} & 0.20 - 0.30 & 0.252 $\pm$ 0.005 & 0.256 $\pm$ 0.005 & 0.299 $\pm$ 0.006 & 0.288 $\pm$ 0.005 & 0.305 $\pm$ 0.006 & 0.287 $\pm$ 0.005  \\
    & & 0.30 - 0.40 & 0.306 $\pm$ 0.009 & 0.314 $\pm$ 0.010 & 0.382 $\pm$ 0.035 & 0.378 $\pm$ 0.014 & 0.358 $\pm$ 0.011 & 0.380 $\pm$ 0.013  \\
    & & 0.40 - 0.60 & 0.372 $\pm$ 0.014 & 0.368 $\pm$ 0.016 & 0.471 $\pm$ 0.032 & 0.446 $\pm$ 0.024 & 0.459 $\pm$ 0.017 & 0.451 $\pm$ 0.262  \\
    & & 0.60 - 0.80 & 0.479 $\pm$ 0.042 & 0.508 $\pm$ 0.054 & 0.499 $\pm$ 0.058 & 0.525 $\pm$ 0.069 & 0.542 $\pm$ 0.041 & 0.521 $\pm$ 2.373  \\
    \cline{2-9}
    
    & \multirow{4}{*}{0.020 - 0.055} & 0.20 - 0.30 & 0.242 $\pm$ 0.002 & 0.248 $\pm$ 0.002 & 0.288 $\pm$ 0.002 & 0.288 $\pm$ 0.002 & 0.288 $\pm$ 0.002 & 0.286 $\pm$ 0.002  \\
    & & 0.30 - 0.40 & 0.298 $\pm$ 0.003 & 0.307 $\pm$ 0.004 & 0.351 $\pm$ 0.004 & 0.358 $\pm$ 0.006 & 0.348 $\pm$ 0.003 & 0.347 $\pm$ 0.004  \\
    & & 0.40 - 0.60 & 0.383 $\pm$ 0.005 & 0.380 $\pm$ 0.006 & 0.459 $\pm$ 0.010 & 0.451 $\pm$ 0.012 & 0.455 $\pm$ 0.006 & 0.442 $\pm$ 0.006  \\
    & & 0.60 - 0.80 & 0.470 $\pm$ 0.014 & 0.452 $\pm$ 0.016 & 0.538 $\pm$ 0.010 & 0.553 $\pm$ 0.021 & 0.552 $\pm$ 0.267 & 0.579 $\pm$ 0.022  \\
    \cline{2-9}
    
    & \multirow{4}{*}{0.055 - 0.100} & 0.20 - 0.30 & 0.233 $\pm$ 0.002 & 0.241 $\pm$ 0.002 & 0.268 $\pm$ 0.003 & 0.275 $\pm$ 0.004 & 0.266 $\pm$ 0.003 & 0.269 $\pm$ 0.003  \\
    & & 0.30 - 0.40 & 0.289 $\pm$ 0.004 & 0.301 $\pm$ 0.005 & 0.340 $\pm$ 0.007 & 0.344 $\pm$ 0.007 & 0.331 $\pm$ 0.005 & 0.337 $\pm$ 0.005  \\
    & & 0.40 - 0.60 & 0.374 $\pm$ 0.007 & 0.386 $\pm$ 0.009 & 0.451 $\pm$ 0.012 & 0.457 $\pm$ 0.026 & 0.451 $\pm$ 0.009 & 0.445 $\pm$ 0.009  \\
    & & 0.60 - 0.80 & 0.520 $\pm$ 0.023 & 0.422 $\pm$ 0.020 & 0.564 $\pm$ 0.035 & 0.543 $\pm$ 0.025 & 0.565 $\pm$ 0.017 & 0.581 $\pm$ 0.039  \\
    \hline
    
    \end{tabular}
    \caption{The value of $\aPtsq$ of the three fits is given For each bin in $\Qsq$, $x$ and $z$ and for positive and negative hadrons separately. $\Qsq$ and $\aPtsq$ are given in units of (GeV/$c$)$^2$. The uncertainties are statistical only.}
        \label{tab:dist_fit}
\end{table}

The linear dependence of $\aPtsq$ on $z^2$, expected from the relation $\aPtsq = z^2\aktsq + \apperpsq$, is only approximately verified at high $x$ and $\Qsq$. Similar trends have been observed in the previous analyses on deuteron \cite{COMPASS:2013bfs,COMPASS:2017mvk}. The deviations from the linear relation does not allow for a direct extraction of $\aktsq$, which would require a linear fit of $\aPtsq$ versus $z^2$, from which $\aktsq$ would be inferred from the slope parameter. However, since the points do not follow a linear trend, this procedure would give different estimates of $\aktsq$ depending on the $z^2$ range considered in the fit. Considering the first two points in $z^2$, the fitted slope would be large, ranging from 0.75 to 1.50 in the various $x$, $\Qsq$ bins. Including the third point, the slope would range from 0.67 to 1.10 and eventually decrease including the fourth point in $z^2$. As it is easy to figure out, the $\chi^2$ of the fit would also degrade by extending the range in $z^2$. Nonetheless, the constant term in the fit, that in this framework would be identified with a $z$-independent $\apperpsq$, does not depend much on the $z^2$ interval: its value ranges from 0.19 to 0.26 regardless of the number of considered points in $z^2$.
We will come back to this point later on. The mean values of some relevant kinematic variables are given, for each $x$, $\Qsq$ and $z$ point, in Appendix~\ref{AppendixC}.

\newpage
\clearpage

\subsection{Comparison with the deuteron results}
\begin{figure}[h!]
\captionsetup{width=\textwidth}
    \centering
	\includegraphics[width=0.75\textwidth]{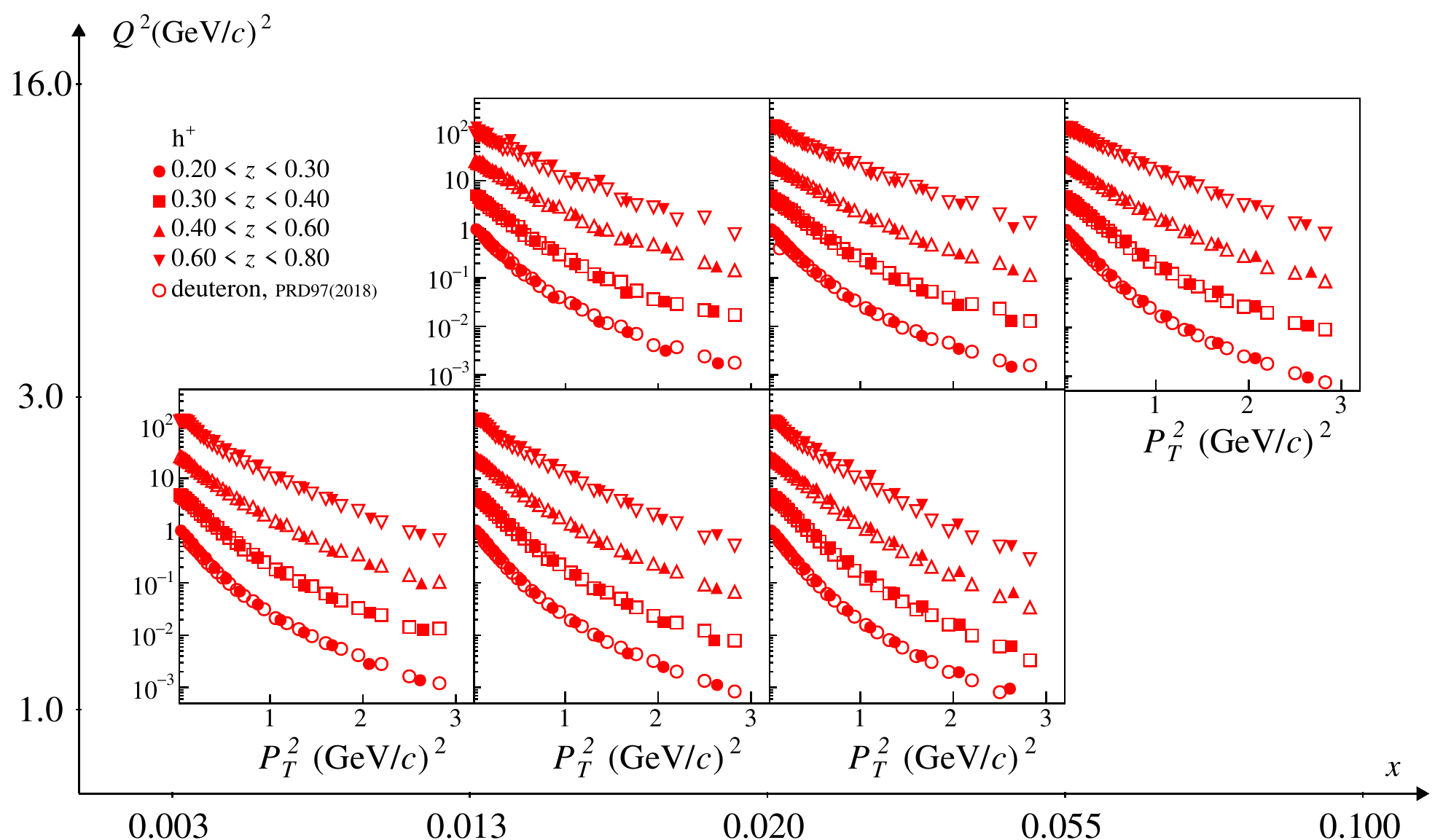}
    \caption{Comparison of the present results for positive hadrons (full markers) with the published ones on deuteron (empty markers).}
   \label{fig:pt2cmp}
\end{figure}

\begin{figure}[h!]
\captionsetup{width=\textwidth}
    \centering
	\includegraphics[width=0.75\textwidth]{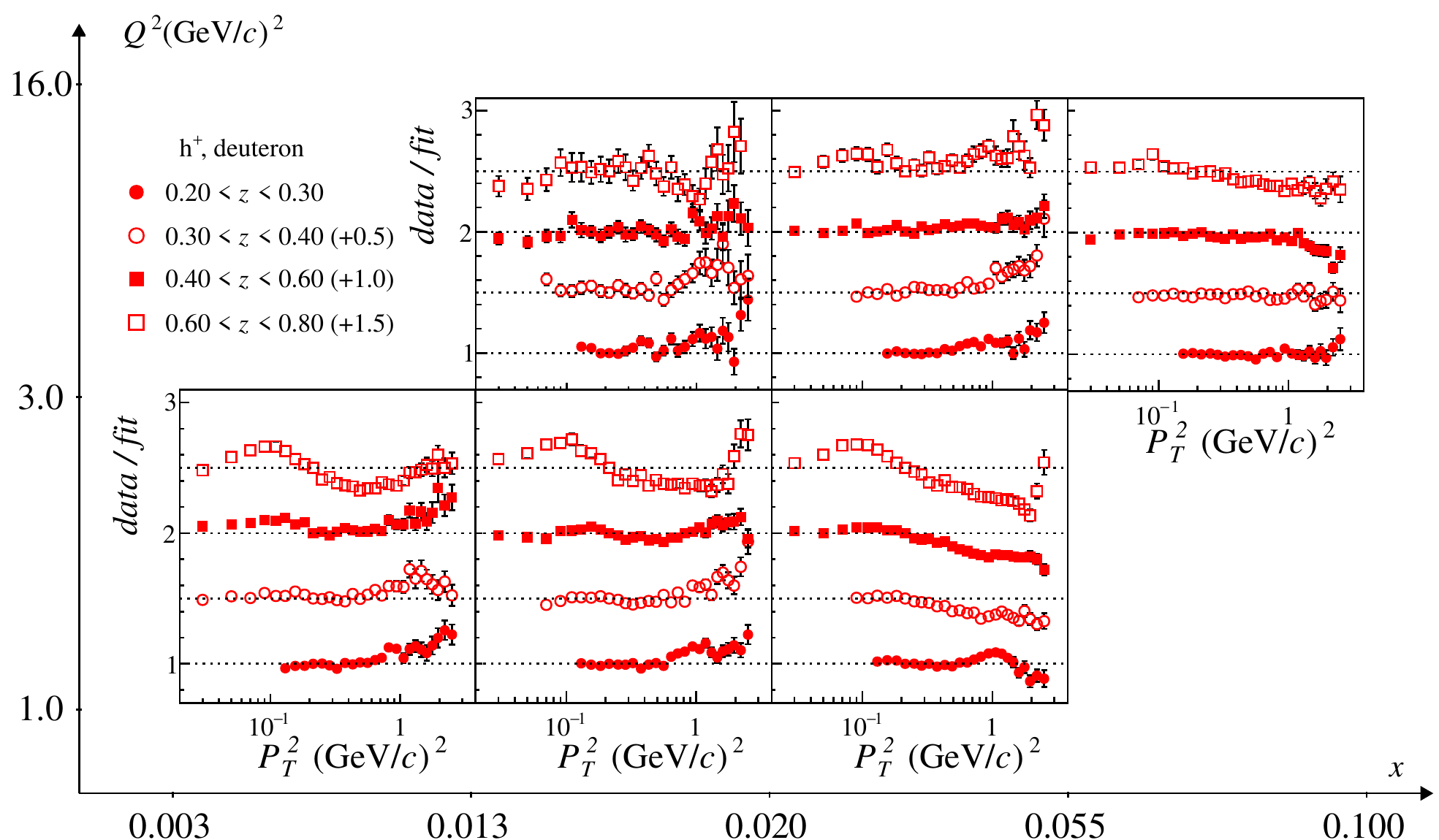}
    \caption{Ratio of the $\Ptsq$-distribution on deuteron over the double-exponential fit function used to describe the proton data, for positive (top) and negative hadrons (bottom) and in bins of $x$, $\Qsq$ and $z$ (staggered).}
   \label{fig:pt2cmp2}
\end{figure}

It is interesting to compare these results with the multiplicities on an isoscalar target published by COMPASS in 2018 \cite{COMPASS:2017mvk}. It must be noted that the measurement of Ref.~\cite{COMPASS:2017mvk} has been performed in the range $0.1<y<0.9$; however, the comparison with the present measurement is correct, since the region $0.1<y<0.2$ is relevant in three only ($x$,$\Qsq$) bins (see Sect.~\ref{sect:ch4_acceptance}) and we have checked that the shape of the $\Ptsq$-distributions in that kinematic region is in very good agreement with those measured at higher $y$. The comparison for positive hadrons is shown in Fig.~\ref{fig:pt2cmp}. The published results have been scaled and averaged in $x$ and $Q^2$ in order to match the current binning, while keeping the binning in $z$ and $\Ptsq$ unchanged. This means that the number of bins in $\Ptsq$ is double in the deuteron case. The four $z$ bins are shown in the same panels, scaled with the factors 1, 5, 25, 125 for better readability. The agreement between the proton and deuteron results can better be evaluated in Fig.~\ref{fig:pt2cmp2}, which shows the ratio of the deuteron data over the double-exponential fit function used to describe the proton data. In each pad, the ratios are shown for the four $z$ bins, which are staggered and added a constant. Clearly, the agreement worsens at large $z$, low $\Pt$ and in the first $\Qsq$ bin, thus indicating that the different procedure for the subtraction of the exclusive hadrons plays an important role. No particular differences can be spotted in the proton-deuteron comparisons, between positive and negative hadrons.

\section{Distributions in $q_T$ and $q_T^2$}
\label{sect:ch4_qt}

Recently, as said in Ch.~\ref{Chapter1_SIDIS}, a lot of interest has risen about the ratio $q_T=P_T/z$. In Ref.~\cite{Scimemi:2019cmh}, the region of validity of the TMD formalism has been selected by requiring $q_T<0.25~Q$, based on the value of the reduced $\chi^2$ obtained from the comparison of the predictions and the data points from SIDIS and Drell-Yan \cite{Scimemi:2017etj}. This cut helps in getting a good global description of SIDIS and Drell-Yan. On the other hand, such strong cut rules out a large part of the available data. The points at high $\Pt$, in particular, are kept only if $z$ and $Q$ are also large. This appears to be in contradiction with the COMPASS results on the $\Pt$-weighted Sivers asymmetries \cite{COMPASS:2018ofp}: there, the Sivers asymmetry, obtained by weighting the hadrons with $\frac{\Pt}{zM}$ (or $\frac{\Pt}{M}$, $M$ being the target nucleon mass), was found to be in good agreement with the standard, unweighted measurement. Also, all the results are in agreement with what expected from TMD formalism and factorization in the whole kinematic domain accessed by COMPASS. Of course, the problem in describing the $\Ptsq$-distributions could be related to effects which cancel in the spin asymmetries.

In addition, a strict cut on $q_T$ could reject the decay products of particles supposed to be in the low-$q_T$ region. Consider for example a  SIDIS process at $\Qsq=1$~(GeV/$c$)$^2$ in which the virtual photon has an energy in the laboratory equal to $E_{\gamma^*}=40$~GeV and in which a $\vmrho$ meson is produced with $E_{\vmrho}=32$~GeV and a transverse momentum $P_T=0.1$~GeV/$c$: this corresponds to a $q_T=0.125$~GeV/$c$, thus in the region of low-$q_T$. Let's suppose that one of its two decay pions has $E_{\pi}=10$~GeV with $P_T=0.5$~GeV/$c$: in this case $q_T=2$~GeV/$c$, thus far from being low-$q_T$. A cut on $q_T$ would then counterintuitively removes the decay products of a particle produced in the supposed TMD region.

\bigskip

\begin{figure}[h!]
\captionsetup{width=\textwidth}
    \centering
    \includegraphics[width=0.85\textwidth]{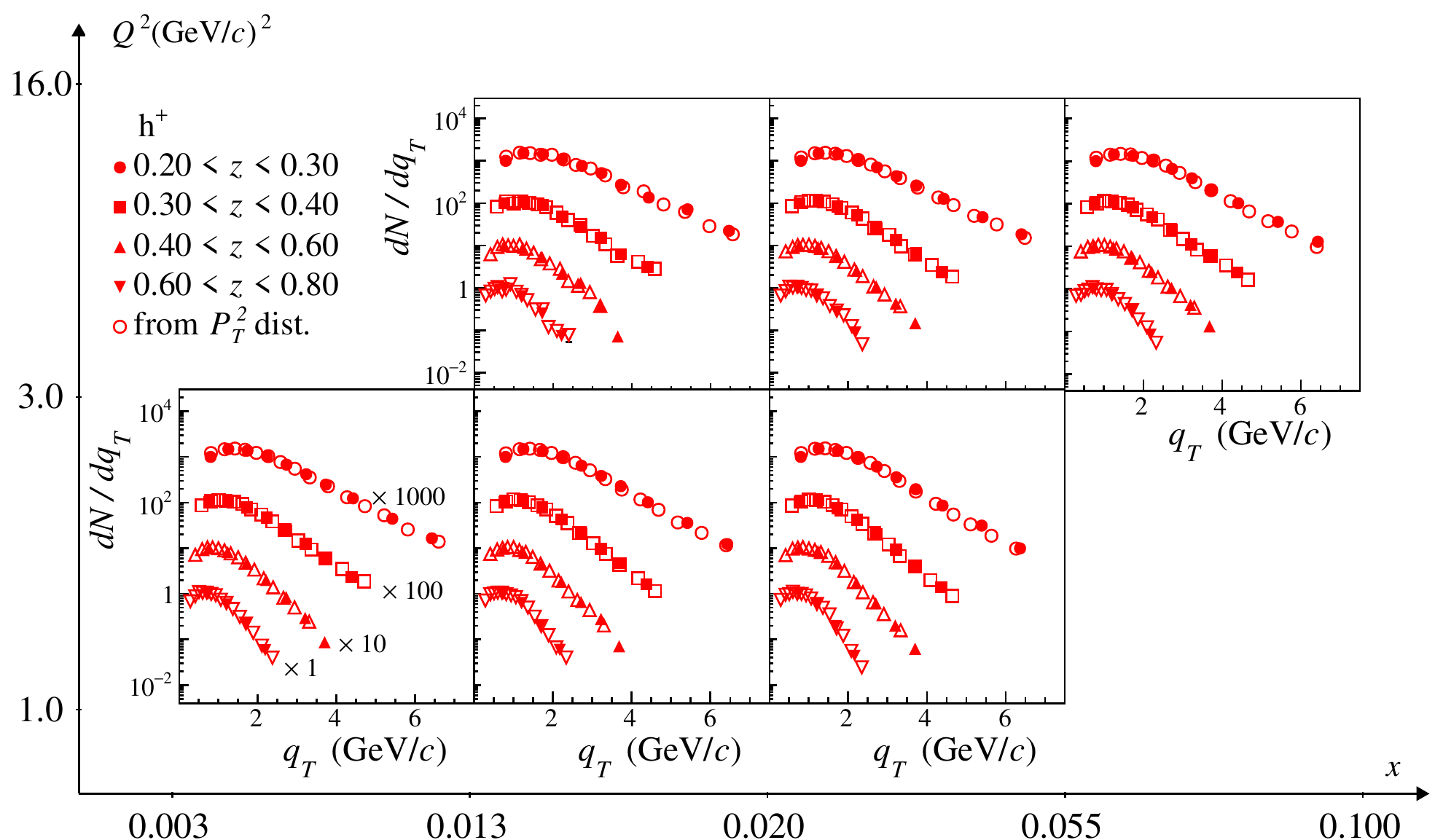}
    \includegraphics[width=0.85\textwidth]{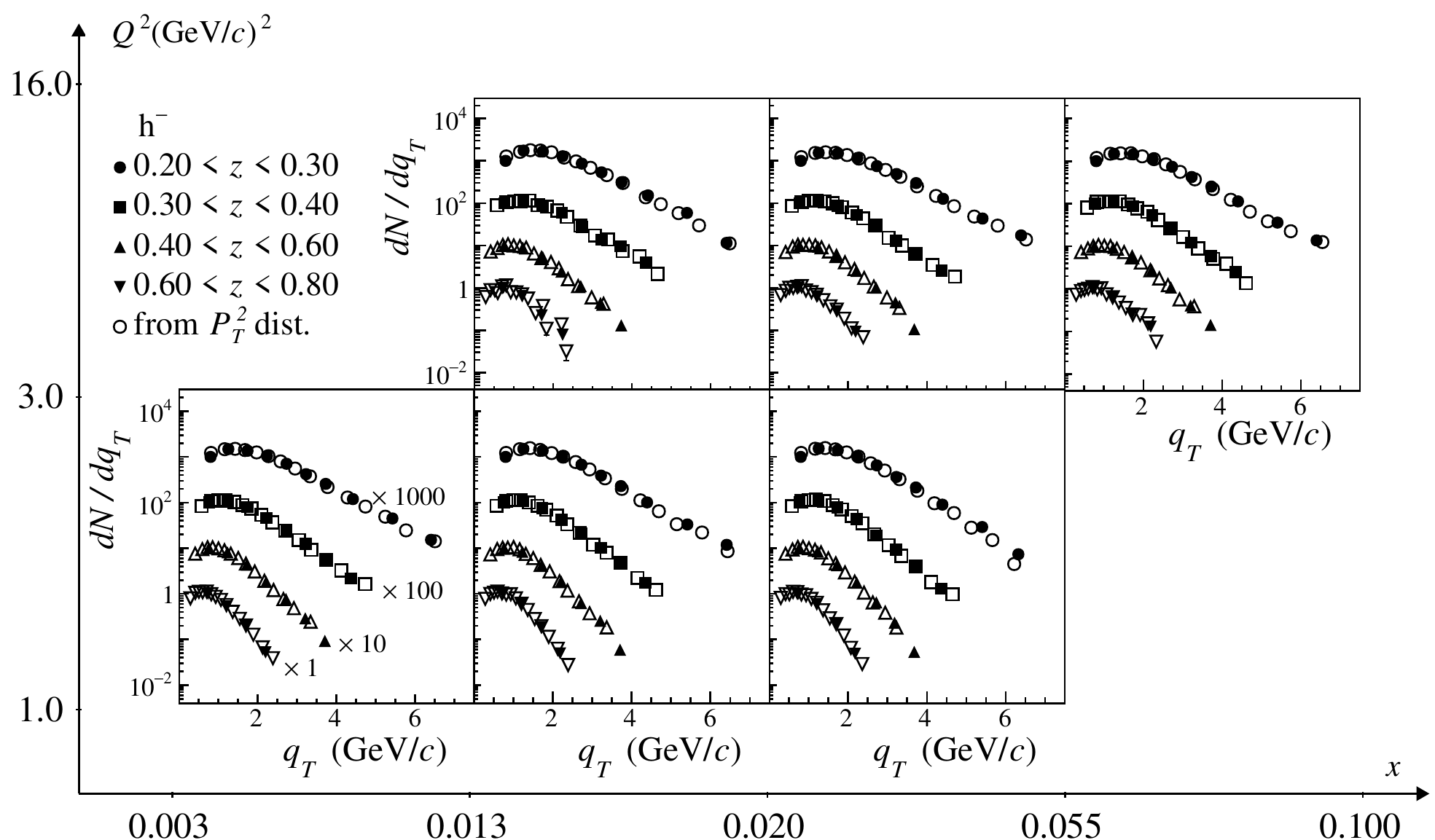}
    \caption{$q_T$-distributions for positive (top) and negative hadrons (bottom), compared to the distributions obtained with the simplified formula, normalized at 1 in the first point and scaled for better readability.}
    \label{fig:qt}
\end{figure}

\begin{figure}[h!]
\captionsetup{width=\textwidth}
    \centering
    \includegraphics[width=0.85\textwidth]{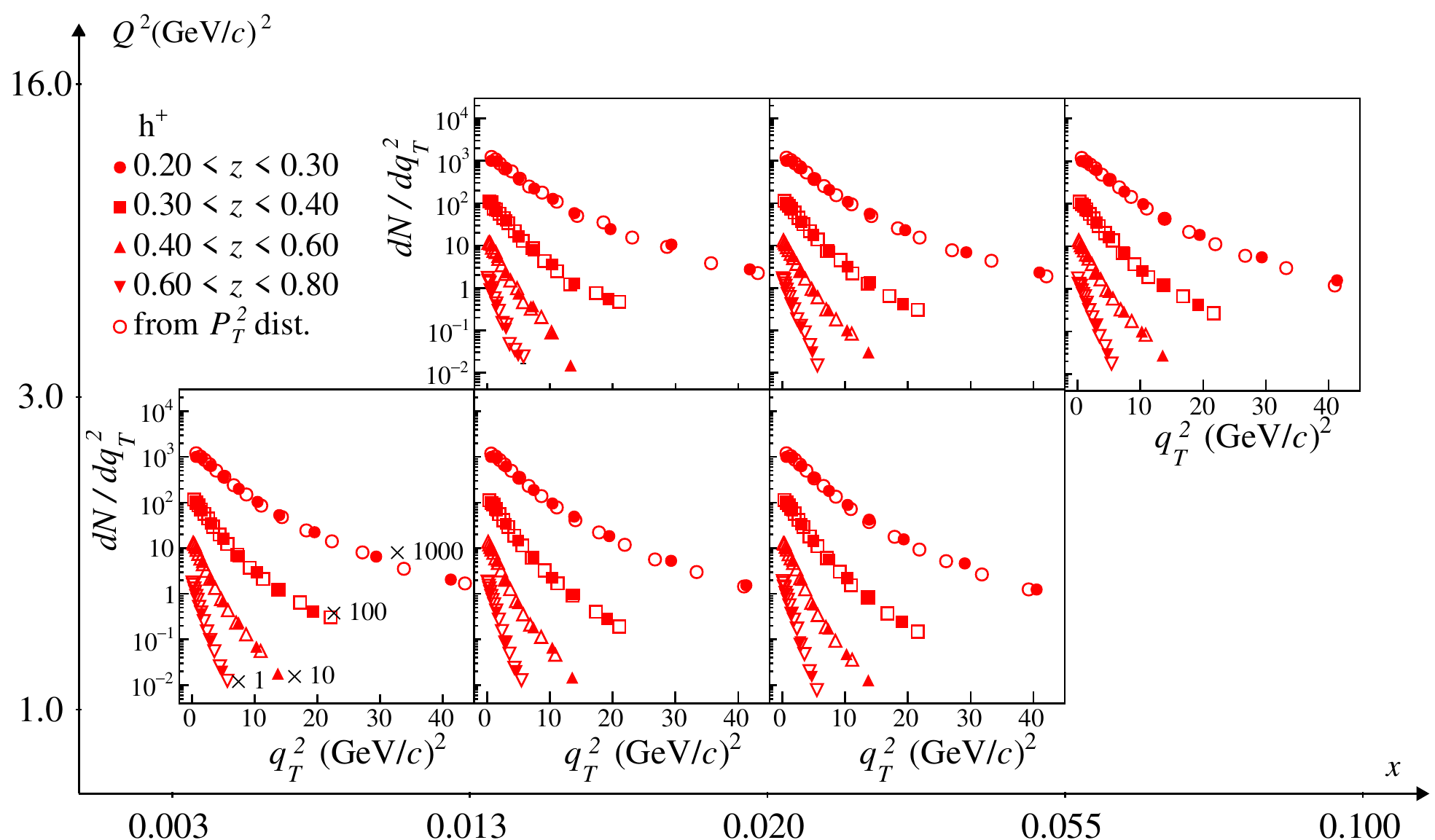}
    \includegraphics[width=0.85\textwidth]{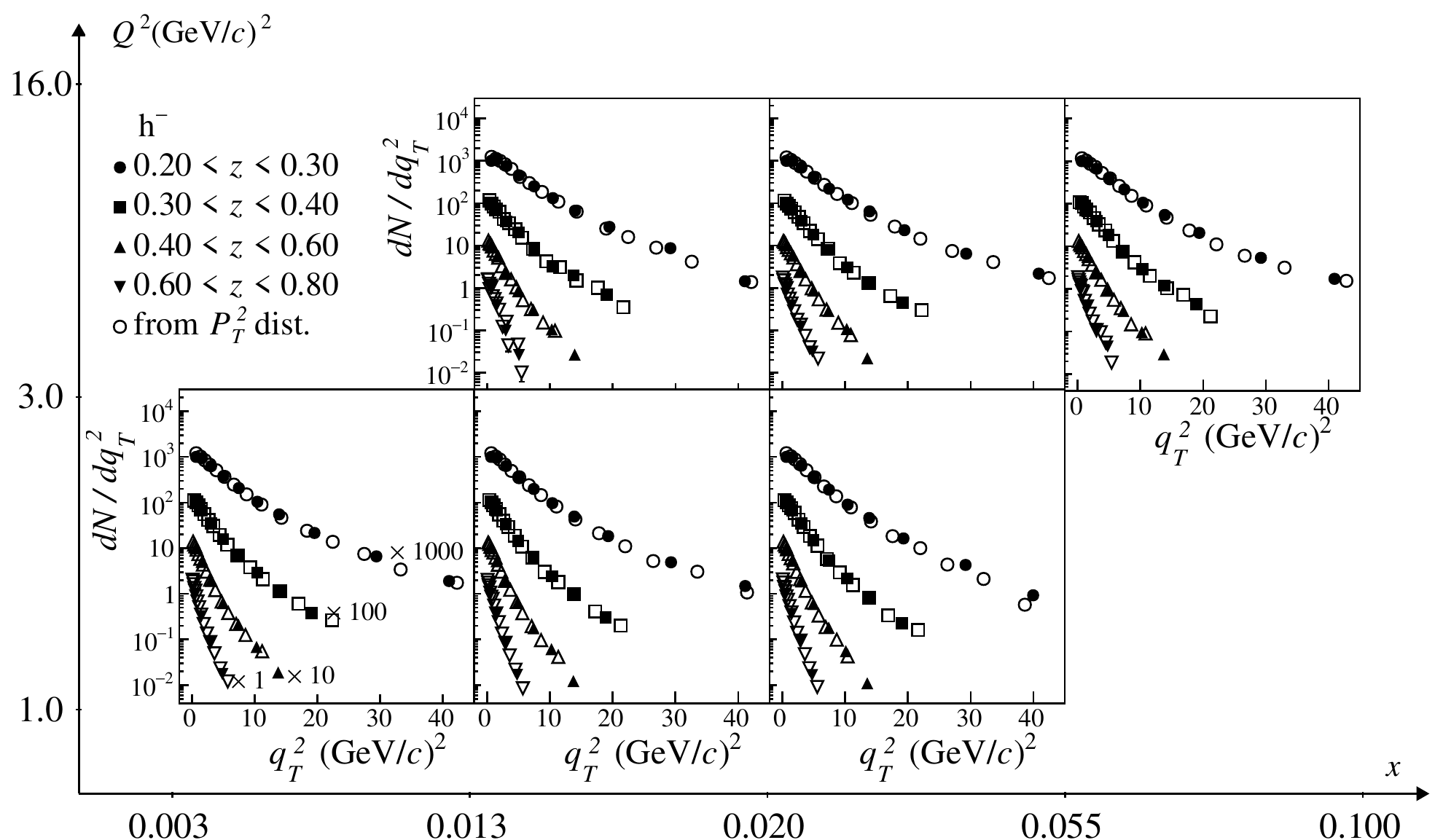}
    \caption{$q_T^2$-distributions for positive (top) and negative hadrons (bottom), compared to the distributions obtained with the simplified formula, normalized at 1 in the first point and scaled for better readability.}
    \label{fig:qt2}
\end{figure}

All this said, the distributions of $q_T$ have also been measured using the data samples used for the measurement of the $\Ptsq$-distributions. The binning in $x$, $\Qsq$ and $z$ has been kept the same as for the $\Ptsq$ distributions, while the $q_T$ range has been divided into 15 bins from $q_T=0.1$~GeV/$c$ to $q_T=17.0$~GeV/$c$. The condition on $\Ptsq$ to be in the same range used for the $\Ptsq$-distributions reduced the number of accessible $q_T$ bins to 10 at low $z$.
The measured distributions, corrected for the exclusive hadron contribution and for acceptance, are shown in Fig.~\ref{fig:qt} for positive and negative hadrons. It can be seen that the range in $q_T$ covered at the different $z$ is different, as a natural consequence of the definition of $q_T$. The distributions are compared in the same Figure to the ones obtained from the $\Ptsq$-distributions, according to the formula:

\begin{equation}
    \frac{\diff^2 N}{\diff z \diff \Ptsq} =  \frac{\diff^2 N}{\diff z~2\Pt \diff \Pt} = \frac{\diff^2 N}{\diff z~2\Pt \frac{\diff \Pt}{z}z} \approx \frac{1}{2z\Pt}\frac{\diff^2 N}{\diff z\diff \qt}
\end{equation}
The qualitative agreement with the approximate method is good. The exponential trend of these distributions can better be observed by looking at the $q_T^2$ distributions: these are given, for the two methods and again with an arbitrary normalization and scaling factor, in Fig.~\ref{fig:qt2}. In general, the distributions are smooth and weakly dependent on $x$ and $\Qsq$, but with a strong dependence on $z$, as expected.

\clearpage
\newpage

\section{Further studies of the kinematic dependences}
\label{sect:ch4_kin_dep}
The dependences of the $\Ptsq$-distribution on the kinematic variables have been further studied, performing their measurements in more bins of the relevant variables. This Section collects the results obtained with

\begin{itemize}
    \item the same two bins in $\Qsq$ and four bins in $x$ (the same used for the 3D measurement of the azimuthal asymmetries, see Ch.~\ref{Chapter5_Azimuthal_asymmetries}) but with seven bins in $z$ instead of four;
    \item the same $x$, $\Qsq$ and $z$ bins as the standard ones, but with two bins in $W$;
    \item the same $x$ and $z$ binning as the standard one, but with four $\Qsq$ bins, both integrated over $W$ and in two bins of $W$.
\end{itemize}

Due to the low statistics of the available Monte Carlo samples, more complete multi-dimensional analyses could not be performed. The goal of these studies was to disentangle the different dependences, in view of global analyses aimed at extracting $\aktsq$. The conclusions are given in Sect.~\ref{sect:ch4_kT2_Q2}.

\subsection{$\Ptsq$-distributions in 7 $z$ bins}
The $\Ptsq$-distributions have been measured in seven $z$ bins, instead of the usual four, while keeping the standard $\Qsq$ and $\Ptsq$ binning, in order to better inspect the trend of the extracted $\aPtsq$ versus $z^2$ and to allow for a simultaneous phenomenological analysis of $\Ptsq$-distributions and azimuthal asymmetries. The binning in $x$ and $z$, given in Tab.~\ref{tab:binspt_7z}, has thus been modified and made similar to the one used for the three-dimensional extraction of the azimuthal asymmetries, treated in the next Chapter, as indicated in Table~\ref{tab:binspt_7z}:

\begin{table}[tbh!]
\centering
\captionsetup{width=\textwidth}
\begin{tabular}{lrrrrrrrrrrrrr}
  \hline
             & 1    & 2    & 3    & 4    & 5     & 6    & 7    & 8    \\
  \hline
    $x$      & 0.003 & 0.012 & 0.020 & 0.038 & 0.130 \\ 
    $z$      & 0.10 & 0.20 & 0.25 & 0.30 & 0.40 & 0.55 & 0.70 & 0.85 \\    \hline
\end{tabular}
\caption{Limits of kinematic bins in $x$ and $z$ for the measurement of the $\Ptsq$ distributions in 7 bins of $z$. The binning in $Q^2$ and $\Ptsq$ has been left unchanged.}
\label{tab:binspt_7z}
\end{table}

\begin{figure}[h!]
\captionsetup{width=\textwidth}
    \centering
    \includegraphics[width=0.95\textwidth]{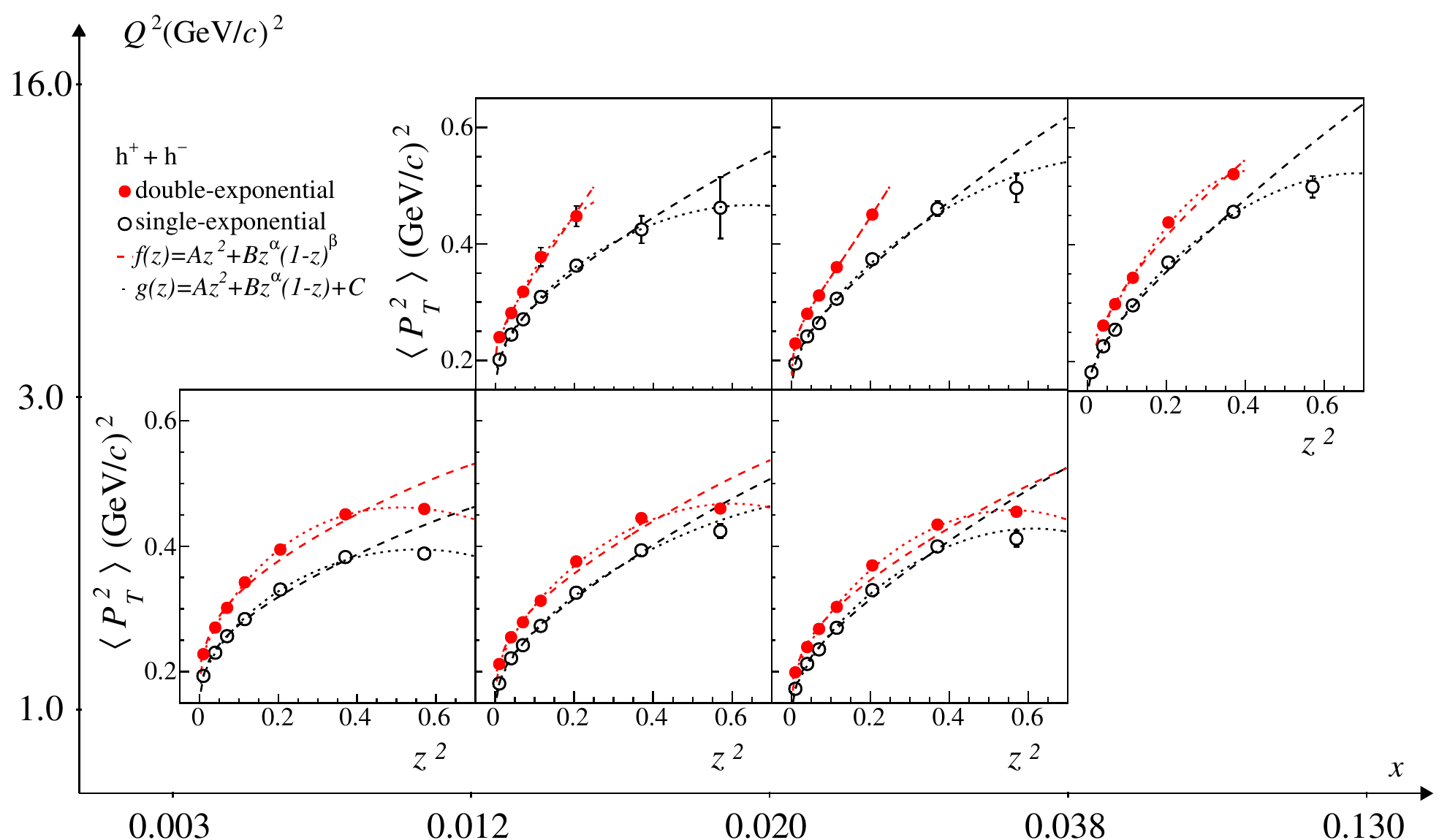}
    \caption{$\aPtsq$ versus $z^2$ as derived from the double-exponential (closed markers) and from the single-exponential fit (open markers) of the $\Ptsq$-distributions, measured in seven $z$ bins and summing over the hadron charge, with the indication of the two fit functions. }
    \label{fig:apt2_7z_fit}
\end{figure}

Apart from the newly-introduced low-$z$ bin, the distributions are affected by large statistical fluctuations that often prevent the double-exponential fit up to $\Ptsq=3$~(GeV/$c$)$^2$ to converge to reasonable values of the two slopes. The distributions for positive and negative hadrons have thus been summed before the fit, increasing the number of bins in which a solid estimate of $\aPtsq$ could be made. Actually, the fit is still unstable in some $z$ bins at large $\Qsq$. The values of $\aPtsq$ estimated from the double-exponential fits are shown in Fig.~\ref{fig:apt2_7z_fit} (closed markers), together with the analogous values from the single-exponential fit (open markers) for comparison. The trend of the two sets of values versus $z^2$ is similar. In particular, the deviation from the linear dependence expected from the relation: $\aPtsq = z^2\aktsq + \apperpsq$, where $\apperpsq$ is assumed to be constant in $z$, is confirmed. For this reason, as it was the case for the analysis in four $z$ bins, an extraction of $\aktsq$ from the slope of $\aPtsq$ versus $z^2$ seems not to be a viable solution. \\

Nonetheless, a description of the observed trend of $\aPtsq$ with respect to $z^2$ has been tried by including the $z$-dependence of $\apperpsq$, assuming it to be of the form: $\apperpsq(z) \propto z^{\alpha}(1-z)^\beta$.
Two functions have been fitted to the data. A first one (the dashed line in Fig.~\ref{fig:apt2_7z_fit}), defined as:
\begin{equation}
    f(z) = Az^2 + Bz^\alpha(1-z)^\beta,
\label{eq:fitfunc}    
\end{equation}
where the parameter $A$ corresponds to $\aktsq$, provides a reasonable description of the data only at low $z$. At intermediate and high $z$, the curve overshoots the data. In the first $\Qsq$ bin, the parameters $A$ and $\alpha$ are found to be compatible in the single- and double-exponential cases, ranging between 0.36 and 0.46 (the former) and between 0.24 and 0.28 (the latter). In the second $\Qsq$ bin, the limited range in $z$ for the double-exponential case prevents a direct comparison of the two sets of results, due to the large variations in the first case.  The $A$-parameter in the single exponential case is found to range between 0.45 (at low $x$) to 0.62 (at high $x$).

A better description of the data at high $z$ can be achieved by adding a constant term, in order to take into account the offset of $\aPtsq$ observed at low $z$. The second fit function (dotted  in Fig.~\ref{fig:apt2_7z_fit}), is defined as:
\begin{equation}
    g(z) = Az^2 + Bz^\alpha(1-z) + C,
\end{equation}
where the parameter $\beta$ (at the exponent of $(1-z)$ in Eq.~\ref{eq:fitfunc}) has been set to one, in order to keep low the number of parameters. In this case, the parameters suffer from a large uncertainty and bin-by-bin variability, even in the first $\Qsq$ bin.

\subsection{$\Ptsq$-distributions in 2 $W$ bins}
\begin{figure}[h!]
    \centering
    \captionsetup{width=\textwidth}
    \includegraphics[width=0.85\textwidth]{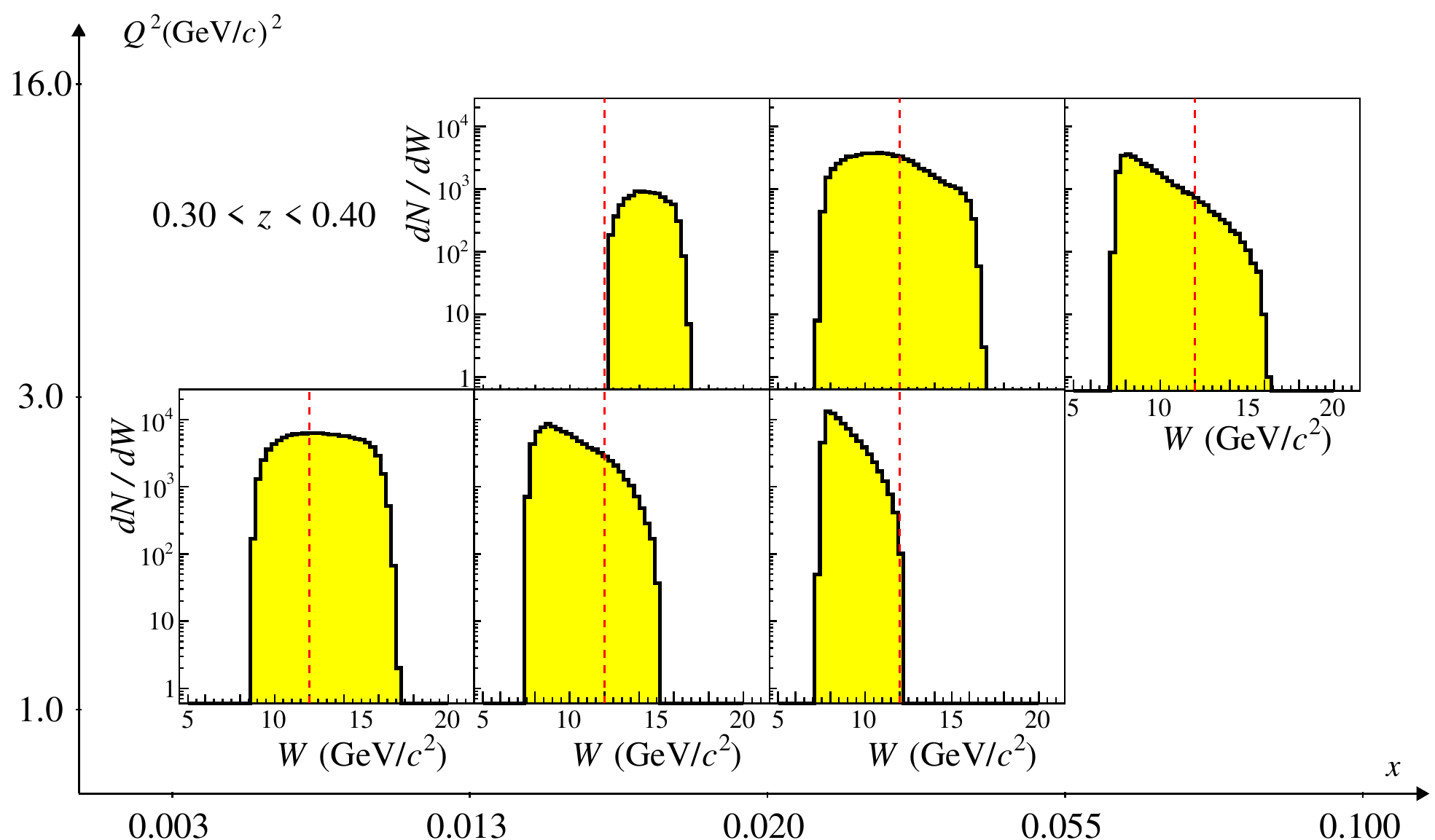}
 \caption{$W$-distributions for all hadrons in bins of $x$ (horizontal axis) and $Q^2$ (vertical axis),for $0.30<z<0.40$. The red, vertical line indicates the cut at $W=12$~GeV/$c^2$.}
    \label{fig:Wdists_0}
\end{figure}

So far, the $\Ptsq$-distributions have been presented in four $x$ bins and two $\Qsq$ bins: no binning in $W$ was applied. It is clear that the DIS process can be fully described by two variables only, and in this respect the dependence of the $\Ptsq$-distribution on $x$ and $\Qsq$ would be sufficient. However, the dependence of the results on different variables is also interesting, as it may give a deeper insight into the kinematics of the process. Here we investigate a possible dependence on $W$, observed e.g. by the EMC Collaboration \cite{Ashman:1991cj}, by extracting the distributions in 2 bins of $W$ ($W<12$~GeV/$c^2$ and $W>12$~GeV/$c^2$). The cut at $W=12$~GeV/$c^2$ is suggested by the measured $W$ distributions for the selected hadrons as shown in Fig.~\ref{fig:Wdists_0} for $0.30<z<0.40$. No $z$-dependence is observed in the shape of the $W$-distributions.

\begin{figure}[h!]
    \centering
    \captionsetup{width=\textwidth}
    \includegraphics[width=0.49\textwidth]{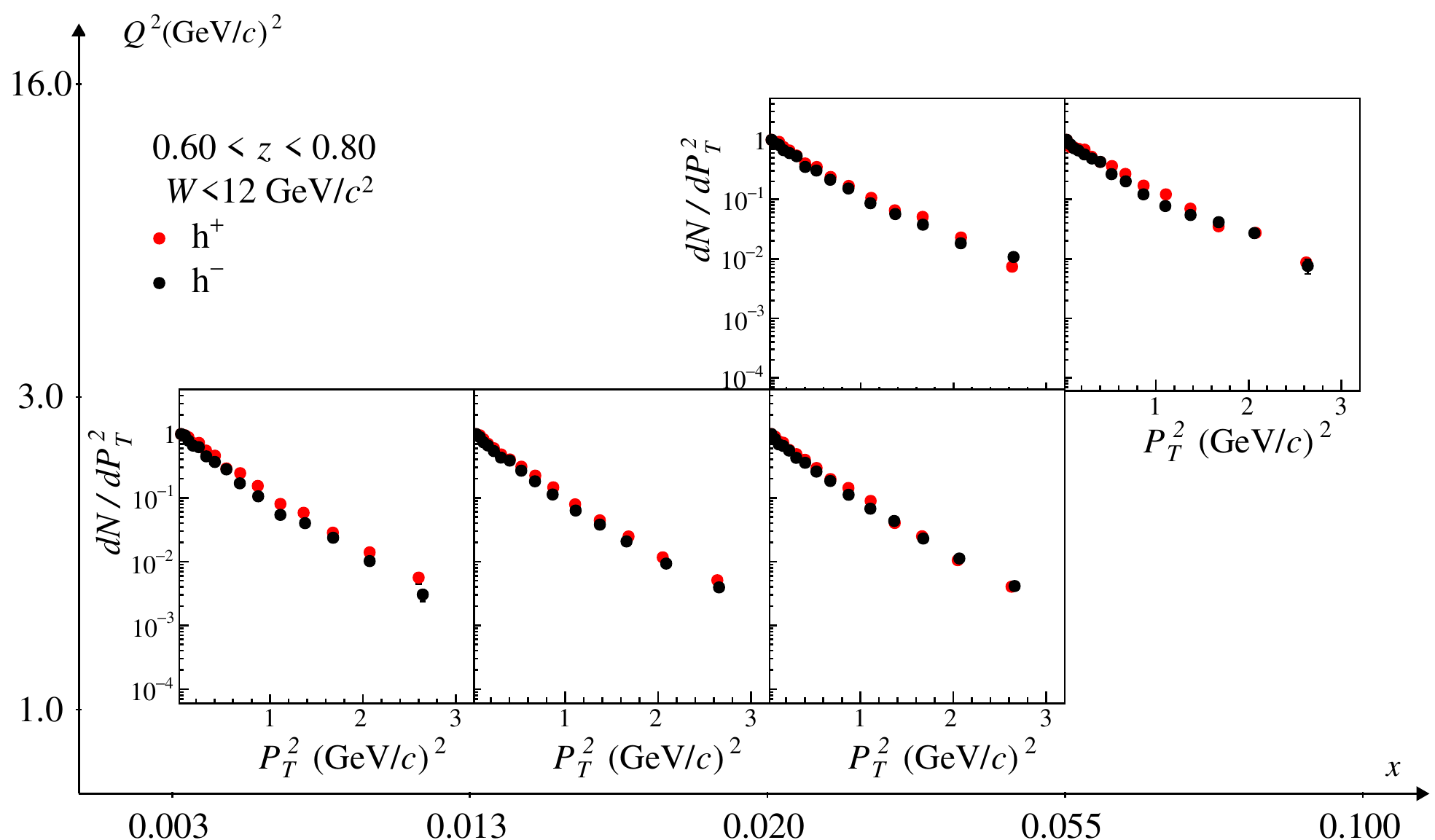}
    \includegraphics[width=0.49\textwidth]{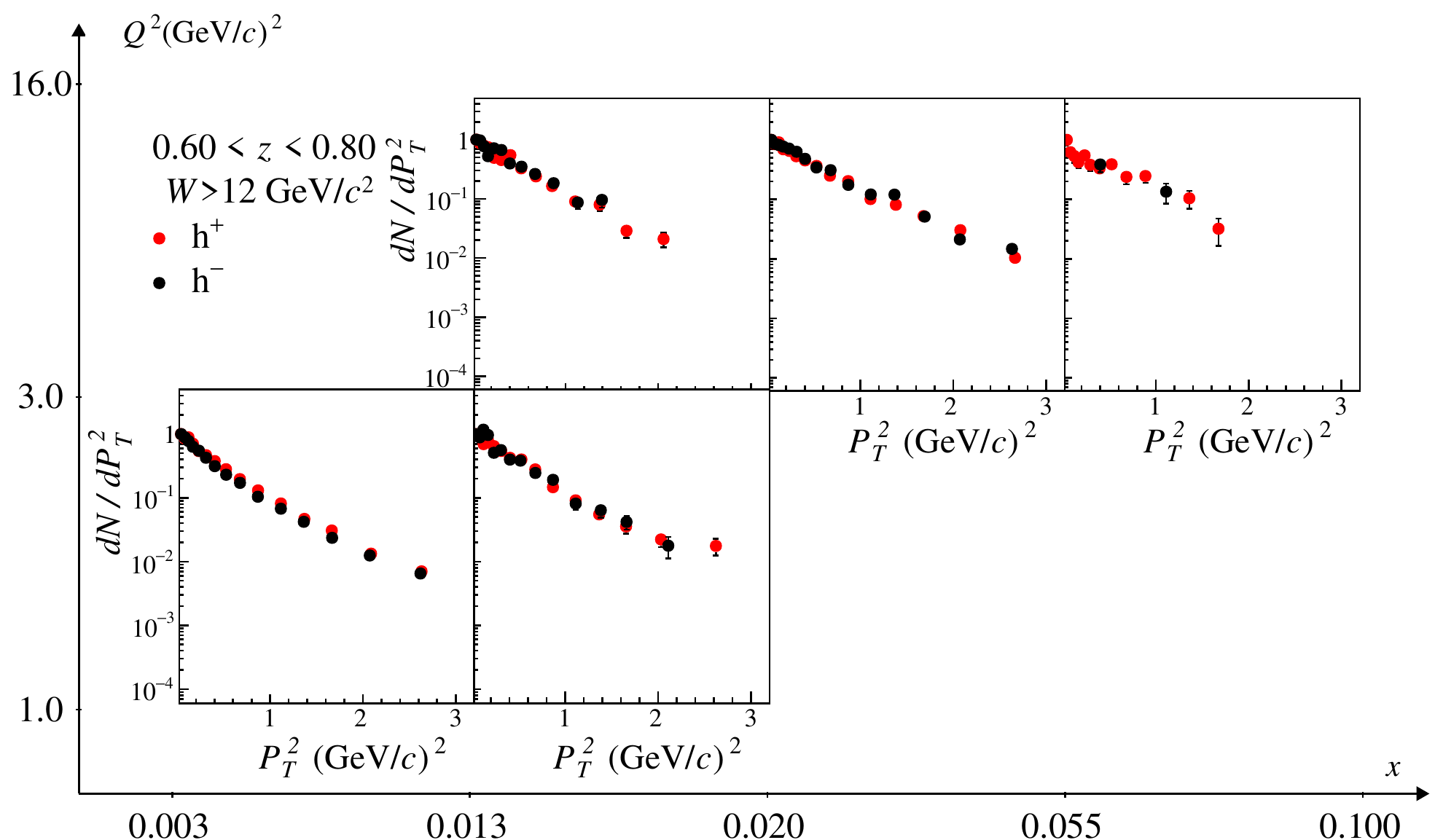}
    
    \includegraphics[width=0.49\textwidth]{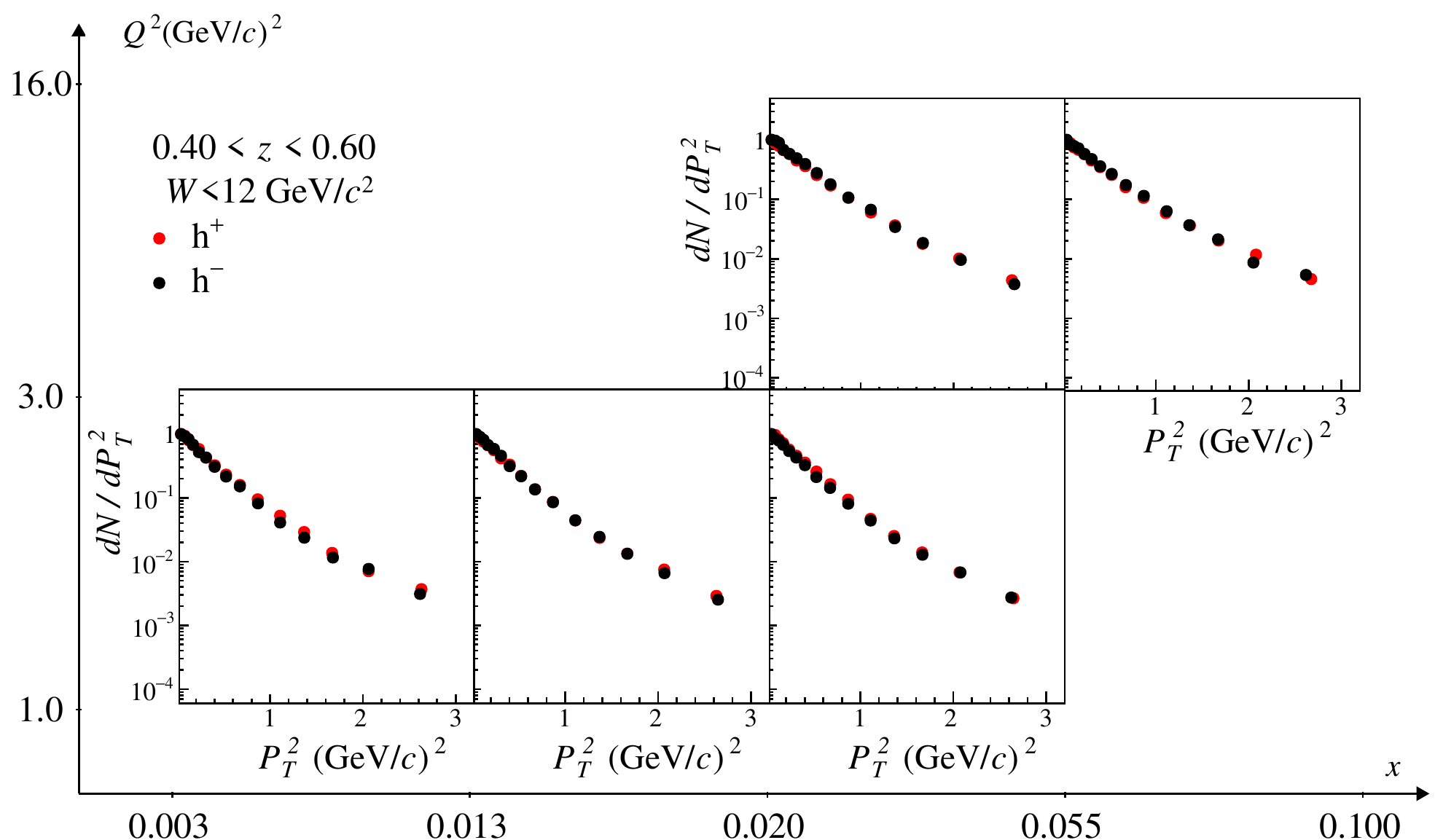}
    \includegraphics[width=0.49\textwidth]{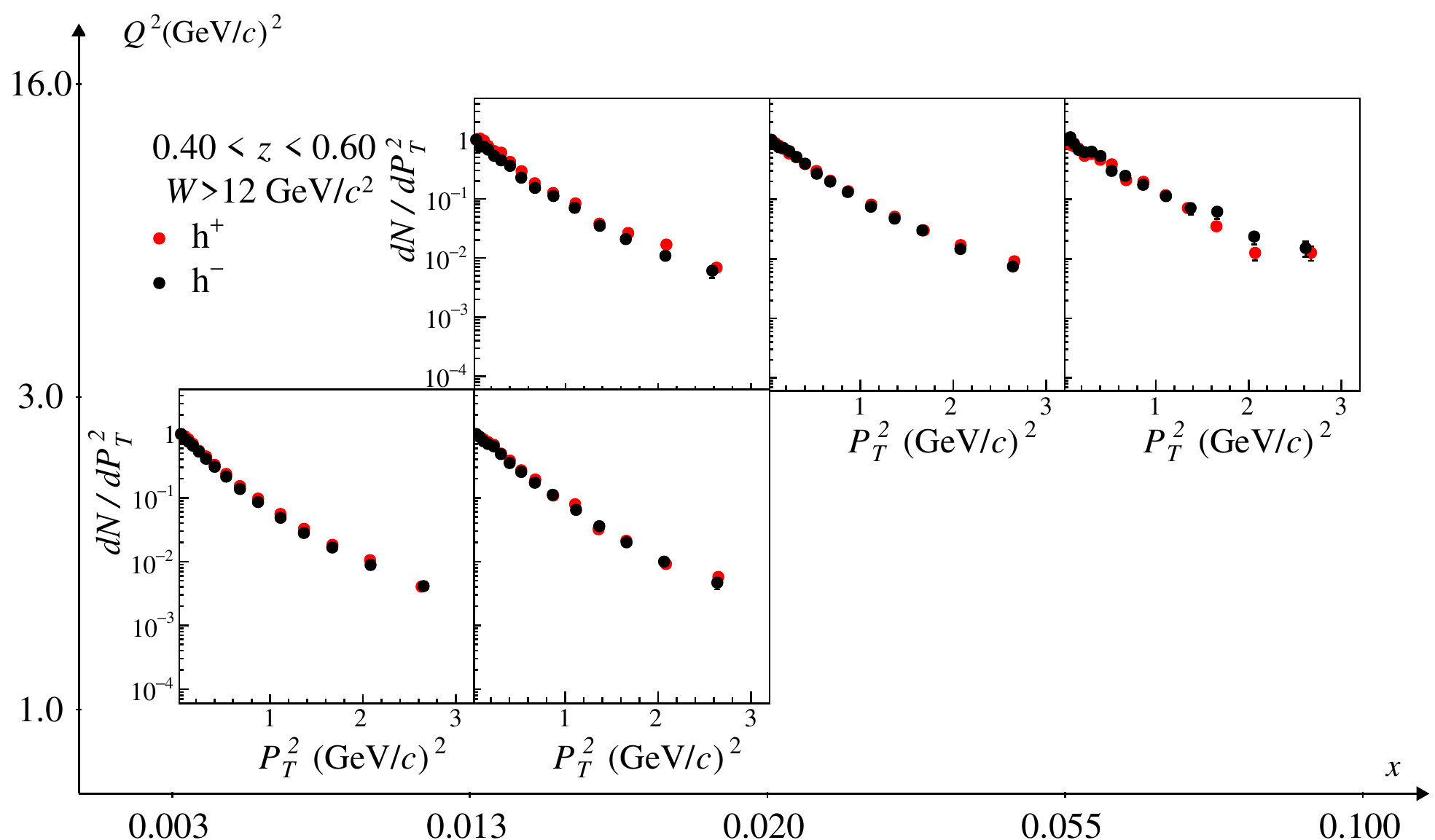}
    
    \includegraphics[width=0.49\textwidth]{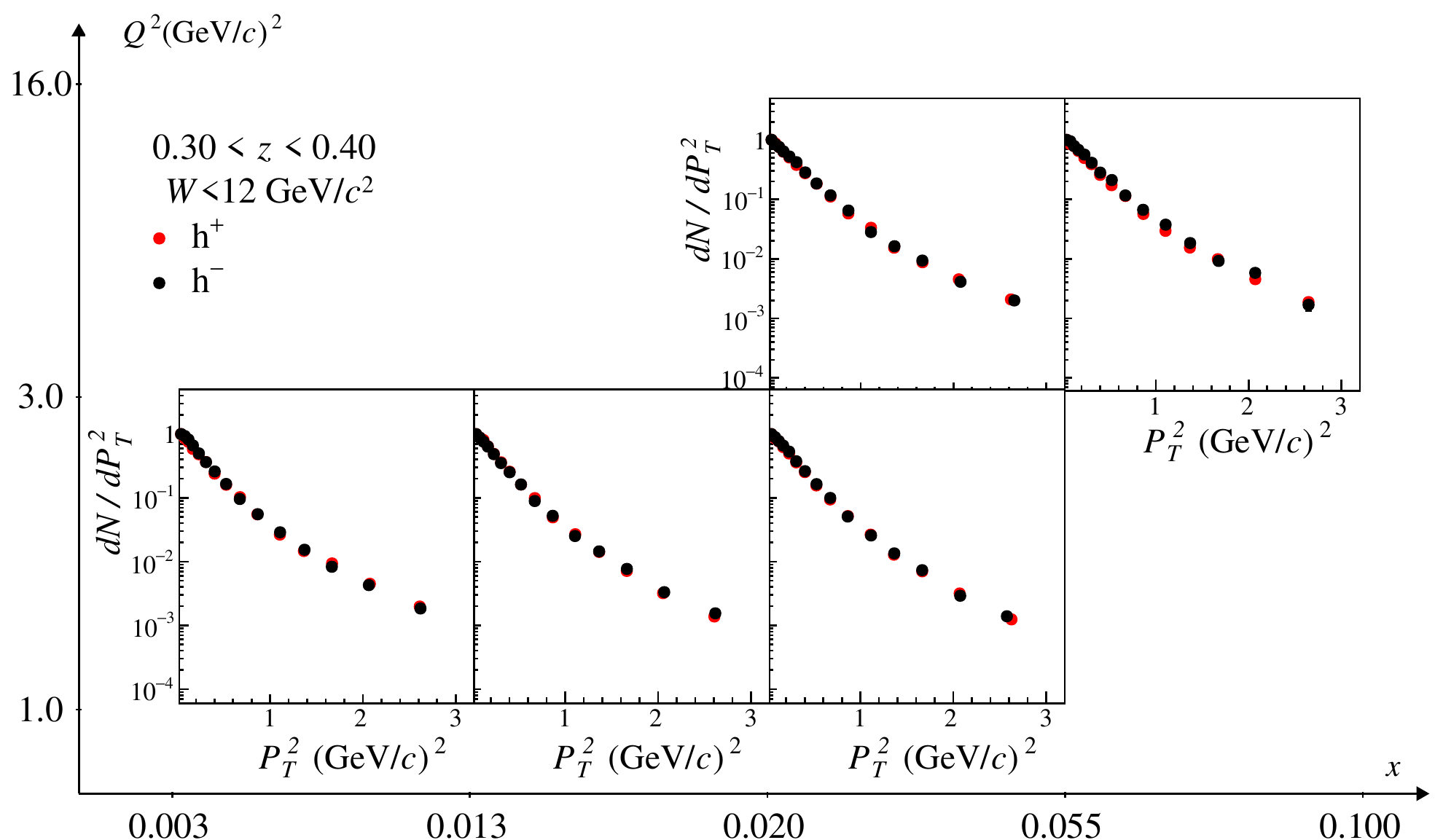}
    \includegraphics[width=0.49\textwidth]{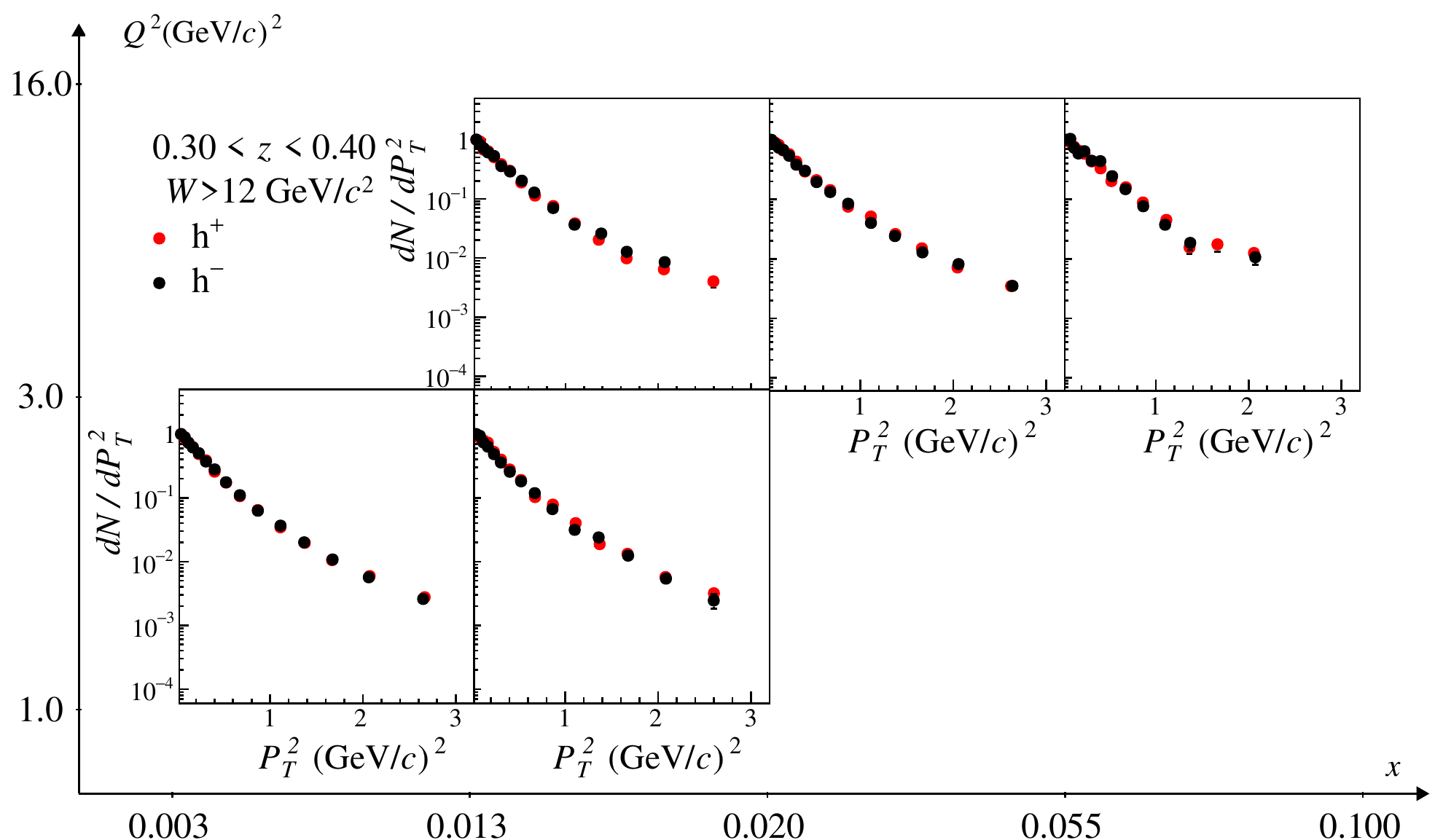}
    
    \includegraphics[width=0.49\textwidth]{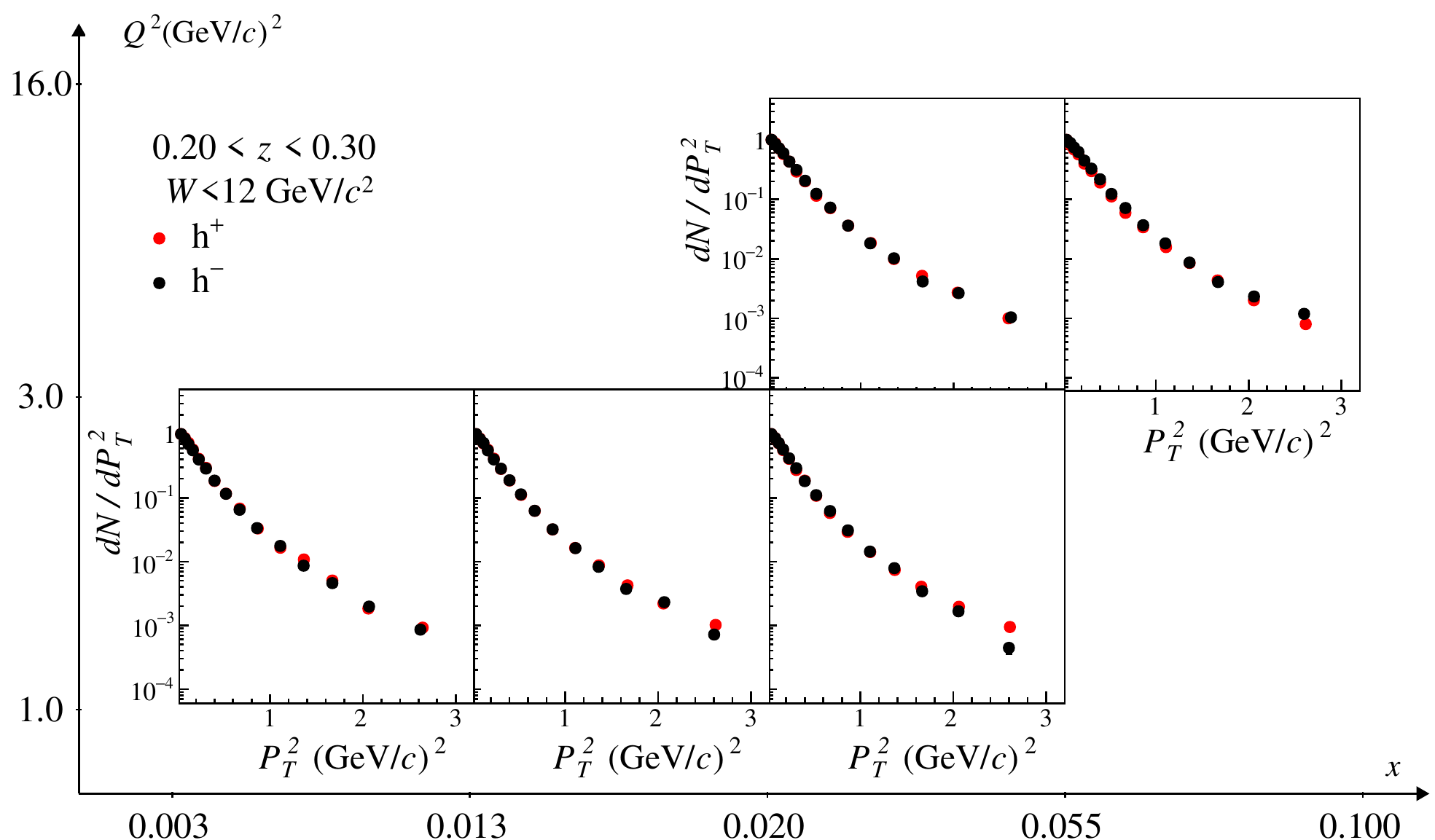}
    \includegraphics[width=0.49\textwidth]{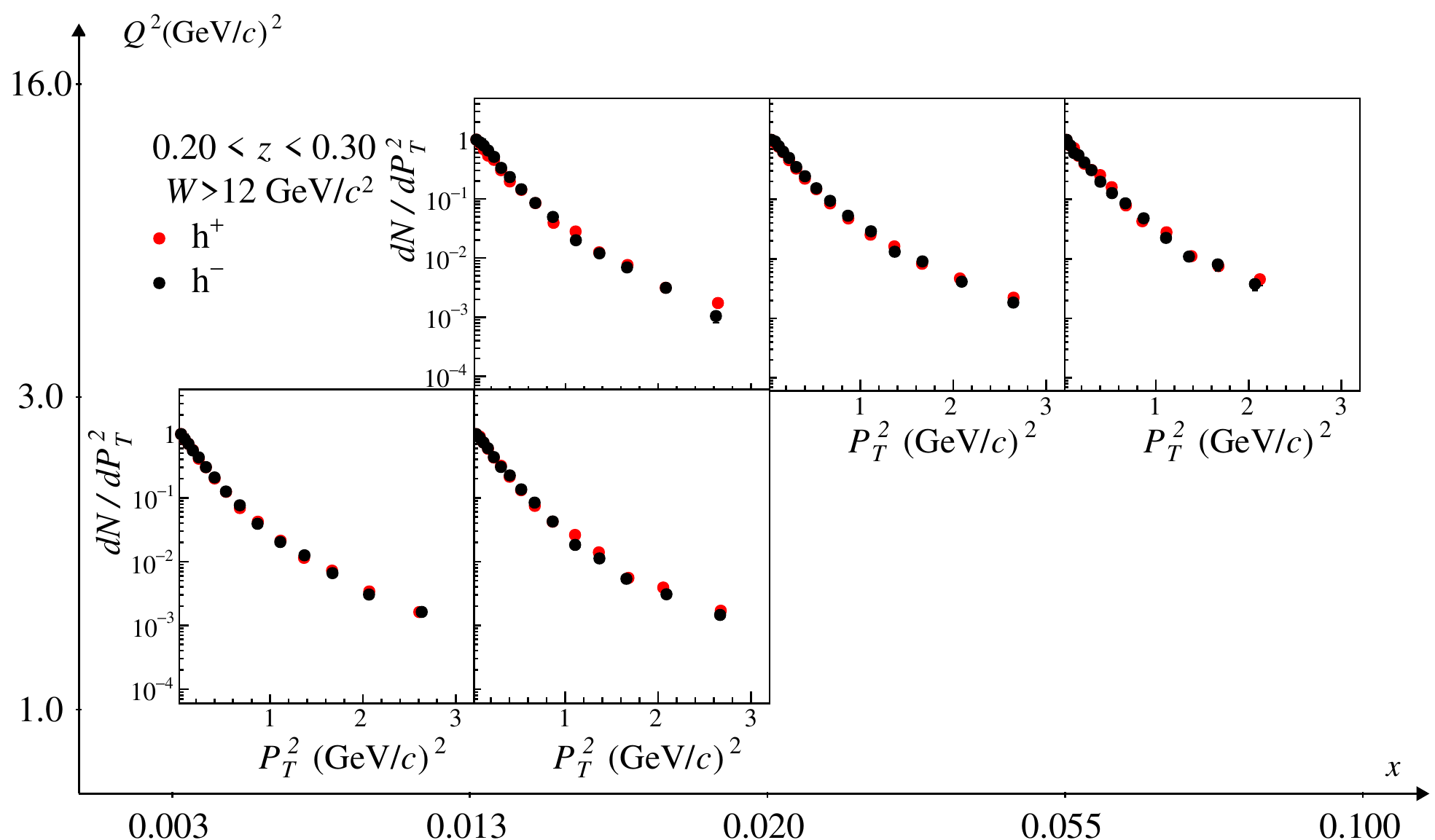}

 \caption{$\Ptsq$-distributions for positive (red) and negative hadrons (black) in bins of $x$ (horizontal axis) and $Q^2$ (vertical axis), in bins of $z$, for $W<12$~GeV/$c^2$ (left) and $W>12$~GeV/$c^2$ (right).}
    \label{fig:pt_2W_2Q2_4z_0}
\end{figure}

\begin{figure}[h!]
    \captionsetup{width=\textwidth}
    \centering
    \includegraphics[width=0.49\textwidth]{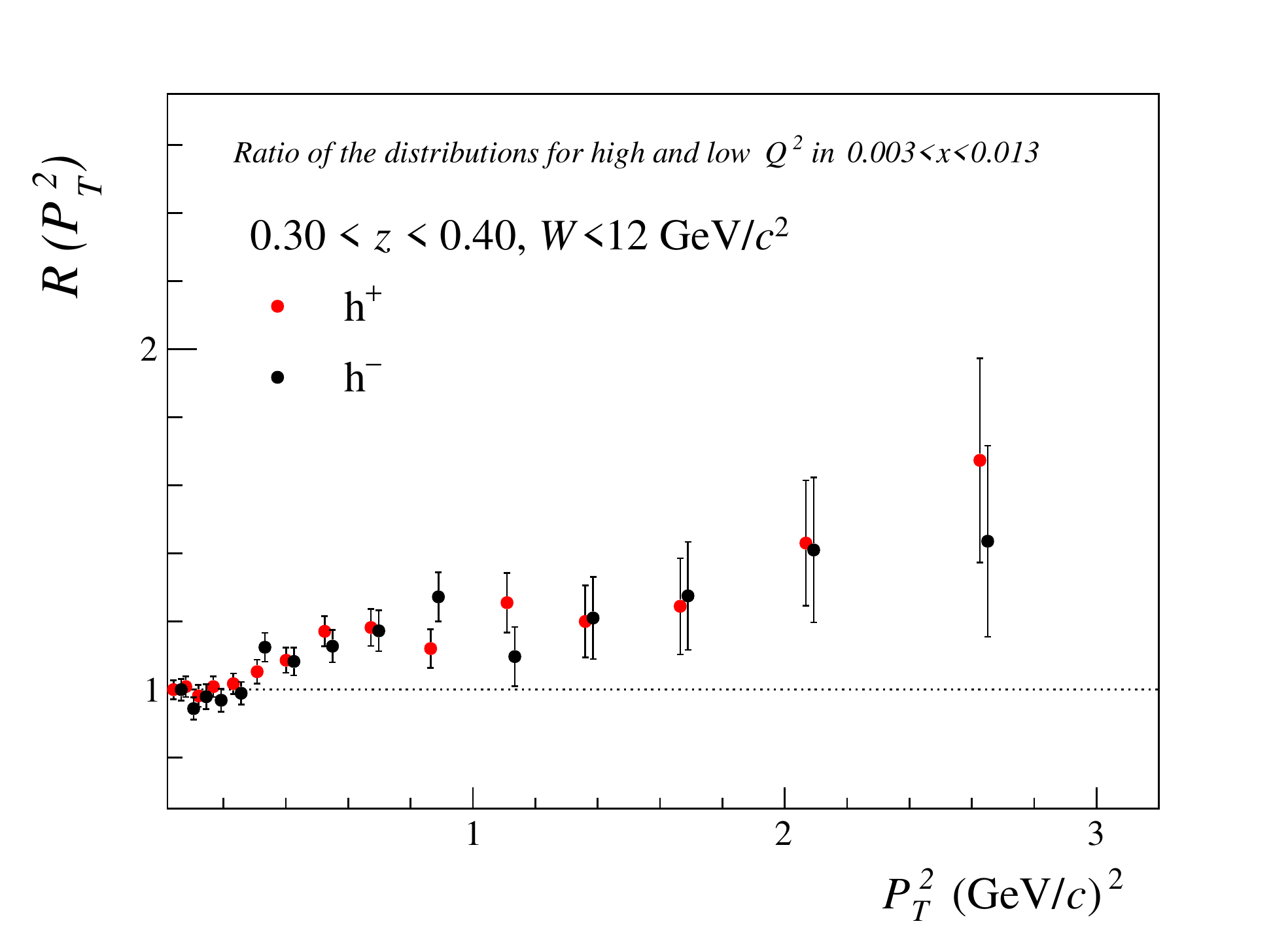}
    \includegraphics[width=0.49\textwidth]{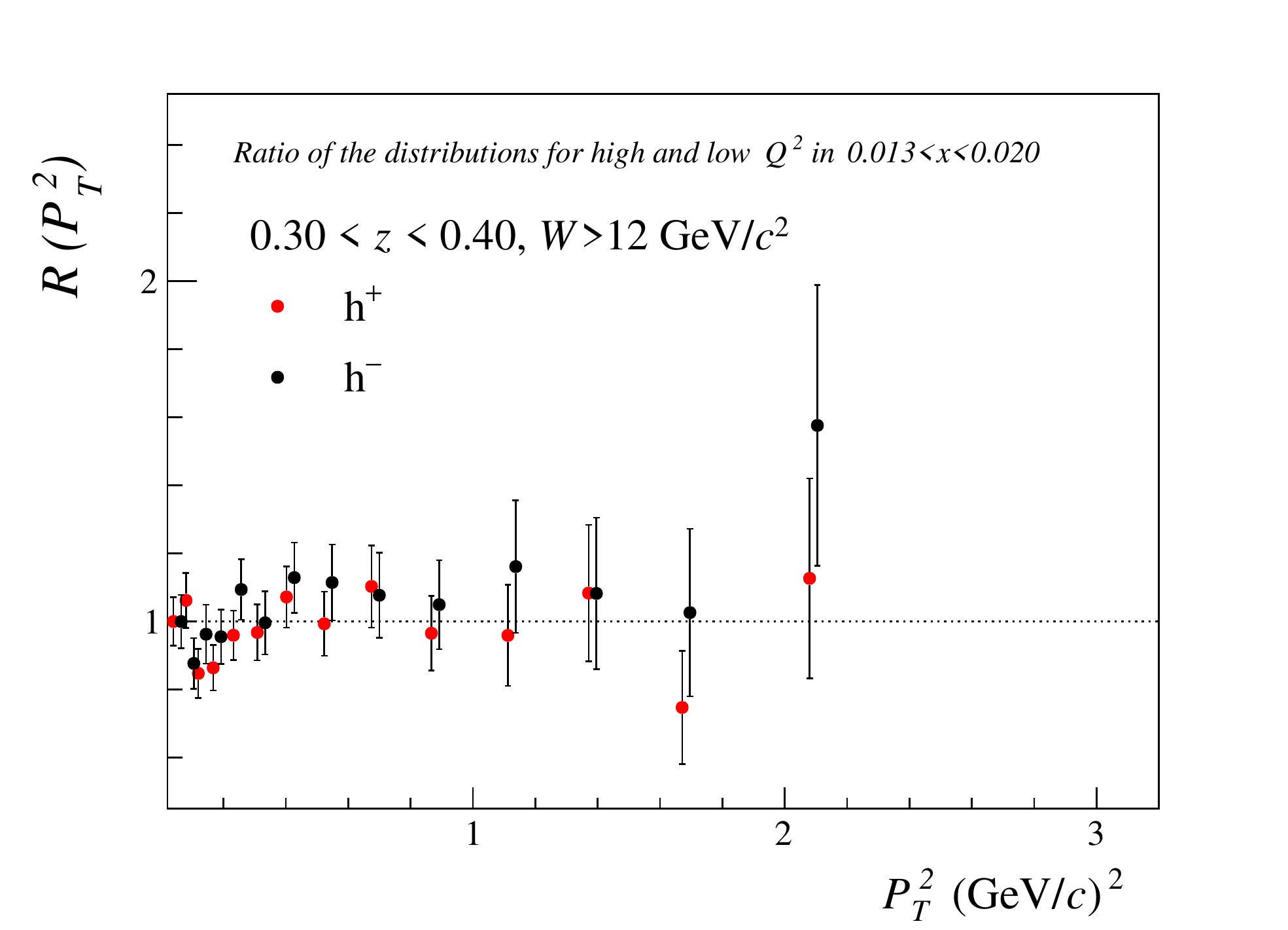}
 \caption{Ratio (high over low $\Qsq$) of the $\Ptsq$-distributions for positive (red) and negative hadrons (black) for $0.30<z<0.40$ in $0.020<x<0.055$ for $W<12$~GeV/$c^2$ and in $0.013<x<0.020$ for $W>12$~GeV/$c^2$.}
    \label{fig:pt_2W_2Q2_4z_rs}
\end{figure}

\begin{figure}
\captionsetup{width=\textwidth}
    \centering
    \includegraphics[width=0.49\textwidth]{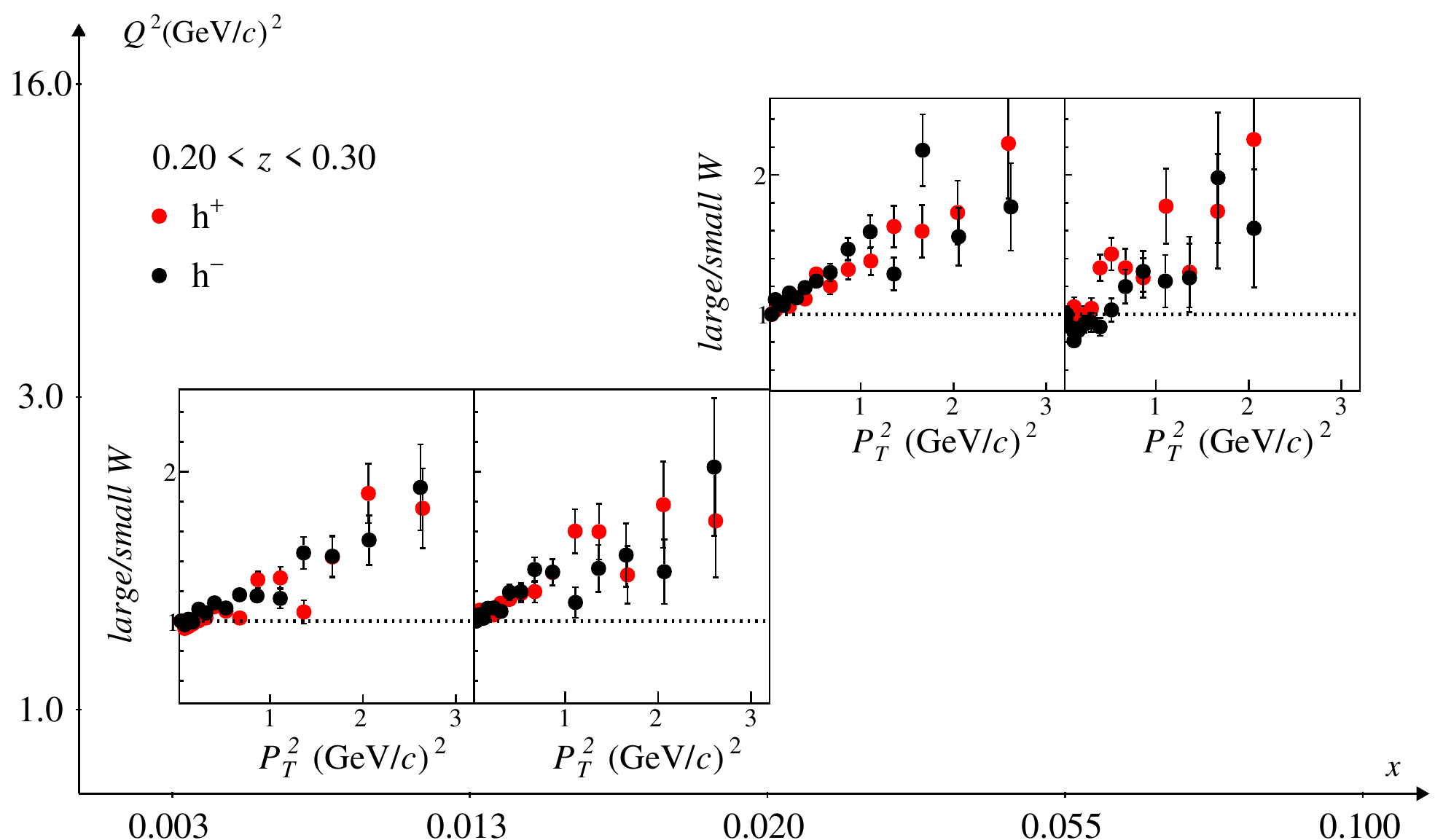}
    \includegraphics[width=0.49\textwidth]{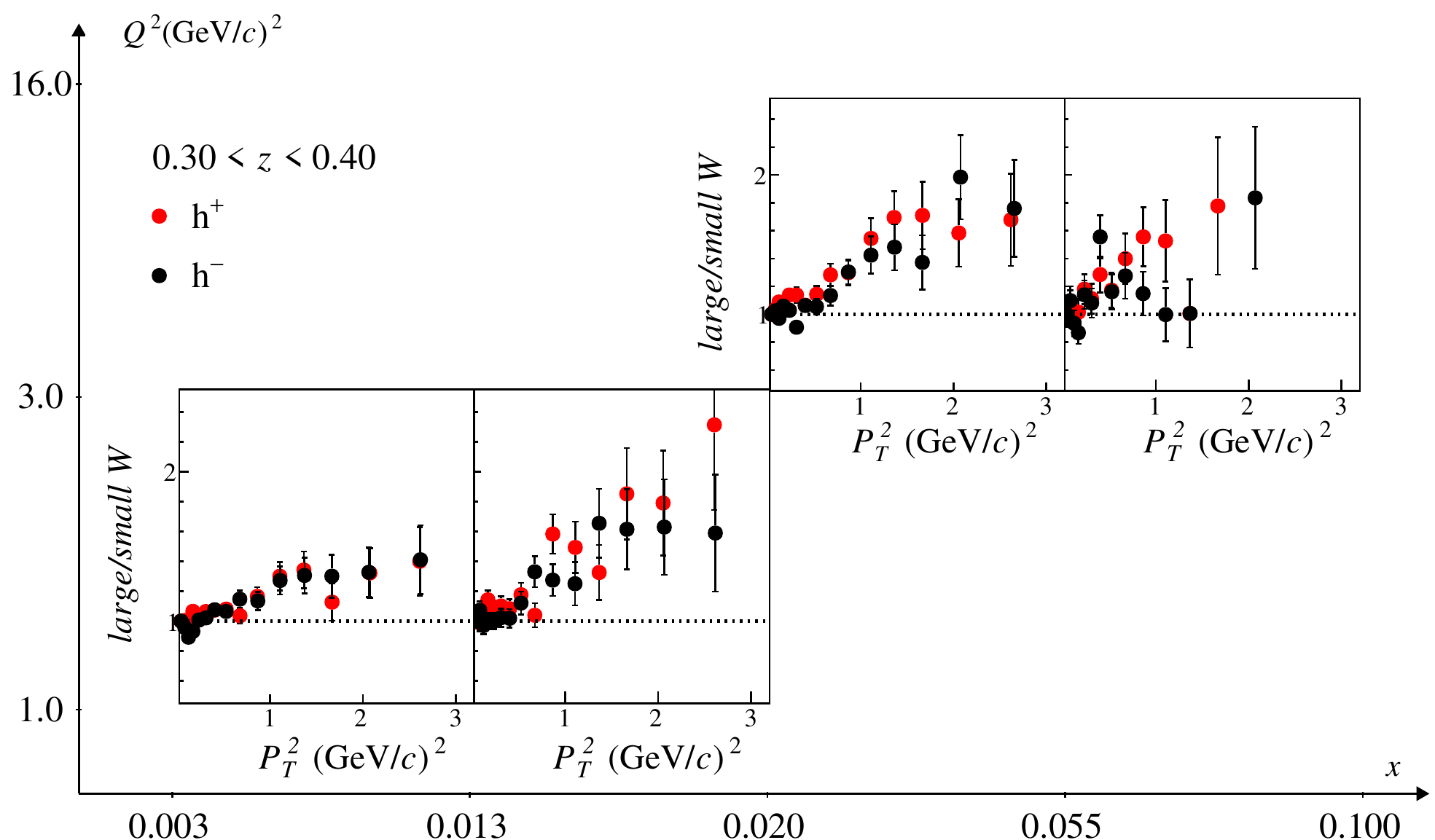}
    \includegraphics[width=0.49\textwidth]{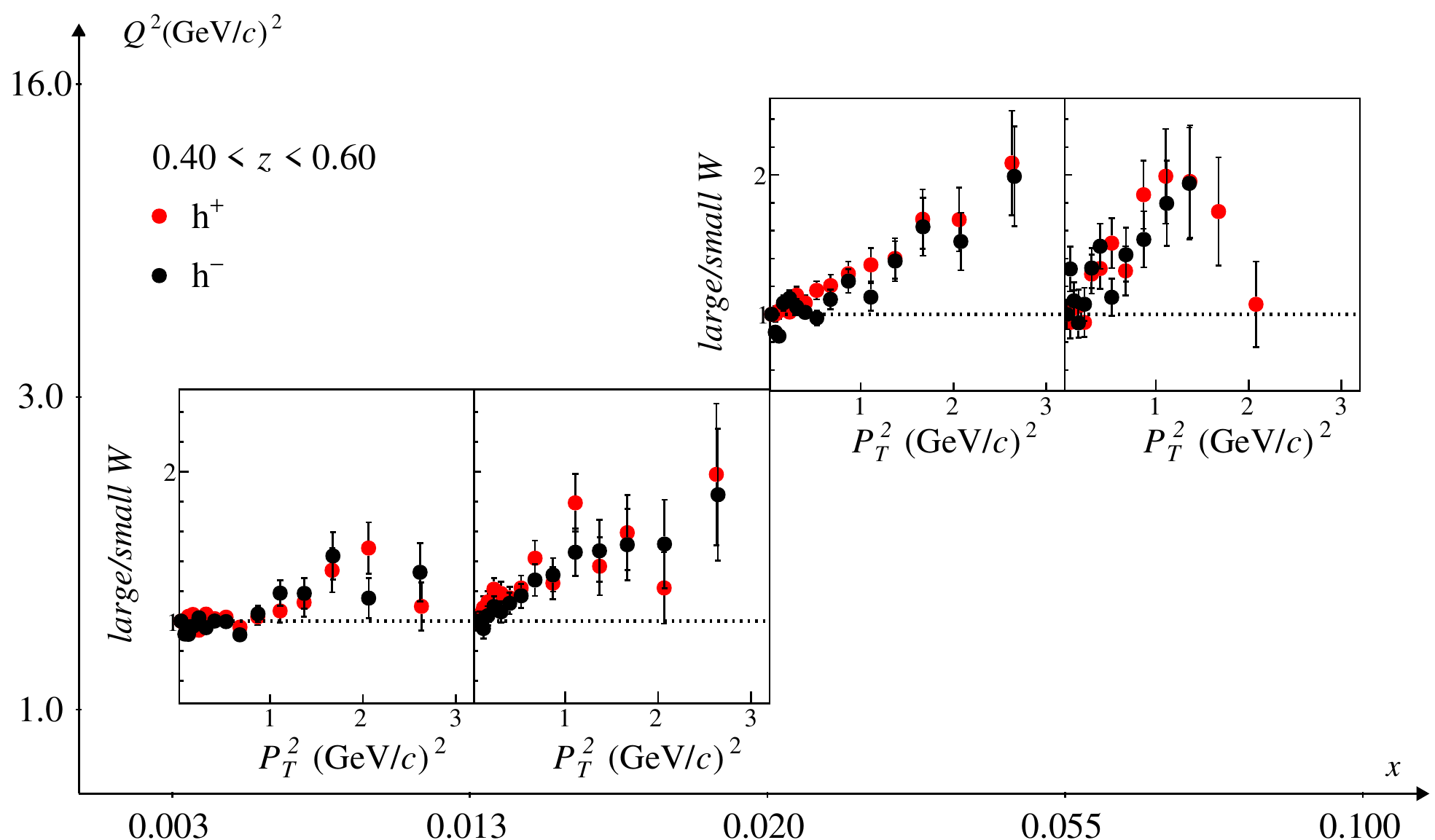}
    \includegraphics[width=0.49\textwidth]{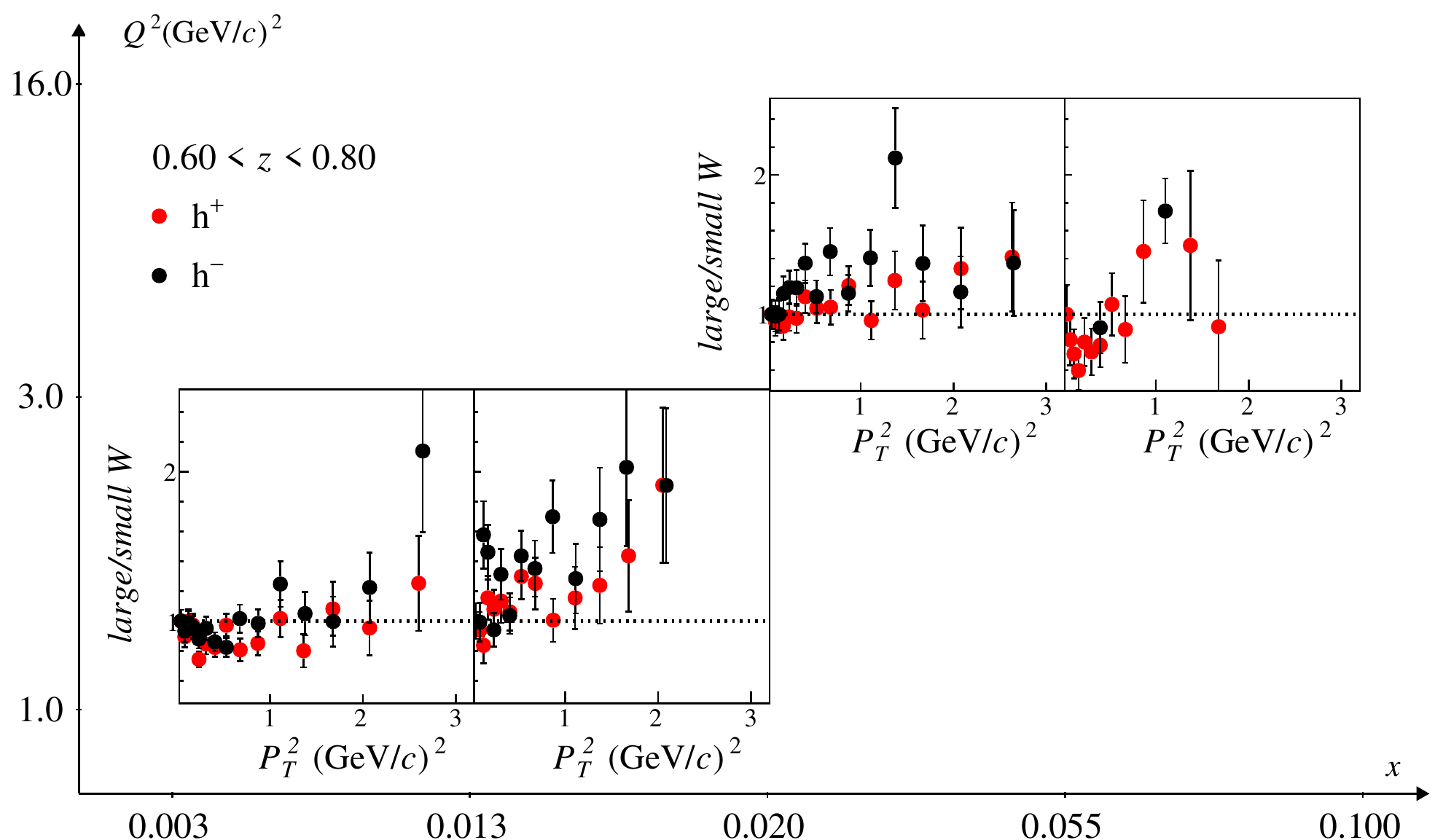}    
 \caption{Ratio (high- over low $W$) of the $\Ptsq$-distributions for positive (red) and negative hadrons (black) in bins of $x$ (horizontal axis) and $Q^2$ (vertical axis).}
    \label{fig:pt_2W_2Q2_4z_rW0}
\end{figure}

The distributions at low $W$ and high $W$ are shown in Fig.~\ref{fig:pt_2W_2Q2_4z_0}. As can clearly be seen, the condition on $W$ limits the number of accessible ($x,\Qsq$) bins, which is reduced by one unit at low $W$ (the second bin in $x$ at high $\Qsq$) and at high $W$ (the bin at high $x$ and low $\Qsq$). In general, as seen before, the dependence on $x$ looks weak, or even negligible. The $\Qsq$ dependence is instead visible at low $W$, as shown in Fig.~\ref{fig:pt_2W_2Q2_4z_rs} (left), where the ratio of the distribution at high and low $\Qsq$ shows, for a representative bin in $z$, a clear linear trend. The slope of the ratio, about 0.15~(GeV/$c$)$^{-2}$, anyway suggest a very limited difference in terms of the $\aPtsq$ of the distributions. Interestingly, the ratio of the distributions at high and low $\Qsq$ is flat at high $W$. It is however not easy to conclude whether the observed trends are a manifestation of a real $\Qsq$ dependence, vanishing at high $W$, or just the combined effect of the $x$-$\Qsq$-$W$ correlation in the considered bins. To study a possible $W$-dependence of the distributions, their ratio (high over low $W$) is taken in the corresponding ($x,\Qsq$) bins, as shown in Fig.~\ref{fig:pt_2W_2Q2_4z_rW0}: except for the first $x$ bin at high $z$, the widening of the distributions with $W$ looks similar in all bins. The inclusion of more data in the analysis would help in reducing the statistical uncertainty affecting the ratios and, as a  consequence, in concluding whether these trends are a manifestation of a real $W$-dependence.

\newpage

\subsection{$\Ptsq$-distributions in 4 $Q^2$ bins}
\begin{figure}[h!]
\captionsetup{width=\textwidth}
    \centering
    \includegraphics[width=0.49\textwidth]{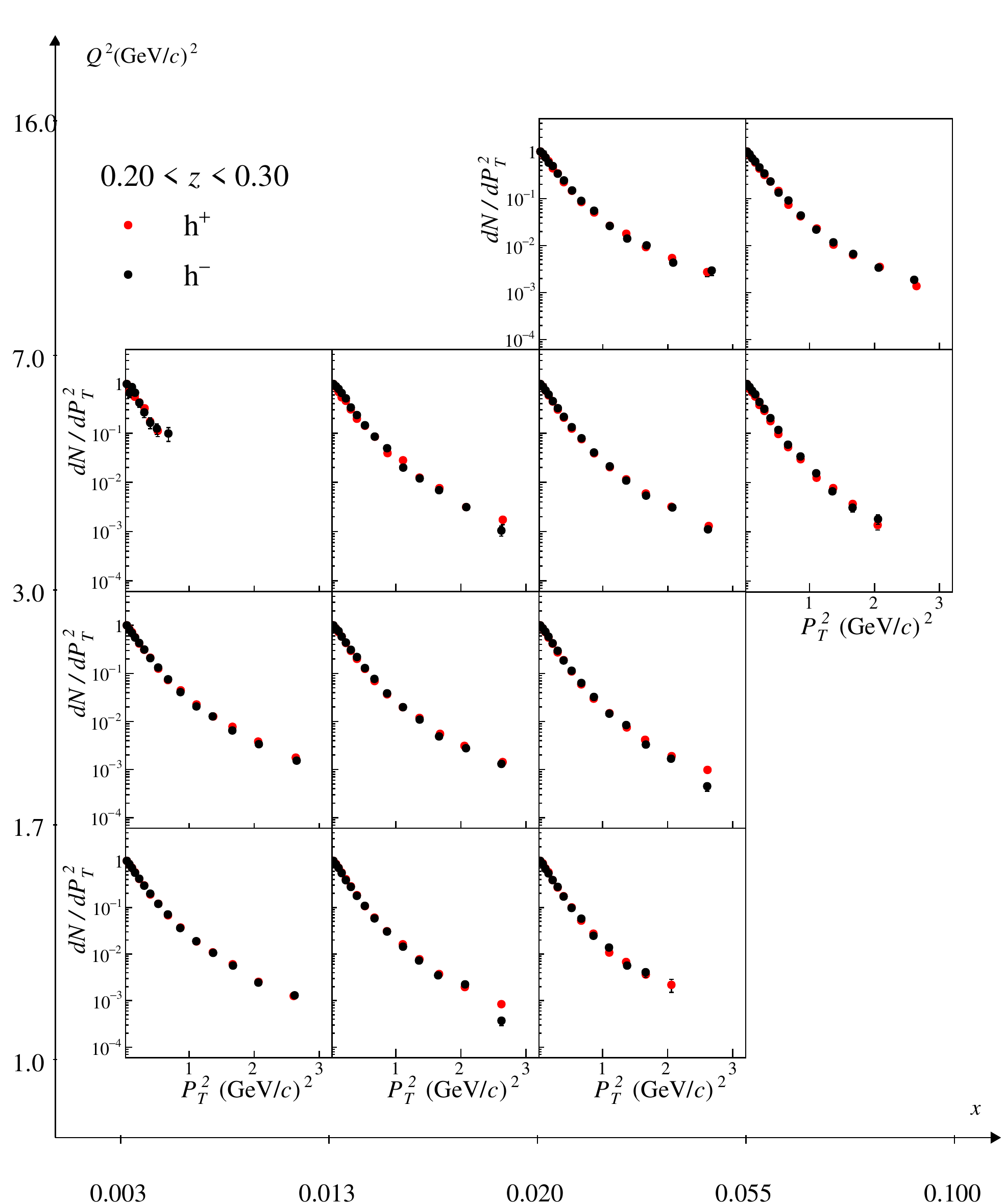}
    \includegraphics[width=0.49\textwidth]{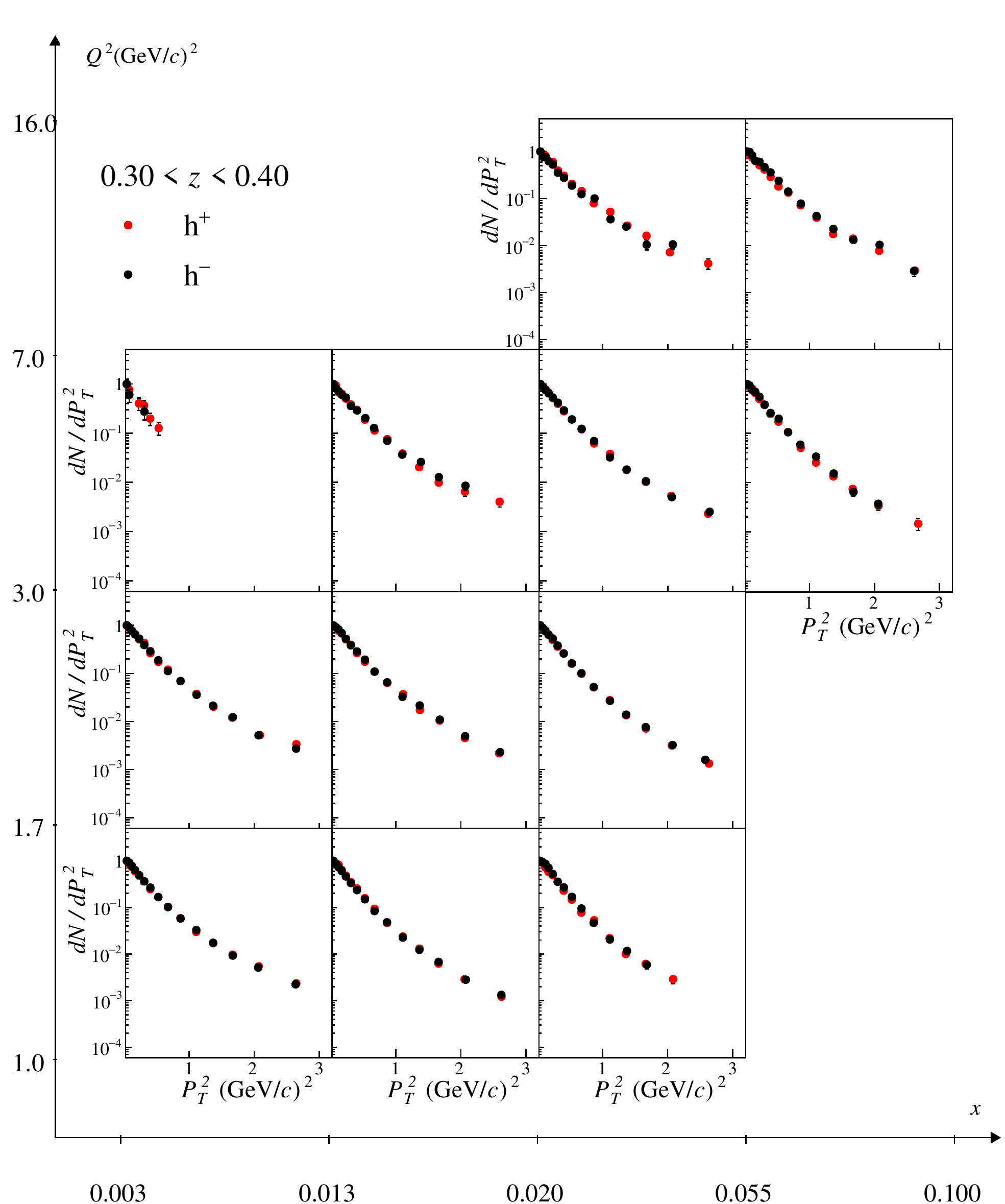}
    \includegraphics[width=0.49\textwidth]{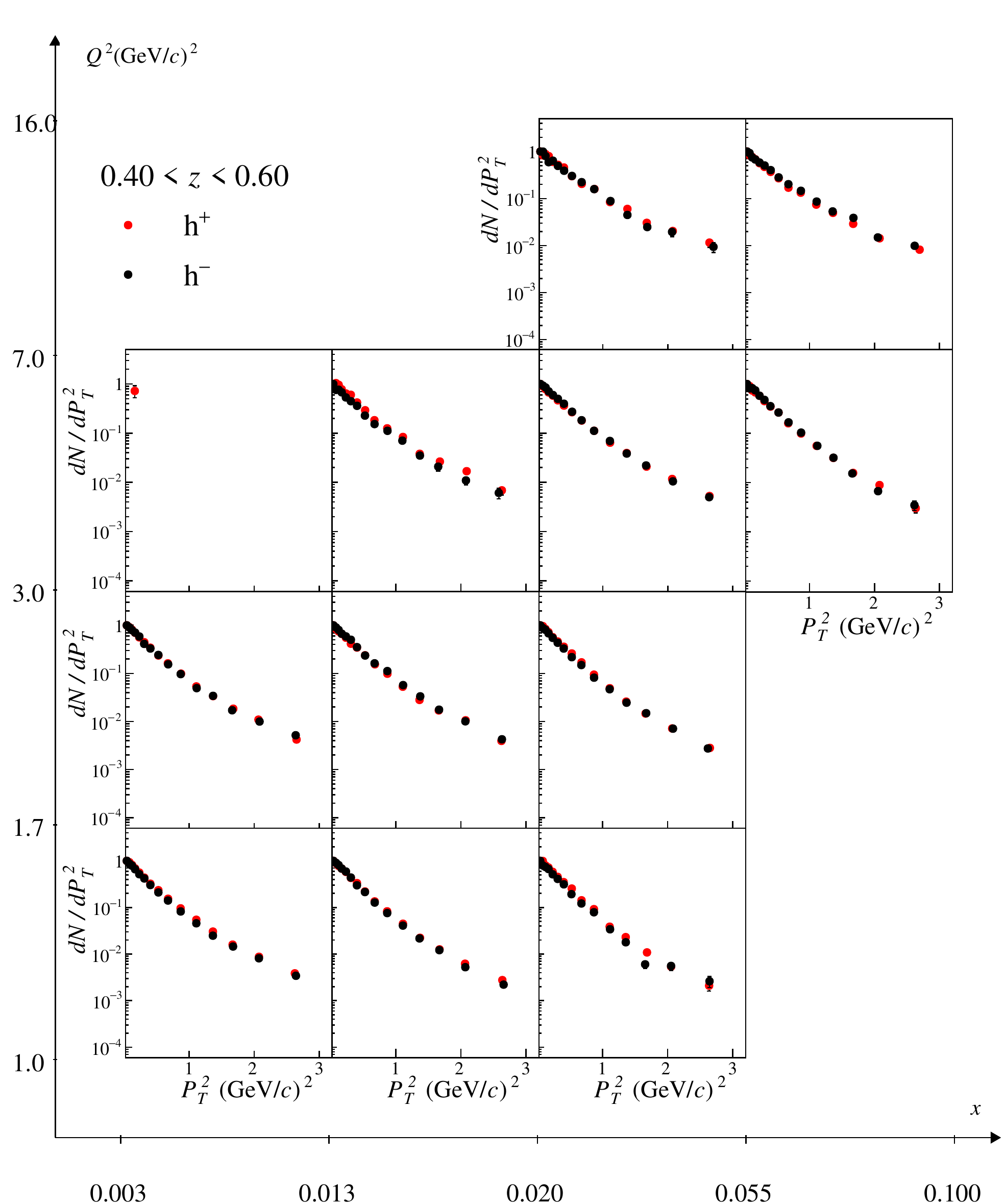}
    \includegraphics[width=0.49\textwidth]{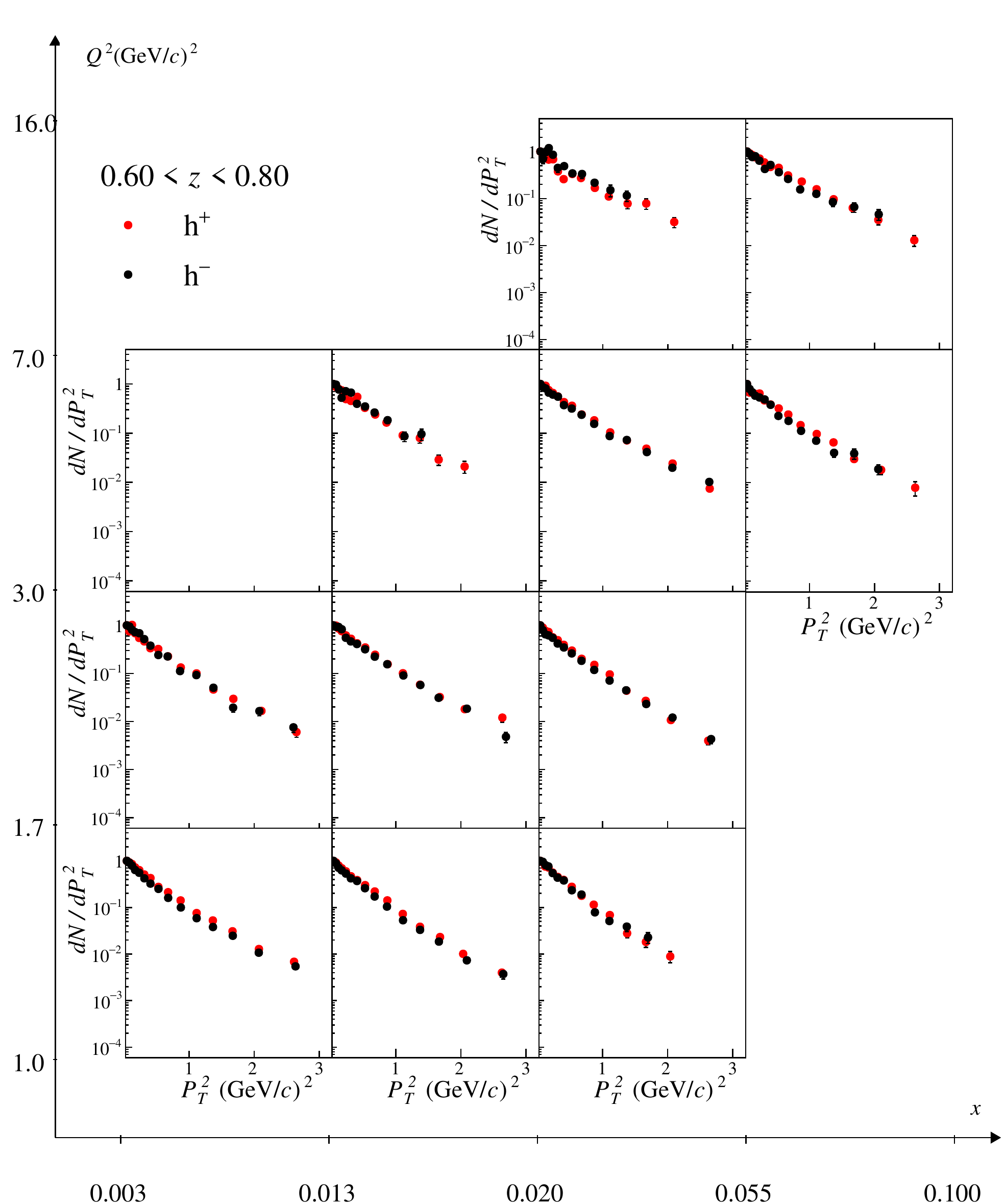}
    \caption{$\Ptsq$-distributions for positive (red) and negative hadrons (black) in bins of $x$ (horizontal axis) and $Q^2$ (vertical axis), in the four $z$ bins.}
    \label{fig:pt_1W_4Q2_4z}
\end{figure}

\begin{figure}[h!]
\captionsetup{width=\textwidth}
    \centering
    \includegraphics[width=0.65\textwidth]{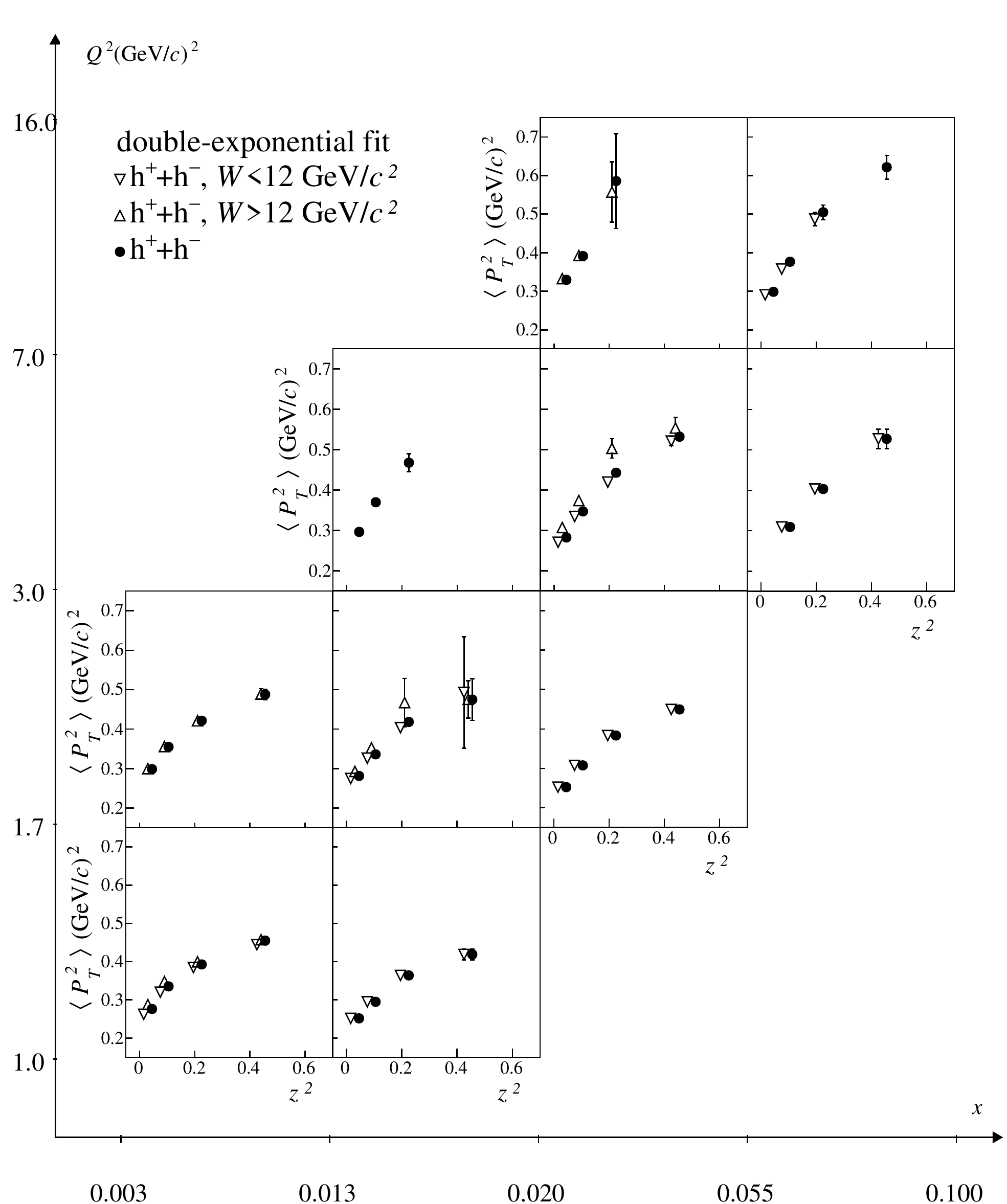}
    \caption{$\aPtsq$ for the sum of positive and negative hadrons shown as a function of $z^2$, in bins of $x$, $\Qsq$ (closed markers) and in two bins of $W$ (open markers).}
    \label{fig:pt_2W_4Q2_4z_1}
\end{figure}

A deeper study of the $Q^2$ dependence of the $\Ptsq$-distributions can be performed by increasing the number of bins. In particular, the two bins in $Q^2$ have been split, so to have four bins in the same $Q^2$ range. The measured distributions in the new ($x$,$\Qsq$) bins are shown in Fig.~\ref{fig:pt_1W_4Q2_4z}: as before, the trend are smooth, the $x$-dependence is weak while the dependences on $\Qsq$ and $z$ are well visible. After summing over positive and negative hadrons, these distributions have been fitted with the usual double-exponential function, from which the corresponding value of $\aPtsq$ has been derived. The dependence of the values of $\aPtsq$ on $z^2$ is shown in  Fig.~\ref{fig:pt_2W_4Q2_4z_1} in all bins where the fit could converge. The same analysis has been performed also splitting the $W$ range in two, obtaining similar indications as for the $x$, $\Qsq$ and $z$ dependences of the $\Ptsq$-distributions. In Fig.~\ref{fig:pt_2W_4Q2_4z_1}, the $\aPtsq$ values obtained from the fits at high and low $W$ are also shown; clearly, because of the limitations in statistics and of the kinematic constraints, the extraction of $\aPtsq$ could not be done in all bins in $x$, $\Qsq$ and $z$. Where the comparison of the low and high $W$ results can be made (along the diagonal in the $x$, $\Qsq$ grid) the $\aPtsq$ results for high $W$ are found slightly larger than the ones at low $W$.

\subsection{Dependence of $\aktsq$ upon $\Qsq$}
\label{sect:ch4_kT2_Q2}
Using the measured values of $\aPtsq$ in bins of $z$ and $\Qsq$, it is possible to derive interesting information on the dependence of $\aktsq$ on $\Qsq$. As will be clear in the following, this requires fixing the $\Qsq$-dependence of $\aPtsq$ in the data. The Leading Order expression for $\aPtsq$ reads:

\begin{equation}
    \aPtsq(x,\Qsq,z) = z^2\aktsq(x,\Qsq) + \apperpsq(z,\Qsq).
\end{equation}
where the kinematic dependences have been written in an explicit way. In particular, $\aktsq$ in general depends on $x$ and $\Qsq$, while $\apperpsq$ can depend on $z$ and $\Qsq$. The dependence of $\aPtsq$ on $x$ is however observed to be small in the data, and it will be neglected in the following. The difference $\Delta_{10}$ of $\aPtsq$, taken at two values of $z$ ($z_0$ and $z_1$), reads:

\begin{equation}
\begin{split}
    \Delta_{10}(\Qsq) & = \aPtsq(\Qsq,z_1) -  \aPtsq(\Qsq,z_0)\\ 
    & = (z_1^2 - z_0^2) \aktsq(\Qsq) + \apperpsq(z_1,\Qsq) - \apperpsq(z_0,\Qsq) 
\end{split}
\end{equation}
Assuming that the dependence of $\apperpsq$ on $\Qsq$ does not change with $z$, one has that:

\begin{equation}
\begin{split}
    \frac{\diff \Delta_{10}(\Qsq)}{\diff \Qsq} & = (z_1^2 - z_0^2) \frac{\diff \aktsq(\Qsq)}{\diff \Qsq} + \cancel{\frac{\diff \apperpsq(z_1,\Qsq)}{\diff \Qsq}} - \cancel{\frac{\diff \apperpsq(z_0,\Qsq)}{\diff \Qsq}} \\
    & = (z_1^2 - z_0^2) \frac{\diff \aktsq(\Qsq)}{\diff \Qsq},
\end{split}
\end{equation}
which means that $\frac{\diff \aktsq(\Qsq)}{\diff \Qsq}$ can be obtained once $\frac{\diff \Delta_{10}(\Qsq)}{\diff \Qsq}$ is known. Two hypotheses have been tested on the data, about the behavior of $\aPtsq$:

\begin{itemize}
    \item \textit{logarithmic hypothesis}. In this case, the average hadron transverse momentum is modeled as $\aPtsq(z,\Qsq) = a(z) + b(z)\log(\Qsq)$, which implies that:        
    \begin{equation}
        \frac{\diff \Delta_{10}(\Qsq)}{\diff \Qsq}  = \frac{b(z_1) - b(z_0)}{\Qsq} = \frac{b_1 - b_0}{\Qsq} \Longrightarrow \frac{\diff \aktsq(\Qsq)}{\diff \Qsq} = \frac{b_1 - b_0}{(z_1^2 - z_0^2)\Qsq}.
    \end{equation}

\item \textit{linear hypothesis}. This time, $\aPtsq(z,\Qsq)$ is written as $ \aPtsq(z,\Qsq) = c(z) + d(z)\Qsq$. As a consequence, 
    \begin{equation}
        \frac{\diff \Delta_{10}(\Qsq)}{\diff \Qsq}  = d(z_1) - d(z_0) = d_1 - d_0 \Longrightarrow \frac{\diff \aktsq(\Qsq)}{\diff \Qsq} = \frac{d_1 - d_0}{z_1^2 - z_0^2}.
    \end{equation}
\end{itemize}

\begin{figure}
\captionsetup{width=\textwidth}
    \centering
    \includegraphics[width=0.75\textwidth]{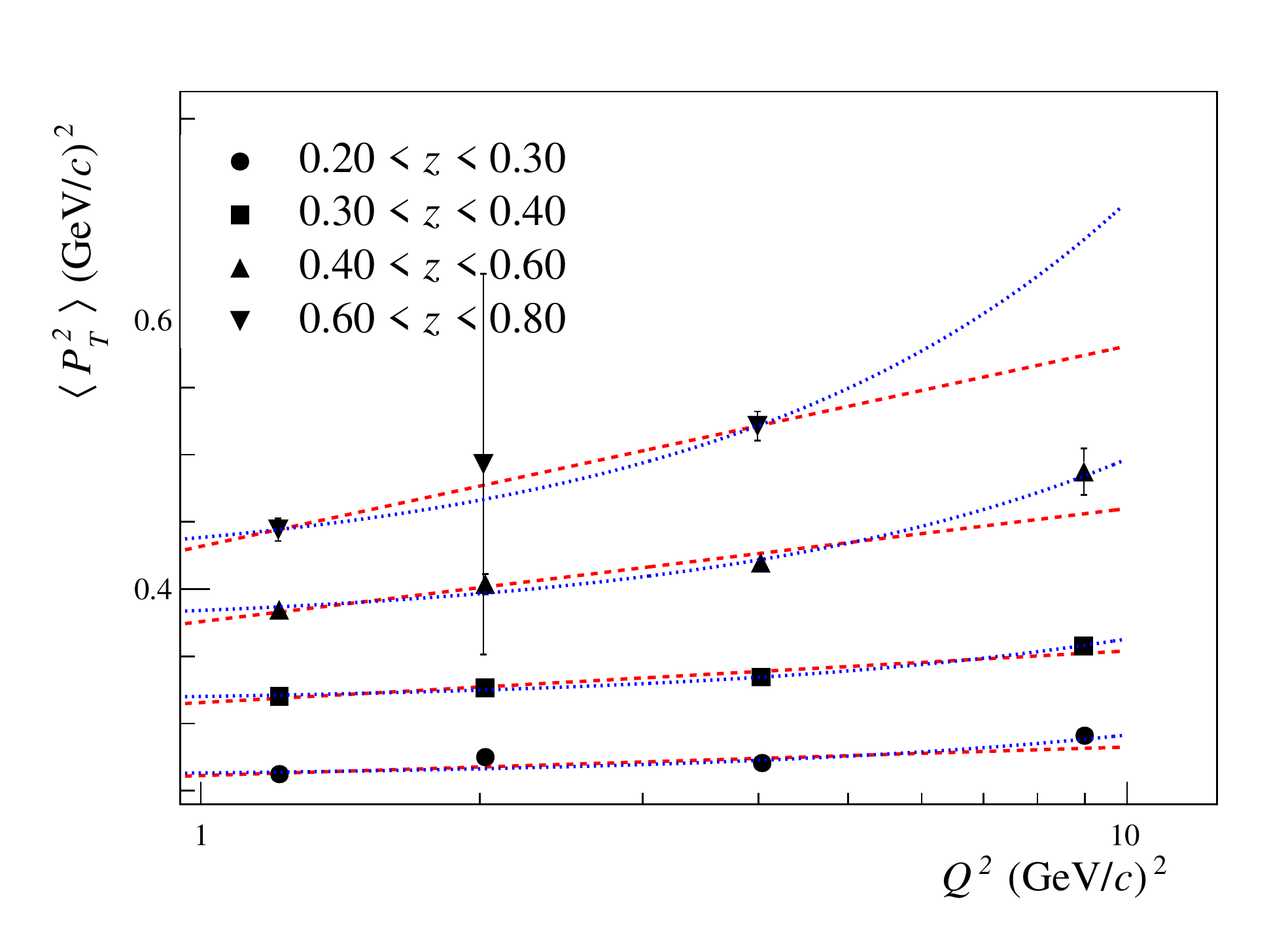}
    \caption{$\aPtsq$ for the sum of positive and negative hadrons, shown as a function of $\Qsq$ in four bins of $z$ and fitted with a logarithmic (red) and linear function (blue).} 
    \label{fig:kt2_evol}
\end{figure}

Figure~\ref{fig:kt2_evol} shows the fit of the measured values of $\aPtsq$ at low $W$ as a function of $\Qsq$, in four $z$ bins, and with the two said hypotheses. At low $z$, both the linear and the logarithmic options describe well the data, while at higher $z$ a better description seems to be offered by the linear options. Within the limited $\Qsq$ range and given the scarce number of points, both options look valid. The results of the fit are presented in Tab.~\ref{tab:dkt2dQ2}: considering for example the first and second bins in $z$, one has that:

\begin{equation}
    \frac{\diff \aktsq(\Qsq)}{\diff \Qsq} = \frac{0.13 \pm 0.04}{\Qsq}
\end{equation}
in the logarithmic case, and
\begin{equation}
    \frac{\diff \aktsq(\Qsq)}{\diff \Qsq} = 0.03 \pm 0.01
\end{equation}
in the linear case. The results are similar, and compatible, if other $z$ bin pairs are considered instead of this, only in the logarithmic case: in the linear case, the derivative would be higher considering bins at higher $z$. \\

In literature (e.g. in Ref.~\cite{Anselmino:2012aa}), the $\Qsq$-dependence of $\aktsq$ is often modeled with a logarithmic function of the form:
\begin{equation}
    \aktsq(\Qsq) = \aktsq(\Qsq_0) + g_2 \log \op \frac{\Qsq}{\Qsq_0}\cp 
\end{equation}
where $g_2$ is a constant to be evaluated from the data. A wide range of values for $g_2$ has been proposed in the recent years. For example, the authors of Ref.~\cite{Anselmino:2012aa} indicate the value $g_2=0.68$; a different study \cite{Aidala:2014hva} proposed the range $0 \leq g_2\leq0.03$; a recent phenomenological work on the extraction of the Boer-Mulders function \cite{Christova:2020ahe} suggests $g_2^{max} = 0.35$. The fit of the $\aPtsq$ values as a function of $\Qsq$, done assuming a logarithmic trend, returns $g_2=0.13\pm0.04$, thus in between the lowest and the highest values suggested so far.

\begin{table}
\captionsetup{width=\textwidth}
\centering
\begin{tabular}{cccc}
  \hline
          $z$ bin & $z^2$   & logarithmic fit   & linear fit \\ 
  \hline
    $0.20<z<0.30$  & 0.06 & $b=0.009 \pm 0.002$ & $d=0.003 \pm 0.001$ \\
    $0.30<z<0.40$  & 0.12 & $b=0.017 \pm 0.003$ & $d=0.005 \pm 0.001$ \\
    $0.40<z<0.60$  & 0.24 & $b=0.037 \pm 0.006$ & $d=0.013 \pm 0.002$ \\
    $0.60<z<0.80$  & 0.47 & $b=0.065 \pm 0.012$ & $d=0.028 \pm 0.005$ \\  \hline
\end{tabular}
\caption{Results of the logarithmic and linear fits of the $\aPtsq$ values as a function of $\Qsq$ in four $z$ bins, at low $W$.} 
\label{tab:dkt2dQ2}
\end{table}

\clearpage 
\newpage

\chapter{Measurement of the azimuthal~asymmetries} 

\label{Chapter5_Azimuthal_asymmetries}

The $A_{UU}^{\cos\fih}$, $A_{UU}^{\cos2\fih}$ and $A_{LU}^{\sin\fih}$ azimuthal asymmetries, defined in Sect.~\ref{sect:ch1_aa_definition}, have been measured using the same event- and hadron samples analyzed for the extraction of the $\Ptsq$-distributions (Ch.~\ref{Chapter4_PT-distributions}). The procedure is also similar. The first step is the definition of the kinematic range and of the binning. In the first place, both a one-dimensional (1D) and a three-dimensional (3D) analyses have been performed, meaning with an alternative or simultaneous binning in $x$, $z$ and $\Pt$. As for the $\Ptsq$-distributions, the \textit{standard} binning has been chosen the same of the published COMPASS deuteron results \cite{COMPASS:2014kcy}. In addition, a binning in $\Qsq$ and on $W$ has been introduced in order to better characterize the kinematic dependences of the measured asymmetries. In each kinematic bin, and separately for positive and negative beams and hadrons, the distribution of the azimuthal angle of the hadron $\fih$ has first been corrected for the residual exclusive background and then for acceptance. Finally, the amplitudes of the three expected modulations have been extracted with a fit and corrected for the corresponding kinematic terms $\eps_i(y)$. \\

In the following Sections (Sect.~\ref{sect:ch5_binning},~\ref{sect:ch5_exclusive},~\ref{sect:ch5_acceptance},~\ref{sect:ch5_fitting}), the specific steps of the analysis are presented in detail. The systematic uncertainties are discussed in Sect.~\ref{sect:ch5_systematics}, while the final results are presented in Sect.~\ref{sect:ch5_results}, where they are also compared with the deuteron measurement. A further investigation about the kinematic dependences of the azimuthal asymmetries is presented in Sect.~\ref{sect:ch5_kinematics}. The final part of this Chapter (\ref{sect:ch5_interpretation}) hosts a possible interpretation of the results, with the definition of an effective transverse momentum $\aktsq_{eff}$ through which the $A_{UU}^{\cos\fih}$ can be described and with the application of the method of the difference asymmetries to $A_{UU}^{\cos2\fih}$.\\

As was the case for the $\Ptsq$-distributions, most of the material presented in this Chapter has been released by the COMPASS Collaboration and shown at international conferences \cite{Matousek:2019dlk,Moretti:2020uee,Moretti:CPHI,Moretti:ICNFP,Matousek:IWHSS,Moretti:DIS}. The data are available upon request.

\section{The \textit{standard} binning}
\label{sect:ch5_binning}
The three azimuthal asymmetries have been measured in bins of $x$, $z$ and $\Pt$. 
In the 1D approach, it is well known that if the acceptance is not flat, some kinematic regions can weigh more than others and the resulting (integrated) asymmetries could be distorted. On the other hand, in the measurements described here the statistics (as for both the real data and Monte Carlo samples) is limited, and in the 1D case it is possible to better investigate the dependence on one of the variables and to have a better overview than in the 3D case. Moreover, as will be shown in the dedicated Section, the acceptance in the selected kinematic region is almost flat.

\begin{figure}[h!]
\captionsetup{width=\textwidth}
\begin{center}
	\includegraphics[width=0.49\textwidth]{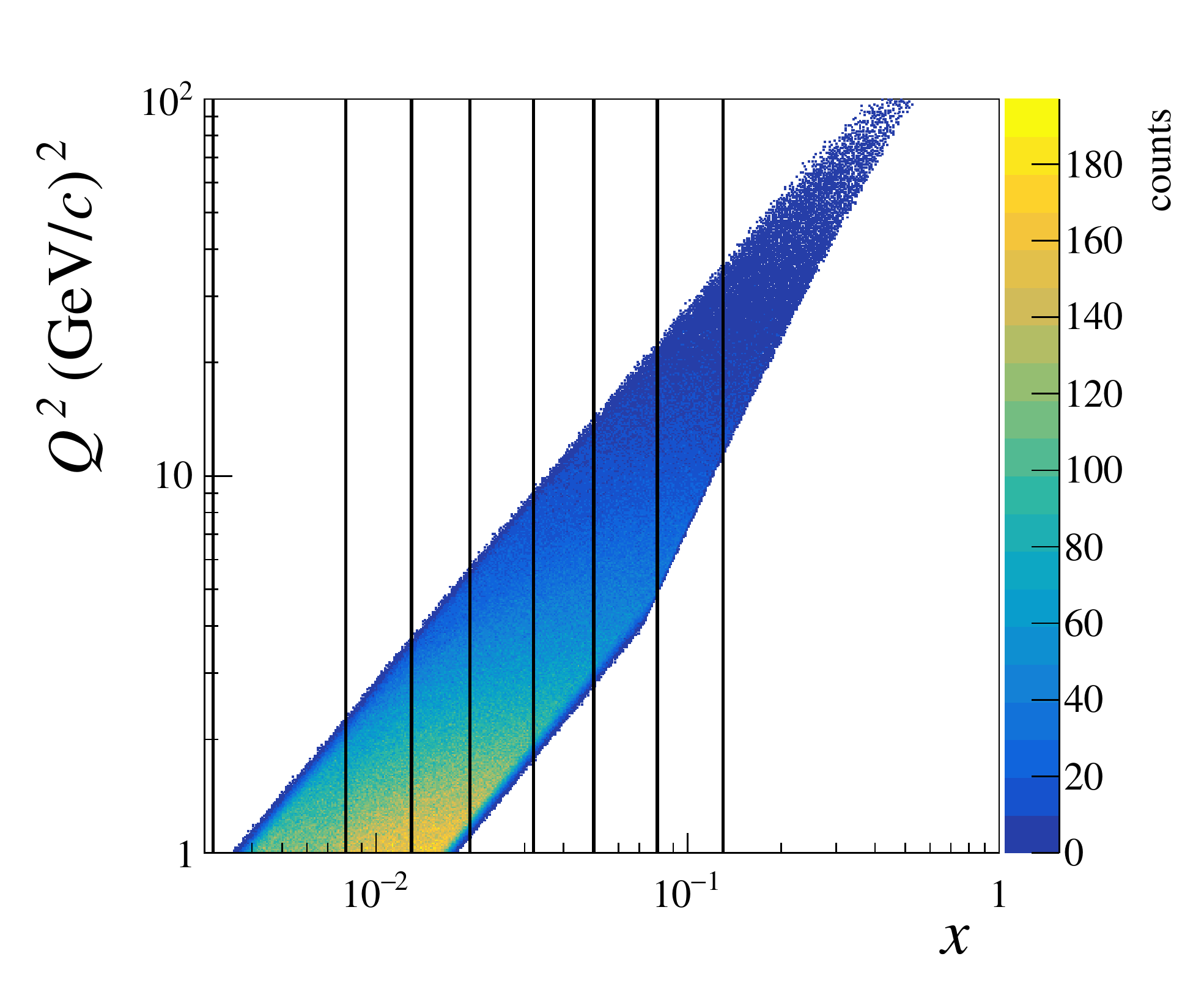}
	\includegraphics[width=0.49\textwidth]{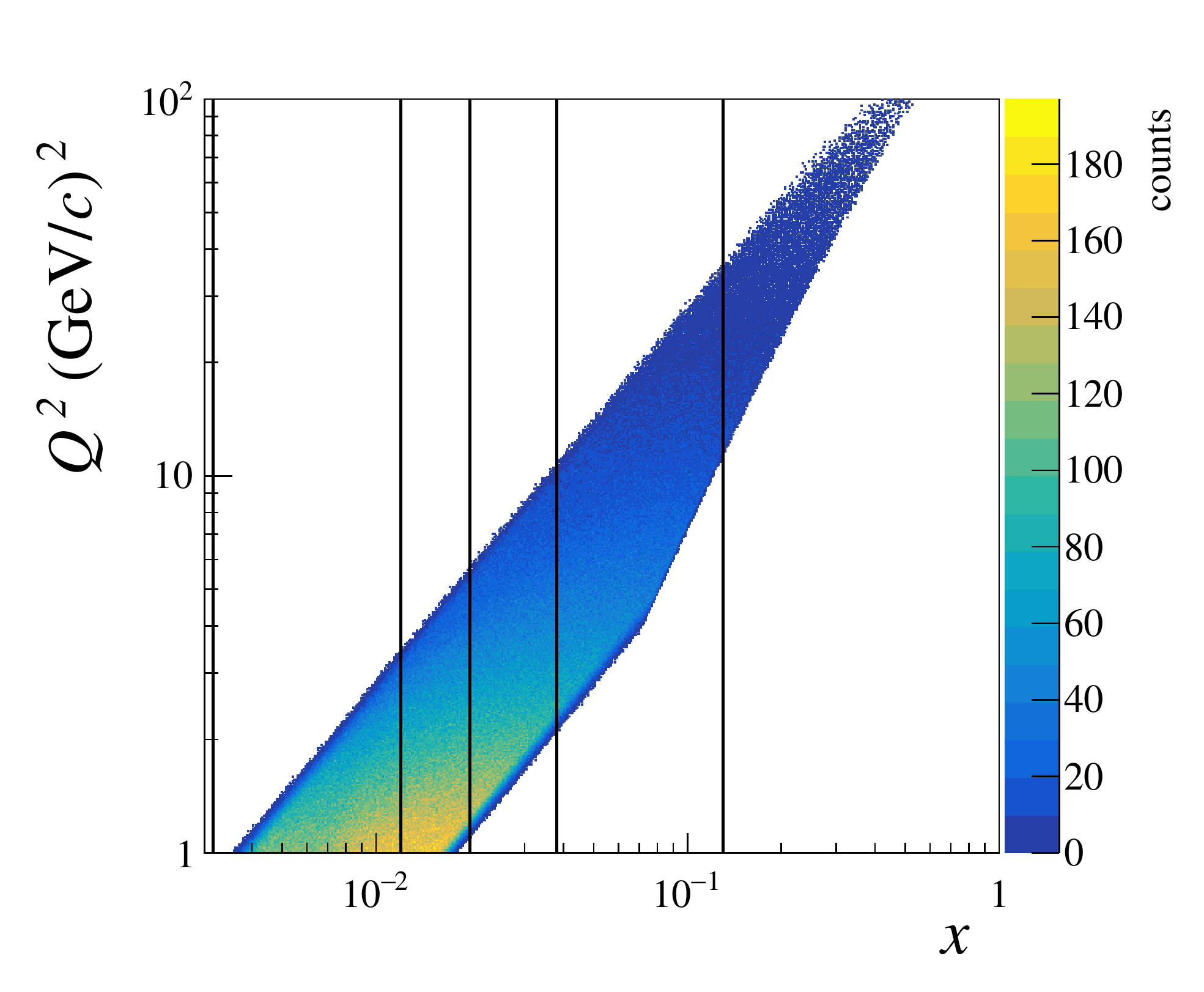}
		\caption{Drawn on top of the $x-\Qsq$ correlation plot are: the standard binning used for the measurement of the azimuthal asymmetries in 1D (left) and in 3D (right).}
\label{fig:xQ2_aa}
\end{center}
\end{figure}

For the 1D analysis, the most natural choice for the binning has been to keep it as in the previous COMPASS analysis on deuteron \cite{COMPASS:2014kcy} (the \textit{standard} binning), corresponding to the conditions: $0.2<y<0.9$, $\Qsq>1$~(GeV/$c$)$^2$, $0.003<x<0.130$. The $x$ range has been divided into 7 bins, as shown in Fig.~\ref{fig:xQ2_aa} (left); the $z$ range ($0.2<z<0.85$) into 8 bins; the $\Pt$ range ($0.1<\Pt~\mathrm{/ (GeV/}c\mathrm{)} <1.0$) into 9 bins. The bin limits were chosen in order to ensure similar statistics in each bin. In addition, two other regions have been studied: the region of small $z$ ($0.1<z<0.2$), where the azimuthal asymmetries have been investigated as a function of $x$ and $\Pt$ using the same binning as in the standard case, and the high $\Pt$ region ($1.00<\Pt~\mathrm{/ (GeV/}c\mathrm{)} <1.73$), where the asymmetries have been studied as a function of $x$ and $z$. These two regions, shown for completeness, show strong kinematic dependences of difficult interpretation.

Also for the 3D analysis, characterized by the same range in $x$, $\Qsq$ and $y$, the binning has been kept the same as in Ref.~\cite{COMPASS:2014kcy}, shown in Fig.~\ref{fig:xQ2_aa} (right). For each of the four bins in $x$, the data have been analyzed in six bins in $z$ ($0.2<z<0.85$) and four bins in $\Pt$ ($0.1<\Pt~\mathrm{/ (GeV/}c\mathrm{)} <1.0$), for a total of 96 bins. In addition, the low $z$ region  ($0.1<z<0.2$) and the high $\Pt$ region ($1.00<\Pt~\mathrm{/ (GeV/}c\mathrm{)}) <1.73$) have been included for completeness.

\section{Exclusive hadron contribution}
\label{sect:ch5_exclusive}
As explained in Ch.~\ref{Chapter3_Data_analysis}, the exclusive vector meson production events observed in the data, in which both exclusive hadrons are reconstructed, have been discarded, and those hadrons not included in the samples used in analysis. On the other hand, the same exclusive events have been used to normalize the HEPGEN Monte Carlo to allow for the subtraction of the hadrons produced in exclusive events not fully reconstructed. In the case of the azimuthal asymmetries, the distribution of the azimuthal angle $\fih$ has been corrected for the non-visible exclusive contribution, by subtracting the corresponding distribution from the normalized Monte Carlo.
This procedure is new with respect to the approach used, e.g., in Ref.~\cite{COMPASS:2019lcm}. There, the correction for the exclusive background to the azimuthal asymmetries was applied at the amplitude level, after the acceptance correction, according to the expression: 

\begin{equation}
    A_{corr} = \frac{A_{tot} - f_{bg}A_{excl}}{1-f_{bg}},
\label{eq:a_corr}
\end{equation}
where the corrected asymmetry $A_{corr}$ was estimated from the total (SIDIS+exclusive) asymmetry $A_{tot}$ by subtracting the exclusive asymmetries $A_{excl}$, after the acceptance correction. Given the time past before realizing that the exclusive vector meson production could have an important impact on the azimuthal asymmetries, this was the only possible choice. Now, with a new analysis ongoing, the new procedure described above could be applied, with the advantage of being less Monte Carlo-dependent. Also, the new method does not rely on the ratio of the cross-sections.  \\

\begin{figure}
\captionsetup{width=\textwidth}
    \centering
	\includegraphics[width=0.8\textwidth]{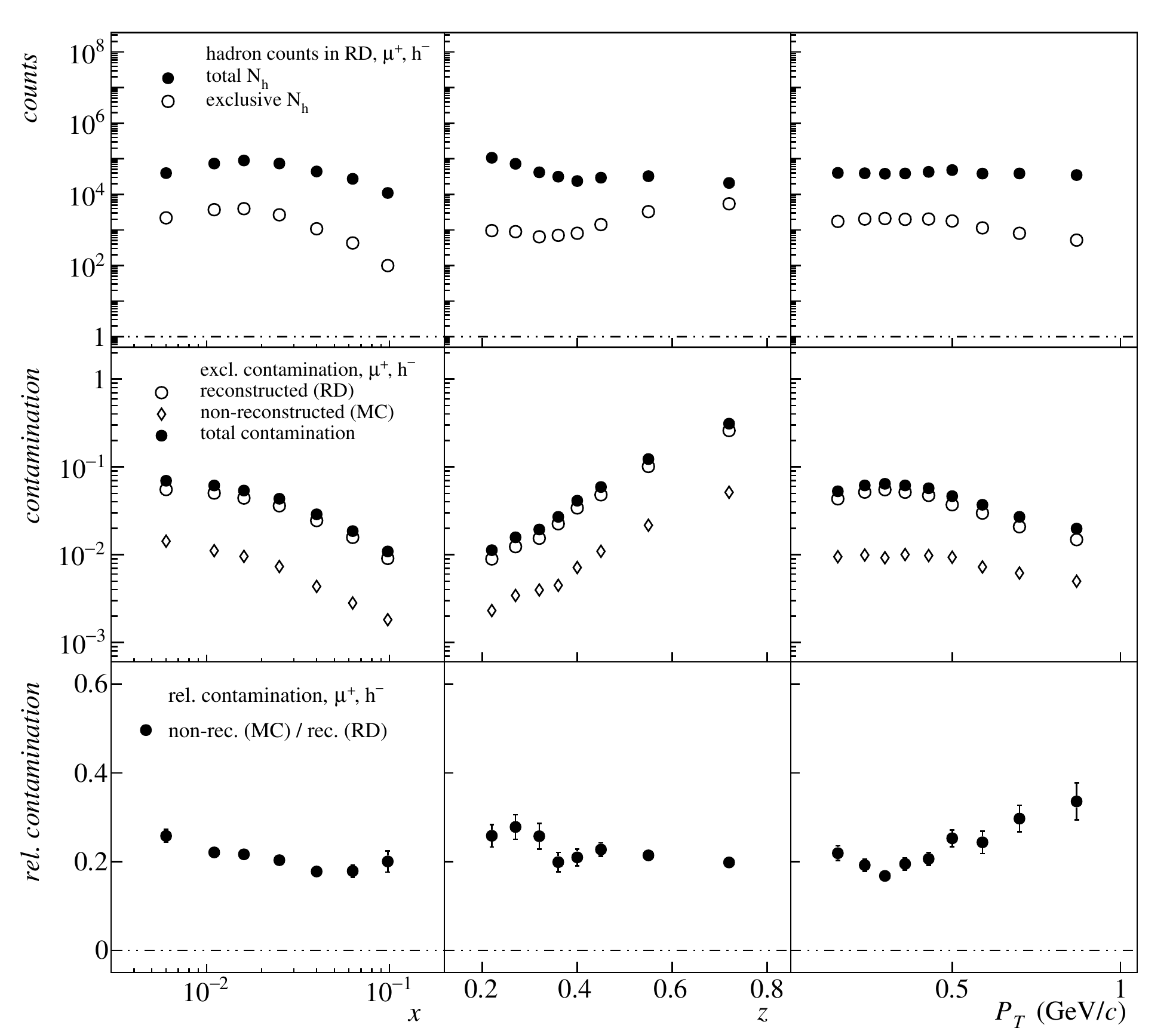}
	\caption{Total number of hadrons in bins of $x$, $z$ and $\Pt$ compared to the reconstructed exclusive component (top row); exclusive hadron contamination and its contribution from data and Monte Carlo (middle row) and ratio between non-reconstructed and reconstructed components (bottom row).}
    \label{fig:counts}
\end{figure}

The relevance of the exclusive hadron contamination is shown in Fig.~\ref{fig:counts} where for the standard 1D case, $\mu^+$ beam and negative hadrons there are:
\begin{itemize}
    \item in the top row: the total number of hadrons (full points) and the number of hadrons from a reconstructed exclusive event (open points) in the real data, not normalized to the bin width. As a function of $x$, the number of hadrons from exclusive events is from one to two orders of magnitudes less than the total number of hadrons. As a function of $z$, it becomes comparable with the total number of hadrons at large $z$. The dependence of the number of exclusive hadrons on $\Pt$ is small.
    \item in the middle row: the percentage of all exclusive hadrons (full points), the percentage of hadron produced in fully reconstructed exclusive events in the data (open circles) and the percentage of hadrons produced in not-reconstructed exclusive events from the normalized Monte Carlo (open diamonds). All the percentages have been calculated on the SIDIS hadron sample. The open circles are the ratios of the numbers of hadrons shown in the first row and the closed points are the sum of the open ones. The total contamination (full points), smaller than 10\% and decreasing as a function of $x$, has the opposite trend when looked as a function of $z$ and can be as high as 40\% at high $z$. Again, the $\Pt$ dependence is weaker.
    \item in the bottom row: the ratio of the non-reconstructed events (from Monte Carlo) over reconstructed (from the data). It ranges between 20 and 40\%, being almost flat as a function of $x$ and $z$ and increasing with $\Pt$.
\end{itemize}
Note that the percentage of exclusive hadrons which are subtracted from the real data are shown as the diamonds in the second row, and ranges from 0.2\% at high $x$ and low $z$ to 8\% at high $z$. This means that the new procedure for the exclusive hadron subtraction is much less model (Monte Carlo) dependent.
The trends are similar for positive hadrons, but with a smaller overall contamination, while they are the same for the $\mu^-$ samples.

\begin{figure}
\captionsetup{width=\textwidth}
    \centering
    \includegraphics[width=0.95\textwidth]{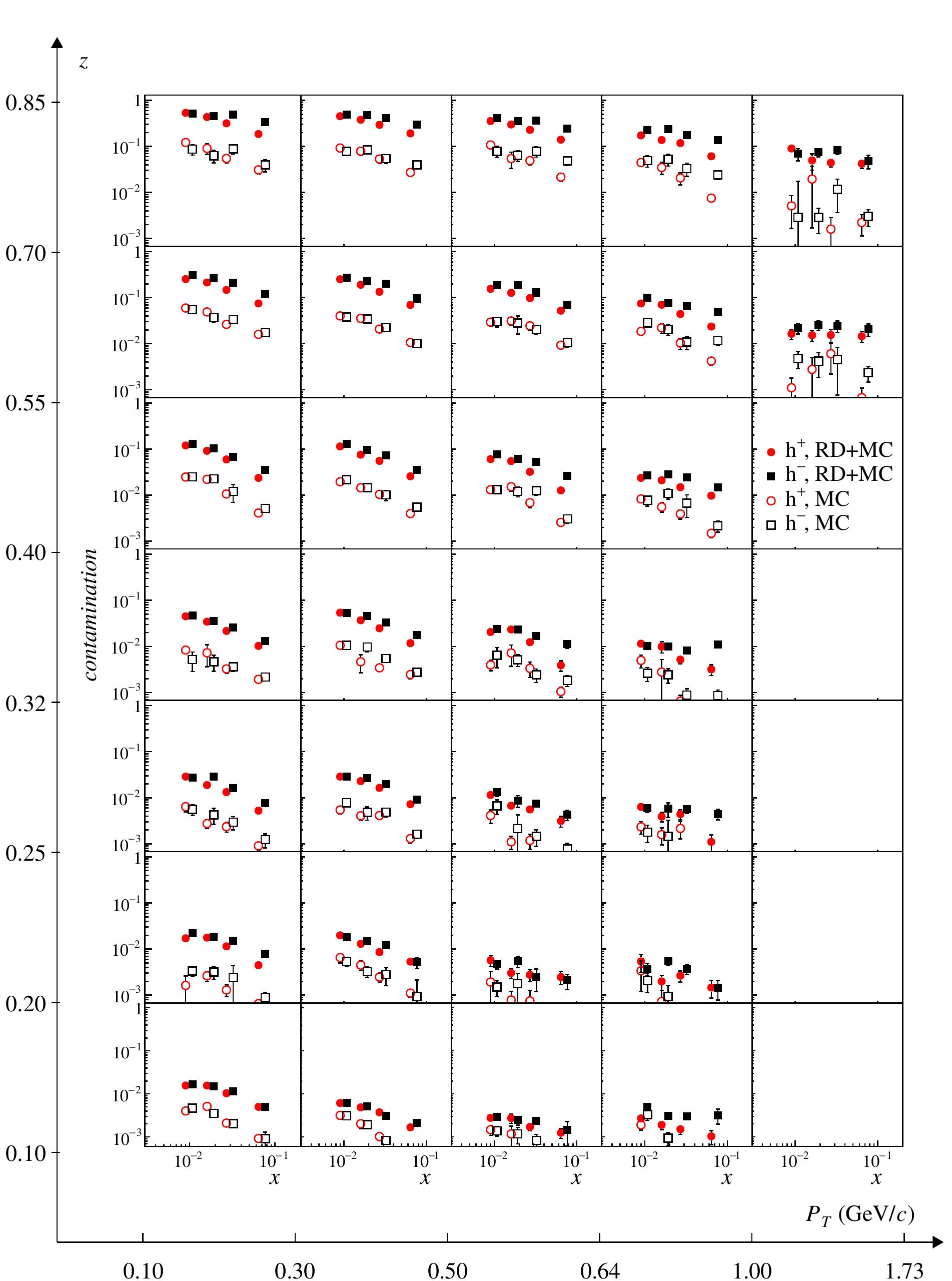}
   	\caption{Total exclusive hadron contamination (RD+MC, full markers) and contamination from exclusive hadrons not reconstructed in the data (MC, open markers).}
  \label{fig:counts_3D}
\end{figure}

As for the 3D analysis, the corresponding information is given in Fig.~\ref{fig:counts_3D}, where the total (real data and Monte Carlo) contamination for positive and negative hadrons (RD+MC, full markers) is given together with the contamination from the non-reconstructed exclusive events (MC, open markers). The last one, which is the relevant one, is of about 10\% at high $z$ dropping to a few per thousand at low $z$. \\

\newpage

\begin{figure}[ht]
\captionsetup{width=\textwidth}
\begin{center}
	\includegraphics[width=0.49\textwidth]{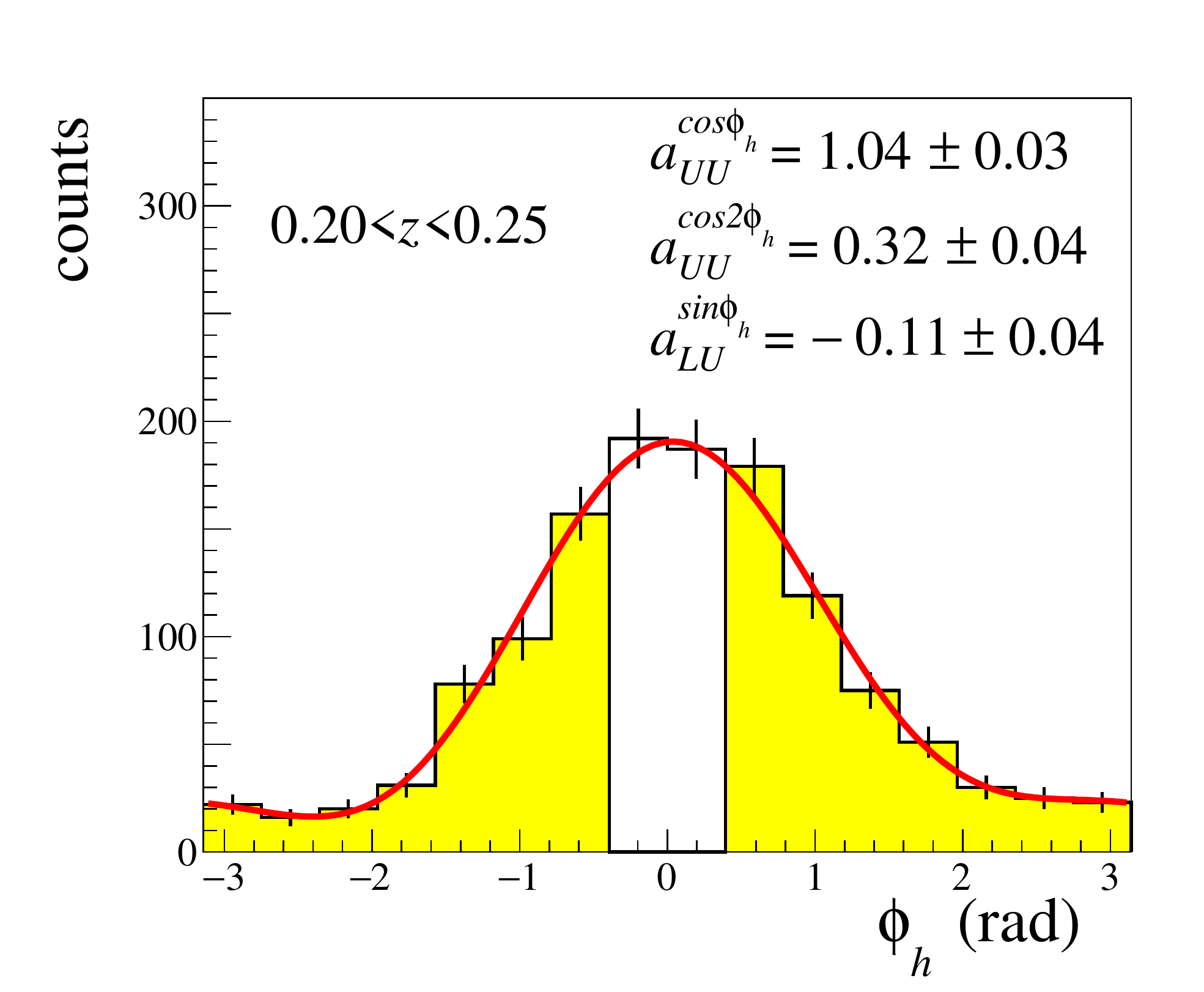} 
	\includegraphics[width=0.49\textwidth]{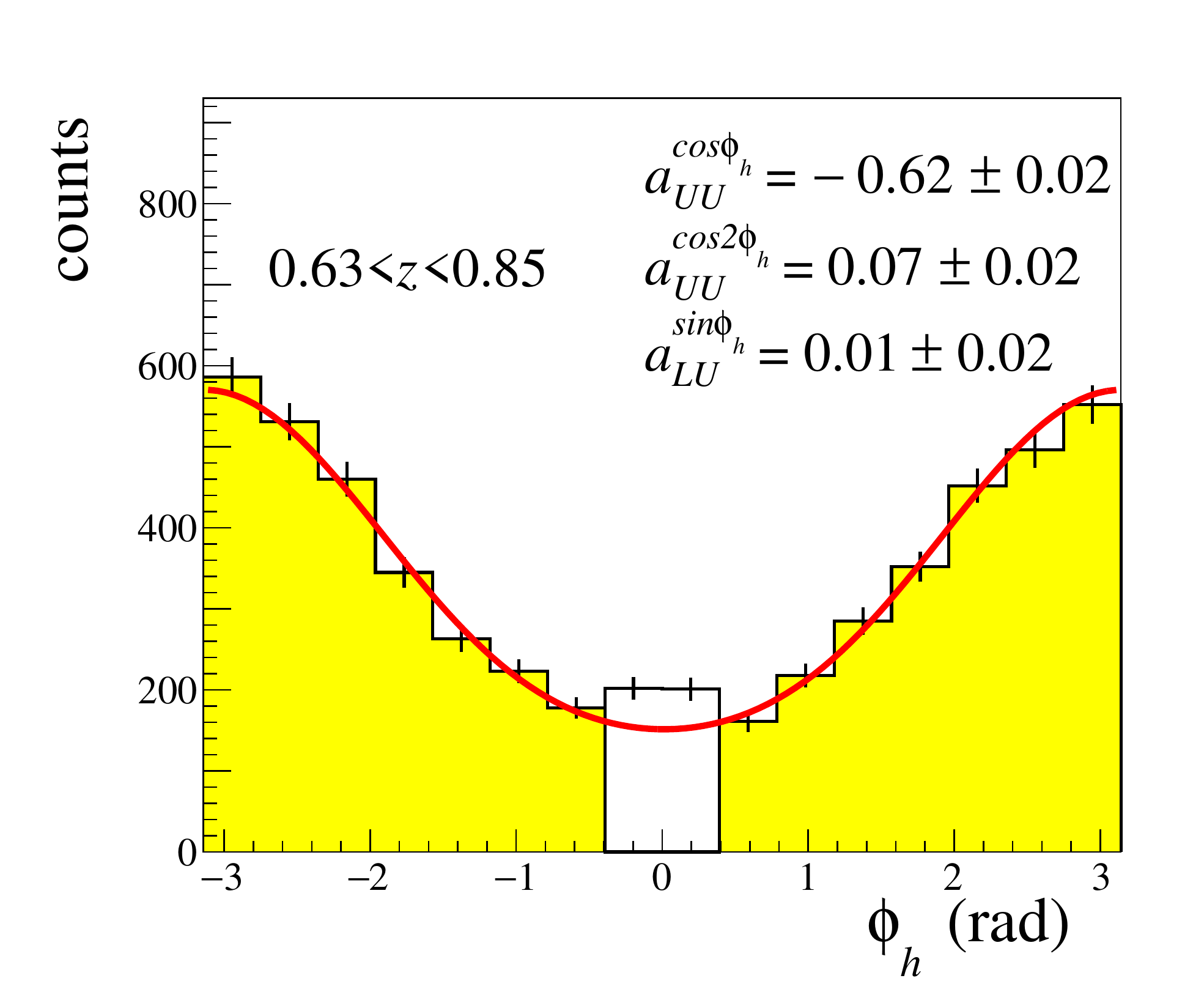} 
    \caption{Raw azimuthal asymmetries for positive exclusive hadrons in the data ($\mu^+$ data) in the first $z$ bin ($0.20<z<0.25$, left) and in the last $z$ bin ($0.63<z<0.85$, right). The two central bins are not considered in the fit.}
\label{fig:fits_example}
\end{center}
\end{figure}

\begin{figure}[ht]
\captionsetup{width=\textwidth}
\begin{center}
	\includegraphics[width=0.8\textwidth]{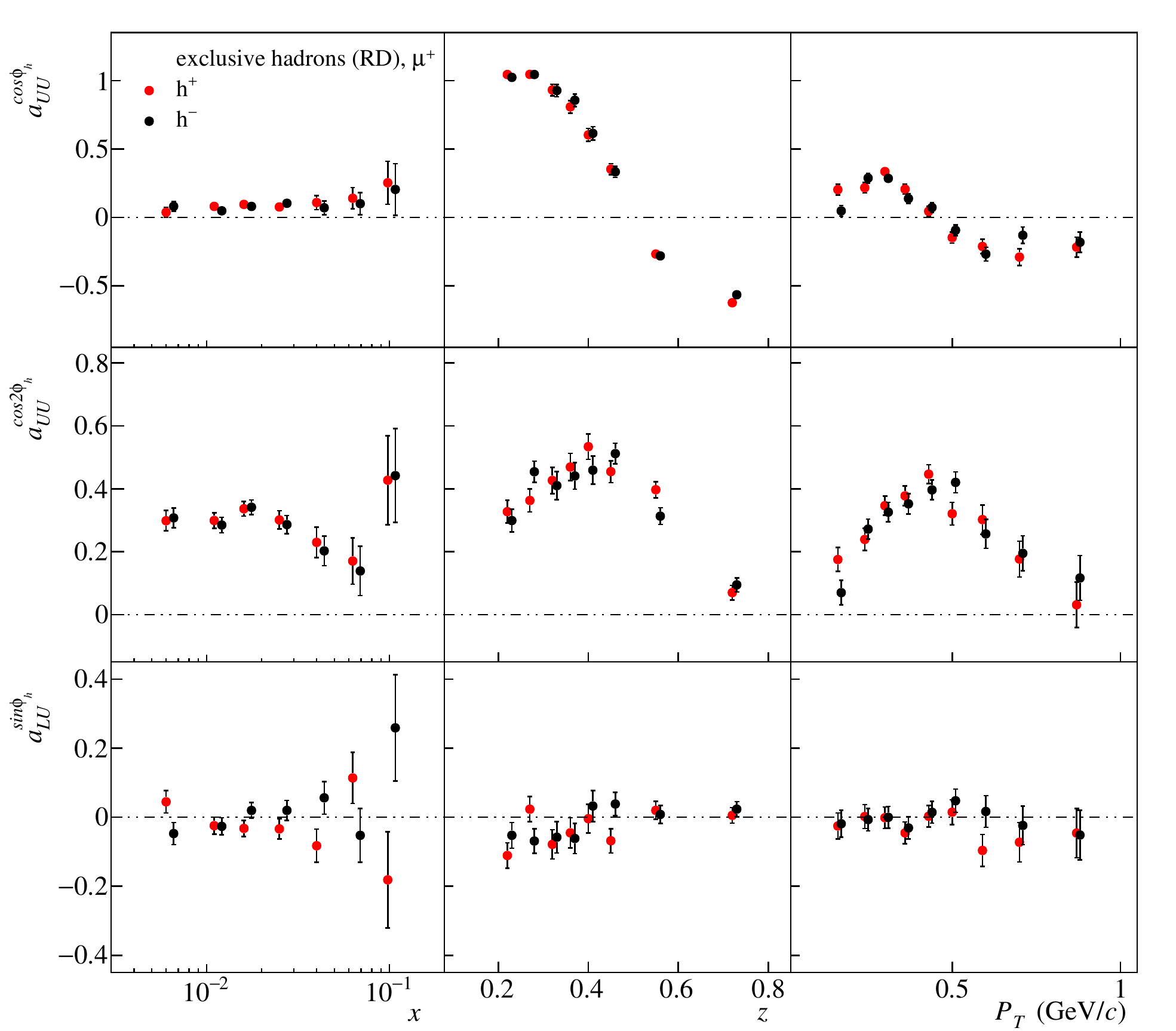}
    \caption{Amplitudes of the raw azimuthal modulations of the exclusive hadrons in the data, in $\cos\fih$ (top row), $\cos2\fih$ (middle row) and $\sin\fih$ (bottom row) for positive beam and positive (red) and negative hadrons (black) in the standard 1D kinematic range. The black points are slightly shifted for a better readability.}
\label{fig:multiplot_aa_excl}
\end{center}
\end{figure}

In order to understand the relevance of the exclusive hadrons rejection/subtraction, it is interesting to inspect some features of the exclusive hadrons in the real data and Monte Carlo samples. As an example, the azimuthal angle distributions of the positive exclusive hadrons in the data sample are given, in the first and last $z$ bins, in Fig.~\ref{fig:fits_example}. The distributions, not corrected for acceptance,  have been fitted with the function:

\begin{equation}
    f(\fih) = a_0 \op 1 + a_{UU}^{\cos\fih} \cos\fih + a_{UU}^{\cos2\fih} \cos2\fih + a_{LU}^{\sin\fih} \sin\fih \cp
\end{equation}
excluding the two central bins in $\fih$ (in white) where the contribution of electrons from photon radiation is generally relevant. As can be seen, the raw $\cos\fih$ modulations are large and with opposite sign in the two $z$ bins. More generally, Fig.~\ref{fig:multiplot_aa_excl} shows, for all the bins of the 1D analysis, the amplitudes of the raw azimuthal modulations of the exclusive hadrons reconstructed in the data (not corrected for acceptance nor for the  kinematic factors $\eps_i(y)$). The amplitudes of the $\cos\fih$ and $\cos2\fih$ modulations show impressive kinematic dependences. In particular, the $\cos\fih$ amplitude, almost flat in $x$ and even larger than one at small $z$, decreases crossing zero at $z\simeq 0.5$ reaching about the value of $-0.6$ at large $z$. The $\cos2\fih$ amplitude is always positive, with maximum value at $z \simeq 0.4$ and $\Pt \simeq$ 0.4~GeV/$c$. shows a peculiar trend in $x$ and a parabolic shape in both $z$ and $\Pt$. The $\sin\fih$ modulation is generally compatible with zero. Almost everywhere, no difference can be seen between positive and negative hadrons, as expected. 

\begin{figure}[ht]
\captionsetup{width=\textwidth}
\begin{center}
	\includegraphics[width=0.8\textwidth]{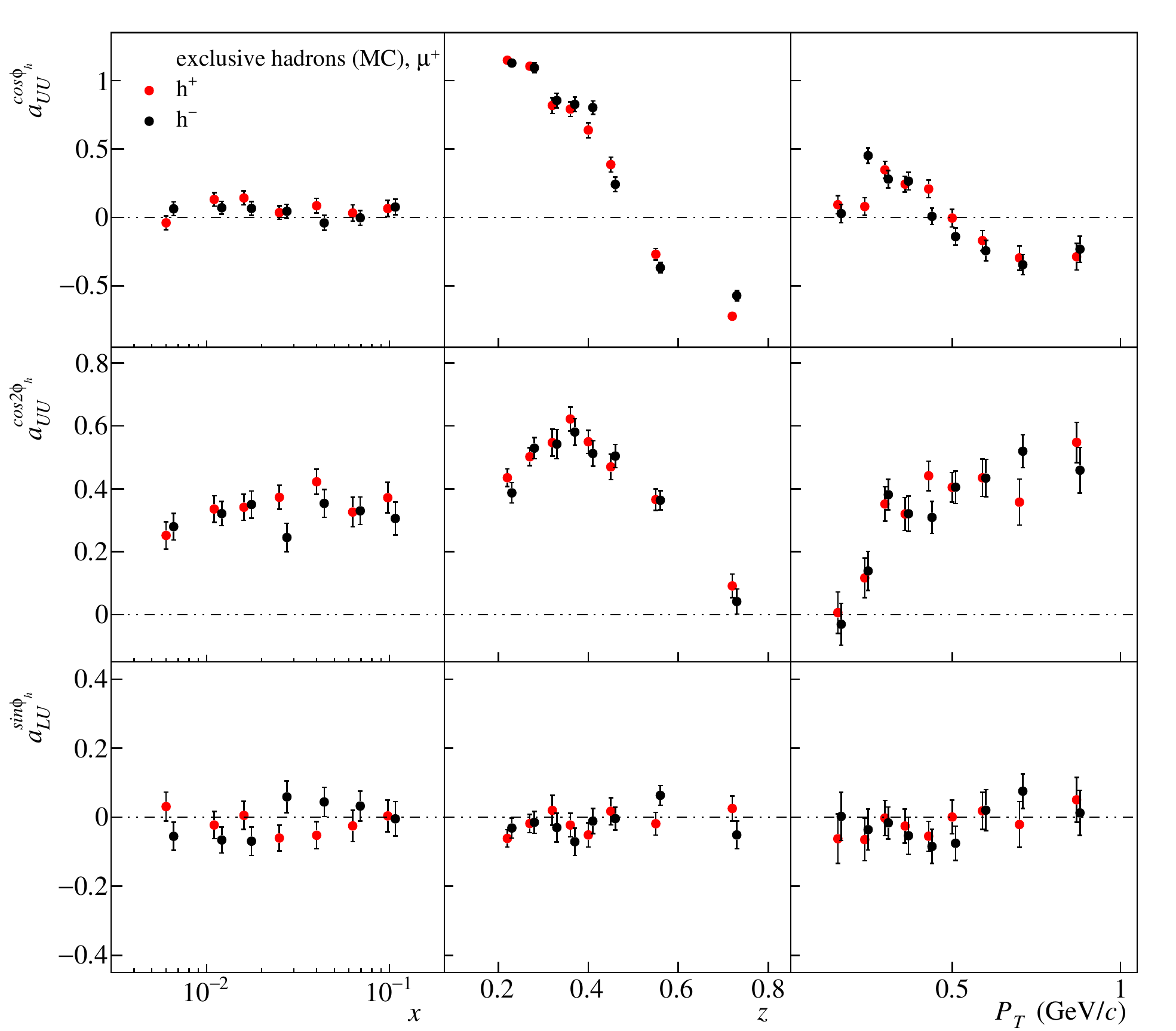}
    \caption{Same as Fig.~\ref{fig:multiplot_aa_excl} but for the HEPGEN Monte Carlo hadrons.}
\label{fig:multiplot_aa_excl_mc}
\end{center}
\end{figure}

The same amplitudes for the Monte Carlo sample are shown in Fig.~\ref{fig:multiplot_aa_excl_mc}. As can be seen, the general features observed in the data are also present in the Monte Carlo. The only exception is the $\Pt$-dependence of the $\cos2\fih$ amplitude. However, such discrepancy appears to be in a region in which the exclusive contamination is small.

\begin{figure}
\captionsetup{width=\textwidth}
\begin{center}
	\includegraphics[angle=0,width=0.95\textwidth]{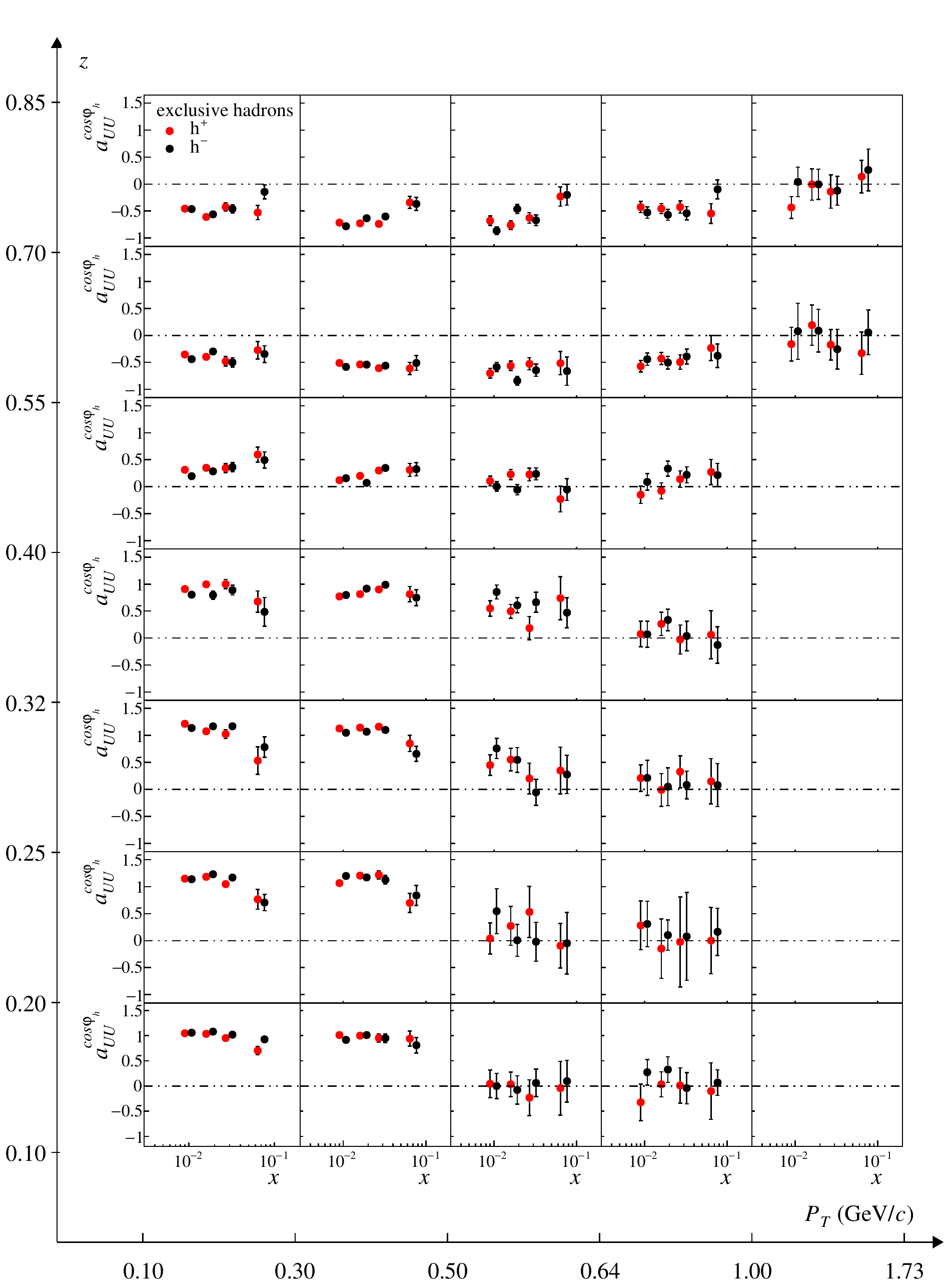}
	\caption{Amplitude of the raw azimuthal modulation in $\cos\fih$ of the exclusive hadrons reconstructed in the data, for positive (red) and negative hadrons (black), as a function of $x$ and in bins of $z$ (vertical axis) and $\Pt$ (horizontal axis). The black points are slightly shifted along $x$ for a better readability.}	
\label{fig:multiplot_aa_excl_3D1}	
\end{center}
\end{figure}

\begin{figure}
\captionsetup{width=\textwidth}
\begin{center}
	\includegraphics[angle=0, width=0.95\textwidth]{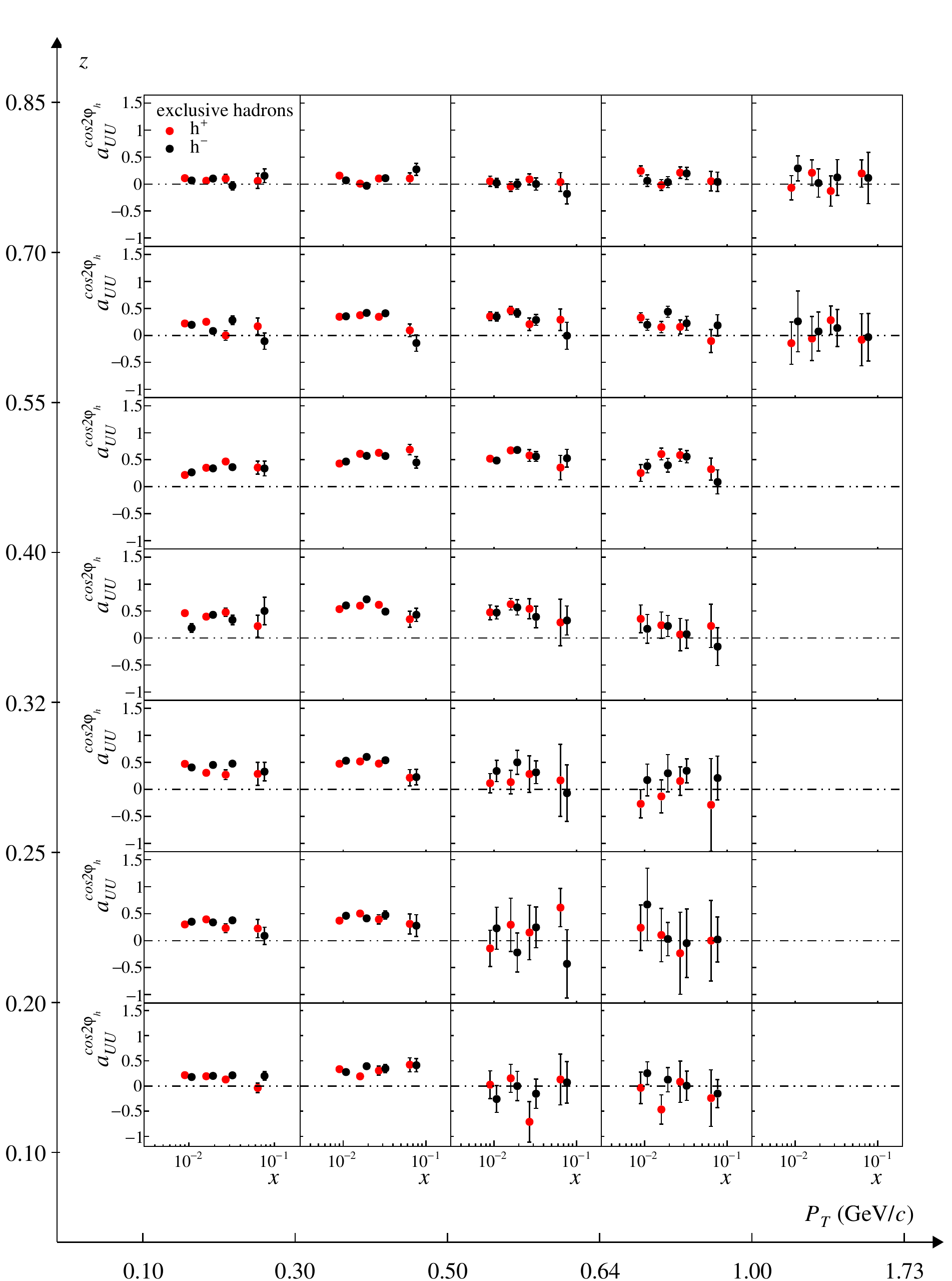}
	\caption{Amplitude of the raw azimuthal modulation in $\cos2\fih$ of the exclusive hadrons reconstructed in the data, for positive (red) and negative hadrons (black), as a function of $x$ and in bins of $z$ (vertical axis) and $\Pt$ (horizontal axis). The points are slightly shifted along $x$ for a better readability.}	
\label{fig:multiplot_aa_excl_3D2}
\end{center}
\end{figure}

More information can be obtained looking at the same amplitudes in the 3D bins. The amplitudes of the $\cos\fih$ and $\cos2\fih$ modulations for the exclusive hadrons reconstructed in the data are shown in Fig.~\ref{fig:multiplot_aa_excl_3D1} and Fig.~\ref{fig:multiplot_aa_excl_3D2} respectively. Some pads at high $\Pt$ and low $z$ are empty because of low statistics. It is interesting to notice how the amplitude of the $\cos\fih$ modulation changes sign along $z$, staying almost flat in $x$ and with a small dependence on $\Pt$. Also, the change of sign as a function of $\Pt$ is due to the $z-\Pt$ correlation. The $\cos2\fih$ amplitude also exhibit a stronger dependence on $z$, as already observed in the 1D results, with an almost negligible dependence on $x$ and $\Pt$. Again, the decrease of the 1D amplitudes at large $\Pt$ is due to the $z-\Pt$ correlation for these hadrons. As before, the $\sin\fih$ modulation is compatible with zero everywhere, and it is thus not shown. 

It can be noticed that the Monte Carlo amplitudes are very similar (as expected) for positive and negative hadrons. Both in 1D and in 3D, also, the Monte Carlo amplitudes for the not-fully reconstructed events (the ones of the hadrons that are subtracted from the data) are similar to the one shown here for the fully reconstructed events.

\newpage
\clearpage

\section{Acceptance correction}
\label{sect:ch5_acceptance}
Analogously to the case of the $\Ptsq$-distributions, the number of hadrons observed in each kinematic bin and in each $\fih$ bin has been corrected for acceptance, according to Eq.~\ref{eq:acc}. The acceptance correction has been estimated for positive and negative hadrons, and for positive and negative muon beams separately. 

\begin{figure}[h!]
\captionsetup{width=\textwidth}
    \centering
   \includegraphics[width=0.75\textwidth]{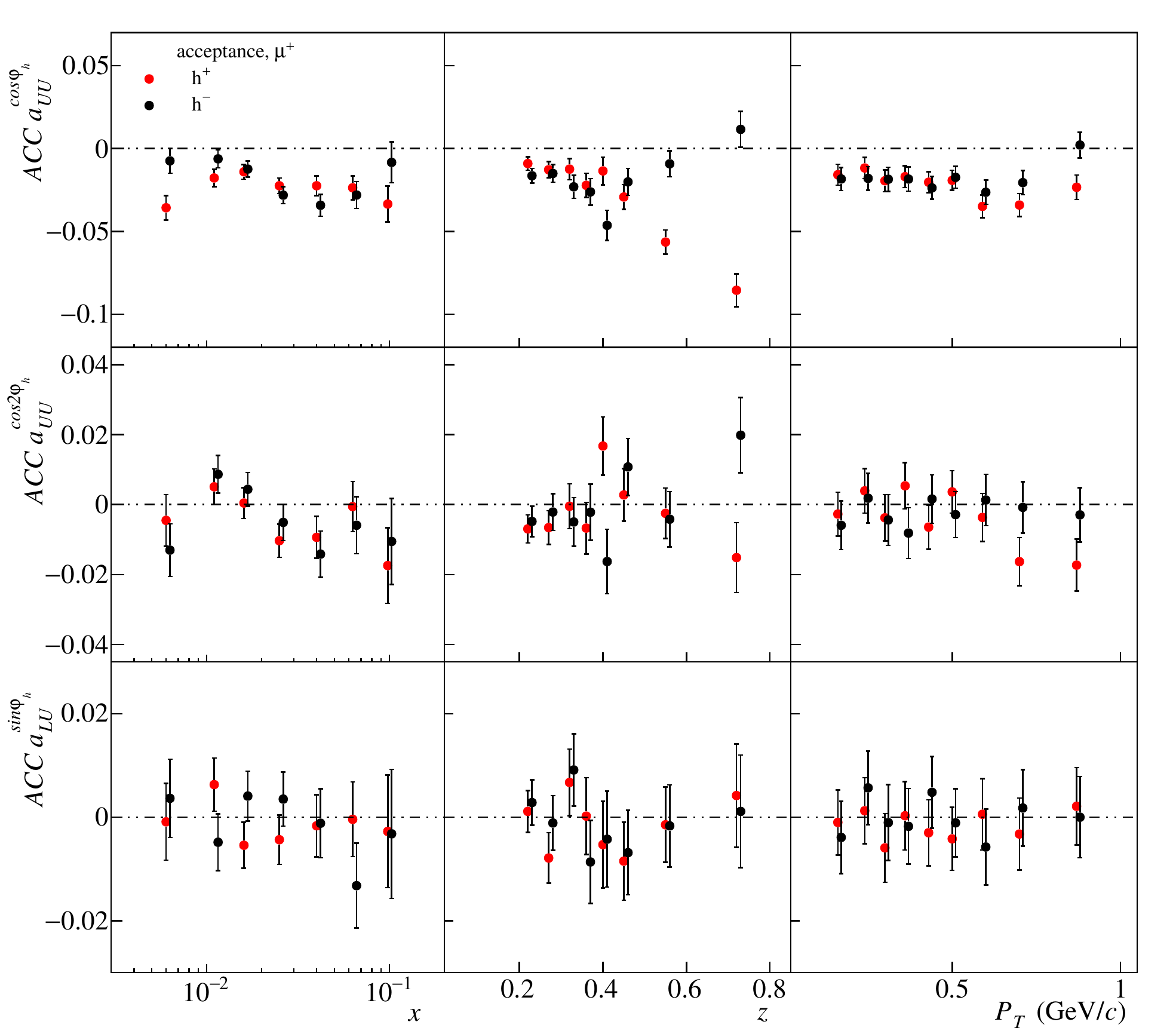}
    \caption{Acceptance amplitudes in $\cos\fih$ (top row), $\cos2\fih$ (middle row) and $\sin\fih$ (bottom row) for positive beam and positive (red) and negative hadrons (black) in the standard 1D kinematic range. The black points are slightly shifted along $x$ for a better readability.}
    \label{fig:accaa1Dmup}
\end{figure}

The kinematic range of the measurements has been chosen in order to have relatively small acceptance corrections. Still, it was important to check \textit{a posteriori} whether this was the case. To this aim only, the amplitudes of the modulations in the azimuthal distributions of the reconstructed hadrons of the LEPTO Monte Carlo sample (the \textit{acceptance amplitudes}) have been evaluated with the same fitting procedure as  for the physics azimuthal asymmetries. Figure~\ref{fig:accaa1Dmup} shows the acceptance amplitudes for positive and negative hadrons and for a positive muon beam as a function of $x$, $z$ and $\Pt$ in the standard 1D binning. While the $\cos2\fih$ and $\sin\fih$ amplitudes are small, with fluctuations due to statistics, the $\cos\fih$ amplitude is generally larger, but never exceeding 10\%. The difference between positive and negative hadrons at high $z$ is due to the cut on the ambiguous events in which particles with the same charge as the beam are reconstructed after the Muon Filters. A symmetric effect arises in the $\mu^-$ case, where positive and negative hadrons exchange their roles. 

\begin{figure}
\captionsetup{width=\textwidth}
    \centering
	\includegraphics[width=0.95\textwidth]{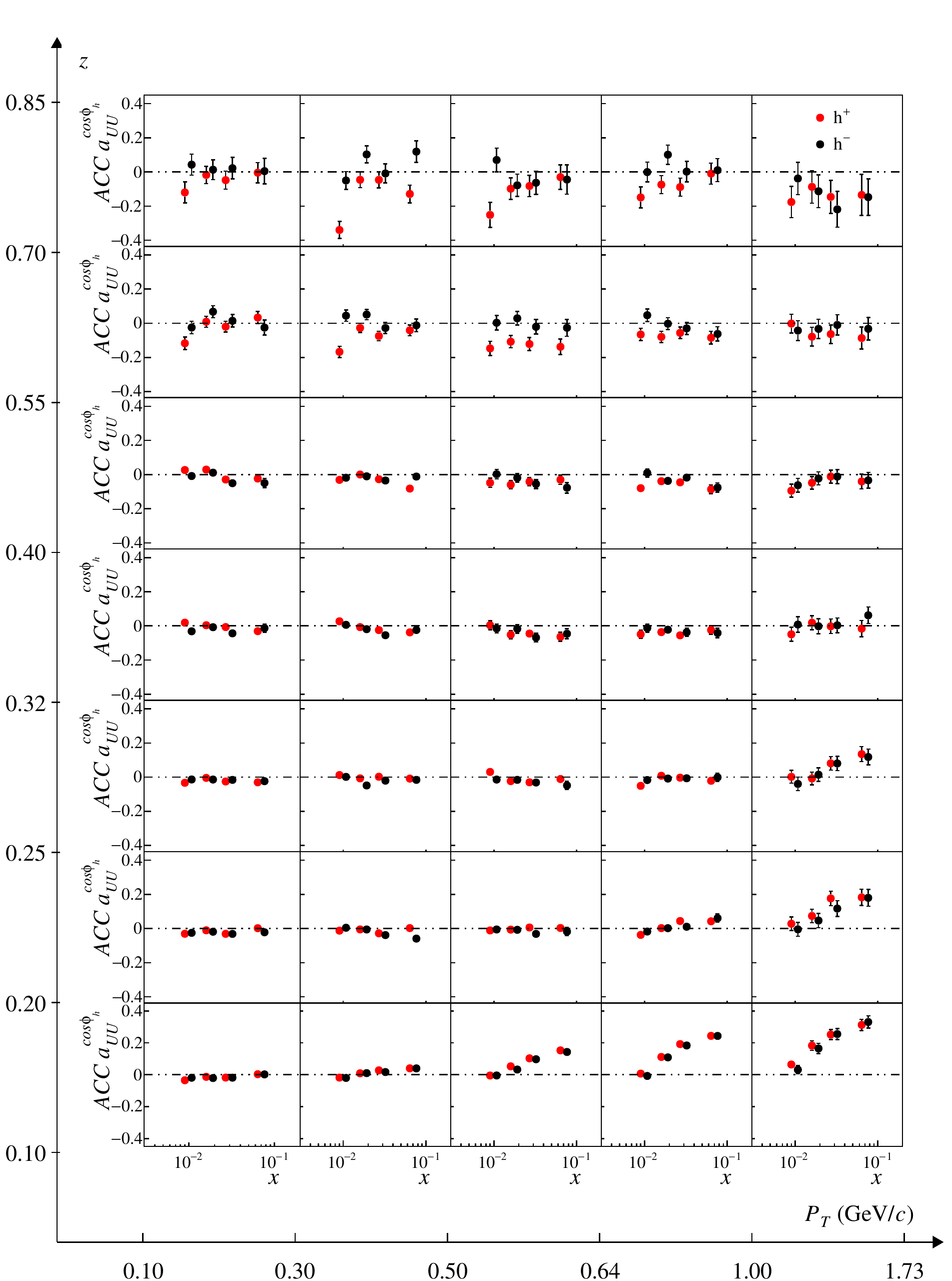}
    \caption{Acceptance amplitudes in $\cos\fih$ for positive (red) and negative (black) hadrons for $\mu^+$ beam in the kinematic bins of the 3D analysis.}
    \label{fig:aaacc3Dcosmup}
\end{figure}

A more detailed overview of the kinematic dependences is offered by the 3D analysis. There, only the $\cos\fih$ amplitude is generally not compatible with zero. It is shown, for a $\mu^+$ beam and for positive and negative hadrons, in Fig.~\ref{fig:aaacc3Dcosmup} with a vertical scale which allows to see the amplitude in all bins, including the lowest $z$ bin and the highest $\Pt$ bin, excluded in the 1D analysis, where they are clearly larger. Excluding the low-$z$ region, the modulation is smaller than, or close to, 10\% in all bins. As in the 1D case, the trend is similar for $\mu^-$, where again positive and negative hadrons exchange their roles. 

\begin{figure}
\captionsetup{width=\textwidth}
    \centering
	\includegraphics[width=0.95\textwidth]{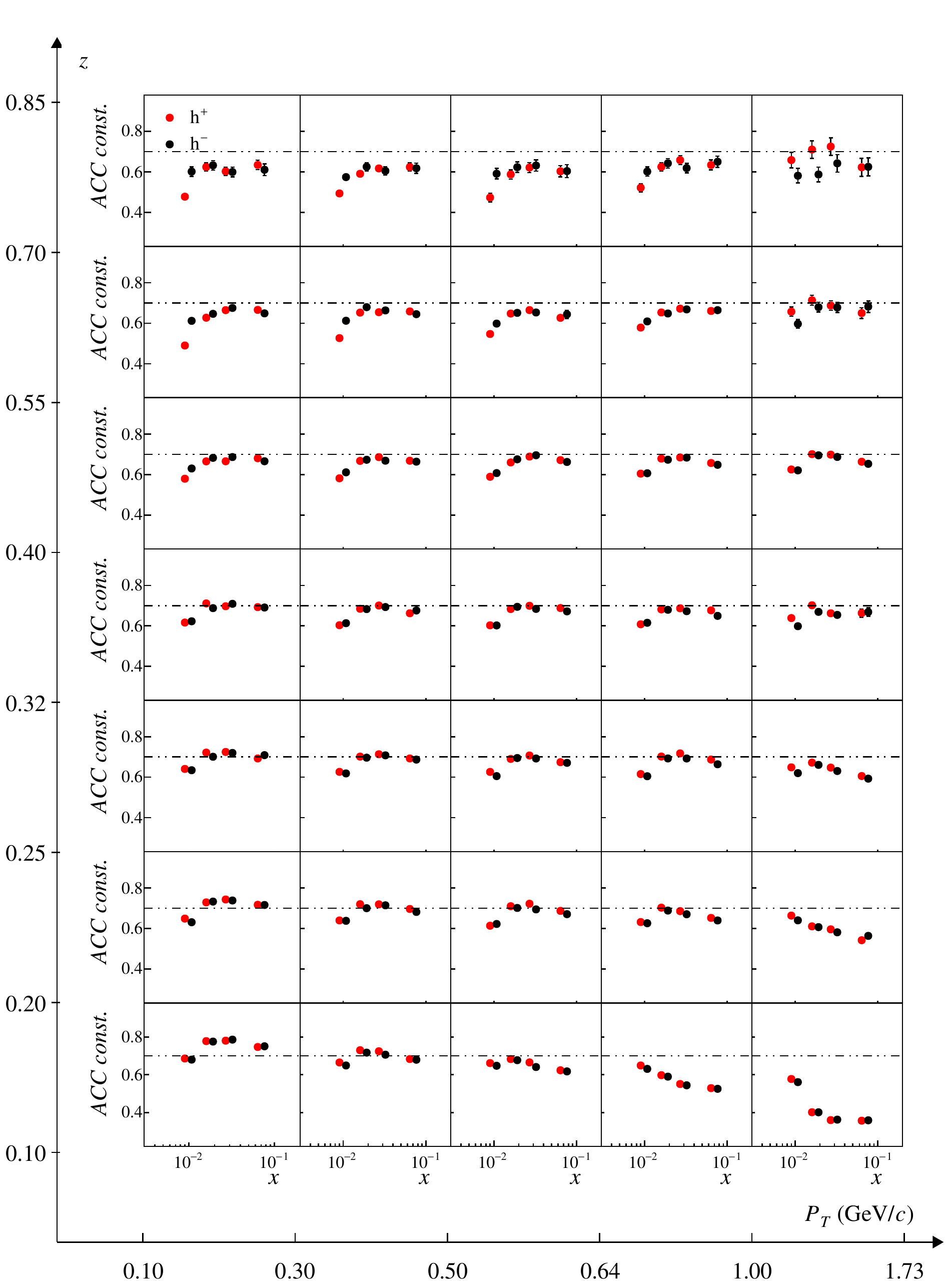}
    \caption{Acceptance for positive (red) and negative hadrons (black) for $\mu^+$ beam, integrated over the azimuthal angle, in the kinematic bins of the 3D analysis.}
    \label{fig:aaacc3Dcost}
\end{figure}

For completeness, the 3D acceptance integrated over the azimuthal angle is given in Fig.~\ref{fig:aaacc3Dcost}, again for positive and negative hadrons in the $\mu^+$ case. A horizontal line at 0.7 is drawn to guide the eye. Generally, the constant term ranges between 0.6 and 0.8, with a moderate dependence on $x$ that becomes clearer at low $z$ and high $\Pt$. As already mentioned, this point is relevant for the interpretation of the 1D measurements.

\clearpage
\newpage

\section{Fitting procedure}
\label{sect:ch5_fitting}

The final results have been obtained from the azimuthal distributions after subtracting the residual exclusive hadron contribution and correcting for acceptance. The procedure has been implemented independently for $\mu^+$ and $\mu^-$ data and the corresponding results have been averaged after checking their compatibility. In each kinematic bin, the distribution of the azimuthal angle $\fih$, ranging ($-\pi$,$\pi$), has been divided into 16 bins of equal width. The two central bins, corresponding to ($-\frac{\pi}{8}, \frac{\pi}{8}$), where electrons from radiative events are present, have been discarded. Then, the amplitudes of the azimuthal modulations have then been extracted by fitting the distributions with the function:
\begin{equation}
    f(\fih) = p_0 \op 1 + p_1 \cos\fih + p_2 \cos2\fih + p_3 \sin\fih \cp
\end{equation}
where $p_i$ ($i=0,\dots,3$) are free parameters and azimuthal asymmetries have been derived from the measured amplitudes according to:
\begin{equation}
    A_{UU}^{\cos\fih} = \frac{p_1}{\langle\eps_1\rangle}, \hspace{1cm} A_{UU}^{\cos2\fih} = \frac{p_2}{\langle\eps_2\rangle},  \hspace{1cm} A_{LU}^{\sin\fih} = \frac{p_3}{\lambda\langle\eps_3\rangle},
\label{eq:asymfit}    
\end{equation}
where the beam polarization $\lambda$ and the kinematic terms $\eps_i$ have already been introduced in Sect.~\ref{sect:ch1_sidis_unpol}. The mean values $\langle \eps_i \rangle$ have been estimated in each kinematic bin separately, using the data on a hadron-by-hadron basis and taking into account the (minimal) corrections due to the subtracted non-visible exclusive component. The beam polarization has been taken as $\lambda=-0.80$ \cite{COMPASS:2007rjf}.

The fits have been performed with the \texttt{MIGRAD} minimizer of the \texttt{MINUIT} package \cite{James:1994vla}. Using the error matrix provided by \texttt{MIGRAD}, the \texttt{MINOS} processor has also been called to accurately refine the uncertainty on the fitted parameters.\\

\begin{figure}[h!]
\captionsetup{width=\textwidth}
    \centering
    \includegraphics[width=0.8\textwidth]{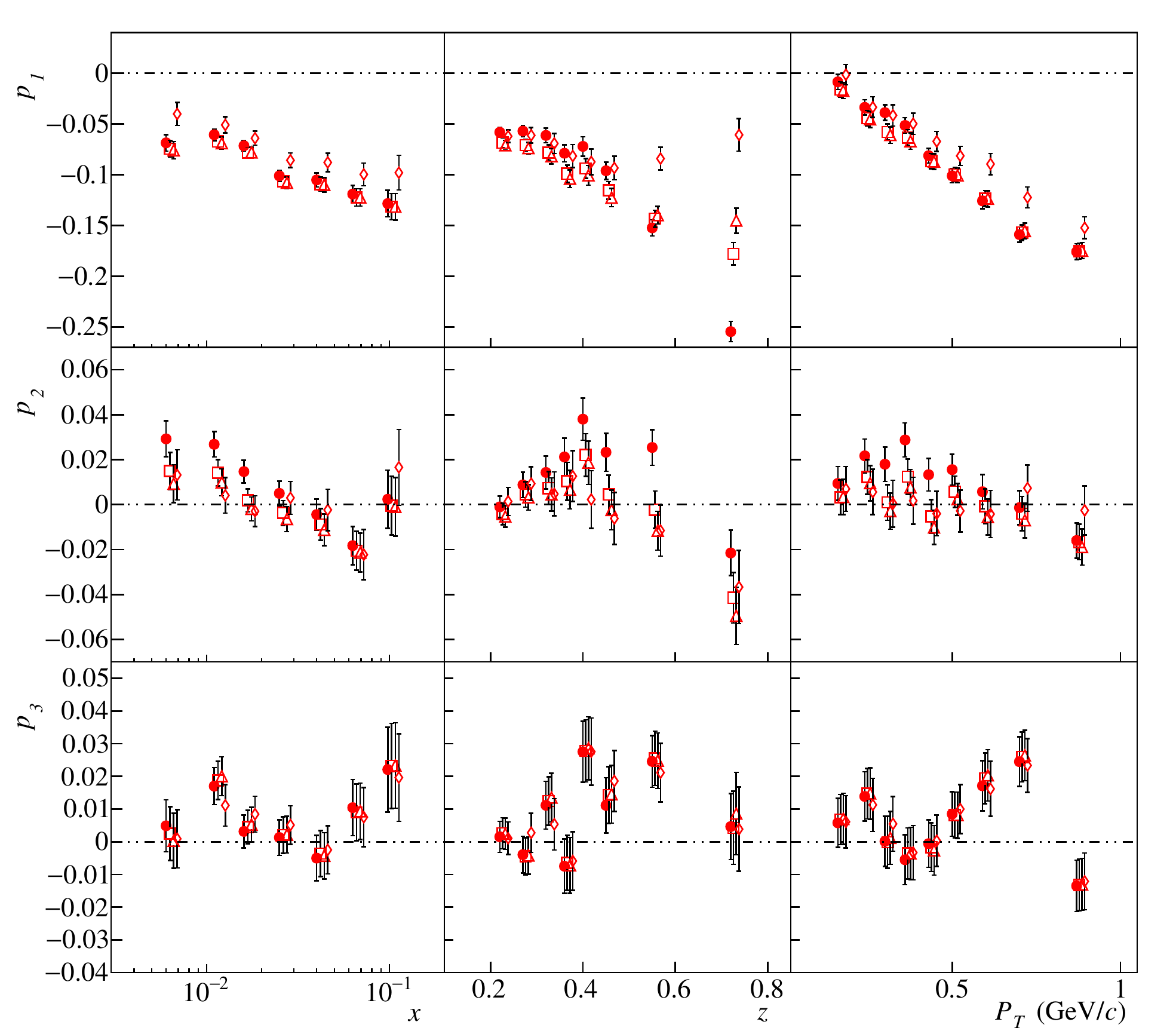}
    \caption{Impact on the fitted parameters $p_1$, $p_2$ and $p_3$ of various corrections applied to the data ($\mu^+$, $h^+$). The raw azimuthal asymmetries (full points) are shown after discarding the reconstructed exclusive events (open squares), after subtracting the non-visible exclusive component (open triangles) and after correcting for acceptance (open diamonds).}
   \label{fig:aaeffectsmup}
\end{figure}

It is interesting to look at the impact of the various corrections applied to the raw data. An example, for positive beam and hadron charges, is shown in Fig.~\ref{fig:aaeffectsmup} where the values of the three parameters $p_1$, $p_2$ and $p_3$ are presented, in the 1D approach with standard binning, as obtained from the fit of the raw data (full points), after discarding the reconstructed exclusive events (open squares), after subtracting the leftover exclusive component with the Monte Carlo (open triangles) and after correcting for acceptance (open diamonds). As expected from the exclusive hadron contamination and azimuthal asymmetries, the removal of the exclusive hadrons has a strong impact on the $p_1$ parameter at high $z$ only; there, the acceptance correction is also strong because the muon and the hadrons have the same charge. As for the $p_2$ parameter, the corrections for exclusive hadrons are important at low $x$ and at intermediate $z$ and $\Pt$ (as expected from Fig.~\ref{fig:multiplot_aa_excl}), while the impact of the acceptance correction is generally small. Lastly, the $p_3$ parameter is almost not impacted by both the exclusive hadron subtraction and the acceptance corrections. \\

\begin{figure}
\captionsetup{width=\textwidth}
    \centering
   	\includegraphics[width=0.8\textwidth]{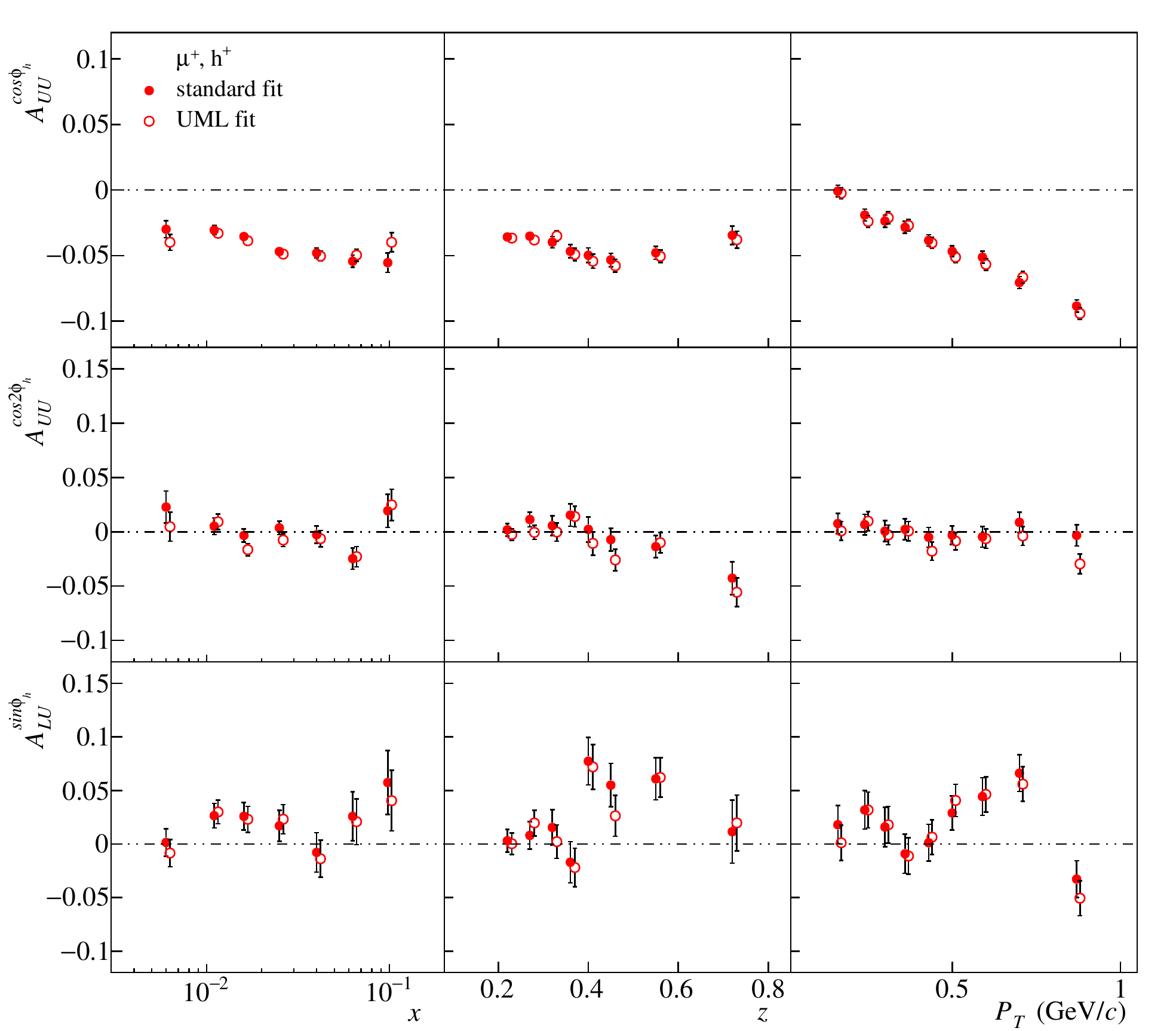}
    \caption{Azimuthal asymmetries $A_{UU}^{\cos\fih}$ (top row), $A_{UU}^{\cos2\fih}$ (middle row) and $A_{LU}^{\sin\fih}$ (bottom row) for positive beam and hadrons. The results of the UML fit (open points) are compared to the ones obtained with the standard binned approach (full points). The open points are slightly shifted horizontally for a better readability.}
    \label{fig:aares1Duml}
\end{figure}

In addition to the fitting procedure described above, the azimuthal asymmetries have been extracted with the Unbinned Maximum Likelihood (UML) method originally implemented for the evaluation of the SDMEs (more details in Ch.~\ref{Chapter6_SDMEs}). In such UML approach, the contribution of the non-visible exclusive events cannot be subtracted before correcting for acceptance. Instead, in  each kinematic bin one needs to know the fraction $f_{bg}$ of hadrons produced in a non-visible exclusive event and the azimuthal asymmetries of the exclusive hadrons. The function $g$ that describes the physical distribution then reads:

\begin{equation}
\begin{split}
    g(\fih,y) & = (1-f_{bg}) \op1 + \eps_1(y)A_{UU}^{\cos\fih}\cos\fih + \eps_2(y)A_{UU}^{\cos2\fih}\cos2\fih + \lambda\eps_3(y)A_{LU}^{\sin\fih}\sin\fih \cp \\
    & + f_{bg} \op1 + \eps_1(y)A_{UU,e}^{\cos\fih}\cos\fih + \eps_2(y)A_{UU,e}^{\cos2\fih}\cos2\fih + \lambda\eps_3(y)A_{LU,e}^{\sin\fih}\sin\fih \cp
\end{split}
\end{equation}
where the dependence on $y$ has been left explicit to indicate that the $\varepsilon$ terms are considered on a hadron-by-hadron basis (instead of taking their mean values in the kinematic bins); the subscript $e$ indicates the asymmetries of the exclusive hadrons, obtained by fitting the visible exclusive component in the data. 
The likelihood $L$ is derived from the function $g$ by simply normalizing it over the useful $\fih$ interval $\Omega_{\fih}$:

\begin{equation}
    L(\fih,y) = \frac{g(\fih,y)}{2 \int_{\Omega_{\fih}} \diff \fih g(\fih,y)}.
\end{equation}
where $\Omega_{\fih} = (\frac{\pi}{8},\pi)$. 
Then, for a given data sample of size $n$ and a reconstructed Monte Carlo of size $m$, used for the acceptance correction, the azimuthal asymmetries are obtained maximizing the quantity:
\begin{equation}
    \sum_{i=1}^n \ln{L(\phi_{h_i},y_i)} - n \ln{\sum_{j=1}^m L(\phi_{h_j},y_j)}.
\end{equation}
In this method it is assumed that $m\gg n$, corresponding to a negligible uncertainty on the acceptance. Due to the limitations in the available Monte Carlo statistics, this requirement prevented an extensive usage of such UML approach and all the results have been produced with the fit of the $\fih$ distributions. The two methods, however, give results in very good agreement, as can be seen in Fig.~\ref{fig:aares1Duml}. Note that for the results from the standard binned method, the statistical uncertainty on the acceptance correction is not taken into account. \\

\begin{figure}[!h]
\captionsetup{width=\textwidth}
    \centering
    \includegraphics[width=0.8\textwidth]{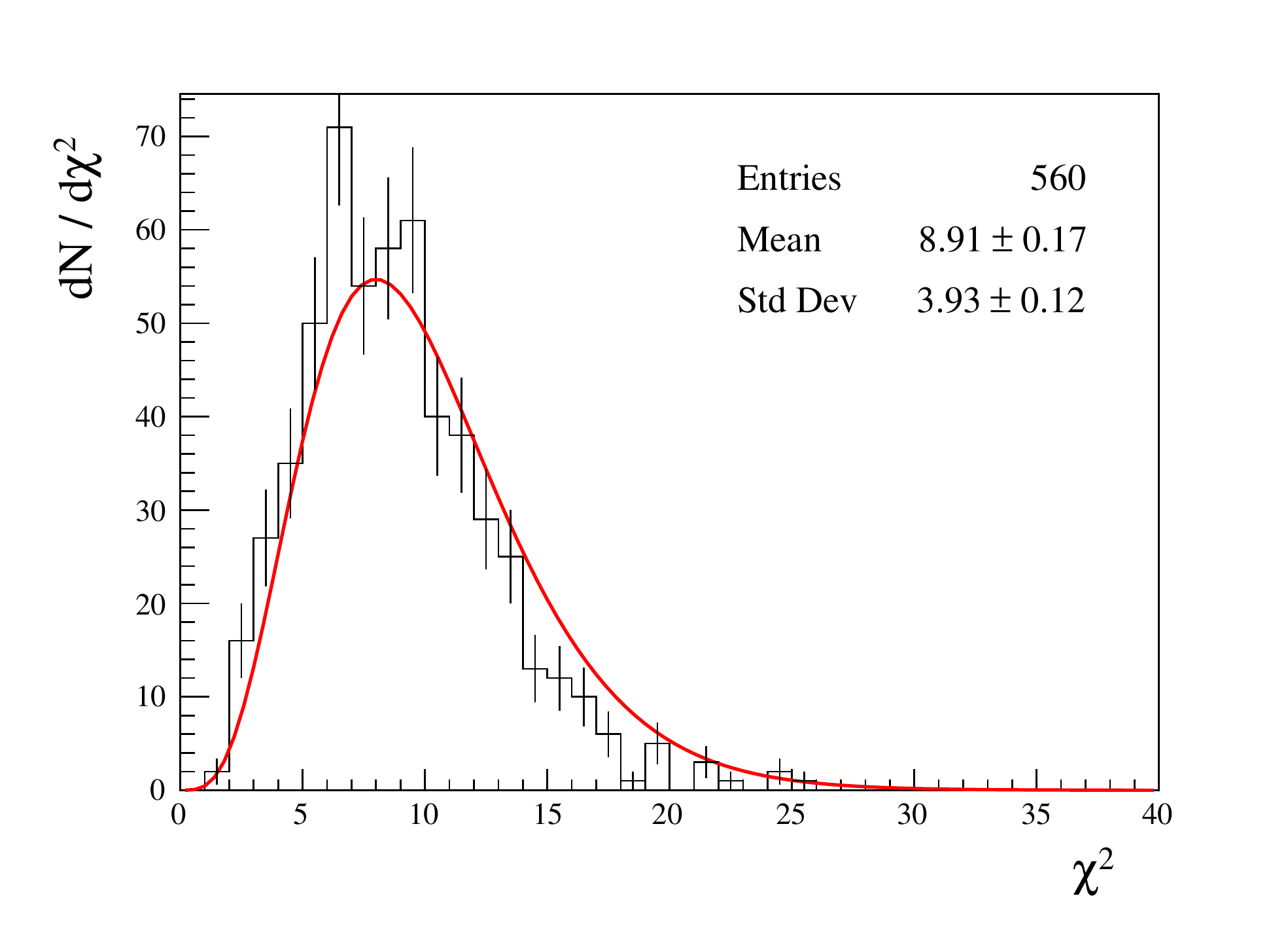}
    \caption{Distribution of the $\chi^2$ values obtained from the fit of the azimuthal asymmetries in 3D bins, compared to the normalized $\chi^2$ distribution with $n.d.f.=10$. }
    \label{fig:chi2_3D}
\end{figure}

The quality of the standard binned fits has been checked looking at the $\chi^2$ values. Figure~\ref{fig:chi2_3D} shows the distribution of the $\chi^2$ values obtained from the fits of the 3D asymmetries, compared to the corresponding $\chi^2$ distribution. The number of degrees of freedom is equal to the number of $\phi_h$ bins considered in the fit (16 minus the two central ones), minus the number of parameters, which makes $n.d.f=10$. The number of considered fits is 560, corresponding to 2 (beam charge) $\times$ 2 (hadron charge) $\times$ 4 ($x$ bins) $\times$ 7 ($z$ bins) $\times$ 5 ($P_T$ bins).

Both the mean value and the standard deviation of the observed $\chi^2$ distribution are smaller than the expected values.

\newpage
\clearpage
\section{Systematic uncertainties}
\label{sect:ch5_systematics}
Several tests have been performed to investigate possible systematic effects. They are described in the following and summarized at the end of this Section. 

\subsection{Period compatibility}
\begin{figure}[h!]
\captionsetup{width=\textwidth}
    \centering
   	\includegraphics[width=\textwidth]{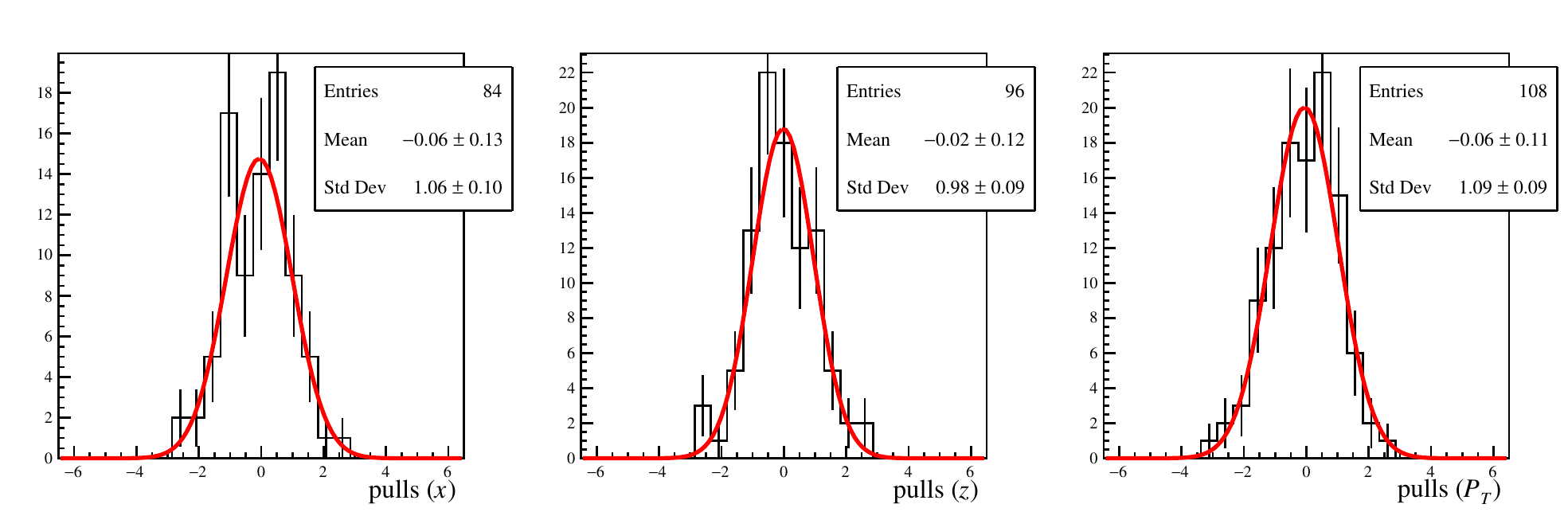}
    \caption{Distributions of the pulls of the asymmetries measured in the three periods of data taking, as a function of $x$ (left), $z$ (center) and $P_T$ (right).}
    \label{fig:period_comp_1D}
\end{figure}

Given the fact that the data taking conditions were the same in the three periods considered in this analysis (P08, P09 and P10), the data have been analyzed in a combined way, building two samples out of them, one per charge of the beam. As done for the $\Ptsq$-distributions, the azimuthal asymmetries have also been measured using the data collected, separately in each period. The compatibility among the results obtained in the three periods has been checked by looking at the \textit{pulls} of the raw asymmetries. For a generic kinematic bin $i$ and period $j$ the pull is defined as:

\begin{equation}
    pull_{ij} = \frac{A_{ij}-\langle A_i \rangle}{\sqrt{\sigma^2_{A_{ij}} - \sigma^2_{\langle A_{i}\rangle}}}
\end{equation}
where $A$ indicates any of the three unpolarized asymmetries and $\langle A \rangle$ is the corresponding mean value calculated as the weighted mean of the results for the three periods. The pulls are expected to be normally distributed: a deviation from this expectation hints at possible systematic effects. Figure~\ref{fig:period_comp_1D} shows the distributions of the pulls for the asymmetries measured as a function of $x$, $z$ and $\Pt$ (left to right). Each plot contains the pulls for positive and negative hadron and beam charges, and for all the three asymmetries. For example, the distribution of the pulls as a function of $x$ has 84 entries, corresponding to the 2 (hadron charge) $\times$ 2 (beam charge) $\times$ 3 (asymmetries) $\times$ 7 (kinematic bins in $x$). Within the uncertainties, no deviation from the normal distribution is observed, neither for the $x$, nor for the $z$ and $\Pt$ cases. The same is true when looking at the pulls of the different asymmetries separately. 

As expected, the same conclusion is reached when inspecting the 3D raw asymmetries, for which the pulls distribution has a mean value $\mu=0.00 \pm 0.01$ and a standard deviation $\sigma = 1.03 \pm 0.01$, thus with no indications of systematic effects.

\subsection{$\mu^{+} - \mu^{-}$ compatibility}
The compatibility between the 1D azimuthal asymmetries measured with $\mu^+$ and $\mu^-$ beam is quite good. It has been studied using the asymmetries corrected for acceptance after the subtraction of the exclusive hadrons and by calculating the $\chi^2$: 
\begin{equation}
    \chi^2_j=\sum_i \frac{(A_{ij}^+-A_{ij}^-)^2}{\sigma_{ij}^{+\,2}+\sigma_{ij}^{-\,2}}
\end{equation}
where the index $j$ indicates a kinematic variable (either $x, \, z$, or $\Pt$) and the hadron charge, the index $i$ a kinematic bin and where $A_{ij}^{+(-)}$ are the asymmetries measured with a $\mu^{+(-)}$ beam. The $\chi^2$ values are given in Tab.~\ref{tab:aasyst_mupm}, for the standard and for the extended kinematic ranges. In all the cases the values of the $\chi^2$ are low, and no indication of relevant systematic effects is seen. The same procedure has been applied to the 3D asymmetries, as can be seen in Tab.~\ref{tab:aasyst3_mupm}, again getting no indication of systematic effects.

\begin{table}[tbh!]
\captionsetup{width=\textwidth}
\scriptsize
\centering
\begin{tabular}{llrrrrrr}
	\hline
\textbf{kinematic range	(1D)} & & \multicolumn{2}{c}{$A_{UU}^{\cos\fih}$} 
		& \multicolumn{2}{c}{$A_{UU}^{\cos2\fih}$} 
		& \multicolumn{2}{c}{$A_{LU}^{\sin\fih}$}  \\
        & & $h^+$ & $h^-$ & $h^+$ & $h^-$ & $h^+$ & $h^-$ \\
\hline
standard & $x$ bins (7) &  6.0 & 8.9 & 3.2 & 9.1 & 6.5 & 1.1  \\
 &$z$ bins (8) &  8.8 & 8.4 & 10.3 & 9.3 & 4.7 & 4.5  \\
 & $\Pt$ bins (9) &  3.5 & 10.1 & 5.0 & 9.2 & 9.4 & 7.7  \\
 &total (48)& \multicolumn{2}{c}{45.7}  &
 \multicolumn{2}{c}{46.1}  &
 \multicolumn{2}{c}{33.9}  \\
    \hline
$0.1<z<0.2$ & $x$ bins (7) & 2.3 & 1.1 & 4.6 & 4.6 & 8.4 & 5.5  \\
 & $z$ bins (1) & 0.4 & 0.1 & 3.1 & 0.8 & 0.3 & 2.3  \\
 & $\Pt$ bins (9) & 4.7 & 5.2 & 4.6 & 6.5 & 6.4 & 4.5  \\
 &total (34)& \multicolumn{2}{c}{13.8}  &
 \multicolumn{2}{c}{24.2}  &
 \multicolumn{2}{c}{27.4}  \\
    \hline 
$1.0<\Pt~\mathrm{(GeV/}c\mathrm{)}<1.73$  & $x$ bins (7) & 5.3 & 4.2 & 3.2   & 4.1 & 3.8 & 4.1  \\
& $z$ bins (8) & 3.1 & 3.3 & 7.6 & 3.4 & 9.1 & 3.1  \\ 
 & $\Pt$ bins (1) & 1.6 & 0.2  & 1.2   & 0.1 & 1.1 & 0.4  \\
 & total (32)& \multicolumn{2}{c}{17.7}  &
 \multicolumn{2}{c}{19.6}  &
 \multicolumn{2}{c}{21.6}  \\
    \hline
\end{tabular}
\caption{$\chi^2$ values for the $\mu^{+}-\mu^{-}$ compatibility for the 1D azimuthal asymmetries: in the $x$, $z$ and $\Pt$ bins for positive and negative hadron, in the three kinematic ranges. The numbers in parentheses are the numbers of bins.}
\label{tab:aasyst_mupm}
\end{table}

\begin{table}[tbh!]
\captionsetup{width=\textwidth}
\scriptsize
\centering
\begin{tabular}{llrrrrrr}
	\hline
\textbf{kinematic range	(3D)} & & \multicolumn{2}{c}{$A_{UU}^{\cos\fih}$} 
		& \multicolumn{2}{c}{$A_{UU}^{\cos2\fih}$} 
		& \multicolumn{2}{c}{$A_{LU}^{\sin\fih}$}  \\
        & & $h^+$ & $h^-$ & $h^+$ & $h^-$ & $h^+$ & $h^-$ \\
\hline
standard &  & 85.5 & 88.8 & 79.3 & 113.3 & 71.3 & 104.3  \\
(4 $x$ bins, 6 $z$ bins, 4 $P_T$ bins)  & total (192) & \multicolumn{2}{c}{174.3}  &
 \multicolumn{2}{c}{192.6}  &
 \multicolumn{2}{c}{175.6}  \\  
    \hline
including low $z$ and high $P_T$ &  & 139.5 & 114.4 & 119.0 & 158.2 & 102.4 & 141.3  \\
(4 $x$ bins, 7 $z$ bins, 5 $P_T$ bins)  & total (280) & \multicolumn{2}{c}{253.9}  &
 \multicolumn{2}{c}{277.2}  &
 \multicolumn{2}{c}{243.7}  \\  
    \hline
\end{tabular}
\caption{$\chi^2$ values for the $\mu^{+}-\mu^{-}$ compatibility for the 3D azimuthal asymmetries, obtained summing over the $x$, $z$ and $\Pt$ bins for positive and negative hadrons, in the standard and extended kinematic ranges. The numbers in parentheses are the numbers of bins.}
\label{tab:aasyst3_mupm}
\end{table}

\newpage
\subsection{Target quarter compatibility}
\begin{figure}[!h]
\captionsetup{width=\textwidth}
    \centering
    \includegraphics[width=\textwidth]{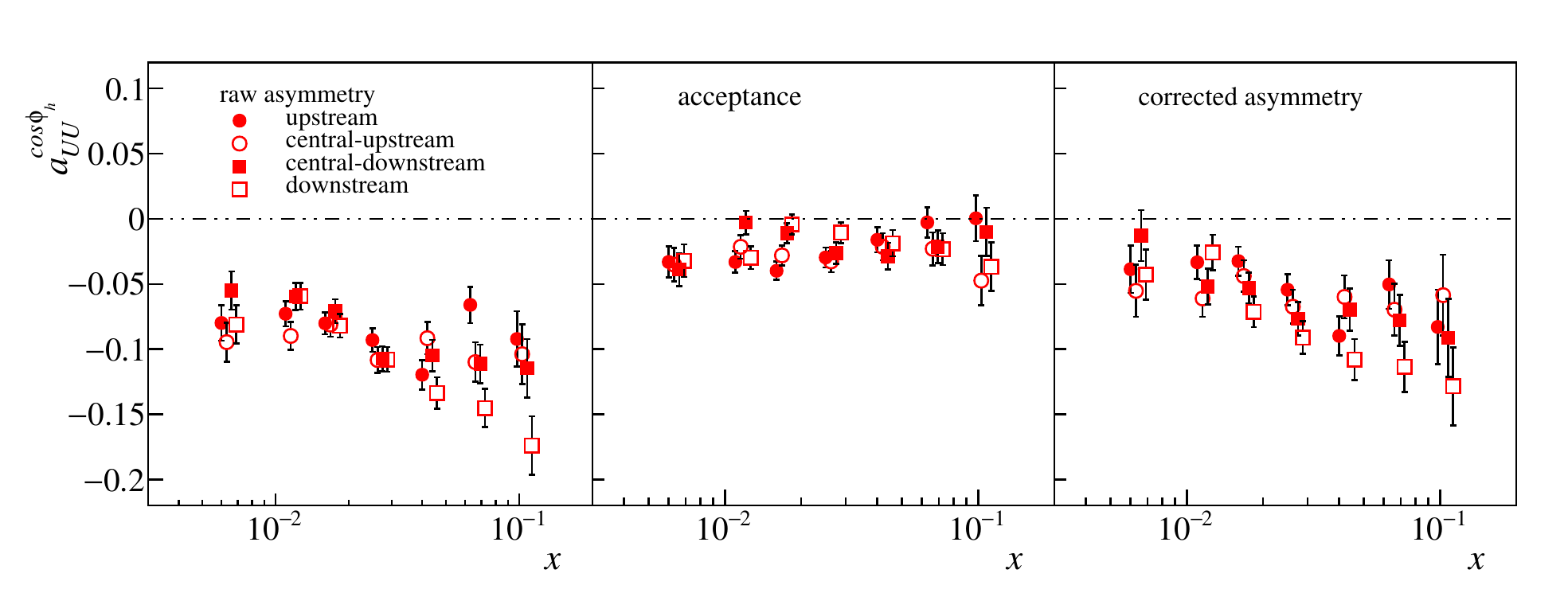}
    \caption{For the four target quarters, the amplitude of the $\cos\fih$ modulation as measured in the raw data (left), in the acceptance (middle) and in the data after the acceptance correction (right).}
    \label{fig:target_effect}
\end{figure}

There is a systematic shift in the acceptances when going from the upstream to the downstream quarter for the 1D $A_{UU}^{\cos \fih}$ asymmetry at small  $x$, which is not expected nor observed in the raw asymmetries. This is shown in Fig.~\ref{fig:target_effect} for positive hadrons and $\mu^+$ beam data. The left panel shows the $\cos\fih$ asymmetry, before the correction for the acceptance, for events  with the primary vertex in the same target quarters used to test the systematic effects for the $\Ptsq$-distributions (Sect.~\ref{sect:ch4_systematics}). As expected, at small $x$, i.e. for hadrons going in the forward direction, where the detection efficiency is expected to be uniform, the measured asymmetries are compatible for the different target quarters. Going at larger $x$, the limited acceptance of the apparatus introduces a positive $\cos\fih$ asymmetry, which is more relevant for the upstream part of the target, and the measured asymmetries become different. The central panel shows the acceptance modulations. While the trend at large $x$ looks as expected, at intermediate $x$ the acceptance amplitudes are shifted. As a consequence, the acceptance-corrected asymmetries for events with primary vertices in the different target regions, shown in the right panel of Fig.~\ref{fig:target_effect}, are different, and there is a systematic shift. The most reasonable conclusion is that there is a problem in the event simulation, which could be either a slightly low efficiency in the central region of the spectrometer, close to the beam line, or a high efficiency in the outer part. It turned out that the $\cos\fih$ modulation is very sensitive to this kind of effects and, in spite of all the work done, the problem in the Monte Carlo could not be identified so far. The half-difference of the final asymmetries, measured from events with the primary vertex in the second and fourth quarters of the target (the most upstream quarter is excluded in this analysis) has been taken as systematic error. It turned out to be approximately equal to the statistical uncertainty in each $x$ bin, both for positive and negative hadrons. Even if this effect is not so strong for the other asymmetries, a common upper limit $\sigma_{syst} = \sigma_{stat}$ for all the 1D asymmetries is considered safe.
As for the 3D asymmetries, the half-difference of the second and fourth quarter of the target shows its largest value (0.007)  for the $\cos\fih$ asymmetry, positive hadrons. The target quarter compatibility for the other asymmetries is better; still, the same, safe upper limit ($\sigma_{syst} = 0.5~\sigma_{stat}$) is taken also here.

\subsection{Uncertainty on the HEPGEN normalization}
The impact of the HEPGEN normalization on the final results has been studied by modifying the normalization values of $\pm$20\%, thus obtaining two new sets of results. The half-difference of the azimuthal asymmetries from the new samples has been considered, to get an estimate of the systematic uncertainty. Except for the one point at high $z$, where it is comparable with the statistical uncertainty, the half-difference is much smaller than 0.1\%. For this reason, this contribution has not been regarded as a source of systematic uncertainty.

\subsection{Summary}
To summarize, out of the several tests on possible systematic effects, only that on the dependence of the measured values of the asymmetries on the primary vertex position turned out to be relevant. The upper limits for the systematic uncertainties are:
\begin{itemize}
    \item 1D asymmetries, standard, low $z$, high $\Pt$ ranges: $\sigma_{syst} = \sigma_{stat}$;
    \item 3D asymmetries: $\sigma_{syst} = 0.5~\sigma_{stat}$ everywhere, except for:
    \begin{itemize}
        \item $0.10<z<0.20,~0.10~\mathrm{GeV/}c<\Pt<0.50~\mathrm{GeV/}c:~~\sigma_{syst} = 1.5~\sigma_{stat}$
        \item $0.10<z<0.20,~0.50~\mathrm{GeV/}c<\Pt<1.00~\mathrm{GeV/}c:~~\sigma_{syst} = 1.0~\sigma_{stat}$
        \item $0.20<z<0.25,~0.10~\mathrm{GeV/}c<\Pt<0.50~\mathrm{GeV/}c:~~\sigma_{syst} = 1.0~\sigma_{stat}$
        \item $0.25<z<0.32,~0.10~\mathrm{GeV/}c<\Pt<0.50~\mathrm{GeV/}c:~~\sigma_{syst} = 1.0~\sigma_{stat}$
    \end{itemize}
\end{itemize}

As said before, the results presented in this work have not been corrected for the radiative effects. Preliminary studies on their possible impact, performed with Monte Carlo simulations based on DJANGOH \cite{Aschenauer:2013iia}, indicate that the $A_{UU}^{\cos\fih}$ asymmetries could get reduced by a factor two at high $\Qsq$. For the asymmetries integrated over $\Qsq$, as the one considered for the standard binning presented here, the overall impact would be dominated by the low-$\Qsq$ region, where they are not much affected by the radiative effects. For the $A_{UU}^{\cos2\fih}$ and $A_{LU}^{\sin\fih}$ asymmetries, the impact has been estimated to be within the statistical uncertainty even at large $\Qsq$, and thus negligible integrating over $\Qsq$.

\section{Results}
\label{sect:ch5_results}
In this Section, the results for the azimuthal asymmetries from the one-dimensional and the three-dimensional results are presented, both with the standard binning and with the extensions at low $z$ and high $\Pt$. These results have been obtained as the weighted average of the asymmetries measured separately for positive and negative muon beam data, merged after checking the $\mu^+-\mu^-$ compatibility, and they are presented for positive and negative hadrons. The comparison with the same asymmetries, measured on a deuteron target at COMPASS, can be found in Sect.~\ref{sect:ch5_results_compD}. In all the plots, the error bars indicate the statistical uncertainties only.

\subsection{One-dimensional results}
\begin{figure}[h!]
\captionsetup{width=\textwidth}
    \centering
   	\includegraphics[width=0.75\textwidth]{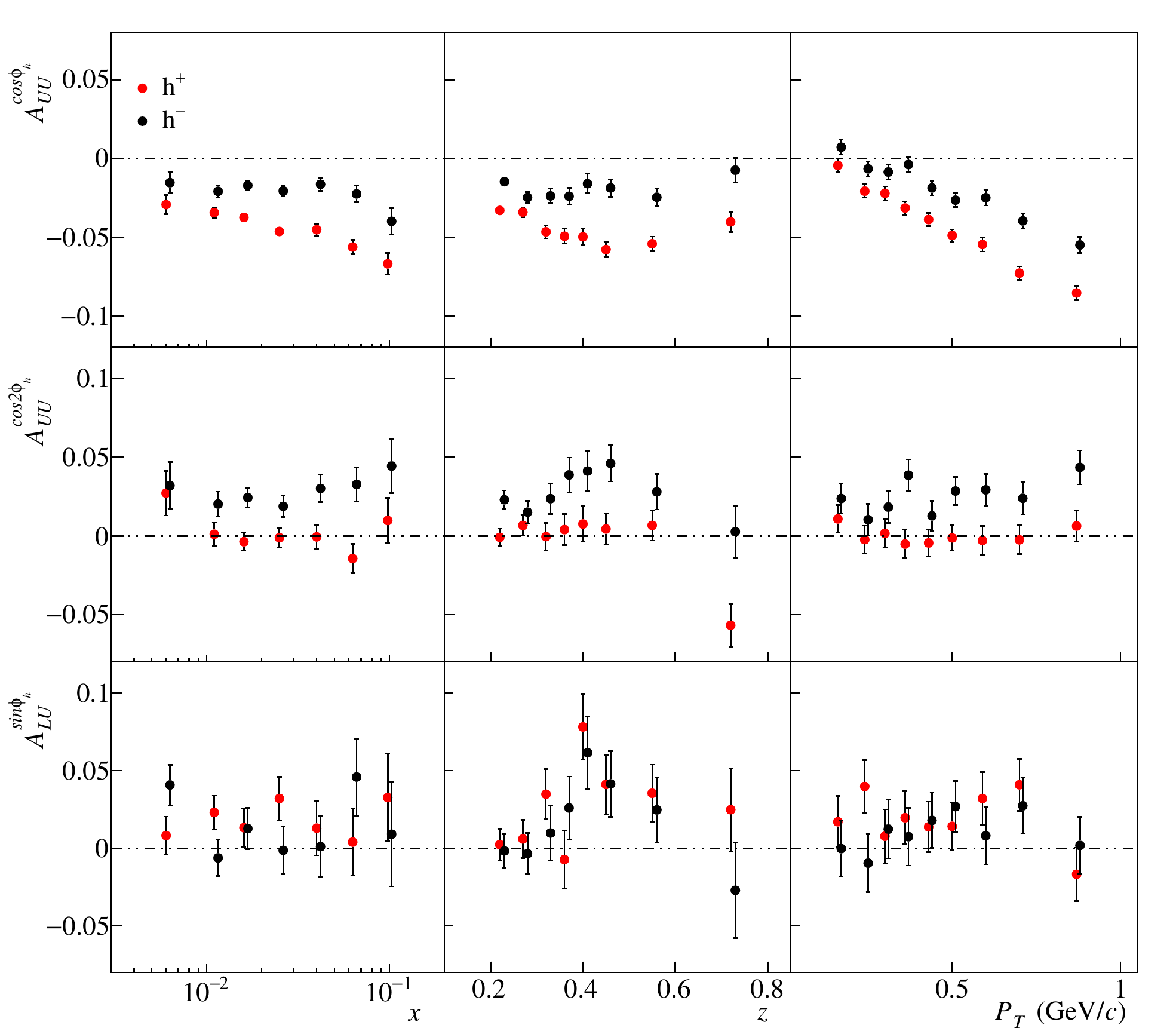}
    \caption{Azimuthal asymmetries $A_{UU}^{\cos\fih}$ (top row), $A_{UU}^{\cos2\fih}$ (middle row) and $A_{LU}^{\sin\fih}$ (bottom row) for positive (red) and negative hadrons (black) in the standard kinematic range. The black points are slightly shifted for a better readability.}
    \label{fig:aares1Dstd}
\end{figure}

\begin{figure}[h!]
\captionsetup{width=\textwidth}
    \centering
   	\includegraphics[width=0.8\textwidth]{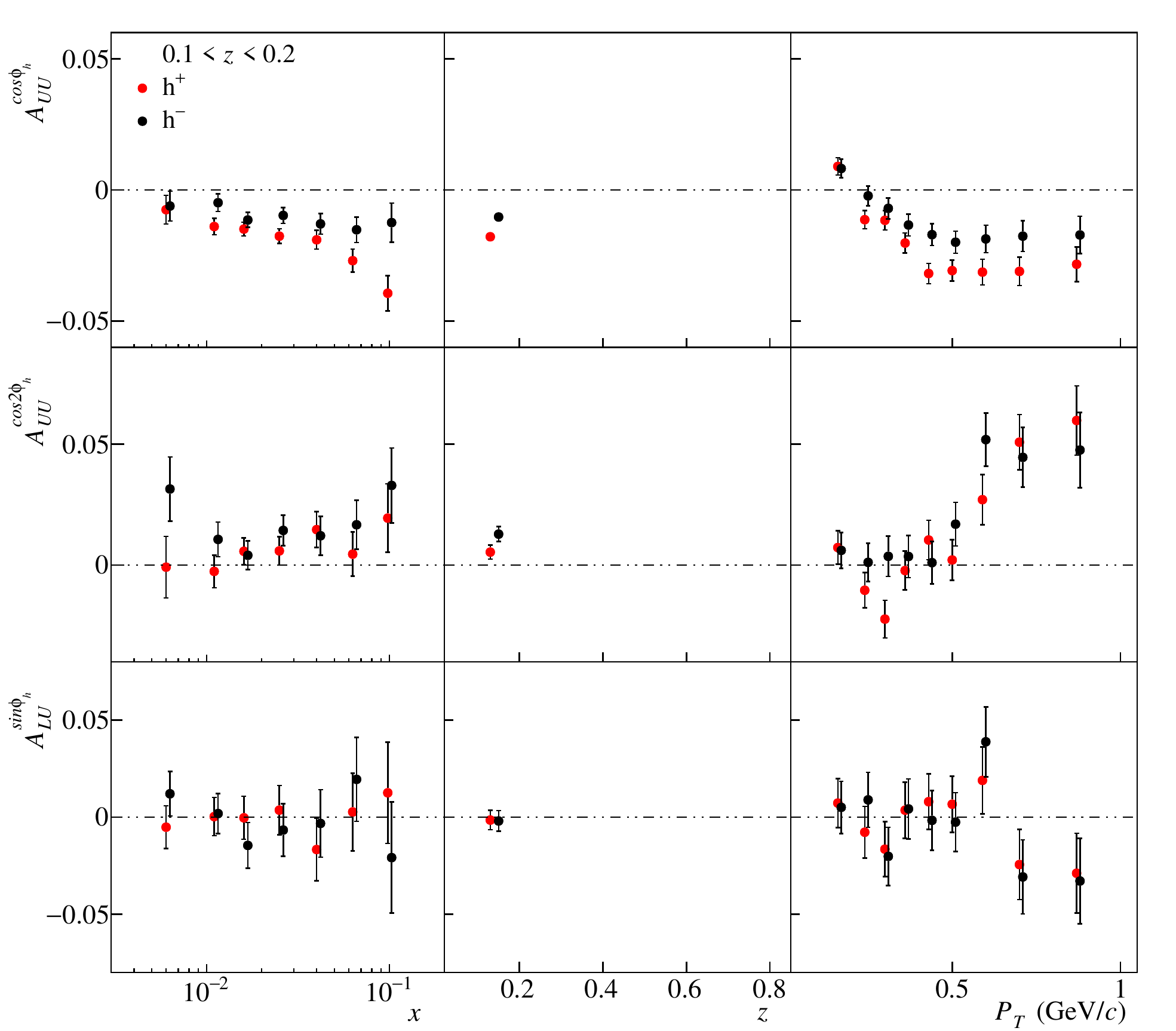}
    \caption{Azimuthal asymmetries $A_{UU}^{\cos\fih}$ (top row), $A_{UU}^{\cos2\fih}$ (middle row) and $A_{LU}^{\sin\fih}$ (bottom row) for positive (red) and negative hadrons (black) in the range $0.1<z<0.2$. The black points are slightly shifted for a better readability.}
    \label{fig:aares1Dlz}
\end{figure}

\begin{figure}[h!]
\captionsetup{width=\textwidth}
    \centering
   	\includegraphics[width=0.8\textwidth]{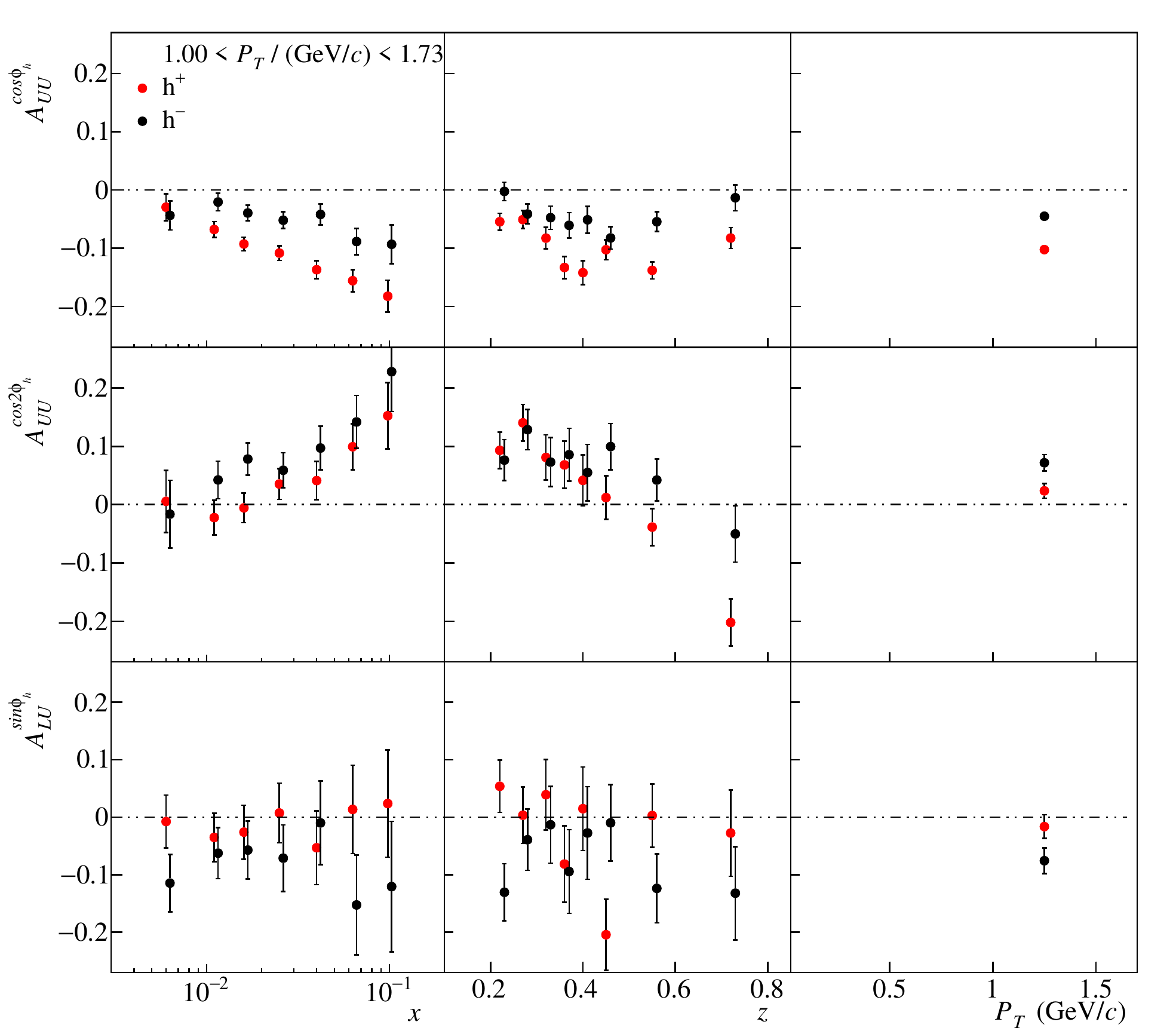}
    \caption{Azimuthal asymmetries $A_{UU}^{\cos\fih}$ (top row), $A_{UU}^{\cos2\fih}$ (middle row) and $A_{LU}^{\sin\fih}$ (bottom row) for positive (red) and negative hadrons (black) in the range $1.0<\Pt~\mathrm{/( GeV/}c\mathrm{)}<1.73$. The black points are slightly shifted for a better readability.} 
    \label{fig:aares1Dhp}
\end{figure}

The 1D results for $A_{UU}^{\cos\fih}$, $A_{UU}^{\cos2\fih}$ and $A_{LU}^{\sin\fih}$ in the standard kinematic range are shown in Fig.~\ref{fig:aares1Dstd} as a function of $x$, $z$ and $\Pt$ for positive and negative hadrons. As can be seen, particularly in the $A_{UU}^{\cos\fih}$ case, the asymmetries are large, clearly different from zero, with a linear trend in $\Pt$ and different between positive and negative hadrons. The $\cos2\fih$ asymmetry is generally smaller, compatible with zero for the positive hadrons and clearly different from zero for the negative hadrons. As for the $\sin\fih$ asymmetry, the uncertainties are larger due to the kinematic terms.  The 1D results in the low-$z$ and high-$\Pt$ kinematic ranges are shown in Fig.~\ref{fig:aares1Dlz} and ~\ref{fig:aares1Dhp} respectively. Also in these cases, despite the large uncertainties, the kinematic dependences are strong and clearly visible.

\subsection{Three-dimensional results}

\begin{figure}
\captionsetup{width=\textwidth}
\centering
	\includegraphics[width=0.95\textwidth]{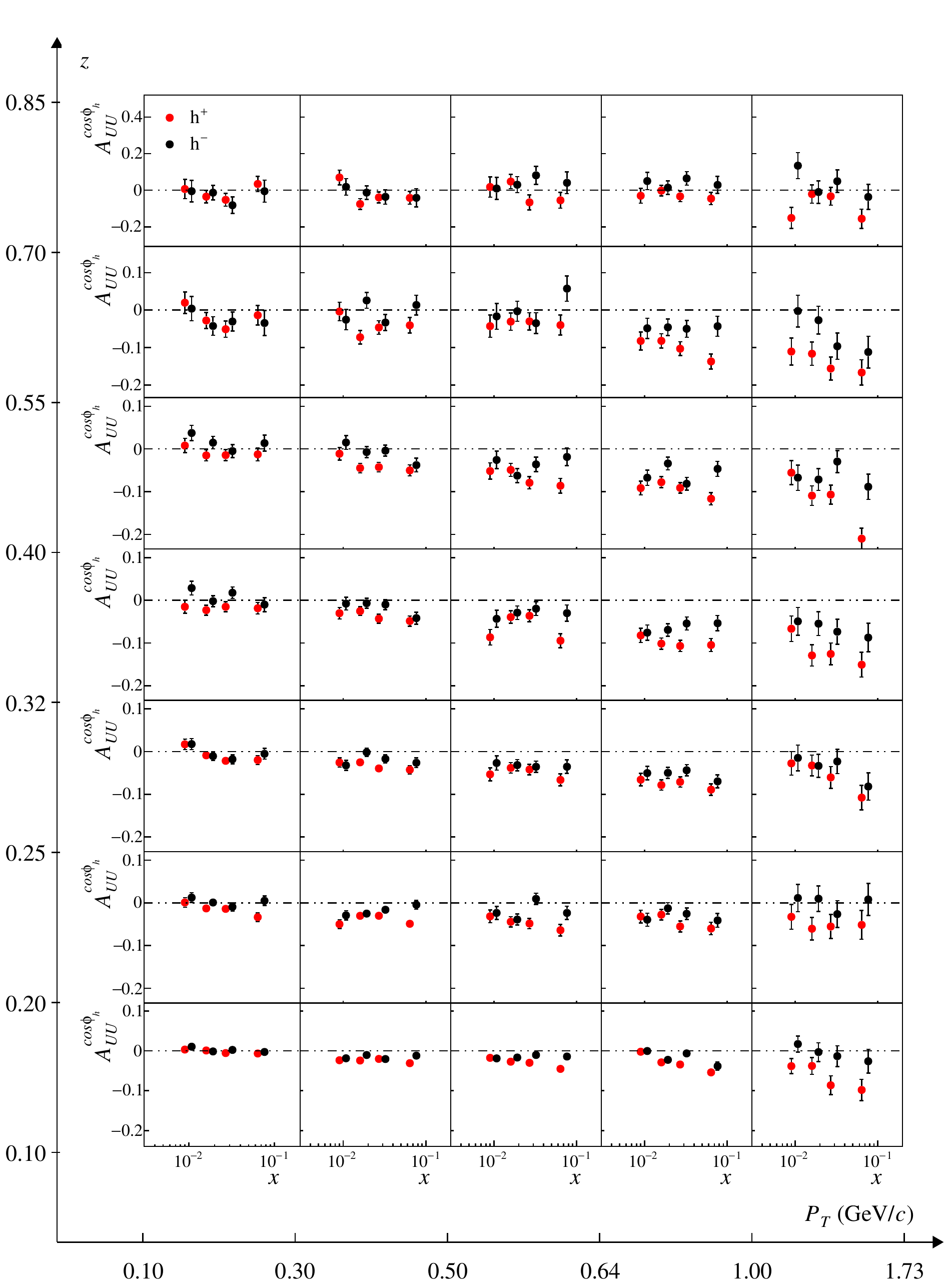}
    \caption{$A_{UU}^{\cos\fih}$ asymmetry for positive (red) and negative hadrons (black), as a function of $x$ and in bins of $z$ (vertical axis) and $\Pt$ (horizontal axis). The black points are slightly shifted for a better readability.}
    \label{fig:aares3Dcos}
\end{figure}

\begin{figure}
\captionsetup{width=\textwidth}
\centering
	\includegraphics[width=0.95\textwidth]{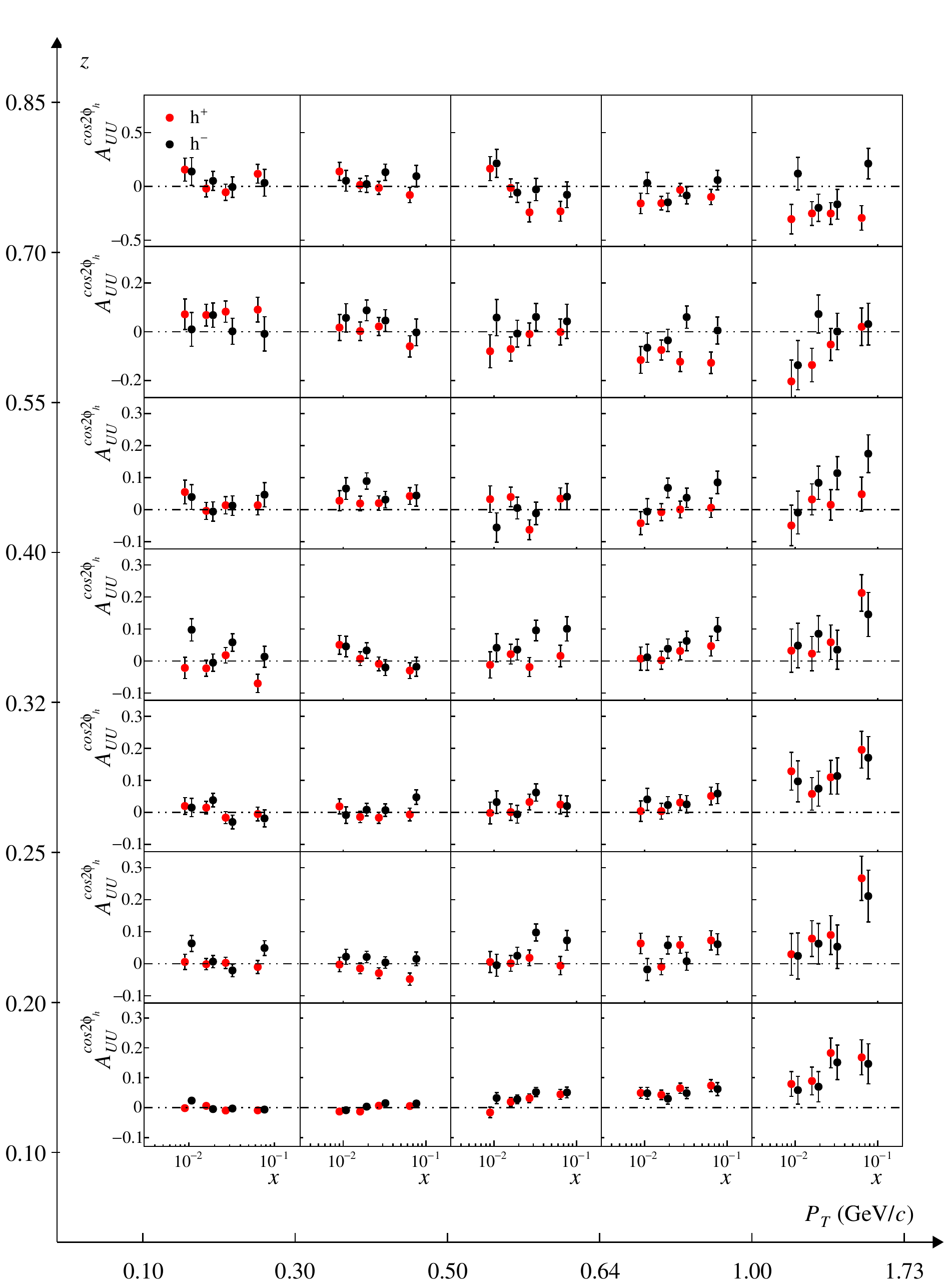}
    \caption{$A_{UU}^{\cos2\fih}$ asymmetry for positive (red) and negative hadrons (black), as a function of $x$ and in bins of $z$ (vertical axis) and $\Pt$ (horizontal axis). The black points are slightly shifted for a better readability.}
    \label{fig:aares3Dcos2}
\end{figure}

\begin{figure}
\captionsetup{width=\textwidth}
\centering
	\includegraphics[width=0.95\textwidth]{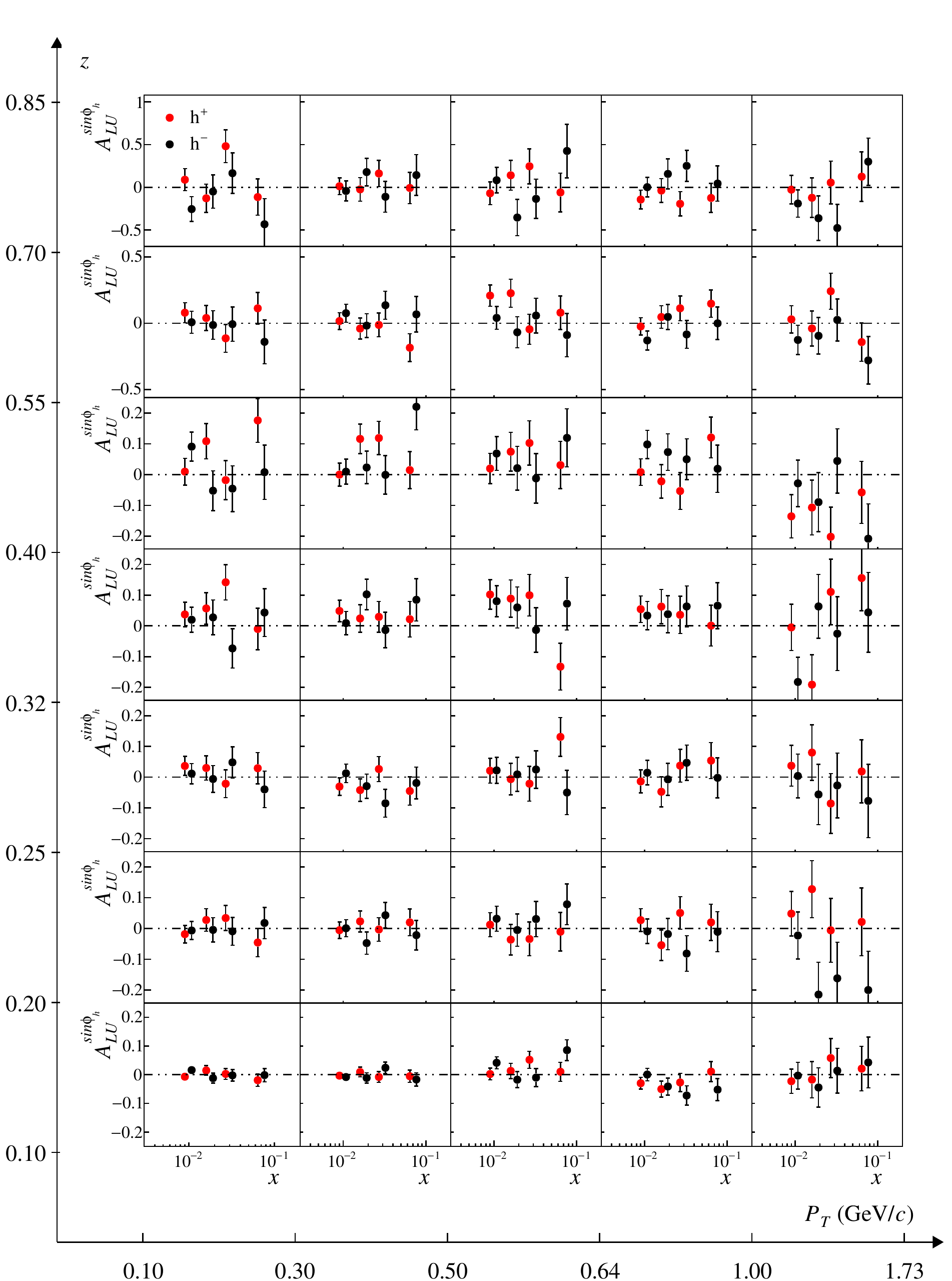}
    \caption{$A_{LU}^{\sin\fih}$ asymmetry for positive (red) and negative hadrons (black), as a function of $x$ and in bins of $z$ (vertical axis) and $\Pt$ (horizontal axis). The black points are slightly shifted for a better readability.}
    \label{fig:aares3Dsin}
\end{figure}

The $A_{UU}^{\cos\fih}$, $A_{UU}^{\cos2\fih}$ and $A_{LU}^{\sin\fih}$ asymmetries have also been measured in 3D bins: the results are shown in Fig.~\ref{fig:aares3Dcos}, \ref{fig:aares3Dcos2} and \ref{fig:aares3Dsin} respectively. 
In particular, the $A_{UU}^{\cos\fih}$ asymmetry shows a clear $x$-dependence in most $z$ bins, increasing with $\Pt$. The dependences on $z$ and $\Pt$ look similar to the ones observed in the 1D case, namely with the same $z$-dependence in all $\Pt$ bins and with the same $\Pt$-dependence at all $z$ (except for the last $z$ bin, where the asymmetry is mostly compatible with zero). The $A_{UU}^{\cos2\fih}$ asymmetry has a more complex dependence on $z$ and $\Pt$. At small $z$, the asymmetry generally increases as a function of $x$ and $\Pt$. For positive hadrons, it changes sign when moving from low to high $z$ at intermediate $\Pt$. As for the $A_{LU}^{\sin\fih}$ asymmetry, no particular trend is observed. 

\begin{figure}[!h]
\captionsetup{width=\textwidth}
    \centering
    \includegraphics[width=\textwidth]{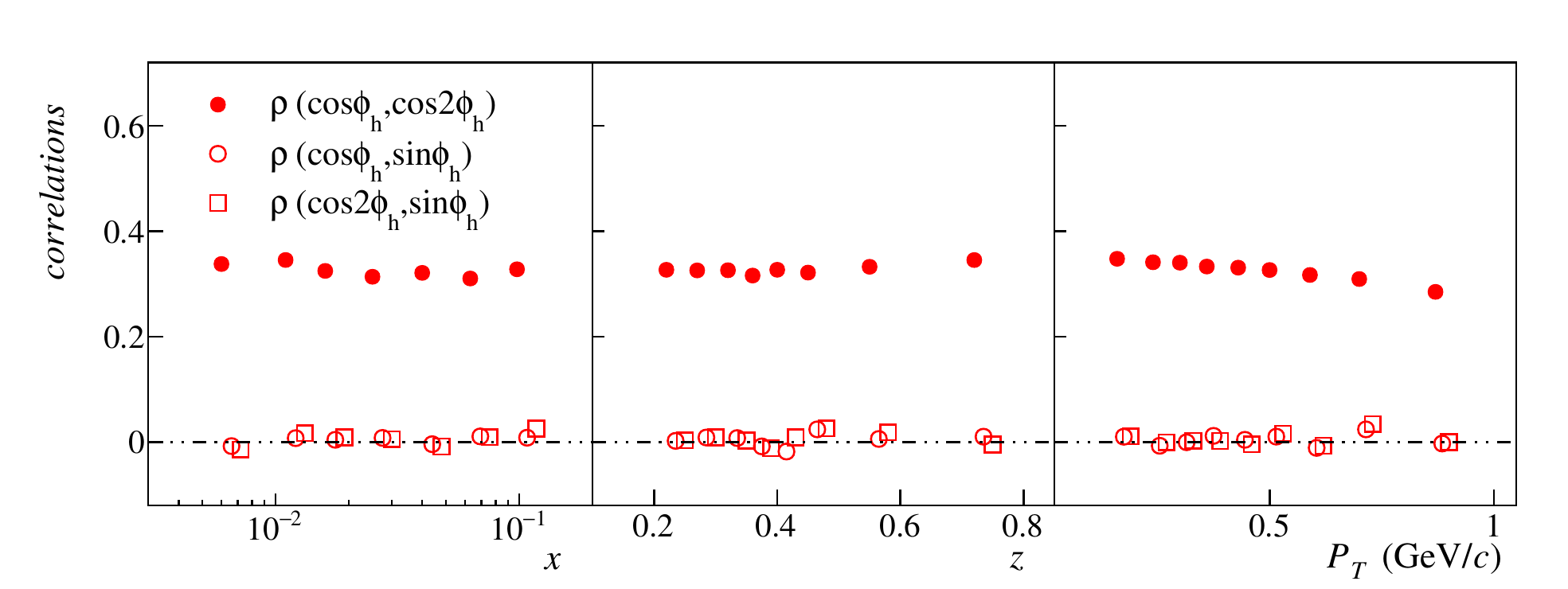}
    \caption{Correlation coefficients  $\rho(\cos\fih,\cos2\fih)$, $\rho(\cos\fih,\sin\fih)$ and $\rho(\cos2\fih,\sin\fih)$ as obtained in the fit of the 1D azimuthal asymmetries. }
    \label{fig:correlations_1D}
\end{figure}

Concerning the correlation among the parameters, the correlation coefficient $\rho(\cos\fih,\cos2\fih)$, $\rho(\cos\fih,\sin\fih)$ and $\rho(\cos2\fih,\sin\fih)$ have been derived from the covariance matrix for the 1D fit. They are shown in Fig.~\ref{fig:correlations_1D} as a function of $x$, $z$ and $P_T$. It is immediate to notice that, while $\rho(\cos\fih,\sin\fih)$ and $\rho(\cos2\fih,\sin\fih)$ are negligible, $\rho(\cos\fih,\cos2\fih)$ is not negligible, its mean value been equal to 0.326$\pm$0.013. Similar results are obtained in the 3D case: in particular, it is $\langle \rho(\cos\fih,\cos2\fih)\rangle^{3D}= 0.316\pm0.003$.

\subsection{Comparisons with the deuteron results}
\label{sect:ch5_results_compD}
Given the different kinematic range covered by the HERMES measurement \cite{Airapetian:2012yg}, it is difficult to compare them with the new COMPASS results. It must however be reminded that the HERMES results for proton and deuteron targets were found compatible. On the contrary, thanks to the choice of the binning, the comparison with the results obtained in COMPASS on a deuteron target should be easy. In Ref.~\cite{COMPASS:2014kcy}, the COMPASS measurements of the azimuthal asymmetries $A_{UU}^{\cos\fih}$ and $A_{UU}^{\cos2\fih}$, both for the 1D and the 3D binning, were not corrected for the contributions of the exclusive hadrons. The impact of the exclusive hadrons on the 3D asymmetries could later be evaluated for the 3D results only and was presented in Ref.~\cite{COMPASS:2019lcm}.

Two comparisons between the COMPASS deuteron and proton results can be performed for:
\begin{enumerate}
    \item the 1D azimuthal asymmetries without the subtraction of the exclusive hadrons contribution;
    \item the 3D azimuthal asymmetries after the correction for the exclusive hadrons.
\end{enumerate}

\begin{figure}[!h]
\captionsetup{width=\textwidth}
    \centering
    \includegraphics[width=0.8\textwidth]{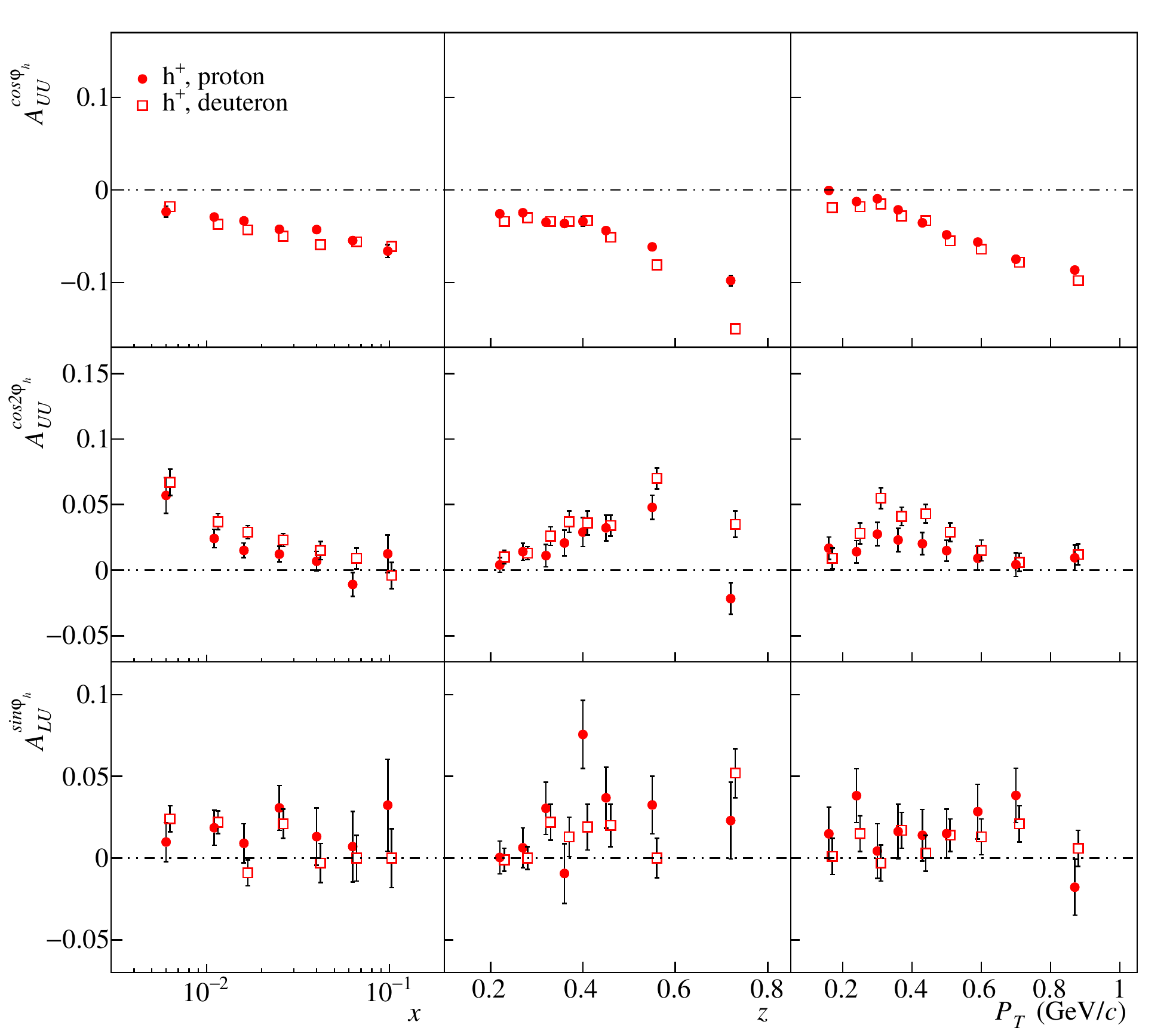}
    \includegraphics[width=0.8\textwidth]{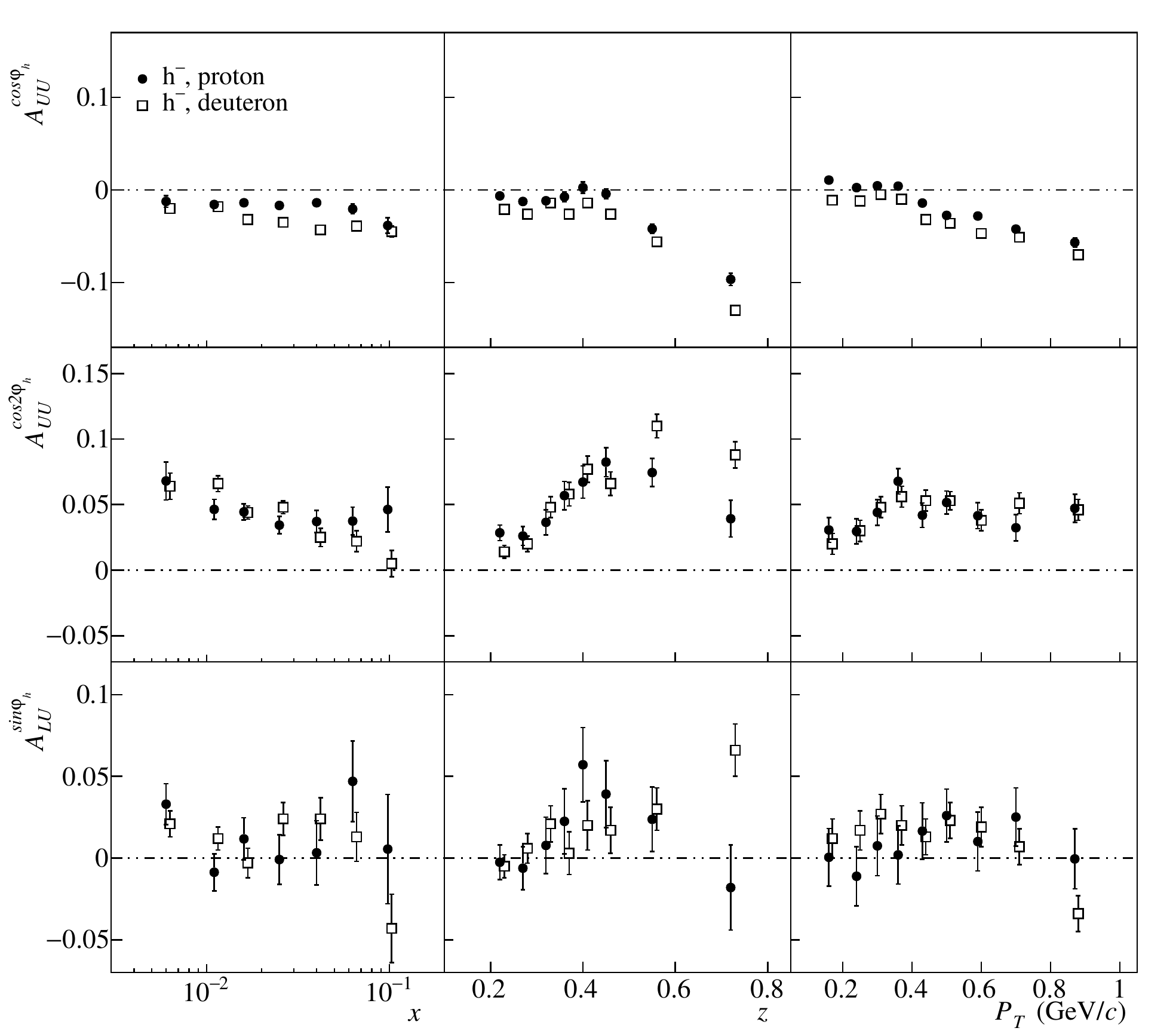}
    \caption{Comparison of the azimuthal asymmetries on a proton and on a deuteron target, separately for a positive (red) and negative hadrons (black), as a function of $x$, $z$ and $P_T$. }
    \label{fig:deut_aa_1D}
\end{figure}

The comparison of the 1D results without the subtraction of the exclusive hadrons is shown in Fig.~\ref{fig:deut_aa_1D}, separately for positive and negative hadrons. While the $A_{LU}^{\sin\fih}$ asymmetries look compatible for the two different targets, some differences can be spotted for $A_{UU}^{\cos\fih}$ and $A_{UU}^{\cos2\fih}$. $A_{UU}^{\cos\fih}$ for positive hadrons on proton looks smaller at intermediate $x$, in the first bin in $\Pt$ and in particular at large $z$; for negative hadrons, the asymmetry on proton looks systematically shifted towards smaller values in all bins of $x$, $z$ and $\Pt$. Conversely, for the $A_{UU}^{\cos2\fih}$ asymmetry a systematic shift towards larger values can be observed for the positive hadrons on deuteron at all $x$ and particularly at high $z$ and small- to intermediate $P_T$, while for negative hadrons the difference is concentrated at high $z$. 

\begin{figure}
\captionsetup{width=\textwidth}
    \centering
    \includegraphics[width=0.49\textwidth]{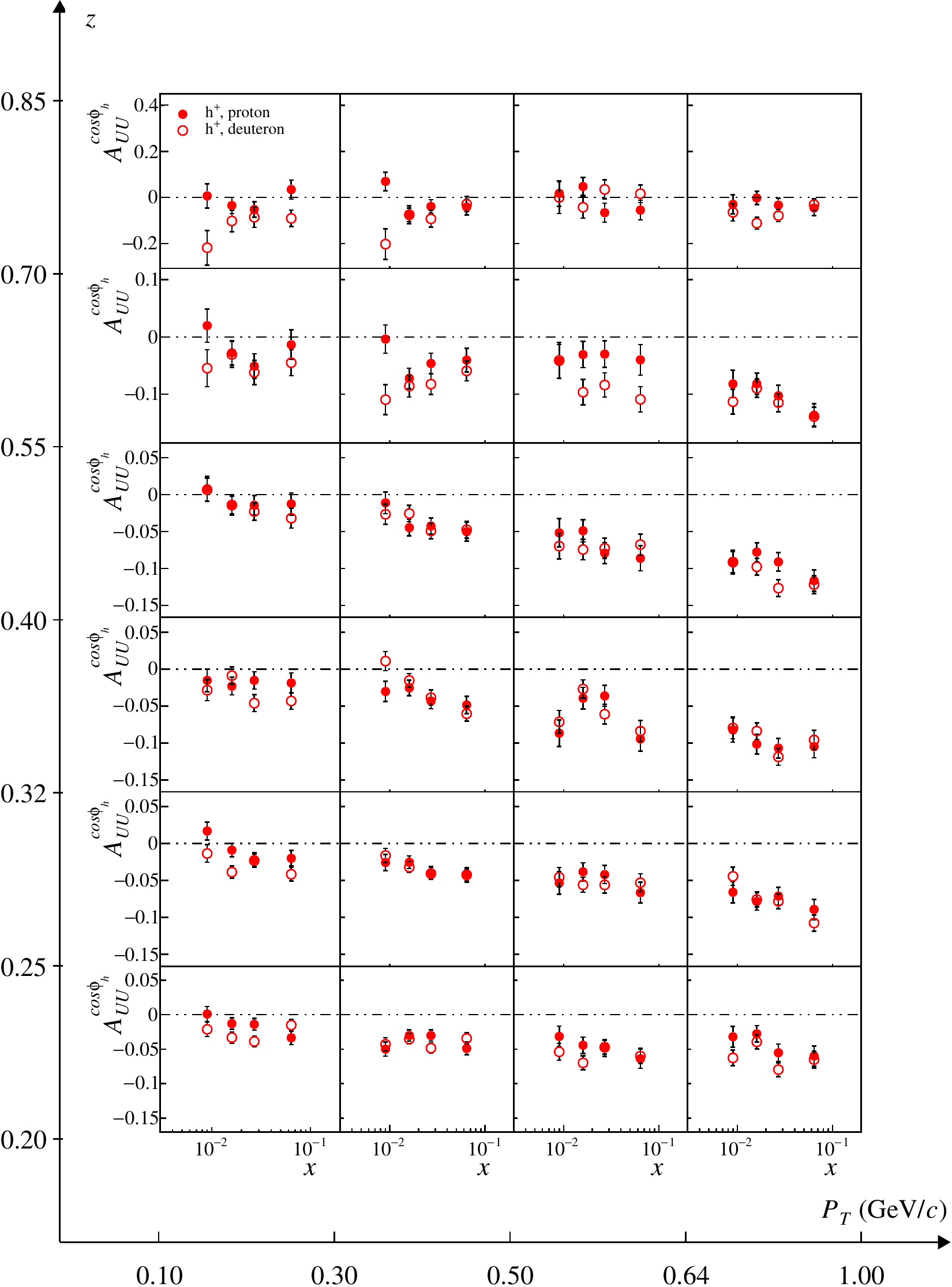}
     \includegraphics[width=0.49\textwidth]{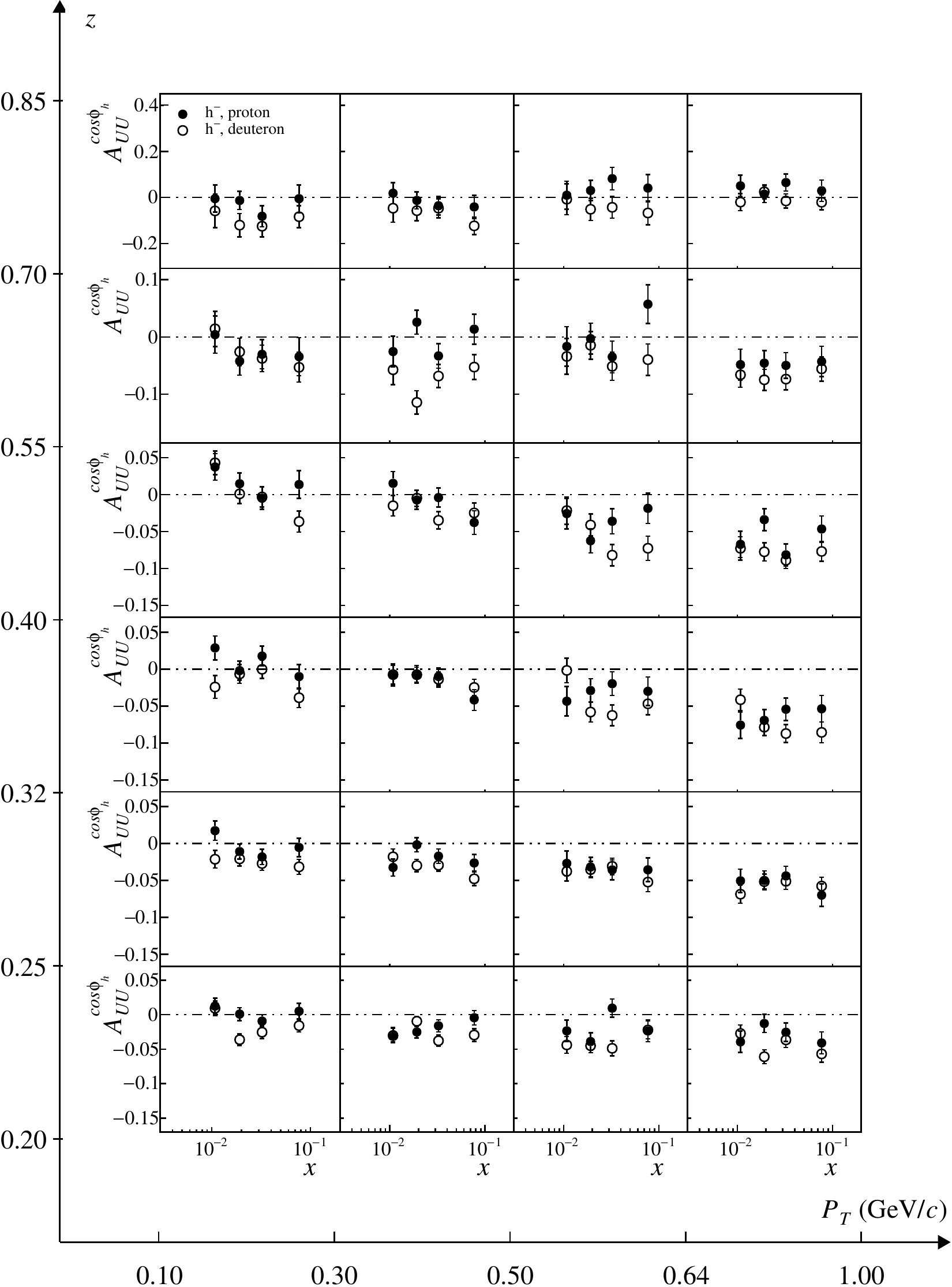}
    \caption{Comparison of the $A_{UU}^{\cos\phi_h}$ asymmetry for positive (left) and negative hadrons (right) between the current results on proton (full points) with the analogous results on deuteron (empty points).}
   \label{fig:f1}
\end{figure}

\begin{figure}
\captionsetup{width=\textwidth}
    \centering
    \includegraphics[width=0.49\textwidth]{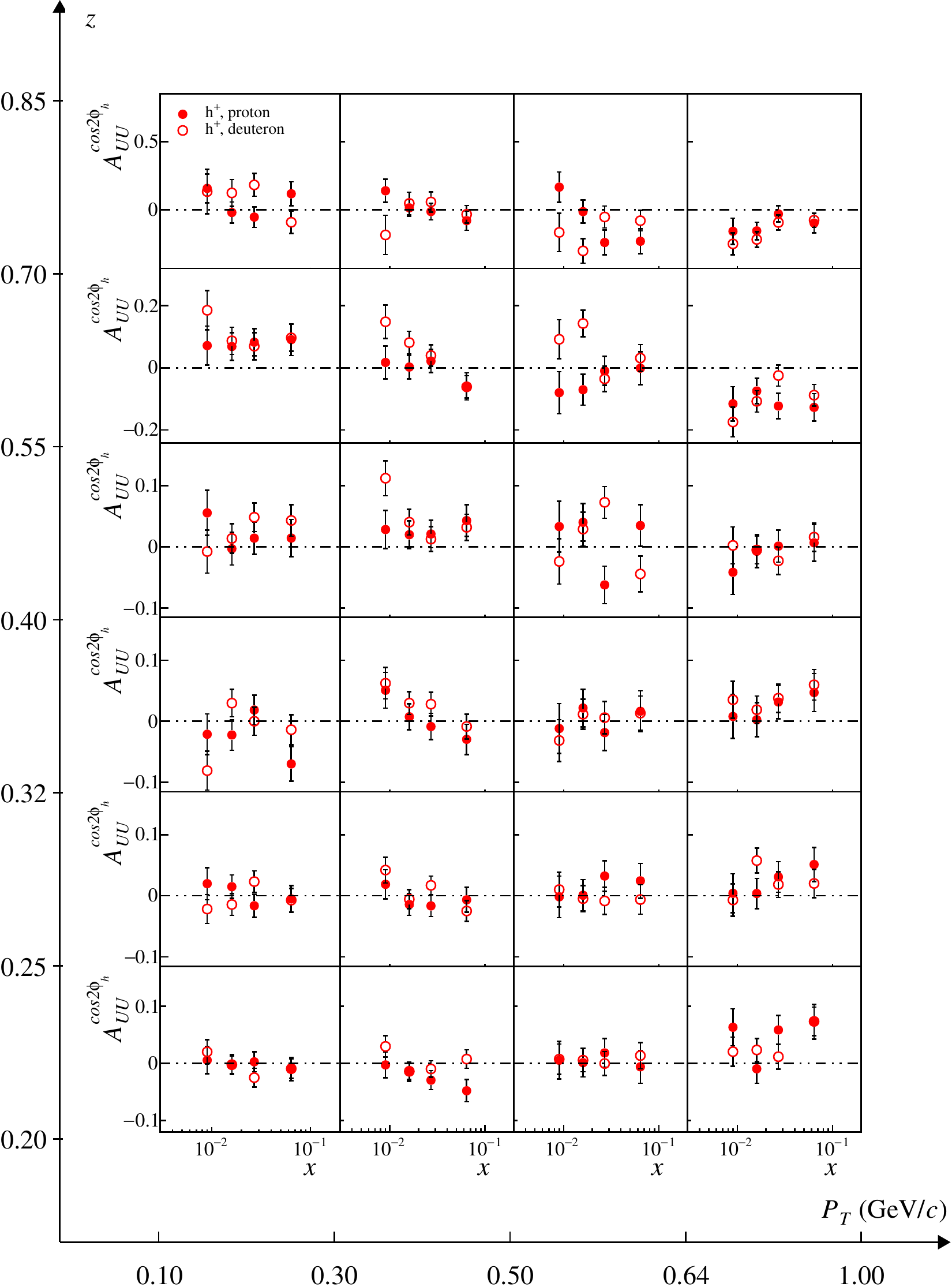}
    \includegraphics[width=0.49\textwidth]{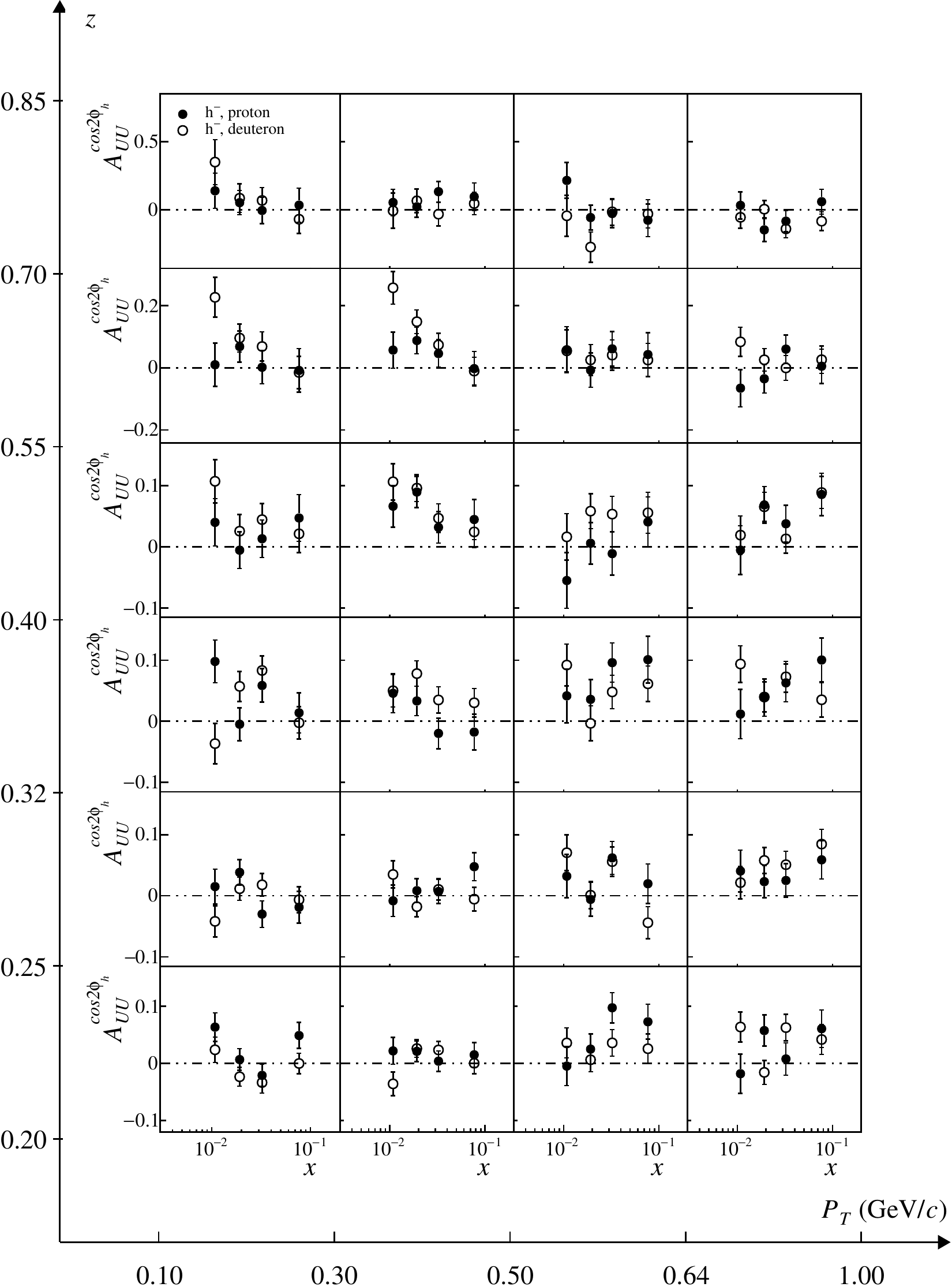}
   \caption{Same as Fig.~\ref{fig:f1}, for the $A_{UU}^{\cos2\fih}$ asymmetry. }
   \label{fig:f2}
\end{figure}

The new 3D results on proton for $A_{UU}^{\cos\fih}$ and $A_{UU}^{\cos2\fih}$, obtained with the correction for the exclusive hadrons, are compared to the ones obtained on deuteron, in the overlapping bins and separately for positive and negative hadrons, in Figg.~\ref{fig:f1} and \ref{fig:f2}. As previously said, the exclusive hadrons have been treated in different ways in the two analyses. The comparison of the $A_{UU}^{\cos\fih}$ is generally good for both positive and negative hadrons; some difference can be seen at high $z$ and at low $x$ and $\Pt$, where the results on proton are closer to zero. As for $A_{UU}^{\cos2\fih}$, the results for proton and deuteron target measurements look in agreement, but the large fluctuations prevent to draw a definite conclusion in absence of a complete phenomenological analysis. 

\begin{figure}[!h]
\captionsetup{width=\textwidth}
    \centering
    \includegraphics[width=0.8\textwidth]{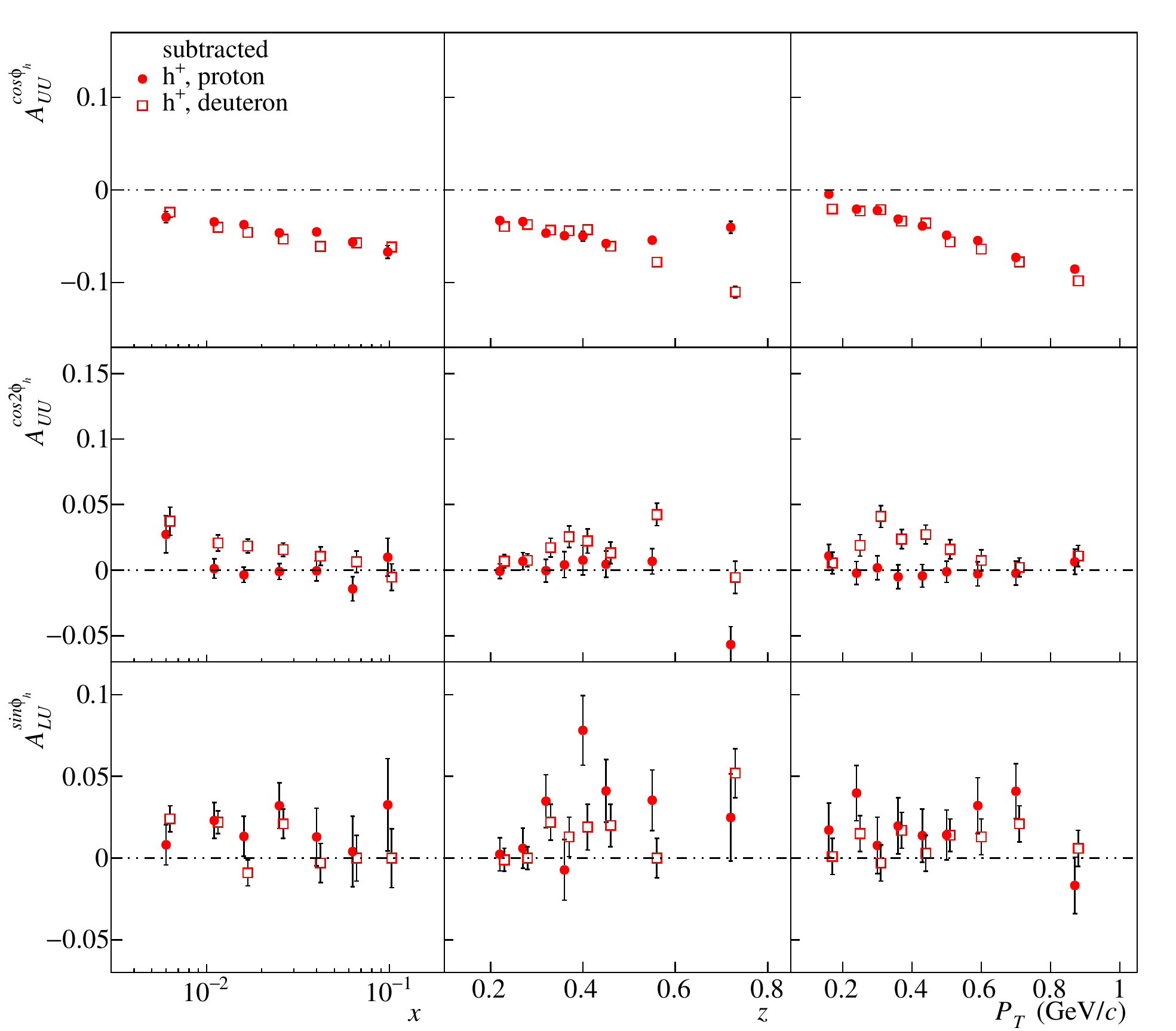}
    \includegraphics[width=0.8\textwidth]{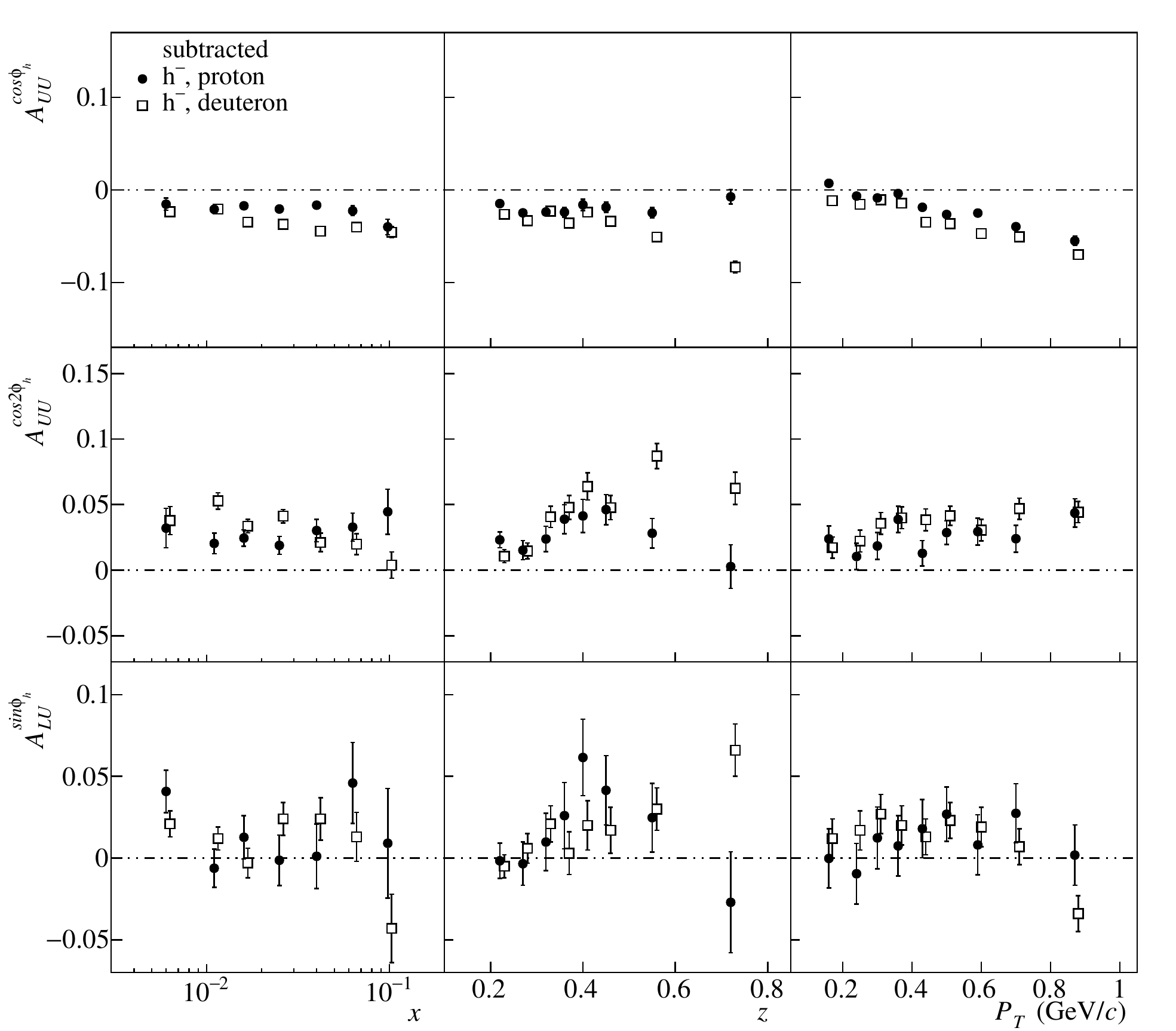}
    \caption{Same as Fig.~\ref{fig:deut_aa_1D}, but for the proton and deuteron results subtracted, or corrected for, the exclusive hadrons contamination. }
    \label{fig:deut_aa_1D_bis}
\end{figure}

The parametrization of the background fraction due to the exclusive hadrons, used in the estimate of the correction to the published 3D deuteron results, has been converted from the 3D treatment to the 1D case by alternatively integrating it over two of the three variables ($x$, $z$ and $\Pt$). In this way, in addition to the corrected 3D asymmetries published in Ref.~\cite{COMPASS:2019lcm}, the 1D corrected asymmetries on deuteron have also been estimated. These new 1D results are compared to the proton results in Fig.~\ref{fig:deut_aa_1D_bis}. It can be observed that, at variance with the proton results, in the deuteron ones the $A_{UU}^{\cos\fih}$ asymmetry shows a more linear trend in $z$, while $A_{UU}^{\cos2\fih}$ as a function of $\Pt$ is not as compatible with zero as in the proton case.

To conclude, the azimuthal asymmetries measured on proton and deuteron results show differences, mainly concentrated at high $z$. Also, both for proton and deuteron the $\Ptsq$-distributions show almost no difference between positive and negative hadrons, while the differences are clear for the azimuthal asymmetries. This appears to be in contrast with the naive interpretation of a strong (flavor-independent) $\aktsq$ contribution to $\aPtsq$ and of the Cahn effect in $A_{UU}^{\cos\fih}$.

\newpage
\clearpage

\section{Further studies of the kinematic dependences}
\label{sect:ch5_kinematics}
\begin{figure}[h!]
\captionsetup{width=\textwidth}
\begin{center}
	\includegraphics[width=0.49\textwidth]{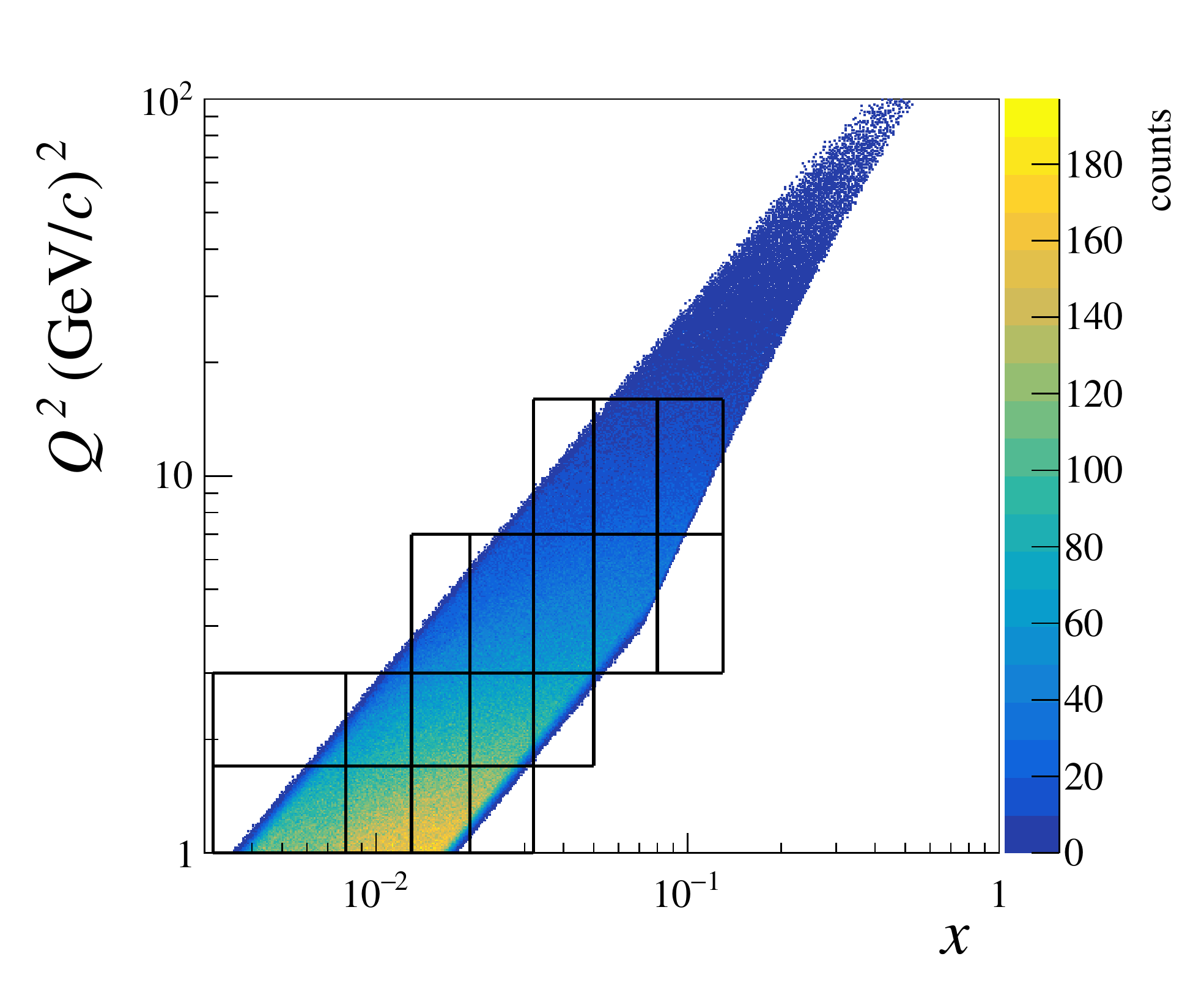}
	\includegraphics[width=0.49\textwidth]{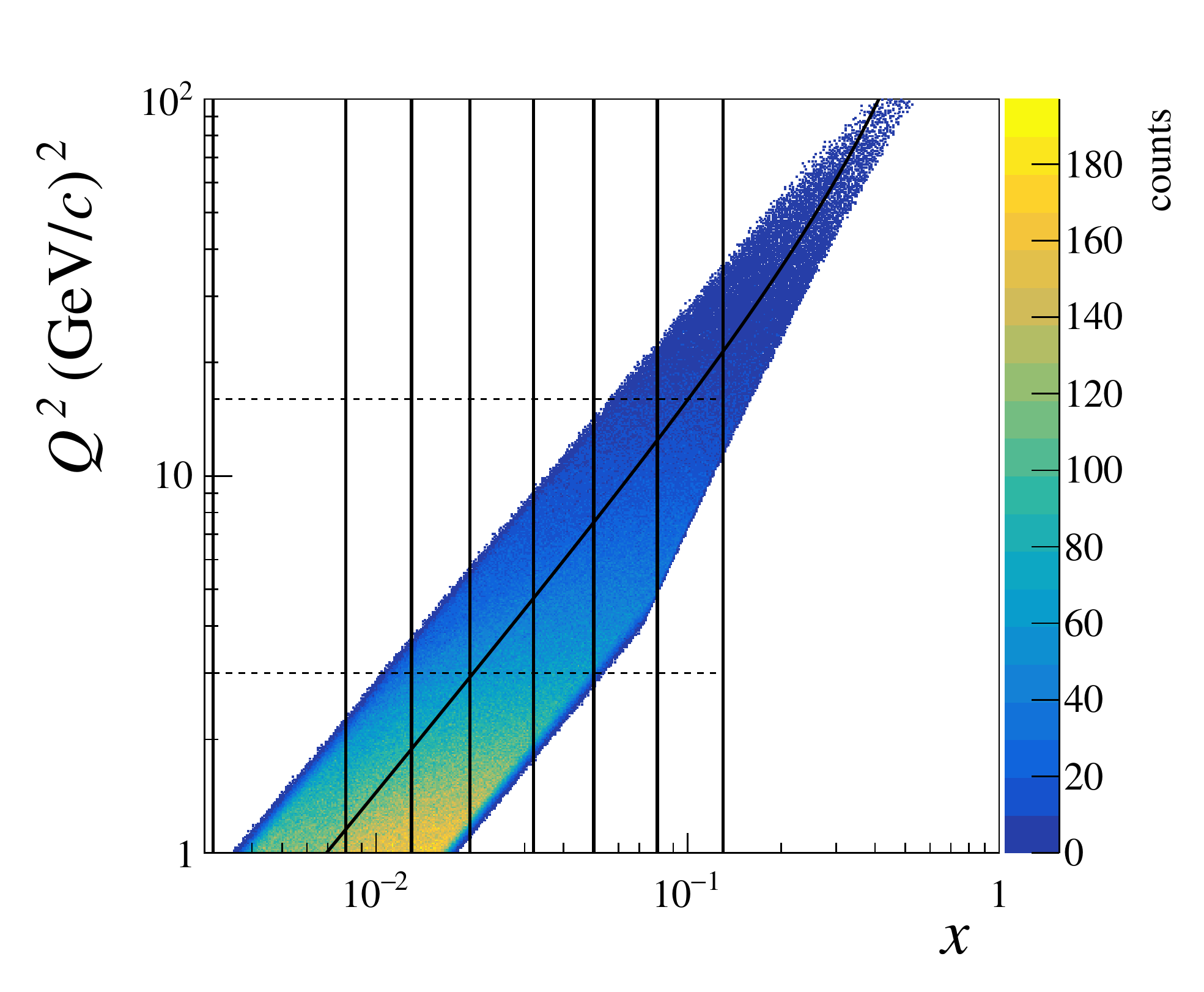}
		\caption{Drawn on top of the $x-\Qsq$ correlation plot: the binning in $x$ and $\Qsq$ used for the measurement of the azimuthal asymmetries in four $\Qsq$ bins (left) and the binning used for the same measurement in two $W$ bins (right), with an inclined line at $W=12$~GeV/$c^2$.}
\label{fig:xQ2_aa2}
\end{center}
\end{figure}

In this Section, we investigate further the kinematic dependences of the azimuthal asymmetries focusing on their $Q^2$- and $W$-dependences. The $\Qsq$-dependence has been investigated measuring the azimuthal asymmetries in four $\Qsq$ bins, limited at 1.0, 1.7, 3.0 and 16.0~(GeV/$c$)$^2$, and in bins of $x$, $z$ or $\Pt$ as in the standard 1D analysis. The bins in $x$ and $\Qsq$ are shown in Fig.~\ref{fig:xQ2_aa2} (left): due to the correlation between these two variables, the binning in $\Qsq$ naturally limits the accessible $x$ range, which is different in each $\Qsq$ bin. Also, the mean value of $x$, for a given $x$ bin, can be different at different $\Qsq$. 

The $W$-dependence has been studied by making two $W$ bins, below and above 12~GeV/$c^2$, with the same $x$, $z$ or $\Pt$ as in the $\Qsq$ case. Due to the statistics limitations, only two bins in $\Qsq$ have been done. The binning in $x$ and $\Qsq$ is shown in Fig.~\ref{fig:xQ2_aa2} (right), where the inclined line corresponds to $W=12$~GeV/$c^2$. Again, the mean values of $x$ in a given bin can change according to the $W$ bin. 

The results for $A_{UU}^{\cos\fih}$ and $A_{UU}^{\cos2\fih}$ are discussed in the next Sections. The corresponding studies for the $A_{LU}^{\sin\fih}$ did not give any particular insight, and the results are not shown here.

\subsection{$\Qsq$-dependence}
\begin{figure}[h!]
\captionsetup{width=\textwidth}
    \centering
    \includegraphics[width=0.95\textwidth]{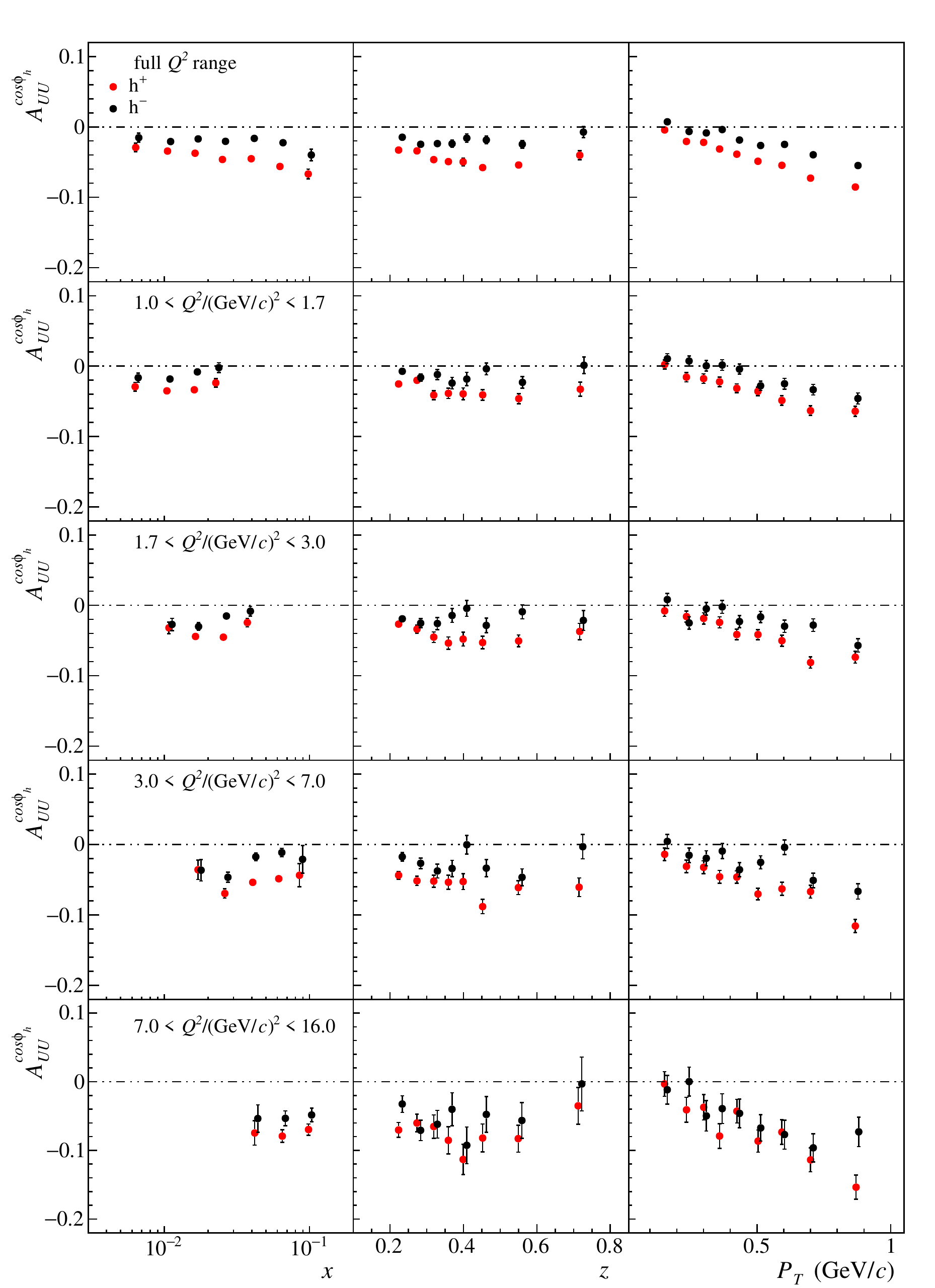}
    \caption{$A_{UU}^{\cos\phi_h}$, as extracted in the 1D approach, for positive (red) and negative hadrons (black) as a function of $x$, $z$ and $P_T$ in the full $Q^2$ range (first row) and in four bins of $Q^2$ (second to fifth row). }
    \label{fig:1D_acos_4Q2}
\end{figure}

The dependence on $Q^2$ of the azimuthal asymmetries $A_{UU}^{\cos\phi_h}$ and $A_{UU}^{\cos2\phi_h}$ has been studied by dividing the final hadron sample into four $Q^2$ bins, while keeping the standard binning in $x$, $z$ and $P_T$. Figure~\ref{fig:1D_acos_4Q2} shows the $A_{UU}^{\cos\fih}$ asymmetry as a function of $x$, $z$ and $\Pt$ (left to right) integrated over $\Qsq$ (first row) and in four $\Qsq$ bins (second to fifth row, for increasing $\Qsq$). It shows an increase with $Q^2$. This is particularly clear when the asymmetry is looked at as a function of $P_T$ or $x$. The increase in the size of the asymmetry appears to be in contrast with the naive expectation from the twist-3 Cahn effect. If the transverse momenta $k_T$ and $p_\perp$ are assumed flavor-independent, using the Gaussian approximation the contribution of the Cahn effect to $A_{UU}^{\cos\phi_h}$ reads:

\begin{equation}
    A_{UU|Cahn}^{\cos\phi_h} = - \frac{2zP_T\langle k_T^2 \rangle}{Q\langle P_T^2 \rangle}
\label{eq:cahn}
\end{equation}
so that a decrease in the size of the asymmetry for increasing $Q^2$ is, in first approximation, expected. The observation that this is not the case unveils a richer kinematic dependence of the various ingredients and, in particular, a strong dependence of $\aktsq$ on $\Qsq$. It is however possible that other effects are dominant. While the Boer-Mulders contribution to the asymmetry is also expected to be decreasing with $Q^2$, other terms at the same- or higher twist could be the reason for the observed trends. The radiative effects, also, are expected to have a larger impact for increasing $\Qsq$: thus, the observed trend in $\Qsq$ could be partially due to them. As already said, no correction for the radiative effects has been applied to the data in the present work. However, preliminary studies on their impact, done with the DJANGOH Monte Carlo, indicate that the $A_{UU}^{\cos\fih}$ could be reduced by a factor two in the last $\Qsq$ bin, while being almost unaffected in the first $\Qsq$ bin. In this case, the radiative effects would be responsible for a non-negligible part of the observed $\Qsq$-dependence, but not for all of it.  \\

\begin{figure}[h!]
\captionsetup{width=\textwidth}
    \centering
    \includegraphics[width=0.3\textwidth]{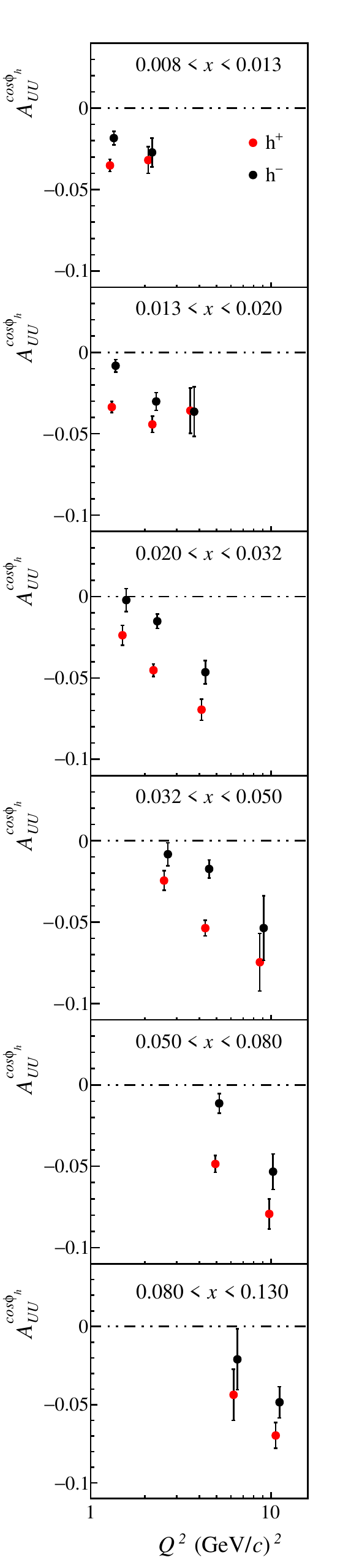}
    \includegraphics[width=0.3\textwidth]{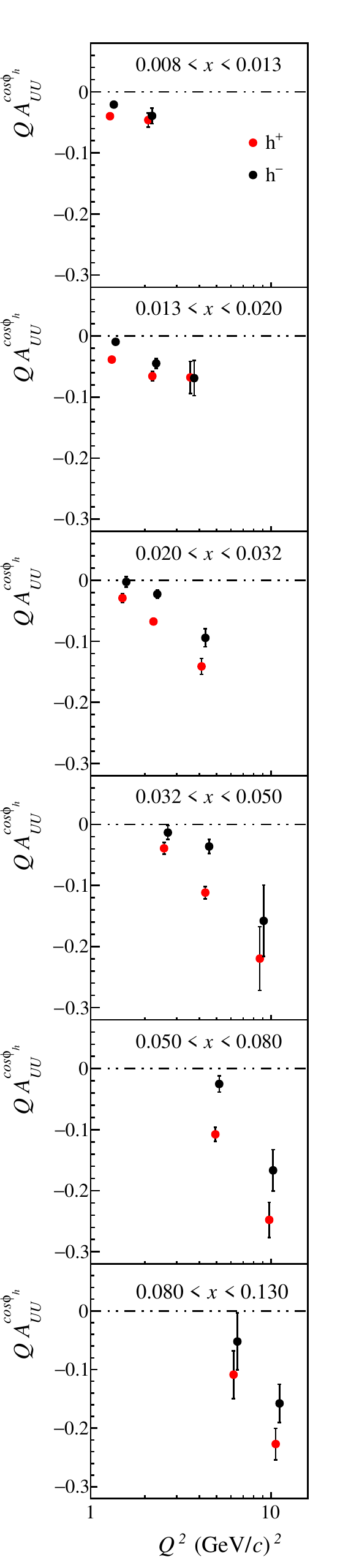}
    \includegraphics[width=0.3\textwidth]{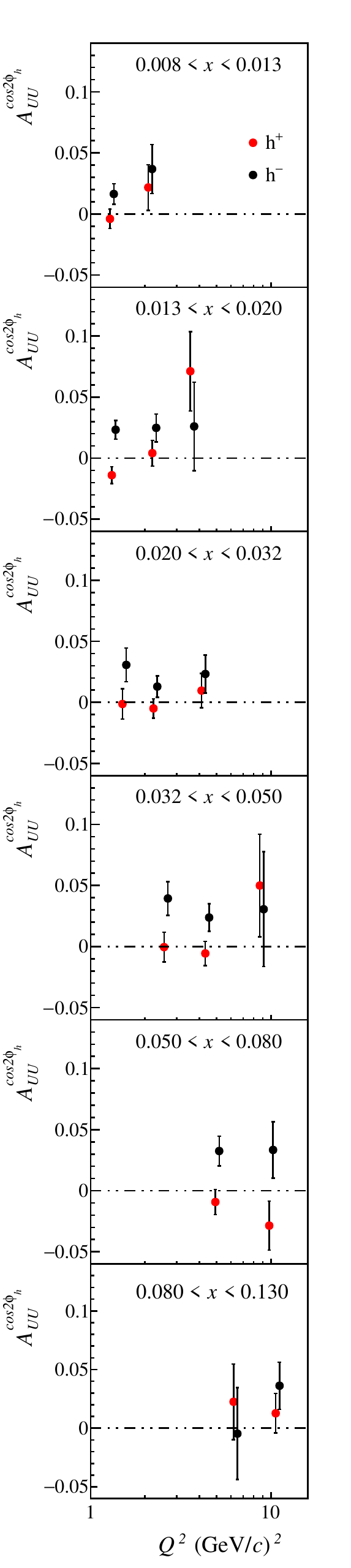}
    \caption{$A_{UU}^{\cos\phi_h}$, $QA_{UU}^{\cos\phi_h}$ and $A_{UU}^{\cos2\phi_h}$ as extracted in the 1D approach, for positive (red) and negative hadrons (black) as a function of $Q^2$ in six bins of $x$. }
    \label{fig:1D_acos_4Q2_Q2}
\end{figure}

Also, the clear difference between positive and negative hadrons at low $Q^2$ (or in the integrated case, dominated by the low $Q^2$ region) is gradually lost moving to higher $Q^2$. This, despite the moderate difference in the values of the $Q^2$ between the first and the last bin in $Q^2$. Note that the QED radiative effects should be the same for positive and negative hadrons, and should not justify the larger increase, in absolute value, of the asymmetry for $h^-$ with respect to the $h^+$. Finally, the dependence on $\Pt$ is almost linear in all $\Qsq$ bins, while the $z$-dependence for $h^-$ changes, becoming similar to that for $h^+$ at high $\Qsq$. The $x$-dependence seems to change in the different bins of $\Qsq$, in particular in the two highest $\Qsq$ bins and for $h^-$. It is also interesting to look at the asymmetry as a function of $\Qsq$ in the different $x$ bins. This has been done rearranging the values of the asymmetries shown in Fig.~\ref{fig:1D_acos_4Q2}, and plotting them as a function of $\langle \Qsq \rangle$ in the different $x$ bins. This is done in Fig.~\ref{fig:1D_acos_4Q2_Q2}, where the left column shows $A_{UU}^{\cos\phi_h}$ as a function of $\Qsq$ in the different $x$ bins. The $A_{UU}^{\cos\phi_h}$ asymmetry, multiplied by $Q \approx \sqrt{Q^2}$, to get rid of the $1/Q$ factor appearing in the definition of this asymmetry at twist-3, is shown in the second column. \\

\begin{figure}[h!]
\captionsetup{width=\textwidth}
    \centering
    \includegraphics[width=0.955\textwidth]{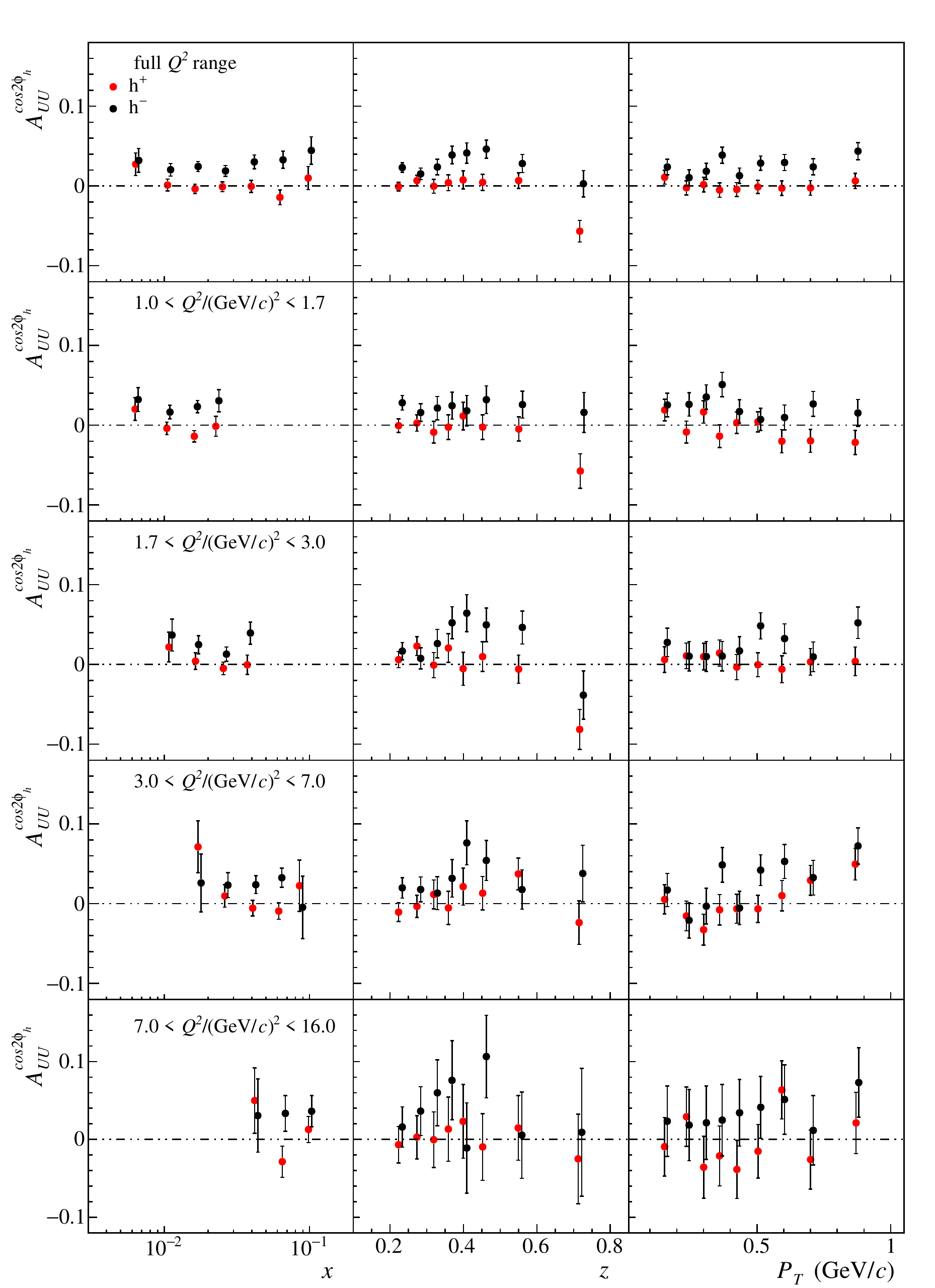}
        \caption{$A_{UU}^{\cos2\phi_h}$, as extracted in the 1D approach, for positive (red) and negative hadrons (black) as a function of $x$, $z$ and $P_T$ in the full $Q^2$ range (first row) and in four bins of $Q^2$ (second to fifth row). }
    \label{fig:1D_acos2_4Q2}
\end{figure}

\begin{figure}[h!]
\captionsetup{width=\textwidth}
    \centering
    \includegraphics[width=0.75\textwidth]{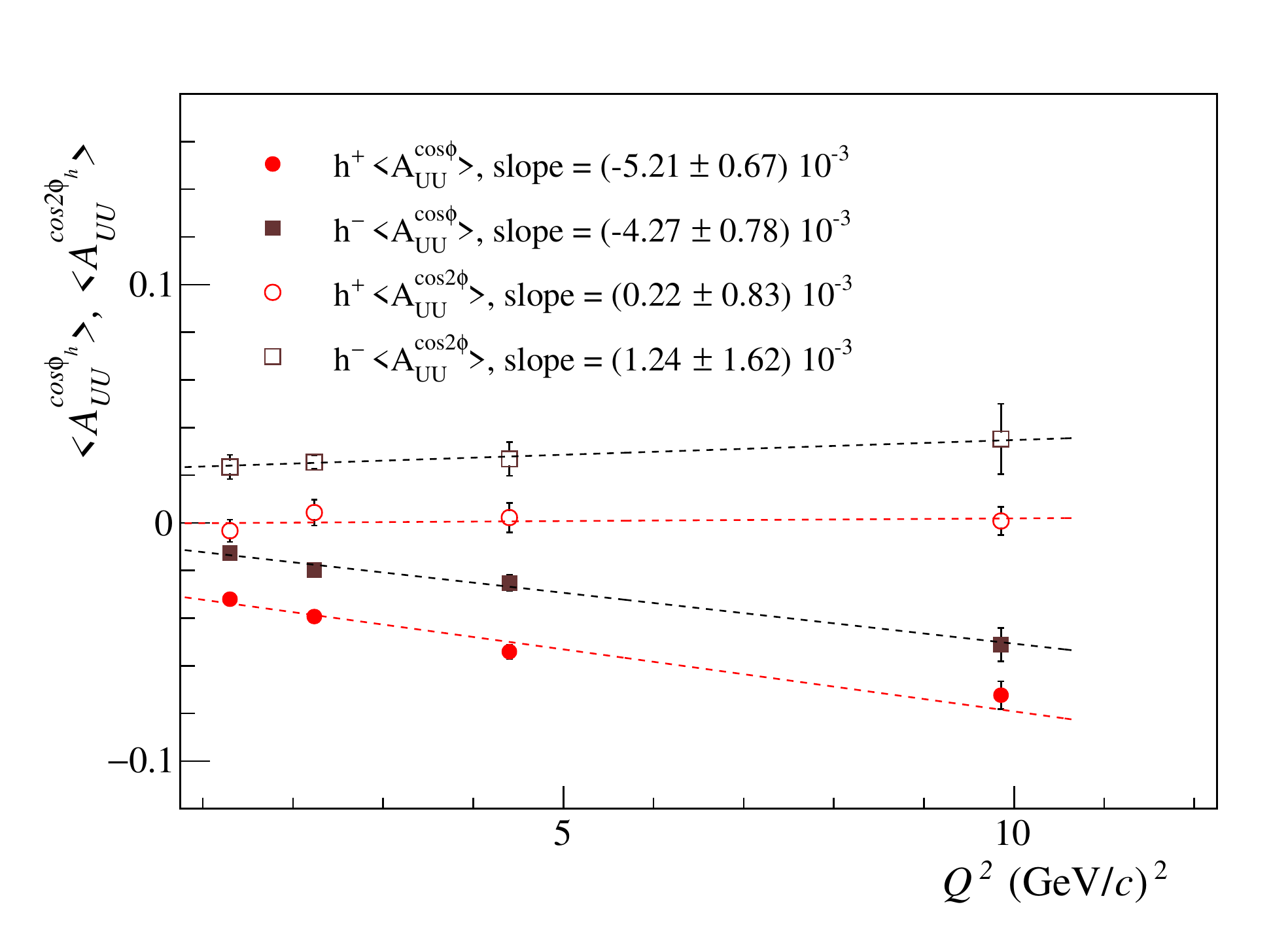}
    \caption{Dependence on $\Qsq$ of the mean azimuthal asymmetries for positive and negative hadrons. The mean values of $x$ in the four $\Qsq$ bins are: $\langle x \rangle$=0.013, $\langle x \rangle$=0.022, $\langle x \rangle$=0.041, $\langle x \rangle$=0.073.}
    \label{fig:Q2dep_intA}
\end{figure}

The case of the $A_{UU}^{\cos2\phi_h}$ is different: as seen in Fig.~\ref{fig:1D_acos2_4Q2}, there is no evidence for changes with $\Qsq$ in the size of the asymmetries nor in the difference between the asymmetries for positive and negative hadrons. The compatibility with zero of this asymmetry for positive hadrons does not look confined to the region at low $Q^2$, where most of the statistics sit. These observations are confirmed looking at the asymmetry as a function of $\Qsq$ in different $x$ bins, shown in Fig.~\ref{fig:1D_acos_4Q2_Q2} (third column). The dependence on $\Qsq$ of the measured asymmetries is presented, in a different way, in Fig.~\ref{fig:Q2dep_intA}: there, the mean values of the asymmetries, obtained from the average of the values in the $z$ bins, are plotted against $\Qsq$ and fitted with a line. Of course, the other dependences, like that on $x$, are hidden is such a representation. In fact, each $\Qsq$ value corresponds to a different mean value of $x$, so that the $\Qsq$-dependence is essentially obtained assuming no $x$-dependence. Still, the linear dependence is impressive. As already observed, the dependence on $\Qsq$ is negligible in the case of the $A_{UU}^{\cos2\fih}$ asymmetry, being well visible for the $A_{UU}^{\cos\fih}$ asymmetry. As for the $A_{UU}^{\cos2\fih}$ asymmetry, the preliminary studies indicate almost no impact from the radiative effects. \\ 

As a final study, the 3D extraction of the azimuthal asymmetries, namely their measurement by simultaneously binning in $x$, $z$ and $\Pt$, has been performed separately in the two $\Qsq$ bins: $1.0~\mathrm{(GeV/}c\mathrm{)}^2 < \Qsq < 3.0~\mathrm{(GeV/}c\mathrm{)}^2$ and $3.0~\mathrm{(GeV/}c\mathrm{)}^2 < \Qsq < 16.0~\mathrm{(GeV/}c\mathrm{)}^2$. Given the low statistics, the results (not shown here) are characterized by large fluctuations, and no definite conclusions could be drawn out of them. 

\subsection{$W$-dependence}
\begin{figure}[h!]
\captionsetup{width=\textwidth}
    \centering
    \includegraphics[width=0.49\textwidth]{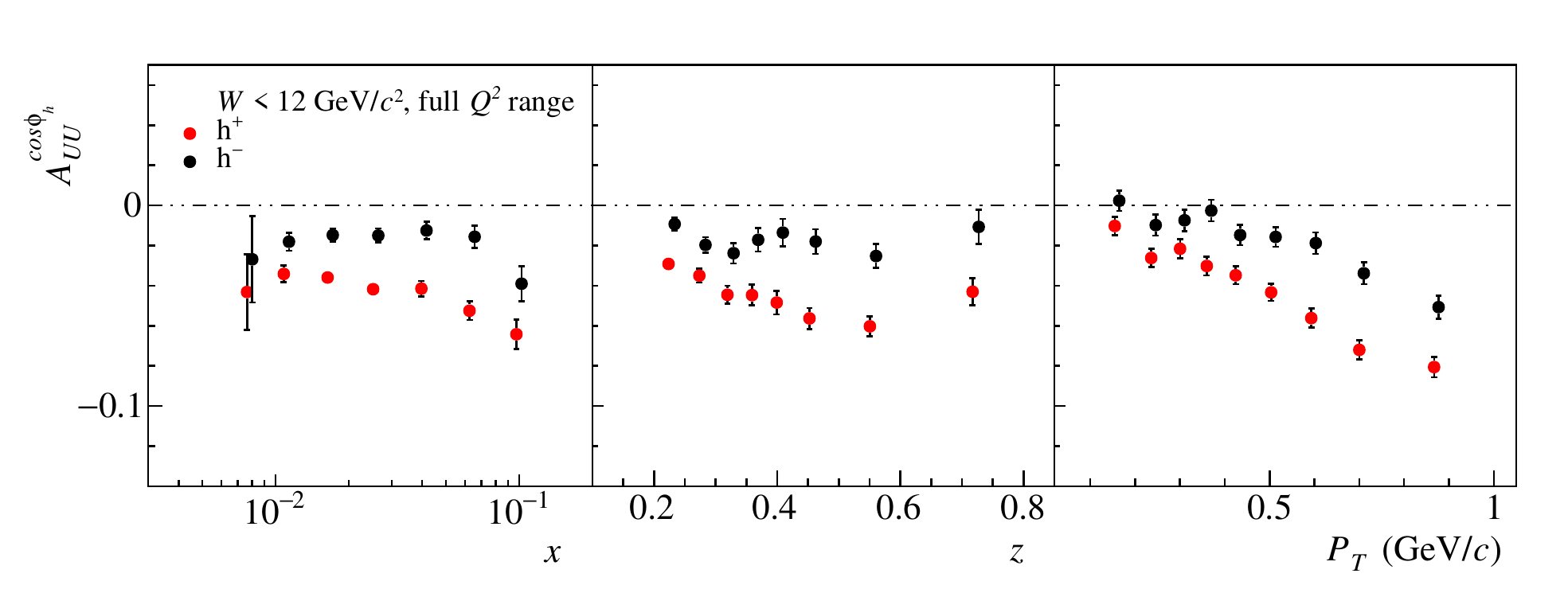}
    \includegraphics[width=0.49\textwidth]{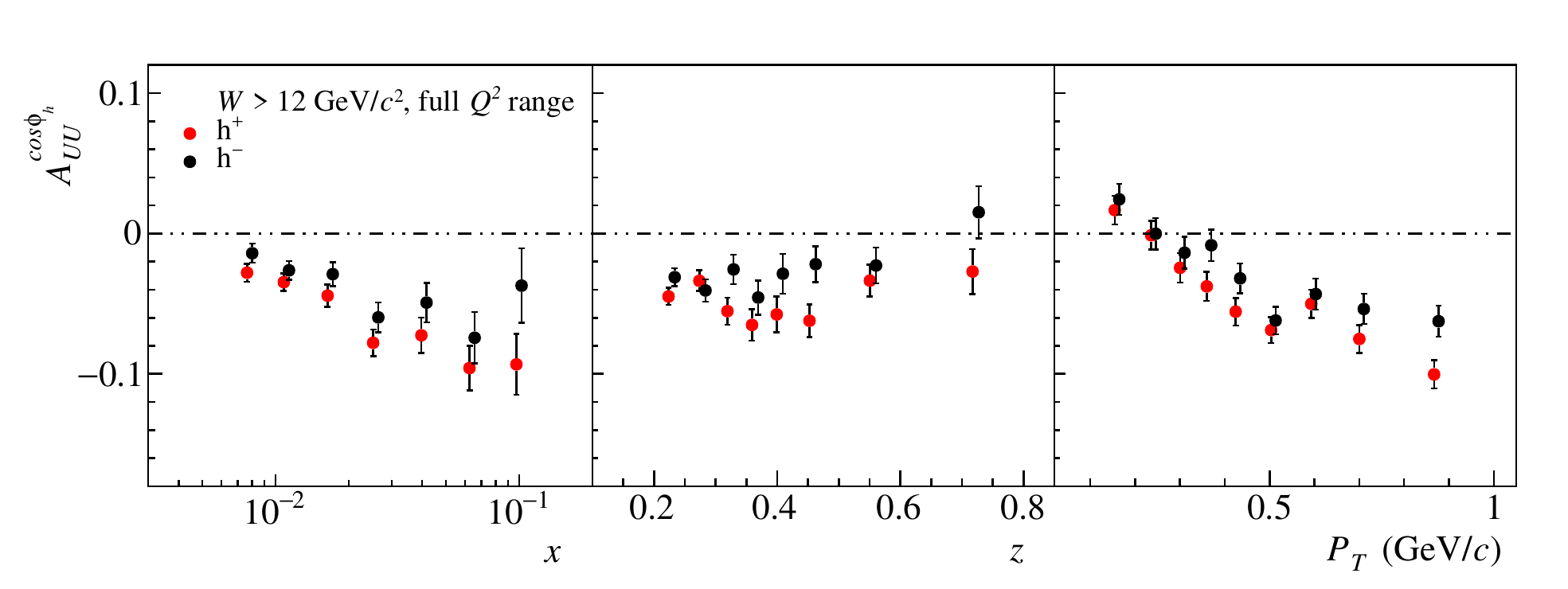}

    \includegraphics[width=0.49\textwidth]{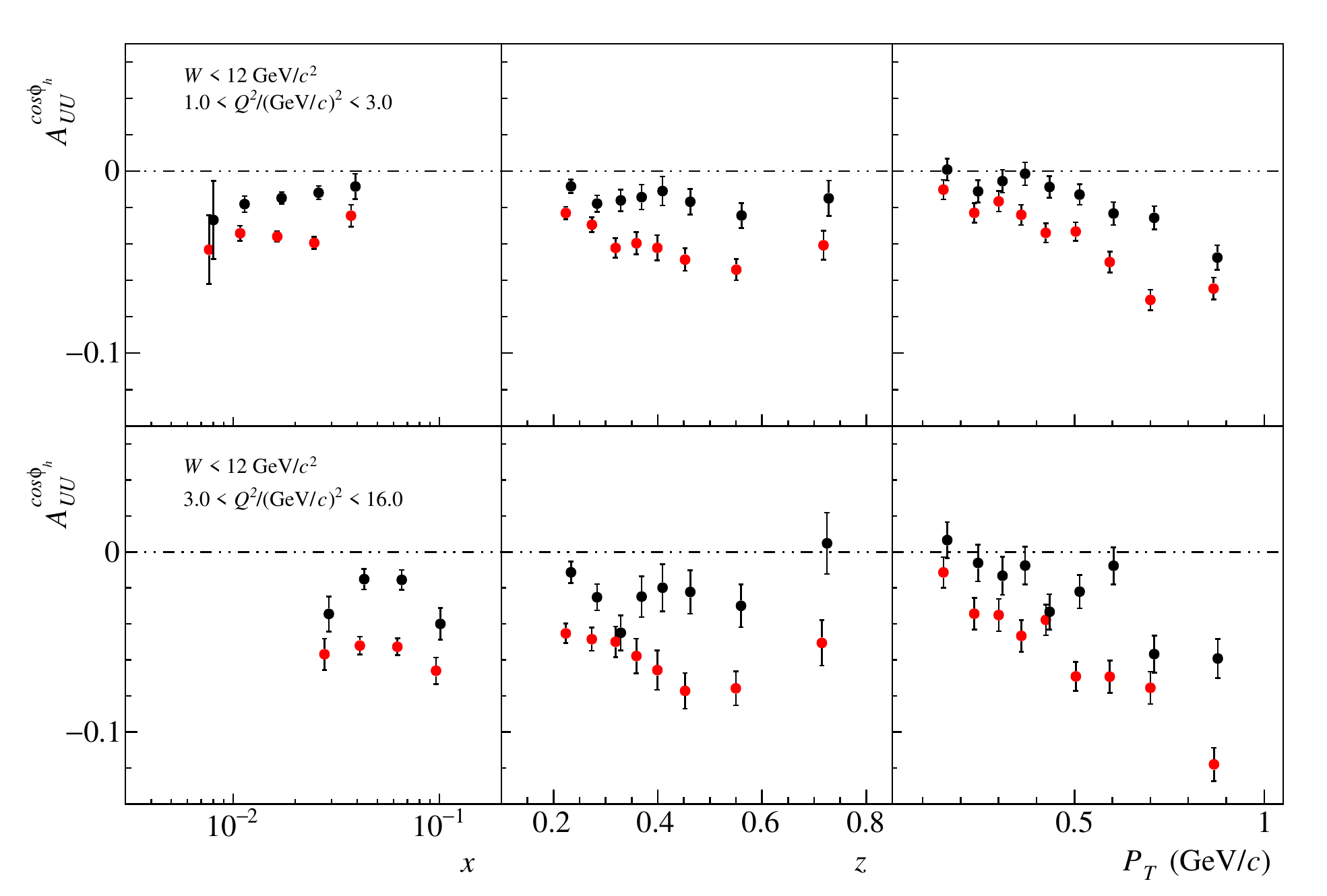}
    \includegraphics[width=0.49\textwidth]{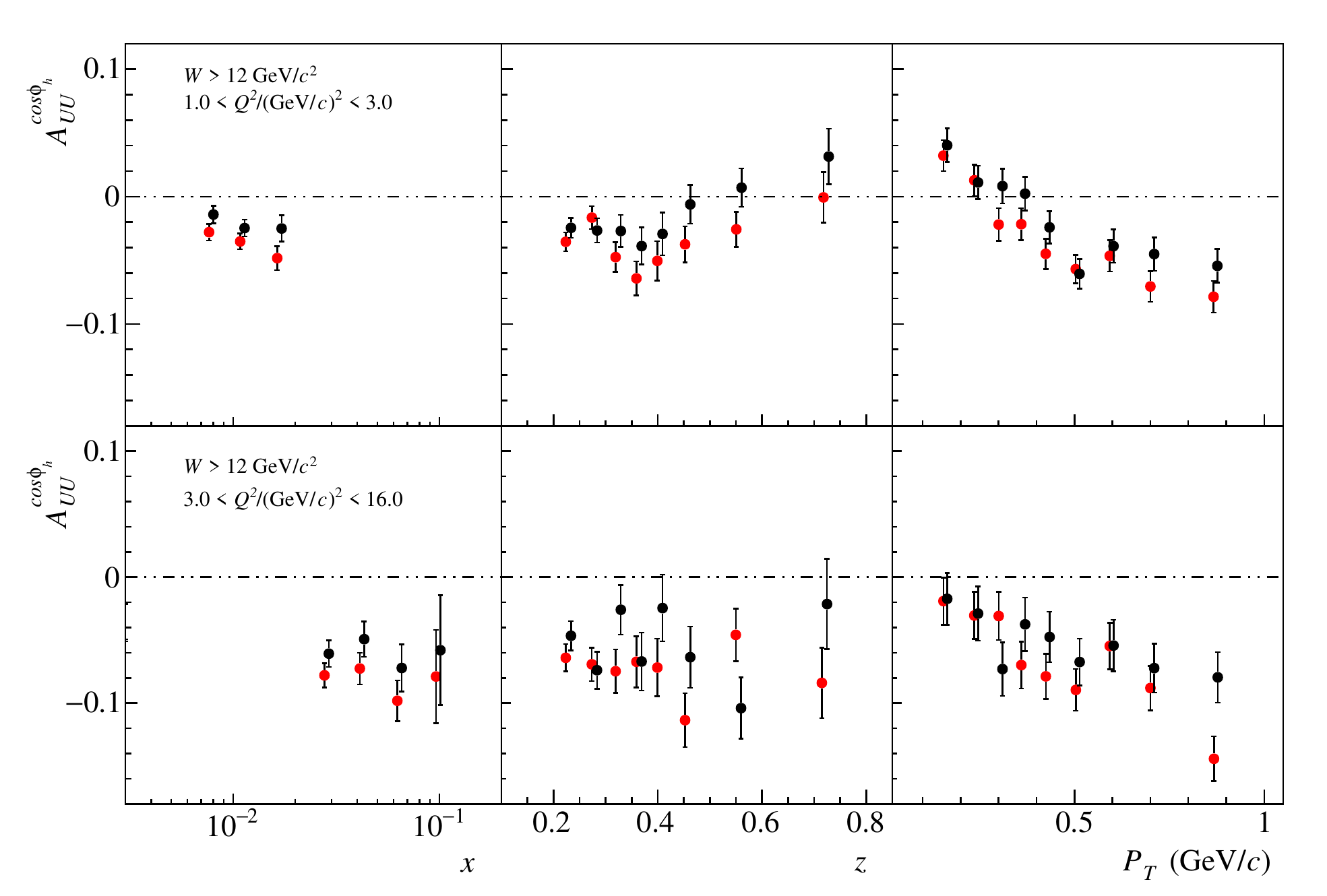}

    \caption{$A_{UU}^{\cos\phi_h}$ at low $W$ (left) and at high $W$, as extracted in the 1D approach, for positive (red) and negative hadrons (black) as a function of $x$, $z$ and $P_T$ in the full $Q^2$ range (top) and in two bins of $Q^2$ (bottom). }
    \label{fig:1D_acos_Wsl12_2Q2}
\end{figure}

It is well known that the DIS process is fully characterized by two variables only: given $x$ and $\Qsq$, $W$ is fixed by the relation $W^2=M^2 + \Qsq (1-x)/x$. Thus, the study of the kinematic dependence of the asymmetries on $x$ and $\Qsq$ exhausts the problem. Nevertheless, looking at the $W$-dependence constitutes a different point of view.
The dependence on $W$ of the azimuthal asymmetries has been studied by making two bins in $W$ ($W$ smaller or larger than 12 GeV/$c^2$), in which the asymmetries have been measured as a function of $x$, $z$ or $\Pt$ using the standard 1D binning. In addition, in both $W$ bins the asymmetries have been measured in two bins of $\Qsq$. Note theta, in the low and high $W$ ranges, the $\Qsq$ binning introduces an effective cut on the accessible $x$ ranges, which turn out to be essentially complementary. 
The $A_{UU}^{\cos\phi_h}$ asymmetry is shown in Fig.~\ref{fig:1D_acos_Wsl12_2Q2} in the small (left) and large $W$ bins (right). In each plot, $A_{UU}^{\cos\phi_h}$ is given for both positive and negative hadrons in the full $Q^2$ range and in the two selected $Q^2$ bins. The dependence on $\Qsq$ is strong at low $W$ but almost negligible at high $W$. Also, the differences between positive and negative hadrons turn out to be sizable at low $W$ while being small at high $W$, where the asymmetries for $h^-$ become almost as large as the ones for $h^+$. The mean values of $\Qsq$ at low $W$ are: $\langle \Qsq \rangle=1.69$~(GeV/$c$)$^2$ in the first $\Qsq$ bin, $\langle \Qsq \rangle=5.39$~(GeV/$c$)$^2$ in the second $\Qsq$ bin and $\langle \Qsq \rangle=2.73$~(GeV/$c$)$^2$ integrating over $\Qsq$. At high $W$, $\langle \Qsq \rangle=1.69$~(GeV/$c$)$^2$ in the first $\Qsq$ bin, $\langle \Qsq \rangle=6.05$~(GeV/$c$)$^2$ in the second $\Qsq$ bin and $\langle \Qsq \rangle=3.08$~(GeV/$c$)$^2$ integrating over $\Qsq$.

In the previous Section, the dependences of the azimuthal asymmetries on $\Qsq$ (Fig.~\ref{fig:Q2dep_intA}) have been obtained, integrating over $W$ and assuming no $x$-dependence. In particular, $A_{UU}^{\cos\fih}$ turned out to decrease linearly with $\Qsq$. It is interesting to check if that dependence is sufficient to explain the measured asymmetries in bins of $W$. If the differences observed in the two bins of $W$ are just a consequence of the $\Qsq-W$ correlation, the $\Qsq$ dependence should in fact be sufficient. This check has been performed by evolving the mean asymmetries from the first bin in $\Qsq$ to the second and comparing the result with the measured asymmetries, according to the expression:

\begin{equation}
    \langle A \rangle (\Qsq_1) = \langle A \rangle (\Qsq_0) + \frac{\diff~ \langle A \rangle}{\diff \Qsq} (\Qsq_1 - \Qsq_0) 
\end{equation}
where $\Qsq_0$ ($\Qsq_1$) is the value of $\Qsq$ in the lowest (highest) $\Qsq$ bin. The values of the measured and expected asymmetries, in bins of $\Qsq$ and $W$, are shown in Tab.~\ref{tab:W_test}. Both at low and high $W$, the measured and expected $A_{UU}^{\cos\fih}$ asymmetries are in fair agreement within the statistical uncertainties, thus giving no clear indication of a pure dependence on $W$.

\begin{table}[]
\captionsetup{width=\textwidth}
    \centering
    \begin{tabular}{c|c|c|c}

    \textbf{\textit{low $W$}} & \textbf{$1<\Qsq$ / (GeV/$c$)$^{2}<3$ } & \textbf{$3<\Qsq$ / (GeV/$c$)$^{2}<16$} & \textbf{estimated $\langle A \rangle_1$} \\ 
     & $\langle \Qsq \rangle_0 = 1.69$  (GeV/$c$)$^{2}$ & $\langle \Qsq \rangle_1 = 5.39$  (GeV/$c$)$^{2}$ &    \\ \hline
    $\langle A_{UU}^{\cos\phi_{h^+}} \rangle$ & --0.036 $\pm$ 0.002 & --0.055 $\pm$ 0.003 & --0.055 $\pm$ 0.003 \\
    $\langle A_{UU}^{\cos\phi_{h^-}} \rangle$ & --0.014 $\pm$ 0.002 & --0.021 $\pm$ 0.003 & --0.030 $\pm$ 0.004 \\ \hline
    & & & \\
    \textbf{\textit{high $W$}} & \textbf{$1<\Qsq$ / (GeV/$c$)$^{2}<3$ } & \textbf{$3<\Qsq$ / (GeV/$c$)$^{2}<16$} & \textbf{estimated $\langle A \rangle_1$} \\ 
     & $\langle \Qsq \rangle_0 = 1.69$  (GeV/$c$)$^{2}$ & $\langle \Qsq \rangle_1 = 6.05$  (GeV/$c$)$^{2}$ &    \\ \hline
    $\langle A_{UU}^{\cos\phi_{h^+}} \rangle$ & --0.035 $\pm$ 0.004 & --0.071 $\pm$ 0.006 & --0.058 $\pm$ 0.007 \\
    $\langle A_{UU}^{\cos\phi_{h^-}} \rangle$ & --0.021 $\pm$ 0.004 & --0.055 $\pm$ 0.007 & --0.040 $\pm$ 0.008 \\ 
      
    \hline
    \end{tabular}
    \caption{Mean value of the $A_{UU}^{\cos\fih}$ asymmetries, measured in bins of $W$ and in the two $\Qsq$ bins, and expected asymmetry in the highest $W$ bin based on the dependence on $\Qsq$ only.}
    \label{tab:W_test}
\end{table}

The same conclusion can be drawn when inspecting the azimuthal asymmetries integrated over $\Qsq$, in the two bins of $W$. In this case, if the observed dependence on $W$ is just due to the $\Qsq-W$ correlation, one expects:

\begin{equation}
\begin{split}
    \langle A \rangle (W_1) & = \langle A \rangle (W_0) + \frac{\diff~ \langle A \rangle}{\diff W} (W_1 - W_0)   \\
    & = \langle A \rangle (W_0) + \frac{2Wx}{1-x} \frac{\diff~ \langle A \rangle}{\diff \Qsq} (W_1 - W_0)   \\
\end{split}
\end{equation}
The estimated mean asymmetries are given in Tab.~\ref{tab:W_test2} together with the measured mean asymmetries in the two $W$ bins: again, the estimated asymmetries are compatible with the measured ones, thus giving no strong indication of a $W$ dependence on top of the one induced by the $\Qsq$ one.

\begin{table}[]
\captionsetup{width=\textwidth}
    \centering
    \begin{tabular}{c|c|c|c}

     & \textbf{$W<12$ GeV/$c^2$} & \textbf{$W>12$ GeV/$c^2$} & \textbf{estimated $\langle A \rangle_1$} \\ 
     & $\langle W \rangle_0 = 9.67$  GeV/$c^2$ & $\langle W \rangle_1 = 13.79$ GeV/$c^2$ &    \\ \hline
    $\langle A_{UU}^{\cos\phi_{h^+}} \rangle$ & --0.041 $\pm$ 0.002 & --0.046 $\pm$ 0.003 & --0.054 $\pm$ 0.004 \\
    $\langle A_{UU}^{\cos\phi_{h^-}} \rangle$ & --0.016 $\pm$ 0.002 & --0.030 $\pm$ 0.004 & --0.027 $\pm$ 0.004 \\   
    \hline
    \end{tabular}
    \caption{Mean value of the asymmetries, measured in bins of $W$ and integrated over $\Qsq$, and expected asymmetry in the highest $\Qsq$ bin based on the dependence on $\Qsq$ only.}
    \label{tab:W_test2}
\end{table}

\section{Interpretation of the results}
\label{sect:ch5_interpretation}
The interpretation of the $A_{UU}^{\cos\fih}$ and $A_{UU}^{\cos2\fih}$ azimuthal asymmetries is a complicated task. The complication arises, e.g., from the difficulties in taking into account, e.g., the possible contribution of higher twists terms, which are expected to contribute, or the TMD evolution, not yet well established. Still, it is interesting to see what the results presented here suggest in a very simple approach, naming assuming the validity of the relation $\vPt = z\vkt + \vpperp$ and of factorization and Gaussian Ansatz, as already done for the interpretation of the $\Ptsq$-distributions. Note that, again, we assume negligible the radiative corrections. Also, no attempt is done to explain the complex kinematic dependences described in the previous Sections, which would require a much deeper phenomenological analysis. Only the 1D results are used here. 

In the first part of the Section, the $A_{UU}^{\cos\fih}$ asymmetry is considered. First, $\aktsq$ is extracted assuming that the only contribution to the asymmetry originates from the Cahn effect and comparing with the extraction from the $\Ptsq$-distributions. The possible impact of the Boer-Mulders effect is then evaluated. For this work, possible constraints on $\aktsq$ due to the phase-space limitations, proven to be sizable particularly at lower beam energies \cite{Boglione:2011wm,Aghasyan:2014zma} have not been taken into account.

The second part of this Section is dedicated to the $A_{UU}^{\cos1\fih}$ asymmetry and to the insight on the Boer-Mulders function. After some general considerations, information on the Boer-Mulders function is obtained using the method of the difference asymmetries.

\subsection{Evaluation of $\aktsq$ from $A_{UU}^{\cos\fih}$}
\begin{figure}
    \centering
    \captionsetup{width=\textwidth}
    \includegraphics[width=\textwidth]{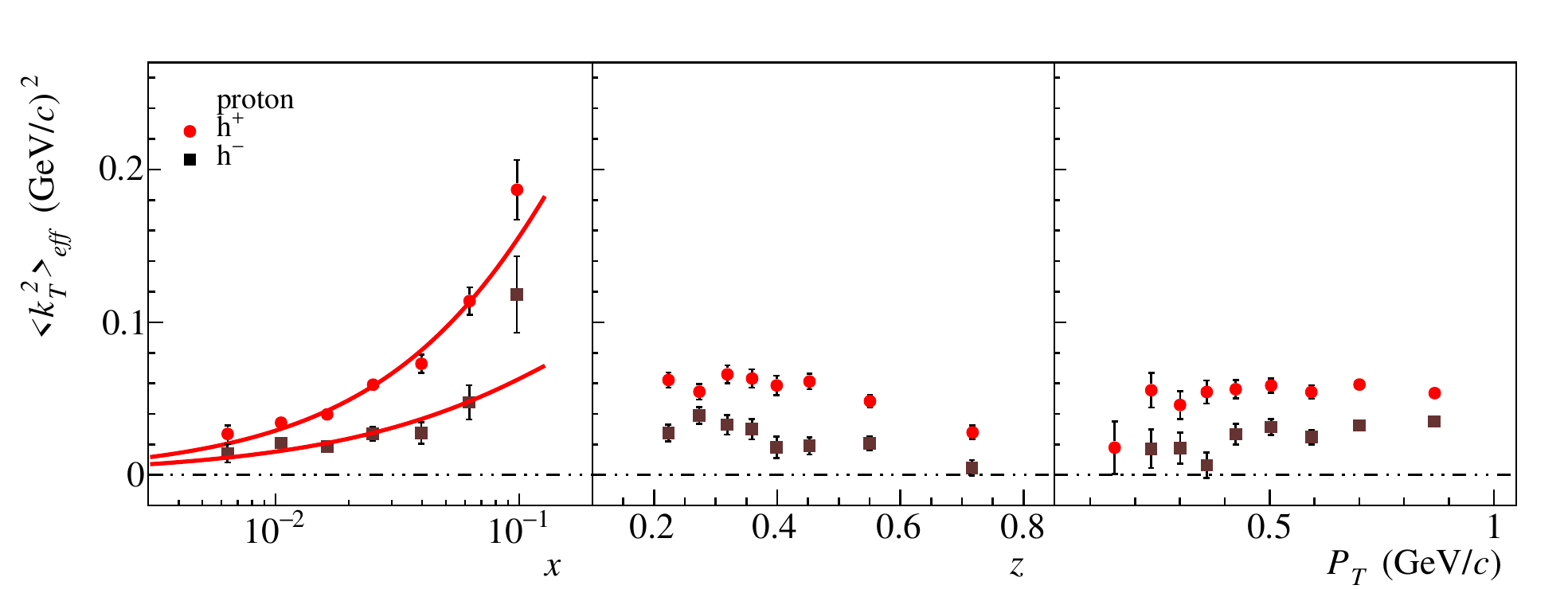}
    \caption{Extraction of $\aktsq_{eff}$ from the 1D $A_{UU}^{\cos\fih}$ asymmetries on proton, as a function of $x$, $z$ and $\Pt$. }
    \label{fig:ckT_proton}
\end{figure}

\begin{figure}
    \centering
    \captionsetup{width=\textwidth}
    \includegraphics[width=\textwidth]{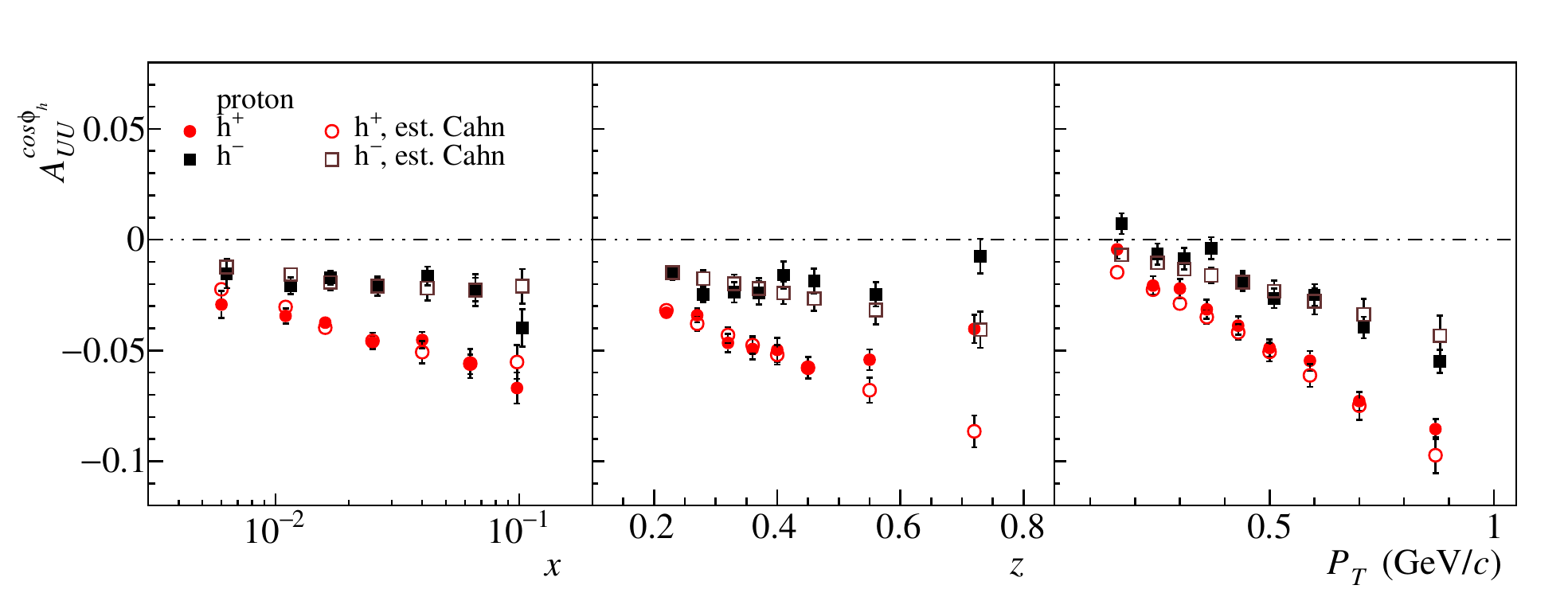}
    \caption{Comparison of the estimated Cahn asymmetry, evaluated using the fitted $\aktsq_{eff}(x)$, for the 1D $A_{UU}^{\cos\fih}$ asymmetries on proton.}
    \label{fig:ctot2_proton}
\end{figure}

As a first step, it is interesting to evaluate to what extent the $A_{UU}^{\cos\fih}$ asymmetries in $\cos\fih$ can be modeled considering only the Cahn mechanism, i.e. through an effective $\aktsq$, hereafter referred to as $\aktsq_{eff}$, such that: 
\begin{equation}
A_{UU}^{\cos\fih} = -\frac{2z\Pt\aktsq_{eff}}{Q\aPtsq} \Longrightarrow \aktsq_{eff}=-\frac{Q\aPtsq}{2z\Pt A_{UU}^{\cos\fih}}.    
\end{equation}
As seen in Ch.~\ref{sect:ch1_aa_definition}, this simple expression for the Cahn asymmetry has been derived assuming flavor-independence. This is the first, very relevant problem: if the $A_{UU}^{\cos\fih}$ is only due to the Cahn effect and if $\aktsq$ and $\apperpsq$ are flavor-independent, the asymmetries should be the same for positive and negative hadrons, while they are not, both on a proton and a deuteron target. Still, it is interesting to extract $\aktsq$ for positive and negative hadrons separately, and see how different they are. Using the above expression and the 1D results for $A_{UU}^{\cos\fih}$, one obtains the values of $\aktsq_{eff}$ shown in Fig.~\ref{fig:ckT_proton}; it is found that $0.00~\mathrm{GeV/}c\mathrm{)}^2<\aktsq_{eff}<0.06~\mathrm{GeV/}c\mathrm{)}^2$, in line with the values found, e.g. in Ref.~\cite{Barone:2015ksa}. A strong trend can be seen for $\aktsq_{eff}$ as a function of $x$, while it is almost flat as a function of $z$ and $\Pt$. The $x$-dependence has been fitted with the function:

\begin{equation}
    f(x) = A \frac{x^\alpha (1-x)}{x_0^\alpha (1-x_0)}
\end{equation}
where $x_0=0.01$; with this choice, $A$ corresponds to $\aktsq_{eff}(x_0=0.01)$. The estimated values of the free parameters are:
\begin{itemize}
    \item for positive hadrons: $A^+=0.029\pm 0.002$, $\alpha^+=0.77\pm 0.05$
    \item for negative hadrons: $A^-=0.015\pm 0.002$, $\alpha^-=0.66\pm 0.15$,
\end{itemize}
which gives about a factor two between positive and negative hadrons.
The next step is to check that the parametrization given for the $x$ dependence of $\aktsq_{eff}$ allows describing the measured asymmetries in bins of $z$ and $\Pt$, in addition to $x$ . This is done by evaluating the estimated Cahn asymmetry using the fitted $\aktsq_{eff}$. Except for the points at high $z$, the level of agreement, shown in Fig.~\ref{fig:ctot2_proton}, is good.

\begin{figure}
    \centering
    \captionsetup{width=\textwidth}
    \includegraphics[width=\textwidth]{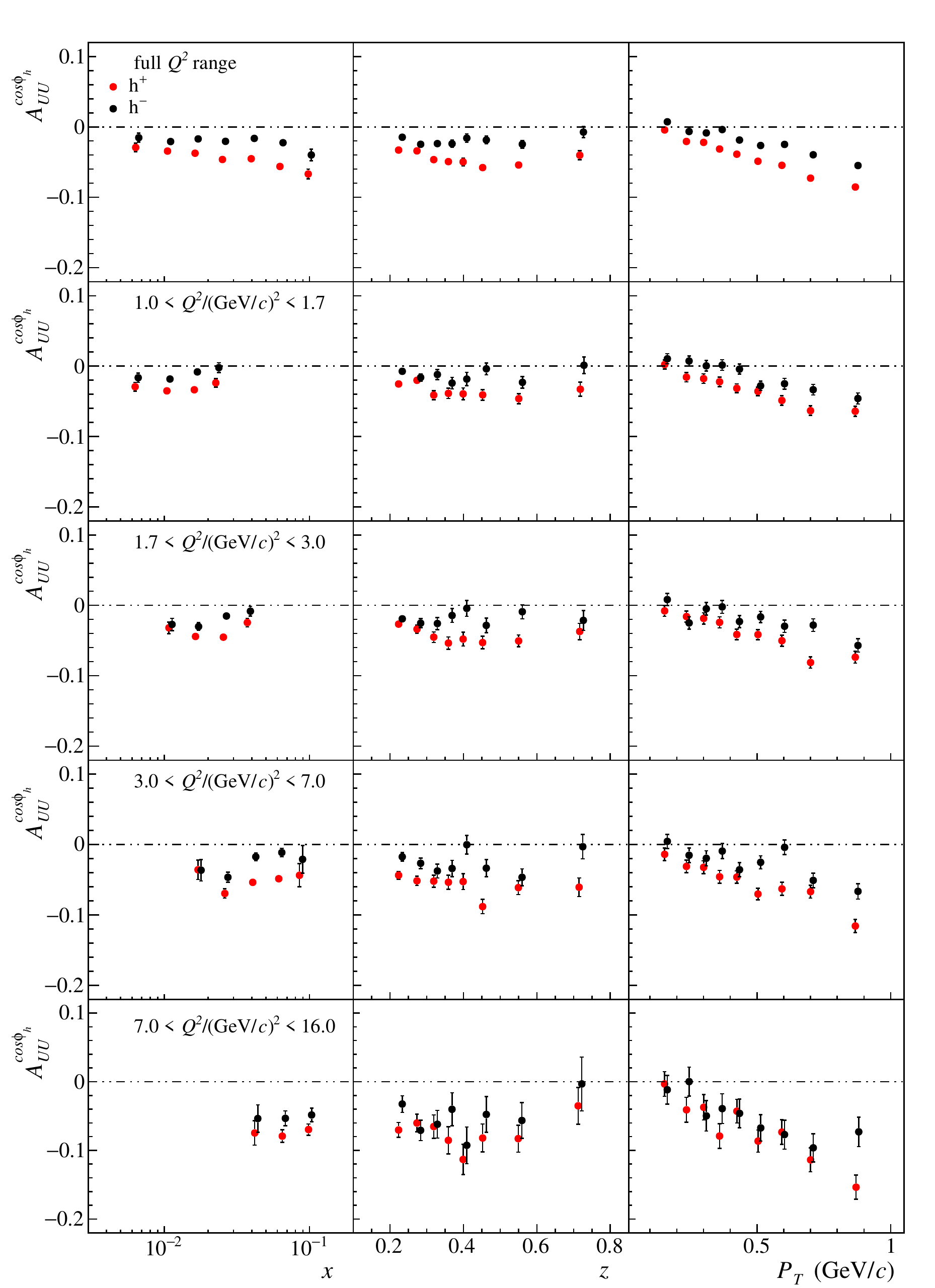}
    \caption{Comparison of the estimated Cahn asymmetry, evaluated using the fitted $\aktsq_{eff}(x)$, with the measured $A_{UU}^{\cos\fih}$ asymmetries on proton (open and closed points respectively), for positive (red) and negative hadrons (black), in four bins in $\Qsq$.}
    \label{fig:cahnQ2}
\end{figure}

\begin{figure}
    \centering
    \captionsetup{width=\textwidth}
    \includegraphics[width=\textwidth]{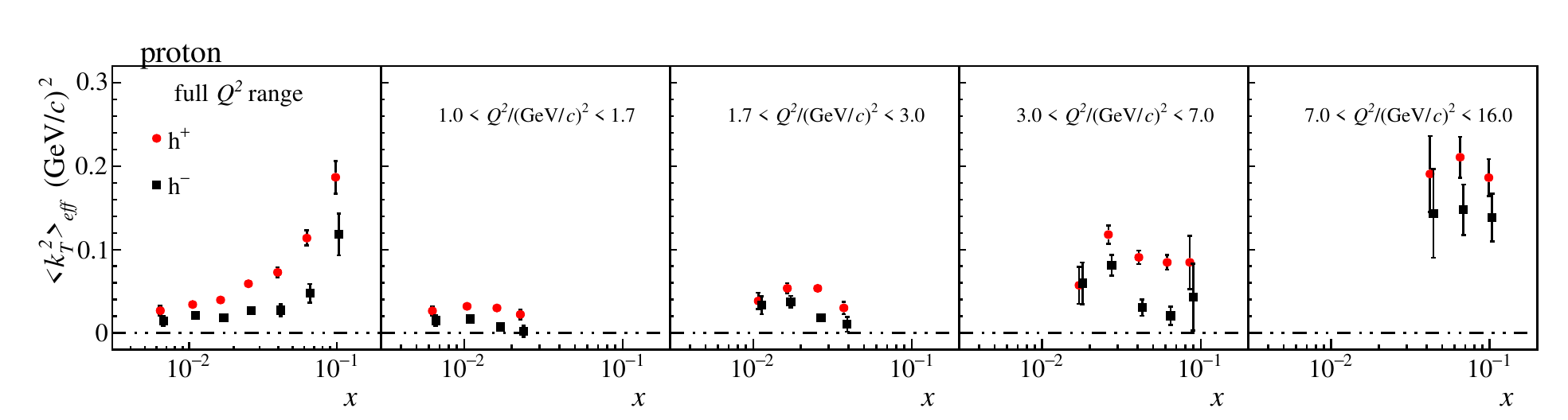}
    \caption{Extraction of $\aktsq_{eff}$ from the 1D azimuthal asymmetries in four $\Qsq$ bins. }
    \label{fig:aktsq_Q2}
\end{figure}

For the new proton results, it is interesting to check whether the level of agreement between measured and estimated asymmetries described above can be reached in bins of $\Qsq$. As can be seen in Fig.\ref{fig:cahnQ2} for the four different $\Qsq$ bins investigated in this work, the agreement is good as low $\Qsq$ and worse at high $\Qsq$. This indicates that, in addition (or instead of, given their correlation) to the $x$ dependence of the $\aktsq_{eff}$, there is a dependence on $\Qsq$. These two options are tested by repeating the extraction of the $\aktsq_{eff}$ in the four $\Qsq$ bins. As shown in Fig.~\ref{fig:aktsq_Q2}, $\aktsq_{eff}$ has almost no dependence on $x$ in each of the $\Qsq$ bins, its overall $x$-dependence arising from the correlation between the two variables when integrating over $\Qsq$. \\

\begin{figure}
    \centering
    \captionsetup{width=\textwidth}
    \includegraphics[width=0.65\textwidth]{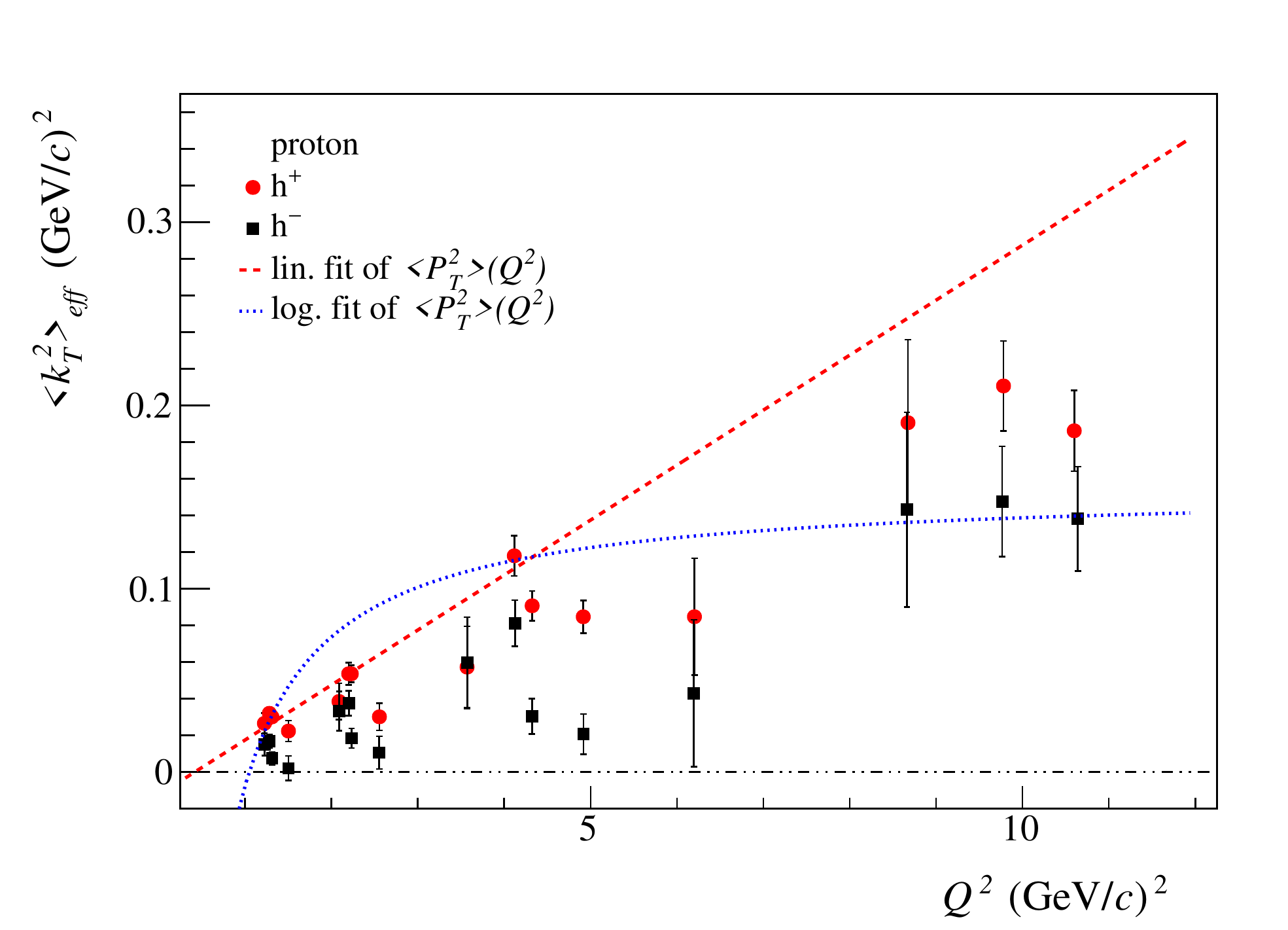}
    \caption{The values of $\aktsq_{eff}$ as in Fig.~\ref{fig:aktsq_Q2}, shown as a function of $\Qsq$ and compared with the expected values from the fits of $\aPtsq$ as a function of $\Qsq$ (Sect.~\ref{sect:ch4_kin_dep}). }
    \label{fig:aktsq_int}
\end{figure}

The values of $\aktsq_{eff}$, extracted in the four $\Qsq$ bins, are shown as a function of $\Qsq$ in Fig.~\ref{fig:aktsq_int}: the trend is almost linear in $\Qsq$ and, as expected since the beginning, different for positive and negative hadrons. The points are compared, in the same Figure, with the values of $\aktsq$ expected from the linear and logarithmic fits of the $\aPtsq$ values as a function of $\Qsq$ (Sect.~\ref{sect:ch4_kin_dep}), where the curves have been adjusted in order to match the value of $\aktsq_{eff}$ at low $\Qsq$ for positive hadrons. The linear fit tends to overshoot the data, while the logarithmic trend is closer to the data at high $\Qsq$. It is to be stressed, however, that the comparison observed here between the $\aktsq$ estimated from the azimuthal asymmetries and from the $\Ptsq$-distributions would be much different if a different pair of $z$ bins was considered for the estimation of the dependence of $\aktsq$ on $\Qsq$ in the linear case. In particular, estimating the derivative at high $z$, the predicted increase of $\aktsq$ with $\Qsq$ would be stronger.

\begin{figure}
    \centering
    \captionsetup{width=\textwidth}
    \includegraphics[width=\textwidth]{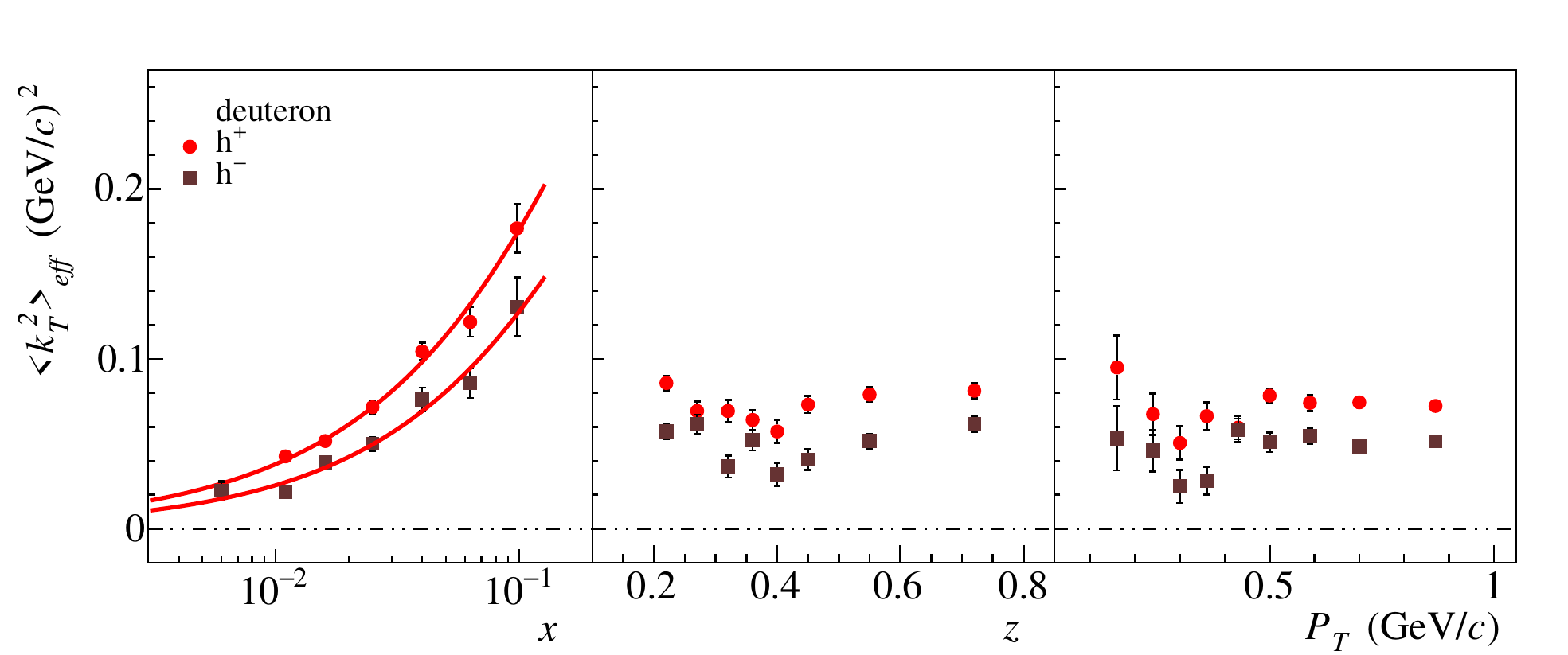}
    \caption{Extraction of $\aktsq_{eff}$ from the 1D $A_{UU}^{\cos\fih}$ asymmetries on deuteron, as a function of $x$, $z$ and $\Pt$. }
    \label{fig:ckT_deuteron}
\end{figure}

A similar exercise has been performed with the asymmetries measured in COMPASS on a deuteron target, after correcting for the exclusive hadron contamination \cite{COMPASS:2014kcy}. The extracted values of $\aktsq_{eff}$ are shown in Fig.~\ref{fig:ckT_deuteron} as a function of $x$, $z$ and $\Pt$. The trend is similar as the one already observed in the proton case. The fitted values of the parameters are in this case:
\begin{itemize}
    \item for positive hadrons: $A^+=0.038\pm 0.002$, $\alpha^+=0.71\pm 0.04$;
    \item for negative hadrons: $A^-=0.026\pm 0.002$, $\alpha^-=0.74\pm 0.06$
    \end{itemize}
thus indicating a similar power trend, but a different mean value of $\aktsq_{eff}$. \\ 

\subsection{Impact of the Boer-Mulders effect in the evaluation of $\aktsq$}
Having derived, presented and discussed the effective quark transverse momentum $\aktsq_{eff}$, thanks to which a reasonable description of the data can be achieved, it remains interesting to investigate by how much the extraction of $\aktsq$ could be affected by the Boer-Mulders contribution. In the following, a general expression for $\aktsq$ will be derived and proposed. Despite the many simplifications, this approach does not require the knowledge of the Boer-Mulders and the Collins functions. The measured $A_{UU}^{\cos\fih}$ and $A_{UU}^{\cos2\fih}$ asymmetries are assumed to be generated by the Cahn and Boer-Mulders mechanisms only, whose contributions to the measured asymmetries can be written as (see \ref{AppendixA} for details):

\begin{equation}
    A_{UU|Cahn}^{\cos\fih} = - \frac{2z\Pt\aktsq}{Q\aPtsq},
\end{equation}

\begin{equation}
    A_{UU|Cahn}^{\cos2\fih} =  \frac{2z^2\Pt^2\aktsq^2}{Q^2\aPtsq^2},
\end{equation}

\begin{equation}
    A_{UU|BM}^{\cos\fih} = - \frac{2\aktsq\apperpsq\Pt}{zQMM_h\aPtsq^3}\op \apperpsq\aPtsq + z^2\aktsq\op \Pt^2 - \aPtsq \cp \cp \underbrace{\frac{\sum_q e_q^2 x \bomq \collq}{\sum_q e_q^2 x f_1^{q} D_1^{h/q}}}_{\Sigma},
\end{equation}

\begin{equation}
    A_{UU|BM}^{\cos2\fih} =  \frac{\Ptsq \aktsq \apperpsq}{MM_h\aPtsq^2} \underbrace{\frac{\sum_q e_q^2 x \bomq \collq}{\sum_q e_q^2 x f_1^{q} D_1^{h/q}})}_{\Sigma}.
\end{equation}
where the $\Sigma$ symbol is introduced for brevity.
Two comments are in order. The first relates to the Cahn and Boer-Mulders contributions to $A_{UU}^{\cos\fih}$, which are given here in the Wandzura-Wilczek approximation, according to which the twist-3 TMDs have been neglected. Their contribution is in general not known, and they could play a relevant role in the extraction of $\aktsq$, as suggested e.g. in Ref.~\cite{Barone:2015ksa}. The second comment is about the Cahn contribution to $A_{UU}^{\cos2\fih}$, which is just one of several twist-4 terms, also not known. Its size is often considered to be small, and it is not always included in phenomenological analyses. To deal with these two considerations in the most generic way, we write the azimuthal asymmetries including four weights $n_i$ ($i=1,\dots,4)$:

\begin{equation}
\begin{split}
    A_{UU}^{\cos\fih} & = n_1 A_{UU|Cahn}^{\cos\fih} + n_2 A_{UU|BM}^{\cos\fih} \\
    A_{UU}^{\cos2\fih} & = n_3 A_{UU|Cahn}^{\cos2\fih} + n_4 A_{UU|BM}^{\cos2\fih}. 
\end{split}
\label{eq:asymm_sum} 
\end{equation}

\paragraph{\underline{The simplest case}:}
If $A_{UU|BM}^{\cos\fih}$ is assumed to be negligible or, in other terms, $n_2=0$, it easily follows that: 

\begin{equation}
    \aktsq_{eff} = - \frac{Q\aPtsq}{2z\Pt}A_{UU}^{\cos\fih} = - \frac{Q\aPtsq}{2z\Pt}n_1 \op - \frac{2z\Pt\aktsq}{Q\aPtsq} \cp = n_1\aktsq
\label{eq:kt2bm}
\end{equation}
The size of $n_1$ is not known, but it can have a very large impact. Assuming, for example \cite{Barone:2015ksa}, that the twist–3 terms reduce by 50\% the Cahn asymmetry ($n_1=0.5)$,  we would conclude that $\aktsq= 2 \cdot \aktsq_{eff}$. 

\paragraph{\underline{Inclusion of the Boer-Mulders contribution:}} 
If $A_{UU|BM}^{\cos\fih}$ is not neglected ($n_2 \neq 0$), the information on $A_{UU}^{\cos2\fih}$ can be used to get an estimate of $\aktsq$ from the $A_{UU}^{\cos\fih}$ asymmetry. Let's consider for simplicity $n_2=1$, thus assuming that the Wandzura-Wilczek approximation is accurate in the Boer-Mulders part of $A_{UU}^{\cos\fih}$, and also $n_4=1$. From $A_{UU}^{\cos2\fih}$, one has that:

\begin{equation}
\begin{split}
    A_{UU}^{\cos2\fih} = n_3\frac{2z^2\Pt^2\aktsq^2}{Q^2\aPtsq^2} + \frac{\Ptsq \aktsq \apperpsq}{MM_h\aPtsq^2}\Sigma \\ 
    \Longrightarrow \Sigma = \frac{MM_h\aPtsq^2}{\Ptsq \aktsq \apperpsq} \op A_{UU}^{\cos2\fih} - n_3\frac{2z^2\Pt^2\aktsq^2}{Q^2\aPtsq^2} \cp    
\end{split}
\end{equation}
Inserting $\Sigma$ into the expression for $A_{UU}^{\cos\fih}$ and substituting $\apperpsq$ with $\aPtsq - z^2\aktsq$  gives:
\begin{equation}
\begin{split}
   A_{UU}^{\cos\fih} & = -n_1 \frac{2z\Pt\aktsq}{Q\aPtsq} - \frac{2\op \apperpsq\aPtsq + z^2\aktsq\op \Pt^2 - \aPtsq \cp\cp}{Qz\Pt\aPtsq} \op A_{UU}^{\cos2\fih} - n_3\frac{2z^2\Pt^2\aktsq^2}{Q^2\aPtsq^2} \cp    \\
   & = -\frac{2z\Pt}{Q\aPtsq} \op n_1\aktsq + \op \frac{\aPtsq^2}{z^2\Ptsq}-\frac{2\aktsq\aPtsq}{\Ptsq}+\aktsq\cp \op A_{UU}^{\cos2\fih} - n_3\frac{2z^2\Pt^2\aktsq^2}{Q^2\aPtsq^2} \cp \cp   
\end{split}   
\end{equation}
so that, introducing the definition of $\aktsq_{eff}$ and neglecting the terms proportional to the third power of $\aktsq$, one has:

\begin{equation}
\begin{split}
    \aktsq_{eff} & =  n_1\aktsq + \op \frac{\aPtsq^2}{z^2\Ptsq}-\frac{2\aktsq\aPtsq}{\Ptsq}+\aktsq\cp \op A_{UU}^{\cos2\fih} - n_3\frac{2z^2\Pt^2\aktsq^2}{Q^2\aPtsq^2} \cp \\
    & \approx n_1\aktsq + \frac{\aPtsq^2 A_{UU}^{\cos2\fih}}{z^2\Ptsq} - \frac{2n_3}{\Qsq}\aktsq^2 - \frac{2\aPtsq A_{UU}^{\cos2\fih}}{\Ptsq}\aktsq + A_{UU}^{\cos2\fih} \aktsq 
\end{split}
\end{equation}
or equivalently:
\begin{equation}
    \frac{2n_3}{\Qsq}\aktsq^2 - \op n_1 - \op\frac{2\aPtsq}{\Ptsq} - 1\cp A_{UU}^{\cos2\fih} \cp \aktsq - \frac{\aPtsq^2 A_{UU}^{\cos2\fih}}{z^2\Ptsq} + \aktsq_{eff} = 0.
\end{equation}

Now, if $n_3=0$ (that is, the Cahn contribution in $A_{UU}^{\cos2\fih}$ is not considered), the final expression for $\aktsq$ reads:
\begin{equation}
      \aktsq |_{n_3=0} =  \frac{\aktsq_{eff} - \frac{\aPtsq^2 A_{UU}^{\cos2\fih}}{z^2\Ptsq}}{ n_1 - \op\frac{2\aPtsq}{\Ptsq} - 1\cp A_{UU}^{\cos2\fih}},
\end{equation}
which reduces to Eq.~\ref{eq:kt2bm} if $A_{UU}^{\cos2\fih}=0$. Even if the $A_{UU}^{\cos2\fih}$ asymmetry has a small size, its contribution could be very relevant at the numerator, where $\aPtsq^2/(z^2\Ptsq)$ can be of order 1 and where $\aktsq_{eff}$ is also generally small. At the denominator, the dominant correction would still be determined by the value of $n_1$. In other words, the inclusion of the Boer-Mulders contribution to the $A_{UU}^{\cos\fih}$ asymmetry would not help much in finding a conclusive estimate of $\aktsq$, if the twist-3 correction to the expression for the Cahn asymmetry is not known. As a further complication, as will be clear in the following, the values of the measured $A_{UU}^{\cos2\fih}$ asymmetries indicate that the Boer-Mulders term cannot be the only contribution to the asymmetry: in our picture, it would mean that $n_3\neq 0$. Thus, the expression for $\aktsq$ would be further complicated: it is given here, for completeness:

\begin{equation}
\begin{split}
    \frac{4n_3}{\Qsq}\aktsq & =     n_1- \op \frac{2\aPtsq}{\Ptsq}-1\cp A_{UU}^{\cos2\fih}  \\
    & ~~~ + \sqrt{\op n_1- \op \frac{2\aPtsq}{\Ptsq}-1\cp A_{UU}^{\cos2\fih}\cp^2 - \frac{8n_3}{\Qsq}\op \aktsq_{eff} -\frac{ \aPtsq^2}{z^2\Ptsq}A_{UU}^{\cos2\fih} \cp}.
\end{split}
\end{equation}

\subsection{Considerations on the $A_{UU}^{\cos2\fih}$ asymmetries}
As for the $A_{UU}^{\cos2\fih}$, there are basic considerations on the measured values which point to more sophisticated mechanisms than those assumed in the simple interpretation of the results discussed here. Let's consider the $A_{UU}^{\cos2\fih}$ asymmetries measured, for positive and negative hadrons, on proton (subscript $p$) and on deuteron (subscript $d$). Their basic features:

\begin{equation}
\begin{cases}
    A_{UU,p}^{\cos2\phi_{h^+}}  \approx 0 \\ 
    A_{UU,p}^{\cos2\phi_{h^-}} > 0 \\ 
    A_{UU,d}^{\cos2\phi_{h^+}} > 0 \\ 
    A_{UU,d}^{\cos2\phi_{h^-}} > 0 
\end{cases}
\end{equation}
are puzzling in view of a possible extraction of the Boer-Mulders function $\bom$. In fact, considering only the Boer-Mulders mechanism at work, and limiting the sum over the flavors to the $u-$ and $d$ quarks, it can be written that:
\begin{equation}
\begin{cases}
    A_{UU,p}^{\cos2\phi_{h^+}} \sim \op 4h_{1}^{\perp u} H_{1,fav}^\perp +  h_{1}^{\perp d} H_{1,unf}^\perp\cp  \approx 0 \\ 
    A_{UU,p}^{\cos2\phi_{h^-}} \sim \op 4h_{1}^{\perp u} H_{1,unf}^\perp +  h_{1}^{\perp d} H_{1,fav}^\perp\cp> 0 \\ 
    A_{UU,d}^{\cos2\phi_{h^+}} \sim \op 4h_{1}^{\perp u} H_{1,fav}^\perp +  h_{1}^{\perp d} H_{1,unf}^\perp + 4h_{1}^{\perp d} H_{1,fav}^\perp +  h_{1}^{\perp u} H_{1,unf}^\perp\cp> 0 \\ 
    A_{UU,d}^{\cos2\phi_{h^-}} \sim \op 4h_{1}^{\perp u} H_{1,unf}^\perp +  h_{1}^{\perp d} H_{1,fav}^\perp + 4h_{1}^{\perp d} H_{1,unf}^\perp +  h_{1}^{\perp u} H_{1,fav}^\perp \cp> 0 
\end{cases}
\end{equation}
where $H_{1,fav}^{\perp}$ ($H_{1,unf}^{\perp}$) is the favored (unfavored) Collins fragmentation function. Assuming that $H_{1,unf}^\perp \approx - H_{1,fav}^\perp >0$, it follows that:
\begin{equation}
\begin{cases}
    \op 4h_{1}^{\perp u} -  h_{1}^{\perp d}\cp H_{1,fav}^\perp  \approx 0 \\ 
    \op 4h_{1}^{\perp u} -  h_{1}^{\perp d} \cp H_{1,fav}^\perp < 0 \\ 
    3 \op h_{1}^{\perp u} +  h_{1}^{\perp d} \cp H_{1,fav}^\perp > 0 \\ 
    3 \op h_{1}^{\perp u} +  h_{1}^{\perp d} \cp H_{1,fav}^\perp < 0 
\end{cases}
\end{equation}
which evidently has no solution. This is a strong indication that the Boer-Mulders mechanism is not sufficient to explain the measured asymmetries. A possible solution would be to have an additional contribution to the asymmetries of positive sign, while the Boer-Mulders contribution would have negative sign and could be at the origin of the difference between positive and negative hadrons. Such positive contribution to $A_{UU}^{\cos2\fih}$ may originate at twist-4 from the Cahn mechanism; however, this would be only one of several possible twist-4 contributions to the asymmetry, presently not known. In this scenario, assuming the flavor-independence of the Cahn contribution, the compatibility of $ A_{UU,p}^{\cos2\phi_{h^+}}$ with zero would give an indication of the size of the Boer-Mulders term.
However, if the unknown contributions to this asymmetry are the same for positive and negative hadrons, information on the Boer-Mulders function can still be extracted, as explained in the following.

\subsection{Difference asymmetries on the $A_{UU}^{\cos2\fih}$ asymmetry}
The method of the \textit{difference asymmetries} was proposed a long time ago \cite{Frankfurt:1989wq,Christova:2000nz,Sissakian:2006vz} as a possible way to access the helicity (later, also the transversity) PDFs. It has been used by the SMC and COMPASS Collaborations to measure the helicity PDFs \cite{Adeva:1995yi,COMPASS:2007esq} and, recently, it has been applied to the COMPASS measurement of the Collins asymmetry on proton and deuteron \cite{Barone:2019yvn} in order to extract the ratio of the transversity functions for the $u_v$- and $d_v$-quarks, with no need for a knowledge of the Collins function. The same method can be applied also to the $A_{UU}^{\cos2\fih}$ asymmetries to gain information on the ratio of the Boer-Mulders function for the $u_v$- and the $d_v$-quarks. Note that this method can only be applied if the asymmetries are measured for positive and negative hadrons and for SIDIS on proton and on deuteron (or neutron) in the same kinematics. Assuming no contribution to the $F_{UU}^{\cos2\fih}$ structure function other than the one occurring at twist-2, related to the Boer-Mulders function, a measurement of the amplitude of the $\cos2\fih$ modulation as a function of $x$ and $z$ allows accessing the quantity:

\begin{equation}
   A_{UU}^{\cos2\fih}(x,z) = \frac{\int \diff \Ptsq~~\mathcal{C}\left[\frac{2\op\hh\cdot\vkt\cp\op\hh\cdot\vpperp\cp-\vkt\cdot\vpperp}{zMM_h}\bom\coll\right]}{\int \diff \Ptsq~~\mathcal{C}\left[f_1D_1\right]} 
\label{eq:axz}   
\end{equation}
where, in the usual Gaussian approximation, the denominator reduces to the products of collinear PDFs and FFs:

\begin{equation}
    \int \diff \Ptsq~~\mathcal{C}\left[f_1D_1\right] = \int \diff \Ptsq~~\sum_q e_q^2 xf_1^q(x)D_1^{h/q}(z) \frac{e^{-\Ptsq/\aPtsq}}{\pi \aPtsq} = \sum_q e_q^2 xf_1^q(x)D_1^{h/q}(z).
\end{equation}

The convolution at the numerator can also be solved in the context of the Gaussian approximation: full details of the calculations are given in Appendix, where it is shown that:
\begin{equation}
   \mathcal{C}\left[\frac{2\op\hh\cdot\vkt\cp\op\hh\cdot\vpperp\cp-\vkt\cdot\vpperp}{zMM_h}\bom\coll\right] =  \sum_q e_q^2 x h_1^{\perp q}(x)H_1^{\perp h/q}(z) \frac{\Ptsq \aktsq \apperpsq}{\pi MM_h\aPtsq^3}e^{-\Ptsq/\aPtsq}
\end{equation}
so that the numerator of Eq.~\ref{eq:axz} reads:
\begin{equation}
\begin{split}
    \int \diff \Ptsq~~\mathcal{C}\left[\frac{2\op\hh\cdot\vkt\cp\op\hh\cdot\vpperp\cp-\vkt\cdot\vpperp}{zMM_h}\bom\coll\right] & = \frac{\aktsq \apperpsq}{ \pi MM_h\aPtsq}\sum_q e_q^2 x h_1^{\perp q}(x)H_1^{\perp h/q}(z) \\
    & = w\sum_q e_q^2 x h_1^{\perp q}(x)H_1^{\perp h/q}(z)
\end{split}
\end{equation}
where $w=\frac{\pi \aktsq\apperpsq}{MM_h\aPtsq}$ is an overall multiplicative factor. The asymmetry finally reads:
\begin{equation}
   A_{UU}^{\cos2\fih}(x,z) = \frac{w\sum_q e_q^2 x h_1^{\perp q}(x)H_1^{\perp h/q}(z)}{\sum_q e_q^2 xf_1^q(x)D_1^{h/q}(z)}. 
\label{eq:axz2}   
\end{equation}

Following the procedure of Ref.~\cite{Barone:2019yvn}, the asymmetry for a generic target $t$ ($t=p,d$ for proton, deuteron) and positive and negative hadrons can be written as:

\begin{equation}
    A_{UU,t}^{\cos2\phi_{h^{\pm}}}(x,z) = \frac{w \sigma_{BM,t}^\pm}{\sigma_{0,t}^\pm}
\end{equation}
where $\sigma_0$ is the cross-section integrated over $\fih$, corresponding to the denominator of Eq. \ref{eq:axz2}, while $\sigma_{BM}$ is the term in the cross-section corresponding to the Boer-Mulders contribution to the $\cos2\fih$ modulation, as in the numerator of Eq. \ref{eq:axz2}. The difference asymmetries read:
\begin{equation}
    A_{D,t} = \frac{w \op \sigma_{BM,t}^+-\sigma_{BM,t}^-\cp}{\sigma_{0,t}^++\sigma_{0,t}^-}.
\end{equation}

Interestingly, as far as the difference of the asymmetries for positive and negative hadrons are considered, the possible contribution to $A_{UU}^{\cos2\fih}$ of other (flavor-independent) terms, like the one due to the Cahn effect at twist-4 or to the Brosdky-Berger mechnism, is canceled out. The same holds, in first approximation, for the systematic uncertainties affecting the measured asymmetries. 

It can be calculated that: 

\begin{equation}
    A_{D,p}(x,z) = \frac{w\op 4h_1^{\perp u_v}(x)-h_1^{\perp d_v}(x)\cp \op H_{1,fav}^{\perp}(z) - H_{1,unf}^{\perp}(z)\cp}{\op 4f_1^u(x)+f_1^{d}(x)+4f_1^{\bar{u}}(x)+f_1^{\bar{d}}(x)\cp \op D_{1,fav}(z)+D_{1,unf}(z)\cp + 2\op f_1^s(x)+f_1^{\bar{s}}(x)\cp D_{1,s}(z) }
\end{equation}

\begin{equation}
    A_{D,d}(x,z) = \frac{3w\op h_1^{\perp u_v}(x)+h_1^{\perp d_v}(x)\cp \op H_{1,fav}^{\perp}(z) - H_{1,unf}^{\perp}(z)\cp}{5\op f_1^u(x)+f_1^{\bar{u}}(x)+f_1^{d}(x)+f_1^{\bar{d}}(x)\cp  \op D_{1,fav}(z)+D_{1,unf}(z)\cp + 4\op f_1^s(x)+f_1^{\bar{s}}(x)\cp  D_{1,s}(z) }
\end{equation}

from which it follows that the difference asymmetries, integrated over $z$, read:
\begin{equation}
\begin{split}
    A_{D,p}(x) & = \frac{\op 4h_1^{\perp u_v}(x)-h_1^{\perp d_v}(x)\cp \int \diff z w \op H_{1,fav}^{\perp}(z) - H_{1,unf}^{\perp}(z)\cp}{\op 4f_1^u(x)+f_1^{d}(x)+4f_1^{\bar{u}}(x)+f_1^{\bar{d}}(x)\cp \int \diff z \op D_{1,fav}(z)+D_{1,unf}(z)\cp + 2\op f_1^s(x)+f_1^{\bar{s}}(x)\cp \int \diff z D_{1,s}(z) } \\
    & = \frac{\op 4h_1^{\perp u_v}(x)-h_1^{\perp d_v}(x)\cp \op \langle wH_{1,fav}^{\perp} \rangle - \langle wH_{1,unf}^{\perp}\rangle\cp}{\op 4f_1^u(x)+f_1^{d}(x)+4f_1^{\bar{u}}(x)+f_1^{\bar{d}}(x)\cp \op \langle D_{1,fav}\rangle + \langle D_{1,unf} \rangle \cp + 2\op f_1^s(x)+f_1^{\bar{s}}(x)\cp \langle D_{1,s} \rangle }. 
\end{split}
\end{equation}

and analogously:
\begin{equation}
    A_{D,d}(x) = \frac{3 \op h_1^{\perp u_v}(x)+h_1^{\perp d_v}(x)\cp \op \langle wH_{1,fav}^{\perp} \rangle - \langle wH_{1,unf}^{\perp}\rangle\cp}{5\op f_1^u(x)+f_1^d(x)+f_1^{\bar{u}}(x)+f_1^{\bar{d}}(x)\cp \op \langle D_{1,fav}\rangle + \langle D_{1,unf} \rangle \cp + 4\op f_1^s(x)+f_1^{\bar{s}}(x)\cp \langle D_{1,s} \rangle }. 
\end{equation}

The ratio of the difference asymmetries for deuteron and proton targets gives:

\begin{equation}
\begin{split}
    \frac{A_{D,d}(x)}{A_{D,p}(x)} & = 3 \frac{ h_1^{\perp u_v}(x)+h_1^{\perp d_v}(x) }{4h_1^{\perp u_v}(x)-h_1^{\perp d_v}(x)} \cdot \\
    & \cdot \underbrace{\frac{\op 4f_1^u(x)+f_1^{\bar{d}}(x)+4f_1^{\bar{u}}(x)+f_1^{\bar{d}}(x)\cp \op \langle D_{1,fav}\rangle + \langle D_{1,unf} \rangle \cp + 2\op f_1^s(x)+f_1^{\bar{s}}(x)\cp \langle D_{1,s} \rangle } {5\op f_1^u(x)+f_1^d(x)+f_1^{\bar{u}}(x)+f_1^{\bar{d}}(x)\cp \op \langle D_{1,fav}\rangle + \langle D_{1,unf} \rangle \cp + 4\op f_1^s(x)+f_1^{\bar{s}}(x)\cp \langle D_{1,s} \rangle }}_{R(x)}
\end{split}
\end{equation}

The quantity $R(x)$ is composed of collinear PDFs and FFs that are generally well known, thus no systematic uncertainty has been considered to affect it. Its value is almost constant along the considered $x$ range, being $R(x)\approx 0.5$. From the ratio of the difference asymmetries, one can easily get the ratio of the Boer-Mulders functions for the $d_v$ and $u_v$-quarks:

\begin{equation}
    \frac{h_1^{\perp d_v}(x)}{h_1^{\perp u_v}(x)} = \frac{4\frac{A_{D,d}(x)}{A_{D,p}(x)} -3R(x)}{\frac{A_{D,d}(x)}{A_{D,p}(x)}+3R(x)}.
\end{equation}

This method has been applied to the $A_{UU}^{\cos2\fih}$ asymmetries measured in COMPASS on the deuteron and the proton target, considering only the statistical uncertainties affecting the measurements. In terms of the measured asymmetries, the difference asymmetries can be expressed as:

\begin{equation}
    A_{D,t} = \frac{\sigma_{0,t}^+}{\sigma_{0,t}^++\sigma_{0,t}^-}A_{UU,t}^{\cos2\phi_{h^+}} - \frac{\sigma_{0,t}^-}{\sigma_{0,t}^++\sigma_{0,t}^-}A_{UU,t}^{\cos2\phi_{h^-}}
\end{equation}
where the $\sigma_0$ terms are proportional to the total number of hadrons entering the fit performed to measure the azimuthal asymmetries. For the proton data, $\sigma_0$ has been replaced with the constant term $N_0$, obtained from the fit of the acceptance-corrected azimuthal distributions:

\begin{equation}
    N(\fih) = N_0\op 1 + \eps_1 A_{UU}^{\cos\fih} + \eps_2 A_{UU}^{\cos2\fih} + \lambda \eps_3 A_{LU}^{\sin\fih}\cp.
\end{equation}
For the deuteron data, for which this information was not available, $\sigma_0$ has been estimated from the uncertainties on the quoted asymmetries following the procedure of Ref.\cite{Barone:2019yvn}, where the integrated acceptance has been proven compatible for positive and negative hadrons. 

\begin{figure}[h!]
    \centering
    \captionsetup{width=\textwidth}
    \includegraphics[width=0.65\textwidth]{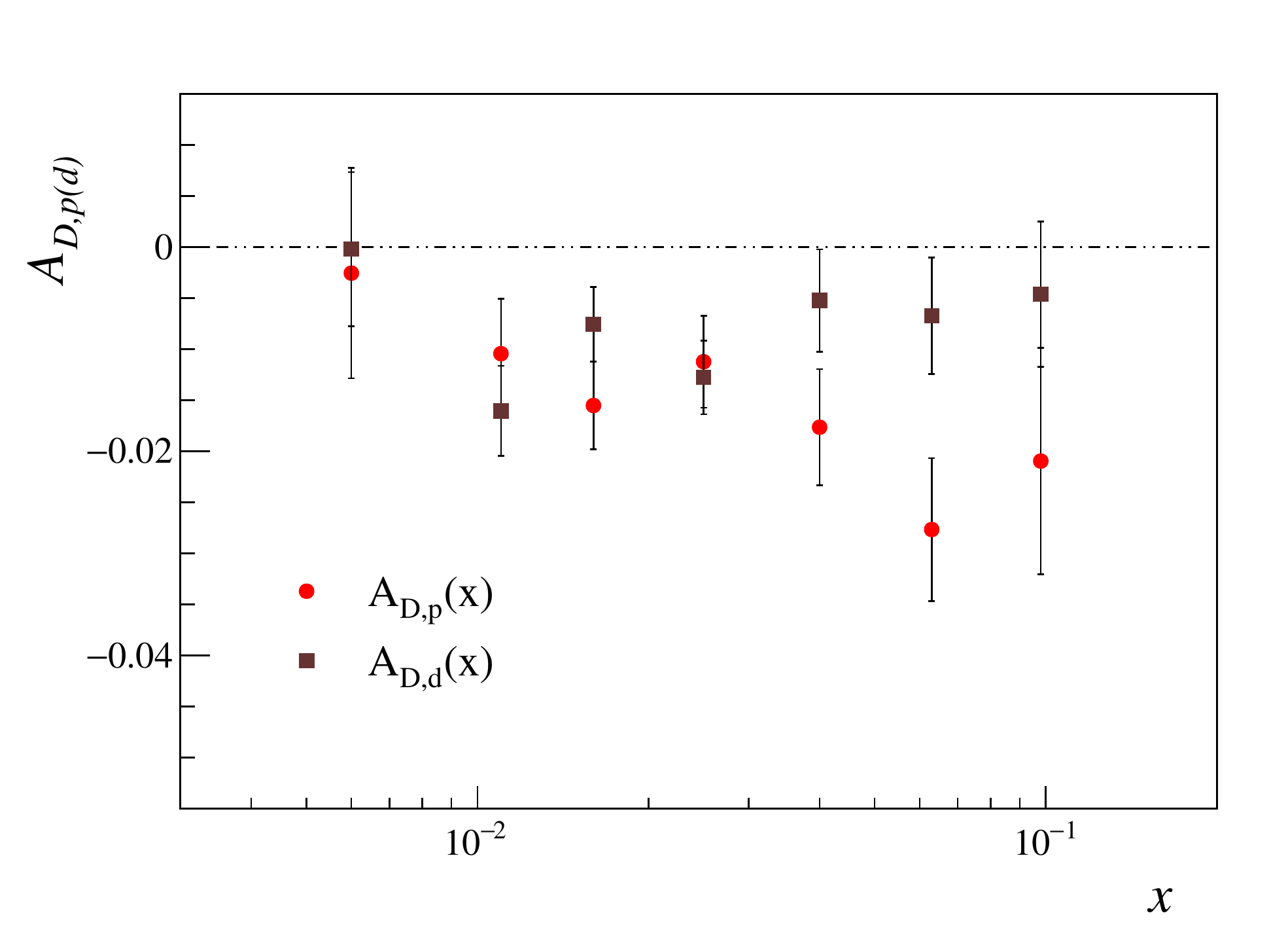}
    \caption{$A_{UU}^{\cos2\fih}$ difference asymmetries for proton and deuteron, as a function of $x$.}
    \label{fig:ADfig1}
\end{figure}

\begin{figure}[h!]
    \centering
    \captionsetup{width=\textwidth}
    \includegraphics[width=0.49\textwidth]{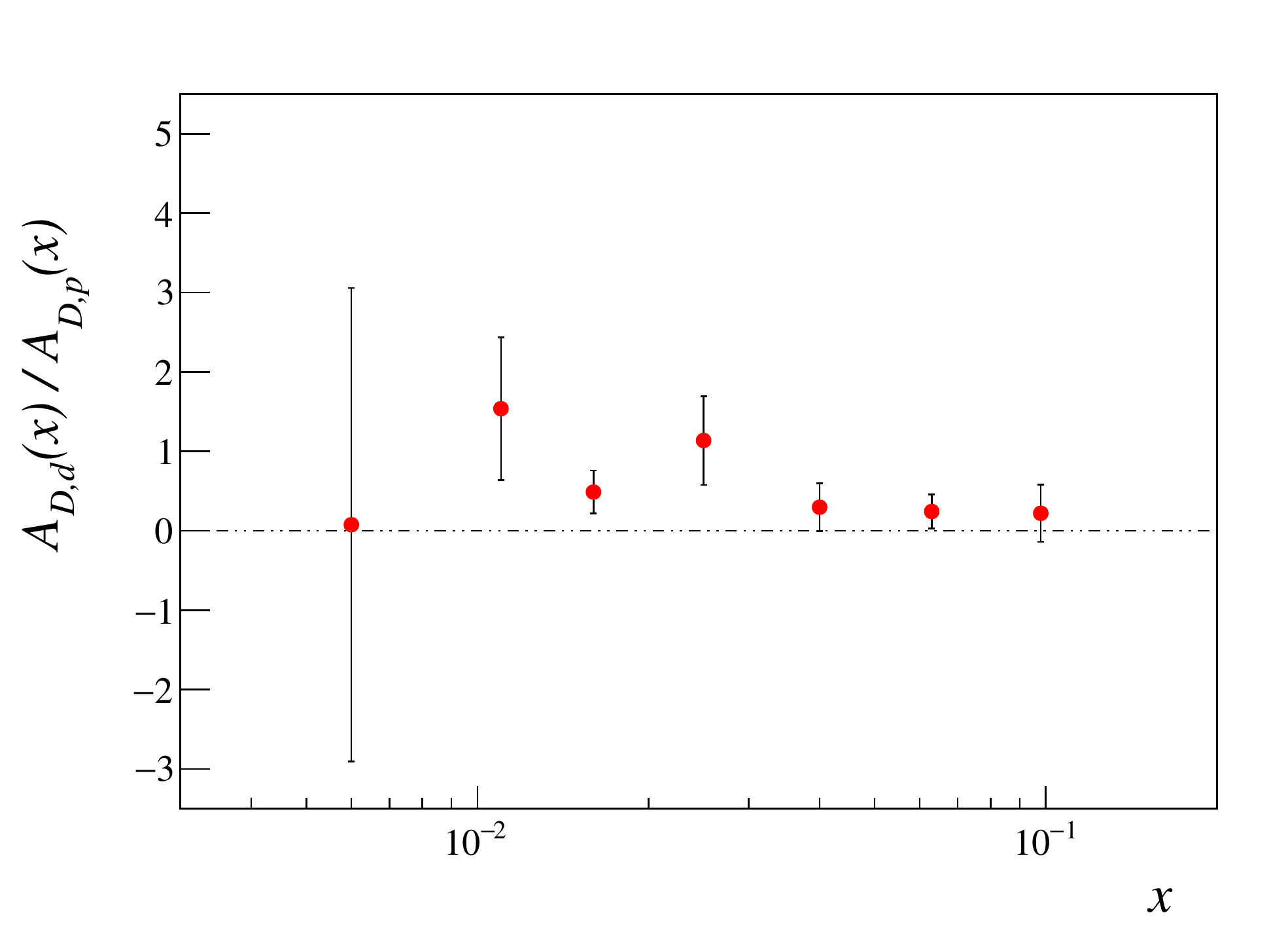}
    \includegraphics[width=0.49\textwidth]{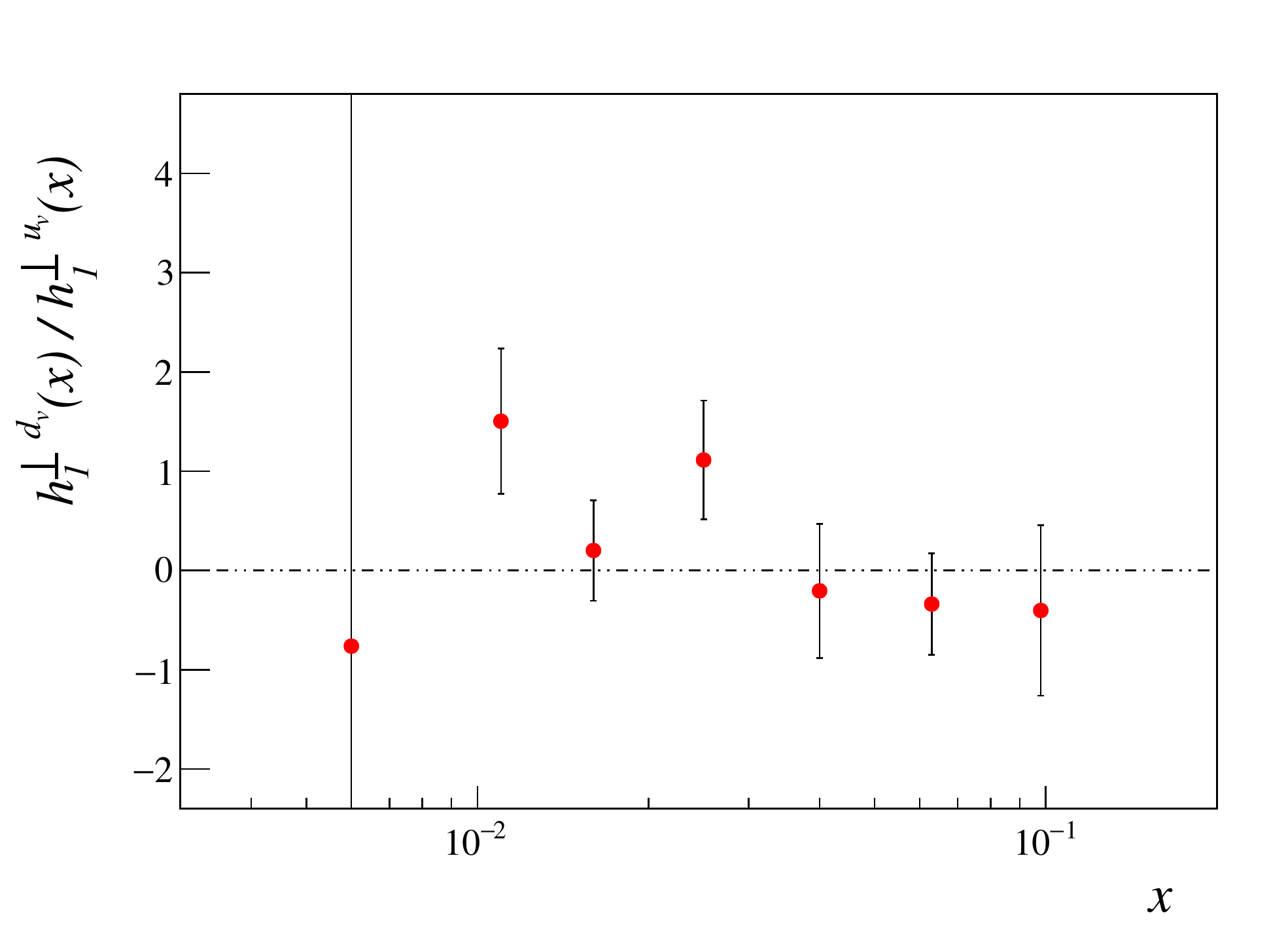}
    \caption{Left: The ratio between deuteron and proton $A_{UU}^{\cos2\fih}$ difference asymmetries as a function of $x$. Right: the corresponding ratio of the $h_1^{\perp~d_v}$ with the $h_1^{\perp~u_v}$ Boer-Mulders function. }
    \label{fig:ADfig2}
\end{figure}

The difference asymmetries for proton and deuteron are shown in Fig.~\ref{fig:ADfig1}: they almost coincide at zero in the first point in $x$, then look similar up to $x=0.03$, their difference being larger at larger $x$. Their ratio and the extracted ratio of the Boer-Mulders functions for the $d_v$- and $u_v$-quarks are shown in Fig.~\ref{fig:ADfig2}. Even with a non-negligible statistical uncertainties, the ratio of the Boer-Mulders functions suggests the same sign for $h_1^{\perp d_v}$ and $h_1^{\perp u_v}$, with a mean value $\langle h_1^{\perp d_v} / h_1^{\perp u_v} \rangle = 0.28 \pm 0.25$ when integrating over $x$ in the range $0.008<x<0.130$. This would be in agreement with the theoretical expectation \cite{Barone:2009hw} that both functions are negative. However at high-$x$, where the Boer-Mulders function is expected to be more sizable, the ratio is found negative, compatible with zero.

\chapter{Diffractive exclusive production of vector mesons} 
\label{Chapter6_SDMEs}

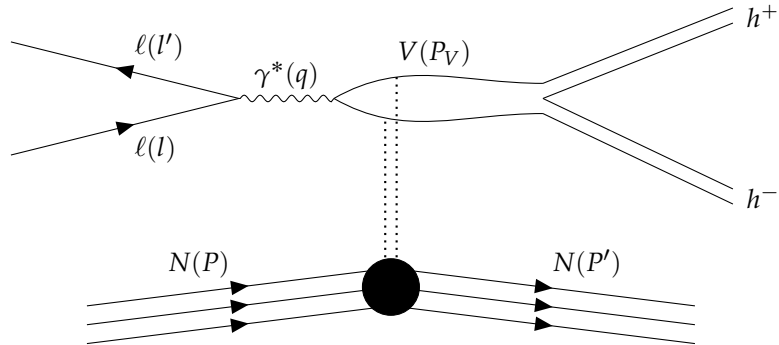
\begin{figure}[h!]
\captionsetup{width=\textwidth}
\centering
\begin{tikzpicture}
    \begin{feynman}
        \vertex[blob,fill=black] (m) at ( 0, -1.5) {};
        \vertex (a) at (-5, 1.75);
        \vertex (b) at (-2,1);
        \vertex (c) at (-5,0.25);
        \vertex (d) at (-0.75, 1);
        
        \vertex (e1) at ( -4, -2.25);
        \vertex (f1) at ( -4, -2.00);
        \vertex (g1) at ( -4, -1.75);
        \vertex (e2) at ( 0, -1.75);
        \vertex (f2) at ( 0, -1.50);
        \vertex (g2) at ( 0, -1.25);
        \vertex (e3) at ( 4, -2.25);
        \vertex (f3) at ( 4, -2.00);
        \vertex (g3) at ( 4, -1.75);

        \vertex (h1) at ( 2, 1.2);        
        \vertex (h2) at ( 4.5, 2.2);        
        \vertex (h3) at ( 2, 1.0);        
        \vertex (h4) at ( 4.5, 2.0);        
        
        \vertex (h5) at ( 2, 1.0);        
        \vertex (h6) at ( 4.5, -0.2);        
        \vertex (h7) at ( 2, 0.8);        
        \vertex (h8) at ( 4.5, -0.4);        
        
        \vertex (p0) at (-0.075, -1.5);
        \vertex (p1) at ( 0.075, -1.5);
        \vertex (p2) at (-0.075, 0.75);
        \vertex (p3) at (0.075, 1.32);
        
        \vertex (b1) at (4.9,2.1) {\(h^{+}\)};
        \vertex (b2) at (4.9,-0.3) {\(h^{-}\)};
     
        \diagram*{
        (c) -- [fermion,edge label'=$\ell(l)$] (b) -- [fermion,edge label'=$\ell(l^\prime)$] (a),
        (b) -- [photon, edge label=$\gamma^{*}(q)$] (d),
        (e1) -- [fermion] (e2) -- [fermion] (e3),
        (f1) -- [fermion] (f2) -- [fermion] (f3),
        (g1) -- [fermion, edge label=$N(P)$] (g2) -- [fermion, edge label=$N(P^\prime)$] (g3),
        
        (h1) -- (h2),
        (h3) -- (h4),
        (h5) -- (h6),
        (h7) -- (h8),
        
        (h1) -- [edge label'=$V(P_V)$, out=180, in=30] (d) -- [out=-30, in=180] (h7),
        (p0) -- [ghost] (p2),
        (p1) -- [ghost] (p3),
        
    };
    
    \end{feynman}
\end{tikzpicture}
\caption{Diffractive production and decay of a vector meson $V$.}
\label{fig:excl_rho_diagram}
\end{figure}

The diffractive lepto-production of vector mesons (Fig.~\ref{fig:excl_rho_diagram}):
\begin{equation}
    \ell(l) \: N(P) \longrightarrow \ell(l^\prime) \: N(P^\prime) \: V(P_V)
\label{eq:exclr}
\end{equation}
is an exclusive, soft process in which a lepton $\ell$ elastically interacts with a target nucleon $N$ to produce a vector meson $V$ in the final state. This production mechanism is characterized by a small value of the momentum transfer squared $t=(P^\prime-P)^2$, where $P$ and $P^\prime$ denote the four-momenta of the nucleon in the initial and in the final state respectively. The cross-section shows an exponential trend in $|t|$: 

\begin{equation}
    \frac{\diff \sigma}{\diff t} \sim e^{-R^2|t|}
\end{equation}
where $R\sim$ 1~fm is the hadron size, which also represents the only scale of the process. 

The vector meson $V$ usually decays into meson pairs of so-called \textit{exclusive hadrons}, where the attribute \textit{exclusive} underlines the exclusive nature of the process that led to their generation. In Ch.~\ref{Chapter3_Data_analysis} the exclusive vector mesons contamination has been analyzed as a source of background for the SIDIS hadron samples. Here, the exclusive vector mesons production is studied in itself. The observed azimuthal modulations of the exclusive hadrons depend on the Spin Density Matrix Elements (SDMEs) of the parent vector meson $V$. In this Chapter we focus on the SDMEs of the $\vmrho$ vector meson: after an introduction to the diffractive production mechanism (Sect.~\ref{sect:ch6_diffr_prod}) and on the SDME formalism (Sect.~\ref{sect:ch6_sdme_form}), the new measurement of the $\vmrho$ SDMEs from the COMPASS data collected with a liquid hydrogen target is presented, and the results discussed (Sect.~\ref{sect:ch6_sdme_measurement} and~\ref{sect:ch6_procedure}).

\section{Diffractive production mechanism}
\label{sect:ch6_diffr_prod}
The diffractive processes are characterized by a large rapidity gap between the recoil nucleon and the decay products on the vector meson \cite{Derrick:1993xh,Ahmed:1994nw}. This is different to what happens in SIDIS, where the rapidity gap between the nucleon remnant and the struck quark is uniformly filled with the produced hadrons.
Also, due to the large scale $R$ of the process, the diffractive production mechanism cannot be explained in perturbative QCD and a dedicated theory of the diffractive processes is needed.

Historically, the soft processes have been described in the context of the Regge theory \cite{Regge:1959mz}, where the soft hadronic phenomena are modeled through the exchange of an object called \textit{Reggeon} at low energies, and through the exchange of a \textit{Pomeron} at higher energies. In modern terms, this interaction can be viewed as an exchange of two quarks in the $t$-channel at low energy and as the exchange of gluons at higher energies. 

Thus, as shown in Fig.~\ref{fig:excl_rho_diagram}, the reaction of Eq.~\ref{eq:exclr} can be thought to be mediated by a virtual photon $\gamma^*$ fluctuating into a $q\bar{q}$ pair (an off-shell vector meson) which then scatters elastically off the nucleon. The (by then on-shell) vector meson finally decays into a hadron pair.

A second approach, alternative to the diffractive one just briefly introduced, describes the production of exclusive vector mesons in terms of the Generalized Parton Distributions (GPDs). If the virtual photon is longitudinally polarized, the amplitude has been proven \cite{Collins:1996fb,Radyushkin:1997ki} to factorize into a calculable hard component and a soft component, related to GPDs. The same cannot be said for transverse photons; however, phenomenological models exists (the Goloskokov-Kroll model being one of the most famous \cite{Goloskokov:2006hr,Goloskokov:2007nt}) that allow one to describe the interaction cross-section for both longitudinal and transverse photon polarizations. In COMPASS, it is not possible to separate the longitudinal and transverse components of the cross-section: a Rosenbluth separation, that would exploit the different $Q^2$ dependence of the two components and tempted elsewhere \cite{Defurne:2016eiy}, is not feasible with a fixed beam energy.

\subsection{Coherence, incoherence and diffractive dissociation}
In a diffractive process, at least two regimes can be identified.
The lower the momentum transfer, the higher the probability that the interaction keeps the target nucleus intact. If this is the case, the nucleon is simply excited to a state of higher mass and the various amplitudes that originate at different parts of the nucleon and contribute to this excitation add up coherently \cite{Goulianos:1982vk}. The slope of the measured cross-section is mostly due to the target nucleus form factor (for an historical example on deuteron, see Ref.~\cite{Glauber:1967zzb}). 

At larger values of $|t|$, the process becomes incoherent and, to a good approximation, can be regarded as the sum of the cross-sections on the target nucleons. Moreover, if $|t|$ is larger (indicatively, $|t|>m_\pi^2$) the probability to have a diffractive dissociation of the target into pions increases. While this might be seen as a complication on the experimental side, as the exclusivity of the process is more difficult to retrieve, it has been found that the ratio of the cross-sections (with or without dissociation) is independent on $Q^2$ and the SDMEs for $\vmrho$ and $\vmphi$ have been found compatible in the two cases \cite{Aaron:2009xp}.

\section{Spin Density Matrix Elements formalism}
\label{sect:ch6_sdme_form}
In this Section we go through the Spin Density Matrix Elements (SDMEs) formalism according to Schilling and Wolf \cite{SCHILLING1973381} \footnote{An alternative treatment has been suggested by Diehl \cite{Diehl:2007jy}.} focusing on the $\vmrho$ vector meson, whose exclusive production can be written as:

\begin{equation}
    \ell N \longrightarrow \ell N \vmrho.
    \label{eq:rho_prod}
\end{equation}
With a branching ratio $BR \sim$ 100\% the $\vmrho$ decays into two charged pions: $\vmrho \longrightarrow \pi^+ \pi^-$. The cross-section for the vector meson production can be conveniently factorized in a part describing the emission of a virtual photon by the incoming lepton and in a part describing the photon oscillation into the $\vmrho$. We focus here on the spin-dependence of the process, which can be understood by using the Spin Density Matrix formalism. \\

The emission of the virtual photon by the incoming lepton is described by the leptonic tensor, also proportional to the photon spin density matrix $\varrho$. If the beam has a longitudinal polarization $P_b$, $\varrho$ reads:
\begin{equation}
    \varrho_{\lambda_\gamma \lambda_\gamma^\prime}^{U+L} = \varrho_{\lambda_\gamma \lambda_\gamma^\prime}^{U} + P_b \varrho_{\lambda_\gamma \lambda_\gamma^\prime}^{L}
\end{equation}
where $U$ ($L$) denotes the unpolarized (polarized) component of the spin density matrix and $\lambda_\gamma$ indicates one of the three possible helicity states of the photon: $\lambda_{\gamma}=-1,0,1$. The transition from the photon spin density matrix $\varrho$ to the vector meson spin density matrix $\rho$ is given by the von Neumann equation: 
\begin{equation}
    \rho_{\lambda_V\lambda_V^\prime} = \frac{1}{2\mathcal{N}} \sum_{\lambda_\gamma \lambda_\gamma^\prime \lambda_N \lambda_N^\prime} F_{\lambda_V \lambda_N^\prime \lambda_\gamma \lambda_N} \: \varrho_{\lambda_\gamma \lambda_\gamma^\prime}^{U+L} \: F^{*}_{\lambda_V^\prime \lambda_N^\prime \lambda_\gamma^\prime \lambda_N}
\label{eq:rho_sdm}    
\end{equation}
where $\mathcal{N}$ is a normalization constant and the sum is performed over the photon and nucleon helicities in the initial and final states. The quantities $F_{\lambda_V \lambda_N^\prime \lambda_\gamma \lambda_N}$, referred to as helicity amplitudes, describe the transition of a virtual photon with helicity $\lambda_\gamma$ to a vector meson with helicity $\lambda_V$ for given helicities of the nucleon ($\lambda_N$ and $\lambda_{N^\prime}$ for the target and recoil nucleon respectively). 

The vector meson spin density matrix $ \rho$ can then be decomposed into nine 3$\times$3 matrices  $\rho^\alpha$, with $\alpha=0,1,\dots,8$, which single out the contributions of the different virtual photon polarizations.  In particular, 

\begin{equation}
    \rho = \sum_{\alpha=0}^{8} \rho^{\alpha} = \frac{1}{2}\sum_{\alpha=0}^{8} \pi_\alpha \Sigma^\alpha
\end{equation}

where $\pi$ is a vector with nine-components (each corresponding to a decomposition constant) and
\begin{equation}
\begin{aligned}
\Sigma^0 & = \begin{pmatrix} 1&0&0 \\ 0&0&0 \\ 0&0&1  \end{pmatrix}; & 
\Sigma^1 & = \begin{pmatrix} 0&0&1 \\ 0&0&0 \\ 1&0&0  \end{pmatrix}; &
\Sigma^2 &= \begin{pmatrix} 0&0&-i \\ 0&0&0 \\ i&0&0  \end{pmatrix}; \\
\Sigma^3 &= \begin{pmatrix} 1&0&0 \\ 0&0&0 \\ 0&0&-1  \end{pmatrix}; &
\Sigma^4 &= 2\begin{pmatrix} 0&0&0 \\ 0&1&0 \\ 0&0&0  \end{pmatrix}; &
\Sigma^5 &= \frac{1}{\sqrt{2}}\begin{pmatrix} 0&1&0 \\ 1&0&-1 \\ 0&-1&0  \end{pmatrix}; \\ 
\Sigma^6 &= \frac{1}{\sqrt{2}}\begin{pmatrix} 0&-i&0 \\ i&0&i \\ 0&-i&0  \end{pmatrix}; &
\Sigma^7 &= \frac{1}{\sqrt{2}}\begin{pmatrix} 0&1&0 \\ 1&0&1 \\ 0&1&0  \end{pmatrix}; & 
\Sigma^8 &= \frac{1}{\sqrt{2}}\begin{pmatrix} 0&-i&0 \\ i&0&-i \\ 0&i&0  \end{pmatrix}. 
\end{aligned}
\end{equation}

The matrices $\Sigma^0-\Sigma^3$ are used to describe photons with transverse polarization, where $\Sigma^0$ gives the unpolarized part; $\Sigma^1$ and $\Sigma^2$ correspond to linear polarization while $\Sigma^3$ represents circular polarization; $\Sigma^4$ describes longitudinally polarized photons; $\Sigma^5-\Sigma^8$ represents interference terms.
The vector components $\pi_\alpha$, with $\alpha=3,7,8$, are different from zero only if the beam is longitudinally polarized. 

The cross-section for the whole process of Eq.~\ref{eq:rho_prod} is proportional to the trace $\mathrm{Tr}(\rho)$. In addition, the spin density matrix of the vector meson in Eq.~\ref{eq:rho_sdm} determines the angular distribution of the decay products in the meson rest frame. 

By studying the angular distributions of the exclusive hadrons, it is possible to extract information on the interaction mechanism encoded in the quantities $F_{\lambda_V \lambda_N^\prime \lambda_\gamma \lambda_N}$. Furthermore, the measurement of the nine $\rho^\alpha$ matrices allows to separate the Natural Parity Exchange (NPE) contributions to $F_{\lambda_V \lambda_N^\prime \lambda_\gamma \lambda_N}$ from the Unnatural Parity Exchange (UPE) ones. By \textit{natural} one means that the exchange is mediated by an object with parity $P=(-1)^J$, corresponding to $J^P = 0^+,1^-,2^+$ etc., and opposite for the \textit{unnatural} case. This separation is possible thanks to a symmetry property of the helicity amplitudes $F=T+U$, according to which the NPE elements $T$ and the UPE terms $U$ behave oppositely:

\begin{equation}
\begin{split}
    T^*_{-\lambda_V \lambda_N^\prime -\lambda_\gamma \lambda_N} = (-1)^{\lambda_V -\lambda_\gamma}T_{\lambda_V \lambda_N^\prime \lambda_\gamma \lambda_N} \\
    U^*_{-\lambda_V \lambda_N^\prime -\lambda_\gamma \lambda_N} = -(-1)^{\lambda_V -\lambda_\gamma}U_{\lambda_V \lambda_N^\prime \lambda_\gamma \lambda_N}  
\end{split}
\end{equation}

Several experimental results (the first from NMC \cite{Arneodo:1994id}) suggest that the helicity of the photon in the GNS system is approximately retained by the vector meson. This observation is referred to as \textit{s-channel helicity conservation} (SCHC). In terms of helicity amplitudes, it means that:
\begin{equation}
    F_{\lambda_V \lambda_N^\prime \lambda_\gamma \lambda_N} = F_{\lambda_V \lambda_N^\prime \lambda_\gamma \lambda_N} \delta_{\lambda_V \lambda_\gamma}\delta_{\lambda_N^\prime \lambda_N}.
\end{equation}
A measurement of the vector meson spin density matrix elements allows one to test the SCHC hypothesis and to understand whether the NPE terms are different from zero and how large they are. However, at fixed beam energy (as it is in the COMPASS case) it is not possible to explicitly separate the contributions coming from longitudinal and transverse photons, and only linear combinations of the $\rho^\alpha$ matrices elements can be accessed. These combinations, hereafter referred to as SDMEs, are:

\begin{equation}
\begin{split}
    & r_{\lambda_V\lambda_V^\prime}^{04} = \frac{\rho_{\lambda_V\lambda_V^\prime}^0 + \op\epsilon+\delta\cp R \rho_{\lambda_V\lambda_V^{\prime}}^4}{1+\op \epsilon+\delta\cp R} \\
    & r_{\lambda_V\lambda_V^\prime}^{\alpha} = \frac{\rho_{\lambda_V\lambda_V^\prime}^{\alpha}}{1+\op \epsilon+\delta\cp R} \hspace{2cm} \op\alpha=1,2,3 \cp \\
    & r_{\lambda_V\lambda_V^\prime}^{\alpha} = \frac{\sqrt{R}\rho_{\lambda_V\lambda_V^\prime}^{\alpha}}{1+\op \epsilon+\delta\cp R}  \hspace{2cm} \op \alpha=5,6,7,8 \cp
\end{split}
\label{eq:sdmes_def}
\end{equation}
where $R=\frac{\sigma_L}{\sigma_T}$ is the ratio of the longitudinal to the transverse virtual photon cross-section for the exclusive $\vmrho$ production and $\epsilon$ is the virtual photon polarization parameter, given by \cite{SCHILLING1973381}:

\begin{equation}
    \epsilon = \op 1 + 2 \frac{Q^2 + \nu^2}{Q^2\op 1-\frac{Q^2_{min}}{Q^2}\cp^2}\tan^2\frac{\Theta}{2}\cp^{-1}
\end{equation}
being $\Theta$ the lepton scattering angle and $Q^2_{min} = -2m_{\ell}^2+2\op E_{\ell}E_{\ell^\prime}- |p_{\ell}||p_{\ell^\prime}|\cp$. The lepton-mass correction factor $\delta$ in Eq.~\ref{eq:sdmes_def} is defined as:

\begin{equation}
    \delta = \frac{2m_{\ell}^2}{Q^2}\op 1-\epsilon\cp.
\end{equation}

In total, 23 SDMEs can be addressed with a longitudinally polarized beam and an unpolarized target. Among them, one of the most interesting is $r_{00}^{04}$. It can be demonstrated that, if SCHC holds, this term is directly related to the ratio $R$ of the longitudinal to transverse cross-section, namely:
\begin{equation}
    r_{00}^{04} \xrightarrow{SCHC} \frac{\op\epsilon+\delta\cp R}{1+\op\epsilon+\delta\cp R }.
\end{equation}

\subsection{Accessing SDMEs through angular distributions}    
The set of SDMEs fully defines the joint ($\vmrho$ and pions) three-dimensional angular distribution $W^{U+L}\op\cos\theta,\Phi,\phi\cp$:
\begin{equation}
    W^{U+L}(\cos\theta,\Phi,\phi) = W^{U}(\cos\theta,\Phi,\phi) + P_b W^{L}(\cos\theta,\Phi,\phi),
\label{eq:WUL}
\end{equation}
where $W^{U}$  is the unpolarized component, namely not dependent on the beam polarization $P_b$, while $W^{L}$ is the polarized component. This distribution is conveniently studied in the $s$-channel helicity frame \cite{Bauer:1977iq}, where the $\vmrho$ direction in the virtual photon-nucleon center of mass system is taken as the quantization axis. The angles $\theta$, $\Phi$ and $\phi$ are defined in Fig.~\ref{fig:excl_rho_angles}, which schematically shows the process. Their meaning is as follows: $\theta$ is the polar angle of the positive decay pion in the $\vmrho$ center of mass system and $\phi$ the azimuthal angle between the production and decay plane. The angle $\Phi$ is that of the $\vmrho$ production plane with respect to the lepton scattering plane. If the $s$-channel helicity conservation (SCHC) holds, the angular distribution $W(\cos\theta,\Phi,\phi)$ reduces to $W(\cos\theta,\psi)$, being $\psi=\phi-\Phi$ the angle of the $\vmrho$ decay plane with respect to the lepton scattering plane.\\

The unpolarized component $W^{U}$ depends on 15 \textit{unpolarized} SDMEs, while $W^{L}$ on 8 \textit{polarized} terms. Explicitly, the two components read:

\begin{figure}
\captionsetup{width=\textwidth}
    \centering
    \includegraphics[scale=0.3]{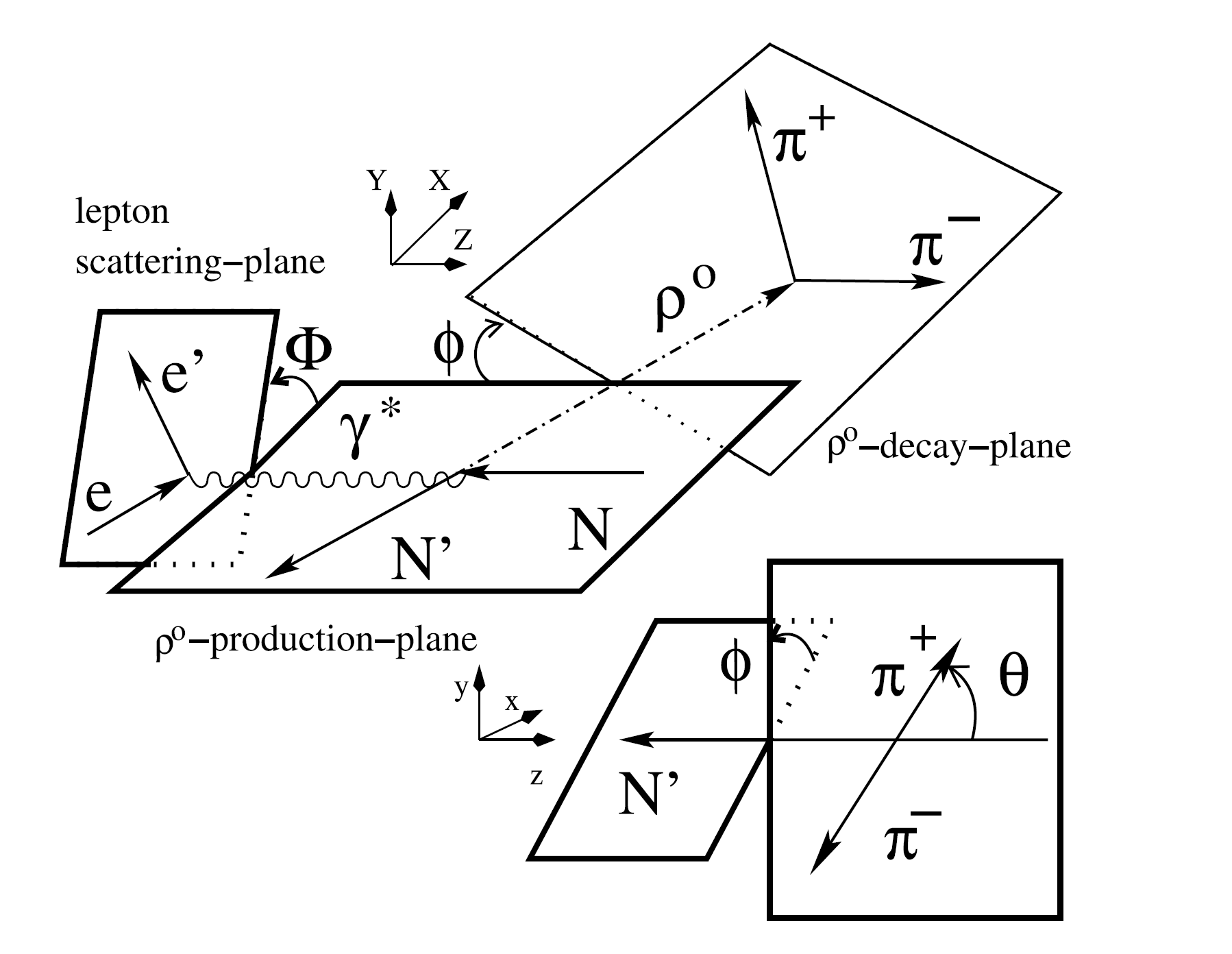}
    \caption{Angular distributions for the electro-production of a $\vmrho$ vector meson and its subsequent decay into a pion pair, from Ref.~\cite{Airapetian:2009af}.}
    \label{fig:excl_rho_angles}
\end{figure}

\begin{equation}
    \begin{split}
        W^{U}&(\cos\theta,\Phi,\phi)  = \frac{3}{8\pi^2} \Biggr[ \frac{1}{2}\op 1-r_{00}^{04}\cp + \frac{1}{2}\op 3r_{00}^{04}-1\cp \cos^2\theta - \sqrt{2}\mathrm{Re}\opc r_{10}^{04}\cpc\sin2\theta\cos\phi - r_{1-1}^{04}\sin^2\theta\cos2\phi \\ 
        & - \epsilon \cos2\Phi \op r_{11}^{1}\sin^2\theta + r_{00}^{1}\cos^2\theta - \sqrt{2}\mathrm{Re}\opc r_{10}^{1}\cpc \sin^2\theta\cos\phi - r_{1-1}^{1}\sin^2\theta\cos2\phi \cp \\
        & - \epsilon \sin2\Phi \op \sqrt{2}\mathrm{Im}\opc r_{10}^{2}\cpc\sin2\theta\sin\phi + \mathrm{Im}\opc r_{1-1}^{2}\cpc\sin^2\theta\sin2\phi \cp \\
        & + \sqrt{2\epsilon\op1+\epsilon\cp} \cos\Phi \op r_{11}^{5}\sin^2\theta + r_{00}^{5}\cos^2\theta - \sqrt{2}\mathrm{Re}\opc r_{10}^{5}\cpc\sin2\theta\cos\phi - r_{1-1}^{5}\sin^2\theta\cos2\phi \cp \\
        & + \sqrt{2\epsilon\op1+\epsilon\cp} \sin\Phi \op \sqrt{2}\mathrm{Im}\opc r_{10}^{6}\cpc\sin2\theta\sin\phi + \mathrm{Im}\opc r_{1-1}^{6}\cpc\sin^2\theta\sin2\phi \cp \Biggr]
    \end{split}
\label{eq:WU}
\end{equation}

\begin{equation}
    \begin{split}
        W^{L}&(\cos\theta,\Phi,\phi) = \frac{3}{8\pi^2} \Biggr[ \sqrt{1-\epsilon^2} \op \sqrt{2}\mathrm{Im}\opc r_{10}^{3}\cpc\sin2\theta\sin\phi + \mathrm{Im}\opc r_{1-1}^{3}\cpc\sin^2\theta\sin2\phi \cp\\
        & +\sqrt{2\epsilon\op1-\epsilon\cp}\cos\Phi\op \sqrt{2}\mathrm{Im}\opc r_{10}^{7}\cpc\sin2\theta\sin\phi + \mathrm{Im}\opc r_{1-1}^{7}\cpc\sin^2\theta\sin2\phi \cp\\
        & +\sqrt{2\epsilon\op1-\epsilon\cp}\sin\Phi\op r_{11}^{8}\sin^2\theta + r_{00}^{8}\cos^2\theta - \sqrt{2}\mathrm{Re}\opc r_{10}^{8}\cpc \sin2\theta\cos\phi - r_{1-1}^{8}\sin^2\theta\cos2\phi\cp \Biggr]
    \end{split}
\label{eq:WL}    
\end{equation}

The angles $\theta$, $\Phi$ and $\phi$ are more precisely defined as follows.\\
The polar angle $\theta$ is defined in the $\vmrho$ center of mass as (minus) the angle between the recoil proton momentum $\vec{P}_{p,rec}^{~\vmrho}$ and the $\pi^+$ momentum $\vec{P}_{\pi^+}^{~\vmrho}$, according to the following expression:

\begin{equation}
        \cos\theta = -\frac{\vec{P}_{p,rec}^{~\vmrho} \cdot \vec{P}_{\pi^+}^{~\vmrho} }{|\vec{P}_{p,rec}^{~\vmrho}| |\vec{P}_{\pi^+}^{~\vmrho}| }
    \end{equation}
where $\vec{P}_{p,rec}$ can be obtained by energy conservation.\\

The azimuthal angle $\Phi$ between the $\vmrho$ production plane and the lepton scattering plane is calculated in the Hadron Center of Mass System (HCMS), defined as the system where the initial state (formed by the virtual photon and the target nucleon) is at rest. To reach the HCMS from the laboratory frame, all the relevant momenta are boosted after applying a rotation, that in the case of the virtual photon makes it aligned to the z-axis. Indicating with the apex $^*$ the momenta in the HCMS, the angle $\Phi$ is given by:
    \begin{equation}
        \Phi = \arctan\op\frac{\sin\Phi}{\cos\Phi}\cp.
    \end{equation}
where:    
    \begin{equation}
        \cos\Phi = \frac{\op\vec{P}_{\gamma}^* \times \vec{P}_{\vmrho}^*\cp \cdot \op\vec{P}_{\mu}^* \times \vec{P}_{\mu^{\prime}}^*\cp }{|\vec{P}_{\gamma}^* \times \vec{P}_{\vmrho}^*| |\vec{P}_{\mu}^* \times \vec{P}_{\mu^{\prime}}^*|}
    \end{equation}
    \begin{equation}
        \sin\Phi = \frac{|\op\vec{P}_{\gamma}^* \times \vec{P}_{\vmrho}^*\cp \times \op\vec{P}_{\mu}^* \times \vec{P}_{\mu^{\prime}}^*\cp| \cdot \vec{P}_{\gamma}^* }{|\vec{P}_{\gamma}^* \times \vec{P}_{\vmrho}^*| |\vec{P}_{\mu}^* \times \vec{P}_{\mu^{\prime}}^*| |\vec{P}_{\gamma}^*|}.
    \end{equation}\\

The azimuthal angle $\phi$ between the production and decay plane, $\phi$, is calculated in the HCMS as:
    \begin{equation}
        \phi = \arctan\op\frac{\sin\phi}{\cos\phi}\cp
    \end{equation}
where: 
    \begin{equation}
        \cos\phi = \frac{\op\vec{P}_{\gamma}^* \times \vec{P}_{\vmrho}^*\cp \cdot \op\vec{P}_{\vmrho}^* \times \vec{P}_{\pi^{+}}^*\cp }{|\vec{P}_{\gamma}^* \times \vec{P}_{\vmrho}^*| |\vec{P}_{\vmrho}^* \times \vec{P}_{\pi^{+}}^*|}
    \end{equation}
    \begin{equation}
        \sin\phi = \frac{\op\vec{P}_{\gamma}^* \times \vec{P}_{\vmrho}^* \times \vec{P}_{\vmrho}^* \cp \cdot \op\vec{P}_{\pi^{+}}^* \times \vec{P}_{\vmrho}^*\cp}{|\vec{P}_{\gamma}^* \times \vec{P}_{\vmrho}^* \times \vec{P}_{\vmrho}^*||\vec{P}_{\pi^{+}}^* \times \vec{P}_{\vmrho}^*|}
    \end{equation}

\bigskip
The 23 SDMEs can be conveniently organized in classes, according to the transition type:
\begin{itemize}
    \item class A: SCHC $\op r_{00}^{04},~r_{1-1}^{1},~\mathrm{Im}\opc r_{1-1}^{2}\cpc,~\mathrm{Re}\opc r_{10}^{5}\cpc\cp$
    
    \item class B: interference $\op \mathrm{Im} \opc r_{10}^{6}\cpc,~\mathrm{Im} \opc r_{10}^{7}\cpc,~\mathrm{Re}\opc r_{10}^{8}\cpc,~\mathrm{Re}\opc r_{10}^{04}\cpc\cp$
    
    \item class C: T$\to$L spin flip $\op \mathrm{Re} \opc r_{10}^{1}\cpc,~\mathrm{Im} \opc r_{10}^{2}\cpc,~r_{00}^{5},~r_{00}^{1},~\mathrm{Im}\opc r_{10}^{3}\cpc,~r_{00}^{8}\cp$
    
    \item class D: L$\to$T spin flip $\op r_{11}^{5},~r_{1-1}^{5},~\mathrm{Im} \opc r_{1-1}^{6} \cpc,~\mathrm{Im}\opc r_{1-1}^{7}\cpc,~r_{11}^{8},~r_{1-1}^{8}\cp$
    
    \item class E: double spin flip $\op r_{1-1}^{04},~r_{11}^{1},~\mathrm{Im} \opc r_{1-1}^{3} \cpc \cp$
\end{itemize}
If SCHC holds, the elements in classes C, D and E are expected to be equal to zero.

\bigskip
\section{Data used in the analysis}
\label{sect:ch6_sdme_measurement}
\subsection{Data and Monte Carlo samples}
The measurement of the $\vmrho$ SDMEs has been performed using the $\mu^+$ and $\mu^-$ data collected in COMPASS during the 2012 pilot run, for which the experimental setup was very similar to the one described in Ch.~\ref{Chapter2_COMPASS} for the 2016 data taking. In particular, compared to the 2016 case, ECAL0 was partially mounted and the recoil proton detector (CAMERA) was not in its final shape. Of the five weeks of data collected in 2012 all have been used here: they are the weeks W44, W45, W46, W47, W48, for which the productions T15, T11, T12, T13, T14 have respectively been considered. A list of bad spills has been taken into account in order to discard the events in which instabilities had been identified. The results presented in the following have been produced without considering the information coming from the CAMERA detector, not to limit the accessible kinematic range, particularly in the transverse momentum. 

Like the SIDIS measurements presented in the previous Chapters, this measurement also requires the use of Monte Carlo simulations to estimate both the acceptance corrections and the background contamination. Also in this case the TGEANT package has been used to produce the simulated events, using the HEPGEN generator \cite{Sandacz:2012at}, in which the relevant angular distributions are generated flat, for the acceptance correction, and the LEPTO generator \cite{Ingelman:1996mq} for both the estimate of the amount of the SIDIS background and of the SDMEs of the SIDIS background, that need to be taken into account when performing the fit of the observed angular distribution.

\subsection{Exclusive events selection}
The selection of exclusive $\vmrho$ events has been performed according to the following list of cuts:
\begin{itemize}
    \item \textbf{Event topology.} The topology of the event has been selected asking for exactly three particles stemming from the vertex, for the scattered muon candidate and the two decay particles; the vertex has been required to be the best \textit{primary}, with one incoming and one outgoing muon track reconstructed in CORAL. 
    \item \textbf{Vertex position.} The position of the primary vertex has been checked to be inside the fiducial target volume: $-311.2~\mathrm{cm}<z_{vtx}<-71.2~\mathrm{cm}$ and within a radius $R=1.9$~cm from the target center. 
    \item \textbf{Beam track properties.} The beam track has been required to have an energy in the range $140~\mathrm{GeV}<E_{beam}< 180~\mathrm{GeV}$ and to cross the full target length within the radius $R$, with at least three hits left in the Beam Momentum Station (BMS), an association probability to the BMS larger than 1\% and a good track quality.
    \item \textbf{Scattered muon track properties.} The quality of the track reconstruction has been required to be good, with its identification as muon being based on the number of crossed radiation lengths ($X/X_0>15$).
    \item \textbf{Trigger.} At least one among Middle Trigger (MT), Ladder Trigger (LT) and Outer Trigger (OT) has been required in the event.
    \item \textbf{Conditions on the two hadrons.} The reconstruction quality of each of the two outgoing tracks (different from the scattered muon) has been required to be good. The hadron identification has been performed asking for a small number of crossed radiation lengths ($X/X_0<10$). For a better track reconstruction, the first hit has been required to be located before SM1 ($Z_{First}<350$~cm). The number of hadrons has been required to be exactly equal to two, with a null total charge.
    \item \textbf{Kinematic range.}  The photon virtuality $Q^2$ has been selected to be in the range $1~\mathrm{(GeV/}c\mathrm{)}^2<Q^2<10~\mathrm{(GeV/}c\mathrm{)}^2$; the invariant mass of the hadronic final state $W$ in the range $5~\mathrm{GeV/}c^2< W < 17~\mathrm{GeV/}c^2$; the inelasticity $y$ in $0.1<y<0.9$ to avoid bad reconstruction precision and high radiative corrections; the virtual photon energy in the laboratory frame $\nu$ has been selected to be $\nu>20$~GeV. The transverse momentum of the reconstructed pair, with respect to the virtual photon has been required to be in the range $0.01~\mathrm{GeV/}c^2<p_T^2<0.5~\mathrm{GeV/}c^2$; the invariant mass of the pair $M_{h^+h^-}$ to satisfy  $0.5~\mathrm{GeV/}c^2<M_{h^+h^-}<1.1~\mathrm{GeV/}c^2 $ and the momentum of the pair above $P_{h^+h^-}>15$~GeV/$c$.  The cut on the invariant mass of the pair has been chosen in order to limit the interference with the non-resonant pair production. The cut on the pair momentum has been introduced to limit the SIDIS background.
\end{itemize}
The number of $\vmrho$ candidates after all selection steps, and summing over the $\mu^+$ and $\mu^-$ samples, amount to 188~844, of which 52~257 in the low missing energy region ($-2.5~\mathrm{GeV}<E_{miss}<2.5~\mathrm{GeV}$). In the same region, the LEPTO and HEPGEN statistics amount to 139~589 and 1~107~827 (unweighted) candidates respectively.

\section{Data analysis}
\label{sect:ch6_procedure}
The $\vmrho$ SDMEs have been measured in one-dimensional bins of $\Qsq$, $p_T^2$ and $W$, according to the following limits:
\begin{itemize}
    \item $Q^2$ binning (GeV/$c$)$^2$: 1.0,~1.3,~2.0,~4.0,~10.0;
    \item $p_T^2$ binning (GeV/$c$)$^2$: 0.01,~0.10,~0.20,~0.30,~0.50;
    \item $W$ binning (GeV/$c^2$): 5.0,~7.3,~9.0,~12.0,~17.0.
\end{itemize}
and in the full phase space. The measurement procedure consists of three steps. The first one, which is the reference for the following two, is the measurement of the SDMEs with no background correction: this means that neither the amount of SIDIS background in the selected exclusive sample nor its possible angular modulations are taken into account. In the second step, the fraction of background is evaluated and the background SDMEs are measured on a sample of SIDIS events. The third and last step consists of the complete approach, in which the fraction of background and its angular modulations are taken into consideration. The SDMEs for the $\vmrho$ vector meson have been extracted with the Unbinned Maximum Likelihood (UML) method by fitting the measured angular distribution to the theoretical expression $W^{U+L}(\cos\theta,\Phi,\phi)$, which is defined in Eq.~\ref{eq:WUL}-\ref{eq:WL} and depends on 23 free parameters. In this UML approach, already applied in the $\omega$ case \cite{COMPASS:2020zre}, real data and data simulated with flat angular distributions are simultaneously fitted. This allows for a simple treatment of the acceptance correction. The disadvantages are that the smearing effect of the detector resolutions is neglected and that the amount of reconstructed Monte Carlo statistics must be large compared to the experimental data. The expression for the likelihood function is derived in Appendix~\ref{AppendixD}.
As introduced in the previous subsection about the Unbinned Maximum Likelihood formulation, the measurement requires a Monte Carlo sample in order to take into account the acceptance of the apparatus. For this purpose, the HEPGEN generator \cite{Sandacz:2012at} has been used, with flat generated distributions of the three quantities $\cos\theta,\Phi,\phi$ and for which the reconstructed events have been selected with the same procedure as the data.

\subsection{Measurement with no background correction}
According to the expression in Eq.~\ref{eq:lnL2} and neglecting the background contribution to the observed angular distributions, the set $\mathcal{R}_0$ of the 23 SDMEs can be measured by minimizing the quantity:
\begin{equation}
\begin{split}
    -\ln L(\mathcal{R}_0) & = - \sum_{i=1}^{N} \ln W^{U+L}(\mathcal{R}_0;\cos\theta_i,\Phi_i,\phi_i) + N \ln \sum_{j=1}^{M} W^{U+L}(\mathcal{R}_0;\cos\theta_j,\Phi_j,\phi_j) \\
    & = - \sum_{i=1}^{N} \ln \frac{W^{U+L}(\mathcal{R}_0;\cos\theta_i,\Phi_i,\phi_i)}{\sum_{j=1}^{M} W^{U+L}(\mathcal{R}_0;\cos\theta_j,\Phi_j,\phi_j)}
\end{split}
\label{eq:lnL4}
\end{equation}
where the first sum runs over the data sample, of size $N$, and the second (the normalization term) over the reconstructed Monte Carlo events, of size $M$. The minimization of the likelihood has been implemented in a dedicated \texttt{C++} minimization software based on \texttt{MINUIT} \cite{James:1994vla}, in which a first call to \texttt{MIGRAD} with the option \texttt{SET STRATEGY 2} (to check that the found minimum is the true one and that the uncertainties are correct) was followed by a call to \texttt{HESSE} for the precise determination of the covariance matrix. The data collected with $\mu^+$ and $\mu^-$ beam were considered altogether in the fit, the polarization of the beam being $P_b=-0.80$ and $P_b=+0.80$ respectively.

\subsection{Background fraction and background SDMEs}

In order to correct for the SIDIS background, one has to measure the set of background SDMEs (indicated as $\mathcal{B}$) and to estimate the background fraction $f_{bg}$ in each kinematic bin.

\begin{figure}
    \centering
    \captionsetup{width=\textwidth}
    \includegraphics[width=\textwidth]{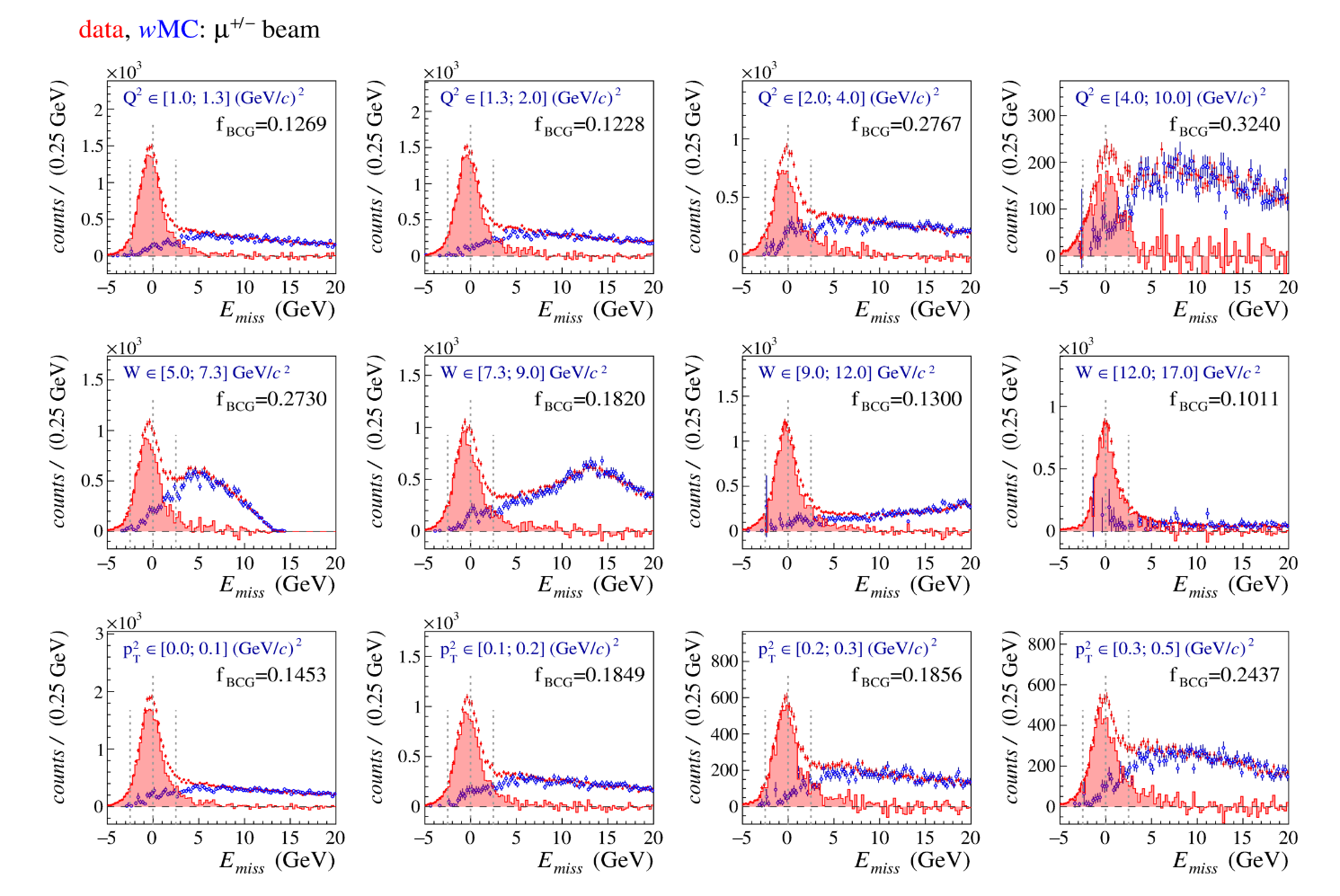}
    \caption{$E_{miss}$ distributions for the data (red points), the reweighted, normalized LEPTO (blue points) and for their difference (filled red histogram). The fraction of background is also indicated. Plot by B.~Parsamyan (COMPASS). }
    \label{fig:fbg_sdmes}
\end{figure}

The first point has been accomplished by applying Eq.~\ref{eq:lnL2} to a reconstructed LEPTO Monte Carlo sample for SIDIS \cite{Ingelman:1996mq}, selected with the same criteria followed for the real data. The same HEPGEN sample has been used, as in the previous step, for the determination of the normalization term. Denoting by $S$ the size of the SIDIS sample, the log-likelihood reads:
\begin{equation}
    -\ln L(\mathcal{B}) = - \sum_{i=1}^{S} \ln \frac{W^{U+L}(\mathcal{B};\cos\theta_i,\Phi_i,\phi_i)}{\sum_{j=1}^{M} W^{U+L}(\mathcal{B};\cos\theta_j,\Phi_j,\phi_j)}
\label{eq:lnL5}
\end{equation}
whose minimum has been found with the same code used in the first step. \\
The estimation of the fraction of background, performed by other members of the COMPASS Collaboration involved in this analysis, has been done by comparing the missing energy distribution of the reconstructed data with the same distribution from LEPTO. To improve the agreement between the two, LEPTO has been reweighted in order to balance the fraction of same-charge and opposite-charge hadron pairs; then, in each kinematic bin, LEPTO has been normalized to the data in the range $7~\mathrm{GeV}<E_{miss}<20~\mathrm{GeV}$. Finally, the fraction of background has been calculated by comparing the number of events in the data and in the reweighted, normalized LEPTO in the range $-2.5\mathrm{GeV}<E_{miss}<2.5\mathrm{GeV}$, giving a fraction of background $f_{bg}$ ranging from 0.10 at high $W$ to 0.32 at high $Q^2$. The $E_{miss}$ distributions from the data and from the normalized LEPTO sample, as well as their difference, are shown in Fig.~\ref{fig:fbg_sdmes} for the various kinematic bins in which the analysis has been performed.

\subsection{Measurement with background subtraction}
Once the fraction of background $f_{bg}$ and the set of background SDMEs $\mathcal{B}$ are known, the full expression for the likelihood reads:
\begin{equation}
    -\ln L(\mathcal{R}) = - \sum_{i=1}^{N} \ln \frac{(1-f_{bg}) W^{U+L}(\mathcal{R};\cos\theta_i,\Phi_i,\phi_i) + f_{bg}W^{U+L}(\mathcal{B};\cos\theta_i,\Phi_i,\phi_i)}{\sum_{j=1}^{M}\op (1-f_{bg}) W^{U+L}(\mathcal{R};\cos\theta_j,\Phi_j,\phi_j) + f_{bg}W^{U+L}(\mathcal{B};\cos\theta_j,\Phi_j,\phi_j)\cp}.
\label{eq:lnL4}
\end{equation}
Namely, for each event $i$ ($i=1,\dots,N$) in the data, the $W^{U+L}$ function is calculated using the known background SDMEs $\mathcal{B}$ (fixed) and the unknown $\vmrho$ SDMEs $\mathcal{R}$ (free parameters), properly scaled with the fraction of background (or its complementary). An identical expression is evaluated for each event $j$ in the HEPGEN Monte Carlo ($j=1,\dots,M$). The minimization algorithm thus allows for a determination of the best set of values $\mathcal{R}$ for given $f_{bg}$ and $\mathcal{B}$.

\section{Results}
\begin{figure}
    \centering
    \captionsetup{width=\textwidth}
    \includegraphics[width=0.9\textwidth]{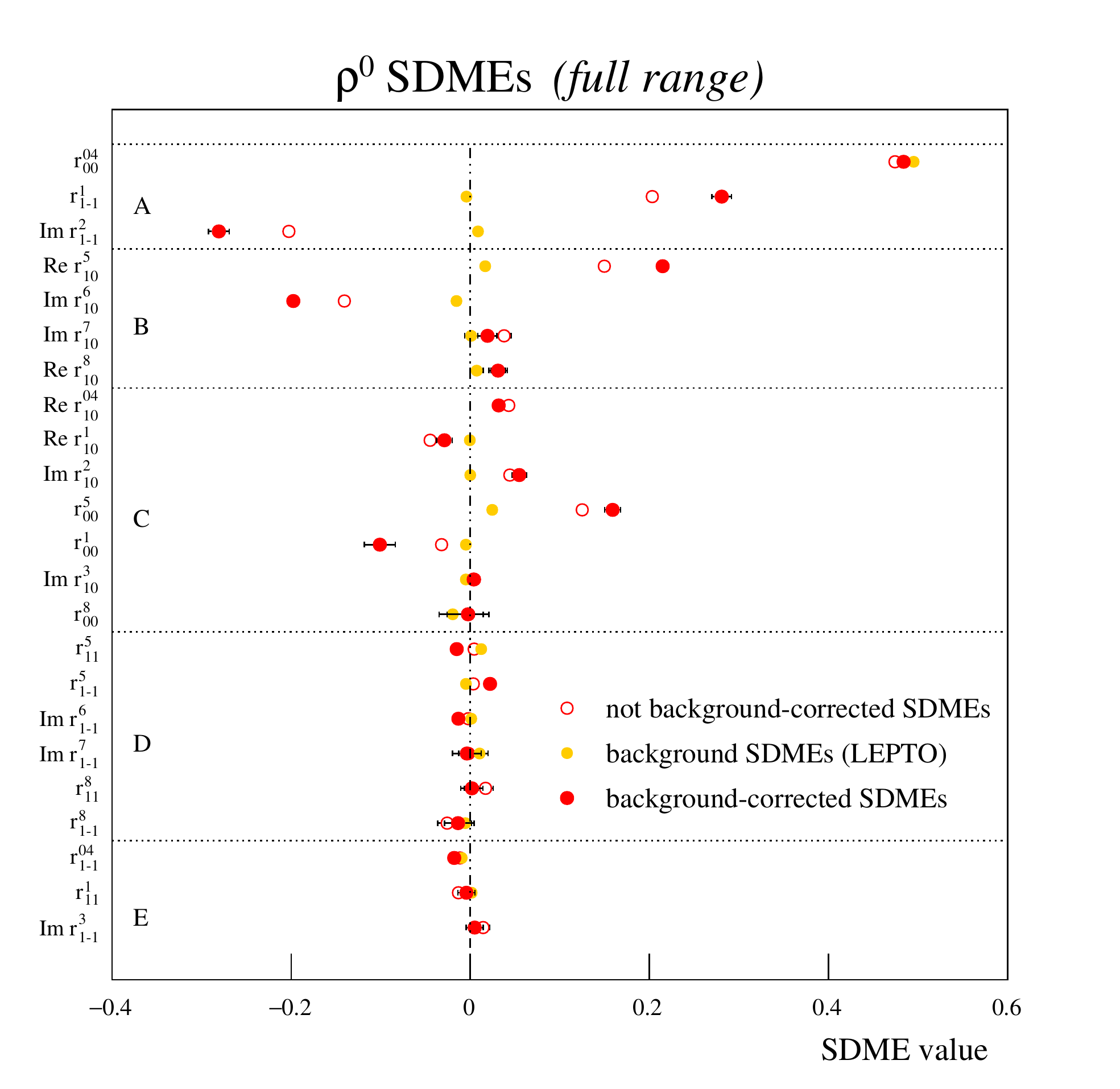} 
    \caption{SDMEs in the full kinematic region not corrected for the background (empty points), of the background (orange points) and of the background-corrected data (closed red points). The uncertainties are statistical only.}
    \label{fig:sdmes_tot}
\end{figure}

\begin{figure}
\captionsetup{width=\textwidth}
    \centering
    \includegraphics[width=\textwidth]{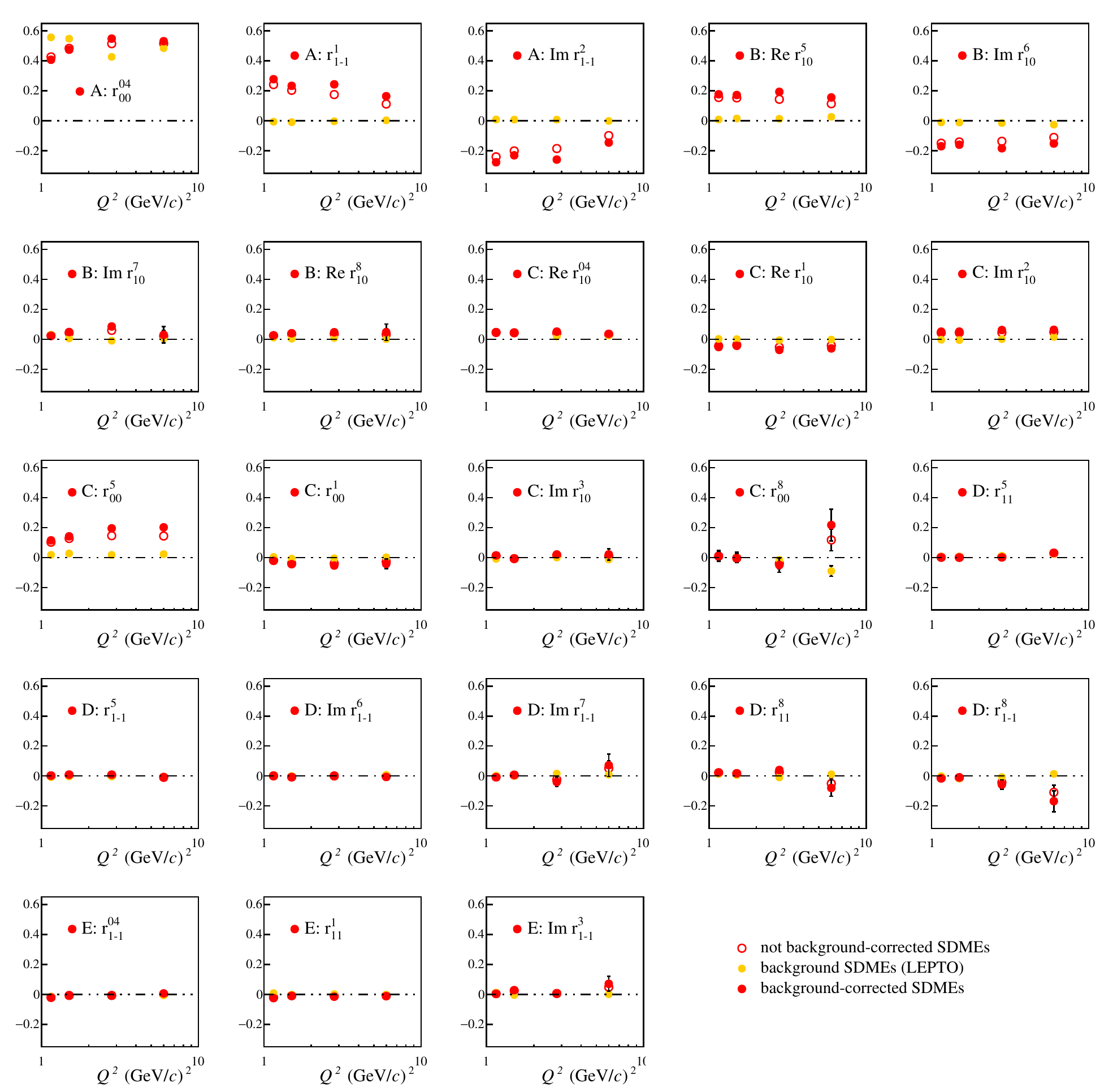} 
    \caption{Dependence on $\Qsq$ of the SDMEs not corrected for the background (empty points), of the background (orange points) and of the background-corrected data (closed red points), for $5~\mathrm{GeV}/c^2\mathrm{}<W<17~\mathrm{GeV}/c^2\mathrm{}$ and $p_T^2<0.50$~(GeV/$c$)$^2$. The uncertainties are statistical only. }
    \label{fig:sdmes_q2}
\end{figure}

\begin{figure}
\captionsetup{width=\textwidth}
    \centering
    \includegraphics[width=\textwidth]{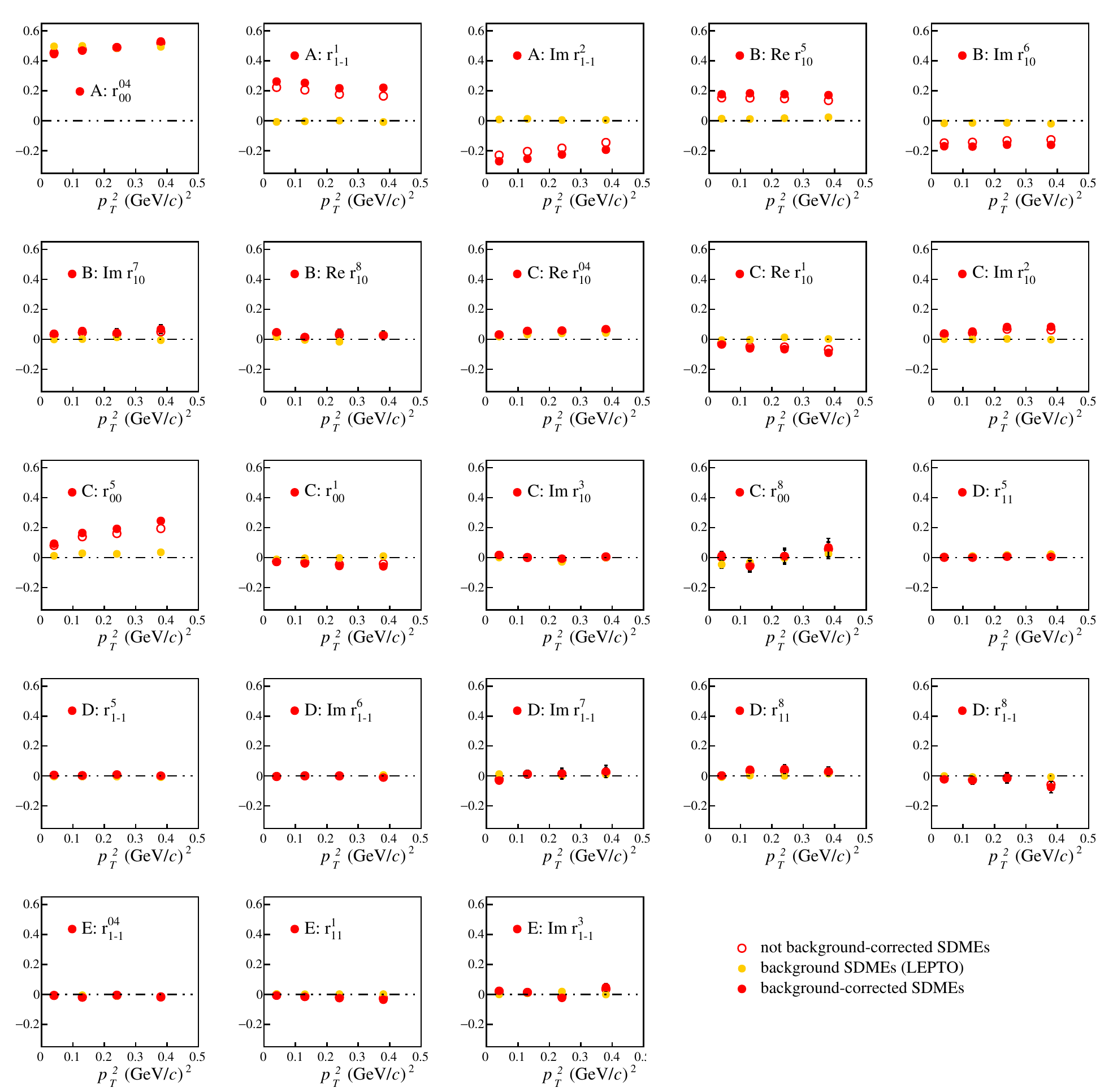} 
    \caption{Dependence on $W$ of the SDMEs not corrected for the background (empty points), of the background (orange points) and of the background-corrected data (closed red points),  for $1~\mathrm{(GeV}/c\mathrm{)}^2<\Qsq<10~\mathrm{(GeV}/c\mathrm{)}^2$ and $p_T^2<0.50$~(GeV/$c$)$^2$. The uncertainties are statistical only.}
    \label{fig:sdmes_w}
\end{figure}

\begin{figure}[h!]
\captionsetup{width=\textwidth}
    \centering
    \includegraphics[width=\textwidth]{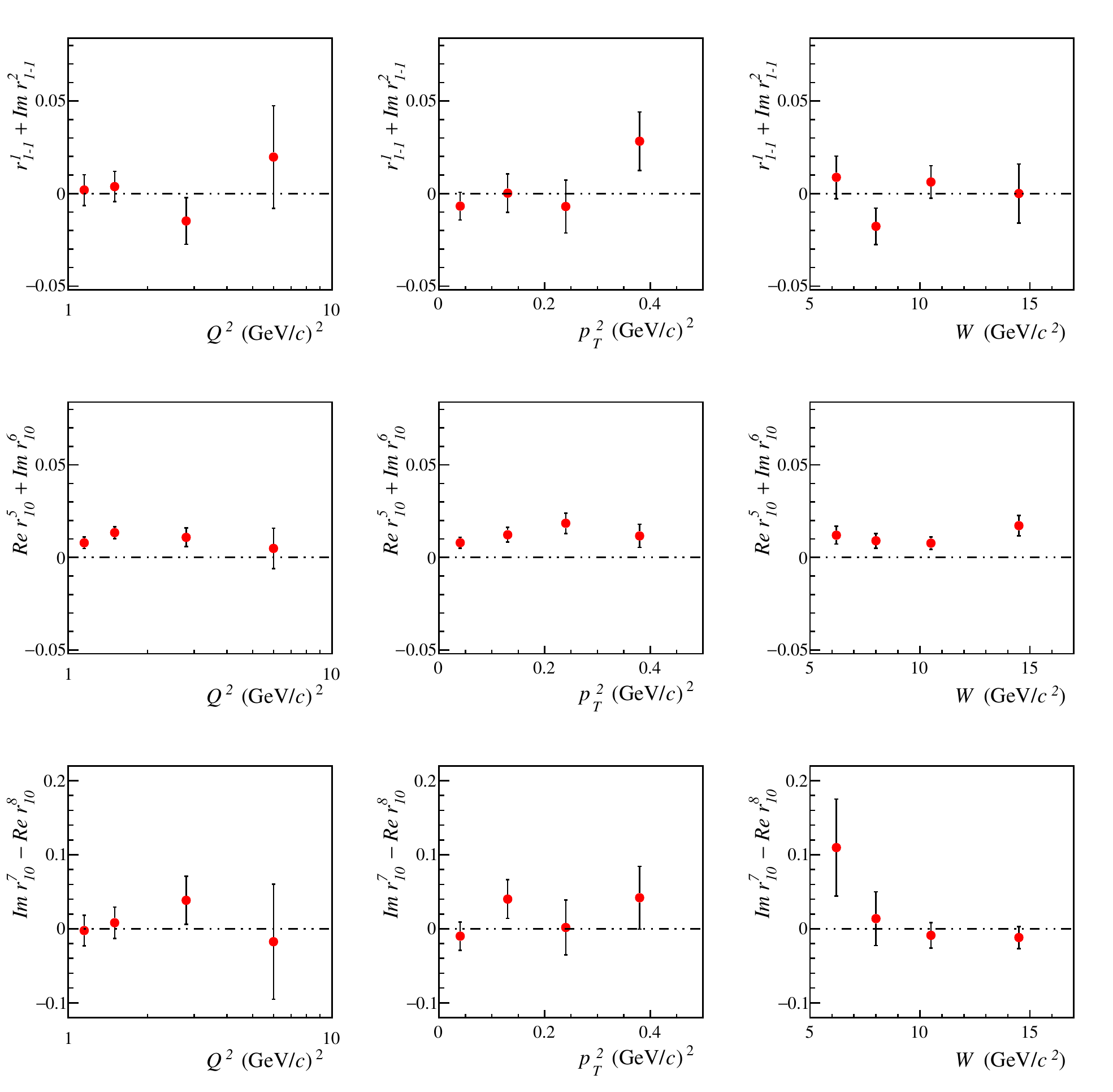} 
    \caption{Dependence on $Q^2$, $p_T^2$ and $W$ of three relations among elements in the classes A and B, expected from the SCHC hypothesis. The uncertainties are statistical only.}
    \label{fig:schc}
\end{figure}

The results for the full kinematic range are presented in Fig.~\ref{fig:sdmes_tot} for the SDMEs not corrected for the background (empty points), for the background (orange points) and for the background-corrected SDMEs (closed red points). The kinematic dependences of the SDMEs are shown in Figg.~\ref{fig:sdmes_q2}-~\ref{fig:sdmes_w}. For all the results, only the statistical uncertainties are shown: the estimate of the systematic uncertainties goes beyond the scope of the preliminary study presented in this Chapter, whose aim is to fix reasonable values for the SDMEs to be included in the HEPGEN Monte Carlo for the angular modulations of the exclusive hadrons.
A significant violation of the SCHC hypothesis can be observed for the elements of the class C (corresponding to the transitions $\gamma_T^*\to\vmrho_L$) which, together with the elements in the classes D and E, are expected to vanish if this hypothesis holds. A milder violation is present for the elements of the classes D and E. Assuming that the SCHC hypothesis holds, three relations can be derived also for some elements in the classes A and B:

\begin{equation}
    \begin{split}
        r_{1-1}^{1} + \mathrm{Im}~r_{1-1}^2 & = 0; \\
        \mathrm{Re}~r_{10}^{5} + \mathrm{Im}~r_{10}^6 & = 0; \\
        \mathrm{Im}~r_{10}^{7} - \mathrm{Re}~r_{10}^8 & = 0.
    \end{split}
\end{equation}
These quantities are shown as a function of $Q^2$, $p_T^2$ and $W$ in Fig.~\ref{fig:schc}. Considering the large uncertainties, only the second relation indicates a violation of the SCHC hypothesis: the violation is however small, being of the order of 1-2\%.

A detailed investigation of these results is beyond the scope of this work. Just as an example, we estimate the contribution to the total cross-section due to the Unnatural Parity Exchange (UPE) transitions, which can be obtained introducing the three quantities $u_1$, $u_2$ and $u_3$, defined as: 

\begin{equation}
    \begin{split}
        u_1 & = 1 - r_{00}^{04} + 2r_{1-1}^{04} -2r_{11}^{1} -2r_{1-1}^{1}; \\
        u_2 & = r_{11}^{5} + r_{1-1}^{5}; \\
        u_3 & = r_{11}^{8} + r_{1-1}^{8}.
        \end{split}
\end{equation}

\begin{figure}[h!]
\captionsetup{width=\textwidth}
    \centering
    \includegraphics[width=\textwidth]{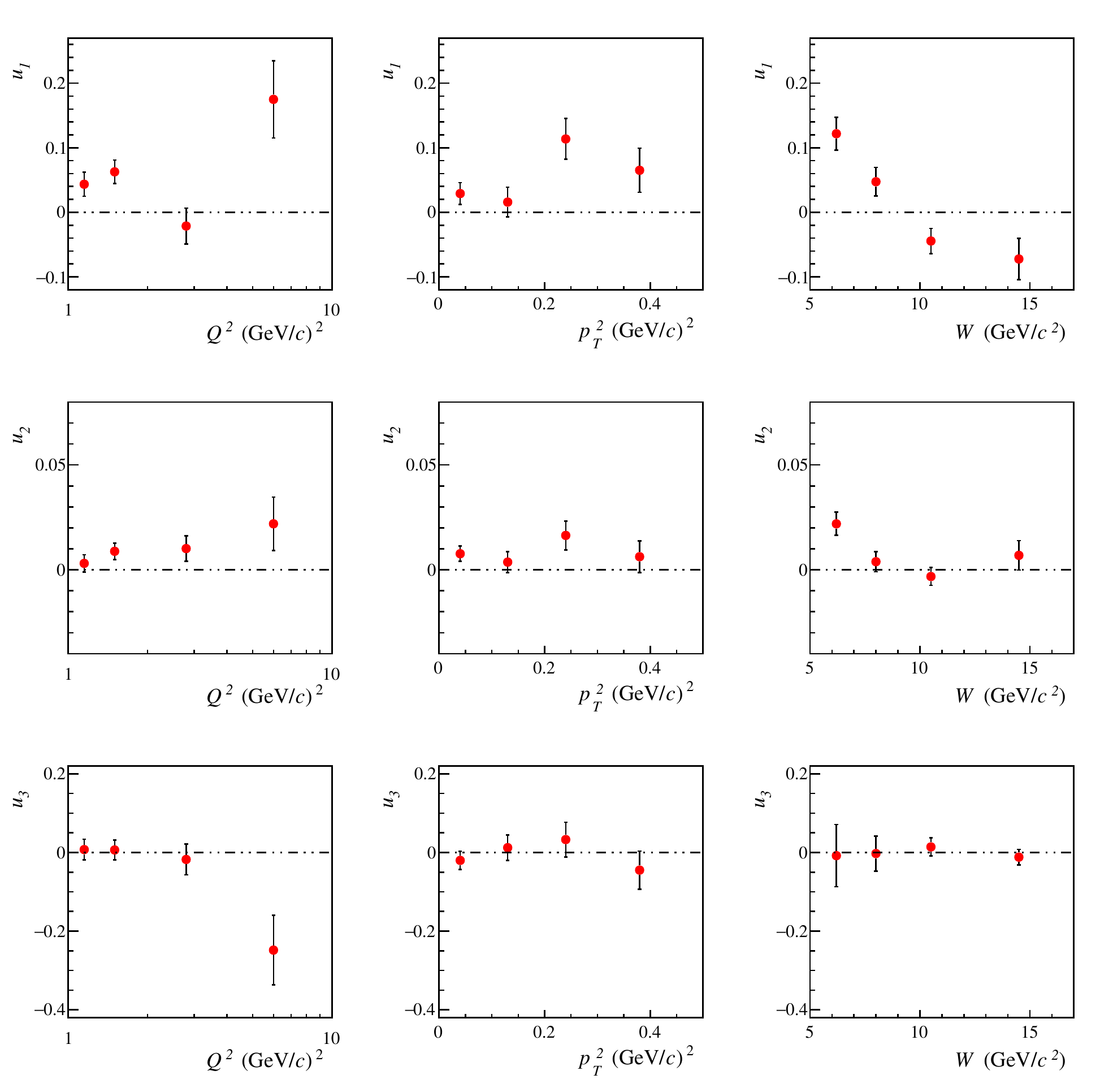} 
    \caption{Dependence on $Q^2$, $p_T^2$ and $W$ of the UPE quantities $u_1$, $u_2$ and $u_3$. The uncertainties are statistical only.}
    \label{fig:upe}
\end{figure}

The kinematic dependences of these quantities are presented in Fig.~\ref{fig:upe}: in particular for $u_1$ and $u_2$, the signals are different from zero, indicating the presence of UPE processes, with a clear decreasing trend as a function of $W$. The value of these quantities is however small, compared to the analogous COMPASS results for the $\omega$ case \cite{COMPASS:2020zre}.

\begin{figure}[h!]
\captionsetup{width=\textwidth}
    \centering
    \includegraphics[width=\textwidth]{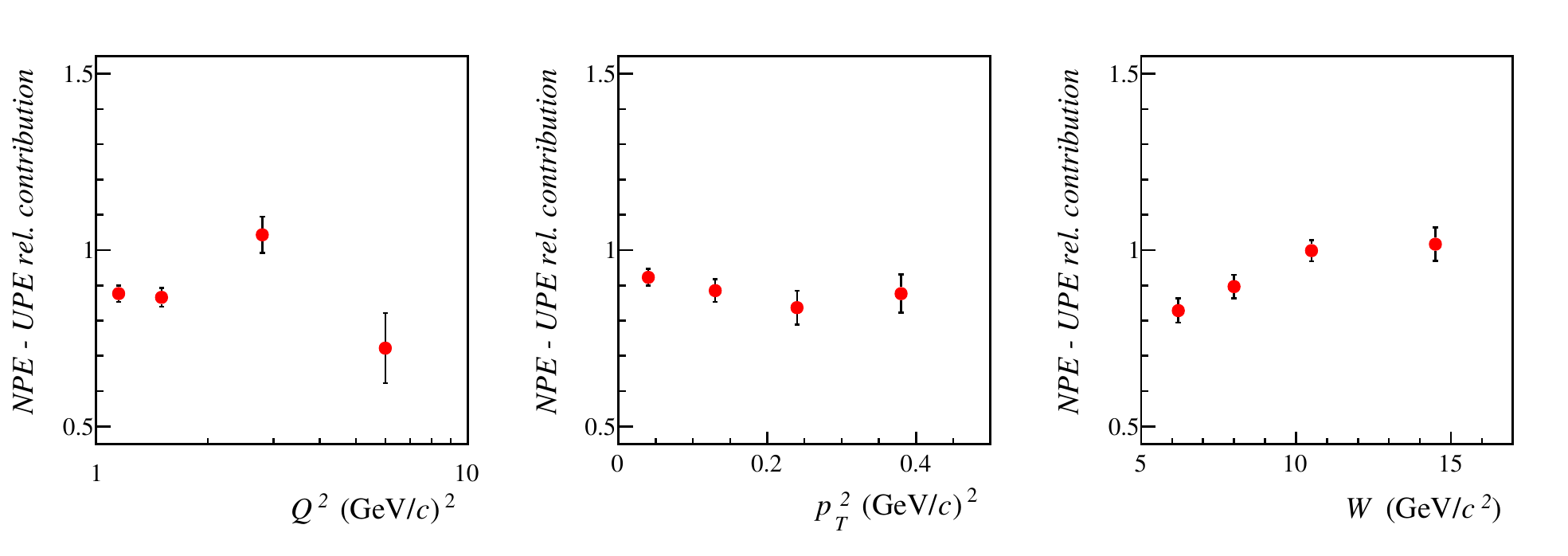} 
    \caption{Dependence on $Q^2$, $p_T^2$ and $W$ of the $P$ asymmetry, related to the NPE-to-UPE relative contribution to the $\gamma_{T}^*\to V_T$ transition. The uncertainties are statistical only.}
    \label{fig:p_asymm}
\end{figure}

Another interesting quantity is the relative contributions of UPE and NPE amplitudes to the transverse differential cross-section for the $\gamma_T^*\to V_T$ transition, defined as $P$-asymmetry:

\begin{equation}
\begin{split}
    P & = \frac{\diff \sigma_T^N(\gamma_T^*\to V_T) - \diff \sigma_T^U(\gamma_T^*\to V_T)}{\diff \sigma_T^N(\gamma_T^*\to V_T) + \diff \sigma_T^U(\gamma_T^*\to V_T)} \\
    & = \frac{2r_{1-1}^{1}}{1-r_{00}^{04}-2r_{1-1}^{04}}
\end{split}
\end{equation}
where the superscripts $N$ and $U$ denote the part of cross-section related to NPE and UPE transitions, respectively. The kinematic dependences of the $P$ asymmetry are shown in Fig.~\ref{fig:p_asymm}: a mild, decreasing trend can be observed as a function of $p_T$, the trend being opposite as a function of $W$.

\begin{figure}
\captionsetup{width=\textwidth}
    \centering
    \includegraphics[width=\textwidth]{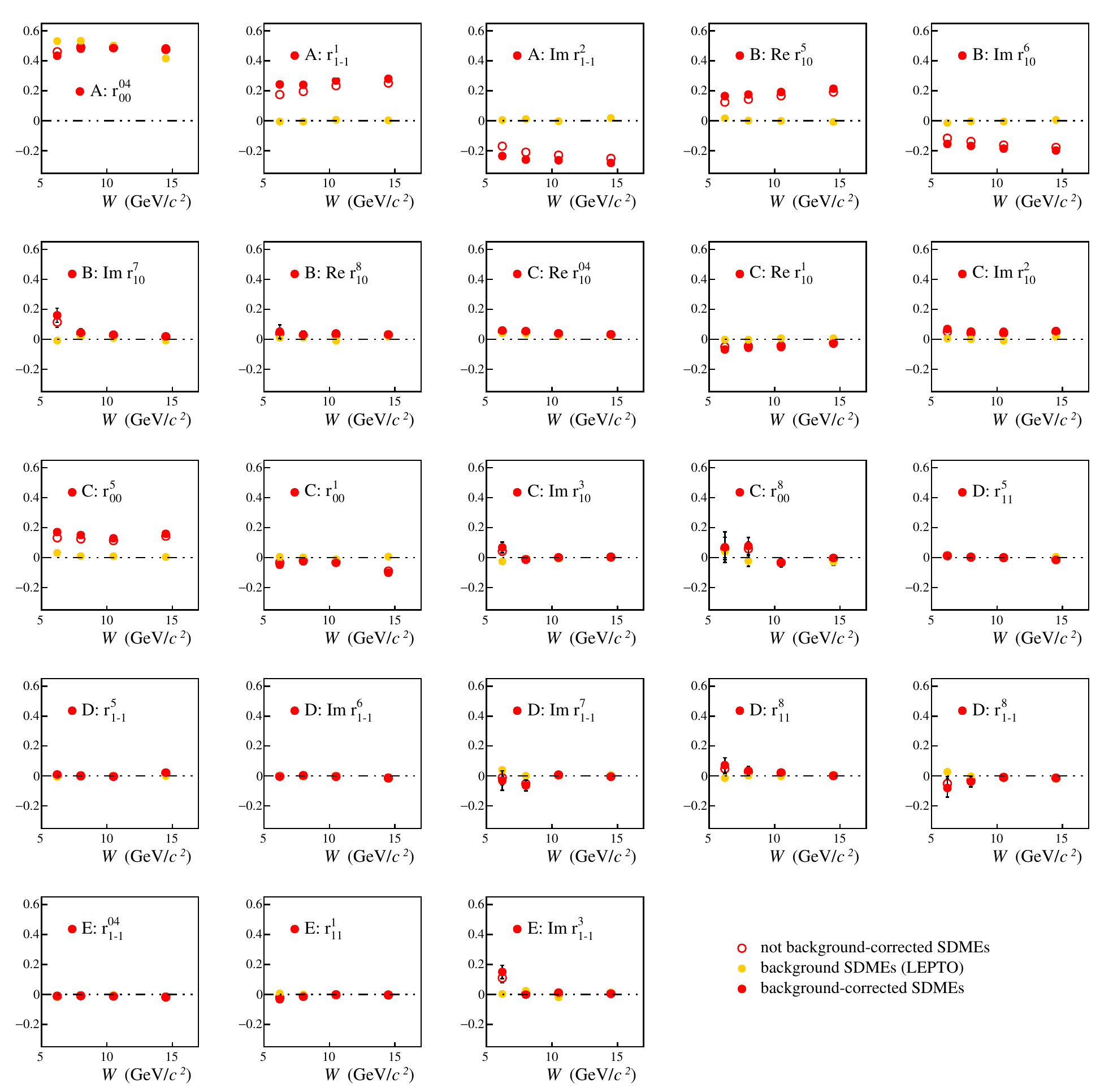} 
    \caption{Dependence on $p_T^2$ of the SDMEs not corrected for the background (empty points), of the background (orange points) and of the background-corrected data (closed red points), for $1~\mathrm{(GeV}/c\mathrm{)}^2<\Qsq<10~\mathrm{(GeV}/c\mathrm{)}^2$ and $5~\mathrm{GeV}/c^2\mathrm{}<W<17~\mathrm{GeV}/c^2\mathrm{}$.  The uncertainties are statistical only.}
    \label{fig:sdmes_pt2}
\end{figure}

\chapter{Conclusions} 

\label{Chapter7_Conclusions}

The study of the transverse-momentum-dependent structure of the nucleon is still today one of the most challenging problems in hadron physics. Both on the theoretical and on the experimental side, a significant effort is leading to a more and more refined understanding of the TMD physics and to a better description of the experimental data.
Together with HERMES and the experiments at the Jefferson Lab and RHIC, the COMPASS Collaboration is giving its contribution in this quest with Semi-Inclusive DIS measurements. Considering the SIDIS process on unpolarized nucleons, two observables are of particular interests: the transverse-momentum distributions, linked to the convolution of the unpolarized PDF $f_1$ and the unpolarized FF $D_1$, and the azimuthal asymmetries, where also the convolution of the Boer-Mulders function $\bom$ with the Collins function $\coll$ appears. \\

In this Thesis, the results for the distributions of the transverse momentum squared $\Ptsq$ and for the $A_{UU}^{\cos\fih}$, $A_{UU}^{\cos2\fih}$ and $A_{LU}^{\sin\fih}$ azimuthal asymmetries in SIDIS on unpolarized protons are presented, discussed and compared to the previous COMPASS results on deuteron. The data analyzed in this work have been collected in COMPASS in 2016 and constitute a small fraction of the full data sample. Particular care has been devoted to the selection of the best kinematic region, to the evaluation of the acceptance correction and to the rejection and subtraction of the background process, namely the diffractive production of exclusive vector mesons, the decay products of which show both peculiar kinematic distributions and azimuthal modulations. All these corrections require precise Monte Carlo simulations with good descriptions of the COMPASS apparatus. The simulations required a very large and coherent effort from many actors in the Collaboration, which unfortunately could not be finished during the time span of this Thesis, so that the systematic uncertainties of the results are still large. More work will be necessary before concluding this analysis. \\

Still, several conclusions can already be drawn from the present analysis. 

The measured $\Ptsq$-distributions show the expected, characteristic exponential trend. The average transverse momentum squared, $\aPtsq$ is found to be the same for positive and negative hadrons. As previously observed on deuteron data, the Leading Order relation among the transverse momenta, $\aPtsq = z^2\aktsq + \apperpsq$, is only approximately verified. From the measurement of $\aPtsq$ in bins of $\Qsq$ and $z$, a description of the $\Qsq$-dependence of $\aktsq$ has been derived and compared to the observed dependence of $\aktsq$ on $\Qsq$ as derived from the $A_{UU}^{\cos\fih}$ azimuthal asymmetry, assuming the Cahn effect to be dominant. The strong and rich kinematic dependences of the $A_{UU}^{\cos\fih}$ and $A_{UU}^{\cos2\fih}$ azimuthal asymmetries, already observed on deuteron, are confirmed. Combining the values of $A_{UU}^{\cos2\fih}$ measured for positive and negative hadrons on proton and on deuteron at COMPASS, the ratio of the $d_v$- and $u_v$- Boer-Mulders functions has been estimated for the first time with the method of the difference asymmetries, giving an indication that $h_1^{\perp u_v}$ and $h_1^{\perp d_v}$ have the same sign integrating over $0.008<x<0.130$, while their ratio is found negative (but compatible with zero) at high-$x$, where the Boer-Mulders function is expected to be more relevant.


\appendix 



\chapter{Structure functions in unpolarized SIDIS}
\label{AppendixA} 

\section{The structure function $F_{UU}$}
Let's calculate the expression for the $F_{UU}$ structure function in Gaussian approximation. From its definition \cite{Bacchetta:2006tn,Anselmino:2011ch} it follows that:
\begin{equation}
\begin{split}
   	F_{UU} & = \mathcal{C}\left[f_1D_1\right] \\
   	& = \sum_{q} e_q^2 xf_1^q(x,\Qsq) D_1^{h/q}(z,\Qsq) \int \diff^2\vkt \diff^2\vpperp \delta^2 \op\vPt-z\vkt-\vpperp\cp \frac{e^{-\ktsq/\Faktsq}}{\pi\Faktsq}\frac{e^{-\pperpsq/\Fapperpsq}}{\pi\Fapperpsq} \\
	 & =  \sum_{q} e_q^2 xf_1^q(x,\Qsq) D_1^{h/q}(z,\Qsq) \frac{1}{\pi^2\Faktsq\Fapperpsq} \underbrace{\int \diff^2\vkt e^{-\left(\frac{\ktsq}{\Faktsq}+\frac{(\vPt-z\vkt)^2}{\Fapperpsq}\right)}}_{I_0}.
\end{split}
\end{equation}
Let's now focus on the integral $I_0$, omitting all flavor indices for the sake of clarity. By expanding the square and rearranging the terms, and inserting the identity $\aPtsq = z^2\aktsq+\apperpsq$, the exponent can be rewritten as:
\begin{equation}
\begin{split}
   \frac{\ktsq}{\aktsq}+\frac{\op\vPt-z\vkt\cp^2}{\apperpsq} & =  \frac{\ktsq}{\aktsq} + \frac{\Ptsq}{\apperpsq}-\frac{2z\vPt\cdot\vkt}{\apperpsq} + \frac{z^2\ktsq}{\apperpsq} \\
   & = \left(\frac{1}{\aktsq}+\frac{z^2}{\apperpsq}\right)\ktsq + \frac{\Ptsq}{\apperpsq} - \frac{2z\vPt\cdot\vkt}{\apperpsq} \\
   & = \frac{\aPtsq}{\aktsq\apperpsq}\left(\ktsq + \frac{\aktsq\Ptsq}{\aPtsq} - \frac{2z\aktsq\vPt\cdot\vkt}{\aPtsq} \right) \\
   & = \frac{\aPtsq}{\aktsq\apperpsq}\left(\vkt-\frac{z\aktsq\vPt}{\aPtsq}\right)^2 + \frac{\Ptsq}{\aPtsq}.
\end{split}
\end{equation}
The integral $I_0$ can then be solved by introducing the vector $\vt = \sqrt{\frac{\aPtsq}{\aktsq\apperpsq}}\op \vkt - \frac{z\aktsq}{\aPtsq}\vPt \cp$: 
\begin{equation}
\begin{split}
    I_0  & = e^{-\frac{\Ptsq}{\aPtsq}} \int \diff^2\vkt e^{-\frac{\aPtsq}{\aktsq\apperpsq}\left(\vkt-\frac{z\aktsq\vPt}{\aPtsq}\right)^2} \\
    & =  e^{-\frac{\Ptsq}{\aPtsq}} \frac{\aktsq\apperpsq}{\aPtsq} \int \diff^2\vt e^{-t^2} = e^{-\frac{\Ptsq}{\aPtsq}} \frac{\pi\aktsq\apperpsq}{\aPtsq},
\end{split}
\end{equation}
so that the flavor-dependent structure function $F_{UU}$ finally reads:
\begin{equation}
\begin{split}
   F_{UU} & = \sum_{q}e_q^2 xf_1^q(x,\Qsq) D_1^{h/q}(z,\Qsq) \frac{e^{-\Ptsq/\FaPtsq}}{\pi\FaPtsq}.
\end{split}
\end{equation}
We thus observe that the structure function is predicted to show an exponential trend in $\Ptsq$, with an \textit{inverse slope} equal to $\aPtsq = z^2\aktsq + \apperpsq$. \\

\section{The Cahn contribution to $F_{UU}^{\cos\fih}$: $F_{UU|Cahn}^{\cos\fih}$}
The Cahn contribution to the fully-differential $F_{UU}^{\cos\fih}$ structure function can be written in terms of the unpolarized PDF $f_1$ and FF $D_1$, according to the expression given in Ref.~\cite{Bacchetta:2006tn,Anselmino:2011ch}. In Gaussian approximation, the convolution $\mathcal{C}$ can be solved as shown in the following.

\begin{equation}
\begin{split}
   F_{UU|Cahn}^{\cos\fih} & = -\frac{2}{Q}\mathcal{C}\left[\hh\cdot\vkt f_1D_1\right] \\
   & = -\frac{2}{Q}\sum_{q} e_q^2 xf_1^q(x,\Qsq) D_1^{h/q}(z,\Qsq)  \cdot \\
   & \hspace{2cm} \cdot \int \diff^2\vkt \diff^2\vpperp \delta^2 \op\vPt-z\vkt-\vpperp\cp \op\hh\cdot\vkt\cp \frac{e^{-\ktsq/\Faktsq}}{\pi\Faktsq}\frac{e^{-\pperpsq/\Fapperpsq}}{\pi\Fapperpsq} \\
   & = -\frac{2}{Q}\sum_{q} e_q^2 xf_1^q(x,\Qsq) D_1^{h/q}(z,\Qsq) \frac{1}{\pi^2\Faktsq\Fapperpsq} \cdot \\
   & \hspace{2cm} \cdot \underbrace{\int \diff^2\vkt \op\hh\cdot\vkt\cp e^{-\left(\frac{\ktsq}{\Faktsq}+\frac{(\vPt-z\vkt)^2}{\Fapperpsq}\right)}}_{I_1}.
\end{split}
\end{equation}
The integral $I_1$ can be solved by introducing the same vector $\vt = \sqrt{\frac{\aPtsq}{\aktsq\apperpsq}}\op \vkt - \frac{z\aktsq}{\aPtsq}\vPt \cp$. Let's first rewrite the $\op\hh\cdot\vkt\cp$ product in terms of $\vt$, naming $\chi$ the angle between $\hh$ and $\vt$, and neglecting the flavor dependence for the sake of clarity:
\begin{equation}
\begin{split}
    \hh\cdot\vkt & = \hh\cdot\left(\sqrt{\frac{\aktsq\apperpsq}{\aPtsq}}\vt+\frac{z\aktsq\vPt}{\aPtsq}\right) \\
    & = \sqrt{\frac{\aktsq\apperpsq}{\aPtsq}}t\cos\chi + \frac{z\aktsq\Pt}{\aPtsq}.
\end{split}
\end{equation}
The integral over $\diff^2\vt$ of $t\cos\chi$ is zero by symmetry, so that $I_1$ reads:
\begin{equation}
   I_1 = e^{-\frac{\Ptsq}{\aPtsq}} \frac{\aktsq\apperpsq}{\aPtsq}\frac{z\aktsq\Pt}{\aPtsq}\int \diff^2\vt  e^{-t^2} = e^{-\frac{\Ptsq}{\aPtsq}}  \frac{\pi z\aktsq^2\apperpsq\Pt}{\aPtsq^2}.
\end{equation}
so that the structure function can be finally derived as:
\begin{equation}
\begin{split}
   F_{UU|Cahn}^{\cos\fih} & = -\frac{2}{Q}\sum_{q} e_q^2 xf_1^q(x,\Qsq) D_1^{h/q}(z,\Qsq) \frac{e^{-\Ptsq/\FaPtsq}}{\pi^2\Faktsq\Fapperpsq}\frac{\pi z\Faktsq^2\Fapperpsq\Pt}{\FaPtsq^2}\\
   & = -\frac{2z\Pt}{Q}\sum_{q} e_q^2 xf_1^q(x,\Qsq) D_1^{h/q}(z,\Qsq) \frac{e^{-\Ptsq/\FaPtsq}}{\pi\FaPtsq}\frac{\Faktsq}{\FaPtsq}.\\
\end{split}
\end{equation}
Under the assumption of flavor-independence of the average transverse momenta, the asymmetry $A_{UU|Cahn}^{\cos\fih}$ can be simplified to give:
\begin{equation}
\begin{split}
   A_{UU|Cahn}^{\cos\fih} & = \frac{F_{UU|Cahn}^{\cos\fih}}{F_{UU}} = -\frac{2z\Pt\aktsq}{Q\aPtsq}.
\end{split}
\end{equation}

\bigskip
\section{The Cahn contribution to $F_{UU}^{\cos2\fih}$: $F_{UU|Cahn}^{\cos2\fih}$}
The Cahn contribution to the $F_{UU}^{\cos2\fih}$ structure function can be written as \cite{Barone:2015ksa}:

\begin{equation}
\begin{split}
   F_{UU|Cahn}^{\cos2\fih} & = \frac{2}{Q^2}\mathcal{C}\left[\op 2\op\hh\cdot\vkt\cp^2-\kt^2 \cp f_1D_1\right] \\
   & = \frac{2}{Q^2}\sum_{q} e_q^2 xf_1^q(x,\Qsq) D_1^{h/q}(z,\Qsq) \cdot \\
   & \hspace{1cm} \cdot \int \diff^2\vkt \diff^2\vpperp \delta^2\op\vPt-z\vkt-\vpperp\cp \op 2\op\hh\cdot\vkt\cp^2-\kt^2\cp \frac{e^{-\ktsq/\Faktsq}}{\pi\Faktsq}\frac{e^{-\pperpsq/\Fapperpsq}}{\pi\Fapperpsq} \\
   & = \frac{2}{Q^2}\sum_{q} e_q^2 xf_1^q(x,\Qsq) D_1^{h/q}(z,\Qsq) \frac{1}{\pi^2\Faktsq\Fapperpsq} \cdot \\
   & \hspace{1cm} \cdot \underbrace{\int \diff^2\vkt \op 2\op\hh\cdot\vkt\cp^2-\kt^2\cp e^{-\left(\frac{\ktsq}{\Faktsq}+\frac{(\vPt-z\vkt)^2}{\Fapperpsq}\right)}}_{I_2}.
\end{split}
\end{equation}
Let's first rewrite the quantity $\op 2\op\hh\cdot\vkt\cp^2-\kt^2\cp$ in terms of $\vt = \sqrt{\frac{\aPtsq}{\aktsq\apperpsq}}\op \vkt - \frac{z\aktsq}{\aPtsq}\vPt \cp$, with $\chi$ the angle between $\hh$ and $\vt$. Neglecting for the moment the flavor indices, one has:
\begin{equation}
\begin{split}
    \op\hh\cdot\vkt\cp^2 & = \frac{\aktsq\apperpsq}{\aPtsq}t^2\cos^2\chi + \frac{z^2\aktsq^2\Pt^2}{\aPtsq^2} + 2\sqrt{\frac{\aktsq\apperpsq}{\aPtsq}}\frac{z\aktsq\Pt}{\aPtsq}t\cos\chi .
\end{split}
\end{equation}

\begin{equation}
\begin{split}
   \kt^2 & = \frac{\aktsq\apperpsq}{\aPtsq}t^2 + 2\frac{z\aktsq\Pt}{\aPtsq}\sqrt{\frac{\aktsq\apperpsq}{\aPtsq}}t\cos\chi + \frac{z^2\aktsq^2\Ptsq}{\aPtsq^2}.
\end{split}
\end{equation}
so that, dropping the terms that would integrate to zero, 
\begin{equation}
   2\op\hh\cdot\vkt\cp^2 - \kt^2 = \frac{z^2\aktsq^2\Pt^2}{\aPtsq^2}
\end{equation}
The integral $I_2$ can now be easily solved:
\begin{equation}
    I_2 = e^{-\frac{\Ptsq}{\aPtsq}} \frac{\aktsq\apperpsq}{\aPtsq} \frac{z^2\aktsq^2\Pt^2}{\aPtsq^2} \int_{0}^{\infty} \diff t~te^{-t^2} \int_0^{2\pi}\diff \chi  = e^{-\frac{\Ptsq}{\aPtsq}} \frac{\pi z^2\aktsq^3\apperpsq\Ptsq}{\aPtsq^3}.
\end{equation}
and the Cahn contribution to the $F_{UU}^{\cos2\fih}$ structure function finally be obtained:
\begin{equation}
\begin{split}
   F_{UU|Cahn}^{\cos2\fih} & = \frac{2}{Q^2}\sum_{q} e_q^2 xf_1^q(x,\Qsq) D_1^{h/q}(z,\Qsq) \frac{e^{-\Ptsq/\FaPtsq}}{\pi^2\Faktsq\Fapperpsq} \frac{\pi z^2\Faktsq^3\Fapperpsq\Ptsq}{\FaPtsq^3} \\
   & = \frac{2z^2\Ptsq}{Q^2}\sum_{q} e_q^2 xf_1^q(x,\Qsq) D_1^{h/q}(z,\Qsq) \frac{e^{-\Ptsq/\FaPtsq}}{\pi\FaPtsq} \frac{\Faktsq^2}{\FaPtsq^2},
\end{split}
\end{equation}
which reduces, in the flavor-independent case, to a contribution to the $A_{UU}^{\cos2\fih}$ as follows:
\begin{equation}
\begin{split}
   A_{UU|Cahn}^{\cos2\fih} & = \frac{F_{UU|Cahn}^{\cos2\fih}}{F_{UU}} = \frac{2z^2\aktsq^2\Ptsq}{Q^2\aPtsq^2} 
\end{split}
\end{equation}

\bigskip

\section{The Boer-Mulders contribution to $F_{UU}^{\cos\fih}$: $F_{UU|BM}^{\cos\fih}$}
The Boer-Mulders contribution to the $F_{UU}^{\cos\fih}$ structure function can be written as \cite{Anselmino:2011ch}:
\begin{equation}
   F_{UU|BM}^{\cos\fih} = -\frac{2}{zQMM_h}\mathcal{C}\left[\op\hh\cdot\vpperp\cp\kt^2 \bom\coll\right] \\
\end{equation}

For the Boer-Mulders and the Collins functions we choose the forms:
\begin{equation}
\begin{split}
    h_1^{\perp q}(x,\ktsq) = h_1^{\perp q}(x)\frac{e^{-k_T^2/\Faktsq}}{\pi\Faktsq} \\
    H_1^{\perp q}(z,\pperpsq) = H_1^{\perp q}(z)\frac{e^{-p_\perp^2/\Fapperpsq}}{\pi\Fapperpsq} \\
\end{split}
\end{equation}
where the transverse momenta $\aktsq$ and $\apperpsq$ have been taken the same as the ones for $f_1$ and $D_1$. With these choices,
\begin{equation}
\begin{split}
   F_{UU|BM}^{\cos\fih} & =  -\frac{2}{zQMM_h} \sum_{q} e_q^2x\bomq(x,\Qsq) \collq(z,\Qsq) \frac{1}{\pi^2\Faktsq\Fapperpsq} \cdot \\
   & ~~~\cdot \int \diff^2\vkt \diff^2\vpperp \delta^2\op\vPt-z\vkt-\vpperp\cp \left[\op\hh\cdot\vpperp\cp\kt^2\right] e^{-\op\frac{\ktsq}{\Faktsq}+\frac{\pperpsq}{\Fapperpsq}\cp} \\
   & = -\frac{2}{zQMM_h} \sum_{q} e_q^2x\bomq(x,\Qsq) \collq(z,\Qsq) \frac{1}{\pi^2\Faktsq\Fapperpsq} \cdot \\
   & ~~~\cdot \underbrace{\int \diff^2\vkt \left[\op\hh\cdot\op\vPt-z\vkt\cp\cp\kt^2\right] e^{-\op\frac{\ktsq}{\Faktsq}+\frac{\pperpsq}{\Fapperpsq}\cp}}_{I_3} 
\end{split}
\end{equation}

Let's now consider the integral $I_3$, neglecting the flavor indices for the sake of clarity. Introducing the usual vector $\vt = \sqrt{\frac{\aPtsq}{\aktsq\apperpsq}}\op \vkt - \frac{z\aktsq}{\aPtsq}\vPt \cp$, with $\chi$ the angle between $\hh$ and $\vt$, one has that:

\begin{equation}
\begin{split}
    \hh\cdot\op\vPt-z\vkt\cp & = \Pt-z\hh\cdot\vkt \\
    & = \Pt-z\hh\cdot\op\sqrt{\frac{\aktsq\apperpsq}{\aPtsq}}\vt + \frac{z\aktsq}{\aPtsq}\vPt\cp\\
    & = \Pt-z\sqrt{\frac{\aktsq\apperpsq}{\aPtsq}}t\cos\chi-\frac{z^2\aktsq\Pt}{\aPtsq} \\
    & = \frac{\apperpsq\Pt}{\aPtsq} -z\sqrt{\frac{\aktsq\apperpsq}{\aPtsq}}t\cos\chi
\end{split}
\end{equation}
while $\kt^2$ can be rewritten as:
\begin{equation}
\begin{split}
    \kt^2 = \frac{\aktsq\apperpsq}{\aPtsq}t^2 + 2\sqrt{\frac{\aktsq\apperpsq}{\aPtsq}}\frac{z\Pt\aktsq}{\aPtsq}t\cos\chi + \frac{z^2\aktsq^2\Ptsq}{\op \aPtsq\cp^2}.
\end{split}
\end{equation}
Dropping the terms that would integrate to zero, the product $\op\Pt-z\hh\cdot\vkt\cp\kt^2$ reads:
\begin{equation}
\begin{split}
    \op\Pt-z\hh\cdot\vkt\cp\kt^2 & = \frac{\aktsq\apperpsq^2\Pt}{\aPtsq^2}t^2 - \frac{2z^2\aktsq^2\apperpsq\Pt}{\aPtsq^2}t^2\cos^2\chi  + \frac{z^2\aktsq^2\apperpsq\Pt^3}{\aPtsq^3}\\
    & = \frac{\aktsq\apperpsq\Pt}{\op\aPtsq\cp^2}\op\apperpsq t^2 - 2z^2\aktsq t^2\cos^2\chi + \frac{z^2\aktsq\Ptsq}{\aPtsq}\cp.
\end{split}
\end{equation}
The integral $I_{3}$ can now be easily solved:
\begin{equation}
\begin{split}
   I_3 & = e^{-\frac{\Ptsq}{\aPtsq}} \frac{\aktsq^2\apperpsq^2\Pt}{\aPtsq^3} \int \diff^2\vt \op\apperpsq t^2 - 2z^2\aktsq t^2\cos^2\chi + \frac{z^2\aktsq\Ptsq}{\aPtsq}\cp e^{-t^2} \\
   & = e^{-\frac{\Ptsq}{\aPtsq}} \frac{\pi\aktsq^2\apperpsq^2\Pt}{\aPtsq^4}\op\apperpsq\aPtsq + z^2\aktsq\op\Ptsq-\aPtsq\cp\cp
\end{split}
\end{equation}
and the $F_{UU|BM}^{\cos\fih}$ structure function reads:
\begin{equation}
\begin{split}
   F_{UU|BM}^{\cos\fih} & = -\frac{2}{zQMM_h} \sum_{q} e_q^2x\bomq(x,\Qsq) \collq(z,\Qsq) \frac{e^{-\frac{\Ptsq}{\FaPtsq}}}{\pi \FaPtsq} \cdot \\
   & ~~~\cdot \frac{\Faktsq\Fapperpsq\Pt}{\FaPtsq^3}\op\Fapperpsq\FaPtsq + z^2\Faktsq\op\Ptsq-\FaPtsq\cp\cp
\end{split}
\end{equation}
which reduces, under the flavor-independent assumption, to the asymmetry:
\begin{equation}
\begin{split}
   A_{UU|BM}^{\cos\fih} = \frac{F_{UU|BM}^{\cos\fih}}{F_{UU}} & = -\frac{2}{zQMM_h} \frac{\aktsq\apperpsq\Pt}{\aPtsq^3}\op\apperpsq\aPtsq + z^2\aktsq\op\Ptsq-\aPtsq\cp\cp \cdot \\
   & ~~~\cdot \frac{\sum_{q} e_q^2x\bomq(x,\Qsq) \collq(z,\Qsq)}{\sum_{q} e_q^2 xf_1^q(x,\Qsq) D_1^{h/q}(z,\Qsq)} 
\end{split}
\end{equation}

\section{The Boer-Mulders contribution to $F_{UU}^{\cos2\fih}$: $F_{UU|BM}^{\cos2\fih}$}
The contribution of the Boer-Mulders term to the $F_{UU}^{\cos2\fih}$ structure function in defined by the convolution \cite{Anselmino:2011ch}:
\begin{equation}
   F_{UU|BM}^{\cos2\fih} = \mathcal{C}\left[\frac{2\op\hh\cdot\vkt\cp\op\hh\cdot\vpperp\cp-\vkt\cdot\vpperp}{zMM_h}\bom\coll\right]. 
\end{equation}
Inserting the parametrizations for $\bom$ and $\coll$ already introduced for the calculation of $F_{UU|BM}^{\cos\fih}$, it can be rewritten as:
\begin{equation}
\begin{split}
   F_{UU|BM}^{\cos2\fih} & = \frac{1}{zMM_h}\sum_{q} e_q^2 x\bomq(x,\Qsq) \collq(z,\Qsq)
   \frac{1}{\pi^2\Faktsq \Fapperpsq} \cdot \\
   &  \cdot \underbrace{\int \diff^2\vkt \diff^2\vpperp \delta^2\op\vPt-z\vkt-\vpperp\cp \left[2\op\hh\cdot\vkt\cp\op\hh\cdot\vpperp\cp-\vkt\cdot\vpperp\right] e^{-\op\frac{\ktsq}{\Faktsq}+\frac{\pperpsq}{\Fapperpsq}\cp}}_{I_4} 
\end{split}
\end{equation}
Let's consider the integral $I_4$. Similarly to the previous calculations, exploiting the relation between the transverse momenta and omitting the flavor indices for the sake of clarity, it can be rearranged in the following way:
\begin{equation}
  I_4 = e^{-\frac{\Ptsq}{\aPtsq}}\int \diff^2\vkt \underbrace{\left[2\op\hh\cdot\vkt\cp\op\hh\cdot\op\vPt-z\vkt\cp\cp-\vkt\cdot\op\vPt-z\vkt\cp\right]}_{\mathcal{T}} e^{-\frac{\aPtsq}{\aktsq\apperpsq}\op\vkt - \frac{z\aktsq\vPt}{\aPtsq}\cp^2}.
\end{equation}
Let's first simplify the quantity in the square bracket, $\mathcal{T}$:
\begin{equation}
\begin{split}
  \mathcal{T} & = 2\op\hh\cdot\vkt\cp\op\hh\cdot\op\vPt-z\vkt\cp\cp-\vkt\cdot\op\vPt-z\vkt\cp \\
  & =  2\op\hh\cdot\vkt\cp \op\Pt-z\hh\cdot\vkt\cp- \Pt\op\hh\cdot\vkt\cp +z\ktsq \\
  & = 2\Pt \op\hh\cdot\vkt\cp - 2z\op\hh\cdot\vkt\cp^2 - \Pt\op\hh\cdot\vkt\cp + z\ktsq \\
  & = \Pt \op\hh\cdot\vkt\cp - 2z\op\hh\cdot\vkt\cp^2 + z\ktsq.
\end{split}
\end{equation}
Now, via the usual vector $\vt = \sqrt{\frac{\aPtsq}{\aktsq\apperpsq}}\op \vkt - \frac{z\aktsq}{\aPtsq}\vPt \cp$, with $\chi$ the angle between $\hh$ and $\vt$, 
\begin{equation}
\begin{split}
  \mathcal{T} & = \sqrt{\frac{\aktsq\apperpsq}{\aPtsq}}\Pt t\cos\chi + \frac{z\aktsq\Ptsq}{\aPtsq}  \\
  & -  \frac{2z\aktsq\apperpsq}{\aPtsq}t^2\cos^2\chi - \frac{2z^3\aktsq^2\Ptsq}{\aPtsq^2} - \frac{4z^2\aktsq\Pt}{\aPtsq}\sqrt{\frac{\aktsq\apperpsq}{\aPtsq}}t\cos\chi \\
  & + \frac{z\aktsq\apperpsq}{\aPtsq}t^2 + \frac{z^3\aktsq^2\Ptsq}{\aPtsq} + \frac{2z^2\aktsq\Pt}{\aPtsq}\sqrt{\frac{\aktsq\apperpsq}{\aPtsq}}t\cos\chi.
\end{split}
\end{equation}
Retaining only the terms that would not integrate to zero, one gets:
\begin{equation}
\begin{split}
  \mathcal{T} & = \frac{z\aktsq\Ptsq}{\aPtsq} - \frac{z^3\aktsq^2\Ptsq}{\aPtsq^2} = \frac{z\aktsq\Ptsq}{\aPtsq^2} \op \aPtsq -z^2\aktsq \cp \\ 
  & = \frac{z\aktsq\Ptsq\apperpsq}{\aPtsq^2}
\end{split}
\end{equation}
The integral $I_4$ then reads:
\begin{equation}
\begin{split}
  I_4 & = e^{-\frac{\Ptsq}{\aPtsq}}\frac{\aktsq\apperpsq}{\aPtsq}\frac{z\aktsq\Ptsq\apperpsq}{\aPtsq^2} \int \diff^2\vt e^{-t^2} \\
  & =  e^{-\frac{\Ptsq}{\aPtsq}}\frac{\pi z \Ptsq \aktsq^2\apperpsq^2}{\aPtsq^3}.
\end{split}
\end{equation}
and, consequently, the structure function can be written as:

\begin{equation}
   F_{UU|BM}^{\cos2\fih} = \frac{1}{zMM_h}\sum_{q} e_q^2 x\bomq(x,\Qsq) \collq(z,\Qsq)
   \frac{e^{-\frac{\Ptsq}{\FaPtsq}}}{\pi\FaPtsq}\frac{ z \Ptsq \Faktsq\Fapperpsq}{\FaPtsq^2}.
\end{equation}
which corresponds, in the flavor-independent case, to the following asymmetry: 
\begin{equation}
   A_{UU|BM}^{\cos2\fih}  = \frac{F_{UU|BM}^{\cos2\fih}}{F_{UU}}= \frac{ \Ptsq \aktsq\apperpsq}{MM_h\aPtsq^2}\frac{\sum_{q} e_q^2 x\bomq(x,\Qsq) \collq(z,\Qsq)}{\sum_{q} e_q^2 xf_1^q(x,\Qsq) D_1^{h/q}(z,\Qsq)} 
\end{equation}

\chapter{Extraction of $\aPtsq$ from the fits}
\label{AppendixB} 

The $\Ptsq$-distributions have been fitted with a single-exponential, a double-exponential and a Tsallis-like function in order to derive the mean value $\aPtsq$. While the single-exponential fit directly gives the best estimate of $\aPtsq$ with its uncertainty, this is not the case for the other two options, where $\aPtsq$ and its uncertainty are obtained by manipulating the fitted parameters and the covariance matrices. In this Appendix, the full calculations and some examples are given for completeness.

\section{Double-exponential case}
The $\Ptsq$-distributions are fitted with the double-exponential function:
\begin{equation}
    f(x) = A e^{-x/a} + B e^{-x/b},
\end{equation}
whose normalization constant $I$ is obtained by integrating $f(x)$ in $[0,+\infty)$:

\begin{equation}
    I = \int_0^{+\infty} \diff x f(x) = A\int_{0}^{+\infty}\diff x~e^{-x/a} + B\int_{0}^{+\infty}\diff x~e^{-x/b} = Aa + Bb.
\end{equation}
Thus, the probability density function associated to $f(x)$, indicated with $p(x)$, reads:

\begin{equation}
      p(x) = \frac{A e^{-x/a} + B e^{-x/b}}{Aa + Bb}.
\end{equation}
The mean value of $x$ can easily be calculated as:
\begin{equation}
    \langle x \rangle = \int_{0}^{+\infty} \diff x~ x p(x) = \frac{A \int_{0}^{+\infty} \diff x~xe^{-x/a} + B \int_{0}^{+\infty} \diff x~xe^{-x/b}}{Aa + Bb} = \frac{Aa^2 + Bb^2}{Aa + Bb}.
\end{equation}
where $A$, $a$, $B$ and $b$ are the best values of the parameters obtained from the fit. Let's now consider the variance on $\langle x \rangle$. It reads:

\begin{equation}
\begin{split}
    \sigma^2(\langle x\rangle) & = \left(\frac{\diff\langle x\rangle}{\diff A}\right)^2 \sigma^2(A) + \left(\frac{\diff\langle x\rangle}{\diff a}\right)^2 \sigma^2(a) + \left(\frac{\diff \langle x\rangle}{\diff B}\right)^2 \sigma^2(B) +  \left(\frac{\diff\langle x\rangle}{\diff b}\right)^2 \sigma^2(b) \\
    & + 2\frac{\diff \langle x\rangle}{\diff A}\frac{\diff \langle x\rangle}{\diff a} cov(A,a) + 2\frac{\diff\langle x\rangle}{\diff A}\frac{\diff\langle x\rangle}{\diff B} cov(A,B) + 2\frac{\diff \langle x\rangle}{\diff A}\frac{\diff \langle x\rangle}{\diff b} cov(A,b) \\
    & + 2\frac{\diff \langle x\rangle}{\diff a}\frac{\diff \langle x\rangle}{\diff B} cov(a,B) + 2\frac{\diff \langle x\rangle}{\diff a}\frac{\diff\langle x\rangle}{\diff b} cov(a,b) + 2\frac{\diff \langle x\rangle}{\diff B}\frac{\diff \langle x\rangle}{\diff b} cov(B,b).
\end{split}
\end{equation}
The values of the covariances $cov(A,a)$, $cov(A,B)$, $cov(A,b)$, $cov(a,B)$, $cov(a,b)$ and $cov(B,b)$ are obtained from MINUIT (with the \texttt{mnemat} command). The derivatives read:

\begin{equation}
    \frac{\diff\langle x\rangle}{\diff A} = \frac{Bab(a-b)}{(Aa+Bb)^2}
\end{equation}
\begin{equation}
    \frac{\diff\langle x\rangle}{\diff B} = \frac{Aab(b-a)}{(Aa+Bb)^2}
\end{equation}
\begin{equation}
    \frac{\diff \langle x\rangle}{\diff a} = \frac{A(Aa^2 +2Bab-Bb^2)}{(Aa+Bb)^2}
\end{equation}
\begin{equation}
    \frac{\diff \langle x\rangle}{\diff b} = \frac{B(Bb^2 +2Aab-Aa^2)}{(Aa+Bb)^2}.
\end{equation}
To get the uncertainty on the estimated $\langle P_T^2 \rangle$, these quantities are computed at the best values of the parameters.

\bigskip
As an example, let's consider the fit of the $\Ptsq$-distribution of positive hadrons in the first bins of $x$, $\Qsq$ and $z$. In this case it is found that:
\begin{equation}
\begin{split}
    A & = (1.06 \pm 0.09)~10^{-1}~(\mathrm{GeV}/c)^{-2} \\
    a & = (5.85 \pm 0.17)~10^{-1}~(\mathrm{GeV}/c)^2\\ 
    B & = (11.19 \pm 0.10)~10^{-1}~(\mathrm{GeV}/c)^{-2} \\ 
    b & = (1.97 \pm 0.03)~10^{-1}~(\mathrm{GeV}/c)^2
\end{split}
\end{equation}
with a covariance matrix $C$ as follows:
\begin{equation}
C = \begin{pmatrix}
81.9 & -148.2 & -40.7 & -20.9\\
-148.2 & 285.9 & 81.5 & 35.6\\
-40.7 & 81.5 & 94.2 & 1.9\\
-20.9 & 35.6 & 1.9 & 7.0
\end{pmatrix}   \cdot 10^{-6}.
\end{equation}

With these ingredients, it is found that:

\begin{equation}
    \aPtsq~/~\mathrm{(GeV/}c\mathrm{)}^2 = \frac{Aa^2 + Bb^2}{Aa+Bb} = \frac{1.06\cdot 5.85^2 + 11.19\cdot1.97^2}{1.06\cdot 5.85 + 11.19\cdot 1.97}10^{-1} = 0.282 
\end{equation}
while the various derivatives take the following values:
\begin{equation}
    \begin{split}
        \frac{\diff \aPtsq}{\diff A} & = \frac{11.19\cdot 5.85 \cdot 1.97 \cdot (5.85-1.97)}{(1.06\cdot5.85 + 11.19\cdot1.97)^2} = 0.62 \\
        \frac{\diff \aPtsq}{\diff a} & = \frac{1.06\cdot(1.06\cdot 5.85^2 +2\cdot 11.19 \cdot 5.85 \cdot 1.97 - 11.19 \cdot 1.97^2)}{(1.06\cdot5.85 + 11.19\cdot1.97)^2} = 0.33 \\
        \frac{\diff \aPtsq}{\diff B} & = \frac{1.06\cdot 5.85 \cdot 1.97 \cdot (1.97-5.85)}{(1.06\cdot5.85 + 11.19\cdot1.97)^2} = -0.06 \\
        \frac{\diff \aPtsq}{\diff b} & = \frac{11.19\cdot (11.19 \cdot 1.97^2 +2\cdot 1.06 \cdot 5.85 \cdot 1.97 - 1.06\cdot 5.85^2)}{(1.06\cdot5.85 + 11.19\cdot1.97)^2} = 0.44
    \end{split}
\end{equation}
so that the variance on $\aPtsq$ is:
\begin{equation}
\begin{split}
    \sigma^2(\aPtsq) /\ 10^{-6} & = 0.62^2 \cdot 81.9 + 0.33^2 \cdot 285.9 + (-0.06)^2\cdot 94.2 + 0.44^2 \cdot 7.0 \\
    & + 2\cdot 0.62 \cdot 0.33 \cdot (-148.2) + 2\cdot 0.62 \cdot (-0.06) \cdot (-40.7) + 2\cdot 0.62 \cdot 0.44 \cdot (-20.9) \\
    & + 2\cdot 0.33 \cdot (-0.06) \cdot 81.5 + 2 \cdot 0.33 \cdot 0.44 \cdot 35.6 + 2 \cdot (-0.06) \cdot 0.44 \cdot 1.9 \\
    & = 64.31 - 62.01 = 2.30
\end{split}
\end{equation}
where, in the last line, the first term in the addition comes from the variances only, while the second one from the covariance-related terms. They almost cancel, so that the uncertainty on $\aPtsq$ is found to be:

\begin{equation}
    \sigma (\aPtsq) = \sqrt{2.30 \cdot 10^{-6}} \approx 0.002,
\end{equation}
thus with a relative uncertainty smaller than 1\%. 

As a side exercise, it is interesting to evaluate the relative contribution of the two exponential to the global $\aPtsq$, so that $\aPtsq$ can be written as:

\begin{equation}
    \aPtsq  = w_a \cdot a + w_b \cdot b.
\end{equation}
For this particular bin, it is found that:
\begin{equation}
    w_{a} = \frac{Aa}{Aa+Bb} = 0.220 \pm 0.013
\end{equation}
and, correspondingly,
\begin{equation}
    w_{b} = \frac{Bb}{Aa+Bb} = 0.780 \pm 0.013.
\end{equation}

\section{Tsallis-like case}
In this case, the $\Ptsq$-distributions are fitted with the function:

\begin{equation}
    g(x) = c_0(1+c_1x)^{-c_2}
\end{equation}
where $c_0$, $c_1$ and $c_2$ are all positive. In general, the normalization integral $J$ is not defined for all values of $c_2$ and may be divergent. In our case, however, the function rapidly falls and no such problem arises.

\begin{equation}
    J = \int_0^{+\infty} \diff x g(x) = \frac{c_0(1+c_1x)^{1-c_2}}{c_1(1-c_2)} \Bigg|_{0}^{+\infty} = \frac{c_0}{c_1(c_2-1)} 
\end{equation}
so that the probability density function associated to $g(x)$ reads: 
\begin{equation}
      q(x) = c_1(c_2-1)(1+c_1x)^{-c_2}.
\end{equation}
As a consequence, the expression for the mean value involves only the two parameters $c_1$ and $c_2$:
\begin{equation}
\begin{split}
    \langle x \rangle & = \int_{0}^{+\infty} \diff x~xq(x) = c_1(c_2-1) \int_{0}^{+\infty} \diff x~x(1+c_1x)^{-c_2} \\
    & = c_1(c_2-1) \frac{-(1+c_1x)^{1-c_2}(c_1(c_2-1)x+1)}{c_1^2(c_2-2)(c_2-1)}\Bigg|_{0}^{+\infty} = \frac{1}{c_1(c_2-2)}
\end{split}
\end{equation}
with a variance $\sigma^2(\langle x\rangle)$ equal to:

\begin{equation}
    \sigma^2(\langle x\rangle) = \frac{\sigma^2(c_1)}{c_1^4(c_2-2)^2} + \frac{\sigma^2(c_2)}{c_1^2(c_2-2)^4} + 2 \frac{cov(c_1,c_2)}{c_1^3(c_2-2)^3}
\end{equation}

As an example, let's consider again the fit of the $\Ptsq$-distribution of positive hadrons in the first bins of $x$, $\Qsq$ and $z$, where it is found that:
\begin{equation}
\begin{split}
    c_0 & = 1.30 \pm 0.01~(\mathrm{GeV}/c)^{-2} \\
    c_1 & = 1.08 \pm 0.03~(\mathrm{GeV}/c)^{-2}\\ 
    c_2 & = 5.26 \pm 0.11 \\ 
\end{split}
\end{equation}
with a covariance matrix $C^\prime$ as follows:
\begin{equation}
C^\prime = \begin{pmatrix}
1.1 & 2.5 & -7.1 \\
2.5 & 10.8 & -35.2\\
-7.1 & -35.2 & 117.6
\end{pmatrix}   \cdot 10^{-4}.
\end{equation}

It is readily found that:

\begin{equation}
    \aPtsq = \frac{1}{1.08 \cdot (5.26-2)} = 0.284
\end{equation}
with a variance:
\begin{equation}
\begin{split}
    \sigma^2(\aPtsq) /\ 10^{-4} & = \frac{10.8}{1.08^4 \cdot (5.26-2)^2} + \frac{117.6}{1.08^2 \cdot (5.26-2)^4} + 2 \frac{-35.2}{1.08^3\cdot(5.26-2)^3} \\
    & = 0.75 + 0.89 - 1.61 = 0.03
\end{split}
\end{equation}
corresponding to an uncertainty $\sigma(\aPtsq) \approx 0.002$. Also in this case, an almost perfect cancellation occurs between the terms related to the variances of the two parameters and the term related to their covariance.

\chapter{Mean values for the $\Ptsq$-distributions (\textit{standard} binning)}
\label{AppendixC} 
In this Appendix we give the mean values of $\Qsq$, $x$, $z$ and $W$ in all the \textit{standard} bins in which the $\Ptsq$-distribution have been fitted, as explained in Sect.~\ref{sec:ptdist_interpr} and shown in Fig.~\ref{fig:apt2_de}.
\bigskip

\begin{table}[tbh!]
\small 
\captionsetup{width=\textwidth}
    \centering
\begin{tabular}{c|c|c|c|c|c|c}
    \hline
    $\Qsq$ & $x$ & $z$ & $\langle \Qsq \rangle$ & $\langle x \rangle$ & $\langle z \rangle$ & $\langle W \rangle$ \\
    (GeV/$c$)$^2$ & & & (GeV/$c$)$^2$ & & & (GeV/$c^2$) \\
    \hline
    \multirow{12}{*}{1.0 - 3.0} & \multirow{4}{*}{0.003 - 0.013} & 0.20 - 0.30 & 1.42 & 0.0091 & 0.24 & 12.7 \\ 
    & & 0.30 - 0.40 & 1.42 & 0.0091 & 0.34 & 12.7 \\
    & & 0.40 - 0.60 & 1.42 & 0.0091 & 0.48 & 12.6 \\
    & & 0.60 - 0.80 & 1.42 & 0.0091 & 0.68 & 12.6 \\
    \cline{2-7}
    
    & \multirow{4}{*}{0.013 - 0.020} & 0.20 - 0.30 & 1.67 & 0.0162 & 0.24 & 10.0 \\ 
    & & 0.30 - 0.40 & 1.65 & 0.0162 & 0.34 & 10.0 \\
    & & 0.40 - 0.60 & 1.63 & 0.0162 & 0.48 & 9.9 \\
    & & 0.60 - 0.80 & 1.62 & 0.0162 & 0.68 & 9.9 \\
    \cline{2-7}
    
    & \multirow{4}{*}{0.020 - 0.055} & 0.20 - 0.30 & 2.15 & 0.0274 & 0.24 & 8.8 \\ 
    & & 0.30 - 0.40 & 2.15 & 0.0275 & 0.34 & 8.8 \\
    & & 0.40 - 0.60 & 2.14 & 0.0275 & 0.48 & 8.8 \\
    & & 0.60 - 0.80 & 2.13 & 0.0275 & 0.68 & 8.8 \\
    \cline{1-7}   
    
    \multirow{12}{*}{3.0 - 16.0} & \multirow{4}{*}{0.013 - 0.020} & 0.20 - 0.30 & 3.17 & 0.0171 & 0.24 & 16.0 \\ 
    & & 0.30 - 0.40 & 3.16 & 0.0170 & 0.34 & 16.0 \\
    & & 0.40 - 0.60 & 3.17 & 0.0171 & 0.48 & 16.0 \\
    & & 0.60 - 0.80 & 3.18 & 0.0171 & 0.68 & 16.0 \\
    \cline{2-7}
    
    & \multirow{4}{*}{0.020 - 0.055} & 0.20 - 0.30 & 4.77 & 0.0369 & 0.24 & 14.4 \\ 
    & & 0.30 - 0.40 & 4.74 & 0.0370 & 0.34 & 14.4 \\
    & & 0.40 - 0.60 & 4.72 & 0.0371 & 0.48 & 14.4 \\
    & & 0.60 - 0.80 & 4.67 & 0.0371 & 0.68 & 14.5 \\
    \cline{2-7}
    
    & \multirow{4}{*}{0.055 - 0.100} & 0.20 - 0.30 & 7.31 & 0.0712 & 0.24 & 9.7 \\ 
    & & 0.30 - 0.40 & 7.23 & 0.0712 & 0.34 & 9.6 \\
    & & 0.40 - 0.60 & 7.18 & 0.0713 & 0.48 & 9.6 \\
    & & 0.60 - 0.80 & 7.01 & 0.0709 & 0.68 & 9.5 \\
    \cline{1-7}       
    \end{tabular}    
\caption{Mean values of $\Qsq$, $x$, $z$ and $W$ in each of the \textit{standard} bins used for the measurement of the $\Ptsq$-distributions.}
\end{table}

\chapter{Unbinned Maximum Likelihood method for measurement of SDMEs}
\label{AppendixD} 

Let's derive the expression for the likelihood, starting from the binned case and then generalizing it to the unbinned case, assuming that the physical process under consideration can be described by the probability density function $f(\vec{x},\vec{p})$, $\vec{x}$ being a vector of kinematic variables and $\vec{p}$ a vector of unknown parameters. For the sake of simplicity, we restrict to the case of only one kinematic variable $x$. The normalization condition reads:

\begin{equation}
    \int_{x_{min}}^{x_{max}}\diff x~f(x,\vec{p}) = 1
\end{equation}
for any choice of $\vec{p}$. Let's divide the $x$ range into $N_x$ bins of identical width, so that:
\begin{equation}
     \Delta x = \frac{x_{max} - x_{min}}{N_x}.
\end{equation}
The acceptance effects (detector efficiency and geometrical coverage, but also track reconstruction) are modelled through the function $a(x)$. The probability $\tilde{p}_j$ associated to the $j-$th bin reads:

\begin{equation}
     \tilde{p}_j = \frac{\int_{x_{j}-\frac{\Delta}{2}}^{x_{j}+\frac{\Delta}{2}}\diff x~a(x)f(x,\vec{p})}{\int_{x_{min}}^{x_{max}}\diff x~a(x)f(x,\vec{p})} = \frac{\int_{x_{j}-\frac{\Delta}{2}}^{x_{j}+\frac{\Delta}{2}}\diff x~a(x)f(x,\vec{p})}{I(\vec{p})}.
\end{equation}
If $K$ is the total number of collected events, the number of expected events in each bin is $\mu_j = K\tilde{p}_j$, while (according to Poissonian statistics) the probability of observing $k_j$ events in each bin reads:
\begin{equation}
     \tilde{P}_j = \frac{\op\mu_j\cp^{k_j}}{k_j!}e^{-\mu_j}.
\end{equation}
We now write the negative log-likelihood as:
\begin{equation}
    -\ln L = - \sum_{j=1}^{N_x} \ln  \tilde{P}_j = \sum_{j=1}^{N_x} \op \mu_j + \ln(k_j!) - k_j\ln \mu_j \cp.
\end{equation}
The terms not depending on $\vec{p}$ can be dropped, as their only effect is to vertically shift the likelihood function, with no effect on the position of the minimum. In this sense, we can write:
\begin{equation}
\begin{split}
    -\ln L & =  K + \sum_{j=1}^{N_x}\ln(k_j!)  -\sum_{j=1}^{N_x} k_j\ln \mu_j \\
    & \approx -\sum_{j=1}^{N_x} k_j \ln \op K a(x_j)f(x_j,\vec{p}) \Delta x /I(\vec{p}) \cp \\
    & = - \sum_{j=1}^{N_x} k_j \ln \op Ka(x_j)\Delta x \cp  -\sum_{j=1}^{N_x} k_j \ln \op f(x_j,\vec{p})/I(\vec{p}) \cp \\
    & = - \sum_{j=1}^{N_x}k_j\ln f(x_j,\vec{p}) + K \ln I(\vec{p}).
\end{split}
\label{eq:lnL1}
\end{equation}

Let's now concentrate on $I(\vec{p})$, the only term where the acceptance correction enters. If $M^{gen}$ is the total number of generated events,  $m_j^{gen} \op m_j^{rec}\cp $  is the number of generated (reconstructed) Monte Carlo events in the $j-$th bin, and if the distribution of the generated values of $x$ is flat and the generated statistics large, we can write:
\begin{equation}
     a_j = \frac{m_j^{rec}}{m_j^{gen}} = \frac{m_j^{rec}}{M^{gen}/N_x},
\end{equation}
so that:
\begin{equation}
\begin{split}
     I(\vec{p}) & \approx  \sum_{j=1}^{N_x}  \frac{m_j^{rec}}{M^{gen}/N_x} f(x_j,\vec{p}) \Delta x = \sum_{j=1}^{N_x}  m_j^{rec}f(x_j,\vec{p}) \cdot  \frac{N_x \Delta x}{M^{gen}} \\
     & = \sum_{j=1}^{N_x}  \sum_{i=0}^{m_j^{rec}}f(x_i,\vec{p}) \cdot \frac{N_x \Delta x}{M^{gen}} \\
     & = \sum_{n=1}^{M^{rec}} f(x_n,\vec{p}) \cdot \frac{N_x \Delta x}{M^{gen}}.
\end{split}
\label{eq:Ivec}
\end{equation}
Thus, the integral $ I(\vec{p})$ can be expressed as the sum over the Monte Carlo reconstructed events of the function $f$, evaluated at $x_n$, times a constant term not dependent on the parameters. From Eq.~\ref{eq:lnL1} and Eq.~\ref{eq:Ivec}, and letting $\Delta x \to 0$, we get to the final formula for the Unbinned Maximum Likelihood:

\begin{equation}
\begin{split}
    -\ln L & = - \sum_{j=1}^{N_x}k_j\ln f(x_j,\vec{p}) + K \ln \sum_{n=1}^{M^{rec}} f(x_n,\vec{p}) +  K \ln \sum_{n=1}^{M^{rec}} \frac{N_x \Delta x}{M^{gen}}\\
    & \xrightarrow[\Delta x \to 0]{} - \sum_{k=1}^{K} \ln f(x_k,\vec{p}) + K \ln \sum_{n=1}^{M^{rec}} f(x_n,\vec{p}).
\end{split}
\label{eq:lnL2}
\end{equation}


\begin{acknowledgements}

This work would not have been possible without the contribution of many people who, in different ways and with various roles, have all been essential. \\

I am very grateful to Prof. Anna Martin for her kind and careful supervision along these years, for her encouragements and patient suggestions, and for all the time she dedicated to me and to this work. Her commitment has been inspirational to me. \\

I sincerely thank Prof. Franco Bradamante, who first introduced me, still a Bachelor student, to the wonders of the Deep Inelastic Scattering: his lessons, advice and suggestions have always been very precious.\\

Thanks to Prof. Mariaelena Boglione and Prof. Harut Avakian for having accepted to referee this Thesis. \\

Thanks to Andrea Bressan for his constant help and friendship, and to all the members of the COMPASS Group in Trieste for their warm company. \\

Thanks to the colleagues with whom I shared the office along the years: Giulio Sbrizzai, Nour Makke, Adam Szabelski and Jan Matousek, for their company, for their help, for the many things they kindly taught me. \\

A special thank to Albi Kerbizi, who first convinced me to join the COMPASS Group, for his friendship, patience, support: his presence has been invaluable. Thanks also to Bakur Parsamyan for his encouragement, enthusiasm and passion. A warm thank you to all the colleagues and friends in the COMPASS Collaboration. \\

The fondest thoughts go to my parents, Sandra and Luciano, to my brother Stefano with Gloria, to Beatrice with Alberto, Emanuele, Maria Gioia and Giosuè, to Elisa with Marco, Davide and Jacopo, to Elena, Marco, Pierpaolo and all my friends. I embrace all of you in my biggest and warmest hug.  \\

\end{acknowledgements}

  \bibliographystyle{ieeetr}
  \bibliography{bibliography.bib}


\end{document}